%% file: ms.tex
\pdfoutput=1
\documentclass[11pt,a4paper]{article}

\usepackage{iftex}
\ifPDFTeX
	\usepackage[utf8]{inputenc}
\fi

\usepackage{natbib}
\usepackage{doi}

\usepackage{microtype}

\usepackage[autostyle]{csquotes}

\usepackage{eufrak}
\usepackage{amssymb}
\usepackage{wasysym}
\usepackage{isabelle,isabellesym}

\hypersetup{
  linkcolor = blue,
  urlcolor = blue,
  citecolor = blue,
  colorlinks = true,
}

\isabellestyle{it}

\newcommand*\sq{\mathbin{\vcenter{\hbox{\rule{.3ex}{.3ex}}}}}

\newcommand{\isactrlitem}{$\sq$}
\newcommand{\isactrlemph}{\isamath{\ast}}

\begin{document}

\title{Mechanization of LAGC Semantics in Isabelle}
\author{Niklas Heidler}



\maketitle

\abstract {
Formal programming language semantics are imperative when trying to verify properties of programs in an automated manner. Using a new approach, \cite{LAGCSemantics} strengthen the ability of reasoning about concurrent programs by proposing a modular trace semantics, which can flexibly adapt to the most prominent imperative programming language paradigms. These semantics decouple the evaluation in the local environments from the evaluation in the global environment by generating abstract, symbolic traces for the individual, local systems. The traces are then composed and concretized, resulting in global traces for the global system. Hence, these semantics are called \textit{Locally Abstract, Globally Concrete~(LAGC)}. \par

In this thesis, we present a formalization of the LAGC semantics in the popular theorem proving environment Isabelle/HOL. The given model is based on the prior work on the theory of LAGC semantics by \cite{LAGCSemantics} and includes formalizations of the basic theorems, the LAGC semantics for the While Language (WL), as well as the LAGC semantics for an extended version of the While Language (WL$_{EXT}$). We furthermore use our Isabelle model in order to provide formal proofs for several advanced properties of the LAGC semantics, which have not been analyzed in the original~paper. \par

Whilst the main goal of the work was to formalize the LAGC semantics in a mathematically rigorous manner, we also achieve a high level of proof automatization and manage to contribute an efficient code-generation for the computation of program traces. As the formalization of the semantics is highly modular, the given theories could in the future be extended with even more sophisticated programming language paradigms. 
}

\tableofcontents

\section{Introduction}

\textit{Locally Abstract, Globally Concrete (LAGC)} semantics as first described by \cite{LAGCSemantics} are modular trace semantics for concurrent programming languages which cleanly decouple the evaluation of expressions and statements in the local systems from the evaluation in the global environment. In order to enforce this level of separation, local evaluation rules are used to generate abstract, symbolic traces for each individual, local system. This implies that states occurring in traces of local systems are allowed to be underspecified (i.e. due to communication with other systems). Using certain composition rules, the local traces are then composed and concretized into global traces for the global~system. In contrast to other semantics, LAGC semantics also smoothly align with deductive verification rules of program logics, thus enabling logical reasoning about concurrent programs via calculi. This objective was part of the main motivation for the definition of LAGC semantics. \par

\textbf{Isabelle.} Isabelle/HOL is a popular proof assistant that uses a higher-order logic theorem proving environment allowing mathematically rigorous and precise formalizations of theories. Within Isabelle, proof tools (e.g. simplifiers, SMT) are used to formally proof given lemmas, so as to ensure their correctness. Additionally, Isabelle contributes the structured proof language Isar, which guarantees that complex proofs can be written in a formal and precise manner without losing a human, natural intuition about the proof structure. Isabelle also comes along with a large, easily usable library consisting of mathematically verified theories (e.g. finite maps, multisets) and includes a compiler, that can produce executable code for SML, OCaml, Haskell, and Scala. \par

\textbf{Contribution.} The prior work on LAGC semantics by \cite{LAGCSemantics} has been purely theoretical, laying the necessary foundation for possible mechanizations in proof assistants. In this thesis, we present a mathematically rigorous and sound formalization of the LAGC semantics using the popular theorem prover Isabelle/HOL. In contrast to the original paper, our mechanization gains a higher precision in its formalizations due to its strict mathematical foundation, thereby also verifying the correctness of the given theory. Whilst our model contains the central designs and schemes of the original paper, we also deviate from given designs in a reasonable margin in order for a correct and sound formalization to be feasible. The justification of these design decisions will be focused on in the next sections. We also provide formal proofs for additional properties (e.g. determinism, concreteness of traces), which have not been given in the original paper. Our model furthermore guarantees a high degree of proof automation and provides a simple and efficient code-generation for the construction of program traces. Note that this report does not describe all auxiliary lemmas, as their length would greatly exceed the scope of this report. However, the complete mechanization, including all supplementary lemmas, is available~on~gitlab\footnote{\url{https://gitlab.com/Niklas_Heidler/mechanization-of-lagc-semantics-in-isabelle}}. \par

\textbf{Difficulties.} Several difficulties during the formalization of the semantics stemmed from implicit assumptions taken in the original paper, which could not be left implicit, so as to ensure the faithfulness of the model. This refers to situations in which syntactically correct specifications are implicitly enforced to have a specific form (e.g. implicit simplifications of expressions in states), as well as occurrences where definitions of supporting functions are not explicitly provided (e.g. the generation of fresh variables). Multiple difficulties also originated from the desire of establishing an efficient code-generation for the program trace construction. 
Considering that Isabelle cannot generate code for quantifications over infinite types, more complex, alternative formalizations deviating from the original paper had to be explored. \par

\textbf{Outline.} The thesis is organized as follows: \textit{Chapter 2} focuses on the formalization of the basics of LAGC Semantics as described in the second section of the original paper. This includes the syntax and semantics of expressions, as well as the definition of states, traces and concretization mappings. Note that this theory will be reusable and independent from concrete instantiations of the LAGC semantics for each specific programming language, thus forming the essence of the LAGC semantics. In \textit{Chapter 3} we formalize the LAGC semantics for the standard While Language (WL) as denoted in the third section of the original paper, whilst also providing proofs for additional properties. \textit{Chapter 4} is based on the fifth section of the original paper and extends the While Language with additional programming language concepts (e.g. concurrency, scheduling). We show that the formalization of the LAGC semantics can be easily adapted to accommodate these new concepts. Whilst \textit{Chapter 5} focuses on related work, \textit{Chapter 6} aims at providing ideas for future extensions of the model and concludes the thesis.

\parindent 0pt\parskip 0.5ex

\input{session}

\newpage

\section{Related Work}

LAGC semantics as first described by \cite{LAGCSemantics} are denotational trace semantics. Although an eclectic variety of trace semantics has been proposed in the past, we only select a few of them for our~discussion. 

\textbf{Transition Traces.} \cite{BrookesA} proposes a denotational semantics, which is considered fully abstract in accordance with the partial correctness property, and builds on the construction of so-called \textit{transition traces}. Contrary to LAGC semantics, transition traces are introduced as sequences of states without events, indicating that communication between systems must be handled in a different manner. The semantics choose to handle parallelism by stitching together the traces corresponding to the different atomic statements of the parallel command using a certain set of rules. Note that the semantics also decide against the use of continuation markers. 

\textbf{Action Traces.} Similar to LAGC semantics, \cite{BrookesB} proposes a denotational trace semantics, which sequentially connects both states and events in its so-called \textit{action traces}. Communication is handled by using corresponding events, whilst parallelism is captured via so-called \textit{mutex-fairmerges}. The semantics focus on locking procedures in order to solve data races, thereby ensuring safe access to shared ressources (e.g. variables). However, note that these semantics cannot handle dynamic process creation and procedure calls, thereby greatly deviating from LAGC semantics.

\textbf{Interaction Trees.} Another recent idea as first proposed by \cite{Xia} is a denotational encoding of program effects as a so-called \textit{interaction tree}. An interaction tree is introduced as a data structure that stores effects of a program, thereby having similarities to our notion of traces. However, note that these semantics abstract away from the notion of states and concretizations, which greatly differs from the core idea of LAGC semantics. Interaction trees also do not allow faithful representations of concurrency, implying that the approach can only be used for sequential programs. Instead, the semantics focus on reasoning about divergence.

\section{Conclusion and Future Work}

\textbf{Conclusion.} The main objective of this thesis was to provide a mathematically rigorous formalization of \textit{Locally Abstract, Globally Concrete (LAGC)} semantics in the higher-order logic theorem proving environment Isabelle/HOL. Whilst Chapter 3 focused on formalizing the semantics for the standard While Language \textit{WL}, Chapter 5 established that our model can be easily adapted in order to accommodate new programming language concepts (e.g. concurrency). Using Isabelle to formalize the theory of the original paper with a higher level of precision also helped us to spot several inconsistencies in the paper, thus assisting the original authors in improving the soundness of~their~theory. \par

During the formalization, we meticulously avoided definitions, which would interfere with Isabelle's code generation, thereby allowing us to automatically generate executable code for the construction of global traces. Note that this was a major difficulty during the formalization procedure, as many definitions needed to be slightly altered without changing the core of the semantics. We also used Isabelle in order to provide a proof automation system, thereby ensuring that we can systematically derive the corresponding set of global traces for any arbitrary program and~initial~state. \par

Note that we also established supplementary theorems, which we have not introduced in the earlier chapters, as their length greatly exceeds the scope of this report. These auxiliary proofs establish the determinism of \textit{WL}, the concreteness of global traces in \textit{WL} and \textit{WL$_{EXT}$}, as well as many more properties. \par

\textbf{Future Work.} Our model only includes formalizations of sections 2, 4 and 5 of the original paper, thus implying that the whole mechanization is still not fully completed. We therefore propose extensions of our mechanized semantics to a shared-memory multiprocessor language, as well as a formalization of the corresponding deductive verification calculus. However, there is also room for improvement in the current state of the model, specifically referring to extensions of finite traces to infinite traces, (speed-related) improvements for our proof automation system, as well as exports of the generated code to other programming languages supported by Isabelle (e.g. Haskell, Scala).

\bibliographystyle{plainnat}
\bibliography{\jobname}

\end{document}

%% file: session.tex
\input{LAGC_Base.tex}

\input{LAGC_WL.tex}

\input{LAGC_WL_Extended.tex}

%% file: LAGC_Base.tex
%
\begin{isabellebody}%
\setisabellecontext{LAGC{\isacharunderscore}{\kern0pt}Base}%
\isadelimdocument
\endisadelimdocument
\isatagdocument
\isamarkupsection{Basics%
}
\isamarkuptrue%
\endisatagdocument
{\isafolddocument}%
\isadelimdocument
\endisadelimdocument
\isadelimtheory
\endisadelimtheory
\isatagtheory
\isacommand{theory}\isamarkupfalse%
\ LAGC{\isacharunderscore}{\kern0pt}Base\isanewline
\ \ \isakeyword{imports}\ Main\ {\isachardoublequoteopen}HOL{\isacharminus}{\kern0pt}Library{\isachardot}{\kern0pt}Finite{\isacharunderscore}{\kern0pt}Map{\isachardoublequoteclose}\isanewline
\isakeyword{begin}%
\endisatagtheory
{\isafoldtheory}%
\isadelimtheory
\endisadelimtheory
\begin{isamarkuptext}%
In this chapter, we formalize the basics of the LAGC semantics as
  described in section~2 of the original paper. This section describes the 
  reusable interface for all concrete instantiations of the LAGC semantics,
  providing the basis for the subsequent chapters.%
\end{isamarkuptext}\isamarkuptrue%
\isadelimdocument
\endisadelimdocument
\isatagdocument
\isamarkupsubsection{Expressions%
}
\isamarkuptrue%
\isamarkupsubsubsection{Primitives%
}
\isamarkuptrue%
\endisatagdocument
{\isafolddocument}%
\isadelimdocument
\endisadelimdocument
\begin{isamarkuptext}%
We model program variables and method names as strings, ensuring that
  the domains for variables and method names are both infinite. This design
  choice also ensures easy usability and readability of variables.%
\end{isamarkuptext}\isamarkuptrue%
\ \ \isacommand{type{\isacharunderscore}{\kern0pt}synonym}\isamarkupfalse%
\ var\ {\isacharequal}{\kern0pt}\ string\isanewline
\ \ \isacommand{type{\isacharunderscore}{\kern0pt}synonym}\isamarkupfalse%
\ method{\isacharunderscore}{\kern0pt}name\ {\isacharequal}{\kern0pt}\ string%
\isadelimdocument
\endisadelimdocument
\isatagdocument
\isamarkupsubsubsection{Syntax%
}
\isamarkuptrue%
\endisatagdocument
{\isafolddocument}%
\isadelimdocument
\endisadelimdocument
\begin{isamarkuptext}%
We now formalize the syntax of expressions for a generic programming language. 
  The formalization is highly modular in order to guarantee that the syntax can be 
  easily adapted to accommodate new syntactical~concepts. \par
  However, the formalization of the syntax in the original paper is not easily 
  transferable to Isabelle due to a lack of type constraints. Without loss of 
  generality, we thus propose a strictly typed grammar that matches the original
  grammar in its core concept. The grammar includes the common operators and
  enforces expressions to be of either arithmetic or Boolean~nature. \par
  We first introduce the standard arithmetic, Boolean and relational operators, 
  which can be used in syntactical derivations of our~expressions.%
\end{isamarkuptext}\isamarkuptrue%
\ \ \isacommand{datatype}\isamarkupfalse%
\ op\isactrlsub a\ {\isacharequal}{\kern0pt}\ \isanewline
\ \ \ \ \ \ add\ %
\isamarkupcmt{Addition%
}\isanewline
\ \ \ \ {\isacharbar}{\kern0pt}\ sub\ %
\isamarkupcmt{Subtraction%
}\isanewline
\ \ \ \ {\isacharbar}{\kern0pt}\ mul\ %
\isamarkupcmt{Multiplication%
}\isanewline
\isanewline
\ \ \isacommand{datatype}\isamarkupfalse%
\ op\isactrlsub b\ {\isacharequal}{\kern0pt}\ \isanewline
\ \ \ \ \ \ conj\ %
\isamarkupcmt{Logical And%
}\isanewline
\ \ \ \ {\isacharbar}{\kern0pt}\ disj\ %
\isamarkupcmt{Logical Or%
}\isanewline
\isanewline
\ \ \isacommand{datatype}\isamarkupfalse%
\ op\isactrlsub r\ {\isacharequal}{\kern0pt}\ \isanewline
\ \ \ \ \ \ leq\ %
\isamarkupcmt{Less Equal%
}\isanewline
\ \ \ \ {\isacharbar}{\kern0pt}\ geq\ %
\isamarkupcmt{Greater Equal%
}\isanewline
\ \ \ \ {\isacharbar}{\kern0pt}\ eq\ %
\isamarkupcmt{Equal%
}%
\begin{isamarkuptext}%
We now provide a formal definition of the grammar, separating between
  the following types of expressions: \begin{description}
  \item[Arithmetic Expressions] Arithmetic expressions consist of arithmetic 
  numerals, variables and binary arithmetic operations using arithmetic 
  operators and arithmetic operands.
  \item[Boolean Expressions] Boolean expressions consist of Boolean truth values, 
  the not operation, binary Boolean operations using Boolean operators and Boolean 
  operands, as well as binary relational operations using relational operators 
  and arithmetic~operands.
  \item[Expressions] Expressions can either be arithmetic expressions, 
  Boolean expressions or method names. The addition of method names to the 
  expression syntax deviates from the original paper. The reason for this
  will be explained at a later point.
  \item[Starred Expressions] Starred Expressions can only be arithmetic expressions 
  or the symbolic value. Note that we do not allow the derivation of Boolean starred 
  expressions, thereby differing from the original paper. This later ensures that 
  all variables have an arithmetic type, thus establishing a simpler model 
  without loss~of~generality.
  \end{description} Note that the syntax can be easily adapted to accommodate new 
  types of~expressions.%
\end{isamarkuptext}\isamarkuptrue%
\ \ \isacommand{datatype}\isamarkupfalse%
\ aexp\ {\isacharequal}{\kern0pt}\ \isanewline
\ \ \ \ \ \ Numeral\ int\ %
\isamarkupcmt{Arithmetic numeral%
}\isanewline
\ \ \ \ {\isacharbar}{\kern0pt}\ Variable\ var\ %
\isamarkupcmt{Variables%
}\isanewline
\ \ \ \ {\isacharbar}{\kern0pt}\ Op\isactrlsub A\ aexp\ op\isactrlsub a\ aexp\ %
\isamarkupcmt{Binary arithmetic operations%
}\isanewline
\isanewline
\ \ \isacommand{datatype}\isamarkupfalse%
\ bexp\ {\isacharequal}{\kern0pt}\ \isanewline
\ \ \ \ \ \ Boolean\ bool\ %
\isamarkupcmt{Boolean truth values%
}\isanewline
\ \ \ \ {\isacharbar}{\kern0pt}\ Not\ bexp\ %
\isamarkupcmt{Not Operation%
}\isanewline
\ \ \ \ {\isacharbar}{\kern0pt}\ Op\isactrlsub B\ bexp\ op\isactrlsub b\ bexp\ %
\isamarkupcmt{Binary Boolean operations%
}\isanewline
\ \ \ \ {\isacharbar}{\kern0pt}\ Op\isactrlsub R\ aexp\ op\isactrlsub r\ aexp\ %
\isamarkupcmt{Binary Relational operations%
}\isanewline
\isanewline
\ \ \isacommand{datatype}\isamarkupfalse%
\ exp\ {\isacharequal}{\kern0pt}\ \isanewline
\ \ \ \ \ \ Aexp\ aexp\ %
\isamarkupcmt{Arithmetic Expressions%
}\isanewline
\ \ \ \ {\isacharbar}{\kern0pt}\ Bexp\ bexp\ %
\isamarkupcmt{Boolean Expressions%
}\isanewline
\ \ \ \ {\isacharbar}{\kern0pt}\ Param\ method{\isacharunderscore}{\kern0pt}name\ %
\isamarkupcmt{Method Names%
}\isanewline
\isanewline
\ \ \isacommand{datatype}\isamarkupfalse%
\ sexp\ {\isacharequal}{\kern0pt}\ \isanewline
\ \ \ \ \ \ Expression\ aexp\ %
\isamarkupcmt{Arithmetic Expressions%
}\isanewline
\ \ \ \ {\isacharbar}{\kern0pt}\ Star\ %
\isamarkupcmt{Symbolic Value%
}%
\begin{isamarkuptext}%
We also add a minimal concrete syntax for expressions, thereby improving 
  the readability of expressions in~programs.%
\end{isamarkuptext}\isamarkuptrue%
\ \ \isacommand{notation}\isamarkupfalse%
\ Numeral\ {\isacharparenleft}{\kern0pt}{\isachardoublequoteopen}Num\ {\isacharunderscore}{\kern0pt}{\isachardoublequoteclose}\ {\isacharbrackleft}{\kern0pt}{\isadigit{1}}{\isadigit{0}}{\isadigit{0}}{\isadigit{0}}{\isacharbrackright}{\kern0pt}\ {\isadigit{6}}{\isadigit{3}}{\isacharparenright}{\kern0pt}\isanewline
\ \ \isacommand{notation}\isamarkupfalse%
\ Variable\ {\isacharparenleft}{\kern0pt}{\isachardoublequoteopen}Var\ {\isacharunderscore}{\kern0pt}{\isachardoublequoteclose}\ {\isacharbrackleft}{\kern0pt}{\isadigit{1}}{\isadigit{0}}{\isadigit{0}}{\isadigit{0}}{\isacharbrackright}{\kern0pt}\ {\isadigit{6}}{\isadigit{3}}{\isacharparenright}{\kern0pt}\isanewline
\ \ \isacommand{notation}\isamarkupfalse%
\ Op\isactrlsub A\ {\isacharparenleft}{\kern0pt}{\isachardoublequoteopen}{\isacharunderscore}{\kern0pt}\ \isactrlsub A{\isacharunderscore}{\kern0pt}\ {\isacharunderscore}{\kern0pt}{\isachardoublequoteclose}\ {\isacharbrackleft}{\kern0pt}{\isadigit{1}}{\isadigit{0}}{\isadigit{0}}{\isadigit{0}}{\isacharcomma}{\kern0pt}\ {\isadigit{1}}{\isadigit{0}}{\isadigit{0}}{\isadigit{0}}{\isacharcomma}{\kern0pt}\ {\isadigit{1}}{\isadigit{0}}{\isadigit{0}}{\isadigit{0}}{\isacharbrackright}{\kern0pt}\ {\isadigit{6}}{\isadigit{2}}{\isacharparenright}{\kern0pt}\isanewline
\isanewline
\ \ \isacommand{notation}\isamarkupfalse%
\ Boolean\ {\isacharparenleft}{\kern0pt}{\isachardoublequoteopen}Bool\ {\isacharunderscore}{\kern0pt}{\isachardoublequoteclose}\ {\isacharbrackleft}{\kern0pt}{\isadigit{1}}{\isadigit{0}}{\isadigit{0}}{\isadigit{0}}{\isacharbrackright}{\kern0pt}\ {\isadigit{6}}{\isadigit{3}}{\isacharparenright}{\kern0pt}\isanewline
\ \ \isacommand{notation}\isamarkupfalse%
\ Not\ {\isacharparenleft}{\kern0pt}{\isachardoublequoteopen}not\ {\isacharunderscore}{\kern0pt}{\isachardoublequoteclose}\ {\isacharbrackleft}{\kern0pt}{\isadigit{1}}{\isadigit{0}}{\isadigit{0}}{\isadigit{0}}{\isacharbrackright}{\kern0pt}\ {\isadigit{6}}{\isadigit{3}}{\isacharparenright}{\kern0pt}\isanewline
\ \ \isacommand{notation}\isamarkupfalse%
\ Op\isactrlsub B\ {\isacharparenleft}{\kern0pt}{\isachardoublequoteopen}{\isacharunderscore}{\kern0pt}\ \isactrlsub B{\isacharunderscore}{\kern0pt}\ {\isacharunderscore}{\kern0pt}{\isachardoublequoteclose}\ {\isacharbrackleft}{\kern0pt}{\isadigit{1}}{\isadigit{0}}{\isadigit{0}}{\isadigit{0}}{\isacharcomma}{\kern0pt}\ {\isadigit{1}}{\isadigit{0}}{\isadigit{0}}{\isadigit{0}}{\isacharcomma}{\kern0pt}\ {\isadigit{1}}{\isadigit{0}}{\isadigit{0}}{\isadigit{0}}{\isacharbrackright}{\kern0pt}\ {\isadigit{6}}{\isadigit{2}}{\isacharparenright}{\kern0pt}\isanewline
\ \ \isacommand{notation}\isamarkupfalse%
\ Op\isactrlsub R\ {\isacharparenleft}{\kern0pt}{\isachardoublequoteopen}{\isacharunderscore}{\kern0pt}\ \isactrlsub R{\isacharunderscore}{\kern0pt}\ {\isacharunderscore}{\kern0pt}{\isachardoublequoteclose}\ {\isacharbrackleft}{\kern0pt}{\isadigit{1}}{\isadigit{0}}{\isadigit{0}}{\isadigit{0}}{\isacharcomma}{\kern0pt}\ {\isadigit{1}}{\isadigit{0}}{\isadigit{0}}{\isadigit{0}}{\isacharcomma}{\kern0pt}\ {\isadigit{1}}{\isadigit{0}}{\isadigit{0}}{\isadigit{0}}{\isacharbrackright}{\kern0pt}\ {\isadigit{6}}{\isadigit{2}}{\isacharparenright}{\kern0pt}\isanewline
\isanewline
\ \ \isacommand{notation}\isamarkupfalse%
\ Aexp\ {\isacharparenleft}{\kern0pt}{\isachardoublequoteopen}A\ {\isacharunderscore}{\kern0pt}{\isachardoublequoteclose}\ {\isacharbrackleft}{\kern0pt}{\isadigit{1}}{\isadigit{0}}{\isadigit{0}}{\isadigit{0}}{\isacharbrackright}{\kern0pt}\ {\isadigit{6}}{\isadigit{3}}{\isacharparenright}{\kern0pt}\isanewline
\ \ \isacommand{notation}\isamarkupfalse%
\ Bexp\ {\isacharparenleft}{\kern0pt}{\isachardoublequoteopen}B\ {\isacharunderscore}{\kern0pt}{\isachardoublequoteclose}\ {\isacharbrackleft}{\kern0pt}{\isadigit{1}}{\isadigit{0}}{\isadigit{0}}{\isadigit{0}}{\isacharbrackright}{\kern0pt}\ {\isadigit{6}}{\isadigit{3}}{\isacharparenright}{\kern0pt}\isanewline
\ \ \isacommand{notation}\isamarkupfalse%
\ Param\ {\isacharparenleft}{\kern0pt}{\isachardoublequoteopen}P\ {\isacharunderscore}{\kern0pt}{\isachardoublequoteclose}\ {\isacharbrackleft}{\kern0pt}{\isadigit{1}}{\isadigit{0}}{\isadigit{0}}{\isadigit{0}}{\isacharbrackright}{\kern0pt}\ {\isadigit{6}}{\isadigit{3}}{\isacharparenright}{\kern0pt}\isanewline
\isanewline
\ \ \isacommand{notation}\isamarkupfalse%
\ Expression\ {\isacharparenleft}{\kern0pt}{\isachardoublequoteopen}Exp\ {\isacharunderscore}{\kern0pt}{\isachardoublequoteclose}\ {\isacharbrackleft}{\kern0pt}{\isadigit{1}}{\isadigit{0}}{\isadigit{0}}{\isadigit{0}}{\isacharbrackright}{\kern0pt}\ {\isadigit{6}}{\isadigit{3}}{\isacharparenright}{\kern0pt}\isanewline
\ \ \isacommand{notation}\isamarkupfalse%
\ Star\ {\isacharparenleft}{\kern0pt}{\isachardoublequoteopen}\isactrlemph {\isachardoublequoteclose}{\isacharparenright}{\kern0pt}%
\begin{isamarkuptext}%
We additionally define helper functions, which project expressions
  onto their encased arithmetic/Boolean expression. Note that both
  projection functions are partial, hence they should only be called if the
  encased type of expression is known~beforehand.%
\end{isamarkuptext}\isamarkuptrue%
\ \ \isacommand{abbreviation}\isamarkupfalse%
\isanewline
\ \ \ \ proj\isactrlsub A\ {\isacharcolon}{\kern0pt}{\isacharcolon}{\kern0pt}\ {\isachardoublequoteopen}exp\ {\isasymRightarrow}\ aexp{\isachardoublequoteclose}\ \isakeyword{where}\isanewline
\ \ \ \ {\isachardoublequoteopen}proj\isactrlsub A\ e\ {\isasymequiv}\ {\isacharparenleft}{\kern0pt}case\ e\ of\ A\ a\ {\isasymRightarrow}\ a{\isacharparenright}{\kern0pt}{\isachardoublequoteclose}\isanewline
\isanewline
\ \ \isacommand{abbreviation}\isamarkupfalse%
\isanewline
\ \ \ \ proj\isactrlsub B\ {\isacharcolon}{\kern0pt}{\isacharcolon}{\kern0pt}\ {\isachardoublequoteopen}exp\ {\isasymRightarrow}\ bexp{\isachardoublequoteclose}\ \isakeyword{where}\isanewline
\ \ \ \ {\isachardoublequoteopen}proj\isactrlsub B\ e\ {\isasymequiv}\ {\isacharparenleft}{\kern0pt}case\ e\ of\ B\ b\ {\isasymRightarrow}\ b{\isacharparenright}{\kern0pt}{\isachardoublequoteclose}%
\begin{isamarkuptext}%
Using our previously defined grammar, we can now derive syntactically correct 
  expressions. We demonstrate this by providing examples for arithmetic and Boolean 
  expressions with our minimal concrete~syntax.%
\end{isamarkuptext}\isamarkuptrue%
\ \ \isacommand{definition}\isamarkupfalse%
\ \isanewline
\ \ \ \ aExp{\isacharunderscore}{\kern0pt}ex\ {\isacharcolon}{\kern0pt}{\isacharcolon}{\kern0pt}\ aexp\ \isakeyword{where}\isanewline
\ \ \ \ {\isachardoublequoteopen}aExp{\isacharunderscore}{\kern0pt}ex\ {\isasymequiv}\ {\isacharparenleft}{\kern0pt}{\isacharparenleft}{\kern0pt}Var\ {\isacharprime}{\kern0pt}{\isacharprime}{\kern0pt}x{\isacharprime}{\kern0pt}{\isacharprime}{\kern0pt}{\isacharparenright}{\kern0pt}\ \isactrlsub Amul\ {\isacharparenleft}{\kern0pt}Var\ {\isacharprime}{\kern0pt}{\isacharprime}{\kern0pt}y{\isacharprime}{\kern0pt}{\isacharprime}{\kern0pt}{\isacharparenright}{\kern0pt}{\isacharparenright}{\kern0pt}\ \isactrlsub Asub\ {\isacharparenleft}{\kern0pt}Var\ {\isacharprime}{\kern0pt}{\isacharprime}{\kern0pt}x{\isacharprime}{\kern0pt}{\isacharprime}{\kern0pt}{\isacharparenright}{\kern0pt}{\isachardoublequoteclose}\ %
\isamarkupcmt{\isa{{\isacharparenleft}{\kern0pt}x\ {\isacharasterisk}{\kern0pt}\ y{\isacharparenright}{\kern0pt}\ {\isacharminus}{\kern0pt}\ x}%
}\isanewline
\isanewline
\ \ \isacommand{definition}\isamarkupfalse%
\ \isanewline
\ \ \ \ bExp{\isacharunderscore}{\kern0pt}ex\ {\isacharcolon}{\kern0pt}{\isacharcolon}{\kern0pt}\ bexp\ \isakeyword{where}\isanewline
\ \ \ \ {\isachardoublequoteopen}bExp{\isacharunderscore}{\kern0pt}ex\ {\isasymequiv}\ {\isacharparenleft}{\kern0pt}{\isacharparenleft}{\kern0pt}Var\ {\isacharprime}{\kern0pt}{\isacharprime}{\kern0pt}x{\isacharprime}{\kern0pt}{\isacharprime}{\kern0pt}{\isacharparenright}{\kern0pt}\ \isactrlsub Req\ {\isacharparenleft}{\kern0pt}Num\ {\isadigit{2}}{\isacharparenright}{\kern0pt}{\isacharparenright}{\kern0pt}\ \isactrlsub Bdisj\ {\isacharparenleft}{\kern0pt}Bool\ False{\isacharparenright}{\kern0pt}{\isachardoublequoteclose}\ \ %
\isamarkupcmt{\isa{x\ {\isacharequal}{\kern0pt}\ {\isadigit{2}}\ {\isasymor}\ False}%
}%
\isadelimdocument
\endisadelimdocument
\isatagdocument
\isamarkupsubsubsection{Variable Mappings%
}
\isamarkuptrue%
\endisatagdocument
{\isafolddocument}%
\isadelimdocument
\endisadelimdocument
\begin{isamarkuptext}%
We now formally introduce variable functions mapping a specific type of expression
  onto a set of their enclosed free variables. We also provide a variable mapping
  for lists of expressions and sets of Boolean expressions, which will be utilized as
  handy abbreviations at a later point. The definitions of these functions 
  are~straightforward.%
\end{isamarkuptext}\isamarkuptrue%
\ \ \isacommand{primrec}\isamarkupfalse%
\isanewline
\ \ \ \ vars\isactrlsub A\ {\isacharcolon}{\kern0pt}{\isacharcolon}{\kern0pt}\ {\isachardoublequoteopen}aexp\ {\isasymRightarrow}\ var\ set{\isachardoublequoteclose}\ \isakeyword{where}\isanewline
\ \ \ \ {\isachardoublequoteopen}vars\isactrlsub A\ {\isacharparenleft}{\kern0pt}Num\ n{\isacharparenright}{\kern0pt}\ {\isacharequal}{\kern0pt}\ {\isacharbraceleft}{\kern0pt}{\isacharbraceright}{\kern0pt}{\isachardoublequoteclose}\ {\isacharbar}{\kern0pt}\isanewline
\ \ \ \ {\isachardoublequoteopen}vars\isactrlsub A\ {\isacharparenleft}{\kern0pt}Var\ x{\isacharparenright}{\kern0pt}\ {\isacharequal}{\kern0pt}\ {\isacharbraceleft}{\kern0pt}x{\isacharbraceright}{\kern0pt}{\isachardoublequoteclose}\ {\isacharbar}{\kern0pt}\isanewline
\ \ \ \ {\isachardoublequoteopen}vars\isactrlsub A\ {\isacharparenleft}{\kern0pt}a\isactrlsub {\isadigit{1}}\ \isactrlsub Aop\ a\isactrlsub {\isadigit{2}}{\isacharparenright}{\kern0pt}\ {\isacharequal}{\kern0pt}\ vars\isactrlsub A{\isacharparenleft}{\kern0pt}a\isactrlsub {\isadigit{1}}{\isacharparenright}{\kern0pt}\ {\isasymunion}\ vars\isactrlsub A{\isacharparenleft}{\kern0pt}a\isactrlsub {\isadigit{2}}{\isacharparenright}{\kern0pt}{\isachardoublequoteclose}\ \isanewline
\isanewline
\ \ \isacommand{primrec}\isamarkupfalse%
\ \isanewline
\ \ \ \ vars\isactrlsub B\ {\isacharcolon}{\kern0pt}{\isacharcolon}{\kern0pt}\ {\isachardoublequoteopen}bexp\ {\isasymRightarrow}\ var\ set{\isachardoublequoteclose}\ \isakeyword{where}\isanewline
\ \ \ \ {\isachardoublequoteopen}vars\isactrlsub B\ {\isacharparenleft}{\kern0pt}Bool\ b{\isacharparenright}{\kern0pt}\ {\isacharequal}{\kern0pt}\ {\isacharbraceleft}{\kern0pt}{\isacharbraceright}{\kern0pt}{\isachardoublequoteclose}\ {\isacharbar}{\kern0pt}\isanewline
\ \ \ \ {\isachardoublequoteopen}vars\isactrlsub B\ {\isacharparenleft}{\kern0pt}not\ b{\isacharparenright}{\kern0pt}\ {\isacharequal}{\kern0pt}\ vars\isactrlsub B{\isacharparenleft}{\kern0pt}b{\isacharparenright}{\kern0pt}{\isachardoublequoteclose}\ {\isacharbar}{\kern0pt}\isanewline
\ \ \ \ {\isachardoublequoteopen}vars\isactrlsub B\ {\isacharparenleft}{\kern0pt}b\isactrlsub {\isadigit{1}}\ \isactrlsub Bop\ b\isactrlsub {\isadigit{2}}{\isacharparenright}{\kern0pt}\ {\isacharequal}{\kern0pt}\ vars\isactrlsub B{\isacharparenleft}{\kern0pt}b\isactrlsub {\isadigit{1}}{\isacharparenright}{\kern0pt}\ {\isasymunion}\ vars\isactrlsub B{\isacharparenleft}{\kern0pt}b\isactrlsub {\isadigit{2}}{\isacharparenright}{\kern0pt}{\isachardoublequoteclose}\ {\isacharbar}{\kern0pt}\isanewline
\ \ \ \ {\isachardoublequoteopen}vars\isactrlsub B\ {\isacharparenleft}{\kern0pt}a\isactrlsub {\isadigit{1}}\ \isactrlsub Rop\ a\isactrlsub {\isadigit{2}}{\isacharparenright}{\kern0pt}\ {\isacharequal}{\kern0pt}\ vars\isactrlsub A{\isacharparenleft}{\kern0pt}a\isactrlsub {\isadigit{1}}{\isacharparenright}{\kern0pt}\ {\isasymunion}\ vars\isactrlsub A{\isacharparenleft}{\kern0pt}a\isactrlsub {\isadigit{2}}{\isacharparenright}{\kern0pt}{\isachardoublequoteclose}\ \isanewline
\isanewline
\ \ \isacommand{primrec}\isamarkupfalse%
\ \isanewline
\ \ \ \ vars\isactrlsub E\ {\isacharcolon}{\kern0pt}{\isacharcolon}{\kern0pt}\ {\isachardoublequoteopen}exp\ {\isasymRightarrow}\ var\ set{\isachardoublequoteclose}\ \isakeyword{where}\isanewline
\ \ \ \ {\isachardoublequoteopen}vars\isactrlsub E\ {\isacharparenleft}{\kern0pt}A\ a{\isacharparenright}{\kern0pt}\ {\isacharequal}{\kern0pt}\ vars\isactrlsub A{\isacharparenleft}{\kern0pt}a{\isacharparenright}{\kern0pt}{\isachardoublequoteclose}\ {\isacharbar}{\kern0pt}\isanewline
\ \ \ \ {\isachardoublequoteopen}vars\isactrlsub E\ {\isacharparenleft}{\kern0pt}B\ b{\isacharparenright}{\kern0pt}\ {\isacharequal}{\kern0pt}\ vars\isactrlsub B{\isacharparenleft}{\kern0pt}b{\isacharparenright}{\kern0pt}{\isachardoublequoteclose}\ {\isacharbar}{\kern0pt}\isanewline
\ \ \ \ {\isachardoublequoteopen}vars\isactrlsub E\ {\isacharparenleft}{\kern0pt}P\ m{\isacharparenright}{\kern0pt}\ {\isacharequal}{\kern0pt}\ {\isacharbraceleft}{\kern0pt}{\isacharbraceright}{\kern0pt}{\isachardoublequoteclose}\isanewline
\isanewline
\ \ \isacommand{primrec}\isamarkupfalse%
\ \isanewline
\ \ \ \ vars\isactrlsub S\ {\isacharcolon}{\kern0pt}{\isacharcolon}{\kern0pt}\ {\isachardoublequoteopen}sexp\ {\isasymRightarrow}\ var\ set{\isachardoublequoteclose}\ \isakeyword{where}\isanewline
\ \ \ \ {\isachardoublequoteopen}vars\isactrlsub S\ {\isacharparenleft}{\kern0pt}Exp\ a{\isacharparenright}{\kern0pt}\ {\isacharequal}{\kern0pt}\ vars\isactrlsub A{\isacharparenleft}{\kern0pt}a{\isacharparenright}{\kern0pt}{\isachardoublequoteclose}\ {\isacharbar}{\kern0pt}\isanewline
\ \ \ \ {\isachardoublequoteopen}vars\isactrlsub S\ \isactrlemph \ {\isacharequal}{\kern0pt}\ {\isacharbraceleft}{\kern0pt}{\isacharbraceright}{\kern0pt}{\isachardoublequoteclose}\isanewline
\isanewline
\ \ \isacommand{primrec}\isamarkupfalse%
\ \isanewline
\ \ \ \ lvars\isactrlsub E\ {\isacharcolon}{\kern0pt}{\isacharcolon}{\kern0pt}\ {\isachardoublequoteopen}exp\ list\ {\isasymRightarrow}\ var\ set{\isachardoublequoteclose}\ \isakeyword{where}\isanewline
\ \ \ \ {\isachardoublequoteopen}lvars\isactrlsub E\ {\isacharbrackleft}{\kern0pt}{\isacharbrackright}{\kern0pt}\ {\isacharequal}{\kern0pt}\ {\isacharbraceleft}{\kern0pt}{\isacharbraceright}{\kern0pt}{\isachardoublequoteclose}\ {\isacharbar}{\kern0pt}\isanewline
\ \ \ \ {\isachardoublequoteopen}lvars\isactrlsub E\ {\isacharparenleft}{\kern0pt}exp\ {\isacharhash}{\kern0pt}\ rest{\isacharparenright}{\kern0pt}\ {\isacharequal}{\kern0pt}\ vars\isactrlsub E{\isacharparenleft}{\kern0pt}exp{\isacharparenright}{\kern0pt}\ {\isasymunion}\ lvars\isactrlsub E{\isacharparenleft}{\kern0pt}rest{\isacharparenright}{\kern0pt}{\isachardoublequoteclose}\isanewline
\isanewline
\ \ \isacommand{fun}\isamarkupfalse%
\ \isanewline
\ \ \ \ svars\isactrlsub B\ {\isacharcolon}{\kern0pt}{\isacharcolon}{\kern0pt}\ {\isachardoublequoteopen}bexp\ set\ {\isasymRightarrow}\ var\ set{\isachardoublequoteclose}\ \isakeyword{where}\isanewline
\ \ \ \ {\isachardoublequoteopen}svars\isactrlsub B\ S\ {\isacharequal}{\kern0pt}\ {\isasymUnion}{\isacharparenleft}{\kern0pt}vars\isactrlsub B\ {\isacharbackquote}{\kern0pt}\ S{\isacharparenright}{\kern0pt}{\isachardoublequoteclose}%
\begin{isamarkuptext}%
In contrast to the original paper we additionally define variable occurrence
  functions mapping an arithmetic/Boolean expression onto a list of their
  enclosed free variables. Note that a variable may occur multiple times inside a 
  returned list, strictly depending on the number of occurrences in the 
  corresponding~expression.%
\end{isamarkuptext}\isamarkuptrue%
\ \ \isacommand{primrec}\isamarkupfalse%
\isanewline
\ \ \ \ occ\isactrlsub A\ {\isacharcolon}{\kern0pt}{\isacharcolon}{\kern0pt}\ {\isachardoublequoteopen}aexp\ {\isasymRightarrow}\ var\ list{\isachardoublequoteclose}\ \isakeyword{where}\isanewline
\ \ \ \ {\isachardoublequoteopen}occ\isactrlsub A\ {\isacharparenleft}{\kern0pt}Num\ n{\isacharparenright}{\kern0pt}\ {\isacharequal}{\kern0pt}\ {\isacharbrackleft}{\kern0pt}{\isacharbrackright}{\kern0pt}{\isachardoublequoteclose}\ {\isacharbar}{\kern0pt}\isanewline
\ \ \ \ {\isachardoublequoteopen}occ\isactrlsub A\ {\isacharparenleft}{\kern0pt}Var\ x{\isacharparenright}{\kern0pt}\ {\isacharequal}{\kern0pt}\ {\isacharbrackleft}{\kern0pt}x{\isacharbrackright}{\kern0pt}{\isachardoublequoteclose}\ {\isacharbar}{\kern0pt}\isanewline
\ \ \ \ {\isachardoublequoteopen}occ\isactrlsub A\ {\isacharparenleft}{\kern0pt}a\isactrlsub {\isadigit{1}}\ \isactrlsub Aop\ a\isactrlsub {\isadigit{2}}{\isacharparenright}{\kern0pt}\ {\isacharequal}{\kern0pt}\ occ\isactrlsub A{\isacharparenleft}{\kern0pt}a\isactrlsub {\isadigit{1}}{\isacharparenright}{\kern0pt}\ {\isacharat}{\kern0pt}\ occ\isactrlsub A{\isacharparenleft}{\kern0pt}a\isactrlsub {\isadigit{2}}{\isacharparenright}{\kern0pt}{\isachardoublequoteclose}\ \isanewline
\isanewline
\ \ \isacommand{primrec}\isamarkupfalse%
\ \isanewline
\ \ \ \ occ\isactrlsub B\ {\isacharcolon}{\kern0pt}{\isacharcolon}{\kern0pt}\ {\isachardoublequoteopen}bexp\ {\isasymRightarrow}\ var\ list{\isachardoublequoteclose}\ \isakeyword{where}\isanewline
\ \ \ \ {\isachardoublequoteopen}occ\isactrlsub B\ {\isacharparenleft}{\kern0pt}Bool\ b{\isacharparenright}{\kern0pt}\ {\isacharequal}{\kern0pt}\ {\isacharbrackleft}{\kern0pt}{\isacharbrackright}{\kern0pt}{\isachardoublequoteclose}\ {\isacharbar}{\kern0pt}\isanewline
\ \ \ \ {\isachardoublequoteopen}occ\isactrlsub B\ {\isacharparenleft}{\kern0pt}not\ b{\isacharparenright}{\kern0pt}\ {\isacharequal}{\kern0pt}\ occ\isactrlsub B{\isacharparenleft}{\kern0pt}b{\isacharparenright}{\kern0pt}{\isachardoublequoteclose}\ {\isacharbar}{\kern0pt}\isanewline
\ \ \ \ {\isachardoublequoteopen}occ\isactrlsub B\ {\isacharparenleft}{\kern0pt}b\isactrlsub {\isadigit{1}}\ \isactrlsub Bop\ b\isactrlsub {\isadigit{2}}{\isacharparenright}{\kern0pt}\ {\isacharequal}{\kern0pt}\ occ\isactrlsub B{\isacharparenleft}{\kern0pt}b\isactrlsub {\isadigit{1}}{\isacharparenright}{\kern0pt}\ {\isacharat}{\kern0pt}\ occ\isactrlsub B{\isacharparenleft}{\kern0pt}b\isactrlsub {\isadigit{2}}{\isacharparenright}{\kern0pt}{\isachardoublequoteclose}\ {\isacharbar}{\kern0pt}\isanewline
\ \ \ \ {\isachardoublequoteopen}occ\isactrlsub B\ {\isacharparenleft}{\kern0pt}a\isactrlsub {\isadigit{1}}\ \isactrlsub Rop\ a\isactrlsub {\isadigit{2}}{\isacharparenright}{\kern0pt}\ {\isacharequal}{\kern0pt}\ occ\isactrlsub A{\isacharparenleft}{\kern0pt}a\isactrlsub {\isadigit{1}}{\isacharparenright}{\kern0pt}\ {\isacharat}{\kern0pt}\ occ\isactrlsub A{\isacharparenleft}{\kern0pt}a\isactrlsub {\isadigit{2}}{\isacharparenright}{\kern0pt}{\isachardoublequoteclose}%
\begin{isamarkuptext}%
Although the variable occurrence functions are similar to the previous variable
  mappings, they differ in their result type. Whilst the variable occurrence
  functions return a finite list, the previous variable mappings return
  a (theoretically) infinite set. Thus, the variable occurrence functions ensure
  that the variables of an arithmetic/Boolean expression can be iterated~over. \par%
\end{isamarkuptext}\isamarkuptrue%
\begin{isamarkuptext}%
We can now provide examples for the use of variable mappings and variable 
  occurrence functions in our proof~system.%
\end{isamarkuptext}\isamarkuptrue%
\ \ \isacommand{lemma}\isamarkupfalse%
\ {\isachardoublequoteopen}vars\isactrlsub A{\isacharparenleft}{\kern0pt}aExp{\isacharunderscore}{\kern0pt}ex{\isacharparenright}{\kern0pt}\ {\isacharequal}{\kern0pt}\ {\isacharbraceleft}{\kern0pt}{\isacharprime}{\kern0pt}{\isacharprime}{\kern0pt}x{\isacharprime}{\kern0pt}{\isacharprime}{\kern0pt}{\isacharcomma}{\kern0pt}\ {\isacharprime}{\kern0pt}{\isacharprime}{\kern0pt}y{\isacharprime}{\kern0pt}{\isacharprime}{\kern0pt}{\isacharbraceright}{\kern0pt}{\isachardoublequoteclose}\ \isanewline
\isadelimproof
\ \ \ \ %
\endisadelimproof
\isatagproof
\isacommand{by}\isamarkupfalse%
\ {\isacharparenleft}{\kern0pt}auto\ simp\ add{\isacharcolon}{\kern0pt}\ aExp{\isacharunderscore}{\kern0pt}ex{\isacharunderscore}{\kern0pt}def{\isacharparenright}{\kern0pt}%
\endisatagproof
{\isafoldproof}%
\isadelimproof
\isanewline
\endisadelimproof
\isanewline
\ \ \isacommand{lemma}\isamarkupfalse%
\ {\isachardoublequoteopen}vars\isactrlsub B{\isacharparenleft}{\kern0pt}bExp{\isacharunderscore}{\kern0pt}ex{\isacharparenright}{\kern0pt}\ {\isacharequal}{\kern0pt}\ {\isacharbraceleft}{\kern0pt}{\isacharprime}{\kern0pt}{\isacharprime}{\kern0pt}x{\isacharprime}{\kern0pt}{\isacharprime}{\kern0pt}{\isacharbraceright}{\kern0pt}{\isachardoublequoteclose}\isanewline
\isadelimproof
\ \ \ \ %
\endisadelimproof
\isatagproof
\isacommand{by}\isamarkupfalse%
\ {\isacharparenleft}{\kern0pt}auto\ simp\ add{\isacharcolon}{\kern0pt}\ bExp{\isacharunderscore}{\kern0pt}ex{\isacharunderscore}{\kern0pt}def{\isacharparenright}{\kern0pt}%
\endisatagproof
{\isafoldproof}%
\isadelimproof
\isanewline
\endisadelimproof
\isanewline
\ \ \isacommand{lemma}\isamarkupfalse%
\ {\isachardoublequoteopen}occ\isactrlsub A{\isacharparenleft}{\kern0pt}aExp{\isacharunderscore}{\kern0pt}ex{\isacharparenright}{\kern0pt}\ {\isacharequal}{\kern0pt}\ {\isacharbrackleft}{\kern0pt}{\isacharprime}{\kern0pt}{\isacharprime}{\kern0pt}x{\isacharprime}{\kern0pt}{\isacharprime}{\kern0pt}{\isacharcomma}{\kern0pt}\ {\isacharprime}{\kern0pt}{\isacharprime}{\kern0pt}y{\isacharprime}{\kern0pt}{\isacharprime}{\kern0pt}{\isacharcomma}{\kern0pt}\ {\isacharprime}{\kern0pt}{\isacharprime}{\kern0pt}x{\isacharprime}{\kern0pt}{\isacharprime}{\kern0pt}{\isacharbrackright}{\kern0pt}{\isachardoublequoteclose}\ \isanewline
\isadelimproof
\ \ \ \ %
\endisadelimproof
\isatagproof
\isacommand{by}\isamarkupfalse%
\ {\isacharparenleft}{\kern0pt}simp\ add{\isacharcolon}{\kern0pt}\ aExp{\isacharunderscore}{\kern0pt}ex{\isacharunderscore}{\kern0pt}def{\isacharparenright}{\kern0pt}%
\endisatagproof
{\isafoldproof}%
\isadelimproof
\isanewline
\endisadelimproof
\isanewline
\ \ \isacommand{lemma}\isamarkupfalse%
\ {\isachardoublequoteopen}occ\isactrlsub B{\isacharparenleft}{\kern0pt}bExp{\isacharunderscore}{\kern0pt}ex{\isacharparenright}{\kern0pt}\ {\isacharequal}{\kern0pt}\ {\isacharbrackleft}{\kern0pt}{\isacharprime}{\kern0pt}{\isacharprime}{\kern0pt}x{\isacharprime}{\kern0pt}{\isacharprime}{\kern0pt}{\isacharbrackright}{\kern0pt}{\isachardoublequoteclose}\isanewline
\isadelimproof
\ \ \ \ %
\endisadelimproof
\isatagproof
\isacommand{by}\isamarkupfalse%
\ {\isacharparenleft}{\kern0pt}simp\ add{\isacharcolon}{\kern0pt}\ bExp{\isacharunderscore}{\kern0pt}ex{\isacharunderscore}{\kern0pt}def{\isacharparenright}{\kern0pt}%
\endisatagproof
{\isafoldproof}%
\isadelimproof
\endisadelimproof
\isadelimdocument
\endisadelimdocument
\isatagdocument
\isamarkupsubsubsection{Variable Substitutions%
}
\isamarkuptrue%
\endisatagdocument
{\isafolddocument}%
\isadelimdocument
\endisadelimdocument
\begin{isamarkuptext}%
We now introduce recursive variable substitution functions, substituting every 
  occurrence of a variable in a specific type of expression with another variable. 
  We again denote abbreviations for lists of expressions and sets of Boolean 
  expressions. The definitions of the substitution functions are~straightforward.%
\end{isamarkuptext}\isamarkuptrue%
\ \ \isacommand{primrec}\isamarkupfalse%
\isanewline
\ \ \ \ substitute\isactrlsub A\ {\isacharcolon}{\kern0pt}{\isacharcolon}{\kern0pt}\ {\isachardoublequoteopen}aexp\ {\isasymRightarrow}\ var\ {\isasymRightarrow}\ var\ {\isasymRightarrow}\ aexp{\isachardoublequoteclose}\ \isakeyword{where}\isanewline
\ \ \ \ {\isachardoublequoteopen}substitute\isactrlsub A\ {\isacharparenleft}{\kern0pt}Num\ n{\isacharparenright}{\kern0pt}\ x\ y\ {\isacharequal}{\kern0pt}\ Num\ n{\isachardoublequoteclose}\ {\isacharbar}{\kern0pt}\isanewline
\ \ \ \ {\isachardoublequoteopen}substitute\isactrlsub A\ {\isacharparenleft}{\kern0pt}Var\ v{\isacharparenright}{\kern0pt}\ x\ y\ {\isacharequal}{\kern0pt}\ {\isacharparenleft}{\kern0pt}if\ x\ {\isacharequal}{\kern0pt}\ v\ then\ Var\ y\ else\ Var\ v{\isacharparenright}{\kern0pt}{\isachardoublequoteclose}\ {\isacharbar}{\kern0pt}\ \isanewline
\ \ \ \ {\isachardoublequoteopen}substitute\isactrlsub A\ {\isacharparenleft}{\kern0pt}a\isactrlsub {\isadigit{1}}\ \isactrlsub Aop\ a\isactrlsub {\isadigit{2}}{\isacharparenright}{\kern0pt}\ x\ y\ {\isacharequal}{\kern0pt}\ {\isacharparenleft}{\kern0pt}substitute\isactrlsub A\ a\isactrlsub {\isadigit{1}}\ x\ y{\isacharparenright}{\kern0pt}\ \isactrlsub Aop\ {\isacharparenleft}{\kern0pt}substitute\isactrlsub A\ a\isactrlsub {\isadigit{2}}\ x\ y{\isacharparenright}{\kern0pt}{\isachardoublequoteclose}\isanewline
\isanewline
\ \ \isacommand{primrec}\isamarkupfalse%
\isanewline
\ \ \ \ substitute\isactrlsub B\ {\isacharcolon}{\kern0pt}{\isacharcolon}{\kern0pt}\ {\isachardoublequoteopen}bexp\ {\isasymRightarrow}\ var\ {\isasymRightarrow}\ var\ {\isasymRightarrow}\ bexp{\isachardoublequoteclose}\ \isakeyword{where}\isanewline
\ \ \ \ {\isachardoublequoteopen}substitute\isactrlsub B\ {\isacharparenleft}{\kern0pt}Bool\ b{\isacharparenright}{\kern0pt}\ x\ y\ {\isacharequal}{\kern0pt}\ Bool\ b{\isachardoublequoteclose}\ {\isacharbar}{\kern0pt}\isanewline
\ \ \ \ {\isachardoublequoteopen}substitute\isactrlsub B\ {\isacharparenleft}{\kern0pt}not\ b{\isacharparenright}{\kern0pt}\ x\ y\ {\isacharequal}{\kern0pt}\ not\ {\isacharparenleft}{\kern0pt}substitute\isactrlsub B\ b\ x\ y{\isacharparenright}{\kern0pt}{\isachardoublequoteclose}\ {\isacharbar}{\kern0pt}\isanewline
\ \ \ \ {\isachardoublequoteopen}substitute\isactrlsub B\ {\isacharparenleft}{\kern0pt}b\isactrlsub {\isadigit{1}}\ \isactrlsub Bop\ b\isactrlsub {\isadigit{2}}{\isacharparenright}{\kern0pt}\ x\ y\ {\isacharequal}{\kern0pt}\ {\isacharparenleft}{\kern0pt}substitute\isactrlsub B\ b\isactrlsub {\isadigit{1}}\ x\ y{\isacharparenright}{\kern0pt}\ \isactrlsub Bop\ {\isacharparenleft}{\kern0pt}substitute\isactrlsub B\ b\isactrlsub {\isadigit{2}}\ x\ y{\isacharparenright}{\kern0pt}{\isachardoublequoteclose}\ {\isacharbar}{\kern0pt}\isanewline
\ \ \ \ {\isachardoublequoteopen}substitute\isactrlsub B\ {\isacharparenleft}{\kern0pt}a\isactrlsub {\isadigit{1}}\ \isactrlsub Rop\ a\isactrlsub {\isadigit{2}}{\isacharparenright}{\kern0pt}\ x\ y\ {\isacharequal}{\kern0pt}\ {\isacharparenleft}{\kern0pt}substitute\isactrlsub A\ a\isactrlsub {\isadigit{1}}\ x\ y{\isacharparenright}{\kern0pt}\ \isactrlsub Rop\ {\isacharparenleft}{\kern0pt}substitute\isactrlsub A\ a\isactrlsub {\isadigit{2}}\ x\ y{\isacharparenright}{\kern0pt}{\isachardoublequoteclose}\isanewline
\isanewline
\ \ \isacommand{primrec}\isamarkupfalse%
\isanewline
\ \ \ \ substitute\isactrlsub E\ {\isacharcolon}{\kern0pt}{\isacharcolon}{\kern0pt}\ {\isachardoublequoteopen}exp\ {\isasymRightarrow}\ var\ {\isasymRightarrow}\ var\ {\isasymRightarrow}\ exp{\isachardoublequoteclose}\ \isakeyword{where}\isanewline
\ \ \ \ {\isachardoublequoteopen}substitute\isactrlsub E\ {\isacharparenleft}{\kern0pt}A\ a{\isacharparenright}{\kern0pt}\ x\ y\ {\isacharequal}{\kern0pt}\ A\ {\isacharparenleft}{\kern0pt}substitute\isactrlsub A\ a\ x\ y{\isacharparenright}{\kern0pt}{\isachardoublequoteclose}\ {\isacharbar}{\kern0pt}\isanewline
\ \ \ \ {\isachardoublequoteopen}substitute\isactrlsub E\ {\isacharparenleft}{\kern0pt}B\ b{\isacharparenright}{\kern0pt}\ x\ y\ {\isacharequal}{\kern0pt}\ B\ {\isacharparenleft}{\kern0pt}substitute\isactrlsub B\ b\ x\ y{\isacharparenright}{\kern0pt}{\isachardoublequoteclose}\ {\isacharbar}{\kern0pt}\isanewline
\ \ \ \ {\isachardoublequoteopen}substitute\isactrlsub E\ {\isacharparenleft}{\kern0pt}P\ m{\isacharparenright}{\kern0pt}\ x\ y\ {\isacharequal}{\kern0pt}\ P\ m{\isachardoublequoteclose}\isanewline
\isanewline
\ \ \isacommand{primrec}\isamarkupfalse%
\isanewline
\ \ \ \ substitute\isactrlsub S\ {\isacharcolon}{\kern0pt}{\isacharcolon}{\kern0pt}\ {\isachardoublequoteopen}sexp\ {\isasymRightarrow}\ var\ {\isasymRightarrow}\ var\ {\isasymRightarrow}\ sexp{\isachardoublequoteclose}\ \isakeyword{where}\isanewline
\ \ \ \ {\isachardoublequoteopen}substitute\isactrlsub S\ {\isacharparenleft}{\kern0pt}Exp\ a{\isacharparenright}{\kern0pt}\ x\ y\ {\isacharequal}{\kern0pt}\ Exp\ {\isacharparenleft}{\kern0pt}substitute\isactrlsub A\ a\ x\ y{\isacharparenright}{\kern0pt}{\isachardoublequoteclose}\ {\isacharbar}{\kern0pt}\isanewline
\ \ \ \ {\isachardoublequoteopen}substitute\isactrlsub S\ \isactrlemph \ x\ y\ {\isacharequal}{\kern0pt}\ \isactrlemph {\isachardoublequoteclose}\isanewline
\isanewline
\ \ \isacommand{primrec}\isamarkupfalse%
\isanewline
\ \ \ \ lsubstitute\isactrlsub E\ {\isacharcolon}{\kern0pt}{\isacharcolon}{\kern0pt}\ {\isachardoublequoteopen}exp\ list\ {\isasymRightarrow}\ var\ {\isasymRightarrow}\ var\ {\isasymRightarrow}\ exp\ list{\isachardoublequoteclose}\ \isakeyword{where}\isanewline
\ \ \ \ {\isachardoublequoteopen}lsubstitute\isactrlsub E\ {\isacharbrackleft}{\kern0pt}{\isacharbrackright}{\kern0pt}\ x\ y\ {\isacharequal}{\kern0pt}\ {\isacharbrackleft}{\kern0pt}{\isacharbrackright}{\kern0pt}{\isachardoublequoteclose}\ {\isacharbar}{\kern0pt}\isanewline
\ \ \ \ {\isachardoublequoteopen}lsubstitute\isactrlsub E\ {\isacharparenleft}{\kern0pt}exp\ {\isacharhash}{\kern0pt}\ rest{\isacharparenright}{\kern0pt}\ x\ y\ {\isacharequal}{\kern0pt}\ {\isacharparenleft}{\kern0pt}substitute\isactrlsub E\ exp\ x\ y{\isacharparenright}{\kern0pt}\ {\isacharhash}{\kern0pt}\ {\isacharparenleft}{\kern0pt}lsubstitute\isactrlsub E\ rest\ x\ y{\isacharparenright}{\kern0pt}{\isachardoublequoteclose}\isanewline
\isanewline
\ \ \isacommand{fun}\isamarkupfalse%
\isanewline
\ \ \ \ ssubstitute\isactrlsub B\ {\isacharcolon}{\kern0pt}{\isacharcolon}{\kern0pt}\ {\isachardoublequoteopen}bexp\ set\ {\isasymRightarrow}\ var\ {\isasymRightarrow}\ var\ {\isasymRightarrow}\ bexp\ set{\isachardoublequoteclose}\ \isakeyword{where}\isanewline
\ \ \ \ {\isachardoublequoteopen}ssubstitute\isactrlsub B\ S\ x\ y\ {\isacharequal}{\kern0pt}\ {\isacharparenleft}{\kern0pt}{\isacharpercent}{\kern0pt}b{\isachardot}{\kern0pt}\ substitute\isactrlsub B\ b\ x\ y{\isacharparenright}{\kern0pt}\ {\isacharbackquote}{\kern0pt}\ S{\isachardoublequoteclose}%
\begin{isamarkuptext}%
We can now observe several example applications of the variable 
  substitution~function.%
\end{isamarkuptext}\isamarkuptrue%
\ \ \isacommand{lemma}\isamarkupfalse%
\ {\isachardoublequoteopen}substitute\isactrlsub A\ aExp{\isacharunderscore}{\kern0pt}ex\ {\isacharprime}{\kern0pt}{\isacharprime}{\kern0pt}x{\isacharprime}{\kern0pt}{\isacharprime}{\kern0pt}\ {\isacharprime}{\kern0pt}{\isacharprime}{\kern0pt}z{\isacharprime}{\kern0pt}{\isacharprime}{\kern0pt}\ {\isacharequal}{\kern0pt}\ {\isacharparenleft}{\kern0pt}{\isacharparenleft}{\kern0pt}Var\ {\isacharprime}{\kern0pt}{\isacharprime}{\kern0pt}z{\isacharprime}{\kern0pt}{\isacharprime}{\kern0pt}{\isacharparenright}{\kern0pt}\ \isactrlsub Amul\ {\isacharparenleft}{\kern0pt}Var\ {\isacharprime}{\kern0pt}{\isacharprime}{\kern0pt}y{\isacharprime}{\kern0pt}{\isacharprime}{\kern0pt}{\isacharparenright}{\kern0pt}{\isacharparenright}{\kern0pt}\ \isactrlsub Asub\ {\isacharparenleft}{\kern0pt}Var\ {\isacharprime}{\kern0pt}{\isacharprime}{\kern0pt}z{\isacharprime}{\kern0pt}{\isacharprime}{\kern0pt}{\isacharparenright}{\kern0pt}{\isachardoublequoteclose}\ \isanewline
\isadelimproof
\ \ \ \ %
\endisadelimproof
\isatagproof
\isacommand{by}\isamarkupfalse%
\ {\isacharparenleft}{\kern0pt}simp\ add{\isacharcolon}{\kern0pt}\ aExp{\isacharunderscore}{\kern0pt}ex{\isacharunderscore}{\kern0pt}def{\isacharparenright}{\kern0pt}%
\endisatagproof
{\isafoldproof}%
\isadelimproof
\isanewline
\endisadelimproof
\isanewline
\ \ \isacommand{lemma}\isamarkupfalse%
\ {\isachardoublequoteopen}substitute\isactrlsub B\ bExp{\isacharunderscore}{\kern0pt}ex\ {\isacharprime}{\kern0pt}{\isacharprime}{\kern0pt}x{\isacharprime}{\kern0pt}{\isacharprime}{\kern0pt}\ {\isacharprime}{\kern0pt}{\isacharprime}{\kern0pt}y{\isacharprime}{\kern0pt}{\isacharprime}{\kern0pt}\ {\isacharequal}{\kern0pt}\ {\isacharparenleft}{\kern0pt}{\isacharparenleft}{\kern0pt}Var\ {\isacharprime}{\kern0pt}{\isacharprime}{\kern0pt}y{\isacharprime}{\kern0pt}{\isacharprime}{\kern0pt}{\isacharparenright}{\kern0pt}\ \isactrlsub Req\ {\isacharparenleft}{\kern0pt}Num\ {\isadigit{2}}{\isacharparenright}{\kern0pt}{\isacharparenright}{\kern0pt}\ \isactrlsub Bdisj\ {\isacharparenleft}{\kern0pt}Bool\ False{\isacharparenright}{\kern0pt}{\isachardoublequoteclose}\isanewline
\isadelimproof
\ \ \ \ %
\endisadelimproof
\isatagproof
\isacommand{by}\isamarkupfalse%
\ {\isacharparenleft}{\kern0pt}simp\ add{\isacharcolon}{\kern0pt}\ bExp{\isacharunderscore}{\kern0pt}ex{\isacharunderscore}{\kern0pt}def{\isacharparenright}{\kern0pt}%
\endisatagproof
{\isafoldproof}%
\isadelimproof
\endisadelimproof
\isadelimdocument
\endisadelimdocument
\isatagdocument
\isamarkupsubsubsection{Concreteness%
}
\isamarkuptrue%
\endisatagdocument
{\isafolddocument}%
\isadelimdocument
\endisadelimdocument
\begin{isamarkuptext}%
We call a specific type of expression concrete iff it contains no
  variables or symbolic values and has already been simplified as much as possible. 
  Similar to the variable mappings, we again provide concreteness
  notions for lists of expressions and sets of Boolean expressions, which will be  
  useful abbreviations at a later point. Although the notion of a concrete
  expression does not exist in the original paper, we propose such a notion 
  in order to later be able to circumvent the implicit simplifications of 
  expressions~in~states.%
\end{isamarkuptext}\isamarkuptrue%
\ \ \isacommand{abbreviation}\isamarkupfalse%
\isanewline
\ \ \ \ concrete\isactrlsub A\ {\isacharcolon}{\kern0pt}{\isacharcolon}{\kern0pt}\ {\isachardoublequoteopen}aexp\ {\isasymRightarrow}\ bool{\isachardoublequoteclose}\ \isakeyword{where}\isanewline
\ \ \ \ {\isachardoublequoteopen}concrete\isactrlsub A\ a\ {\isasymequiv}\ {\isacharparenleft}{\kern0pt}case\ a\ of\ {\isacharparenleft}{\kern0pt}Num\ n{\isacharparenright}{\kern0pt}\ {\isasymRightarrow}\ True\ {\isacharbar}{\kern0pt}\ {\isacharunderscore}{\kern0pt}\ {\isasymRightarrow}\ False{\isacharparenright}{\kern0pt}{\isachardoublequoteclose}\ \isanewline
\isanewline
\ \ \isacommand{abbreviation}\isamarkupfalse%
\isanewline
\ \ \ \ concrete\isactrlsub B\ {\isacharcolon}{\kern0pt}{\isacharcolon}{\kern0pt}\ {\isachardoublequoteopen}bexp\ {\isasymRightarrow}\ bool{\isachardoublequoteclose}\ \isakeyword{where}\isanewline
\ \ \ \ {\isachardoublequoteopen}concrete\isactrlsub B\ b\ {\isasymequiv}\ {\isacharparenleft}{\kern0pt}case\ b\ of\ {\isacharparenleft}{\kern0pt}Bool\ b{\isacharparenright}{\kern0pt}\ {\isasymRightarrow}\ True\ {\isacharbar}{\kern0pt}\ {\isacharunderscore}{\kern0pt}\ {\isasymRightarrow}\ False{\isacharparenright}{\kern0pt}{\isachardoublequoteclose}\isanewline
\isanewline
\ \ \isacommand{abbreviation}\isamarkupfalse%
\isanewline
\ \ \ \ concrete\isactrlsub E\ {\isacharcolon}{\kern0pt}{\isacharcolon}{\kern0pt}\ {\isachardoublequoteopen}exp\ {\isasymRightarrow}\ bool{\isachardoublequoteclose}\ \isakeyword{where}\isanewline
\ \ \ \ {\isachardoublequoteopen}concrete\isactrlsub E\ e\ {\isasymequiv}\ {\isacharparenleft}{\kern0pt}case\ e\ of\ {\isacharparenleft}{\kern0pt}A\ a{\isacharparenright}{\kern0pt}\ {\isasymRightarrow}\ concrete\isactrlsub A{\isacharparenleft}{\kern0pt}a{\isacharparenright}{\kern0pt}\ {\isacharbar}{\kern0pt}\ {\isacharparenleft}{\kern0pt}B\ b{\isacharparenright}{\kern0pt}\ {\isasymRightarrow}\ concrete\isactrlsub B{\isacharparenleft}{\kern0pt}b{\isacharparenright}{\kern0pt}\ {\isacharbar}{\kern0pt}\ {\isacharparenleft}{\kern0pt}P\ m{\isacharparenright}{\kern0pt}\ {\isasymRightarrow}\ True{\isacharparenright}{\kern0pt}{\isachardoublequoteclose}\ \isanewline
\isanewline
\ \ \isacommand{abbreviation}\isamarkupfalse%
\isanewline
\ \ \ \ concrete\isactrlsub S\ {\isacharcolon}{\kern0pt}{\isacharcolon}{\kern0pt}\ {\isachardoublequoteopen}sexp\ {\isasymRightarrow}\ bool{\isachardoublequoteclose}\ \isakeyword{where}\isanewline
\ \ \ \ {\isachardoublequoteopen}concrete\isactrlsub S\ s\ {\isasymequiv}\ {\isacharparenleft}{\kern0pt}case\ s\ of\ {\isacharparenleft}{\kern0pt}Exp\ a{\isacharparenright}{\kern0pt}\ {\isasymRightarrow}\ concrete\isactrlsub A{\isacharparenleft}{\kern0pt}a{\isacharparenright}{\kern0pt}\ {\isacharbar}{\kern0pt}\ \isactrlemph \ {\isasymRightarrow}\ False{\isacharparenright}{\kern0pt}{\isachardoublequoteclose}\isanewline
\isanewline
\ \ \isacommand{primrec}\isamarkupfalse%
\isanewline
\ \ \ \ lconcrete\isactrlsub E\ {\isacharcolon}{\kern0pt}{\isacharcolon}{\kern0pt}\ {\isachardoublequoteopen}exp\ list\ {\isasymRightarrow}\ bool{\isachardoublequoteclose}\ \isakeyword{where}\isanewline
\ \ \ \ {\isachardoublequoteopen}lconcrete\isactrlsub E\ {\isacharbrackleft}{\kern0pt}{\isacharbrackright}{\kern0pt}\ {\isacharequal}{\kern0pt}\ True{\isachardoublequoteclose}\ {\isacharbar}{\kern0pt}\isanewline
\ \ \ \ {\isachardoublequoteopen}lconcrete\isactrlsub E\ {\isacharparenleft}{\kern0pt}exp\ {\isacharhash}{\kern0pt}\ rest{\isacharparenright}{\kern0pt}\ {\isacharequal}{\kern0pt}\ {\isacharparenleft}{\kern0pt}concrete\isactrlsub E{\isacharparenleft}{\kern0pt}exp{\isacharparenright}{\kern0pt}\ {\isasymand}\ lconcrete\isactrlsub E{\isacharparenleft}{\kern0pt}rest{\isacharparenright}{\kern0pt}{\isacharparenright}{\kern0pt}{\isachardoublequoteclose}\isanewline
\isanewline
\ \ \isacommand{abbreviation}\isamarkupfalse%
\isanewline
\ \ \ \ sconcrete\isactrlsub B\ {\isacharcolon}{\kern0pt}{\isacharcolon}{\kern0pt}\ {\isachardoublequoteopen}bexp\ set\ {\isasymRightarrow}\ bool{\isachardoublequoteclose}\ \isakeyword{where}\isanewline
\ \ \ \ {\isachardoublequoteopen}sconcrete\isactrlsub B\ S\ {\isasymequiv}\ {\isacharparenleft}{\kern0pt}{\isasymforall}b\ {\isasymin}\ S{\isachardot}{\kern0pt}\ concrete\isactrlsub B{\isacharparenleft}{\kern0pt}b{\isacharparenright}{\kern0pt}{\isacharparenright}{\kern0pt}{\isachardoublequoteclose}%
\begin{isamarkuptext}%
We can now establish a connection between the notion of concrete
  expressions and the variable mappings. By making use of structural induction,
  we show that a concrete expression is always variable-free. This property is
  obvious, as it trivially holds due to the definition of the concreteness notion.%
\end{isamarkuptext}\isamarkuptrue%
\ \ \isacommand{lemma}\isamarkupfalse%
\ concrete{\isacharunderscore}{\kern0pt}vars{\isacharunderscore}{\kern0pt}imp\isactrlsub A{\isacharcolon}{\kern0pt}\ {\isachardoublequoteopen}concrete\isactrlsub A{\isacharparenleft}{\kern0pt}a{\isacharparenright}{\kern0pt}\ {\isasymlongrightarrow}\ vars\isactrlsub A{\isacharparenleft}{\kern0pt}a{\isacharparenright}{\kern0pt}\ {\isacharequal}{\kern0pt}\ {\isacharbraceleft}{\kern0pt}{\isacharbraceright}{\kern0pt}{\isachardoublequoteclose}\ \isanewline
\isadelimproof
\ \ \ \ %
\endisadelimproof
\isatagproof
\isacommand{by}\isamarkupfalse%
\ {\isacharparenleft}{\kern0pt}induct\ a{\isacharsemicolon}{\kern0pt}\ simp{\isacharparenright}{\kern0pt}%
\endisatagproof
{\isafoldproof}%
\isadelimproof
\isanewline
\endisadelimproof
\isanewline
\ \ \isacommand{lemma}\isamarkupfalse%
\ concrete{\isacharunderscore}{\kern0pt}vars{\isacharunderscore}{\kern0pt}imp\isactrlsub B{\isacharcolon}{\kern0pt}\ {\isachardoublequoteopen}concrete\isactrlsub B{\isacharparenleft}{\kern0pt}b{\isacharparenright}{\kern0pt}\ {\isasymlongrightarrow}\ vars\isactrlsub B{\isacharparenleft}{\kern0pt}b{\isacharparenright}{\kern0pt}\ {\isacharequal}{\kern0pt}\ {\isacharbraceleft}{\kern0pt}{\isacharbraceright}{\kern0pt}{\isachardoublequoteclose}\ \isanewline
\isadelimproof
\ \ \ \ %
\endisadelimproof
\isatagproof
\isacommand{by}\isamarkupfalse%
\ {\isacharparenleft}{\kern0pt}induct\ b{\isacharsemicolon}{\kern0pt}\ simp{\isacharparenright}{\kern0pt}%
\endisatagproof
{\isafoldproof}%
\isadelimproof
\isanewline
\endisadelimproof
\isanewline
\ \ \isacommand{lemma}\isamarkupfalse%
\ concrete{\isacharunderscore}{\kern0pt}vars{\isacharunderscore}{\kern0pt}imp\isactrlsub E{\isacharcolon}{\kern0pt}\ {\isachardoublequoteopen}concrete\isactrlsub E{\isacharparenleft}{\kern0pt}e{\isacharparenright}{\kern0pt}\ {\isasymlongrightarrow}\ vars\isactrlsub E{\isacharparenleft}{\kern0pt}e{\isacharparenright}{\kern0pt}\ {\isacharequal}{\kern0pt}\ {\isacharbraceleft}{\kern0pt}{\isacharbraceright}{\kern0pt}{\isachardoublequoteclose}\isanewline
\isadelimproof
\ \ \ \ %
\endisadelimproof
\isatagproof
\isacommand{using}\isamarkupfalse%
\ concrete{\isacharunderscore}{\kern0pt}vars{\isacharunderscore}{\kern0pt}imp\isactrlsub A\ concrete{\isacharunderscore}{\kern0pt}vars{\isacharunderscore}{\kern0pt}imp\isactrlsub B\ \isacommand{by}\isamarkupfalse%
\ {\isacharparenleft}{\kern0pt}induct\ e{\isacharsemicolon}{\kern0pt}\ simp{\isacharparenright}{\kern0pt}%
\endisatagproof
{\isafoldproof}%
\isadelimproof
\isanewline
\endisadelimproof
\isanewline
\ \ \isacommand{lemma}\isamarkupfalse%
\ concrete{\isacharunderscore}{\kern0pt}vars{\isacharunderscore}{\kern0pt}imp\isactrlsub S{\isacharcolon}{\kern0pt}\ {\isachardoublequoteopen}concrete\isactrlsub S{\isacharparenleft}{\kern0pt}s{\isacharparenright}{\kern0pt}\ {\isasymlongrightarrow}\ vars\isactrlsub S{\isacharparenleft}{\kern0pt}s{\isacharparenright}{\kern0pt}\ {\isacharequal}{\kern0pt}\ {\isacharbraceleft}{\kern0pt}{\isacharbraceright}{\kern0pt}{\isachardoublequoteclose}\isanewline
\isadelimproof
\ \ \ \ %
\endisadelimproof
\isatagproof
\isacommand{using}\isamarkupfalse%
\ concrete{\isacharunderscore}{\kern0pt}vars{\isacharunderscore}{\kern0pt}imp\isactrlsub A\ \isacommand{by}\isamarkupfalse%
\ {\isacharparenleft}{\kern0pt}induct\ s{\isacharsemicolon}{\kern0pt}\ simp{\isacharparenright}{\kern0pt}%
\endisatagproof
{\isafoldproof}%
\isadelimproof
\isanewline
\endisadelimproof
\isanewline
\ \ \isacommand{lemma}\isamarkupfalse%
\ l{\isacharunderscore}{\kern0pt}concrete{\isacharunderscore}{\kern0pt}vars{\isacharunderscore}{\kern0pt}imp\isactrlsub E{\isacharcolon}{\kern0pt}\ {\isachardoublequoteopen}lconcrete\isactrlsub E{\isacharparenleft}{\kern0pt}l{\isacharparenright}{\kern0pt}\ {\isasymlongrightarrow}\ lvars\isactrlsub E{\isacharparenleft}{\kern0pt}l{\isacharparenright}{\kern0pt}\ {\isacharequal}{\kern0pt}\ {\isacharbraceleft}{\kern0pt}{\isacharbraceright}{\kern0pt}{\isachardoublequoteclose}\isanewline
\isadelimproof
\ \ \ \ %
\endisadelimproof
\isatagproof
\isacommand{using}\isamarkupfalse%
\ concrete{\isacharunderscore}{\kern0pt}vars{\isacharunderscore}{\kern0pt}imp\isactrlsub E\ \isacommand{by}\isamarkupfalse%
\ {\isacharparenleft}{\kern0pt}induct\ l{\isacharsemicolon}{\kern0pt}\ simp{\isacharparenright}{\kern0pt}%
\endisatagproof
{\isafoldproof}%
\isadelimproof
\isanewline
\endisadelimproof
\isanewline
\ \ \isacommand{lemma}\isamarkupfalse%
\ s{\isacharunderscore}{\kern0pt}concrete{\isacharunderscore}{\kern0pt}vars{\isacharunderscore}{\kern0pt}imp\isactrlsub B{\isacharcolon}{\kern0pt}\ {\isachardoublequoteopen}sconcrete\isactrlsub B{\isacharparenleft}{\kern0pt}S{\isacharparenright}{\kern0pt}\ {\isasymlongrightarrow}\ svars\isactrlsub B{\isacharparenleft}{\kern0pt}S{\isacharparenright}{\kern0pt}\ {\isacharequal}{\kern0pt}\ {\isacharbraceleft}{\kern0pt}{\isacharbraceright}{\kern0pt}{\isachardoublequoteclose}\isanewline
\isadelimproof
\ \ \ \ %
\endisadelimproof
\isatagproof
\isacommand{using}\isamarkupfalse%
\ concrete{\isacharunderscore}{\kern0pt}vars{\isacharunderscore}{\kern0pt}imp\isactrlsub B\ \isacommand{by}\isamarkupfalse%
\ simp%
\endisatagproof
{\isafoldproof}%
\isadelimproof
\endisadelimproof
\isadelimdocument
\endisadelimdocument
\isatagdocument
\isamarkupsubsection{States%
}
\isamarkuptrue%
\isamarkupsubsubsection{Definition of States%
}
\isamarkuptrue%
\endisatagdocument
{\isafolddocument}%
\isadelimdocument
\endisadelimdocument
\begin{isamarkuptext}%
Similar to the original paper, we define a symbolic state as a 
  partial mapping from the set of program variables to the set of starred 
  expressions. However, this approach also allows symbolic states
  to be total functions, thereby resulting in symbolic states possibly having 
  an infinite domain. Due to this fact, Isabelle will not be able to generate
  code when trying to compute the domain of a state, which clearly violates
  our objective of contributing an efficient code generation. \par
  In order to solve this problem, we enforce that the domain of a symbolic state
  must be of finite nature. The \emph{Finite Map} theory of the HOL-Library builds 
  on top of the partial function theory and already implements this exact concept. 
  Whilst a finite map guarantees the domain of the function to be finite, 
  it otherwise behaves very similar to a normal explicit partial function. Using
  this approach, we can guarantee the computability of~state~domains. \par
  We denote the set of all symbolic states as \isa{{\isasymSigma}}.%
\end{isamarkuptext}\isamarkuptrue%
\ \ \isacommand{type{\isacharunderscore}{\kern0pt}synonym}\isamarkupfalse%
\ {\isasymSigma}\ {\isacharequal}{\kern0pt}\ {\isachardoublequoteopen}{\isacharparenleft}{\kern0pt}var{\isacharcomma}{\kern0pt}\ sexp{\isacharparenright}{\kern0pt}\ fmap{\isachardoublequoteclose}%
\begin{isamarkuptext}%
We now introduce several notational abbreviations in order to ease the 
  handling of finite maps. We abbreviate the state that does not
   define any program variables as \isa{{\isasymcircle}}. Additionally we introduce 
  \isa{fm} as an abbreviation for the translation from a tuple list to a corresponding
  finite map. Last but not least, we also denote an easier readable abbreviation for 
  the finite~map~update.%
\end{isamarkuptext}\isamarkuptrue%
\ \ \isacommand{notation}\isamarkupfalse%
\ fmempty\ {\isacharparenleft}{\kern0pt}{\isachardoublequoteopen}{\isasymcircle}{\isachardoublequoteclose}{\isacharparenright}{\kern0pt}\isanewline
\ \ \isacommand{notation}\isamarkupfalse%
\ fmap{\isacharunderscore}{\kern0pt}of{\isacharunderscore}{\kern0pt}list\ {\isacharparenleft}{\kern0pt}{\isachardoublequoteopen}fm{\isachardoublequoteclose}{\isacharparenright}{\kern0pt}\isanewline
\ \ \isacommand{notation}\isamarkupfalse%
\ fmupd\ {\isacharparenleft}{\kern0pt}{\isachardoublequoteopen}{\isacharbrackleft}{\kern0pt}{\isacharunderscore}{\kern0pt}\ {\isasymlongmapsto}\ {\isacharunderscore}{\kern0pt}{\isacharbrackright}{\kern0pt}\ {\isacharunderscore}{\kern0pt}{\isachardoublequoteclose}\ {\isadigit{7}}{\isadigit{0}}{\isacharparenright}{\kern0pt}%
\begin{isamarkuptext}%
Using the newly defined concepts and notations, we can now provide
  several examples for symbolic~states. Note that we always denote a state
  as a tuple list that is translated to a corresponding finite map using~$fm$.%
\end{isamarkuptext}\isamarkuptrue%
\ \ \isacommand{definition}\isamarkupfalse%
\ \isanewline
\ \ \ \ {\isasymsigma}\isactrlsub {\isadigit{1}}\ {\isacharcolon}{\kern0pt}{\isacharcolon}{\kern0pt}\ {\isasymSigma}\ \isakeyword{where}\isanewline
\ \ \ \ {\isachardoublequoteopen}{\isasymsigma}\isactrlsub {\isadigit{1}}\ {\isasymequiv}\ fm{\isacharbrackleft}{\kern0pt}{\isacharparenleft}{\kern0pt}{\isacharprime}{\kern0pt}{\isacharprime}{\kern0pt}x{\isacharprime}{\kern0pt}{\isacharprime}{\kern0pt}{\isacharcomma}{\kern0pt}\ Exp\ {\isacharparenleft}{\kern0pt}{\isacharparenleft}{\kern0pt}Var\ {\isacharprime}{\kern0pt}{\isacharprime}{\kern0pt}y{\isacharprime}{\kern0pt}{\isacharprime}{\kern0pt}{\isacharparenright}{\kern0pt}\ \isactrlsub Amul\ {\isacharparenleft}{\kern0pt}Num\ {\isadigit{4}}{\isacharparenright}{\kern0pt}{\isacharparenright}{\kern0pt}{\isacharparenright}{\kern0pt}{\isacharcomma}{\kern0pt}\ {\isacharparenleft}{\kern0pt}{\isacharprime}{\kern0pt}{\isacharprime}{\kern0pt}y{\isacharprime}{\kern0pt}{\isacharprime}{\kern0pt}{\isacharcomma}{\kern0pt}\ \isactrlemph {\isacharparenright}{\kern0pt}{\isacharbrackright}{\kern0pt}{\isachardoublequoteclose}\isanewline
\isanewline
\ \ \isacommand{definition}\isamarkupfalse%
\ \isanewline
\ \ \ \ {\isasymsigma}\isactrlsub {\isadigit{2}}\ {\isacharcolon}{\kern0pt}{\isacharcolon}{\kern0pt}\ {\isasymSigma}\ \isakeyword{where}\isanewline
\ \ \ \ {\isachardoublequoteopen}{\isasymsigma}\isactrlsub {\isadigit{2}}\ {\isasymequiv}\ fm{\isacharbrackleft}{\kern0pt}{\isacharparenleft}{\kern0pt}{\isacharprime}{\kern0pt}{\isacharprime}{\kern0pt}x{\isacharprime}{\kern0pt}{\isacharprime}{\kern0pt}{\isacharcomma}{\kern0pt}\ Exp\ {\isacharparenleft}{\kern0pt}Num\ {\isadigit{8}}{\isacharparenright}{\kern0pt}{\isacharparenright}{\kern0pt}{\isacharcomma}{\kern0pt}\ {\isacharparenleft}{\kern0pt}{\isacharprime}{\kern0pt}{\isacharprime}{\kern0pt}y{\isacharprime}{\kern0pt}{\isacharprime}{\kern0pt}{\isacharcomma}{\kern0pt}\ Exp\ {\isacharparenleft}{\kern0pt}Num\ {\isadigit{2}}{\isacharparenright}{\kern0pt}{\isacharparenright}{\kern0pt}{\isacharbrackright}{\kern0pt}{\isachardoublequoteclose}%
\isadelimdocument
\endisadelimdocument
\isatagdocument
\isamarkupsubsubsection{Symbolic Variables%
}
\isamarkuptrue%
\endisatagdocument
{\isafolddocument}%
\isadelimdocument
\endisadelimdocument
\begin{isamarkuptext}%
A symbolic variable of a state is a variable that maps to the symbolic value. 
  The symbolic variable function maps a given state to the set of all 
  its symbolic variables. The definition of the function is straightforward. 
  Note that the nature of a symbolic state enforces the set of symbolic
  variables to~be~finite.%
\end{isamarkuptext}\isamarkuptrue%
\ \ \isacommand{definition}\isamarkupfalse%
\ \isanewline
\ \ \ \ symb\isactrlsub {\isasymSigma}\ {\isacharcolon}{\kern0pt}{\isacharcolon}{\kern0pt}\ {\isachardoublequoteopen}{\isasymSigma}\ {\isasymRightarrow}\ var\ set{\isachardoublequoteclose}\ \isakeyword{where}\isanewline
\ \ \ \ {\isachardoublequoteopen}symb\isactrlsub {\isasymSigma}\ {\isasymsigma}\ {\isasymequiv}\ {\isacharbraceleft}{\kern0pt}X\ {\isasymin}\ fmdom{\isacharprime}{\kern0pt}{\isacharparenleft}{\kern0pt}{\isasymsigma}{\isacharparenright}{\kern0pt}{\isachardot}{\kern0pt}\ fmlookup\ {\isasymsigma}\ X\ {\isacharequal}{\kern0pt}\ Some\ \isactrlemph {\isacharbraceright}{\kern0pt}{\isachardoublequoteclose}%
\begin{isamarkuptext}%
We can now provide various examples for the use of the symbolic 
  variable~function.%
\end{isamarkuptext}\isamarkuptrue%
\ \ \isacommand{lemma}\isamarkupfalse%
\ {\isachardoublequoteopen}symb\isactrlsub {\isasymSigma}\ {\isasymsigma}\isactrlsub {\isadigit{1}}\ {\isacharequal}{\kern0pt}\ {\isacharbraceleft}{\kern0pt}{\isacharprime}{\kern0pt}{\isacharprime}{\kern0pt}y{\isacharprime}{\kern0pt}{\isacharprime}{\kern0pt}{\isacharbraceright}{\kern0pt}{\isachardoublequoteclose}\ \isanewline
\isadelimproof
\ \ \ \ %
\endisadelimproof
\isatagproof
\isacommand{by}\isamarkupfalse%
\ {\isacharparenleft}{\kern0pt}auto\ simp\ add{\isacharcolon}{\kern0pt}\ {\isasymsigma}\isactrlsub {\isadigit{1}}{\isacharunderscore}{\kern0pt}def\ symb\isactrlsub {\isasymSigma}{\isacharunderscore}{\kern0pt}def{\isacharparenright}{\kern0pt}%
\endisatagproof
{\isafoldproof}%
\isadelimproof
\isanewline
\endisadelimproof
\isanewline
\ \ \isacommand{lemma}\isamarkupfalse%
\ {\isachardoublequoteopen}symb\isactrlsub {\isasymSigma}\ {\isasymsigma}\isactrlsub {\isadigit{2}}\ {\isacharequal}{\kern0pt}\ {\isacharbraceleft}{\kern0pt}{\isacharbraceright}{\kern0pt}{\isachardoublequoteclose}\ \isanewline
\isadelimproof
\ \ \ \ %
\endisadelimproof
\isatagproof
\isacommand{by}\isamarkupfalse%
\ {\isacharparenleft}{\kern0pt}auto\ simp\ add{\isacharcolon}{\kern0pt}\ {\isasymsigma}\isactrlsub {\isadigit{2}}{\isacharunderscore}{\kern0pt}def\ symb\isactrlsub {\isasymSigma}{\isacharunderscore}{\kern0pt}def{\isacharparenright}{\kern0pt}%
\endisatagproof
{\isafoldproof}%
\isadelimproof
\endisadelimproof
\isadelimdocument
\endisadelimdocument
\isatagdocument
\isamarkupsubsubsection{Wellformedness%
}
\isamarkuptrue%
\endisatagdocument
{\isafolddocument}%
\isadelimdocument
\endisadelimdocument
\begin{isamarkuptext}%
We consider a state wellformed iff all variables occurring
  in mapped expressions of a state are symbolic variables. The definition
  of this predicate is straightforward and smoothly aligns with the definition in 
  the original paper.%
\end{isamarkuptext}\isamarkuptrue%
\ \ \isacommand{definition}\isamarkupfalse%
\ \isanewline
\ \ \ \ wf\isactrlsub {\isasymSigma}\ {\isacharcolon}{\kern0pt}{\isacharcolon}{\kern0pt}\ {\isachardoublequoteopen}{\isasymSigma}\ {\isasymRightarrow}\ bool{\isachardoublequoteclose}\ \isakeyword{where}\isanewline
\ \ \ \ {\isachardoublequoteopen}wf\isactrlsub {\isasymSigma}\ {\isasymsigma}\ {\isasymequiv}\ {\isacharbraceleft}{\kern0pt}v{\isachardot}{\kern0pt}\ {\isasymexists}v{\isacharprime}{\kern0pt}\ {\isasymin}\ fmdom{\isacharprime}{\kern0pt}{\isacharparenleft}{\kern0pt}{\isasymsigma}{\isacharparenright}{\kern0pt}{\isachardot}{\kern0pt}\ v\ {\isasymin}\ vars\isactrlsub S{\isacharparenleft}{\kern0pt}the{\isacharparenleft}{\kern0pt}fmlookup\ {\isasymsigma}\ v{\isacharprime}{\kern0pt}{\isacharparenright}{\kern0pt}{\isacharparenright}{\kern0pt}{\isacharbraceright}{\kern0pt}\ {\isasymsubseteq}\ symb\isactrlsub {\isasymSigma}{\isacharparenleft}{\kern0pt}{\isasymsigma}{\isacharparenright}{\kern0pt}{\isachardoublequoteclose}%
\begin{isamarkuptext}%
As can be easily observed, \isa{{\isasymsigma}\isactrlsub {\isadigit{1}}} and \isa{{\isasymsigma}\isactrlsub {\isadigit{2}}} are both wellformed states, which
  can also be inferred~by~Isabelle.%
\end{isamarkuptext}\isamarkuptrue%
\ \ \isacommand{lemma}\isamarkupfalse%
\ {\isachardoublequoteopen}wf\isactrlsub {\isasymSigma}\ {\isasymsigma}\isactrlsub {\isadigit{1}}\ {\isasymand}\ wf\isactrlsub {\isasymSigma}\ {\isasymsigma}\isactrlsub {\isadigit{2}}{\isachardoublequoteclose}\ \isanewline
\isadelimproof
\ \ \ \ %
\endisadelimproof
\isatagproof
\isacommand{by}\isamarkupfalse%
\ {\isacharparenleft}{\kern0pt}simp\ add{\isacharcolon}{\kern0pt}\ {\isasymsigma}\isactrlsub {\isadigit{1}}{\isacharunderscore}{\kern0pt}def\ {\isasymsigma}\isactrlsub {\isadigit{2}}{\isacharunderscore}{\kern0pt}def\ symb\isactrlsub {\isasymSigma}{\isacharunderscore}{\kern0pt}def\ wf\isactrlsub {\isasymSigma}{\isacharunderscore}{\kern0pt}def{\isacharparenright}{\kern0pt}%
\endisatagproof
{\isafoldproof}%
\isadelimproof
\endisadelimproof
\isadelimdocument
\endisadelimdocument
\isatagdocument
\isamarkupsubsubsection{Concreteness%
}
\isamarkuptrue%
\endisatagdocument
{\isafolddocument}%
\isadelimdocument
\endisadelimdocument
\begin{isamarkuptext}%
We now propose a notion of concrete states that greatly differs
  from the original paper. A state is considered concrete iff it is wellformed 
  and contains no symbolic variables. However, the original paper additionally 
  assumes states to be implicitly simplified, which cannot be reasonably 
  modeled~in~Isabelle. \par
  In order to solve this problem, we desire to provide an explicit condition that requires 
  concrete states to be fully simplified. Note that we have already introduced concrete 
  expressions as fully simplified non-symbolic variable-free expressions. Consequently, 
  the concreteness notion of the original paper is equivalent to confirming that all 
  domain variables map to concrete expressions. This design choice ensures a feasible
  formalization of concreteness, whilst keeping the property~simple.%
\end{isamarkuptext}\isamarkuptrue%
\ \ \isacommand{definition}\isamarkupfalse%
\isanewline
\ \ \ \ concrete\isactrlsub {\isasymSigma}\ {\isacharcolon}{\kern0pt}{\isacharcolon}{\kern0pt}\ {\isachardoublequoteopen}{\isasymSigma}\ {\isasymRightarrow}\ bool{\isachardoublequoteclose}\ \isakeyword{where}\isanewline
\ \ \ \ {\isachardoublequoteopen}concrete\isactrlsub {\isasymSigma}\ {\isasymsigma}\ {\isasymequiv}\ {\isasymforall}v\ {\isasymin}\ fmdom{\isacharprime}{\kern0pt}{\isacharparenleft}{\kern0pt}{\isasymsigma}{\isacharparenright}{\kern0pt}{\isachardot}{\kern0pt}\ concrete\isactrlsub S{\isacharparenleft}{\kern0pt}the{\isacharparenleft}{\kern0pt}fmlookup\ {\isasymsigma}\ v{\isacharparenright}{\kern0pt}{\isacharparenright}{\kern0pt}{\isachardoublequoteclose}%
\begin{isamarkuptext}%
We can now establish that the concreteness property of a 
  state is preserved if it is updated with a concrete arithmetic expression. This
  ensures a speedup of automatic proofs as it is faster to show the concreteness 
  of a singular arithmetic expression instead of reproofing the concreteness of 
  the whole state after each update. Thus, we will later prefer the usage of derived 
  lemmas instead of the original concreteness definition when trying to prove that the
  concreteness of a state has been preserved.%
\end{isamarkuptext}\isamarkuptrue%
\ \ \isacommand{lemma}\isamarkupfalse%
\ concrete{\isacharunderscore}{\kern0pt}upd{\isacharunderscore}{\kern0pt}pr\isactrlsub S{\isacharcolon}{\kern0pt}\ {\isachardoublequoteopen}concrete\isactrlsub {\isasymSigma}{\isacharparenleft}{\kern0pt}{\isasymsigma}{\isacharparenright}{\kern0pt}\ {\isasymand}\ concrete\isactrlsub S{\isacharparenleft}{\kern0pt}e{\isacharparenright}{\kern0pt}\ {\isasymlongrightarrow}\ concrete\isactrlsub {\isasymSigma}{\isacharparenleft}{\kern0pt}{\isacharbrackleft}{\kern0pt}x\ {\isasymlongmapsto}\ e{\isacharbrackright}{\kern0pt}\ {\isasymsigma}{\isacharparenright}{\kern0pt}{\isachardoublequoteclose}\isanewline
\isadelimproof
\ \ \ \ %
\endisadelimproof
\isatagproof
\isacommand{by}\isamarkupfalse%
\ {\isacharparenleft}{\kern0pt}simp\ add{\isacharcolon}{\kern0pt}\ concrete\isactrlsub {\isasymSigma}{\isacharunderscore}{\kern0pt}def{\isacharparenright}{\kern0pt}%
\endisatagproof
{\isafoldproof}%
\isadelimproof
\isanewline
\endisadelimproof
\isanewline
\ \ \isacommand{lemma}\isamarkupfalse%
\ concrete{\isacharunderscore}{\kern0pt}upd{\isacharunderscore}{\kern0pt}pr\isactrlsub A{\isacharcolon}{\kern0pt}\ {\isachardoublequoteopen}concrete\isactrlsub {\isasymSigma}\ {\isasymsigma}\ {\isasymand}\ concrete\isactrlsub A\ a\ {\isasymlongrightarrow}\ concrete\isactrlsub {\isasymSigma}{\isacharparenleft}{\kern0pt}{\isacharbrackleft}{\kern0pt}x\ {\isasymlongmapsto}\ Exp\ a{\isacharbrackright}{\kern0pt}\ {\isasymsigma}{\isacharparenright}{\kern0pt}{\isachardoublequoteclose}\ \isanewline
\isadelimproof
\ \ \ \ %
\endisadelimproof
\isatagproof
\isacommand{using}\isamarkupfalse%
\ concrete{\isacharunderscore}{\kern0pt}upd{\isacharunderscore}{\kern0pt}pr\isactrlsub S\ \isacommand{by}\isamarkupfalse%
\ simp%
\endisatagproof
{\isafoldproof}%
\isadelimproof
\endisadelimproof
\begin{isamarkuptext}%
In the following lemmas we can now establish a connection between
  wellformed states and concrete states. We trivially prove that a concrete
  state has no symbolic variables. Additionally, we deduce that concreteness
  implies wellformedness, but not vice versa.%
\end{isamarkuptext}\isamarkuptrue%
\ \ \isacommand{lemma}\isamarkupfalse%
\ concrete{\isacharunderscore}{\kern0pt}symb{\isacharunderscore}{\kern0pt}imp\isactrlsub {\isasymSigma}{\isacharcolon}{\kern0pt}\ {\isachardoublequoteopen}concrete\isactrlsub {\isasymSigma}{\isacharparenleft}{\kern0pt}{\isasymsigma}{\isacharparenright}{\kern0pt}\ {\isasymlongrightarrow}\ symb\isactrlsub {\isasymSigma}{\isacharparenleft}{\kern0pt}{\isasymsigma}{\isacharparenright}{\kern0pt}\ {\isacharequal}{\kern0pt}\ {\isacharbraceleft}{\kern0pt}{\isacharbraceright}{\kern0pt}{\isachardoublequoteclose}\isanewline
\isadelimproof
\ \ \ \ %
\endisadelimproof
\isatagproof
\isacommand{by}\isamarkupfalse%
\ {\isacharparenleft}{\kern0pt}auto\ simp\ add{\isacharcolon}{\kern0pt}\ symb\isactrlsub {\isasymSigma}{\isacharunderscore}{\kern0pt}def\ concrete\isactrlsub {\isasymSigma}{\isacharunderscore}{\kern0pt}def{\isacharparenright}{\kern0pt}%
\endisatagproof
{\isafoldproof}%
\isadelimproof
\ \isanewline
\endisadelimproof
\isanewline
\ \ \isacommand{lemma}\isamarkupfalse%
\ concrete{\isacharunderscore}{\kern0pt}wf{\isacharunderscore}{\kern0pt}imp\isactrlsub {\isasymSigma}{\isacharcolon}{\kern0pt}\ {\isachardoublequoteopen}concrete\isactrlsub {\isasymSigma}{\isacharparenleft}{\kern0pt}{\isasymsigma}{\isacharparenright}{\kern0pt}\ {\isasymlongrightarrow}\ wf\isactrlsub {\isasymSigma}{\isacharparenleft}{\kern0pt}{\isasymsigma}{\isacharparenright}{\kern0pt}{\isachardoublequoteclose}\isanewline
\isadelimproof
\ \ \ \ %
\endisadelimproof
\isatagproof
\isacommand{using}\isamarkupfalse%
\ concrete{\isacharunderscore}{\kern0pt}vars{\isacharunderscore}{\kern0pt}imp\isactrlsub S\ \isacommand{by}\isamarkupfalse%
\ {\isacharparenleft}{\kern0pt}simp\ add{\isacharcolon}{\kern0pt}\ wf\isactrlsub {\isasymSigma}{\isacharunderscore}{\kern0pt}def\ concrete\isactrlsub {\isasymSigma}{\isacharunderscore}{\kern0pt}def{\isacharparenright}{\kern0pt}%
\endisatagproof
{\isafoldproof}%
\isadelimproof
\isanewline
\endisadelimproof
\isanewline
\ \ \isacommand{lemma}\isamarkupfalse%
\ {\isachardoublequoteopen}{\isasymexists}{\isasymsigma}{\isachardot}{\kern0pt}\ {\isasymnot}{\isacharparenleft}{\kern0pt}wf\isactrlsub {\isasymSigma}{\isacharparenleft}{\kern0pt}{\isasymsigma}{\isacharparenright}{\kern0pt}\ {\isasymlongrightarrow}\ concrete\isactrlsub {\isasymSigma}{\isacharparenleft}{\kern0pt}{\isasymsigma}{\isacharparenright}{\kern0pt}{\isacharparenright}{\kern0pt}{\isachardoublequoteclose}\isanewline
\isadelimproof
\ \ \ \ %
\endisadelimproof
\isatagproof
\isacommand{apply}\isamarkupfalse%
\ {\isacharparenleft}{\kern0pt}rule{\isacharunderscore}{\kern0pt}tac\ x\ {\isacharequal}{\kern0pt}\ {\isachardoublequoteopen}{\isacharbrackleft}{\kern0pt}{\isacharprime}{\kern0pt}{\isacharprime}{\kern0pt}x{\isadigit{0}}{\isacharprime}{\kern0pt}{\isacharprime}{\kern0pt}\ {\isasymlongmapsto}\ \isactrlemph {\isacharbrackright}{\kern0pt}\ {\isasymcircle}{\isachardoublequoteclose}\ \isakeyword{in}\ exI{\isacharparenright}{\kern0pt}\isanewline
\ \ \ \ \isacommand{by}\isamarkupfalse%
\ {\isacharparenleft}{\kern0pt}simp\ add{\isacharcolon}{\kern0pt}\ wf\isactrlsub {\isasymSigma}{\isacharunderscore}{\kern0pt}def\ concrete\isactrlsub {\isasymSigma}{\isacharunderscore}{\kern0pt}def{\isacharparenright}{\kern0pt}%
\endisatagproof
{\isafoldproof}%
\isadelimproof
\endisadelimproof
\begin{isamarkuptext}%
Let us take another look at our example states. \isa{{\isasymsigma}\isactrlsub {\isadigit{1}}} is not concrete, as 
  x1 and x2 do not map to concrete expressions. However, \isa{{\isasymsigma}\isactrlsub {\isadigit{2}}} is concrete, as
  8 and 2 are both (concrete) arithmetic numerals. We prove this reasoning with a 
  corresponding Isabelle~lemma.%
\end{isamarkuptext}\isamarkuptrue%
\ \ \isacommand{lemma}\isamarkupfalse%
\ {\isachardoublequoteopen}{\isasymnot}concrete\isactrlsub {\isasymSigma}\ {\isasymsigma}\isactrlsub {\isadigit{1}}\ {\isasymand}\ concrete\isactrlsub {\isasymSigma}\ {\isasymsigma}\isactrlsub {\isadigit{2}}{\isachardoublequoteclose}\ \isanewline
\isadelimproof
\ \ \ \ %
\endisadelimproof
\isatagproof
\isacommand{by}\isamarkupfalse%
\ {\isacharparenleft}{\kern0pt}simp\ add{\isacharcolon}{\kern0pt}\ {\isasymsigma}\isactrlsub {\isadigit{1}}{\isacharunderscore}{\kern0pt}def\ {\isasymsigma}\isactrlsub {\isadigit{2}}{\isacharunderscore}{\kern0pt}def\ concrete\isactrlsub {\isasymSigma}{\isacharunderscore}{\kern0pt}def{\isacharparenright}{\kern0pt}%
\endisatagproof
{\isafoldproof}%
\isadelimproof
\endisadelimproof
\isadelimdocument
\endisadelimdocument
\isatagdocument
\isamarkupsubsubsection{Variable Generation%
}
\isamarkuptrue%
\endisatagdocument
{\isafolddocument}%
\isadelimdocument
\endisadelimdocument
\begin{isamarkuptext}%
Let a fresh variable denote a variable that does not occur in the domain 
  of a provided state. We now setup a variable generation that deterministically
  returns a fresh variable for any given state. The design choice of enforcing
  determinism ensures an easier handling of future definitions and proofs without 
  loss of generality. It also guarantees an automatic generation of executable 
  code for the variable generator, which would have not have been easily feasible 
  without imposing~determinism. \par
  Our core idea for the variable generation is repeatedly attaching the
  character $c$ in front of a given variable until we eventually find a variable name, 
  which is not an element of the given state domain. This approach is easy and
  fast, but can possibly result in very long variable names. However, this is not
  a big problem for most programs, as reusing the same variable name more
  than a few times is extremely uncommon. \par
  We first provide a recursive helper function that attaches the character $c$ 
  n-times in front of a given variable, before finally returning~it.%
\end{isamarkuptext}\isamarkuptrue%
\ \ \isacommand{primrec}\isamarkupfalse%
\isanewline
\ \ \ \ vargen\isactrlsub N\ {\isacharcolon}{\kern0pt}{\isacharcolon}{\kern0pt}\ {\isachardoublequoteopen}nat\ {\isasymRightarrow}\ var\ {\isasymRightarrow}\ var{\isachardoublequoteclose}\ \isakeyword{where}\isanewline
\ \ \ \ {\isachardoublequoteopen}vargen\isactrlsub N\ {\isadigit{0}}\ v\ {\isacharequal}{\kern0pt}\ v{\isachardoublequoteclose}\ {\isacharbar}{\kern0pt}\isanewline
\ \ \ \ {\isachardoublequoteopen}vargen\isactrlsub N\ {\isacharparenleft}{\kern0pt}Suc\ n{\isacharparenright}{\kern0pt}\ v\ {\isacharequal}{\kern0pt}\ vargen\isactrlsub N\ n\ {\isacharparenleft}{\kern0pt}CHR\ {\isacharprime}{\kern0pt}{\isacharprime}{\kern0pt}c{\isacharprime}{\kern0pt}{\isacharprime}{\kern0pt}\ {\isacharhash}{\kern0pt}\ v{\isacharparenright}{\kern0pt}{\isachardoublequoteclose}%
\begin{isamarkuptext}%
Using the prepared helper function, we can now define the variable generation 
  selecting a fresh variable for any given state. The generator is given
  a state \isa{{\isasymsigma}}, a variable length n, a maximum bound b and a variable v. It then
  checks the freshness of variable v prepended with n characters. If the variable
  is fresh, we can return it, otherwise we recursively call our variable generation
  with an increased variable length and a reduced bound. If our bound becomes 0, we
  return a standardized variable implying that something~went~wrong. \par
  The notion of a maximal bound ensures that the variable generation terminates
  in finite time. However, this design choice also violates the freshness property of the
  chosen variable, considering that the returned variable is standardized when the bound 
  is exceeded (i.e. it may map to a variable, which already occurs in the state domain). 
  Hence, we have to accept that the generated variable is only fresh as long 
  as the corresponding bound is not exceeded during a program execution. This is 
  not a problem however, as long as a sufficiently high bound~is~provided.%
\end{isamarkuptext}\isamarkuptrue%
\ \ \isacommand{primrec}\isamarkupfalse%
\isanewline
\ \ \ \ vargen\ {\isacharcolon}{\kern0pt}{\isacharcolon}{\kern0pt}\ {\isachardoublequoteopen}{\isasymSigma}\ {\isasymRightarrow}\ nat\ {\isasymRightarrow}\ nat\ {\isasymRightarrow}\ var\ {\isasymRightarrow}\ var{\isachardoublequoteclose}\ \isakeyword{where}\isanewline
\ \ \ \ {\isachardoublequoteopen}vargen\ {\isasymsigma}\ n\ {\isadigit{0}}\ v\ {\isacharequal}{\kern0pt}\ {\isacharprime}{\kern0pt}{\isacharprime}{\kern0pt}{\isachardollar}{\kern0pt}BOUND{\isacharunderscore}{\kern0pt}EXCEEDED{\isacharcolon}{\kern0pt}{\isacharcolon}{\kern0pt}{\isacharprime}{\kern0pt}{\isacharprime}{\kern0pt}\ {\isacharat}{\kern0pt}\ v{\isachardoublequoteclose}\ {\isacharbar}{\kern0pt}\isanewline
\ \ \ \ {\isachardoublequoteopen}vargen\ {\isasymsigma}\ n\ {\isacharparenleft}{\kern0pt}Suc\ b{\isacharparenright}{\kern0pt}\ v\ {\isacharequal}{\kern0pt}\ {\isacharparenleft}{\kern0pt}if\ vargen\isactrlsub N\ n\ v\ {\isasymnotin}\ fmdom{\isacharprime}{\kern0pt}{\isacharparenleft}{\kern0pt}{\isasymsigma}{\isacharparenright}{\kern0pt}\ then\ vargen\isactrlsub N\ n\ v\ else\ {\isacharparenleft}{\kern0pt}vargen\ {\isasymsigma}\ {\isacharparenleft}{\kern0pt}Suc\ n{\isacharparenright}{\kern0pt}\ b\ v{\isacharparenright}{\kern0pt}{\isacharparenright}{\kern0pt}{\isachardoublequoteclose}%
\begin{isamarkuptext}%
We introduce the convention that the variable input of the
  variable generator must have the standardized form \isa{{\isachardollar}{\kern0pt}{\isacharless}{\kern0pt}var{\isachargreater}{\kern0pt}{\isacharless}{\kern0pt}sf{\isachargreater}{\kern0pt}},
  where \isa{var} matches the original variable and \isa{sf} corresponds to a suffix. 
  This later ensures an easier readability of generated variables.%
\end{isamarkuptext}\isamarkuptrue%
\isadelimdocument
\endisadelimdocument
\isatagdocument
\isamarkupsubsubsection{Initial States%
}
\isamarkuptrue%
\endisatagdocument
{\isafolddocument}%
\isadelimdocument
\endisadelimdocument
\begin{isamarkuptext}%
An initial state of a program is a state that maps all program variables
  onto the arithmetic numeral 0. We are now interested in the automatic construction 
  of such initial~states. \par 
  We first define a helper function that receives a list of program variables
  and constructs a corresponding list of (variable, starred expression) tuples,
  assigning each variable the initial numeral 0. In order for a traversion of all
  program variables to be feasible, the input of the function must be a list instead 
  of a set. This motivates the previously defined variable occurrence~functions.%
\end{isamarkuptext}\isamarkuptrue%
\ \ \isacommand{fun}\isamarkupfalse%
\isanewline
\ \ \ \ init\isactrlsub {\isasymSigma}\ {\isacharcolon}{\kern0pt}{\isacharcolon}{\kern0pt}\ {\isachardoublequoteopen}var\ list\ {\isasymRightarrow}\ {\isacharparenleft}{\kern0pt}var\ {\isacharasterisk}{\kern0pt}\ sexp{\isacharparenright}{\kern0pt}\ list{\isachardoublequoteclose}\ \isakeyword{where}\isanewline
\ \ \ \ {\isachardoublequoteopen}init\isactrlsub {\isasymSigma}\ {\isacharbrackleft}{\kern0pt}{\isacharbrackright}{\kern0pt}\ {\isacharequal}{\kern0pt}\ {\isacharbrackleft}{\kern0pt}{\isacharbrackright}{\kern0pt}{\isachardoublequoteclose}\ {\isacharbar}{\kern0pt}\isanewline
\ \ \ \ {\isachardoublequoteopen}init\isactrlsub {\isasymSigma}\ {\isacharparenleft}{\kern0pt}v\ {\isacharhash}{\kern0pt}\ rest{\isacharparenright}{\kern0pt}\ {\isacharequal}{\kern0pt}\ {\isacharparenleft}{\kern0pt}v{\isacharcomma}{\kern0pt}\ Exp\ {\isacharparenleft}{\kern0pt}Num\ {\isadigit{0}}{\isacharparenright}{\kern0pt}{\isacharparenright}{\kern0pt}\ {\isacharhash}{\kern0pt}\ init\isactrlsub {\isasymSigma}\ rest{\isachardoublequoteclose}%
\begin{isamarkuptext}%
It is now possible to construct an initial state from a given list
  of variables. In order to achieve this, we start by eliminating all duplicate 
  variables in the given list. We then use the init function to construct a tuple
  list assigning each variable the initial value, before finally transforming
  the list into a finite map. Note that the computation of the input variable list 
  depends on the overlying programming~language.%
\end{isamarkuptext}\isamarkuptrue%
\ \ \isacommand{fun}\isamarkupfalse%
\isanewline
\ \ \ \ get{\isacharunderscore}{\kern0pt}initial\isactrlsub {\isasymSigma}\ {\isacharcolon}{\kern0pt}{\isacharcolon}{\kern0pt}\ {\isachardoublequoteopen}var\ list\ {\isasymRightarrow}\ {\isasymSigma}{\isachardoublequoteclose}\ \isakeyword{where}\isanewline
\ \ \ \ {\isachardoublequoteopen}get{\isacharunderscore}{\kern0pt}initial\isactrlsub {\isasymSigma}\ vars\ {\isacharequal}{\kern0pt}\ fm\ {\isacharparenleft}{\kern0pt}init\isactrlsub {\isasymSigma}\ {\isacharparenleft}{\kern0pt}remdups\ vars{\isacharparenright}{\kern0pt}{\isacharparenright}{\kern0pt}{\isachardoublequoteclose}%
\isadelimdocument
\endisadelimdocument
\isatagdocument
\isamarkupsubsection{Evaluation%
}
\isamarkuptrue%
\isamarkupsubsubsection{Operator Interpretations%
}
\isamarkuptrue%
\endisatagdocument
{\isafolddocument}%
\isadelimdocument
\endisadelimdocument
\begin{isamarkuptext}%
In order to provide formal semantics for expressions, it is of  
  crucial importance to give faithful interpretations of the syntactic 
  operator~symbols.%
\end{isamarkuptext}\isamarkuptrue%
\ \ \isacommand{primrec}\isamarkupfalse%
\ \isanewline
\ \ \ \ valop\isactrlsub a\ {\isacharcolon}{\kern0pt}{\isacharcolon}{\kern0pt}\ {\isachardoublequoteopen}op\isactrlsub a\ {\isasymRightarrow}\ int\ {\isasymRightarrow}\ int\ {\isasymRightarrow}\ int{\isachardoublequoteclose}\ \isakeyword{where}\isanewline
\ \ \ \ {\isachardoublequoteopen}valop\isactrlsub a\ add\ x\ y\ {\isacharequal}{\kern0pt}\ x\ {\isacharplus}{\kern0pt}\ y{\isachardoublequoteclose}\ {\isacharbar}{\kern0pt}\isanewline
\ \ \ \ {\isachardoublequoteopen}valop\isactrlsub a\ sub\ x\ y\ {\isacharequal}{\kern0pt}\ x\ {\isacharminus}{\kern0pt}\ y{\isachardoublequoteclose}\ {\isacharbar}{\kern0pt}\isanewline
\ \ \ \ {\isachardoublequoteopen}valop\isactrlsub a\ mul\ x\ y\ {\isacharequal}{\kern0pt}\ x\ {\isacharasterisk}{\kern0pt}\ y{\isachardoublequoteclose}\isanewline
\isanewline
\ \ \isacommand{primrec}\isamarkupfalse%
\ \isanewline
\ \ \ \ valop\isactrlsub b\ {\isacharcolon}{\kern0pt}{\isacharcolon}{\kern0pt}\ {\isachardoublequoteopen}op\isactrlsub b\ {\isasymRightarrow}\ bool\ {\isasymRightarrow}\ bool\ {\isasymRightarrow}\ bool{\isachardoublequoteclose}\ \isakeyword{where}\isanewline
\ \ \ \ {\isachardoublequoteopen}valop\isactrlsub b\ conj\ x\ y\ {\isacharequal}{\kern0pt}\ {\isacharparenleft}{\kern0pt}x\ {\isasymand}\ y{\isacharparenright}{\kern0pt}{\isachardoublequoteclose}\ {\isacharbar}{\kern0pt}\isanewline
\ \ \ \ {\isachardoublequoteopen}valop\isactrlsub b\ disj\ x\ y\ {\isacharequal}{\kern0pt}\ {\isacharparenleft}{\kern0pt}x\ {\isasymor}\ y{\isacharparenright}{\kern0pt}{\isachardoublequoteclose}\ \isanewline
\isanewline
\ \ \isacommand{primrec}\isamarkupfalse%
\ \isanewline
\ \ \ \ valop\isactrlsub r\ {\isacharcolon}{\kern0pt}{\isacharcolon}{\kern0pt}\ {\isachardoublequoteopen}op\isactrlsub r\ {\isasymRightarrow}\ int\ {\isasymRightarrow}\ int\ {\isasymRightarrow}\ bool{\isachardoublequoteclose}\ \isakeyword{where}\isanewline
\ \ \ \ {\isachardoublequoteopen}valop\isactrlsub r\ leq\ x\ y\ {\isacharequal}{\kern0pt}\ {\isacharparenleft}{\kern0pt}x\ {\isasymle}\ y{\isacharparenright}{\kern0pt}{\isachardoublequoteclose}\ {\isacharbar}{\kern0pt}\isanewline
\ \ \ \ {\isachardoublequoteopen}valop\isactrlsub r\ geq\ x\ y\ {\isacharequal}{\kern0pt}\ {\isacharparenleft}{\kern0pt}x\ {\isasymge}\ y{\isacharparenright}{\kern0pt}{\isachardoublequoteclose}\ {\isacharbar}{\kern0pt}\isanewline
\ \ \ \ {\isachardoublequoteopen}valop\isactrlsub r\ eq\ x\ y\ {\isacharequal}{\kern0pt}\ {\isacharparenleft}{\kern0pt}x\ {\isacharequal}{\kern0pt}\ y{\isacharparenright}{\kern0pt}{\isachardoublequoteclose}%
\isadelimdocument
\endisadelimdocument
\isatagdocument
\isamarkupsubsubsection{Evaluation Functions%
}
\isamarkuptrue%
\endisatagdocument
{\isafolddocument}%
\isadelimdocument
\endisadelimdocument
\begin{isamarkuptext}%
Before formalizing the evaluation of expressions, we will first have to
  introduce several necessary helper functions. The first function maps an
  arithmetic expression onto its encased numeral while the second function 
  maps a Boolean expression onto its encased truth value. The mappings are
  undefined iff the arithmetic/Boolean expression is not of concrete nature,
  indicating that these are partial functions. This also implies that the
  functions should only be called if the arithmetic/Boolean expression is known
  to~be~concrete.%
\end{isamarkuptext}\isamarkuptrue%
\ \ \isacommand{abbreviation}\isamarkupfalse%
\ \isanewline
\ \ \ \ getNum\ {\isacharcolon}{\kern0pt}{\isacharcolon}{\kern0pt}\ {\isachardoublequoteopen}aexp\ {\isasymRightarrow}\ int{\isachardoublequoteclose}\ \isakeyword{where}\isanewline
\ \ \ \ {\isachardoublequoteopen}getNum\ a\ {\isasymequiv}\ {\isacharparenleft}{\kern0pt}case\ a\ of\ {\isacharparenleft}{\kern0pt}Num\ n{\isacharparenright}{\kern0pt}\ {\isasymRightarrow}\ n{\isacharparenright}{\kern0pt}{\isachardoublequoteclose}\isanewline
\isanewline
\ \ \isacommand{abbreviation}\isamarkupfalse%
\ \isanewline
\ \ \ \ getBool\ {\isacharcolon}{\kern0pt}{\isacharcolon}{\kern0pt}\ {\isachardoublequoteopen}bexp\ {\isasymRightarrow}\ bool{\isachardoublequoteclose}\ \isakeyword{where}\isanewline
\ \ \ \ {\isachardoublequoteopen}getBool\ b\ {\isasymequiv}\ {\isacharparenleft}{\kern0pt}case\ b\ of\ {\isacharparenleft}{\kern0pt}Bool\ v{\isacharparenright}{\kern0pt}\ {\isasymRightarrow}\ v{\isacharparenright}{\kern0pt}{\isachardoublequoteclose}%
\begin{isamarkuptext}%
We can now formalize the evaluation of expressions in a very similar fashion to
  the original paper. Whilst the given paper only needed one singular evaluation
  function, we need to define an evaluation function for each expression type due
  to our additional type constraints. In contrast to the paper, ill-typed expressions
  can not be derived via our grammar, hence ensuring more compact and easier
  readable semantics, as no type checks need to take place. \par
  The evaluation of a specific type of expression in a given state generally
  works~as~follows: \begin{description}
  \item[Primitives] Numerals, truth values and method names are already
  considered concrete, hence they are not simplified~any~further.
  \item[Variables] Variables are simplified by mapping them to their corresponding 
    arithmetic expression in the given state. If the state maps the variable to
    the symbolic value, the variable will not be simplified. As a variable can 
    only map to arithmetic expressions or the symbolic value, no further type 
    checks need to take place. Note that we, similar to the paper, assume all 
    occurring variables to be located in the state domain, otherwise the evaluation 
    function will be~undefined. 
  \item[Operations] Operations are simplified by evaluating them iff all 
    subexpressions can be simplified to a concrete expression. Otherwise both 
    subexpressions will be simplified as much as possible, whilst syntactically 
    preserving the~operation.
  \end{description} We also provide an evaluation function for lists of expressions 
  and sets of Boolean expressions, as these will later serve as useful~abbreviations.%
\end{isamarkuptext}\isamarkuptrue%
\ \ \isacommand{primrec}\isamarkupfalse%
\ \isanewline
\ \ \ \ val\isactrlsub A\ {\isacharcolon}{\kern0pt}{\isacharcolon}{\kern0pt}\ {\isachardoublequoteopen}aexp\ {\isasymRightarrow}\ {\isasymSigma}\ {\isasymRightarrow}\ aexp{\isachardoublequoteclose}\ \isakeyword{where}\isanewline
\ \ \ \ {\isachardoublequoteopen}val\isactrlsub A\ {\isacharparenleft}{\kern0pt}Num\ n{\isacharparenright}{\kern0pt}\ {\isasymsigma}\ {\isacharequal}{\kern0pt}\ Num\ n{\isachardoublequoteclose}\ {\isacharbar}{\kern0pt}\isanewline
\ \ \ \ {\isachardoublequoteopen}val\isactrlsub A\ {\isacharparenleft}{\kern0pt}Var\ x{\isacharparenright}{\kern0pt}\ {\isasymsigma}\ {\isacharequal}{\kern0pt}\ {\isacharparenleft}{\kern0pt}case\ fmlookup\ {\isasymsigma}\ x\ of\ Some\ \isactrlemph \ {\isasymRightarrow}\ Var\ x\ {\isacharbar}{\kern0pt}\ Some\ {\isacharparenleft}{\kern0pt}Exp\ e{\isacharparenright}{\kern0pt}\ {\isasymRightarrow}\ e{\isacharparenright}{\kern0pt}{\isachardoublequoteclose}\ {\isacharbar}{\kern0pt}\isanewline
\ \ \ \ {\isachardoublequoteopen}val\isactrlsub A\ {\isacharparenleft}{\kern0pt}Op\isactrlsub A\ a\isactrlsub {\isadigit{1}}\ op\ a\isactrlsub {\isadigit{2}}{\isacharparenright}{\kern0pt}\ {\isasymsigma}\ {\isacharequal}{\kern0pt}\ {\isacharparenleft}{\kern0pt}if\ concrete\isactrlsub A\ {\isacharparenleft}{\kern0pt}val\isactrlsub A\ a\isactrlsub {\isadigit{1}}\ {\isasymsigma}{\isacharparenright}{\kern0pt}\ {\isasymand}\ concrete\isactrlsub A\ {\isacharparenleft}{\kern0pt}val\isactrlsub A\ a\isactrlsub {\isadigit{2}}\ {\isasymsigma}{\isacharparenright}{\kern0pt}\ \isanewline
\ \ \ \ \ \ \ \ \ \ then\ Num\ {\isacharparenleft}{\kern0pt}valop\isactrlsub a\ op\ {\isacharparenleft}{\kern0pt}getNum\ {\isacharparenleft}{\kern0pt}val\isactrlsub A\ a\isactrlsub {\isadigit{1}}\ {\isasymsigma}{\isacharparenright}{\kern0pt}{\isacharparenright}{\kern0pt}\ {\isacharparenleft}{\kern0pt}getNum\ {\isacharparenleft}{\kern0pt}val\isactrlsub A\ a\isactrlsub {\isadigit{2}}\ {\isasymsigma}{\isacharparenright}{\kern0pt}{\isacharparenright}{\kern0pt}{\isacharparenright}{\kern0pt}\isanewline
\ \ \ \ \ \ \ \ \ \ else\ Op\isactrlsub A\ {\isacharparenleft}{\kern0pt}val\isactrlsub A\ a\isactrlsub {\isadigit{1}}\ {\isasymsigma}{\isacharparenright}{\kern0pt}\ op\ {\isacharparenleft}{\kern0pt}val\isactrlsub A\ a\isactrlsub {\isadigit{2}}\ {\isasymsigma}{\isacharparenright}{\kern0pt}{\isacharparenright}{\kern0pt}{\isachardoublequoteclose}\isanewline
\isanewline
\ \ \isacommand{primrec}\isamarkupfalse%
\ \isanewline
\ \ \ \ val\isactrlsub B\ {\isacharcolon}{\kern0pt}{\isacharcolon}{\kern0pt}\ {\isachardoublequoteopen}bexp\ {\isasymRightarrow}\ {\isasymSigma}\ {\isasymRightarrow}\ bexp{\isachardoublequoteclose}\ \isakeyword{where}\isanewline
\ \ \ \ {\isachardoublequoteopen}val\isactrlsub B\ {\isacharparenleft}{\kern0pt}Bool\ b{\isacharparenright}{\kern0pt}\ {\isasymsigma}\ {\isacharequal}{\kern0pt}\ Bool\ b{\isachardoublequoteclose}\ {\isacharbar}{\kern0pt}\isanewline
\ \ \ \ {\isachardoublequoteopen}val\isactrlsub B\ {\isacharparenleft}{\kern0pt}Not\ b{\isacharparenright}{\kern0pt}\ {\isasymsigma}\ {\isacharequal}{\kern0pt}\ {\isacharparenleft}{\kern0pt}if\ concrete\isactrlsub B\ {\isacharparenleft}{\kern0pt}val\isactrlsub B\ b\ {\isasymsigma}{\isacharparenright}{\kern0pt}\ then\ Bool\ {\isacharparenleft}{\kern0pt}{\isasymnot}{\isacharparenleft}{\kern0pt}getBool\ {\isacharparenleft}{\kern0pt}val\isactrlsub B\ b\ {\isasymsigma}{\isacharparenright}{\kern0pt}{\isacharparenright}{\kern0pt}{\isacharparenright}{\kern0pt}\ \isanewline
\ \ \ \ \ \ \ \ \ \ else\ {\isacharparenleft}{\kern0pt}Not\ {\isacharparenleft}{\kern0pt}val\isactrlsub B\ b\ {\isasymsigma}{\isacharparenright}{\kern0pt}{\isacharparenright}{\kern0pt}{\isacharparenright}{\kern0pt}{\isachardoublequoteclose}\ {\isacharbar}{\kern0pt}\isanewline
\ \ \ \ {\isachardoublequoteopen}val\isactrlsub B\ {\isacharparenleft}{\kern0pt}Op\isactrlsub B\ b\isactrlsub {\isadigit{1}}\ op\ b\isactrlsub {\isadigit{2}}{\isacharparenright}{\kern0pt}\ {\isasymsigma}\ {\isacharequal}{\kern0pt}\ {\isacharparenleft}{\kern0pt}if\ concrete\isactrlsub B\ {\isacharparenleft}{\kern0pt}val\isactrlsub B\ b\isactrlsub {\isadigit{1}}\ {\isasymsigma}{\isacharparenright}{\kern0pt}\ {\isasymand}\ concrete\isactrlsub B\ {\isacharparenleft}{\kern0pt}val\isactrlsub B\ b\isactrlsub {\isadigit{2}}\ {\isasymsigma}{\isacharparenright}{\kern0pt}\ \isanewline
\ \ \ \ \ \ \ \ \ \ then\ Bool\ {\isacharparenleft}{\kern0pt}valop\isactrlsub b\ op\ {\isacharparenleft}{\kern0pt}getBool\ {\isacharparenleft}{\kern0pt}val\isactrlsub B\ b\isactrlsub {\isadigit{1}}\ {\isasymsigma}{\isacharparenright}{\kern0pt}{\isacharparenright}{\kern0pt}\ {\isacharparenleft}{\kern0pt}getBool\ {\isacharparenleft}{\kern0pt}val\isactrlsub B\ b\isactrlsub {\isadigit{2}}\ {\isasymsigma}{\isacharparenright}{\kern0pt}{\isacharparenright}{\kern0pt}{\isacharparenright}{\kern0pt}\isanewline
\ \ \ \ \ \ \ \ \ \ else\ Op\isactrlsub B\ {\isacharparenleft}{\kern0pt}val\isactrlsub B\ b\isactrlsub {\isadigit{1}}\ {\isasymsigma}{\isacharparenright}{\kern0pt}\ op\ {\isacharparenleft}{\kern0pt}val\isactrlsub B\ b\isactrlsub {\isadigit{2}}\ {\isasymsigma}{\isacharparenright}{\kern0pt}{\isacharparenright}{\kern0pt}{\isachardoublequoteclose}\ {\isacharbar}{\kern0pt}\isanewline
\ \ \ \ {\isachardoublequoteopen}val\isactrlsub B\ {\isacharparenleft}{\kern0pt}Op\isactrlsub R\ a\isactrlsub {\isadigit{1}}\ op\ a\isactrlsub {\isadigit{2}}{\isacharparenright}{\kern0pt}\ {\isasymsigma}\ {\isacharequal}{\kern0pt}\ {\isacharparenleft}{\kern0pt}if\ concrete\isactrlsub A\ {\isacharparenleft}{\kern0pt}val\isactrlsub A\ a\isactrlsub {\isadigit{1}}\ {\isasymsigma}{\isacharparenright}{\kern0pt}\ {\isasymand}\ concrete\isactrlsub A\ {\isacharparenleft}{\kern0pt}val\isactrlsub A\ a\isactrlsub {\isadigit{2}}\ {\isasymsigma}{\isacharparenright}{\kern0pt}\ \isanewline
\ \ \ \ \ \ \ \ \ \ then\ Bool\ {\isacharparenleft}{\kern0pt}valop\isactrlsub r\ op\ {\isacharparenleft}{\kern0pt}getNum\ {\isacharparenleft}{\kern0pt}val\isactrlsub A\ a\isactrlsub {\isadigit{1}}\ {\isasymsigma}{\isacharparenright}{\kern0pt}{\isacharparenright}{\kern0pt}\ {\isacharparenleft}{\kern0pt}getNum\ {\isacharparenleft}{\kern0pt}val\isactrlsub A\ a\isactrlsub {\isadigit{2}}\ {\isasymsigma}{\isacharparenright}{\kern0pt}{\isacharparenright}{\kern0pt}{\isacharparenright}{\kern0pt}\isanewline
\ \ \ \ \ \ \ \ \ \ else\ Op\isactrlsub R\ {\isacharparenleft}{\kern0pt}val\isactrlsub A\ a\isactrlsub {\isadigit{1}}\ {\isasymsigma}{\isacharparenright}{\kern0pt}\ op\ {\isacharparenleft}{\kern0pt}val\isactrlsub A\ a\isactrlsub {\isadigit{2}}\ {\isasymsigma}{\isacharparenright}{\kern0pt}{\isacharparenright}{\kern0pt}{\isachardoublequoteclose}\isanewline
\isanewline
\ \ \isacommand{primrec}\isamarkupfalse%
\ \isanewline
\ \ \ \ val\isactrlsub E\ {\isacharcolon}{\kern0pt}{\isacharcolon}{\kern0pt}\ {\isachardoublequoteopen}exp\ {\isasymRightarrow}\ {\isasymSigma}\ {\isasymRightarrow}\ exp{\isachardoublequoteclose}\ \isakeyword{where}\isanewline
\ \ \ \ {\isachardoublequoteopen}val\isactrlsub E\ {\isacharparenleft}{\kern0pt}A\ a{\isacharparenright}{\kern0pt}\ {\isasymsigma}\ {\isacharequal}{\kern0pt}\ A\ {\isacharparenleft}{\kern0pt}val\isactrlsub A\ a\ {\isasymsigma}{\isacharparenright}{\kern0pt}{\isachardoublequoteclose}\ {\isacharbar}{\kern0pt}\isanewline
\ \ \ \ {\isachardoublequoteopen}val\isactrlsub E\ {\isacharparenleft}{\kern0pt}B\ b{\isacharparenright}{\kern0pt}\ {\isasymsigma}\ {\isacharequal}{\kern0pt}\ B\ {\isacharparenleft}{\kern0pt}val\isactrlsub B\ b\ {\isasymsigma}{\isacharparenright}{\kern0pt}{\isachardoublequoteclose}\ {\isacharbar}{\kern0pt}\isanewline
\ \ \ \ {\isachardoublequoteopen}val\isactrlsub E\ {\isacharparenleft}{\kern0pt}P\ m{\isacharparenright}{\kern0pt}\ {\isasymsigma}\ {\isacharequal}{\kern0pt}\ P\ m{\isachardoublequoteclose}\isanewline
\isanewline
\ \ \isacommand{primrec}\isamarkupfalse%
\ \isanewline
\ \ \ \ val\isactrlsub S\ {\isacharcolon}{\kern0pt}{\isacharcolon}{\kern0pt}\ {\isachardoublequoteopen}sexp\ {\isasymRightarrow}\ {\isasymSigma}\ {\isasymRightarrow}\ sexp{\isachardoublequoteclose}\ \isakeyword{where}\isanewline
\ \ \ \ {\isachardoublequoteopen}val\isactrlsub S\ {\isacharparenleft}{\kern0pt}Exp\ a{\isacharparenright}{\kern0pt}\ {\isasymsigma}\ {\isacharequal}{\kern0pt}\ Exp\ {\isacharparenleft}{\kern0pt}val\isactrlsub A\ a\ {\isasymsigma}{\isacharparenright}{\kern0pt}{\isachardoublequoteclose}\ {\isacharbar}{\kern0pt}\isanewline
\ \ \ \ {\isachardoublequoteopen}val\isactrlsub S\ \isactrlemph \ {\isasymsigma}\ {\isacharequal}{\kern0pt}\ \isactrlemph {\isachardoublequoteclose}\ \isanewline
\isanewline
\ \ \isacommand{primrec}\isamarkupfalse%
\ \isanewline
\ \ \ \ lval\isactrlsub E\ {\isacharcolon}{\kern0pt}{\isacharcolon}{\kern0pt}\ {\isachardoublequoteopen}exp\ list\ {\isasymRightarrow}\ {\isasymSigma}\ {\isasymRightarrow}\ exp\ list{\isachardoublequoteclose}\ \isakeyword{where}\isanewline
\ \ \ \ {\isachardoublequoteopen}lval\isactrlsub E\ {\isacharbrackleft}{\kern0pt}{\isacharbrackright}{\kern0pt}\ {\isasymsigma}\ {\isacharequal}{\kern0pt}\ {\isacharbrackleft}{\kern0pt}{\isacharbrackright}{\kern0pt}{\isachardoublequoteclose}\ {\isacharbar}{\kern0pt}\isanewline
\ \ \ \ {\isachardoublequoteopen}lval\isactrlsub E\ {\isacharparenleft}{\kern0pt}exp\ {\isacharhash}{\kern0pt}\ rest{\isacharparenright}{\kern0pt}\ {\isasymsigma}\ {\isacharequal}{\kern0pt}\ {\isacharparenleft}{\kern0pt}val\isactrlsub E\ exp\ {\isasymsigma}{\isacharparenright}{\kern0pt}\ {\isacharhash}{\kern0pt}\ {\isacharparenleft}{\kern0pt}lval\isactrlsub E\ rest\ {\isasymsigma}{\isacharparenright}{\kern0pt}{\isachardoublequoteclose}\isanewline
\isanewline
\ \ \isacommand{fun}\isamarkupfalse%
\isanewline
\ \ \ \ sval\isactrlsub B\ {\isacharcolon}{\kern0pt}{\isacharcolon}{\kern0pt}\ {\isachardoublequoteopen}bexp\ set\ {\isasymRightarrow}\ {\isasymSigma}\ {\isasymRightarrow}\ bexp\ set{\isachardoublequoteclose}\ \isakeyword{where}\isanewline
\ \ \ \ {\isachardoublequoteopen}sval\isactrlsub B\ S\ {\isasymsigma}\ {\isacharequal}{\kern0pt}\ {\isacharparenleft}{\kern0pt}{\isacharpercent}{\kern0pt}e{\isachardot}{\kern0pt}\ {\isacharparenleft}{\kern0pt}val\isactrlsub B\ e\ {\isasymsigma}{\isacharparenright}{\kern0pt}{\isacharparenright}{\kern0pt}\ {\isacharbackquote}{\kern0pt}\ S{\isachardoublequoteclose}%
\begin{isamarkuptext}%
The following proofs establish that a concrete expression preserves
  its concreteness when evaluated under an arbitrary state. This is obvious,
  considering that a concrete expression cannot be simplified any further.
  The proofs are trivial by making use of structural induction over
  the construction~of~expressions.%
\end{isamarkuptext}\isamarkuptrue%
\ \ \isacommand{lemma}\isamarkupfalse%
\ concrete{\isacharunderscore}{\kern0pt}pr\isactrlsub A{\isacharcolon}{\kern0pt}\ {\isachardoublequoteopen}concrete\isactrlsub A{\isacharparenleft}{\kern0pt}a{\isacharparenright}{\kern0pt}\ {\isasymlongrightarrow}\ concrete\isactrlsub A{\isacharparenleft}{\kern0pt}val\isactrlsub A\ a\ {\isasymsigma}{\isacharparenright}{\kern0pt}{\isachardoublequoteclose}\ \isanewline
\isadelimproof
\ \ \ \ %
\endisadelimproof
\isatagproof
\isacommand{by}\isamarkupfalse%
\ {\isacharparenleft}{\kern0pt}induct\ a{\isacharcomma}{\kern0pt}\ auto{\isacharparenright}{\kern0pt}%
\endisatagproof
{\isafoldproof}%
\isadelimproof
\isanewline
\endisadelimproof
\isanewline
\ \ \isacommand{lemma}\isamarkupfalse%
\ concrete{\isacharunderscore}{\kern0pt}pr\isactrlsub B{\isacharcolon}{\kern0pt}\ {\isachardoublequoteopen}concrete\isactrlsub B{\isacharparenleft}{\kern0pt}b{\isacharparenright}{\kern0pt}\ {\isasymlongrightarrow}\ concrete\isactrlsub B{\isacharparenleft}{\kern0pt}val\isactrlsub B\ b\ {\isasymsigma}{\isacharparenright}{\kern0pt}{\isachardoublequoteclose}\ \isanewline
\isadelimproof
\ \ \ \ %
\endisadelimproof
\isatagproof
\isacommand{by}\isamarkupfalse%
\ {\isacharparenleft}{\kern0pt}induct\ b{\isacharcomma}{\kern0pt}\ auto{\isacharparenright}{\kern0pt}%
\endisatagproof
{\isafoldproof}%
\isadelimproof
\isanewline
\endisadelimproof
\isanewline
\ \ \isacommand{lemma}\isamarkupfalse%
\ concrete{\isacharunderscore}{\kern0pt}pr\isactrlsub E{\isacharcolon}{\kern0pt}\ {\isachardoublequoteopen}concrete\isactrlsub E{\isacharparenleft}{\kern0pt}e{\isacharparenright}{\kern0pt}\ {\isasymlongrightarrow}\ concrete\isactrlsub E{\isacharparenleft}{\kern0pt}val\isactrlsub E\ e\ {\isasymsigma}{\isacharparenright}{\kern0pt}{\isachardoublequoteclose}\isanewline
\isadelimproof
\ \ \ \ %
\endisadelimproof
\isatagproof
\isacommand{using}\isamarkupfalse%
\ concrete{\isacharunderscore}{\kern0pt}pr\isactrlsub A\ concrete{\isacharunderscore}{\kern0pt}pr\isactrlsub B\ \isacommand{by}\isamarkupfalse%
\ {\isacharparenleft}{\kern0pt}induct\ e{\isacharcomma}{\kern0pt}\ auto{\isacharparenright}{\kern0pt}%
\endisatagproof
{\isafoldproof}%
\isadelimproof
\isanewline
\endisadelimproof
\isanewline
\ \ \isacommand{lemma}\isamarkupfalse%
\ concrete{\isacharunderscore}{\kern0pt}pr\isactrlsub S{\isacharcolon}{\kern0pt}\ {\isachardoublequoteopen}concrete\isactrlsub S{\isacharparenleft}{\kern0pt}s{\isacharparenright}{\kern0pt}\ {\isasymlongrightarrow}\ concrete\isactrlsub S{\isacharparenleft}{\kern0pt}val\isactrlsub S\ s\ {\isasymsigma}{\isacharparenright}{\kern0pt}{\isachardoublequoteclose}\isanewline
\isadelimproof
\ \ \ \ %
\endisadelimproof
\isatagproof
\isacommand{using}\isamarkupfalse%
\ concrete{\isacharunderscore}{\kern0pt}pr\isactrlsub A\ \isacommand{by}\isamarkupfalse%
\ {\isacharparenleft}{\kern0pt}induct\ s{\isacharcomma}{\kern0pt}\ auto{\isacharparenright}{\kern0pt}%
\endisatagproof
{\isafoldproof}%
\isadelimproof
\isanewline
\endisadelimproof
\isanewline
\ \ \isacommand{lemma}\isamarkupfalse%
\ l{\isacharunderscore}{\kern0pt}concrete{\isacharunderscore}{\kern0pt}pr\isactrlsub E{\isacharcolon}{\kern0pt}\ {\isachardoublequoteopen}lconcrete\isactrlsub E{\isacharparenleft}{\kern0pt}l{\isacharparenright}{\kern0pt}\ {\isasymlongrightarrow}\ lconcrete\isactrlsub E{\isacharparenleft}{\kern0pt}lval\isactrlsub E\ l\ {\isasymsigma}{\isacharparenright}{\kern0pt}{\isachardoublequoteclose}\isanewline
\isadelimproof
\ \ \ \ %
\endisadelimproof
\isatagproof
\isacommand{using}\isamarkupfalse%
\ concrete{\isacharunderscore}{\kern0pt}pr\isactrlsub E\ \isacommand{by}\isamarkupfalse%
\ {\isacharparenleft}{\kern0pt}induct\ l{\isacharcomma}{\kern0pt}\ auto{\isacharparenright}{\kern0pt}%
\endisatagproof
{\isafoldproof}%
\isadelimproof
\isanewline
\endisadelimproof
\isanewline
\ \ \isacommand{lemma}\isamarkupfalse%
\ s{\isacharunderscore}{\kern0pt}concrete{\isacharunderscore}{\kern0pt}pr\isactrlsub B{\isacharcolon}{\kern0pt}\ {\isachardoublequoteopen}sconcrete\isactrlsub B{\isacharparenleft}{\kern0pt}S{\isacharparenright}{\kern0pt}\ {\isasymlongrightarrow}\ sconcrete\isactrlsub B{\isacharparenleft}{\kern0pt}sval\isactrlsub B\ S\ {\isasymsigma}{\isacharparenright}{\kern0pt}{\isachardoublequoteclose}\isanewline
\isadelimproof
\ \ \ \ %
\endisadelimproof
\isatagproof
\isacommand{using}\isamarkupfalse%
\ concrete{\isacharunderscore}{\kern0pt}pr\isactrlsub B\ \isacommand{by}\isamarkupfalse%
\ simp%
\endisatagproof
{\isafoldproof}%
\isadelimproof
\endisadelimproof
\begin{isamarkuptext}%
We now infer that the value of a concrete expression does not change 
  after an evaluation under an arbitrary state. Similar to the above proof,
  we use structural induction to reason that a concrete expression cannot
  be simplified~any~further.%
\end{isamarkuptext}\isamarkuptrue%
\ \ \isacommand{lemma}\isamarkupfalse%
\ value{\isacharunderscore}{\kern0pt}pr\isactrlsub A{\isacharcolon}{\kern0pt}\ {\isachardoublequoteopen}concrete\isactrlsub A{\isacharparenleft}{\kern0pt}a{\isacharparenright}{\kern0pt}\ {\isasymlongrightarrow}\ val\isactrlsub A\ a\ {\isasymsigma}\ {\isacharequal}{\kern0pt}\ a{\isachardoublequoteclose}\isanewline
\isadelimproof
\ \ \ \ %
\endisadelimproof
\isatagproof
\isacommand{by}\isamarkupfalse%
\ {\isacharparenleft}{\kern0pt}induct\ a{\isacharsemicolon}{\kern0pt}\ auto{\isacharparenright}{\kern0pt}%
\endisatagproof
{\isafoldproof}%
\isadelimproof
\isanewline
\endisadelimproof
\isanewline
\ \ \isacommand{lemma}\isamarkupfalse%
\ value{\isacharunderscore}{\kern0pt}pr\isactrlsub B{\isacharcolon}{\kern0pt}\ {\isachardoublequoteopen}concrete\isactrlsub B{\isacharparenleft}{\kern0pt}b{\isacharparenright}{\kern0pt}\ {\isasymlongrightarrow}\ val\isactrlsub B\ b\ {\isasymsigma}\ {\isacharequal}{\kern0pt}\ b{\isachardoublequoteclose}\isanewline
\isadelimproof
\ \ \ \ %
\endisadelimproof
\isatagproof
\isacommand{by}\isamarkupfalse%
\ {\isacharparenleft}{\kern0pt}induct\ b{\isacharsemicolon}{\kern0pt}\ auto{\isacharparenright}{\kern0pt}%
\endisatagproof
{\isafoldproof}%
\isadelimproof
\isanewline
\endisadelimproof
\isanewline
\ \ \isacommand{lemma}\isamarkupfalse%
\ value{\isacharunderscore}{\kern0pt}pr\isactrlsub E{\isacharcolon}{\kern0pt}\ {\isachardoublequoteopen}concrete\isactrlsub E{\isacharparenleft}{\kern0pt}e{\isacharparenright}{\kern0pt}\ {\isasymlongrightarrow}\ val\isactrlsub E\ e\ {\isasymsigma}\ {\isacharequal}{\kern0pt}\ e{\isachardoublequoteclose}\isanewline
\isadelimproof
\ \ \ \ %
\endisadelimproof
\isatagproof
\isacommand{using}\isamarkupfalse%
\ value{\isacharunderscore}{\kern0pt}pr\isactrlsub A\ value{\isacharunderscore}{\kern0pt}pr\isactrlsub B\ \isacommand{by}\isamarkupfalse%
\ {\isacharparenleft}{\kern0pt}induct\ e{\isacharsemicolon}{\kern0pt}\ auto{\isacharparenright}{\kern0pt}%
\endisatagproof
{\isafoldproof}%
\isadelimproof
\isanewline
\endisadelimproof
\isanewline
\ \ \isacommand{lemma}\isamarkupfalse%
\ value{\isacharunderscore}{\kern0pt}pr\isactrlsub S{\isacharcolon}{\kern0pt}\ {\isachardoublequoteopen}concrete\isactrlsub S{\isacharparenleft}{\kern0pt}s{\isacharparenright}{\kern0pt}\ {\isasymlongrightarrow}\ val\isactrlsub S\ s\ {\isasymsigma}\ {\isacharequal}{\kern0pt}\ s{\isachardoublequoteclose}\isanewline
\isadelimproof
\ \ \ \ %
\endisadelimproof
\isatagproof
\isacommand{using}\isamarkupfalse%
\ value{\isacharunderscore}{\kern0pt}pr\isactrlsub A\ \isacommand{by}\isamarkupfalse%
\ {\isacharparenleft}{\kern0pt}induct\ s{\isacharsemicolon}{\kern0pt}\ auto{\isacharparenright}{\kern0pt}%
\endisatagproof
{\isafoldproof}%
\isadelimproof
\isanewline
\endisadelimproof
\isanewline
\ \ \isacommand{lemma}\isamarkupfalse%
\ l{\isacharunderscore}{\kern0pt}value{\isacharunderscore}{\kern0pt}pr\isactrlsub E{\isacharcolon}{\kern0pt}\ {\isachardoublequoteopen}lconcrete\isactrlsub E{\isacharparenleft}{\kern0pt}l{\isacharparenright}{\kern0pt}\ {\isasymlongrightarrow}\ lval\isactrlsub E\ l\ {\isasymsigma}\ {\isacharequal}{\kern0pt}\ l{\isachardoublequoteclose}\isanewline
\isadelimproof
\ \ \ \ %
\endisadelimproof
\isatagproof
\isacommand{using}\isamarkupfalse%
\ value{\isacharunderscore}{\kern0pt}pr\isactrlsub E\ \isacommand{by}\isamarkupfalse%
\ {\isacharparenleft}{\kern0pt}induct\ l{\isacharsemicolon}{\kern0pt}\ auto{\isacharparenright}{\kern0pt}%
\endisatagproof
{\isafoldproof}%
\isadelimproof
\isanewline
\endisadelimproof
\isanewline
\ \ \isacommand{lemma}\isamarkupfalse%
\ s{\isacharunderscore}{\kern0pt}value{\isacharunderscore}{\kern0pt}pr\isactrlsub B{\isacharcolon}{\kern0pt}\ {\isachardoublequoteopen}sconcrete\isactrlsub B{\isacharparenleft}{\kern0pt}S{\isacharparenright}{\kern0pt}\ {\isasymlongrightarrow}\ sval\isactrlsub B\ S\ {\isasymsigma}\ {\isacharequal}{\kern0pt}\ S{\isachardoublequoteclose}\isanewline
\isadelimproof
\ \ \ \ %
\endisadelimproof
\isatagproof
\isacommand{using}\isamarkupfalse%
\ value{\isacharunderscore}{\kern0pt}pr\isactrlsub B\ \isacommand{by}\isamarkupfalse%
\ auto%
\endisatagproof
{\isafoldproof}%
\isadelimproof
\endisadelimproof
\begin{isamarkuptext}%
The following proofs establish that a variable-free expression 
  always becomes concrete after an evaluation under an arbitrary state. This is
  obvious, considering that the evaluation of variable-free expressions is guaranteed 
  to be independent of the state. We again prove the property using 
  structural~induction. Note that we have to make an exception for the symbolic
  value, as it is not considered~concrete.%
\end{isamarkuptext}\isamarkuptrue%
\ \ \isacommand{lemma}\isamarkupfalse%
\ vars{\isacharunderscore}{\kern0pt}concrete{\isacharunderscore}{\kern0pt}imp\isactrlsub A{\isacharcolon}{\kern0pt}\ {\isachardoublequoteopen}vars\isactrlsub A{\isacharparenleft}{\kern0pt}a{\isacharparenright}{\kern0pt}\ {\isacharequal}{\kern0pt}\ {\isacharbraceleft}{\kern0pt}{\isacharbraceright}{\kern0pt}\ {\isasymlongrightarrow}\ concrete\isactrlsub A{\isacharparenleft}{\kern0pt}val\isactrlsub A\ a\ {\isasymsigma}{\isacharparenright}{\kern0pt}{\isachardoublequoteclose}\isanewline
\isadelimproof
\ \ \ \ %
\endisadelimproof
\isatagproof
\isacommand{by}\isamarkupfalse%
\ {\isacharparenleft}{\kern0pt}induct\ a{\isacharcomma}{\kern0pt}\ auto{\isacharparenright}{\kern0pt}%
\endisatagproof
{\isafoldproof}%
\isadelimproof
\isanewline
\endisadelimproof
\isanewline
\ \ \isacommand{lemma}\isamarkupfalse%
\ vars{\isacharunderscore}{\kern0pt}concrete{\isacharunderscore}{\kern0pt}imp\isactrlsub B{\isacharcolon}{\kern0pt}\ {\isachardoublequoteopen}vars\isactrlsub B{\isacharparenleft}{\kern0pt}b{\isacharparenright}{\kern0pt}\ {\isacharequal}{\kern0pt}\ {\isacharbraceleft}{\kern0pt}{\isacharbraceright}{\kern0pt}\ {\isasymlongrightarrow}\ concrete\isactrlsub B{\isacharparenleft}{\kern0pt}val\isactrlsub B\ b\ {\isasymsigma}{\isacharparenright}{\kern0pt}{\isachardoublequoteclose}\isanewline
\isadelimproof
\ \ \ \ %
\endisadelimproof
\isatagproof
\isacommand{using}\isamarkupfalse%
\ vars{\isacharunderscore}{\kern0pt}concrete{\isacharunderscore}{\kern0pt}imp\isactrlsub A\ \isacommand{by}\isamarkupfalse%
\ {\isacharparenleft}{\kern0pt}induct\ b{\isacharcomma}{\kern0pt}\ auto{\isacharparenright}{\kern0pt}%
\endisatagproof
{\isafoldproof}%
\isadelimproof
\ \isanewline
\endisadelimproof
\isanewline
\ \ \isacommand{lemma}\isamarkupfalse%
\ vars{\isacharunderscore}{\kern0pt}concrete{\isacharunderscore}{\kern0pt}imp\isactrlsub E{\isacharcolon}{\kern0pt}\ {\isachardoublequoteopen}vars\isactrlsub E{\isacharparenleft}{\kern0pt}e{\isacharparenright}{\kern0pt}\ {\isacharequal}{\kern0pt}\ {\isacharbraceleft}{\kern0pt}{\isacharbraceright}{\kern0pt}\ {\isasymlongrightarrow}\ concrete\isactrlsub E{\isacharparenleft}{\kern0pt}val\isactrlsub E\ e\ {\isasymsigma}{\isacharparenright}{\kern0pt}{\isachardoublequoteclose}\isanewline
\isadelimproof
\ \ \ \ %
\endisadelimproof
\isatagproof
\isacommand{using}\isamarkupfalse%
\ vars{\isacharunderscore}{\kern0pt}concrete{\isacharunderscore}{\kern0pt}imp\isactrlsub A\ vars{\isacharunderscore}{\kern0pt}concrete{\isacharunderscore}{\kern0pt}imp\isactrlsub B\ \isacommand{by}\isamarkupfalse%
\ {\isacharparenleft}{\kern0pt}induct\ e{\isacharcomma}{\kern0pt}\ auto{\isacharparenright}{\kern0pt}%
\endisatagproof
{\isafoldproof}%
\isadelimproof
\ \isanewline
\endisadelimproof
\isanewline
\ \ \isacommand{lemma}\isamarkupfalse%
\ vars{\isacharunderscore}{\kern0pt}concrete{\isacharunderscore}{\kern0pt}imp\isactrlsub S{\isacharcolon}{\kern0pt}\ {\isachardoublequoteopen}vars\isactrlsub S{\isacharparenleft}{\kern0pt}s{\isacharparenright}{\kern0pt}\ {\isacharequal}{\kern0pt}\ {\isacharbraceleft}{\kern0pt}{\isacharbraceright}{\kern0pt}\ {\isasymlongrightarrow}\ s\ {\isacharequal}{\kern0pt}\ \isactrlemph \ {\isasymor}\ concrete\isactrlsub S{\isacharparenleft}{\kern0pt}val\isactrlsub S\ s\ {\isasymsigma}{\isacharparenright}{\kern0pt}{\isachardoublequoteclose}\isanewline
\isadelimproof
\ \ \ \ %
\endisadelimproof
\isatagproof
\isacommand{using}\isamarkupfalse%
\ vars{\isacharunderscore}{\kern0pt}concrete{\isacharunderscore}{\kern0pt}imp\isactrlsub A\ \isacommand{by}\isamarkupfalse%
\ {\isacharparenleft}{\kern0pt}induct\ s{\isacharcomma}{\kern0pt}\ auto{\isacharparenright}{\kern0pt}%
\endisatagproof
{\isafoldproof}%
\isadelimproof
\isanewline
\endisadelimproof
\isanewline
\ \ \isacommand{lemma}\isamarkupfalse%
\ l{\isacharunderscore}{\kern0pt}vars{\isacharunderscore}{\kern0pt}concrete{\isacharunderscore}{\kern0pt}imp\isactrlsub E{\isacharcolon}{\kern0pt}\ {\isachardoublequoteopen}lvars\isactrlsub E{\isacharparenleft}{\kern0pt}l{\isacharparenright}{\kern0pt}\ {\isacharequal}{\kern0pt}\ {\isacharbraceleft}{\kern0pt}{\isacharbraceright}{\kern0pt}\ {\isasymlongrightarrow}\ lconcrete\isactrlsub E{\isacharparenleft}{\kern0pt}lval\isactrlsub E\ l\ {\isasymsigma}{\isacharparenright}{\kern0pt}{\isachardoublequoteclose}\isanewline
\isadelimproof
\ \ \ \ %
\endisadelimproof
\isatagproof
\isacommand{using}\isamarkupfalse%
\ vars{\isacharunderscore}{\kern0pt}concrete{\isacharunderscore}{\kern0pt}imp\isactrlsub E\ \isacommand{by}\isamarkupfalse%
\ {\isacharparenleft}{\kern0pt}induct\ l{\isacharcomma}{\kern0pt}\ auto{\isacharparenright}{\kern0pt}%
\endisatagproof
{\isafoldproof}%
\isadelimproof
\ \isanewline
\endisadelimproof
\isanewline
\ \ \isacommand{lemma}\isamarkupfalse%
\ s{\isacharunderscore}{\kern0pt}vars{\isacharunderscore}{\kern0pt}concrete{\isacharunderscore}{\kern0pt}imp\isactrlsub B{\isacharcolon}{\kern0pt}\ {\isachardoublequoteopen}svars\isactrlsub B{\isacharparenleft}{\kern0pt}S{\isacharparenright}{\kern0pt}\ {\isacharequal}{\kern0pt}\ {\isacharbraceleft}{\kern0pt}{\isacharbraceright}{\kern0pt}\ {\isasymlongrightarrow}\ sconcrete\isactrlsub B{\isacharparenleft}{\kern0pt}sval\isactrlsub B\ S\ {\isasymsigma}{\isacharparenright}{\kern0pt}{\isachardoublequoteclose}\isanewline
\isadelimproof
\ \ \ \ %
\endisadelimproof
\isatagproof
\isacommand{using}\isamarkupfalse%
\ vars{\isacharunderscore}{\kern0pt}concrete{\isacharunderscore}{\kern0pt}imp\isactrlsub B\ \isacommand{by}\isamarkupfalse%
\ simp%
\endisatagproof
{\isafoldproof}%
\isadelimproof
\endisadelimproof
\begin{isamarkuptext}%
Last but not least, we prove that any expression turns concrete 
  after an evaluation under a concrete state, provided that all variables occurring 
  in the expression are located in the domain of the state. This property holds, since
  all variables occurring in the expression can be replaced with concrete numerals
  due to the concreteness of the state. The proofs are conducted by 
  making use of structural induction. Note that we again have to make an exception 
  for the symbolic~value.%
\end{isamarkuptext}\isamarkuptrue%
\ \ \isacommand{lemma}\isamarkupfalse%
\ concrete{\isacharunderscore}{\kern0pt}imp\isactrlsub A{\isacharcolon}{\kern0pt}\ {\isachardoublequoteopen}concrete\isactrlsub {\isasymSigma}{\isacharparenleft}{\kern0pt}{\isasymsigma}{\isacharparenright}{\kern0pt}\ {\isasymand}\ vars\isactrlsub A{\isacharparenleft}{\kern0pt}a{\isacharparenright}{\kern0pt}\ {\isasymsubseteq}\ fmdom{\isacharprime}{\kern0pt}{\isacharparenleft}{\kern0pt}{\isasymsigma}{\isacharparenright}{\kern0pt}\ {\isasymlongrightarrow}\ concrete\isactrlsub A{\isacharparenleft}{\kern0pt}val\isactrlsub A\ a\ {\isasymsigma}{\isacharparenright}{\kern0pt}{\isachardoublequoteclose}\isanewline
\isadelimproof
\ \ %
\endisadelimproof
\isatagproof
\isacommand{proof}\isamarkupfalse%
\ {\isacharparenleft}{\kern0pt}induct\ a{\isacharparenright}{\kern0pt}\isanewline
\ \ \ \ \isacommand{case}\isamarkupfalse%
\ {\isacharparenleft}{\kern0pt}Numeral\ n{\isacharparenright}{\kern0pt}\isanewline
\ \ \ \ \isacommand{show}\isamarkupfalse%
\ {\isacharquery}{\kern0pt}case\ \isacommand{by}\isamarkupfalse%
\ auto\ \isanewline
\ \ \ \ %
\isamarkupcmt{The evaluation of numerals is independent of the state, thus trivially
       concrete.%
}\isanewline
\ \ \isacommand{next}\isamarkupfalse%
\isanewline
\ \ \ \ \isacommand{case}\isamarkupfalse%
\ {\isacharparenleft}{\kern0pt}Variable\ x{\isacharparenright}{\kern0pt}\isanewline
\ \ \ \ \isacommand{show}\isamarkupfalse%
\ {\isachardoublequoteopen}concrete\isactrlsub {\isasymSigma}{\isacharparenleft}{\kern0pt}{\isasymsigma}{\isacharparenright}{\kern0pt}\ {\isasymand}\ vars\isactrlsub A{\isacharparenleft}{\kern0pt}Var\ x{\isacharparenright}{\kern0pt}\ {\isasymsubseteq}\ fmdom{\isacharprime}{\kern0pt}{\isacharparenleft}{\kern0pt}{\isasymsigma}{\isacharparenright}{\kern0pt}\ {\isasymlongrightarrow}\ concrete\isactrlsub A{\isacharparenleft}{\kern0pt}val\isactrlsub A\ {\isacharparenleft}{\kern0pt}Var\ x{\isacharparenright}{\kern0pt}\ {\isasymsigma}{\isacharparenright}{\kern0pt}{\isachardoublequoteclose}\isanewline
\ \ \ \ %
\isamarkupcmt{The evaluation of variables depends on the state, thus we have to analyze
       this case further.%
}\isanewline
\ \ \ \ \isacommand{proof}\isamarkupfalse%
\ {\isacharparenleft}{\kern0pt}rule\ impI{\isacharparenright}{\kern0pt}\isanewline
\ \ \ \ \ \ \isacommand{assume}\isamarkupfalse%
\ premise{\isacharcolon}{\kern0pt}\ {\isachardoublequoteopen}concrete\isactrlsub {\isasymSigma}{\isacharparenleft}{\kern0pt}{\isasymsigma}{\isacharparenright}{\kern0pt}\ {\isasymand}\ vars\isactrlsub A{\isacharparenleft}{\kern0pt}Var\ x{\isacharparenright}{\kern0pt}\ {\isasymsubseteq}\ fmdom{\isacharprime}{\kern0pt}{\isacharparenleft}{\kern0pt}{\isasymsigma}{\isacharparenright}{\kern0pt}{\isachardoublequoteclose}\isanewline
\ \ \ \ \ \ %
\isamarkupcmt{We assume \isa{{\isasymsigma}} is concrete and x is located in the domain of \isa{{\isasymsigma}}.%
}\isanewline
\ \ \ \ \ \ \isacommand{hence}\isamarkupfalse%
\ {\isachardoublequoteopen}concrete\isactrlsub S{\isacharparenleft}{\kern0pt}the{\isacharparenleft}{\kern0pt}fmlookup\ {\isasymsigma}\ x{\isacharparenright}{\kern0pt}{\isacharparenright}{\kern0pt}{\isachardoublequoteclose}\isanewline
\ \ \ \ \ \ \ \ \isacommand{by}\isamarkupfalse%
\ {\isacharparenleft}{\kern0pt}simp\ add{\isacharcolon}{\kern0pt}\ concrete\isactrlsub {\isasymSigma}{\isacharunderscore}{\kern0pt}def{\isacharparenright}{\kern0pt}\ \isanewline
\ \ \ \ \ \ %
\isamarkupcmt{Hence the value of x in \isa{{\isasymsigma}} must also be of concrete nature.%
}\isanewline
\ \ \ \ \ \ \isacommand{hence}\isamarkupfalse%
\ {\isachardoublequoteopen}{\isasymexists}a{\isachardot}{\kern0pt}\ the{\isacharparenleft}{\kern0pt}fmlookup\ {\isasymsigma}\ x{\isacharparenright}{\kern0pt}\ {\isacharequal}{\kern0pt}\ Exp\ a\ {\isasymand}\ concrete\isactrlsub A\ a{\isachardoublequoteclose}\ \isanewline
\ \ \ \ \ \ \ \ \isacommand{by}\isamarkupfalse%
\ {\isacharparenleft}{\kern0pt}metis\ sexp{\isachardot}{\kern0pt}exhaust\ sexp{\isachardot}{\kern0pt}simps{\isacharparenleft}{\kern0pt}{\isadigit{4}}{\isacharparenright}{\kern0pt}\ sexp{\isachardot}{\kern0pt}simps{\isacharparenleft}{\kern0pt}{\isadigit{5}}{\isacharparenright}{\kern0pt}{\isacharparenright}{\kern0pt}\ \isanewline
\ \ \ \ \ \ %
\isamarkupcmt{This means that the value of x in \isa{{\isasymsigma}} must be a concrete arithmetic expression a.%
}\isanewline
\ \ \ \ \ \ \isacommand{hence}\isamarkupfalse%
\ {\isachardoublequoteopen}{\isasymexists}n{\isachardot}{\kern0pt}\ the{\isacharparenleft}{\kern0pt}fmlookup\ {\isasymsigma}\ x{\isacharparenright}{\kern0pt}\ {\isacharequal}{\kern0pt}\ Exp\ {\isacharparenleft}{\kern0pt}Num\ n{\isacharparenright}{\kern0pt}{\isachardoublequoteclose}\isanewline
\ \ \ \ \ \ \ \ \isacommand{by}\isamarkupfalse%
\ {\isacharparenleft}{\kern0pt}metis\ aexp{\isachardot}{\kern0pt}exhaust\ aexp{\isachardot}{\kern0pt}simps{\isacharparenleft}{\kern0pt}{\isadigit{1}}{\isadigit{1}}{\isacharparenright}{\kern0pt}\ aexp{\isachardot}{\kern0pt}simps{\isacharparenleft}{\kern0pt}{\isadigit{1}}{\isadigit{2}}{\isacharparenright}{\kern0pt}{\isacharparenright}{\kern0pt}\isanewline
\ \ \ \ \ \ %
\isamarkupcmt{The concreteness of $a$ in turn implies that the value of x in \isa{{\isasymsigma}} must be a 
         numeral n.%
}\isanewline
\ \ \ \ \ \ \isacommand{with}\isamarkupfalse%
\ premise\ \isacommand{show}\isamarkupfalse%
\ {\isachardoublequoteopen}concrete\isactrlsub A{\isacharparenleft}{\kern0pt}val\isactrlsub A\ {\isacharparenleft}{\kern0pt}Var\ x{\isacharparenright}{\kern0pt}\ {\isasymsigma}{\isacharparenright}{\kern0pt}{\isachardoublequoteclose}\ \isanewline
\ \ \ \ \ \ \ \ \isacommand{using}\isamarkupfalse%
\ fmdom{\isacharprime}{\kern0pt}{\isacharunderscore}{\kern0pt}notI\ \isacommand{by}\isamarkupfalse%
\ force\isanewline
\ \ \ \ \ \ %
\isamarkupcmt{Knowing that x maps to a numeral, we infer that the evaluation of x must
         be concrete.%
}\isanewline
\ \ \ \ \isacommand{qed}\isamarkupfalse%
\isanewline
\ \ \isacommand{next}\isamarkupfalse%
\isanewline
\ \ \ \ \isacommand{case}\isamarkupfalse%
\ {\isacharparenleft}{\kern0pt}Op\isactrlsub A\ a\isactrlsub {\isadigit{1}}\ op\ a\isactrlsub {\isadigit{2}}{\isacharparenright}{\kern0pt}\isanewline
\ \ \ \ \isacommand{show}\isamarkupfalse%
\ {\isacharquery}{\kern0pt}case\ \isanewline
\ \ \ \ \ \ \isacommand{using}\isamarkupfalse%
\ Op\isactrlsub A{\isachardot}{\kern0pt}hyps{\isacharparenleft}{\kern0pt}{\isadigit{1}}{\isacharparenright}{\kern0pt}\ Op\isactrlsub A{\isachardot}{\kern0pt}hyps{\isacharparenleft}{\kern0pt}{\isadigit{2}}{\isacharparenright}{\kern0pt}\ \isacommand{by}\isamarkupfalse%
\ simp\ \isanewline
\ \ \ \ %
\isamarkupcmt{The induction step can be trivially closed via the induction hypothesis.%
}\isanewline
\ \ \isacommand{qed}\isamarkupfalse%
\endisatagproof
{\isafoldproof}%
\isadelimproof
\isanewline
\endisadelimproof
\isanewline
\ \ \isacommand{lemma}\isamarkupfalse%
\ concrete{\isacharunderscore}{\kern0pt}imp\isactrlsub B{\isacharcolon}{\kern0pt}\ {\isachardoublequoteopen}concrete\isactrlsub {\isasymSigma}{\isacharparenleft}{\kern0pt}{\isasymsigma}{\isacharparenright}{\kern0pt}\ {\isasymand}\ vars\isactrlsub B{\isacharparenleft}{\kern0pt}b{\isacharparenright}{\kern0pt}\ {\isasymsubseteq}\ fmdom{\isacharprime}{\kern0pt}{\isacharparenleft}{\kern0pt}{\isasymsigma}{\isacharparenright}{\kern0pt}\ {\isasymlongrightarrow}\ concrete\isactrlsub B{\isacharparenleft}{\kern0pt}val\isactrlsub B\ b\ {\isasymsigma}{\isacharparenright}{\kern0pt}{\isachardoublequoteclose}\isanewline
\isadelimproof
\ \ \ \ %
\endisadelimproof
\isatagproof
\isacommand{using}\isamarkupfalse%
\ concrete{\isacharunderscore}{\kern0pt}imp\isactrlsub A\ \isacommand{by}\isamarkupfalse%
\ {\isacharparenleft}{\kern0pt}induct\ b{\isacharcomma}{\kern0pt}\ auto{\isacharparenright}{\kern0pt}%
\endisatagproof
{\isafoldproof}%
\isadelimproof
\isanewline
\endisadelimproof
\isanewline
\ \ \isacommand{lemma}\isamarkupfalse%
\ concrete{\isacharunderscore}{\kern0pt}imp\isactrlsub E{\isacharcolon}{\kern0pt}\ {\isachardoublequoteopen}concrete\isactrlsub {\isasymSigma}{\isacharparenleft}{\kern0pt}{\isasymsigma}{\isacharparenright}{\kern0pt}\ {\isasymand}\ vars\isactrlsub E{\isacharparenleft}{\kern0pt}e{\isacharparenright}{\kern0pt}\ {\isasymsubseteq}\ fmdom{\isacharprime}{\kern0pt}{\isacharparenleft}{\kern0pt}{\isasymsigma}{\isacharparenright}{\kern0pt}\ {\isasymlongrightarrow}\ concrete\isactrlsub E{\isacharparenleft}{\kern0pt}val\isactrlsub E\ e\ {\isasymsigma}{\isacharparenright}{\kern0pt}{\isachardoublequoteclose}\isanewline
\isadelimproof
\ \ \ \ %
\endisadelimproof
\isatagproof
\isacommand{using}\isamarkupfalse%
\ concrete{\isacharunderscore}{\kern0pt}imp\isactrlsub A\ concrete{\isacharunderscore}{\kern0pt}imp\isactrlsub B\ \isacommand{by}\isamarkupfalse%
\ {\isacharparenleft}{\kern0pt}induct\ e{\isacharcomma}{\kern0pt}\ auto{\isacharparenright}{\kern0pt}%
\endisatagproof
{\isafoldproof}%
\isadelimproof
\isanewline
\endisadelimproof
\isanewline
\ \ \isacommand{lemma}\isamarkupfalse%
\ concrete{\isacharunderscore}{\kern0pt}imp\isactrlsub S{\isacharcolon}{\kern0pt}\ {\isachardoublequoteopen}concrete\isactrlsub {\isasymSigma}{\isacharparenleft}{\kern0pt}{\isasymsigma}{\isacharparenright}{\kern0pt}\ {\isasymand}\ vars\isactrlsub S{\isacharparenleft}{\kern0pt}s{\isacharparenright}{\kern0pt}\ {\isasymsubseteq}\ fmdom{\isacharprime}{\kern0pt}{\isacharparenleft}{\kern0pt}{\isasymsigma}{\isacharparenright}{\kern0pt}\ {\isasymlongrightarrow}\ s\ {\isacharequal}{\kern0pt}\ \isactrlemph \ {\isasymor}\ concrete\isactrlsub S{\isacharparenleft}{\kern0pt}val\isactrlsub S\ s\ {\isasymsigma}{\isacharparenright}{\kern0pt}{\isachardoublequoteclose}\isanewline
\isadelimproof
\ \ \ \ %
\endisadelimproof
\isatagproof
\isacommand{using}\isamarkupfalse%
\ concrete{\isacharunderscore}{\kern0pt}imp\isactrlsub A\ \isacommand{by}\isamarkupfalse%
\ {\isacharparenleft}{\kern0pt}induct\ s{\isacharcomma}{\kern0pt}\ auto{\isacharparenright}{\kern0pt}%
\endisatagproof
{\isafoldproof}%
\isadelimproof
\isanewline
\endisadelimproof
\isanewline
\ \ \isacommand{lemma}\isamarkupfalse%
\ l{\isacharunderscore}{\kern0pt}concrete{\isacharunderscore}{\kern0pt}imp\isactrlsub E{\isacharcolon}{\kern0pt}\ {\isachardoublequoteopen}concrete\isactrlsub {\isasymSigma}{\isacharparenleft}{\kern0pt}{\isasymsigma}{\isacharparenright}{\kern0pt}\ {\isasymand}\ lvars\isactrlsub E{\isacharparenleft}{\kern0pt}l{\isacharparenright}{\kern0pt}\ {\isasymsubseteq}\ fmdom{\isacharprime}{\kern0pt}{\isacharparenleft}{\kern0pt}{\isasymsigma}{\isacharparenright}{\kern0pt}\ {\isasymlongrightarrow}\ lconcrete\isactrlsub E{\isacharparenleft}{\kern0pt}lval\isactrlsub E\ l\ {\isasymsigma}{\isacharparenright}{\kern0pt}{\isachardoublequoteclose}\isanewline
\isadelimproof
\ \ \ \ %
\endisadelimproof
\isatagproof
\isacommand{using}\isamarkupfalse%
\ concrete{\isacharunderscore}{\kern0pt}imp\isactrlsub E\ \isacommand{by}\isamarkupfalse%
\ {\isacharparenleft}{\kern0pt}induct\ l{\isacharcomma}{\kern0pt}\ auto{\isacharparenright}{\kern0pt}%
\endisatagproof
{\isafoldproof}%
\isadelimproof
\isanewline
\endisadelimproof
\isanewline
\ \ \isacommand{lemma}\isamarkupfalse%
\ s{\isacharunderscore}{\kern0pt}concrete{\isacharunderscore}{\kern0pt}imp\isactrlsub B{\isacharcolon}{\kern0pt}\ {\isachardoublequoteopen}concrete\isactrlsub {\isasymSigma}{\isacharparenleft}{\kern0pt}{\isasymsigma}{\isacharparenright}{\kern0pt}\ {\isasymand}\ svars\isactrlsub B{\isacharparenleft}{\kern0pt}S{\isacharparenright}{\kern0pt}\ {\isasymsubseteq}\ fmdom{\isacharprime}{\kern0pt}{\isacharparenleft}{\kern0pt}{\isasymsigma}{\isacharparenright}{\kern0pt}\ {\isasymlongrightarrow}\ sconcrete\isactrlsub B{\isacharparenleft}{\kern0pt}sval\isactrlsub B\ S\ {\isasymsigma}{\isacharparenright}{\kern0pt}{\isachardoublequoteclose}\isanewline
\isadelimproof
\ \ \ \ %
\endisadelimproof
\isatagproof
\isacommand{using}\isamarkupfalse%
\ concrete{\isacharunderscore}{\kern0pt}imp\isactrlsub B\ \isacommand{by}\isamarkupfalse%
\ {\isacharparenleft}{\kern0pt}simp\ add{\isacharcolon}{\kern0pt}\ UN{\isacharunderscore}{\kern0pt}subset{\isacharunderscore}{\kern0pt}iff{\isacharparenright}{\kern0pt}%
\endisatagproof
{\isafoldproof}%
\isadelimproof
\endisadelimproof
\begin{isamarkuptext}%
The evaluation of expressions can be demonstrated using the following
  practical examples.%
\end{isamarkuptext}\isamarkuptrue%
\ \ \isacommand{lemma}\isamarkupfalse%
\ {\isachardoublequoteopen}val\isactrlsub A\ aExp{\isacharunderscore}{\kern0pt}ex\ {\isasymsigma}\isactrlsub {\isadigit{2}}\ {\isacharequal}{\kern0pt}\ Num\ {\isadigit{8}}{\isachardoublequoteclose}\isanewline
\isadelimproof
\ \ \ \ %
\endisadelimproof
\isatagproof
\isacommand{by}\isamarkupfalse%
\ {\isacharparenleft}{\kern0pt}simp\ add{\isacharcolon}{\kern0pt}\ aExp{\isacharunderscore}{\kern0pt}ex{\isacharunderscore}{\kern0pt}def\ {\isasymsigma}\isactrlsub {\isadigit{2}}{\isacharunderscore}{\kern0pt}def{\isacharparenright}{\kern0pt}%
\endisatagproof
{\isafoldproof}%
\isadelimproof
\isanewline
\endisadelimproof
\isanewline
\ \ \isacommand{lemma}\isamarkupfalse%
\ {\isachardoublequoteopen}val\isactrlsub B\ bExp{\isacharunderscore}{\kern0pt}ex\ {\isasymsigma}\isactrlsub {\isadigit{1}}\ {\isacharequal}{\kern0pt}\ {\isacharparenleft}{\kern0pt}{\isacharparenleft}{\kern0pt}{\isacharparenleft}{\kern0pt}Var\ {\isacharprime}{\kern0pt}{\isacharprime}{\kern0pt}y{\isacharprime}{\kern0pt}{\isacharprime}{\kern0pt}{\isacharparenright}{\kern0pt}\ \isactrlsub Amul\ {\isacharparenleft}{\kern0pt}Num\ {\isadigit{4}}{\isacharparenright}{\kern0pt}{\isacharparenright}{\kern0pt}\ \isactrlsub Req\ {\isacharparenleft}{\kern0pt}Num\ {\isadigit{2}}{\isacharparenright}{\kern0pt}{\isacharparenright}{\kern0pt}\ \isactrlsub Bdisj\ {\isacharparenleft}{\kern0pt}Bool\ False{\isacharparenright}{\kern0pt}{\isachardoublequoteclose}\isanewline
\isadelimproof
\ \ \ \ %
\endisadelimproof
\isatagproof
\isacommand{by}\isamarkupfalse%
\ {\isacharparenleft}{\kern0pt}simp\ add{\isacharcolon}{\kern0pt}\ bExp{\isacharunderscore}{\kern0pt}ex{\isacharunderscore}{\kern0pt}def\ {\isasymsigma}\isactrlsub {\isadigit{1}}{\isacharunderscore}{\kern0pt}def{\isacharparenright}{\kern0pt}%
\endisatagproof
{\isafoldproof}%
\isadelimproof
\endisadelimproof
\isadelimdocument
\endisadelimdocument
\isatagdocument
\isamarkupsubsubsection{State Simplifications%
}
\isamarkuptrue%
\endisatagdocument
{\isafolddocument}%
\isadelimdocument
\endisadelimdocument
\begin{isamarkuptext}%
In the original paper, states are always implicitly assumed to be simplified
  as much as possible. However, such a formalization is not reasonably feasible in 
  Isabelle. We therefore decide to additionally propose a state simplification function 
  that explicitly simplifies all expressions in the image of a given state.  This
  function can be used to further simplify non-simplified states. It can also be 
  utilized to simplify a state if the simplification status of the given 
  state~is~unknown. \par
  Note that the already predefined \isa{fmmap{\isacharminus}{\kern0pt}keys} function of the \isa{finite{\isacharminus}{\kern0pt}map} theory 
  can be used to update all key-value tuples of the state given as the second argument 
  with the function provided as the first~argument.%
\end{isamarkuptext}\isamarkuptrue%
\ \ \isacommand{definition}\isamarkupfalse%
\isanewline
\ \ \ \ sim\isactrlsub {\isasymSigma}\ {\isacharcolon}{\kern0pt}{\isacharcolon}{\kern0pt}\ {\isachardoublequoteopen}{\isasymSigma}\ {\isasymRightarrow}\ {\isasymSigma}{\isachardoublequoteclose}\ \isakeyword{where}\isanewline
\ \ \ \ {\isachardoublequoteopen}sim\isactrlsub {\isasymSigma}\ {\isasymsigma}\ {\isasymequiv}\ fmmap{\isacharunderscore}{\kern0pt}keys\ {\isacharparenleft}{\kern0pt}{\isasymlambda}v\ e{\isachardot}{\kern0pt}\ val\isactrlsub S\ e\ {\isasymsigma}{\isacharparenright}{\kern0pt}\ {\isasymsigma}{\isachardoublequoteclose}%
\begin{isamarkuptext}%
We can now show that the execution of the state simplification function
  preserves the domain of the state, implying that no variables are added 
  or~removed.%
\end{isamarkuptext}\isamarkuptrue%
\ \ \isacommand{lemma}\isamarkupfalse%
\ sim\isactrlsub {\isasymSigma}{\isacharunderscore}{\kern0pt}dom{\isacharunderscore}{\kern0pt}pr{\isacharcolon}{\kern0pt}\ {\isachardoublequoteopen}fmdom{\isacharprime}{\kern0pt}{\isacharparenleft}{\kern0pt}{\isasymsigma}{\isacharparenright}{\kern0pt}\ {\isacharequal}{\kern0pt}\ fmdom{\isacharprime}{\kern0pt}{\isacharparenleft}{\kern0pt}sim\isactrlsub {\isasymSigma}\ {\isasymsigma}{\isacharparenright}{\kern0pt}{\isachardoublequoteclose}\isanewline
\isadelimproof
\ \ \ \ %
\endisadelimproof
\isatagproof
\isacommand{by}\isamarkupfalse%
\ {\isacharparenleft}{\kern0pt}simp\ add{\isacharcolon}{\kern0pt}\ sim\isactrlsub {\isasymSigma}{\isacharunderscore}{\kern0pt}def\ fmdom{\isacharprime}{\kern0pt}{\isacharunderscore}{\kern0pt}alt{\isacharunderscore}{\kern0pt}def{\isacharparenright}{\kern0pt}%
\endisatagproof
{\isafoldproof}%
\isadelimproof
\endisadelimproof
\begin{isamarkuptext}%
Using the previous property, we can now establish that a concrete
  state preserves its concreteness after performing a state~simplification.
  This holds, because the concreteness of starred expressions is preserved under
  arbitrary state evaluations, which we have already shown~earlier.%
\end{isamarkuptext}\isamarkuptrue%
\ \ \isacommand{lemma}\isamarkupfalse%
\ concrete{\isacharunderscore}{\kern0pt}sim{\isacharunderscore}{\kern0pt}pr{\isacharcolon}{\kern0pt}\ {\isachardoublequoteopen}concrete\isactrlsub {\isasymSigma}{\isacharparenleft}{\kern0pt}{\isasymsigma}{\isacharparenright}{\kern0pt}\ {\isasymlongrightarrow}\ concrete\isactrlsub {\isasymSigma}{\isacharparenleft}{\kern0pt}sim\isactrlsub {\isasymSigma}\ {\isasymsigma}{\isacharparenright}{\kern0pt}{\isachardoublequoteclose}\ \isanewline
\isadelimproof
\ \ \ \ %
\endisadelimproof
\isatagproof
\isacommand{apply}\isamarkupfalse%
\ {\isacharparenleft}{\kern0pt}simp\ add{\isacharcolon}{\kern0pt}\ concrete\isactrlsub {\isasymSigma}{\isacharunderscore}{\kern0pt}def\ sim\isactrlsub {\isasymSigma}{\isacharunderscore}{\kern0pt}def{\isacharparenright}{\kern0pt}\isanewline
\ \ \ \ \isacommand{using}\isamarkupfalse%
\ concrete{\isacharunderscore}{\kern0pt}pr\isactrlsub S\ sim\isactrlsub {\isasymSigma}{\isacharunderscore}{\kern0pt}dom{\isacharunderscore}{\kern0pt}pr\ sim\isactrlsub {\isasymSigma}{\isacharunderscore}{\kern0pt}def\ \isacommand{by}\isamarkupfalse%
\ {\isacharparenleft}{\kern0pt}metis\ fmdom{\isacharprime}{\kern0pt}{\isacharunderscore}{\kern0pt}notI\ option{\isachardot}{\kern0pt}map{\isacharunderscore}{\kern0pt}sel{\isacharparenright}{\kern0pt}%
\endisatagproof
{\isafoldproof}%
\isadelimproof
\endisadelimproof
\begin{isamarkuptext}%
We close this section by proving that a wellformed state without symbolic 
  variables is guaranteed to turn concrete after a state simplification. This holds,
  because any expression in a wellformed state without symbolic variables must be
  variable-free, and therefore can always be fully~simplified.%
\end{isamarkuptext}\isamarkuptrue%
\ \ \isacommand{lemma}\isamarkupfalse%
\ sim{\isacharunderscore}{\kern0pt}concrete{\isacharunderscore}{\kern0pt}imp{\isacharcolon}{\kern0pt}\ \isanewline
\ \ \ \ \isakeyword{assumes}\ {\isachardoublequoteopen}symb\isactrlsub {\isasymSigma}{\isacharparenleft}{\kern0pt}{\isasymsigma}{\isacharparenright}{\kern0pt}\ {\isacharequal}{\kern0pt}\ {\isacharbraceleft}{\kern0pt}{\isacharbraceright}{\kern0pt}\ {\isasymand}\ wf\isactrlsub {\isasymSigma}{\isacharparenleft}{\kern0pt}{\isasymsigma}{\isacharparenright}{\kern0pt}{\isachardoublequoteclose}\ \isanewline
\ \ \ \ \isakeyword{shows}\ {\isachardoublequoteopen}concrete\isactrlsub {\isasymSigma}{\isacharparenleft}{\kern0pt}sim\isactrlsub {\isasymSigma}\ {\isasymsigma}{\isacharparenright}{\kern0pt}{\isachardoublequoteclose}\isanewline
\isadelimproof
\ \ %
\endisadelimproof
\isatagproof
\isacommand{proof}\isamarkupfalse%
\ {\isacharminus}{\kern0pt}\isanewline
\ \ \ \ \isacommand{have}\isamarkupfalse%
\ {\isachardoublequoteopen}{\isasymforall}v\ {\isasymin}\ fmdom{\isacharprime}{\kern0pt}{\isacharparenleft}{\kern0pt}{\isasymsigma}{\isacharparenright}{\kern0pt}{\isachardot}{\kern0pt}\ vars\isactrlsub S{\isacharparenleft}{\kern0pt}the{\isacharparenleft}{\kern0pt}fmlookup\ {\isasymsigma}\ v{\isacharparenright}{\kern0pt}{\isacharparenright}{\kern0pt}\ {\isacharequal}{\kern0pt}\ {\isacharbraceleft}{\kern0pt}{\isacharbraceright}{\kern0pt}{\isachardoublequoteclose}\ \isanewline
\ \ \ \ \ \ \isacommand{using}\isamarkupfalse%
\ assms\ \isacommand{by}\isamarkupfalse%
\ {\isacharparenleft}{\kern0pt}auto\ simp\ add{\isacharcolon}{\kern0pt}\ wf\isactrlsub {\isasymSigma}{\isacharunderscore}{\kern0pt}def{\isacharparenright}{\kern0pt}\isanewline
\ \ \ \ %
\isamarkupcmt{Due to the wellformedness of the state and the empty set of symbolic variables, 
       no variable of any kind can occur in mapped expressions of the state.%
}\isanewline
\ \ \ \ \isacommand{thus}\isamarkupfalse%
\ {\isachardoublequoteopen}concrete\isactrlsub {\isasymSigma}{\isacharparenleft}{\kern0pt}sim\isactrlsub {\isasymSigma}\ {\isasymsigma}{\isacharparenright}{\kern0pt}{\isachardoublequoteclose}\ \isanewline
\ \ \ \ \ \ \isacommand{using}\isamarkupfalse%
\ assms\ vars{\isacharunderscore}{\kern0pt}concrete{\isacharunderscore}{\kern0pt}imp\isactrlsub S\ \isacommand{apply}\isamarkupfalse%
\ {\isacharparenleft}{\kern0pt}simp\ add{\isacharcolon}{\kern0pt}\ symb\isactrlsub {\isasymSigma}{\isacharunderscore}{\kern0pt}def\ concrete\isactrlsub {\isasymSigma}{\isacharunderscore}{\kern0pt}def\ sim\isactrlsub {\isasymSigma}{\isacharunderscore}{\kern0pt}def{\isacharparenright}{\kern0pt}\ \isanewline
\ \ \ \ \ \ \isacommand{by}\isamarkupfalse%
\ {\isacharparenleft}{\kern0pt}metis\ {\isacharparenleft}{\kern0pt}no{\isacharunderscore}{\kern0pt}types{\isacharcomma}{\kern0pt}\ lifting{\isacharparenright}{\kern0pt}\ None{\isacharunderscore}{\kern0pt}eq{\isacharunderscore}{\kern0pt}map{\isacharunderscore}{\kern0pt}option{\isacharunderscore}{\kern0pt}iff\ fmlookup{\isacharunderscore}{\kern0pt}dom{\isacharprime}{\kern0pt}{\isacharunderscore}{\kern0pt}iff\ fmlookup{\isacharunderscore}{\kern0pt}fmmap{\isacharunderscore}{\kern0pt}keys\ is{\isacharunderscore}{\kern0pt}none{\isacharunderscore}{\kern0pt}code\ option{\isachardot}{\kern0pt}collapse\ the{\isacharunderscore}{\kern0pt}map{\isacharunderscore}{\kern0pt}option{\isacharparenright}{\kern0pt}\isanewline
\ \ \ \ %
\isamarkupcmt{Hence our state must be concrete after a singular state simplification.%
}\isanewline
\ \ \isacommand{qed}\isamarkupfalse%
\endisatagproof
{\isafoldproof}%
\isadelimproof
\endisadelimproof
\isadelimdocument
\endisadelimdocument
\isatagdocument
\isamarkupsubsection{Traces%
}
\isamarkuptrue%
\isamarkupsubsubsection{Symbolic Traces%
}
\isamarkuptrue%
\endisatagdocument
{\isafolddocument}%
\isadelimdocument
\endisadelimdocument
\begin{isamarkuptext}%
We are now interested in introducing a notion of symbolic traces, similar
  to the one provided in the original paper. We begin by proposing a new type
  consisting of all possible events that can occur during program executions. An
  input event represents the receival of a temporary unknown expression from another 
  system. A method invocation event symbolizes the invocation of another method, whilst
  a method invocation reaction event matches the corresponding reaction of~the~callee. 
  \par
  Instead of providing a type for events, it is also possible to denote a constructor 
  for each kind of event in the trace syntax. However, this would cause problems
  when adding or removing specific events, as all inductions over traces would
  need to be adapted. We have therefore decided to outsource the events
  into an own event type, thereby increasing the modularity of the model. Note that
  this implies that the event type can be freely modified without having to reprove
  basic trace~properties.%
\end{isamarkuptext}\isamarkuptrue%
\ \ \isacommand{datatype}\isamarkupfalse%
\ event{\isacharunderscore}{\kern0pt}marker\ {\isacharequal}{\kern0pt}\ \isanewline
\ \ \ \ \ \ inpEv\ %
\isamarkupcmt{Input Event%
}\isanewline
\ \ \ \ {\isacharbar}{\kern0pt}\ invEv\ %
\isamarkupcmt{Method Invocation Event%
}\isanewline
\ \ \ \ {\isacharbar}{\kern0pt}\ invREv\ %
\isamarkupcmt{Method Invocation Reaction Event%
}%
\begin{isamarkuptext}%
Singular elements of a trace are called trace atoms. A trace atom can either 
  be a state or an event with a list of expressions as~its~arguments.%
\end{isamarkuptext}\isamarkuptrue%
\ \ \isacommand{datatype}\isamarkupfalse%
\ trace{\isacharunderscore}{\kern0pt}atom\ {\isacharequal}{\kern0pt}\ \isanewline
\ \ \ \ \ \ State\ {\isasymSigma}\ %
\isamarkupcmt{State%
}\isanewline
\ \ \ \ {\isacharbar}{\kern0pt}\ Event\ event{\isacharunderscore}{\kern0pt}marker\ {\isachardoublequoteopen}exp\ list{\isachardoublequoteclose}\ %
\isamarkupcmt{Event%
}%
\begin{isamarkuptext}%
We can now consider a symbolic trace as a sequence of trace atoms starting
  at the empty trace. Note that this definition forces all of our traces to be
  finite, implying that infinite traces cannot be formalized in this model. This
  problem could be circumvented by defining the set of traces as the set of all 
  partial functions mapping from natural numbers to trace atoms. We propose this as
  an idea for further expansions of~this~model.%
\end{isamarkuptext}\isamarkuptrue%
\ \ \isacommand{datatype}\isamarkupfalse%
\ {\isasymT}\ {\isacharequal}{\kern0pt}\ \isanewline
\ \ \ \ \ \ Epsilon\ %
\isamarkupcmt{Empty Trace%
}\isanewline
\ \ \ \ {\isacharbar}{\kern0pt}\ Transition\ {\isasymT}\ trace{\isacharunderscore}{\kern0pt}atom\ %
\isamarkupcmt{Trace Transition%
}%
\begin{isamarkuptext}%
We also add a minimal concrete syntax for traces, thereby improving 
  the readability of trace atoms and traces.%
\end{isamarkuptext}\isamarkuptrue%
\ \ \isacommand{notation}\isamarkupfalse%
\ State\ {\isacharparenleft}{\kern0pt}{\isachardoublequoteopen}State\ {\isasymllangle}{\isacharunderscore}{\kern0pt}{\isasymrrangle}{\isachardoublequoteclose}\ {\isadigit{6}}{\isadigit{1}}{\isacharparenright}{\kern0pt}\isanewline
\ \ \isacommand{notation}\isamarkupfalse%
\ Event\ {\isacharparenleft}{\kern0pt}{\isachardoublequoteopen}Event\ {\isasymllangle}{\isacharunderscore}{\kern0pt}{\isacharcomma}{\kern0pt}{\isacharunderscore}{\kern0pt}{\isasymrrangle}{\isachardoublequoteclose}\ {\isadigit{6}}{\isadigit{1}}{\isacharparenright}{\kern0pt}\isanewline
\ \ \isacommand{notation}\isamarkupfalse%
\ Epsilon\ {\isacharparenleft}{\kern0pt}{\isachardoublequoteopen}{\isasymepsilon}{\isachardoublequoteclose}{\isacharparenright}{\kern0pt}\ \isanewline
\ \ \isacommand{notation}\isamarkupfalse%
\ Transition\ {\isacharparenleft}{\kern0pt}\isakeyword{infix}\ {\isachardoublequoteopen}{\isasymleadsto}{\isachardoublequoteclose}\ {\isadigit{6}}{\isadigit{0}}{\isacharparenright}{\kern0pt}%
\begin{isamarkuptext}%
We call a trace consisting of a singular state a singleton trace. Similar
  to the paper, we denote the singleton trace of state \isa{{\isasymsigma}} as \isa{{\isasymlangle}{\isasymsigma}{\isasymrangle}}. This
  abbreviation will be used as a condensed notation when formalizing~traces.%
\end{isamarkuptext}\isamarkuptrue%
\ \ \isacommand{abbreviation}\isamarkupfalse%
\ \isanewline
\ \ \ \ singleton{\isacharunderscore}{\kern0pt}trace\ {\isacharcolon}{\kern0pt}{\isacharcolon}{\kern0pt}\ {\isachardoublequoteopen}{\isasymSigma}\ {\isasymRightarrow}\ {\isasymT}{\isachardoublequoteclose}\ {\isacharparenleft}{\kern0pt}{\isachardoublequoteopen}{\isasymlangle}{\isacharunderscore}{\kern0pt}{\isasymrangle}{\isachardoublequoteclose}{\isacharparenright}{\kern0pt}\ \isakeyword{where}\isanewline
\ \ \ \ {\isachardoublequoteopen}{\isasymlangle}{\isasymsigma}{\isasymrangle}\ {\isasymequiv}\ {\isasymepsilon}\ {\isasymleadsto}\ State{\isasymllangle}{\isasymsigma}{\isasymrrangle}{\isachardoublequoteclose}%
\begin{isamarkuptext}%
In the following we provide multiple examples for traces that can be
  syntactically derived in the given trace~syntax.%
\end{isamarkuptext}\isamarkuptrue%
\ \ \isacommand{definition}\isamarkupfalse%
\isanewline
\ \ \ \ {\isasymtau}\isactrlsub {\isadigit{1}}\ {\isacharcolon}{\kern0pt}{\isacharcolon}{\kern0pt}\ {\isasymT}\ \isakeyword{where}\isanewline
\ \ \ \ {\isachardoublequoteopen}{\isasymtau}\isactrlsub {\isadigit{1}}\ {\isasymequiv}\ {\isacharparenleft}{\kern0pt}{\isasymlangle}{\isasymsigma}\isactrlsub {\isadigit{1}}{\isasymrangle}\ {\isasymleadsto}\ Event{\isasymllangle}inpEv{\isacharcomma}{\kern0pt}\ {\isacharbrackleft}{\kern0pt}{\isacharbrackright}{\kern0pt}{\isasymrrangle}{\isacharparenright}{\kern0pt}\ {\isasymleadsto}\ State{\isasymllangle}{\isasymsigma}\isactrlsub {\isadigit{1}}{\isasymrrangle}{\isachardoublequoteclose}\isanewline
\isanewline
\ \ \isacommand{definition}\isamarkupfalse%
\isanewline
\ \ \ \ {\isasymtau}\isactrlsub {\isadigit{2}}\ {\isacharcolon}{\kern0pt}{\isacharcolon}{\kern0pt}\ {\isasymT}\ \isakeyword{where}\isanewline
\ \ \ \ {\isachardoublequoteopen}{\isasymtau}\isactrlsub {\isadigit{2}}\ {\isasymequiv}\ {\isacharparenleft}{\kern0pt}{\isasymlangle}{\isasymsigma}\isactrlsub {\isadigit{1}}{\isasymrangle}\ {\isasymleadsto}\ Event{\isasymllangle}invREv{\isacharcomma}{\kern0pt}\ {\isacharbrackleft}{\kern0pt}P\ {\isacharprime}{\kern0pt}{\isacharprime}{\kern0pt}foo{\isacharprime}{\kern0pt}{\isacharprime}{\kern0pt}{\isacharcomma}{\kern0pt}\ A\ {\isacharparenleft}{\kern0pt}Num\ {\isadigit{2}}{\isacharparenright}{\kern0pt}{\isacharbrackright}{\kern0pt}{\isasymrrangle}{\isacharparenright}{\kern0pt}\ {\isasymleadsto}\ State{\isasymllangle}{\isasymsigma}\isactrlsub {\isadigit{2}}{\isasymrrangle}{\isachardoublequoteclose}\isanewline
\isanewline
\ \ \isacommand{definition}\isamarkupfalse%
\isanewline
\ \ \ \ {\isasymtau}\isactrlsub {\isadigit{3}}\ {\isacharcolon}{\kern0pt}{\isacharcolon}{\kern0pt}\ {\isasymT}\ \isakeyword{where}\isanewline
\ \ \ \ {\isachardoublequoteopen}{\isasymtau}\isactrlsub {\isadigit{3}}\ {\isasymequiv}\ {\isasymlangle}{\isasymsigma}\isactrlsub {\isadigit{2}}{\isasymrangle}\ {\isasymleadsto}\ State{\isasymllangle}{\isasymcircle}{\isasymrrangle}{\isachardoublequoteclose}%
\isadelimdocument
\endisadelimdocument
\isatagdocument
\isamarkupsubsubsection{Conditioned Symbolic Traces%
}
\isamarkuptrue%
\endisatagdocument
{\isafolddocument}%
\isadelimdocument
\endisadelimdocument
\begin{isamarkuptext}%
Resembling the definition of the original paper, we introduce a path condition
  as a set of Boolean~expressions.%
\end{isamarkuptext}\isamarkuptrue%
\ \ \isacommand{type{\isacharunderscore}{\kern0pt}synonym}\isamarkupfalse%
\ path{\isacharunderscore}{\kern0pt}condition\ {\isacharequal}{\kern0pt}\ {\isachardoublequoteopen}bexp\ set{\isachardoublequoteclose}%
\begin{isamarkuptext}%
A path condition is consistent iff the path condition is concrete and does 
  not contain~false.%
\end{isamarkuptext}\isamarkuptrue%
\ \ \isacommand{definition}\isamarkupfalse%
\ \isanewline
\ \ \ \ consistent\ {\isacharcolon}{\kern0pt}{\isacharcolon}{\kern0pt}\ {\isachardoublequoteopen}path{\isacharunderscore}{\kern0pt}condition\ {\isasymRightarrow}\ bool{\isachardoublequoteclose}\ \isakeyword{where}\isanewline
\ \ \ \ {\isachardoublequoteopen}consistent\ pc\ {\isasymequiv}\ sconcrete\isactrlsub B{\isacharparenleft}{\kern0pt}pc{\isacharparenright}{\kern0pt}\ {\isasymand}\ {\isacharparenleft}{\kern0pt}{\isasymforall}b\ {\isasymin}\ pc{\isachardot}{\kern0pt}\ getBool{\isacharparenleft}{\kern0pt}b{\isacharparenright}{\kern0pt}{\isacharparenright}{\kern0pt}{\isachardoublequoteclose}%
\begin{isamarkuptext}%
We define a conditioned symbolic trace as a combination of a path condition 
  and a symbolic trace. Similar to the paper, we use \isa{{\isasymtriangleright}} as an infix~symbol.%
\end{isamarkuptext}\isamarkuptrue%
\ \ \isacommand{datatype}\isamarkupfalse%
\ {\isasymCC}{\isasymT}\ {\isacharequal}{\kern0pt}\ \isanewline
\ \ \ \ Trace\ path{\isacharunderscore}{\kern0pt}condition\ {\isasymT}\ {\isacharparenleft}{\kern0pt}\isakeyword{infix}\ {\isachardoublequoteopen}{\isasymtriangleright}{\isachardoublequoteclose}\ {\isadigit{5}}{\isadigit{9}}{\isacharparenright}{\kern0pt}%
\begin{isamarkuptext}%
We can now denote projections mapping a conditioned symbolic trace onto its 
  specific components. These functions will serve as handy abbreviations at a
  later~point.%
\end{isamarkuptext}\isamarkuptrue%
\ \ \isacommand{fun}\isamarkupfalse%
\isanewline
\ \ \ \ pc{\isacharunderscore}{\kern0pt}projection\ {\isacharcolon}{\kern0pt}{\isacharcolon}{\kern0pt}\ {\isachardoublequoteopen}{\isasymCC}{\isasymT}\ {\isasymRightarrow}\ path{\isacharunderscore}{\kern0pt}condition{\isachardoublequoteclose}\ {\isacharparenleft}{\kern0pt}{\isachardoublequoteopen}{\isasymDown}\isactrlsub p{\isachardoublequoteclose}{\isacharparenright}{\kern0pt}\ \isakeyword{where}\isanewline
\ \ \ \ {\isachardoublequoteopen}{\isasymDown}\isactrlsub p\ {\isacharparenleft}{\kern0pt}pc\ {\isasymtriangleright}\ {\isasymtau}{\isacharparenright}{\kern0pt}\ {\isacharequal}{\kern0pt}\ pc{\isachardoublequoteclose}\isanewline
\isanewline
\ \ \isacommand{fun}\isamarkupfalse%
\isanewline
\ \ \ \ trace{\isacharunderscore}{\kern0pt}projection\ {\isacharcolon}{\kern0pt}{\isacharcolon}{\kern0pt}\ {\isachardoublequoteopen}{\isasymCC}{\isasymT}\ {\isasymRightarrow}\ {\isasymT}{\isachardoublequoteclose}\ {\isacharparenleft}{\kern0pt}{\isachardoublequoteopen}{\isasymDown}\isactrlsub {\isasymT}{\isachardoublequoteclose}{\isacharparenright}{\kern0pt}\ \isakeyword{where}\isanewline
\ \ \ \ {\isachardoublequoteopen}{\isasymDown}\isactrlsub {\isasymT}\ {\isacharparenleft}{\kern0pt}pc\ {\isasymtriangleright}\ {\isasymtau}{\isacharparenright}{\kern0pt}\ {\isacharequal}{\kern0pt}\ {\isasymtau}{\isachardoublequoteclose}%
\begin{isamarkuptext}%
We show that a consistent path condition must always be concrete. Note that 
  this can be trivially inferred using the consistency~definition.%
\end{isamarkuptext}\isamarkuptrue%
\ \ \isacommand{lemma}\isamarkupfalse%
\ consistent{\isacharunderscore}{\kern0pt}concrete{\isacharunderscore}{\kern0pt}imp{\isacharcolon}{\kern0pt}\ {\isachardoublequoteopen}consistent{\isacharparenleft}{\kern0pt}pc{\isacharparenright}{\kern0pt}\ {\isasymlongrightarrow}\ sconcrete\isactrlsub B{\isacharparenleft}{\kern0pt}pc{\isacharparenright}{\kern0pt}{\isachardoublequoteclose}\isanewline
\isadelimproof
\ \ \ \ %
\endisadelimproof
\isatagproof
\isacommand{by}\isamarkupfalse%
\ {\isacharparenleft}{\kern0pt}simp\ add{\isacharcolon}{\kern0pt}\ consistent{\isacharunderscore}{\kern0pt}def{\isacharparenright}{\kern0pt}%
\endisatagproof
{\isafoldproof}%
\isadelimproof
\endisadelimproof
\begin{isamarkuptext}%
Similar to the properties of the evaluation functions, we establish
  that the consistency of a path condition is preserved when evaluated under
  an arbitrary~state.%
\end{isamarkuptext}\isamarkuptrue%
\ \ \isacommand{lemma}\isamarkupfalse%
\ consistent{\isacharunderscore}{\kern0pt}pr{\isacharcolon}{\kern0pt}\ {\isachardoublequoteopen}consistent\ pc\ {\isasymlongrightarrow}\ consistent\ {\isacharparenleft}{\kern0pt}sval\isactrlsub B\ pc\ {\isasymsigma}{\isacharparenright}{\kern0pt}{\isachardoublequoteclose}\isanewline
\isadelimproof
\ \ \ \ %
\endisadelimproof
\isatagproof
\isacommand{using}\isamarkupfalse%
\ consistent{\isacharunderscore}{\kern0pt}def\ s{\isacharunderscore}{\kern0pt}value{\isacharunderscore}{\kern0pt}pr\isactrlsub B\ \isacommand{by}\isamarkupfalse%
\ auto%
\endisatagproof
{\isafoldproof}%
\isadelimproof
\endisadelimproof
\begin{isamarkuptext}%
We furthermore derive that the consistency of a path condition constructed 
  from an evaluated property \textit{b} also implies the concreteness of the path 
  condition constructed from the evaluated negated property \textit{not b}. This follows
  from the facts that a consistent path condition is guaranteed to be concrete 
  and that the negation of a concrete Boolean expression preserves its concreteness
  if simplified.%
\end{isamarkuptext}\isamarkuptrue%
\ \ \isacommand{lemma}\isamarkupfalse%
\ conc{\isacharunderscore}{\kern0pt}pc{\isacharunderscore}{\kern0pt}pr\isactrlsub B{\isacharcolon}{\kern0pt}\ {\isachardoublequoteopen}consistent\ {\isacharbraceleft}{\kern0pt}val\isactrlsub B\ b\ {\isasymsigma}{\isacharbraceright}{\kern0pt}\ {\isasymlongrightarrow}\ sconcrete\isactrlsub B\ {\isacharbraceleft}{\kern0pt}val\isactrlsub B\ {\isacharparenleft}{\kern0pt}not\ b{\isacharparenright}{\kern0pt}\ {\isasymsigma}{\isacharbraceright}{\kern0pt}{\isachardoublequoteclose}\isanewline
\isadelimproof
\ \ \ \ %
\endisadelimproof
\isatagproof
\isacommand{by}\isamarkupfalse%
\ {\isacharparenleft}{\kern0pt}simp\ add{\isacharcolon}{\kern0pt}\ consistent{\isacharunderscore}{\kern0pt}def{\isacharparenright}{\kern0pt}%
\endisatagproof
{\isafoldproof}%
\isadelimproof
\isanewline
\endisadelimproof
\isanewline
\ \ \isacommand{lemma}\isamarkupfalse%
\ conc{\isacharunderscore}{\kern0pt}pc{\isacharunderscore}{\kern0pt}pr\isactrlsub B\isactrlsub N{\isacharcolon}{\kern0pt}\ {\isachardoublequoteopen}consistent\ {\isacharbraceleft}{\kern0pt}val\isactrlsub B\ {\isacharparenleft}{\kern0pt}not\ b{\isacharparenright}{\kern0pt}\ {\isasymsigma}{\isacharbraceright}{\kern0pt}\ {\isasymlongrightarrow}\ sconcrete\isactrlsub B\ {\isacharbraceleft}{\kern0pt}val\isactrlsub B\ b\ {\isasymsigma}{\isacharbraceright}{\kern0pt}{\isachardoublequoteclose}\isanewline
\isadelimproof
\ \ \ \ %
\endisadelimproof
\isatagproof
\isacommand{by}\isamarkupfalse%
\ {\isacharparenleft}{\kern0pt}simp\ add{\isacharcolon}{\kern0pt}\ consistent{\isacharunderscore}{\kern0pt}def{\isacharparenright}{\kern0pt}%
\endisatagproof
{\isafoldproof}%
\isadelimproof
\endisadelimproof
\begin{isamarkuptext}%
We can now take a look at several examples for conditioned symbolic traces in
  our~model.%
\end{isamarkuptext}\isamarkuptrue%
\ \ \isacommand{definition}\isamarkupfalse%
\isanewline
\ \ \ \ {\isasympi}\isactrlsub {\isadigit{1}}\ {\isacharcolon}{\kern0pt}{\isacharcolon}{\kern0pt}\ {\isasymCC}{\isasymT}\ \isakeyword{where}\isanewline
\ \ \ \ {\isachardoublequoteopen}{\isasympi}\isactrlsub {\isadigit{1}}\ {\isasymequiv}\ {\isacharbraceleft}{\kern0pt}{\isacharbraceright}{\kern0pt}\ {\isasymtriangleright}\ {\isasymtau}\isactrlsub {\isadigit{1}}{\isachardoublequoteclose}\isanewline
\isanewline
\ \ \isacommand{definition}\isamarkupfalse%
\isanewline
\ \ \ \ {\isasympi}\isactrlsub {\isadigit{2}}\ {\isacharcolon}{\kern0pt}{\isacharcolon}{\kern0pt}\ {\isasymCC}{\isasymT}\ \isakeyword{where}\isanewline
\ \ \ \ {\isachardoublequoteopen}{\isasympi}\isactrlsub {\isadigit{2}}\ {\isasymequiv}\ {\isacharbraceleft}{\kern0pt}Bool\ True{\isacharbraceright}{\kern0pt}\ {\isasymtriangleright}\ {\isasymtau}\isactrlsub {\isadigit{2}}{\isachardoublequoteclose}\isanewline
\isanewline
\ \ \isacommand{definition}\isamarkupfalse%
\isanewline
\ \ \ \ {\isasympi}\isactrlsub {\isadigit{3}}\ {\isacharcolon}{\kern0pt}{\isacharcolon}{\kern0pt}\ {\isasymCC}{\isasymT}\ \isakeyword{where}\isanewline
\ \ \ \ {\isachardoublequoteopen}{\isasympi}\isactrlsub {\isadigit{3}}\ {\isasymequiv}\ {\isacharbraceleft}{\kern0pt}Bool\ False{\isacharbraceright}{\kern0pt}\ {\isasymtriangleright}\ {\isasymtau}\isactrlsub {\isadigit{3}}{\isachardoublequoteclose}%
\isadelimdocument
\endisadelimdocument
\isatagdocument
\isamarkupsubsubsection{Trace Modifications%
}
\isamarkuptrue%
\endisatagdocument
{\isafolddocument}%
\isadelimdocument
\endisadelimdocument
\begin{isamarkuptext}%
We provide a straightforward, formal definition of trace concatenation.%
\end{isamarkuptext}\isamarkuptrue%
\ \ \isacommand{primrec}\isamarkupfalse%
\ \isanewline
\ \ \ \ concat\ {\isacharcolon}{\kern0pt}{\isacharcolon}{\kern0pt}\ {\isachardoublequoteopen}{\isasymT}\ {\isasymRightarrow}\ {\isasymT}\ {\isasymRightarrow}\ {\isasymT}{\isachardoublequoteclose}\ {\isacharparenleft}{\kern0pt}\isakeyword{infix}\ {\isachardoublequoteopen}{\isasymcdot}{\isachardoublequoteclose}\ {\isadigit{6}}{\isadigit{0}}{\isacharparenright}{\kern0pt}\ \isakeyword{where}\isanewline
\ \ \ \ {\isachardoublequoteopen}{\isasymtau}\ {\isasymcdot}\ {\isasymepsilon}\ {\isacharequal}{\kern0pt}\ {\isasymtau}{\isachardoublequoteclose}\ {\isacharbar}{\kern0pt}\isanewline
\ \ \ \ {\isachardoublequoteopen}{\isasymtau}\ {\isasymcdot}\ {\isacharparenleft}{\kern0pt}{\isasymtau}{\isacharprime}{\kern0pt}\ {\isasymleadsto}\ t{\isacharparenright}{\kern0pt}\ {\isacharequal}{\kern0pt}\ {\isacharparenleft}{\kern0pt}{\isasymtau}\ {\isasymcdot}\ {\isasymtau}{\isacharprime}{\kern0pt}{\isacharparenright}{\kern0pt}\ {\isasymleadsto}\ t{\isachardoublequoteclose}%
\begin{isamarkuptext}%
We next introduce recursive partial functions that map a symbolic trace onto 
  its first and last occurring state. In order for the function mapping a trace onto 
  its first state to be defined, we demand the given trace to have a state as its 
  initial trace atom. In contrast, the function mapping a trace onto its last element 
  requires the trace to end with a state. This behaviour is not a problem however, as 
  we will later ensure that any wellformed trace starts and ends with~a~state.%
\end{isamarkuptext}\isamarkuptrue%
\ \ \isacommand{fun}\isamarkupfalse%
\ \isanewline
\ \ \ \ first\isactrlsub {\isasymT}\ {\isacharcolon}{\kern0pt}{\isacharcolon}{\kern0pt}\ {\isachardoublequoteopen}{\isasymT}\ {\isasymRightarrow}\ {\isasymSigma}{\isachardoublequoteclose}\ \isakeyword{where}\isanewline
\ \ \ \ {\isachardoublequoteopen}first\isactrlsub {\isasymT}\ {\isasymlangle}{\isasymsigma}{\isasymrangle}\ {\isacharequal}{\kern0pt}\ {\isasymsigma}{\isachardoublequoteclose}\ {\isacharbar}{\kern0pt}\isanewline
\ \ \ \ {\isachardoublequoteopen}first\isactrlsub {\isasymT}\ {\isacharparenleft}{\kern0pt}{\isasymtau}\ {\isasymleadsto}\ t{\isacharparenright}{\kern0pt}\ {\isacharequal}{\kern0pt}\ first\isactrlsub {\isasymT}{\isacharparenleft}{\kern0pt}{\isasymtau}{\isacharparenright}{\kern0pt}{\isachardoublequoteclose}\ {\isacharbar}{\kern0pt}\isanewline
\ \ \ \ {\isachardoublequoteopen}first\isactrlsub {\isasymT}\ {\isasymepsilon}\ {\isacharequal}{\kern0pt}\ undefined{\isachardoublequoteclose}\isanewline
\isanewline
\ \ \isacommand{fun}\isamarkupfalse%
\ \isanewline
\ \ \ \ last\isactrlsub {\isasymT}\ {\isacharcolon}{\kern0pt}{\isacharcolon}{\kern0pt}\ {\isachardoublequoteopen}{\isasymT}\ {\isasymRightarrow}\ {\isasymSigma}{\isachardoublequoteclose}\ \isakeyword{where}\isanewline
\ \ \ \ {\isachardoublequoteopen}last\isactrlsub {\isasymT}\ {\isacharparenleft}{\kern0pt}{\isasymtau}\ {\isasymleadsto}\ State{\isasymllangle}{\isasymsigma}{\isasymrrangle}{\isacharparenright}{\kern0pt}\ {\isacharequal}{\kern0pt}\ {\isasymsigma}{\isachardoublequoteclose}\ {\isacharbar}{\kern0pt}\isanewline
\ \ \ \ {\isachardoublequoteopen}last\isactrlsub {\isasymT}\ {\isacharunderscore}{\kern0pt}\ {\isacharequal}{\kern0pt}\ undefined{\isachardoublequoteclose}%
\begin{isamarkuptext}%
We next give a definition for the semantic chop of two symbolic traces, 
  while slightly deviating from the definition in the original paper.
  In order to perform a semantic chop on two traces, it is vital that both traces align,
  meaning that the last state of the first trace must match with the first state of 
  the second trace. The semantic chop then concatenates both traces, whilst removing 
  one of the duplicates. In contrast to the original paper however, we will not 
  explicitly ensure that both traces align with each other. Instead, the alignment 
  will be established in the formalization of the overlying LAGC semantics. The 
  outsourcing of this property makes sure that Isabelle can generate code for the 
  semantic chop~operation. \par
  Note that we require the first trace to end with a state in order for this 
  function to be defined, thus making this a partial~function.%
\end{isamarkuptext}\isamarkuptrue%
\ \ \isacommand{fun}\isamarkupfalse%
\ \isanewline
\ \ \ \ semantic{\isacharunderscore}{\kern0pt}chop\isactrlsub {\isasymT}\ {\isacharcolon}{\kern0pt}{\isacharcolon}{\kern0pt}\ {\isachardoublequoteopen}{\isasymT}\ {\isasymRightarrow}\ {\isasymT}\ {\isasymRightarrow}\ {\isasymT}{\isachardoublequoteclose}\ {\isacharparenleft}{\kern0pt}\isakeyword{infix}\ {\isachardoublequoteopen}\isactrlemph \isactrlemph {\isachardoublequoteclose}\ {\isadigit{6}}{\isadigit{1}}{\isacharparenright}{\kern0pt}\ \isakeyword{where}\isanewline
\ \ \ \ {\isachardoublequoteopen}{\isacharparenleft}{\kern0pt}{\isasymtau}\ {\isasymleadsto}\ State{\isasymllangle}{\isasymsigma}{\isasymrrangle}{\isacharparenright}{\kern0pt}\ \isactrlemph \isactrlemph \ {\isasymtau}{\isacharprime}{\kern0pt}\ {\isacharequal}{\kern0pt}\ {\isasymtau}\ {\isasymcdot}\ {\isasymtau}{\isacharprime}{\kern0pt}{\isachardoublequoteclose}\ {\isacharbar}{\kern0pt}\isanewline
\ \ \ \ {\isachardoublequoteopen}{\isacharunderscore}{\kern0pt}\ \isactrlemph \isactrlemph \ {\isasymtau}{\isacharprime}{\kern0pt}\ {\isacharequal}{\kern0pt}\ undefined{\isachardoublequoteclose}%
\begin{isamarkuptext}%
We can now heighten the previous definition to the semantic chop between 
  two conditioned symbolic traces. In this scenario, the semantic chop denotes
  the union of both path conditions and the semantic chop of both encased
  symbolic~traces.%
\end{isamarkuptext}\isamarkuptrue%
\ \ \isacommand{fun}\isamarkupfalse%
\ \isanewline
\ \ \ \ semantic{\isacharunderscore}{\kern0pt}chop\isactrlsub {\isasympi}\ {\isacharcolon}{\kern0pt}{\isacharcolon}{\kern0pt}\ {\isachardoublequoteopen}{\isasymCC}{\isasymT}\ {\isasymRightarrow}\ {\isasymCC}{\isasymT}\ {\isasymRightarrow}\ {\isasymCC}{\isasymT}{\isachardoublequoteclose}\ {\isacharparenleft}{\kern0pt}\isakeyword{infix}\ {\isachardoublequoteopen}\isactrlemph \isactrlemph \isactrlsub {\isasympi}{\isachardoublequoteclose}\ {\isadigit{6}}{\isadigit{1}}{\isacharparenright}{\kern0pt}\ \isakeyword{where}\isanewline
\ \ \ \ {\isachardoublequoteopen}{\isacharparenleft}{\kern0pt}pc\isactrlsub {\isadigit{1}}\ {\isasymtriangleright}\ {\isasymtau}{\isacharparenright}{\kern0pt}\ \isactrlemph \isactrlemph \isactrlsub {\isasympi}\ {\isacharparenleft}{\kern0pt}pc\isactrlsub {\isadigit{2}}\ {\isasymtriangleright}\ {\isasymtau}{\isacharprime}{\kern0pt}{\isacharparenright}{\kern0pt}\ {\isacharequal}{\kern0pt}\ {\isacharparenleft}{\kern0pt}pc\isactrlsub {\isadigit{1}}\ {\isasymunion}\ pc\isactrlsub {\isadigit{2}}{\isacharparenright}{\kern0pt}\ {\isasymtriangleright}\ {\isacharparenleft}{\kern0pt}{\isasymtau}\ \isactrlemph \isactrlemph \ {\isasymtau}{\isacharprime}{\kern0pt}{\isacharparenright}{\kern0pt}{\isachardoublequoteclose}%
\begin{isamarkuptext}%
In the following, we provide an example for the semantic chop~operation.%
\end{isamarkuptext}\isamarkuptrue%
\ \ \isacommand{lemma}\isamarkupfalse%
\ {\isachardoublequoteopen}{\isasympi}\isactrlsub {\isadigit{1}}\ \isactrlemph \isactrlemph \isactrlsub {\isasympi}\ {\isasympi}\isactrlsub {\isadigit{3}}\ {\isacharequal}{\kern0pt}\ {\isacharbraceleft}{\kern0pt}Bool\ False{\isacharbraceright}{\kern0pt}\ {\isasymtriangleright}\ {\isacharparenleft}{\kern0pt}{\isacharparenleft}{\kern0pt}{\isasymlangle}{\isasymsigma}\isactrlsub {\isadigit{1}}{\isasymrangle}\ {\isasymleadsto}\ Event{\isasymllangle}inpEv{\isacharcomma}{\kern0pt}\ {\isacharbrackleft}{\kern0pt}{\isacharbrackright}{\kern0pt}{\isasymrrangle}{\isacharparenright}{\kern0pt}\ {\isasymleadsto}\ State{\isasymllangle}{\isasymsigma}\isactrlsub {\isadigit{2}}{\isasymrrangle}{\isacharparenright}{\kern0pt}\ {\isasymleadsto}\ State{\isasymllangle}{\isasymcircle}{\isasymrrangle}{\isachardoublequoteclose}\isanewline
\isadelimproof
\ \ \ \ %
\endisadelimproof
\isatagproof
\isacommand{by}\isamarkupfalse%
\ {\isacharparenleft}{\kern0pt}auto\ simp\ add{\isacharcolon}{\kern0pt}\ {\isasymtau}\isactrlsub {\isadigit{1}}{\isacharunderscore}{\kern0pt}def\ {\isasymtau}\isactrlsub {\isadigit{3}}{\isacharunderscore}{\kern0pt}def\ {\isasympi}\isactrlsub {\isadigit{1}}{\isacharunderscore}{\kern0pt}def\ {\isasympi}\isactrlsub {\isadigit{3}}{\isacharunderscore}{\kern0pt}def{\isacharparenright}{\kern0pt}%
\endisatagproof
{\isafoldproof}%
\isadelimproof
\endisadelimproof
\isadelimdocument
\endisadelimdocument
\isatagdocument
\isamarkupsubsubsection{Event Insertions%
}
\isamarkuptrue%
\endisatagdocument
{\isafolddocument}%
\isadelimdocument
\endisadelimdocument
\begin{isamarkuptext}%
The wellformedness notion for traces will require each event occurring in a
  trace to be surrounded by identical states. In fashion to the original paper, 
  we can now introduce a function that automatically surrounds a given
  event with a particular state. This function can later be used to 
  abbreviate the insertion of specific events into given traces whilst 
  preserving~wellformedness. Note that the inserted event is simplified before
  the actual~insertion.%
\end{isamarkuptext}\isamarkuptrue%
\ \ \isacommand{definition}\isamarkupfalse%
\ {\isacharbrackleft}{\kern0pt}simp{\isacharbrackright}{\kern0pt}{\isacharcolon}{\kern0pt}\isanewline
\ \ \ \ {\isachardoublequoteopen}gen{\isacharunderscore}{\kern0pt}event\ ev\ {\isasymsigma}\ e\ {\isasymequiv}\ {\isacharparenleft}{\kern0pt}{\isasymlangle}{\isasymsigma}{\isasymrangle}\ {\isasymleadsto}\ Event{\isasymllangle}ev{\isacharcomma}{\kern0pt}\ lval\isactrlsub E\ e\ {\isasymsigma}{\isasymrrangle}{\isacharparenright}{\kern0pt}\ {\isasymleadsto}\ State{\isasymllangle}{\isasymsigma}{\isasymrrangle}{\isachardoublequoteclose}%
\begin{isamarkuptext}%
We can now establish the equality of the first and last state of any symbolic trace 
  generated by the previous definition, implying that the states surrounding the 
  event~match.%
\end{isamarkuptext}\isamarkuptrue%
\ \ \isacommand{lemma}\isamarkupfalse%
\ gen{\isacharunderscore}{\kern0pt}event{\isacharunderscore}{\kern0pt}match\isactrlsub {\isasymSigma}{\isacharcolon}{\kern0pt}\ {\isachardoublequoteopen}first\isactrlsub {\isasymT}\ {\isacharparenleft}{\kern0pt}gen{\isacharunderscore}{\kern0pt}event\ ev\ {\isasymsigma}\ e{\isacharparenright}{\kern0pt}\ {\isacharequal}{\kern0pt}\ last\isactrlsub {\isasymT}\ {\isacharparenleft}{\kern0pt}gen{\isacharunderscore}{\kern0pt}event\ ev\ {\isasymsigma}\ e{\isacharparenright}{\kern0pt}{\isachardoublequoteclose}\isanewline
\isadelimproof
\ \ \ \ %
\endisadelimproof
\isatagproof
\isacommand{by}\isamarkupfalse%
\ simp%
\endisatagproof
{\isafoldproof}%
\isadelimproof
\endisadelimproof
\isadelimdocument
\endisadelimdocument
\isatagdocument
\isamarkupsubsubsection{Wellformedness%
}
\isamarkuptrue%
\endisatagdocument
{\isafolddocument}%
\isadelimdocument
\endisadelimdocument
\begin{isamarkuptext}%
Before we begin with the formalization of the wellformedness predicate,
  we first denote an inductive function that maps a trace onto a set of all its 
  symbolic variables, i.e. all variables that occur symbolic in at least one state 
  of the trace. The definition of the function is straightforward. We unfold the
  trace from the back, whilst recursively collecting its symbolic~variables.%
\end{isamarkuptext}\isamarkuptrue%
\ \ \isacommand{fun}\isamarkupfalse%
\isanewline
\ \ \ \ symb\isactrlsub {\isasymT}\ {\isacharcolon}{\kern0pt}{\isacharcolon}{\kern0pt}\ {\isachardoublequoteopen}{\isasymT}\ {\isasymRightarrow}\ var\ set{\isachardoublequoteclose}\ \isakeyword{where}\isanewline
\ \ \ \ {\isachardoublequoteopen}symb\isactrlsub {\isasymT}\ {\isasymepsilon}\ {\isacharequal}{\kern0pt}\ {\isacharbraceleft}{\kern0pt}{\isacharbraceright}{\kern0pt}{\isachardoublequoteclose}\ {\isacharbar}{\kern0pt}\isanewline
\ \ \ \ {\isachardoublequoteopen}symb\isactrlsub {\isasymT}\ {\isacharparenleft}{\kern0pt}{\isasymtau}\ {\isasymleadsto}\ Event{\isasymllangle}ev{\isacharcomma}{\kern0pt}\ e{\isasymrrangle}{\isacharparenright}{\kern0pt}\ {\isacharequal}{\kern0pt}\ symb\isactrlsub {\isasymT}{\isacharparenleft}{\kern0pt}{\isasymtau}{\isacharparenright}{\kern0pt}{\isachardoublequoteclose}\ {\isacharbar}{\kern0pt}\isanewline
\ \ \ \ {\isachardoublequoteopen}symb\isactrlsub {\isasymT}\ {\isacharparenleft}{\kern0pt}{\isasymtau}\ {\isasymleadsto}\ State{\isasymllangle}{\isasymsigma}{\isasymrrangle}{\isacharparenright}{\kern0pt}\ {\isacharequal}{\kern0pt}\ symb\isactrlsub {\isasymSigma}{\isacharparenleft}{\kern0pt}{\isasymsigma}{\isacharparenright}{\kern0pt}\ {\isasymunion}\ symb\isactrlsub {\isasymT}{\isacharparenleft}{\kern0pt}{\isasymtau}{\isacharparenright}{\kern0pt}{\isachardoublequoteclose}%
\begin{isamarkuptext}%
Using the previous function definition, we can now establish a wellformedness
  notion for conditioned symbolic traces. Similar to the paper, we demand the 
  following five properties~to~hold: \begin{description} \par

  \item[Condition 1] All variables that occur non-symbolic in any state of the given 
  trace are not allowed to occur symbolic anywhere in the same~trace. \par
  In order to formalize this property, it is conceivable to define a function 
  that collects all states of a trace and then quantifies over all these states, 
  thereby closely following the formalization in the paper. However, we instead 
  decide to provide an inductive definition of the property, as this ensures that we 
  can later provide simple inductive proofs without another layer of~complexity. \par
  \item[Condition 2] All variables occurring in the path condition of the conditioned
  symbolic trace must be symbolic variables of the~trace. \par
  The formalization of this property is trivial, matching the definition
  in the~paper. \par
  \item[Condition 3] All variables occurring in expressions of trace events must be 
  symbolic variables of the~trace. \par
  Similar to the first condition, we again do not quantify over all events, 
  but provide an inductive property instead, thus also ensuring that our 
  formalizations are consistent in its design~choices. \par
  \item[Condition 4] All events occurring in a trace must be surrounded by 
  identical~states. \par
  The formalization of this property strongly deviates from the paper, as we
  again choose to give an inductive formalization instead of quantifying over all
  events in order to uphold consistency. \par
  We split the formalization of this property into two separate functions. The first
  function uses trace pattern matching to check that any two states, which occur next
  to trace events, match. The second function encases the first function, whilst
  additionally verifying that the trace does not start or end with~an~event. \par
  \item[Condition 5] All states of the trace must be wellformed. \par
  The formalization of this property is also straightforward by making use
  of~recursion. \par
  \end{description} We can now call a conditioned symbolic trace wellformed iff 
  all five wellformedness conditions are~satisfied.%
\end{isamarkuptext}\isamarkuptrue%
\ \ \isacommand{fun}\isamarkupfalse%
\ %
\isamarkupcmt{Condition (1)%
}\isanewline
\ \ \ \ non{\isacharunderscore}{\kern0pt}symb{\isacharunderscore}{\kern0pt}disjunct\isactrlsub {\isasymT}\ {\isacharcolon}{\kern0pt}{\isacharcolon}{\kern0pt}\ {\isachardoublequoteopen}{\isasymT}\ {\isasymRightarrow}\ var\ set\ {\isasymRightarrow}\ bool{\isachardoublequoteclose}\ {\isacharparenleft}{\kern0pt}\isakeyword{infix}\ {\isachardoublequoteopen}{\isasymdiamondop}{\isasymdiamondop}\isactrlsub {\isasymT}{\isachardoublequoteclose}\ {\isadigit{6}}{\isadigit{4}}{\isacharparenright}{\kern0pt}\ \isakeyword{where}\isanewline
\ \ \ \ {\isachardoublequoteopen}{\isasymepsilon}\ {\isasymdiamondop}{\isasymdiamondop}\isactrlsub {\isasymT}\ V\ {\isacharequal}{\kern0pt}\ True{\isachardoublequoteclose}\ {\isacharbar}{\kern0pt}\isanewline
\ \ \ \ {\isachardoublequoteopen}{\isacharparenleft}{\kern0pt}{\isasymtau}\ {\isasymleadsto}\ Event{\isasymllangle}ev{\isacharcomma}{\kern0pt}\ e{\isasymrrangle}{\isacharparenright}{\kern0pt}\ {\isasymdiamondop}{\isasymdiamondop}\isactrlsub {\isasymT}\ V\ {\isacharequal}{\kern0pt}\ {\isasymtau}\ {\isasymdiamondop}{\isasymdiamondop}\isactrlsub {\isasymT}\ V{\isachardoublequoteclose}\ {\isacharbar}{\kern0pt}\isanewline
\ \ \ \ {\isachardoublequoteopen}{\isacharparenleft}{\kern0pt}{\isasymtau}\ {\isasymleadsto}\ State{\isasymllangle}{\isasymsigma}{\isasymrrangle}{\isacharparenright}{\kern0pt}\ {\isasymdiamondop}{\isasymdiamondop}\isactrlsub {\isasymT}\ V\ {\isacharequal}{\kern0pt}\ {\isacharparenleft}{\kern0pt}{\isacharparenleft}{\kern0pt}fmdom{\isacharprime}{\kern0pt}{\isacharparenleft}{\kern0pt}{\isasymsigma}{\isacharparenright}{\kern0pt}\ {\isacharminus}{\kern0pt}\ symb\isactrlsub {\isasymSigma}{\isacharparenleft}{\kern0pt}{\isasymsigma}{\isacharparenright}{\kern0pt}{\isacharparenright}{\kern0pt}\ {\isasyminter}\ V\ {\isacharequal}{\kern0pt}\ {\isacharbraceleft}{\kern0pt}{\isacharbraceright}{\kern0pt}\ {\isasymand}\ {\isasymtau}\ {\isasymdiamondop}{\isasymdiamondop}\isactrlsub {\isasymT}\ V{\isacharparenright}{\kern0pt}{\isachardoublequoteclose}\isanewline
\isanewline
\ \ \isacommand{fun}\isamarkupfalse%
\ %
\isamarkupcmt{Condition (2)%
}\isanewline
\ \ \ \ wf{\isacharunderscore}{\kern0pt}pc\isactrlsub {\isasympi}\ {\isacharcolon}{\kern0pt}{\isacharcolon}{\kern0pt}\ {\isachardoublequoteopen}path{\isacharunderscore}{\kern0pt}condition\ {\isasymRightarrow}\ var\ set\ {\isasymRightarrow}\ bool{\isachardoublequoteclose}\ {\isacharparenleft}{\kern0pt}\isakeyword{infix}\ {\isachardoublequoteopen}{\isasymsubseteq}\isactrlsub p{\isachardoublequoteclose}\ {\isadigit{6}}{\isadigit{4}}{\isacharparenright}{\kern0pt}\ \isakeyword{where}\isanewline
\ \ \ \ {\isachardoublequoteopen}pc\ {\isasymsubseteq}\isactrlsub p\ V\ {\isacharequal}{\kern0pt}\ {\isacharparenleft}{\kern0pt}svars\isactrlsub B{\isacharparenleft}{\kern0pt}pc{\isacharparenright}{\kern0pt}\ {\isasymsubseteq}\ V{\isacharparenright}{\kern0pt}{\isachardoublequoteclose}\isanewline
\isanewline
\ \ \isacommand{fun}\isamarkupfalse%
\ %
\isamarkupcmt{Condition (3)%
}\isanewline
\ \ \ \ wf{\isacharunderscore}{\kern0pt}events\isactrlsub {\isasymT}\ {\isacharcolon}{\kern0pt}{\isacharcolon}{\kern0pt}\ {\isachardoublequoteopen}{\isasymT}\ {\isasymRightarrow}\ var\ set\ {\isasymRightarrow}\ bool{\isachardoublequoteclose}\ {\isacharparenleft}{\kern0pt}\isakeyword{infix}\ {\isachardoublequoteopen}{\isasymsubseteq}\isactrlsub E{\isachardoublequoteclose}\ {\isadigit{6}}{\isadigit{4}}{\isacharparenright}{\kern0pt}\ \isakeyword{where}\ \isanewline
\ \ \ \ {\isachardoublequoteopen}{\isasymepsilon}\ {\isasymsubseteq}\isactrlsub E\ V\ {\isacharequal}{\kern0pt}\ True{\isachardoublequoteclose}\ {\isacharbar}{\kern0pt}\isanewline
\ \ \ \ {\isachardoublequoteopen}{\isacharparenleft}{\kern0pt}{\isasymtau}\ {\isasymleadsto}\ State{\isasymllangle}{\isasymsigma}{\isasymrrangle}{\isacharparenright}{\kern0pt}\ {\isasymsubseteq}\isactrlsub E\ V\ {\isacharequal}{\kern0pt}\ {\isasymtau}\ {\isasymsubseteq}\isactrlsub E\ V{\isachardoublequoteclose}\ {\isacharbar}{\kern0pt}\isanewline
\ \ \ \ {\isachardoublequoteopen}{\isacharparenleft}{\kern0pt}{\isasymtau}\ {\isasymleadsto}\ Event{\isasymllangle}ev{\isacharcomma}{\kern0pt}\ e{\isasymrrangle}{\isacharparenright}{\kern0pt}\ {\isasymsubseteq}\isactrlsub E\ V\ {\isacharequal}{\kern0pt}\ {\isacharparenleft}{\kern0pt}lvars\isactrlsub E{\isacharparenleft}{\kern0pt}e{\isacharparenright}{\kern0pt}\ {\isasymsubseteq}\ V\ {\isasymand}\ {\isasymtau}\ {\isasymsubseteq}\isactrlsub E\ V{\isacharparenright}{\kern0pt}{\isachardoublequoteclose}\isanewline
\isanewline
\ \ \isacommand{fun}\isamarkupfalse%
\ %
\isamarkupcmt{Condition (4a)%
}\isanewline
\ \ \ \ wf{\isacharunderscore}{\kern0pt}surround\isactrlsub {\isasymT}\ {\isacharcolon}{\kern0pt}{\isacharcolon}{\kern0pt}\ {\isachardoublequoteopen}{\isasymT}\ {\isasymRightarrow}\ bool{\isachardoublequoteclose}\ \isakeyword{where}\isanewline
\ \ \ \ {\isachardoublequoteopen}wf{\isacharunderscore}{\kern0pt}surround\isactrlsub {\isasymT}\ {\isasymepsilon}\ {\isacharequal}{\kern0pt}\ True{\isachardoublequoteclose}\ {\isacharbar}{\kern0pt}\isanewline
\ \ \ \ {\isachardoublequoteopen}wf{\isacharunderscore}{\kern0pt}surround\isactrlsub {\isasymT}\ {\isacharparenleft}{\kern0pt}{\isacharparenleft}{\kern0pt}{\isacharparenleft}{\kern0pt}{\isasymtau}\ {\isasymleadsto}\ State{\isasymllangle}{\isasymsigma}{\isacharprime}{\kern0pt}{\isasymrrangle}{\isacharparenright}{\kern0pt}\ {\isasymleadsto}\ Event{\isasymllangle}ev{\isacharcomma}{\kern0pt}\ e{\isasymrrangle}{\isacharparenright}{\kern0pt}\ {\isasymleadsto}\ State{\isasymllangle}{\isasymsigma}{\isasymrrangle}{\isacharparenright}{\kern0pt}\ {\isacharequal}{\kern0pt}\ {\isacharparenleft}{\kern0pt}{\isasymsigma}\ {\isacharequal}{\kern0pt}\ {\isasymsigma}{\isacharprime}{\kern0pt}\ {\isasymand}\ wf{\isacharunderscore}{\kern0pt}surround\isactrlsub {\isasymT}{\isacharparenleft}{\kern0pt}{\isasymtau}\ {\isasymleadsto}\ State{\isasymllangle}{\isasymsigma}{\isacharprime}{\kern0pt}{\isasymrrangle}{\isacharparenright}{\kern0pt}{\isacharparenright}{\kern0pt}{\isachardoublequoteclose}\ {\isacharbar}{\kern0pt}\isanewline
\ \ \ \ {\isachardoublequoteopen}wf{\isacharunderscore}{\kern0pt}surround\isactrlsub {\isasymT}\ {\isacharparenleft}{\kern0pt}{\isasymtau}\ {\isasymleadsto}\ ta{\isacharparenright}{\kern0pt}\ {\isacharequal}{\kern0pt}\ wf{\isacharunderscore}{\kern0pt}surround\isactrlsub {\isasymT}{\isacharparenleft}{\kern0pt}{\isasymtau}{\isacharparenright}{\kern0pt}{\isachardoublequoteclose}\isanewline
\isanewline
\ \ \isacommand{fun}\isamarkupfalse%
\ %
\isamarkupcmt{Condition (4b)%
}\isanewline
\ \ \ \ wf{\isacharunderscore}{\kern0pt}seq\isactrlsub {\isasymT}\ {\isacharcolon}{\kern0pt}{\isacharcolon}{\kern0pt}\ {\isachardoublequoteopen}{\isasymT}\ {\isasymRightarrow}\ bool{\isachardoublequoteclose}\ \isakeyword{where}\isanewline
\ \ \ \ {\isachardoublequoteopen}wf{\isacharunderscore}{\kern0pt}seq\isactrlsub {\isasymT}\ {\isasymtau}\ {\isacharequal}{\kern0pt}\ {\isacharparenleft}{\kern0pt}wf{\isacharunderscore}{\kern0pt}surround\isactrlsub {\isasymT}{\isacharparenleft}{\kern0pt}{\isasymtau}{\isacharparenright}{\kern0pt}\ {\isasymand}\ {\isacharparenleft}{\kern0pt}{\isasymexists}{\isasymsigma}\ {\isasymsigma}{\isacharprime}{\kern0pt}{\isachardot}{\kern0pt}\ first\isactrlsub {\isasymT}{\isacharparenleft}{\kern0pt}{\isasymtau}{\isacharparenright}{\kern0pt}\ {\isacharequal}{\kern0pt}\ {\isasymsigma}\ {\isasymand}\ last\isactrlsub {\isasymT}{\isacharparenleft}{\kern0pt}{\isasymtau}{\isacharparenright}{\kern0pt}\ {\isacharequal}{\kern0pt}\ {\isasymsigma}{\isacharprime}{\kern0pt}{\isacharparenright}{\kern0pt}{\isacharparenright}{\kern0pt}{\isachardoublequoteclose}\isanewline
\isanewline
\ \ \isacommand{fun}\isamarkupfalse%
\ %
\isamarkupcmt{Condition (5)%
}\isanewline
\ \ \ \ wf{\isacharunderscore}{\kern0pt}states\isactrlsub {\isasymT}\ {\isacharcolon}{\kern0pt}{\isacharcolon}{\kern0pt}\ {\isachardoublequoteopen}{\isasymT}\ {\isasymRightarrow}\ bool{\isachardoublequoteclose}\ \isakeyword{where}\isanewline
\ \ \ \ {\isachardoublequoteopen}wf{\isacharunderscore}{\kern0pt}states\isactrlsub {\isasymT}\ {\isasymepsilon}\ {\isacharequal}{\kern0pt}\ True{\isachardoublequoteclose}\ {\isacharbar}{\kern0pt}\isanewline
\ \ \ \ {\isachardoublequoteopen}wf{\isacharunderscore}{\kern0pt}states\isactrlsub {\isasymT}\ {\isacharparenleft}{\kern0pt}{\isasymtau}\ {\isasymleadsto}\ Event{\isasymllangle}ev{\isacharcomma}{\kern0pt}\ e{\isasymrrangle}{\isacharparenright}{\kern0pt}\ {\isacharequal}{\kern0pt}\ wf{\isacharunderscore}{\kern0pt}states\isactrlsub {\isasymT}{\isacharparenleft}{\kern0pt}{\isasymtau}{\isacharparenright}{\kern0pt}{\isachardoublequoteclose}\ {\isacharbar}{\kern0pt}\isanewline
\ \ \ \ {\isachardoublequoteopen}wf{\isacharunderscore}{\kern0pt}states\isactrlsub {\isasymT}\ {\isacharparenleft}{\kern0pt}{\isasymtau}\ {\isasymleadsto}\ State{\isasymllangle}{\isasymsigma}{\isasymrrangle}{\isacharparenright}{\kern0pt}\ {\isacharequal}{\kern0pt}\ {\isacharparenleft}{\kern0pt}wf\isactrlsub {\isasymSigma}{\isacharparenleft}{\kern0pt}{\isasymsigma}{\isacharparenright}{\kern0pt}\ {\isasymand}\ wf{\isacharunderscore}{\kern0pt}states\isactrlsub {\isasymT}{\isacharparenleft}{\kern0pt}{\isasymtau}{\isacharparenright}{\kern0pt}{\isacharparenright}{\kern0pt}{\isachardoublequoteclose}\isanewline
\isanewline
\ \ \isacommand{definition}\isamarkupfalse%
\ %
\isamarkupcmt{Wellformedness%
}\isanewline
\ \ \ \ wf\isactrlsub {\isasympi}\ {\isacharcolon}{\kern0pt}{\isacharcolon}{\kern0pt}\ {\isachardoublequoteopen}{\isasymCC}{\isasymT}\ {\isasymRightarrow}\ bool{\isachardoublequoteclose}\ \isakeyword{where}\isanewline
\ \ \ \ {\isachardoublequoteopen}wf\isactrlsub {\isasympi}\ {\isasympi}\ {\isasymequiv}\ {\isacharparenleft}{\kern0pt}{\isasymDown}\isactrlsub {\isasymT}\ {\isasympi}\ {\isasymdiamondop}{\isasymdiamondop}\isactrlsub {\isasymT}\ symb\isactrlsub {\isasymT}\ {\isacharparenleft}{\kern0pt}{\isasymDown}\isactrlsub {\isasymT}\ {\isasympi}{\isacharparenright}{\kern0pt}\ {\isasymand}\ \isanewline
\ \ \ \ \ \ \ \ \ \ \ \ \ \ {\isasymDown}\isactrlsub p\ {\isasympi}\ {\isasymsubseteq}\isactrlsub p\ symb\isactrlsub {\isasymT}\ {\isacharparenleft}{\kern0pt}{\isasymDown}\isactrlsub {\isasymT}\ {\isasympi}{\isacharparenright}{\kern0pt}\ {\isasymand}\ \isanewline
\ \ \ \ \ \ \ \ \ \ \ \ \ \ {\isasymDown}\isactrlsub {\isasymT}\ {\isasympi}\ {\isasymsubseteq}\isactrlsub E\ symb\isactrlsub {\isasymT}\ {\isacharparenleft}{\kern0pt}{\isasymDown}\isactrlsub {\isasymT}\ {\isasympi}{\isacharparenright}{\kern0pt}\ {\isasymand}\ \isanewline
\ \ \ \ \ \ \ \ \ \ \ \ \ \ wf{\isacharunderscore}{\kern0pt}seq\isactrlsub {\isasymT}\ {\isacharparenleft}{\kern0pt}{\isasymDown}\isactrlsub {\isasymT}\ {\isasympi}{\isacharparenright}{\kern0pt}\ {\isasymand}\ \isanewline
\ \ \ \ \ \ \ \ \ \ \ \ \ \ wf{\isacharunderscore}{\kern0pt}states\isactrlsub {\isasymT}\ {\isacharparenleft}{\kern0pt}{\isasymDown}\isactrlsub {\isasymT}\ {\isasympi}{\isacharparenright}{\kern0pt}{\isacharparenright}{\kern0pt}{\isachardoublequoteclose}%
\begin{isamarkuptext}%
We establish that the symbolic variables of two arbitrary traces
  match the symbolic variables of the concatenation of both traces, thereby showing that
  the symbolic function for traces is homomorphic w.r.t. union~and~concatenation.%
\end{isamarkuptext}\isamarkuptrue%
\ \ \isacommand{lemma}\isamarkupfalse%
\ symb{\isacharunderscore}{\kern0pt}union{\isacharcolon}{\kern0pt}\ {\isachardoublequoteopen}symb\isactrlsub {\isasymT}\ {\isasymtau}\ {\isasymunion}\ symb\isactrlsub {\isasymT}\ {\isasymtau}{\isacharprime}{\kern0pt}\ {\isacharequal}{\kern0pt}\ symb\isactrlsub {\isasymT}{\isacharparenleft}{\kern0pt}{\isasymtau}\ {\isasymcdot}\ {\isasymtau}{\isacharprime}{\kern0pt}{\isacharparenright}{\kern0pt}{\isachardoublequoteclose}\isanewline
\isadelimproof
\ \ %
\endisadelimproof
\isatagproof
\isacommand{proof}\isamarkupfalse%
\ {\isacharparenleft}{\kern0pt}induct\ {\isasymtau}{\isacharprime}{\kern0pt}{\isacharparenright}{\kern0pt}\isanewline
\ \ %
\isamarkupcmt{We prove the theorem via induction over trace \isa{{\isasymtau}{\isacharprime}{\kern0pt}}.%
}\isanewline
\ \ \ \ \isacommand{case}\isamarkupfalse%
\ Epsilon\isanewline
\ \ \ \ \isacommand{thus}\isamarkupfalse%
\ {\isacharquery}{\kern0pt}case\ \isacommand{by}\isamarkupfalse%
\ simp\isanewline
\ \ \ \ %
\isamarkupcmt{If \isa{{\isasymtau}{\isacharprime}{\kern0pt}} is \isa{{\isasymepsilon}}, the case is trivial, because \isa{symb\isactrlsub {\isasymT}{\isacharparenleft}{\kern0pt}{\isasymepsilon}{\isacharparenright}{\kern0pt}\ {\isacharequal}{\kern0pt}\ {\isacharbraceleft}{\kern0pt}{\isacharbraceright}{\kern0pt}}.%
}\isanewline
\ \ \isacommand{next}\isamarkupfalse%
\isanewline
\ \ \ \ \isacommand{case}\isamarkupfalse%
\ {\isacharparenleft}{\kern0pt}Transition\ {\isasymtau}{\isacharprime}{\kern0pt}{\isacharprime}{\kern0pt}\ ta{\isacharparenright}{\kern0pt}\isanewline
\ \ \ \ \isacommand{thus}\isamarkupfalse%
\ {\isacharquery}{\kern0pt}case\isanewline
\ \ \ \ \ \ \isacommand{using}\isamarkupfalse%
\ Transition{\isachardot}{\kern0pt}hyps\ \isacommand{by}\isamarkupfalse%
\ {\isacharparenleft}{\kern0pt}induct\ ta{\isacharsemicolon}{\kern0pt}\ auto{\isacharparenright}{\kern0pt}\isanewline
\ \ \ \ %
\isamarkupcmt{If \isa{{\isasymtau}{\isacharprime}{\kern0pt}} is \isa{{\isacharparenleft}{\kern0pt}{\isasymtau}{\isacharprime}{\kern0pt}{\isacharprime}{\kern0pt}\ {\isasymleadsto}\ ta{\isacharparenright}{\kern0pt}}, we use our induction hypothesis and a case
       distinction regarding the nature of trace atom \isa{ta} to 
       automatically close the case.%
}\isanewline
\ \ \isacommand{qed}\isamarkupfalse%
\endisatagproof
{\isafoldproof}%
\isadelimproof
\endisadelimproof
\begin{isamarkuptext}%
We take another look at our example traces. As can be reasoned, 
  \isa{{\isasymtau}\isactrlsub {\isadigit{2}}} is not wellformed, as it already violates the first wellformedness
  condition. $y$ occurs non-symbolic in \isa{{\isasymsigma}\isactrlsub {\isadigit{2}}}, but symbolic in \isa{{\isasymsigma}\isactrlsub {\isadigit{1}}}, which is a state 
  of the same trace. However, \isa{{\isasymtau}\isactrlsub {\isadigit{1}}} and \isa{{\isasymtau}\isactrlsub {\isadigit{3}}} are wellformed, as all five conditions 
  are satisfied. This conclusion can also be inferred using the Isabelle~system.%
\end{isamarkuptext}\isamarkuptrue%
\ \ \isacommand{lemma}\isamarkupfalse%
\ {\isachardoublequoteopen}wf\isactrlsub {\isasympi}\ {\isasympi}\isactrlsub {\isadigit{1}}\ {\isasymand}\ {\isasymnot}wf\isactrlsub {\isasympi}\ {\isasympi}\isactrlsub {\isadigit{2}}\ {\isasymand}\ wf\isactrlsub {\isasympi}\ {\isasympi}\isactrlsub {\isadigit{3}}{\isachardoublequoteclose}\isanewline
\isadelimproof
\ \ \ \ %
\endisadelimproof
\isatagproof
\isacommand{by}\isamarkupfalse%
\ {\isacharparenleft}{\kern0pt}auto\ simp\ add{\isacharcolon}{\kern0pt}\ {\isasymsigma}\isactrlsub {\isadigit{1}}{\isacharunderscore}{\kern0pt}def\ {\isasymsigma}\isactrlsub {\isadigit{2}}{\isacharunderscore}{\kern0pt}def\ {\isasymtau}\isactrlsub {\isadigit{1}}{\isacharunderscore}{\kern0pt}def\ {\isasymtau}\isactrlsub {\isadigit{2}}{\isacharunderscore}{\kern0pt}def\ {\isasymtau}\isactrlsub {\isadigit{3}}{\isacharunderscore}{\kern0pt}def\ {\isasympi}\isactrlsub {\isadigit{1}}{\isacharunderscore}{\kern0pt}def\ {\isasympi}\isactrlsub {\isadigit{2}}{\isacharunderscore}{\kern0pt}def\ {\isasympi}\isactrlsub {\isadigit{3}}{\isacharunderscore}{\kern0pt}def\ symb\isactrlsub {\isasymSigma}{\isacharunderscore}{\kern0pt}def\ wf\isactrlsub {\isasymSigma}{\isacharunderscore}{\kern0pt}def\ wf\isactrlsub {\isasympi}{\isacharunderscore}{\kern0pt}def{\isacharparenright}{\kern0pt}%
\endisatagproof
{\isafoldproof}%
\isadelimproof
\endisadelimproof
\isadelimdocument
\endisadelimdocument
\isatagdocument
\isamarkupsubsubsection{Concreteness%
}
\isamarkuptrue%
\endisatagdocument
{\isafolddocument}%
\isadelimdocument
\endisadelimdocument
\begin{isamarkuptext}%
We will now provide a definition for the concreteness of traces.
  Note that we are going to deviate from the original paper by clearly separating 
  the concreteness notion of symbolic traces from the concreteness notion of 
  conditioned symbolic~traces. \par
  We consider a symbolic trace concrete iff all states and event expressions
  occurring in the trace are concrete. In contrast to the original paper, we do not 
  require a concrete symbolic trace to satisfy any wellformedness condition. This design 
  choice strongly simplifies the original property of the paper, thus later ensuring 
  a significant reduction of the size of trace concreteness proofs. We propose an
  extension with wellformedness guarantees as a possible idea to further 
  develop~this~model.%
\end{isamarkuptext}\isamarkuptrue%
\ \ \isacommand{fun}\isamarkupfalse%
\isanewline
\ \ \ \ concrete\isactrlsub {\isasymT}\ {\isacharcolon}{\kern0pt}{\isacharcolon}{\kern0pt}\ {\isachardoublequoteopen}{\isasymT}\ {\isasymRightarrow}\ bool{\isachardoublequoteclose}\ \isakeyword{where}\isanewline
\ \ \ \ {\isachardoublequoteopen}concrete\isactrlsub {\isasymT}\ {\isasymepsilon}\ {\isacharequal}{\kern0pt}\ True{\isachardoublequoteclose}\ {\isacharbar}{\kern0pt}\isanewline
\ \ \ \ {\isachardoublequoteopen}concrete\isactrlsub {\isasymT}\ {\isacharparenleft}{\kern0pt}{\isasymtau}\ {\isasymleadsto}\ State{\isasymllangle}{\isasymsigma}{\isasymrrangle}{\isacharparenright}{\kern0pt}\ {\isacharequal}{\kern0pt}\ {\isacharparenleft}{\kern0pt}concrete\isactrlsub {\isasymSigma}\ {\isasymsigma}\ {\isasymand}\ concrete\isactrlsub {\isasymT}\ {\isasymtau}{\isacharparenright}{\kern0pt}{\isachardoublequoteclose}\ {\isacharbar}{\kern0pt}\isanewline
\ \ \ \ {\isachardoublequoteopen}concrete\isactrlsub {\isasymT}\ {\isacharparenleft}{\kern0pt}{\isasymtau}\ {\isasymleadsto}\ Event{\isasymllangle}ev{\isacharcomma}{\kern0pt}\ e{\isasymrrangle}{\isacharparenright}{\kern0pt}\ {\isacharequal}{\kern0pt}\ {\isacharparenleft}{\kern0pt}lconcrete\isactrlsub E\ e\ {\isasymand}\ concrete\isactrlsub {\isasymT}\ {\isasymtau}{\isacharparenright}{\kern0pt}{\isachardoublequoteclose}%
\begin{isamarkuptext}%
Similar to the original paper, we consider a conditioned symbolic trace as 
  concrete iff it is wellformed, and consists of a concrete path condition, as well as 
  a concrete symbolic trace. Note that this strongly differs from our concreteness
  notion of symbolic traces, as we require all wellformedness conditions to
  be~satisfied.%
\end{isamarkuptext}\isamarkuptrue%
\ \ \isacommand{definition}\isamarkupfalse%
\isanewline
\ \ \ \ concrete\isactrlsub {\isasympi}\ {\isacharcolon}{\kern0pt}{\isacharcolon}{\kern0pt}\ {\isachardoublequoteopen}{\isasymCC}{\isasymT}\ {\isasymRightarrow}\ bool{\isachardoublequoteclose}\ \isakeyword{where}\isanewline
\ \ \ \ {\isachardoublequoteopen}concrete\isactrlsub {\isasympi}\ {\isasympi}\ {\isacharequal}{\kern0pt}\ {\isacharparenleft}{\kern0pt}wf\isactrlsub {\isasympi}\ {\isasympi}\ {\isasymand}\ sconcrete\isactrlsub B\ {\isacharparenleft}{\kern0pt}{\isasymDown}\isactrlsub p\ {\isasympi}{\isacharparenright}{\kern0pt}\ {\isasymand}\ concrete\isactrlsub {\isasymT}\ {\isacharparenleft}{\kern0pt}{\isasymDown}\isactrlsub {\isasymT}\ {\isasympi}{\isacharparenright}{\kern0pt}{\isacharparenright}{\kern0pt}{\isachardoublequoteclose}%
\begin{isamarkuptext}%
We establish that a concrete symbolic trace only consists of wellformed
  states. This proof is trivial, since all states occurring in concrete traces
  must be of concrete nature. Concrete states in turn are always wellformed, as we
  have already proven in an earlier~lemma.%
\end{isamarkuptext}\isamarkuptrue%
\ \ \isacommand{lemma}\isamarkupfalse%
\ conc{\isacharunderscore}{\kern0pt}wf{\isacharunderscore}{\kern0pt}imp\isactrlsub {\isasymT}{\isacharcolon}{\kern0pt}\ {\isachardoublequoteopen}concrete\isactrlsub {\isasymT}\ {\isasymtau}\ {\isasymlongrightarrow}\ wf{\isacharunderscore}{\kern0pt}states\isactrlsub {\isasymT}\ {\isasymtau}{\isachardoublequoteclose}\isanewline
\isadelimproof
\ \ %
\endisadelimproof
\isatagproof
\isacommand{proof}\isamarkupfalse%
\ {\isacharparenleft}{\kern0pt}induct\ {\isasymtau}{\isacharparenright}{\kern0pt}\isanewline
\ \ %
\isamarkupcmt{We conduct a structural induction over the trace \isa{{\isasymtau}}.%
}\isanewline
\ \ \ \ \isacommand{case}\isamarkupfalse%
\ Epsilon\isanewline
\ \ \ \ \isacommand{thus}\isamarkupfalse%
\ {\isacharquery}{\kern0pt}case\ \isacommand{by}\isamarkupfalse%
\ simp\isanewline
\ \ \ \ %
\isamarkupcmt{If \isa{{\isasymtau}} is \isa{{\isasymepsilon}}, the case can be trivially closed, as it consists of no states.%
}\isanewline
\ \ \isacommand{next}\isamarkupfalse%
\isanewline
\ \ \ \ \isacommand{case}\isamarkupfalse%
\ {\isacharparenleft}{\kern0pt}Transition\ {\isasymtau}{\isacharprime}{\kern0pt}\ ta{\isacharparenright}{\kern0pt}\isanewline
\ \ \ \ \isacommand{thus}\isamarkupfalse%
\ {\isacharquery}{\kern0pt}case\ \isacommand{using}\isamarkupfalse%
\ concrete{\isacharunderscore}{\kern0pt}wf{\isacharunderscore}{\kern0pt}imp\isactrlsub {\isasymSigma}\ \isacommand{by}\isamarkupfalse%
\ {\isacharparenleft}{\kern0pt}induct\ ta{\isacharsemicolon}{\kern0pt}\ simp{\isacharparenright}{\kern0pt}\isanewline
\ \ \ \ %
\isamarkupcmt{If \isa{{\isasymtau}} is \isa{{\isacharparenleft}{\kern0pt}{\isasymtau}{\isacharprime}{\kern0pt}\ {\isasymleadsto}\ ta{\isacharparenright}{\kern0pt}}, the case can be closed via a case distinction
       over trace atom \isa{ta} and the knowledge that every concrete state is
       also~wellformed.%
}\isanewline
\ \ \isacommand{qed}\isamarkupfalse%
\endisatagproof
{\isafoldproof}%
\isadelimproof
\endisadelimproof
\begin{isamarkuptext}%
Every concrete conditioned symbolic trace is also guaranteed to be wellformed.
  Note that this directly follows from the definition of conditioned symbolic~traces.%
\end{isamarkuptext}\isamarkuptrue%
\ \ \isacommand{lemma}\isamarkupfalse%
\ conc{\isacharunderscore}{\kern0pt}wf{\isacharunderscore}{\kern0pt}imp\isactrlsub {\isasympi}{\isacharcolon}{\kern0pt}\ {\isachardoublequoteopen}concrete\isactrlsub {\isasympi}{\isacharparenleft}{\kern0pt}{\isasymtau}{\isacharparenright}{\kern0pt}\ {\isasymlongrightarrow}\ wf\isactrlsub {\isasympi}{\isacharparenleft}{\kern0pt}{\isasymtau}{\isacharparenright}{\kern0pt}{\isachardoublequoteclose}\isanewline
\isadelimproof
\ \ \ \ %
\endisadelimproof
\isatagproof
\isacommand{by}\isamarkupfalse%
\ {\isacharparenleft}{\kern0pt}simp\ add{\isacharcolon}{\kern0pt}\ concrete\isactrlsub {\isasympi}{\isacharunderscore}{\kern0pt}def{\isacharparenright}{\kern0pt}%
\endisatagproof
{\isafoldproof}%
\isadelimproof
\endisadelimproof
\begin{isamarkuptext}%
We prove the concreteness of the minimal event trace generated from 
  arbitrary events, arbitrary concrete symbolic states and an empty expression
  list. The concreteness holds, because all path conditions and event expressions
  of arbitrary concrete traces are guaranteed to be~variable-free.%
\end{isamarkuptext}\isamarkuptrue%
\ \ \isacommand{lemma}\isamarkupfalse%
\ {\isachardoublequoteopen}concrete\isactrlsub {\isasymSigma}\ {\isasymsigma}\ {\isasymlongrightarrow}\ concrete\isactrlsub {\isasympi}\ {\isacharparenleft}{\kern0pt}{\isacharbraceleft}{\kern0pt}{\isacharbraceright}{\kern0pt}\ {\isasymtriangleright}\ {\isacharparenleft}{\kern0pt}gen{\isacharunderscore}{\kern0pt}event\ ev\ {\isasymsigma}\ {\isacharbrackleft}{\kern0pt}{\isacharbrackright}{\kern0pt}{\isacharparenright}{\kern0pt}{\isacharparenright}{\kern0pt}{\isachardoublequoteclose}\isanewline
\isadelimproof
\ \ \ \ %
\endisadelimproof
\isatagproof
\isacommand{using}\isamarkupfalse%
\ concrete{\isacharunderscore}{\kern0pt}wf{\isacharunderscore}{\kern0pt}imp\isactrlsub {\isasymSigma}\ \isacommand{by}\isamarkupfalse%
\ {\isacharparenleft}{\kern0pt}auto\ simp\ add{\isacharcolon}{\kern0pt}\ consistent{\isacharunderscore}{\kern0pt}def\ wf\isactrlsub {\isasympi}{\isacharunderscore}{\kern0pt}def\ concrete\isactrlsub {\isasympi}{\isacharunderscore}{\kern0pt}def{\isacharparenright}{\kern0pt}%
\endisatagproof
{\isafoldproof}%
\isadelimproof
\endisadelimproof
\begin{isamarkuptext}%
A concrete trace only contains concrete states, which in turn are not 
  allowed to map variables to symbolic values. Hence, we can use Isabelle to
  derive that all concrete traces contain no symbolic~variables.%
\end{isamarkuptext}\isamarkuptrue%
\ \ \isacommand{lemma}\isamarkupfalse%
\ concrete{\isacharunderscore}{\kern0pt}symb{\isacharunderscore}{\kern0pt}imp\isactrlsub {\isasymT}{\isacharcolon}{\kern0pt}\ {\isachardoublequoteopen}concrete\isactrlsub {\isasymT}{\isacharparenleft}{\kern0pt}{\isasymtau}{\isacharparenright}{\kern0pt}\ {\isasymlongrightarrow}\ symb\isactrlsub {\isasymT}{\isacharparenleft}{\kern0pt}{\isasymtau}{\isacharparenright}{\kern0pt}\ {\isacharequal}{\kern0pt}\ {\isacharbraceleft}{\kern0pt}{\isacharbraceright}{\kern0pt}{\isachardoublequoteclose}\isanewline
\isadelimproof
\ \ %
\endisadelimproof
\isatagproof
\isacommand{proof}\isamarkupfalse%
\ {\isacharparenleft}{\kern0pt}induct\ {\isasymtau}{\isacharparenright}{\kern0pt}\isanewline
\ \ %
\isamarkupcmt{We conduct a structural induction over \isa{{\isasymtau}}.%
}\isanewline
\ \ \ \ \isacommand{case}\isamarkupfalse%
\ Epsilon\isanewline
\ \ \ \ \isacommand{show}\isamarkupfalse%
\ {\isacharquery}{\kern0pt}case\ \isacommand{by}\isamarkupfalse%
\ simp\isanewline
\ \ \ \ %
\isamarkupcmt{If \isa{{\isasymtau}} is \isa{{\isasymepsilon}}, the case is trivial, because \isa{symb\isactrlsub {\isasymT}{\isacharparenleft}{\kern0pt}{\isasymepsilon}{\isacharparenright}{\kern0pt}\ {\isacharequal}{\kern0pt}\ {\isacharbraceleft}{\kern0pt}{\isacharbraceright}{\kern0pt}}.%
}\isanewline
\ \ \isacommand{next}\isamarkupfalse%
\isanewline
\ \ \ \ \isacommand{case}\isamarkupfalse%
\ {\isacharparenleft}{\kern0pt}Transition\ {\isasymtau}{\isacharprime}{\kern0pt}\ ta{\isacharparenright}{\kern0pt}\isanewline
\ \ \ \ \isacommand{show}\isamarkupfalse%
\ {\isacharquery}{\kern0pt}case\isanewline
\ \ \ \ \ \ \isacommand{using}\isamarkupfalse%
\ Transition{\isachardot}{\kern0pt}hyps\ concrete{\isacharunderscore}{\kern0pt}symb{\isacharunderscore}{\kern0pt}imp\isactrlsub {\isasymSigma}\ \isacommand{by}\isamarkupfalse%
\ {\isacharparenleft}{\kern0pt}induct\ ta{\isacharsemicolon}{\kern0pt}\ simp{\isacharparenright}{\kern0pt}\isanewline
\ \ \ \ \ \ %
\isamarkupcmt{If \isa{{\isasymtau}} is \isa{{\isacharparenleft}{\kern0pt}{\isasymtau}{\isacharprime}{\kern0pt}\ {\isasymleadsto}\ ta{\isacharparenright}{\kern0pt}}, we can close the case using the induction
         hypothesis and the information that every concrete state contains
         no symbolic variables.%
}\isanewline
\ \ \isacommand{qed}\isamarkupfalse%
\endisatagproof
{\isafoldproof}%
\isadelimproof
\endisadelimproof
\begin{isamarkuptext}%
We can now take another look at our earlier examples of conditioned
  symbolic traces and analyze their concreteness. \isa{{\isasympi}\isactrlsub {\isadigit{1}}} is not concrete as 
  state \isa{{\isasymsigma}\isactrlsub {\isadigit{1}}} is already not concrete. Given that \isa{{\isasympi}\isactrlsub {\isadigit{2}}} is not even wellformed, 
  its concreteness is trivially violated. However, \isa{{\isasympi}\isactrlsub {\isadigit{3}}} satisfies all concreteness 
  conditions and is therefore a concrete conditioned symbolic trace.%
\end{isamarkuptext}\isamarkuptrue%
\ \ \isacommand{lemma}\isamarkupfalse%
\ {\isachardoublequoteopen}{\isasymnot}concrete\isactrlsub {\isasympi}\ {\isasympi}\isactrlsub {\isadigit{1}}\ {\isasymand}\ {\isasymnot}concrete\isactrlsub {\isasympi}\ {\isasympi}\isactrlsub {\isadigit{2}}\ {\isasymand}\ concrete\isactrlsub {\isasympi}\ {\isasympi}\isactrlsub {\isadigit{3}}{\isachardoublequoteclose}\isanewline
\isadelimproof
\ \ \ \ %
\endisadelimproof
\isatagproof
\isacommand{by}\isamarkupfalse%
\ {\isacharparenleft}{\kern0pt}simp\ add{\isacharcolon}{\kern0pt}\ {\isasymsigma}\isactrlsub {\isadigit{1}}{\isacharunderscore}{\kern0pt}def\ {\isasymsigma}\isactrlsub {\isadigit{2}}{\isacharunderscore}{\kern0pt}def\ {\isasymtau}\isactrlsub {\isadigit{1}}{\isacharunderscore}{\kern0pt}def\ {\isasymtau}\isactrlsub {\isadigit{2}}{\isacharunderscore}{\kern0pt}def\ {\isasymtau}\isactrlsub {\isadigit{3}}{\isacharunderscore}{\kern0pt}def\ {\isasympi}\isactrlsub {\isadigit{1}}{\isacharunderscore}{\kern0pt}def\ {\isasympi}\isactrlsub {\isadigit{2}}{\isacharunderscore}{\kern0pt}def\ {\isasympi}\isactrlsub {\isadigit{3}}{\isacharunderscore}{\kern0pt}def\ symb\isactrlsub {\isasymSigma}{\isacharunderscore}{\kern0pt}def\ wf\isactrlsub {\isasymSigma}{\isacharunderscore}{\kern0pt}def\ concrete\isactrlsub {\isasymSigma}{\isacharunderscore}{\kern0pt}def\ wf\isactrlsub {\isasympi}{\isacharunderscore}{\kern0pt}def\ concrete\isactrlsub {\isasympi}{\isacharunderscore}{\kern0pt}def{\isacharparenright}{\kern0pt}%
\endisatagproof
{\isafoldproof}%
\isadelimproof
\endisadelimproof
\isadelimdocument
\endisadelimdocument
\isatagdocument
\isamarkupsubsection{Concretization Mappings%
}
\isamarkuptrue%
\isamarkupsubsubsection{Concretization of States%
}
\isamarkuptrue%
\endisatagdocument
{\isafolddocument}%
\isadelimdocument
\endisadelimdocument
\begin{isamarkuptext}%
States with symbolic variables are eventually concretized,
  meaning that all previously symbolic variables are reassigned to concrete 
  values. This motivates a notion of concretization mappings. A concretization 
  mapping \isa{{\isasymrho}} for a state \isa{{\isasymsigma}} is supposed to map all symbolic variables of \isa{{\isasymsigma}} to 
  arithmetic numerals, thereby concretizing them. Note that \isa{{\isasymrho}} may additionally 
  define variables, which are not in the domain~of~\isa{{\isasymsigma}}. \par
  We can therefore call \isa{{\isasymrho}} a concretization mapping for \isa{{\isasymsigma}} if the domains of
  \isa{{\isasymrho}} and \isa{{\isasymsigma}} match exactly in the symbolic variables of \isa{{\isasymsigma}}. We additionally
  require \isa{{\isasymrho}} to be concrete, such that all variables of \isa{{\isasymrho}} map to arithmetic
  numerals. \par
  In the original paper, concretization mappings are not explicitly denoted
  as states. Their concreteness is ensured by choosing their image set to be atomic 
  values. However, this idea does not synergize well with the actual concretization 
  process in Isabelle, as these concretization mappings cannot be easily combined with 
  states due to their differing~datatypes. \par
  In order to solve this problem, we instead choose to explicitly model 
  concretization mappings as states and enforce their concreteness via the state 
  concreteness notion. Both modeling choices match in their core ideas. However, our 
  design choice ensures that states and concretization mappings originate from the 
  same datatype, thus simplifying the actual concretization process.%
\end{isamarkuptext}\isamarkuptrue%
\ \ \isacommand{definition}\isamarkupfalse%
\isanewline
\ \ \ \ is{\isacharunderscore}{\kern0pt}conc{\isacharunderscore}{\kern0pt}map\isactrlsub {\isasymSigma}\ {\isacharcolon}{\kern0pt}{\isacharcolon}{\kern0pt}\ {\isachardoublequoteopen}{\isasymSigma}\ {\isasymRightarrow}\ {\isasymSigma}\ {\isasymRightarrow}\ bool{\isachardoublequoteclose}\ \isakeyword{where}\isanewline
\ \ \ \ {\isachardoublequoteopen}is{\isacharunderscore}{\kern0pt}conc{\isacharunderscore}{\kern0pt}map\isactrlsub {\isasymSigma}\ {\isasymrho}\ {\isasymsigma}\ {\isasymequiv}\ fmdom{\isacharprime}{\kern0pt}{\isacharparenleft}{\kern0pt}{\isasymrho}{\isacharparenright}{\kern0pt}\ {\isasyminter}\ fmdom{\isacharprime}{\kern0pt}{\isacharparenleft}{\kern0pt}{\isasymsigma}{\isacharparenright}{\kern0pt}\ {\isacharequal}{\kern0pt}\ symb\isactrlsub {\isasymSigma}{\isacharparenleft}{\kern0pt}{\isasymsigma}{\isacharparenright}{\kern0pt}\ {\isasymand}\ concrete\isactrlsub {\isasymSigma}{\isacharparenleft}{\kern0pt}{\isasymrho}{\isacharparenright}{\kern0pt}{\isachardoublequoteclose}%
\begin{isamarkuptext}%
We can now define a function representing the state concretization. 
  For this purpose, we aim to formalize that a concretization mapping 
  \isa{{\isasymrho}} concretizes all symbolic variables of the original state \isa{{\isasymsigma}}, whilst optionally
  introducing new~variables. \par
  Our formalization closely resembles the definition of the original paper. The predefined
  function \isa{{\isacharplus}{\kern0pt}{\isacharplus}{\kern0pt}\isactrlsub f} of the \isa{finite{\isacharminus}{\kern0pt}map} theory guarantees that the resulting state
  matches the concretization mapping \isa{{\isasymrho}} (provided as the second argument) in all 
  its defined variables and the simplified state \isa{{\isasymsigma}} (provided as the first argument) 
  in all other (i.e. non-symbolic) variables. Note that \isa{{\isasymsigma}} is additionally evaluated 
  under \isa{{\isasymrho}} in order to simplify the resulting state, thereby ensuring the 
  concreteness of states after corresponding state~concretizations.%
\end{isamarkuptext}\isamarkuptrue%
\ \ \isacommand{fun}\isamarkupfalse%
\isanewline
\ \ \ \ conc{\isacharunderscore}{\kern0pt}map\isactrlsub {\isasymSigma}\ {\isacharcolon}{\kern0pt}{\isacharcolon}{\kern0pt}\ {\isachardoublequoteopen}{\isasymSigma}\ {\isasymRightarrow}\ {\isasymSigma}\ {\isasymRightarrow}\ {\isasymSigma}{\isachardoublequoteclose}\ {\isacharparenleft}{\kern0pt}\isakeyword{infix}\ {\isachardoublequoteopen}{\isasymbullet}{\isachardoublequoteclose}\ {\isadigit{6}}{\isadigit{1}}{\isacharparenright}{\kern0pt}\ \isakeyword{where}\isanewline
\ \ \ \ {\isachardoublequoteopen}{\isasymrho}\ {\isasymbullet}\ {\isasymsigma}\ {\isacharequal}{\kern0pt}\ fmmap{\isacharunderscore}{\kern0pt}keys\ {\isacharparenleft}{\kern0pt}{\isasymlambda}v\ e{\isachardot}{\kern0pt}\ val\isactrlsub S\ e\ {\isasymrho}{\isacharparenright}{\kern0pt}\ {\isasymsigma}\ {\isacharplus}{\kern0pt}{\isacharplus}{\kern0pt}\isactrlsub f\ {\isasymrho}{\isachardoublequoteclose}%
\begin{isamarkuptext}%
In order to analyze how the empty state connects to state concretizations, we
  first establish that the empty state is a concretization mapping of itself. This
  is trivial, considering that the empty state~is~concrete.%
\end{isamarkuptext}\isamarkuptrue%
\ \ \isacommand{lemma}\isamarkupfalse%
\ empty{\isacharunderscore}{\kern0pt}conc{\isacharunderscore}{\kern0pt}map\isactrlsub {\isasymSigma}{\isacharcolon}{\kern0pt}\ {\isachardoublequoteopen}is{\isacharunderscore}{\kern0pt}conc{\isacharunderscore}{\kern0pt}map\isactrlsub {\isasymSigma}\ {\isasymcircle}\ {\isasymcircle}{\isachardoublequoteclose}\isanewline
\isadelimproof
\ \ \ \ %
\endisadelimproof
\isatagproof
\isacommand{by}\isamarkupfalse%
\ {\isacharparenleft}{\kern0pt}simp\ add{\isacharcolon}{\kern0pt}\ symb\isactrlsub {\isasymSigma}{\isacharunderscore}{\kern0pt}def\ concrete\isactrlsub {\isasymSigma}{\isacharunderscore}{\kern0pt}def\ is{\isacharunderscore}{\kern0pt}conc{\isacharunderscore}{\kern0pt}map\isactrlsub {\isasymSigma}{\isacharunderscore}{\kern0pt}def{\isacharparenright}{\kern0pt}%
\endisatagproof
{\isafoldproof}%
\isadelimproof
\endisadelimproof
\begin{isamarkuptext}%
We now prove that the empty state is a concretization mapping for every 
  concrete state. This holds, because no concrete state contains any symbolic 
  variables. Thus, the concretization mapping conditions are trivially~satisfied.%
\end{isamarkuptext}\isamarkuptrue%
\ \ \isacommand{lemma}\isamarkupfalse%
\ empty{\isacharunderscore}{\kern0pt}conc{\isacharunderscore}{\kern0pt}map{\isacharunderscore}{\kern0pt}imp\isactrlsub {\isasymSigma}{\isacharcolon}{\kern0pt}\ {\isachardoublequoteopen}concrete\isactrlsub {\isasymSigma}{\isacharparenleft}{\kern0pt}{\isasymsigma}{\isacharparenright}{\kern0pt}\ {\isasymlongrightarrow}\ is{\isacharunderscore}{\kern0pt}conc{\isacharunderscore}{\kern0pt}map\isactrlsub {\isasymSigma}\ {\isasymcircle}\ {\isasymsigma}{\isachardoublequoteclose}\isanewline
\isadelimproof
\ \ \ \ %
\endisadelimproof
\isatagproof
\isacommand{using}\isamarkupfalse%
\ concrete{\isacharunderscore}{\kern0pt}symb{\isacharunderscore}{\kern0pt}imp\isactrlsub {\isasymSigma}\ \isacommand{by}\isamarkupfalse%
\ {\isacharparenleft}{\kern0pt}simp\ add{\isacharcolon}{\kern0pt}\ concrete\isactrlsub {\isasymSigma}{\isacharunderscore}{\kern0pt}def\ is{\isacharunderscore}{\kern0pt}conc{\isacharunderscore}{\kern0pt}map\isactrlsub {\isasymSigma}{\isacharunderscore}{\kern0pt}def{\isacharparenright}{\kern0pt}%
\endisatagproof
{\isafoldproof}%
\isadelimproof
\endisadelimproof
\begin{isamarkuptext}%
If we concretize the empty state, the resulting state always matches with the 
  applied concretization mapping. This property is also trivially provable using the
  definition of state~concretizations.%
\end{isamarkuptext}\isamarkuptrue%
\ \ \isacommand{lemma}\isamarkupfalse%
\ empty{\isacharunderscore}{\kern0pt}conc{\isacharunderscore}{\kern0pt}map{\isacharunderscore}{\kern0pt}pr\isactrlsub {\isasymSigma}{\isacharcolon}{\kern0pt}\ {\isachardoublequoteopen}{\isasymrho}\ {\isasymbullet}\ {\isasymcircle}\ {\isacharequal}{\kern0pt}\ {\isasymrho}{\isachardoublequoteclose}\isanewline
\isadelimproof
\ \ \ \ %
\endisadelimproof
\isatagproof
\isacommand{by}\isamarkupfalse%
\ {\isacharparenleft}{\kern0pt}auto{\isacharsemicolon}{\kern0pt}\ metis\ fmadd{\isacharunderscore}{\kern0pt}empty{\isacharparenleft}{\kern0pt}{\isadigit{1}}{\isacharparenright}{\kern0pt}\ fmrestrict{\isacharunderscore}{\kern0pt}set{\isacharunderscore}{\kern0pt}fmmap{\isacharunderscore}{\kern0pt}keys\ fmrestrict{\isacharunderscore}{\kern0pt}set{\isacharunderscore}{\kern0pt}null{\isacharparenright}{\kern0pt}%
\endisatagproof
{\isafoldproof}%
\isadelimproof
\endisadelimproof
\begin{isamarkuptext}%
Finally, we establish that any concretization mapping is guaranteed to be
  concrete, which obviously holds due to the definition of the 
  corresponding~predicate.%
\end{isamarkuptext}\isamarkuptrue%
\ \ \isacommand{lemma}\isamarkupfalse%
\ conc{\isacharunderscore}{\kern0pt}map{\isacharunderscore}{\kern0pt}concrete\isactrlsub {\isasymSigma}{\isacharcolon}{\kern0pt}\ {\isachardoublequoteopen}is{\isacharunderscore}{\kern0pt}conc{\isacharunderscore}{\kern0pt}map\isactrlsub {\isasymSigma}\ {\isasymrho}\ {\isasymsigma}\ {\isasymlongrightarrow}\ concrete\isactrlsub {\isasymSigma}{\isacharparenleft}{\kern0pt}{\isasymrho}{\isacharparenright}{\kern0pt}{\isachardoublequoteclose}\isanewline
\isadelimproof
\ \ \ \ %
\endisadelimproof
\isatagproof
\isacommand{by}\isamarkupfalse%
\ {\isacharparenleft}{\kern0pt}simp\ add{\isacharcolon}{\kern0pt}\ is{\isacharunderscore}{\kern0pt}conc{\isacharunderscore}{\kern0pt}map\isactrlsub {\isasymSigma}{\isacharunderscore}{\kern0pt}def{\isacharparenright}{\kern0pt}%
\endisatagproof
{\isafoldproof}%
\isadelimproof
\endisadelimproof
\begin{isamarkuptext}%
We demonstrate the state concretization with an example. The previously
  defined state \isa{{\isasymsigma}\isactrlsub {\isadigit{1}}} can be concretized with any concrete state that includes
  y, but not x, in its domain. If y is mapped onto the arithmetic numeral 2, 
  we result in state \isa{{\isasymsigma}\isactrlsub {\isadigit{2}}}. This intuitive conclusion can also be inferred by 
  Isabelle using the following~lemma.%
\end{isamarkuptext}\isamarkuptrue%
\ \ \isacommand{lemma}\isamarkupfalse%
\ {\isachardoublequoteopen}fm{\isacharbrackleft}{\kern0pt}{\isacharparenleft}{\kern0pt}{\isacharprime}{\kern0pt}{\isacharprime}{\kern0pt}y{\isacharprime}{\kern0pt}{\isacharprime}{\kern0pt}{\isacharcomma}{\kern0pt}\ Exp\ {\isacharparenleft}{\kern0pt}Num\ {\isadigit{2}}{\isacharparenright}{\kern0pt}{\isacharparenright}{\kern0pt}{\isacharbrackright}{\kern0pt}\ {\isasymbullet}\ {\isasymsigma}\isactrlsub {\isadigit{1}}\ {\isacharequal}{\kern0pt}\ {\isasymsigma}\isactrlsub {\isadigit{2}}{\isachardoublequoteclose}\isanewline
\isadelimproof
\ \ \ \ %
\endisadelimproof
\isatagproof
\isacommand{by}\isamarkupfalse%
\ {\isacharparenleft}{\kern0pt}simp\ add{\isacharcolon}{\kern0pt}\ {\isasymsigma}\isactrlsub {\isadigit{1}}{\isacharunderscore}{\kern0pt}def\ {\isasymsigma}\isactrlsub {\isadigit{2}}{\isacharunderscore}{\kern0pt}def\ fmap{\isacharunderscore}{\kern0pt}ext\ fmupd{\isachardot}{\kern0pt}rep{\isacharunderscore}{\kern0pt}eq\ map{\isacharunderscore}{\kern0pt}upd{\isacharunderscore}{\kern0pt}def{\isacharparenright}{\kern0pt}%
\endisatagproof
{\isafoldproof}%
\isadelimproof
\endisadelimproof
\isadelimdocument
\endisadelimdocument
\isatagdocument
\isamarkupsubsubsection{Minimal State Concretization%
}
\isamarkuptrue%
\endisatagdocument
{\isafolddocument}%
\isadelimdocument
\endisadelimdocument
\begin{isamarkuptext}%
In contrast to the original paper, we additionally propose a minimal state 
  concretization that will later prove useful when automatically constructing
  deterministic state concretization mappings. A minimal state concretization mapping 
  \isa{{\isasymrho}} for a given state \isa{{\isasymsigma}} is characterized by ensuring that \isa{dom{\isacharparenleft}{\kern0pt}{\isasymrho}{\isacharparenright}{\kern0pt}\ {\isacharequal}{\kern0pt}\ symb{\isacharparenleft}{\kern0pt}{\isasymsigma}{\isacharparenright}{\kern0pt}}, 
  implying that the concretization mapping is not allowed to introduce new variables. 
  Also note that all variables of a minimal state concretization are mapped onto a 
  singular, deterministic arithmetic~numeral. \par
  The deterministic minimal state concretization for a state \isa{{\isasymsigma}} is constructed as 
  follows: At the start, all key-value tuples of \isa{{\isasymsigma}} are modified, such that all 
  variables switch their symbolic nature. This implies that all previously symbolic
  variables afterwards map to a deterministic arithmetic numeral $n$, while all
  previously non-symbolic variables become symbolic. In the second step all 
  symbolic variables are filtered out of the modified state, resulting
  in a concrete state, which aligns with the intuition of our desired minimal 
  state~concretization.%
\end{isamarkuptext}\isamarkuptrue%
\ \ \isacommand{fun}\isamarkupfalse%
\isanewline
\ \ \ \ min{\isacharunderscore}{\kern0pt}conc{\isacharunderscore}{\kern0pt}map\isactrlsub {\isasymSigma}\ {\isacharcolon}{\kern0pt}{\isacharcolon}{\kern0pt}\ {\isachardoublequoteopen}{\isasymSigma}\ {\isasymRightarrow}\ int\ {\isasymRightarrow}\ {\isasymSigma}{\isachardoublequoteclose}\ \isakeyword{where}\isanewline
\ \ \ \ {\isachardoublequoteopen}min{\isacharunderscore}{\kern0pt}conc{\isacharunderscore}{\kern0pt}map\isactrlsub {\isasymSigma}\ {\isasymsigma}\ n\ {\isacharequal}{\kern0pt}\ fmfilter\ {\isacharparenleft}{\kern0pt}{\isacharpercent}{\kern0pt}v{\isachardot}{\kern0pt}\ fmlookup\ {\isasymsigma}\ v\ {\isacharequal}{\kern0pt}\ Some\ \isactrlemph {\isacharparenright}{\kern0pt}\ \isanewline
\ \ \ \ \ \ \ \ \ \ \ \ \ \ \ \ \ \ \ \ \ \ \ {\isacharparenleft}{\kern0pt}fmmap{\isacharunderscore}{\kern0pt}keys\ {\isacharparenleft}{\kern0pt}{\isasymlambda}v\ e{\isachardot}{\kern0pt}\ {\isacharparenleft}{\kern0pt}if\ e\ {\isacharequal}{\kern0pt}\ \isactrlemph \ then\ Exp\ {\isacharparenleft}{\kern0pt}Num\ n{\isacharparenright}{\kern0pt}\ else\ \isactrlemph {\isacharparenright}{\kern0pt}{\isacharparenright}{\kern0pt}\ {\isasymsigma}{\isacharparenright}{\kern0pt}{\isachardoublequoteclose}%
\begin{isamarkuptext}%
It can now be inferred that the domain of any minimal concretization mapping
  for \isa{{\isasymsigma}} contains exactly the symbolic variables of \isa{{\isasymsigma}}. This directly follows
  from the construction algorithm of minimal state concretizations as described~above.%
\end{isamarkuptext}\isamarkuptrue%
\ \ \isacommand{lemma}\isamarkupfalse%
\ only{\isacharunderscore}{\kern0pt}symb{\isacharunderscore}{\kern0pt}conc\isactrlsub {\isasymSigma}{\isacharcolon}{\kern0pt}\ {\isachardoublequoteopen}fmdom{\isacharprime}{\kern0pt}{\isacharparenleft}{\kern0pt}min{\isacharunderscore}{\kern0pt}conc{\isacharunderscore}{\kern0pt}map\isactrlsub {\isasymSigma}\ {\isasymsigma}\ n{\isacharparenright}{\kern0pt}\ {\isacharequal}{\kern0pt}\ symb\isactrlsub {\isasymSigma}{\isacharparenleft}{\kern0pt}{\isasymsigma}{\isacharparenright}{\kern0pt}{\isachardoublequoteclose}\isanewline
\isadelimproof
\ \ \ \ %
\endisadelimproof
\isatagproof
\isacommand{apply}\isamarkupfalse%
\ {\isacharparenleft}{\kern0pt}auto\ simp\ add{\isacharcolon}{\kern0pt}\ symb\isactrlsub {\isasymSigma}{\isacharunderscore}{\kern0pt}def{\isacharparenright}{\kern0pt}\isanewline
\ \ \ \ \isacommand{apply}\isamarkupfalse%
\ {\isacharparenleft}{\kern0pt}metis\ {\isacharparenleft}{\kern0pt}mono{\isacharunderscore}{\kern0pt}tags{\isacharcomma}{\kern0pt}\ lifting{\isacharparenright}{\kern0pt}\ fmdom{\isacharprime}{\kern0pt}{\isacharunderscore}{\kern0pt}notI\ fmfilter{\isacharunderscore}{\kern0pt}fmmap{\isacharunderscore}{\kern0pt}keys\ fmlookup{\isacharunderscore}{\kern0pt}filter\ fmlookup{\isacharunderscore}{\kern0pt}dom{\isacharprime}{\kern0pt}{\isacharunderscore}{\kern0pt}iff{\isacharparenright}{\kern0pt}\isanewline
\ \ \ \ \isacommand{apply}\isamarkupfalse%
\ {\isacharparenleft}{\kern0pt}metis\ {\isacharparenleft}{\kern0pt}mono{\isacharunderscore}{\kern0pt}tags{\isacharcomma}{\kern0pt}\ lifting{\isacharparenright}{\kern0pt}\ fmdom{\isacharprime}{\kern0pt}{\isacharunderscore}{\kern0pt}notI\ fmfilter{\isacharunderscore}{\kern0pt}fmmap{\isacharunderscore}{\kern0pt}keys\ fmlookup{\isacharunderscore}{\kern0pt}filter{\isacharparenright}{\kern0pt}\isanewline
\ \ \ \ \isacommand{using}\isamarkupfalse%
\ fmlookup{\isacharunderscore}{\kern0pt}dom{\isacharprime}{\kern0pt}{\isacharunderscore}{\kern0pt}iff\ \isacommand{by}\isamarkupfalse%
\ fastforce%
\endisatagproof
{\isafoldproof}%
\isadelimproof
\endisadelimproof
\begin{isamarkuptext}%
The minimal state concretization of a concrete state is always the empty
  state. This is obvious, considering that a concrete state is not allowed to
  contain any symbolic variables, implying that no concretization can take~place.%
\end{isamarkuptext}\isamarkuptrue%
\ \ \isacommand{lemma}\isamarkupfalse%
\ min{\isacharunderscore}{\kern0pt}conc{\isacharunderscore}{\kern0pt}map{\isacharunderscore}{\kern0pt}of{\isacharunderscore}{\kern0pt}concrete\isactrlsub {\isasymSigma}{\isacharcolon}{\kern0pt}\ {\isachardoublequoteopen}concrete\isactrlsub {\isasymSigma}{\isacharparenleft}{\kern0pt}{\isasymsigma}{\isacharparenright}{\kern0pt}\ {\isasymlongrightarrow}\ min{\isacharunderscore}{\kern0pt}conc{\isacharunderscore}{\kern0pt}map\isactrlsub {\isasymSigma}\ {\isasymsigma}\ n\ {\isacharequal}{\kern0pt}\ {\isasymcircle}{\isachardoublequoteclose}\isanewline
\isadelimproof
\ \ \ \ %
\endisadelimproof
\isatagproof
\isacommand{using}\isamarkupfalse%
\ only{\isacharunderscore}{\kern0pt}symb{\isacharunderscore}{\kern0pt}conc\isactrlsub {\isasymSigma}\ concrete{\isacharunderscore}{\kern0pt}symb{\isacharunderscore}{\kern0pt}imp\isactrlsub {\isasymSigma}\ \isanewline
\ \ \ \ \isacommand{by}\isamarkupfalse%
\ {\isacharparenleft}{\kern0pt}metis\ fmrestrict{\isacharunderscore}{\kern0pt}set{\isacharunderscore}{\kern0pt}dom\ fmrestrict{\isacharunderscore}{\kern0pt}set{\isacharunderscore}{\kern0pt}null{\isacharparenright}{\kern0pt}%
\endisatagproof
{\isafoldproof}%
\isadelimproof
\endisadelimproof
\begin{isamarkuptext}%
We can now take a look at the minimal concretization mapping for our
  previously defined state \isa{{\isasymsigma}\isactrlsub {\isadigit{1}}}. As only y is symbolic in \isa{{\isasymsigma}\isactrlsub {\isadigit{1}}}, its minimal
  concretization mapping will only specify y. We demonstrate this fact using
  the following~lemma.%
\end{isamarkuptext}\isamarkuptrue%
\ \ \isacommand{lemma}\isamarkupfalse%
\ {\isachardoublequoteopen}min{\isacharunderscore}{\kern0pt}conc{\isacharunderscore}{\kern0pt}map\isactrlsub {\isasymSigma}\ {\isasymsigma}\isactrlsub {\isadigit{1}}\ {\isadigit{0}}\ {\isacharequal}{\kern0pt}\ fm{\isacharbrackleft}{\kern0pt}{\isacharparenleft}{\kern0pt}{\isacharprime}{\kern0pt}{\isacharprime}{\kern0pt}y{\isacharprime}{\kern0pt}{\isacharprime}{\kern0pt}{\isacharcomma}{\kern0pt}\ Exp\ {\isacharparenleft}{\kern0pt}Num\ {\isadigit{0}}{\isacharparenright}{\kern0pt}{\isacharparenright}{\kern0pt}{\isacharbrackright}{\kern0pt}{\isachardoublequoteclose}\isanewline
\isadelimproof
\ \ \ \ %
\endisadelimproof
\isatagproof
\isacommand{by}\isamarkupfalse%
\ {\isacharparenleft}{\kern0pt}simp\ add{\isacharcolon}{\kern0pt}\ {\isasymsigma}\isactrlsub {\isadigit{1}}{\isacharunderscore}{\kern0pt}def\ fmap{\isacharunderscore}{\kern0pt}ext{\isacharparenright}{\kern0pt}%
\endisatagproof
{\isafoldproof}%
\isadelimproof
\endisadelimproof
\isadelimdocument
\endisadelimdocument
\isatagdocument
\isamarkupsubsubsection{Concretization of Traces%
}
\isamarkuptrue%
\endisatagdocument
{\isafolddocument}%
\isadelimdocument
\endisadelimdocument
\begin{isamarkuptext}%
We now aim to strengthen the notion of state concretization mappings to 
  concretization mappings for symbolic traces and conditioned symbolic~traces. \par
  Similar to the original paper, we call a state \isa{{\isasymrho}} a valid concretization mapping 
  for a given symbolic trace \isa{{\isasymtau}} iff \isa{{\isasymrho}} is a concretization mapping for all states 
  occurring in \isa{{\isasymtau}}. The definition of the predicate is straightforward by making
  use of recursion. Note that we enforce the concreteness of \isa{{\isasymrho}}, even if
  the argument trace is the empty~trace.%
\end{isamarkuptext}\isamarkuptrue%
\ \ \isacommand{fun}\isamarkupfalse%
\isanewline
\ \ \ \ is{\isacharunderscore}{\kern0pt}conc{\isacharunderscore}{\kern0pt}map\isactrlsub {\isasymT}\ {\isacharcolon}{\kern0pt}{\isacharcolon}{\kern0pt}\ {\isachardoublequoteopen}{\isasymSigma}\ {\isasymRightarrow}\ {\isasymT}\ {\isasymRightarrow}\ bool{\isachardoublequoteclose}\ \isakeyword{where}\isanewline
\ \ \ \ {\isachardoublequoteopen}is{\isacharunderscore}{\kern0pt}conc{\isacharunderscore}{\kern0pt}map\isactrlsub {\isasymT}\ {\isasymrho}\ {\isasymepsilon}\ {\isacharequal}{\kern0pt}\ concrete\isactrlsub {\isasymSigma}{\isacharparenleft}{\kern0pt}{\isasymrho}{\isacharparenright}{\kern0pt}{\isachardoublequoteclose}\ {\isacharbar}{\kern0pt}\isanewline
\ \ \ \ {\isachardoublequoteopen}is{\isacharunderscore}{\kern0pt}conc{\isacharunderscore}{\kern0pt}map\isactrlsub {\isasymT}\ {\isasymrho}\ {\isacharparenleft}{\kern0pt}{\isasymtau}\ {\isasymleadsto}\ Event{\isasymllangle}ev{\isacharcomma}{\kern0pt}\ e{\isasymrrangle}{\isacharparenright}{\kern0pt}\ {\isacharequal}{\kern0pt}\ is{\isacharunderscore}{\kern0pt}conc{\isacharunderscore}{\kern0pt}map\isactrlsub {\isasymT}\ {\isasymrho}\ {\isasymtau}{\isachardoublequoteclose}\ {\isacharbar}{\kern0pt}\isanewline
\ \ \ \ {\isachardoublequoteopen}is{\isacharunderscore}{\kern0pt}conc{\isacharunderscore}{\kern0pt}map\isactrlsub {\isasymT}\ {\isasymrho}\ {\isacharparenleft}{\kern0pt}{\isasymtau}\ {\isasymleadsto}\ State{\isasymllangle}{\isasymsigma}{\isasymrrangle}{\isacharparenright}{\kern0pt}\ {\isacharequal}{\kern0pt}\ {\isacharparenleft}{\kern0pt}is{\isacharunderscore}{\kern0pt}conc{\isacharunderscore}{\kern0pt}map\isactrlsub {\isasymSigma}\ {\isasymrho}\ {\isasymsigma}\ {\isasymand}\ is{\isacharunderscore}{\kern0pt}conc{\isacharunderscore}{\kern0pt}map\isactrlsub {\isasymT}\ {\isasymrho}\ {\isasymtau}{\isacharparenright}{\kern0pt}{\isachardoublequoteclose}%
\begin{isamarkuptext}%
We can now formalize that a concretization mapping \isa{{\isasymrho}} concretizes a
  symbolic trace \isa{{\isasymtau}} by concretizing all states of \isa{{\isasymtau}} with \isa{{\isasymrho}} and 
  simplifying all event expressions occurring in events of \isa{{\isasymtau}} with \isa{{\isasymrho}}.
  Note that we require trace \isa{{\isasymtau}} to be wellformed in order to ensure 
  that the application of this function results in a concrete symbolic~trace
  (e.g. such that we can guarantee that all event expressions are concretized).%
\end{isamarkuptext}\isamarkuptrue%
\ \ \isacommand{fun}\isamarkupfalse%
\isanewline
\ \ \ \ trace{\isacharunderscore}{\kern0pt}conc\ {\isacharcolon}{\kern0pt}{\isacharcolon}{\kern0pt}\ {\isachardoublequoteopen}{\isasymSigma}\ {\isasymRightarrow}\ {\isasymT}\ {\isasymRightarrow}\ {\isasymT}{\isachardoublequoteclose}\ \isakeyword{where}\isanewline
\ \ \ \ {\isachardoublequoteopen}trace{\isacharunderscore}{\kern0pt}conc\ {\isasymrho}\ {\isasymepsilon}\ {\isacharequal}{\kern0pt}\ {\isasymepsilon}{\isachardoublequoteclose}\ {\isacharbar}{\kern0pt}\isanewline
\ \ \ \ {\isachardoublequoteopen}trace{\isacharunderscore}{\kern0pt}conc\ {\isasymrho}\ {\isacharparenleft}{\kern0pt}{\isasymtau}\ {\isasymleadsto}\ State{\isasymllangle}{\isasymsigma}{\isasymrrangle}{\isacharparenright}{\kern0pt}\ {\isacharequal}{\kern0pt}\ {\isacharparenleft}{\kern0pt}trace{\isacharunderscore}{\kern0pt}conc\ {\isasymrho}\ {\isasymtau}{\isacharparenright}{\kern0pt}\ {\isasymleadsto}\ State{\isasymllangle}{\isacharparenleft}{\kern0pt}{\isasymrho}\ {\isasymbullet}\ {\isasymsigma}{\isacharparenright}{\kern0pt}{\isasymrrangle}{\isachardoublequoteclose}\ {\isacharbar}{\kern0pt}\isanewline
\ \ \ \ {\isachardoublequoteopen}trace{\isacharunderscore}{\kern0pt}conc\ {\isasymrho}\ {\isacharparenleft}{\kern0pt}{\isasymtau}\ {\isasymleadsto}\ Event{\isasymllangle}ev{\isacharcomma}{\kern0pt}\ e{\isasymrrangle}{\isacharparenright}{\kern0pt}\ {\isacharequal}{\kern0pt}\ {\isacharparenleft}{\kern0pt}trace{\isacharunderscore}{\kern0pt}conc\ {\isasymrho}\ {\isasymtau}{\isacharparenright}{\kern0pt}\ {\isasymleadsto}\ Event{\isasymllangle}ev{\isacharcomma}{\kern0pt}\ {\isacharparenleft}{\kern0pt}lval\isactrlsub E\ e\ {\isasymrho}{\isacharparenright}{\kern0pt}{\isasymrrangle}{\isachardoublequoteclose}%
\begin{isamarkuptext}%
A concretization mapping \isa{{\isasymrho}} concretizes a conditioned symbolic trace \isa{{\isasympi}} by 
  concretizing the symbolic trace of \isa{{\isasympi}} with \isa{{\isasymrho}} and simplifying the path condition 
  of \isa{{\isasympi}} with \isa{{\isasymrho}}. Note that we again require \isa{{\isasympi}} to be wellformed if we want to 
  guarantee the concreteness of the resulting conditioned symbolic~trace.%
\end{isamarkuptext}\isamarkuptrue%
\ \ \isacommand{fun}\isamarkupfalse%
\isanewline
\ \ \ \ conc{\isacharunderscore}{\kern0pt}map\isactrlsub {\isasymT}\ {\isacharcolon}{\kern0pt}{\isacharcolon}{\kern0pt}\ {\isachardoublequoteopen}{\isasymSigma}\ {\isasymRightarrow}\ {\isasymCC}{\isasymT}\ {\isasymRightarrow}\ {\isasymCC}{\isasymT}{\isachardoublequoteclose}\ {\isacharparenleft}{\kern0pt}\isakeyword{infix}\ {\isachardoublequoteopen}{\isasymsqdot}{\isachardoublequoteclose}\ {\isadigit{6}}{\isadigit{1}}{\isacharparenright}{\kern0pt}\ \isakeyword{where}\isanewline
\ \ \ \ {\isachardoublequoteopen}{\isasymrho}\ {\isasymsqdot}\ {\isacharparenleft}{\kern0pt}pc\ {\isasymtriangleright}\ {\isasymtau}{\isacharparenright}{\kern0pt}\ {\isacharequal}{\kern0pt}\ {\isacharparenleft}{\kern0pt}sval\isactrlsub B\ pc\ {\isasymrho}{\isacharparenright}{\kern0pt}\ {\isasymtriangleright}\ {\isacharparenleft}{\kern0pt}trace{\isacharunderscore}{\kern0pt}conc\ {\isasymrho}\ {\isasymtau}{\isacharparenright}{\kern0pt}{\isachardoublequoteclose}%
\begin{isamarkuptext}%
We next establish that a valid trace concretization mapping is always concrete.
  This property trivially holds and can be inferred by Isabelle using the 
  definition of the corresponding~predicate.%
\end{isamarkuptext}\isamarkuptrue%
\ \ \isacommand{lemma}\isamarkupfalse%
\ conc{\isacharunderscore}{\kern0pt}map{\isacharunderscore}{\kern0pt}concrete\isactrlsub {\isasymT}{\isacharcolon}{\kern0pt}\ {\isachardoublequoteopen}is{\isacharunderscore}{\kern0pt}conc{\isacharunderscore}{\kern0pt}map\isactrlsub {\isasymT}\ {\isasymrho}\ {\isasymtau}\ {\isasymlongrightarrow}\ concrete\isactrlsub {\isasymSigma}{\isacharparenleft}{\kern0pt}{\isasymrho}{\isacharparenright}{\kern0pt}{\isachardoublequoteclose}\isanewline
\isadelimproof
\ \ \ \ %
\endisadelimproof
\isatagproof
\isacommand{apply}\isamarkupfalse%
\ {\isacharparenleft}{\kern0pt}induct\ {\isasymtau}{\isacharparenright}{\kern0pt}\isanewline
\ \ \ \ \isacommand{using}\isamarkupfalse%
\ is{\isacharunderscore}{\kern0pt}conc{\isacharunderscore}{\kern0pt}map\isactrlsub {\isasymT}{\isachardot}{\kern0pt}elims{\isacharparenleft}{\kern0pt}{\isadigit{1}}{\isacharparenright}{\kern0pt}\ \isacommand{by}\isamarkupfalse%
\ auto%
\endisatagproof
{\isafoldproof}%
\isadelimproof
\endisadelimproof
\begin{isamarkuptext}%
The following lemma provides an example for a trace concretization. 
  As we have previously reasoned, applying a specific concretization mapping on \isa{{\isasymsigma}\isactrlsub {\isadigit{1}}}
  results in \isa{{\isasymsigma}\isactrlsub {\isadigit{2}}}. Hence, we can also concretize the conditioned symbolic trace \isa{{\isasympi}\isactrlsub {\isadigit{1}}} 
  with the same concretization mapping in order to result in a conditioned symbolic 
  trace, in which the state \isa{{\isasymsigma}\isactrlsub {\isadigit{1}}} is replaced~with~\isa{{\isasymsigma}\isactrlsub {\isadigit{2}}}.%
\end{isamarkuptext}\isamarkuptrue%
\ \ \isacommand{lemma}\isamarkupfalse%
\ {\isachardoublequoteopen}fm{\isacharbrackleft}{\kern0pt}{\isacharparenleft}{\kern0pt}{\isacharprime}{\kern0pt}{\isacharprime}{\kern0pt}y{\isacharprime}{\kern0pt}{\isacharprime}{\kern0pt}{\isacharcomma}{\kern0pt}\ Exp\ {\isacharparenleft}{\kern0pt}Num\ {\isadigit{2}}{\isacharparenright}{\kern0pt}{\isacharparenright}{\kern0pt}{\isacharbrackright}{\kern0pt}\ {\isasymsqdot}\ {\isasympi}\isactrlsub {\isadigit{1}}\ {\isacharequal}{\kern0pt}\ {\isacharbraceleft}{\kern0pt}{\isacharbraceright}{\kern0pt}\ {\isasymtriangleright}\ {\isacharparenleft}{\kern0pt}{\isasymlangle}{\isasymsigma}\isactrlsub {\isadigit{2}}{\isasymrangle}\ {\isasymleadsto}\ Event{\isasymllangle}inpEv{\isacharcomma}{\kern0pt}\ {\isacharbrackleft}{\kern0pt}{\isacharbrackright}{\kern0pt}{\isasymrrangle}{\isacharparenright}{\kern0pt}\ {\isasymleadsto}\ State{\isasymllangle}{\isasymsigma}\isactrlsub {\isadigit{2}}{\isasymrrangle}{\isachardoublequoteclose}\isanewline
\isadelimproof
\ \ \ \ %
\endisadelimproof
\isatagproof
\isacommand{by}\isamarkupfalse%
\ {\isacharparenleft}{\kern0pt}auto\ simp\ add{\isacharcolon}{\kern0pt}\ {\isasymsigma}\isactrlsub {\isadigit{1}}{\isacharunderscore}{\kern0pt}def\ {\isasymsigma}\isactrlsub {\isadigit{2}}{\isacharunderscore}{\kern0pt}def\ {\isasymtau}\isactrlsub {\isadigit{1}}{\isacharunderscore}{\kern0pt}def\ {\isasympi}\isactrlsub {\isadigit{1}}{\isacharunderscore}{\kern0pt}def\ fmap{\isacharunderscore}{\kern0pt}ext\ fmupd{\isachardot}{\kern0pt}rep{\isacharunderscore}{\kern0pt}eq\ map{\isacharunderscore}{\kern0pt}upd{\isacharunderscore}{\kern0pt}def{\isacharparenright}{\kern0pt}%
\endisatagproof
{\isafoldproof}%
\isadelimproof
\endisadelimproof
\isadelimdocument
\endisadelimdocument
\isatagdocument
\isamarkupsubsubsection{Minimal Trace Concretization%
}
\isamarkuptrue%
\endisatagdocument
{\isafolddocument}%
\isadelimdocument
\endisadelimdocument
\begin{isamarkuptext}%
Given that we have already defined minimal state concretizations, we are now 
  interested in heightening this definition to traces. For this purpose, we 
  define a minimal trace concretization as a minimal concretization mapping for 
  all of the states occurring in the trace. The choice of this concretization mapping 
  will again be deterministic, thus ensuring that the Isabelle compiler can generate 
  efficient code for their~construction. \par
  The construction of the minimal trace concretization is simple. We unfold
  the trace from the back, whilst recursively combining the minimal concretization 
  mappings of all traversed states. Note that the constructed concretization
  mapping is therefore only a valid trace concretization mapping if all states 
  occurring in the trace agree on their symbolic variables, which is guaranteed by 
  the wellformedness~notion.%
\end{isamarkuptext}\isamarkuptrue%
\ \ \isacommand{fun}\isamarkupfalse%
\isanewline
\ \ \ \ min{\isacharunderscore}{\kern0pt}conc{\isacharunderscore}{\kern0pt}map\isactrlsub {\isasymT}\ {\isacharcolon}{\kern0pt}{\isacharcolon}{\kern0pt}\ {\isachardoublequoteopen}{\isasymT}\ {\isasymRightarrow}\ int\ {\isasymRightarrow}\ {\isasymSigma}{\isachardoublequoteclose}\ \isakeyword{where}\isanewline
\ \ \ \ {\isachardoublequoteopen}min{\isacharunderscore}{\kern0pt}conc{\isacharunderscore}{\kern0pt}map\isactrlsub {\isasymT}\ {\isasymepsilon}\ n\ {\isacharequal}{\kern0pt}\ {\isasymcircle}{\isachardoublequoteclose}\ {\isacharbar}{\kern0pt}\isanewline
\ \ \ \ {\isachardoublequoteopen}min{\isacharunderscore}{\kern0pt}conc{\isacharunderscore}{\kern0pt}map\isactrlsub {\isasymT}\ {\isacharparenleft}{\kern0pt}{\isasymtau}\ {\isasymleadsto}\ State{\isasymllangle}{\isasymsigma}{\isasymrrangle}{\isacharparenright}{\kern0pt}\ n\ {\isacharequal}{\kern0pt}\ min{\isacharunderscore}{\kern0pt}conc{\isacharunderscore}{\kern0pt}map\isactrlsub {\isasymT}\ {\isasymtau}\ n\ {\isacharplus}{\kern0pt}{\isacharplus}{\kern0pt}\isactrlsub f\ min{\isacharunderscore}{\kern0pt}conc{\isacharunderscore}{\kern0pt}map\isactrlsub {\isasymSigma}\ {\isasymsigma}\ n{\isachardoublequoteclose}\ {\isacharbar}{\kern0pt}\isanewline
\ \ \ \ {\isachardoublequoteopen}min{\isacharunderscore}{\kern0pt}conc{\isacharunderscore}{\kern0pt}map\isactrlsub {\isasymT}\ {\isacharparenleft}{\kern0pt}{\isasymtau}\ {\isasymleadsto}\ Event{\isasymllangle}ev{\isacharcomma}{\kern0pt}\ e{\isasymrrangle}{\isacharparenright}{\kern0pt}\ n\ {\isacharequal}{\kern0pt}\ min{\isacharunderscore}{\kern0pt}conc{\isacharunderscore}{\kern0pt}map\isactrlsub {\isasymT}\ {\isasymtau}\ n{\isachardoublequoteclose}%
\begin{isamarkuptext}%
The domain of any minimal concretization mapping for \isa{{\isasymtau}} contains exactly the 
  symbolic variables of \isa{{\isasymtau}}. This directly follows from the construction algorithm 
  of minimal trace concretizations as described~above.%
\end{isamarkuptext}\isamarkuptrue%
\ \ \isacommand{lemma}\isamarkupfalse%
\ only{\isacharunderscore}{\kern0pt}symb{\isacharunderscore}{\kern0pt}conc\isactrlsub {\isasymT}{\isacharcolon}{\kern0pt}\ {\isachardoublequoteopen}fmdom{\isacharprime}{\kern0pt}{\isacharparenleft}{\kern0pt}min{\isacharunderscore}{\kern0pt}conc{\isacharunderscore}{\kern0pt}map\isactrlsub {\isasymT}\ {\isasymtau}\ n{\isacharparenright}{\kern0pt}\ {\isacharequal}{\kern0pt}\ symb\isactrlsub {\isasymT}{\isacharparenleft}{\kern0pt}{\isasymtau}{\isacharparenright}{\kern0pt}{\isachardoublequoteclose}\isanewline
\isadelimproof
\ \ %
\endisadelimproof
\isatagproof
\isacommand{proof}\isamarkupfalse%
\ {\isacharparenleft}{\kern0pt}induct\ {\isasymtau}{\isacharparenright}{\kern0pt}\isanewline
\ \ %
\isamarkupcmt{We perform a structural induction over the construction of \isa{{\isasymtau}}.%
}\isanewline
\ \ \ \ \isacommand{case}\isamarkupfalse%
\ Epsilon\isanewline
\ \ \ \ \isacommand{thus}\isamarkupfalse%
\ {\isacharquery}{\kern0pt}case\ \isacommand{by}\isamarkupfalse%
\ simp\isanewline
\ \ \ \ %
\isamarkupcmt{If \isa{{\isasymtau}} is \isa{{\isasymepsilon}}, the case is trivial.%
}\isanewline
\ \ \isacommand{next}\isamarkupfalse%
\isanewline
\ \ \ \ \isacommand{case}\isamarkupfalse%
\ {\isacharparenleft}{\kern0pt}Transition\ {\isasymtau}{\isacharprime}{\kern0pt}\ ta{\isacharparenright}{\kern0pt}\isanewline
\ \ \ \ %
\isamarkupcmt{In the induction step, we assume \isa{{\isasymtau}} is \isa{{\isacharparenleft}{\kern0pt}{\isasymtau}{\isacharprime}{\kern0pt}\ {\isasymleadsto}\ ta{\isacharparenright}{\kern0pt}}.%
}\isanewline
\ \ \ \ \isacommand{thus}\isamarkupfalse%
\ {\isacharquery}{\kern0pt}case\isanewline
\ \ \ \ \isacommand{proof}\isamarkupfalse%
\ {\isacharparenleft}{\kern0pt}induct\ ta{\isacharparenright}{\kern0pt}\isanewline
\ \ \ \ %
\isamarkupcmt{We perform a case distinction over trace atom \isa{ta}.%
}\isanewline
\ \ \ \ \ \ \isacommand{case}\isamarkupfalse%
\ {\isacharparenleft}{\kern0pt}Event\ ev\ e{\isacharparenright}{\kern0pt}\isanewline
\ \ \ \ \ \ \isacommand{thus}\isamarkupfalse%
\ {\isacharquery}{\kern0pt}case\ \isacommand{by}\isamarkupfalse%
\ simp\isanewline
\ \ \ \ \ \ %
\isamarkupcmt{We assume \isa{ta} is an event \isa{e}. Then the case trivially holds in combination
         with the induction hypothesis, because the domain of the minimal trace 
         concretization mapping is independent of the events ocurring in the trace.%
}\isanewline
\ \ \ \ \isacommand{next}\isamarkupfalse%
\isanewline
\ \ \ \ \ \ \isacommand{case}\isamarkupfalse%
\ {\isacharparenleft}{\kern0pt}State\ {\isasymsigma}{\isacharparenright}{\kern0pt}\isanewline
\ \ \ \ \ \ \isacommand{thus}\isamarkupfalse%
\ {\isacharquery}{\kern0pt}case\ \isacommand{using}\isamarkupfalse%
\ only{\isacharunderscore}{\kern0pt}symb{\isacharunderscore}{\kern0pt}conc\isactrlsub {\isasymSigma}\ \isacommand{by}\isamarkupfalse%
\ auto\isanewline
\ \ \ \ \ \ %
\isamarkupcmt{We assume \isa{ta} is a state \isa{{\isasymsigma}}. Due to an earlier proof, we know that the 
         domain of the minimal state concretization mapping of \isa{{\isasymsigma}} contains exactly 
         the symbolic variables of \isa{{\isasymsigma}}. Combined with the induction hypothesis, this
         closes the case.%
}\isanewline
\ \ \ \ \isacommand{qed}\isamarkupfalse%
\isanewline
\ \ \isacommand{qed}\isamarkupfalse%
\endisatagproof
{\isafoldproof}%
\isadelimproof
\endisadelimproof
\begin{isamarkuptext}%
The minimal trace concretization of a concrete trace is always the empty
  state. This is obvious, considering that a concrete trace is not allowed to
  contain any symbolic variables, implying that no concretization can take~place.%
\end{isamarkuptext}\isamarkuptrue%
\isacommand{lemma}\isamarkupfalse%
\ min{\isacharunderscore}{\kern0pt}conc{\isacharunderscore}{\kern0pt}map{\isacharunderscore}{\kern0pt}of{\isacharunderscore}{\kern0pt}concrete\isactrlsub {\isasymT}{\isacharcolon}{\kern0pt}\ {\isachardoublequoteopen}concrete\isactrlsub {\isasymT}{\isacharparenleft}{\kern0pt}{\isasymtau}{\isacharparenright}{\kern0pt}\ {\isasymlongrightarrow}\ min{\isacharunderscore}{\kern0pt}conc{\isacharunderscore}{\kern0pt}map\isactrlsub {\isasymT}\ {\isasymtau}\ n\ {\isacharequal}{\kern0pt}\ {\isasymcircle}{\isachardoublequoteclose}\isanewline
\isadelimproof
\ \ %
\endisadelimproof
\isatagproof
\isacommand{using}\isamarkupfalse%
\ only{\isacharunderscore}{\kern0pt}symb{\isacharunderscore}{\kern0pt}conc\isactrlsub {\isasymT}\ concrete{\isacharunderscore}{\kern0pt}symb{\isacharunderscore}{\kern0pt}imp\isactrlsub {\isasymT}\ \isanewline
\ \ \isacommand{by}\isamarkupfalse%
\ {\isacharparenleft}{\kern0pt}metis\ fmrestrict{\isacharunderscore}{\kern0pt}set{\isacharunderscore}{\kern0pt}dom\ fmrestrict{\isacharunderscore}{\kern0pt}set{\isacharunderscore}{\kern0pt}null{\isacharparenright}{\kern0pt}%
\endisatagproof
{\isafoldproof}%
\isadelimproof
\endisadelimproof
\begin{isamarkuptext}%
We can now take another look at our earlier example. Considering that \isa{{\isasymtau}\isactrlsub {\isadigit{1}}} 
  only contains the state \isa{{\isasymsigma}\isactrlsub {\isadigit{1}}}, its minimal trace concretization mapping will match 
  the minimal state concretization mapping of \isa{{\isasymsigma}\isactrlsub {\isadigit{1}}} from~earlier.%
\end{isamarkuptext}\isamarkuptrue%
\ \ \isacommand{lemma}\isamarkupfalse%
\ {\isachardoublequoteopen}min{\isacharunderscore}{\kern0pt}conc{\isacharunderscore}{\kern0pt}map\isactrlsub {\isasymT}\ {\isasymtau}\isactrlsub {\isadigit{1}}\ {\isadigit{0}}\ {\isacharequal}{\kern0pt}\ fm{\isacharbrackleft}{\kern0pt}{\isacharparenleft}{\kern0pt}{\isacharprime}{\kern0pt}{\isacharprime}{\kern0pt}y{\isacharprime}{\kern0pt}{\isacharprime}{\kern0pt}{\isacharcomma}{\kern0pt}\ Exp\ {\isacharparenleft}{\kern0pt}Num\ {\isadigit{0}}{\isacharparenright}{\kern0pt}{\isacharparenright}{\kern0pt}{\isacharbrackright}{\kern0pt}{\isachardoublequoteclose}\isanewline
\isadelimproof
\ \ \ \ %
\endisadelimproof
\isatagproof
\isacommand{by}\isamarkupfalse%
\ {\isacharparenleft}{\kern0pt}simp\ add{\isacharcolon}{\kern0pt}\ {\isasymsigma}\isactrlsub {\isadigit{1}}{\isacharunderscore}{\kern0pt}def\ {\isasymtau}\isactrlsub {\isadigit{1}}{\isacharunderscore}{\kern0pt}def\ fmap{\isacharunderscore}{\kern0pt}ext{\isacharparenright}{\kern0pt}%
\endisatagproof
{\isafoldproof}%
\isadelimproof
\endisadelimproof
\isadelimdocument
\endisadelimdocument
\isatagdocument
\isamarkupsubsubsection{Proof Automation%
}
\isamarkuptrue%
\endisatagdocument
{\isafolddocument}%
\isadelimdocument
\endisadelimdocument
\begin{isamarkuptext}%
During a state concretization, the \isa{fmmap{\isacharminus}{\kern0pt}keys} function of the finite map
  theory is used to simplify all expressions in the image of a given state 
  with a provided concretization mapping. However, under several circumstances
  the Isabelle simplifier is not able to automatically compute this function, 
  hence obstructing the proof automation for state concretizations. In order to
  circumvent this problem, we propose additional supporting theorems for the 
  simplifier, which can be used to resolve these problematic~situations. \par
  We first establish that the simplification of the empty state with an arbitrary
  state preserves the empty state. This property is trivial, considering that
  there exists no key-value pair in the empty~state.%
\end{isamarkuptext}\isamarkuptrue%
\ \ \isacommand{lemma}\isamarkupfalse%
\ fmmap{\isacharunderscore}{\kern0pt}keys{\isacharunderscore}{\kern0pt}empty{\isacharcolon}{\kern0pt}\ {\isachardoublequoteopen}fmmap{\isacharunderscore}{\kern0pt}keys\ {\isacharparenleft}{\kern0pt}{\isasymlambda}v\ e{\isachardot}{\kern0pt}\ val\isactrlsub S\ e\ {\isasymrho}{\isacharparenright}{\kern0pt}\ {\isasymcircle}\ {\isacharequal}{\kern0pt}\ {\isasymcircle}{\isachardoublequoteclose}\isanewline
\isadelimproof
\ \ \ \ %
\endisadelimproof
\isatagproof
\isacommand{by}\isamarkupfalse%
\ {\isacharparenleft}{\kern0pt}metis\ fmrestrict{\isacharunderscore}{\kern0pt}set{\isacharunderscore}{\kern0pt}fmmap{\isacharunderscore}{\kern0pt}keys\ fmrestrict{\isacharunderscore}{\kern0pt}set{\isacharunderscore}{\kern0pt}null{\isacharparenright}{\kern0pt}%
\endisatagproof
{\isafoldproof}%
\isadelimproof
\endisadelimproof
\begin{isamarkuptext}%
We next prove that any concrete state \isa{{\isasymsigma}} simplified with another concrete 
  state \isa{{\isasymrho}} preserves~\isa{{\isasymsigma}}. Considering that \isa{{\isasymsigma}} is already concrete, no update 
  can take place. This implies that \isa{{\isasymrho}} is not used during the simplification process,
  thus trivially proving the~lemma.%
\end{isamarkuptext}\isamarkuptrue%
\ \ \isacommand{lemma}\isamarkupfalse%
\ fmmap{\isacharunderscore}{\kern0pt}keys{\isacharunderscore}{\kern0pt}conc{\isacharcolon}{\kern0pt}\ {\isachardoublequoteopen}concrete\isactrlsub {\isasymSigma}{\isacharparenleft}{\kern0pt}{\isasymsigma}{\isacharparenright}{\kern0pt}\ {\isasymlongrightarrow}\ fmmap{\isacharunderscore}{\kern0pt}keys\ {\isacharparenleft}{\kern0pt}{\isasymlambda}v\ e{\isachardot}{\kern0pt}\ val\isactrlsub S\ e\ {\isasymrho}{\isacharparenright}{\kern0pt}\ {\isasymsigma}\ {\isacharequal}{\kern0pt}\ {\isasymsigma}{\isachardoublequoteclose}\isanewline
\isadelimproof
\ \ %
\endisadelimproof
\isatagproof
\isacommand{proof}\isamarkupfalse%
\ {\isacharparenleft}{\kern0pt}rule\ impI{\isacharparenright}{\kern0pt}\isanewline
\ \ \ \ \isacommand{assume}\isamarkupfalse%
\ premise{\isacharcolon}{\kern0pt}\ {\isachardoublequoteopen}concrete\isactrlsub {\isasymSigma}{\isacharparenleft}{\kern0pt}{\isasymsigma}{\isacharparenright}{\kern0pt}{\isachardoublequoteclose}\isanewline
\ \ \ \ %
\isamarkupcmt{We assume that \isa{{\isasymsigma}} is of concrete nature.%
}\isanewline
\ \ \ \ \isacommand{hence}\isamarkupfalse%
\ {\isachardoublequoteopen}{\isasymforall}v\ {\isasymin}\ fmdom{\isacharprime}{\kern0pt}{\isacharparenleft}{\kern0pt}{\isasymsigma}{\isacharparenright}{\kern0pt}{\isachardot}{\kern0pt}\ val\isactrlsub S{\isacharparenleft}{\kern0pt}the{\isacharparenleft}{\kern0pt}fmlookup\ {\isasymsigma}\ v{\isacharparenright}{\kern0pt}{\isacharparenright}{\kern0pt}\ {\isasymrho}\ {\isacharequal}{\kern0pt}\ the{\isacharparenleft}{\kern0pt}fmlookup\ {\isasymsigma}\ v{\isacharparenright}{\kern0pt}{\isachardoublequoteclose}\isanewline
\ \ \ \ \ \ \isacommand{by}\isamarkupfalse%
\ {\isacharparenleft}{\kern0pt}simp\ add{\isacharcolon}{\kern0pt}\ concrete\isactrlsub {\isasymSigma}{\isacharunderscore}{\kern0pt}def\ value{\isacharunderscore}{\kern0pt}pr\isactrlsub S{\isacharparenright}{\kern0pt}\isanewline
\ \ \ \ %
\isamarkupcmt{This in turn implies that all expressions in the image of \isa{{\isasymsigma}} will be 
       preserved when simplified, which can be inferred by using our earlier expression 
       preservation~theorem~\isa{value{\isacharunderscore}{\kern0pt}pr\isactrlsub S}.%
}\isanewline
\ \ \ \ \isacommand{hence}\isamarkupfalse%
\ {\isachardoublequoteopen}{\isasymforall}v{\isachardot}{\kern0pt}\ fmlookup\ {\isacharparenleft}{\kern0pt}fmmap{\isacharunderscore}{\kern0pt}keys\ {\isacharparenleft}{\kern0pt}{\isasymlambda}v\ e{\isachardot}{\kern0pt}\ val\isactrlsub S\ e\ {\isasymrho}{\isacharparenright}{\kern0pt}\ {\isasymsigma}{\isacharparenright}{\kern0pt}\ v\ {\isacharequal}{\kern0pt}\ fmlookup\ {\isasymsigma}\ v{\isachardoublequoteclose}\isanewline
\ \ \ \ \ \ \isacommand{by}\isamarkupfalse%
\ {\isacharparenleft}{\kern0pt}smt\ {\isacharparenleft}{\kern0pt}z{\isadigit{3}}{\isacharparenright}{\kern0pt}\ fmdom{\isacharprime}{\kern0pt}{\isacharunderscore}{\kern0pt}notD\ fmlookup{\isacharunderscore}{\kern0pt}dom{\isacharprime}{\kern0pt}{\isacharunderscore}{\kern0pt}iff\ fmlookup{\isacharunderscore}{\kern0pt}fmmap{\isacharunderscore}{\kern0pt}keys\ map{\isacharunderscore}{\kern0pt}option{\isacharunderscore}{\kern0pt}eq{\isacharunderscore}{\kern0pt}Some\ option{\isachardot}{\kern0pt}sel{\isacharparenright}{\kern0pt}\isanewline
\ \ \ \ %
\isamarkupcmt{We can now use the SMT solver in order to establish that the simplified state 
       and the original state match in all their key-value~tuples.%
}\isanewline
\ \ \ \ \isacommand{thus}\isamarkupfalse%
\ {\isachardoublequoteopen}fmmap{\isacharunderscore}{\kern0pt}keys\ {\isacharparenleft}{\kern0pt}{\isasymlambda}v\ e{\isachardot}{\kern0pt}\ val\isactrlsub S\ e\ {\isasymrho}{\isacharparenright}{\kern0pt}\ {\isasymsigma}\ {\isacharequal}{\kern0pt}\ {\isasymsigma}{\isachardoublequoteclose}\ \isanewline
\ \ \ \ \ \ \isacommand{using}\isamarkupfalse%
\ fmap{\isacharunderscore}{\kern0pt}ext\ \isacommand{by}\isamarkupfalse%
\ blast\isanewline
\ \ \ \ %
\isamarkupcmt{Finally, we apply the equivalence definition of finite maps to deduce their 
       equality, which is what needed to be proven in the first~place.%
}\isanewline
\ \ \isacommand{qed}\isamarkupfalse%
\endisatagproof
{\isafoldproof}%
\isadelimproof
\endisadelimproof
\begin{isamarkuptext}%
Let us assume that \isa{{\isasymsigma}{\isacharprime}{\kern0pt}} is the result of simplifying \isa{{\isasymsigma}} with a state \isa{{\isasymrho}}. 
  Let us furthermore assume that we update \isa{{\isasymsigma}}, such that an arbitrary variable \isa{y}
  is now mapped onto an arbitrary expression \isa{e}. If we simplify this updated state 
  using \isa{{\isasymrho}}, we will then result in state \isa{{\isasymsigma}{\isacharprime}{\kern0pt}}, in which \isa{y} is updated with the 
  under \isa{{\isasymrho}} evaluated version of \isa{e}. \par
  This property directly follows from the definition of the \isa{fmmap{\isacharminus}{\kern0pt}keys} function. 
  Note that this lemma is crucial in our proof system, as it ensures that 
  we can later remove the \isa{fmmap{\isacharminus}{\kern0pt}keys} function step by step out of the corresponding 
  proof~obligations.%
\end{isamarkuptext}\isamarkuptrue%
\ \ \isacommand{lemma}\isamarkupfalse%
\ fmmap{\isacharunderscore}{\kern0pt}keys{\isacharunderscore}{\kern0pt}dom{\isacharunderscore}{\kern0pt}upd{\isacharcolon}{\kern0pt}\isanewline
\ \ \ \ \isakeyword{assumes}\ {\isachardoublequoteopen}{\isasymsigma}{\isacharprime}{\kern0pt}\ {\isacharequal}{\kern0pt}\ fmmap{\isacharunderscore}{\kern0pt}keys\ {\isacharparenleft}{\kern0pt}{\isasymlambda}v\ e{\isachardot}{\kern0pt}\ val\isactrlsub S\ e\ {\isasymrho}{\isacharparenright}{\kern0pt}\ {\isasymsigma}{\isachardoublequoteclose}\isanewline
\ \ \ \ \isakeyword{shows}\ {\isachardoublequoteopen}fmmap{\isacharunderscore}{\kern0pt}keys\ {\isacharparenleft}{\kern0pt}{\isasymlambda}v\ e{\isachardot}{\kern0pt}\ val\isactrlsub S\ e\ {\isasymrho}{\isacharparenright}{\kern0pt}\ {\isacharparenleft}{\kern0pt}{\isacharbrackleft}{\kern0pt}y\ {\isasymlongmapsto}\ e{\isacharbrackright}{\kern0pt}\ {\isasymsigma}{\isacharparenright}{\kern0pt}\ {\isacharequal}{\kern0pt}\ {\isacharbrackleft}{\kern0pt}y\ {\isasymlongmapsto}\ val\isactrlsub S\ e\ {\isasymrho}{\isacharbrackright}{\kern0pt}\ {\isasymsigma}{\isacharprime}{\kern0pt}{\isachardoublequoteclose}\isanewline
\isadelimproof
\ \ %
\endisadelimproof
\isatagproof
\isacommand{proof}\isamarkupfalse%
\ {\isacharminus}{\kern0pt}\isanewline
\ \ \ \ \isacommand{have}\isamarkupfalse%
\ {\isachardoublequoteopen}{\isasymforall}v{\isachardot}{\kern0pt}\ fmlookup\ {\isacharparenleft}{\kern0pt}fmmap{\isacharunderscore}{\kern0pt}keys\ {\isacharparenleft}{\kern0pt}{\isasymlambda}v\ e{\isachardot}{\kern0pt}\ val\isactrlsub S\ e\ {\isasymrho}{\isacharparenright}{\kern0pt}\ {\isacharparenleft}{\kern0pt}{\isacharbrackleft}{\kern0pt}y\ {\isasymlongmapsto}\ e{\isacharbrackright}{\kern0pt}\ {\isasymsigma}{\isacharparenright}{\kern0pt}{\isacharparenright}{\kern0pt}\ v\ {\isacharequal}{\kern0pt}\ fmlookup\ {\isacharparenleft}{\kern0pt}{\isacharbrackleft}{\kern0pt}y\ {\isasymlongmapsto}\ val\isactrlsub S\ e\ {\isasymrho}{\isacharbrackright}{\kern0pt}\ {\isasymsigma}{\isacharprime}{\kern0pt}{\isacharparenright}{\kern0pt}\ v{\isachardoublequoteclose}\isanewline
\ \ \ \ \ \ \isacommand{using}\isamarkupfalse%
\ assms\ \isacommand{by}\isamarkupfalse%
\ fastforce\ \ \ \ \isanewline
\ \ \ \ %
\isamarkupcmt{Due to the assumptions, we know that simplifying \isa{{\isasymsigma}} with \isa{{\isasymrho}} results in \isa{{\isasymsigma}{\isacharprime}{\kern0pt}}.
       The definition of the \isa{fmmap{\isacharminus}{\kern0pt}keys} function furthermore guarantees that the
       lookup value of the updated variable y matches \isa{e} simplified with \isa{{\isasymrho}}. This
       establishes that both states of the equation align in all their lookup~values.%
}\isanewline
\ \ \ \ \isacommand{thus}\isamarkupfalse%
\ {\isacharquery}{\kern0pt}thesis\ \isanewline
\ \ \ \ \ \ \isacommand{using}\isamarkupfalse%
\ fmap{\isacharunderscore}{\kern0pt}ext\ \isacommand{by}\isamarkupfalse%
\ blast\isanewline
\ \ \ \ %
\isamarkupcmt{We can then deduce that both states are equivalent using the equivalence
       notion of finite maps. This concludes the~lemma.%
}\isanewline
\ \ \isacommand{qed}\isamarkupfalse%
\endisatagproof
{\isafoldproof}%
\isadelimproof
\isanewline
\endisadelimproof
\isadelimtheory
\isanewline
\endisadelimtheory
\isatagtheory
\isacommand{end}\isamarkupfalse%
\endisatagtheory
{\isafoldtheory}%
\isadelimtheory
\endisadelimtheory
\end{isabellebody}%

%% file: LAGC_WL.tex
\begin{isabellebody}%
\setisabellecontext{LAGC{\isacharunderscore}{\kern0pt}WL}%
\isadelimdocument
\endisadelimdocument
\isatagdocument
\isamarkupsection{LAGC Semantics for WL%
}
\isamarkuptrue%
\endisatagdocument
{\isafolddocument}%
\isadelimdocument
\endisadelimdocument
\isadelimtheory
\endisadelimtheory
\isatagtheory
\isacommand{theory}\isamarkupfalse%
\ LAGC{\isacharunderscore}{\kern0pt}WL\isanewline
\ \ \isakeyword{imports}\ {\isachardoublequoteopen}{\isachardot}{\kern0pt}{\isachardot}{\kern0pt}{\isacharslash}{\kern0pt}basics{\isacharslash}{\kern0pt}LAGC{\isacharunderscore}{\kern0pt}Base{\isachardoublequoteclose}\isanewline
\isakeyword{begin}%
\endisatagtheory
{\isafoldtheory}%
\isadelimtheory
\endisadelimtheory
\begin{isamarkuptext}%
In this chapter we formalize the LAGC semantics for the standard
  While Language \isa{WL} as described in section~3 of the original paper.
  One of our main objectives is establishing a proof automation system for
  the trace construction, implying that we will later be able to systematically 
  derive the set of traces generated for any specific program within Isabelle. The
  second goal is setting up an efficient code generation for the construction of
  traces, thereby ensuring that we can later dynamically output the set of traces for any
  particular program in the~console.%
\end{isamarkuptext}\isamarkuptrue%
\isadelimdocument
\endisadelimdocument
\isatagdocument
\isamarkupsubsection{While Language (WL)%
}
\isamarkuptrue%
\isamarkupsubsubsection{Syntax%
}
\isamarkuptrue%
\endisatagdocument
{\isafolddocument}%
\isadelimdocument
\endisadelimdocument
\begin{isamarkuptext}%
We begin by defining the syntax of the standard imperative While Language (WL), 
  which we will later use as the underlying programming language of our LAGC
  semantics. Statements of WL can be categorized~as~follows: \begin{description}
  \item[Skip Statement] This command performs no operation.
  \item[Assignment] This command assigns a variable the value of an 
      arithmetic~expression. Note that a variable cannot be assigned to a Boolean 
      value, thereby aligning with the definition~of~states.
  \item[If-Branch] An If-Branch hides a statement behind a Boolean guard. If the
      Boolean guard evaluates to true, the statement is executed. Otherwise,
      the statement is~skipped.
  \item[While-Loop] A While-Loop consists of a Boolean guard and a statement. If the
      Boolean guard evaluates to true, the statement is executed and the loop
      preserved. Otherwise, nothing happens and the loop~exits.
  \item[Sequential Statement] This command executes two statements in~sequence.
  \end{description}%
\end{isamarkuptext}\isamarkuptrue%
\ \ \isacommand{datatype}\isamarkupfalse%
\ stmt\ {\isacharequal}{\kern0pt}\ \isanewline
\ \ \ \ \ \ SKIP\ %
\isamarkupcmt{No-Op%
}\isanewline
\ \ \ \ {\isacharbar}{\kern0pt}\ Assign\ var\ aexp\ %
\isamarkupcmt{Assignment of a variable%
}\isanewline
\ \ \ \ {\isacharbar}{\kern0pt}\ If\ bexp\ stmt\ %
\isamarkupcmt{If-Branch%
}\isanewline
\ \ \ \ {\isacharbar}{\kern0pt}\ While\ bexp\ stmt\ %
\isamarkupcmt{While-Loop%
}\isanewline
\ \ \ \ {\isacharbar}{\kern0pt}\ Seq\ stmt\ stmt\ %
\isamarkupcmt{Sequential Statement%
}%
\begin{isamarkuptext}%
We also add a minimal concrete syntax for WL statements, so as to
  improve the readability of programs.%
\end{isamarkuptext}\isamarkuptrue%
\ \ \isacommand{notation}\isamarkupfalse%
\ Assign\ {\isacharparenleft}{\kern0pt}\isakeyword{infix}\ {\isachardoublequoteopen}{\isacharcolon}{\kern0pt}{\isacharequal}{\kern0pt}{\isachardoublequoteclose}\ {\isadigit{6}}{\isadigit{1}}{\isacharparenright}{\kern0pt}\isanewline
\ \ \isacommand{notation}\isamarkupfalse%
\ If\ {\isacharparenleft}{\kern0pt}{\isachardoublequoteopen}{\isacharparenleft}{\kern0pt}IF\ {\isacharunderscore}{\kern0pt}{\isacharslash}{\kern0pt}\ THEN\ {\isacharunderscore}{\kern0pt}{\isacharslash}{\kern0pt}\ FI{\isacharparenright}{\kern0pt}{\isachardoublequoteclose}\ {\isacharbrackleft}{\kern0pt}{\isadigit{1}}{\isadigit{0}}{\isadigit{0}}{\isadigit{0}}{\isacharcomma}{\kern0pt}\ {\isadigit{0}}{\isacharbrackright}{\kern0pt}\ {\isadigit{6}}{\isadigit{1}}{\isacharparenright}{\kern0pt}\isanewline
\ \ \isacommand{notation}\isamarkupfalse%
\ While\ {\isacharparenleft}{\kern0pt}{\isachardoublequoteopen}{\isacharparenleft}{\kern0pt}WHILE\ {\isacharunderscore}{\kern0pt}{\isacharslash}{\kern0pt}\ DO\ {\isacharunderscore}{\kern0pt}{\isacharslash}{\kern0pt}\ OD{\isacharparenright}{\kern0pt}{\isachardoublequoteclose}\ {\isacharbrackleft}{\kern0pt}{\isadigit{1}}{\isadigit{0}}{\isadigit{0}}{\isadigit{0}}{\isacharcomma}{\kern0pt}\ {\isadigit{0}}{\isacharbrackright}{\kern0pt}\ {\isadigit{6}}{\isadigit{1}}{\isacharparenright}{\kern0pt}\isanewline
\ \ \isacommand{notation}\isamarkupfalse%
\ Seq\ {\isacharparenleft}{\kern0pt}\isakeyword{infix}\ {\isachardoublequoteopen}{\isacharsemicolon}{\kern0pt}{\isacharsemicolon}{\kern0pt}{\isachardoublequoteclose}\ {\isadigit{6}}{\isadigit{0}}{\isacharparenright}{\kern0pt}%
\begin{isamarkuptext}%
Using our previously defined grammar, it is now possible to derive 
  syntactically correct WL statements. We demonstrate this by presenting short 
  examples for programs utilizing our minimal concrete syntax. The first presented 
  program switches the program variables x and y, whilst the second program computes 
  the factorial~of~6.%
\end{isamarkuptext}\isamarkuptrue%
\ \ \isacommand{definition}\isamarkupfalse%
\isanewline
\ \ \ \ WL{\isacharunderscore}{\kern0pt}ex\isactrlsub {\isadigit{1}}\ {\isacharcolon}{\kern0pt}{\isacharcolon}{\kern0pt}\ stmt\ \isakeyword{where}\isanewline
\ \ \ \ {\isachardoublequoteopen}WL{\isacharunderscore}{\kern0pt}ex\isactrlsub {\isadigit{1}}\ {\isasymequiv}\ IF\ {\isacharparenleft}{\kern0pt}not\ {\isacharparenleft}{\kern0pt}{\isacharparenleft}{\kern0pt}Var\ {\isacharprime}{\kern0pt}{\isacharprime}{\kern0pt}x{\isacharprime}{\kern0pt}{\isacharprime}{\kern0pt}{\isacharparenright}{\kern0pt}\ \isactrlsub Req\ {\isacharparenleft}{\kern0pt}Var\ {\isacharprime}{\kern0pt}{\isacharprime}{\kern0pt}y{\isacharprime}{\kern0pt}{\isacharprime}{\kern0pt}{\isacharparenright}{\kern0pt}{\isacharparenright}{\kern0pt}{\isacharparenright}{\kern0pt}\isanewline
\ \ \ \ \ \ \ \ \ \ \ \ \ \ \ \ \ \ \ \ \ \ \ \ \ \ THEN\ {\isacharparenleft}{\kern0pt}{\isacharprime}{\kern0pt}{\isacharprime}{\kern0pt}z{\isacharprime}{\kern0pt}{\isacharprime}{\kern0pt}\ {\isacharcolon}{\kern0pt}{\isacharequal}{\kern0pt}\ Var\ {\isacharprime}{\kern0pt}{\isacharprime}{\kern0pt}y{\isacharprime}{\kern0pt}{\isacharprime}{\kern0pt}{\isacharsemicolon}{\kern0pt}{\isacharsemicolon}{\kern0pt}\ {\isacharprime}{\kern0pt}{\isacharprime}{\kern0pt}y{\isacharprime}{\kern0pt}{\isacharprime}{\kern0pt}\ {\isacharcolon}{\kern0pt}{\isacharequal}{\kern0pt}\ Var\ {\isacharprime}{\kern0pt}{\isacharprime}{\kern0pt}x{\isacharprime}{\kern0pt}{\isacharprime}{\kern0pt}{\isacharparenright}{\kern0pt}{\isacharsemicolon}{\kern0pt}{\isacharsemicolon}{\kern0pt}\ {\isacharprime}{\kern0pt}{\isacharprime}{\kern0pt}x{\isacharprime}{\kern0pt}{\isacharprime}{\kern0pt}\ {\isacharcolon}{\kern0pt}{\isacharequal}{\kern0pt}\ Var\ {\isacharprime}{\kern0pt}{\isacharprime}{\kern0pt}z{\isacharprime}{\kern0pt}{\isacharprime}{\kern0pt}\ \isanewline
\ \ \ \ \ \ \ \ \ \ \ \ \ \ \ \ \ \ \ \ FI{\isachardoublequoteclose}\isanewline
\isanewline
\ \ \isacommand{definition}\isamarkupfalse%
\isanewline
\ \ \ \ WL{\isacharunderscore}{\kern0pt}ex\isactrlsub {\isadigit{2}}\ {\isacharcolon}{\kern0pt}{\isacharcolon}{\kern0pt}\ stmt\ \isakeyword{where}\isanewline
\ \ \ \ {\isachardoublequoteopen}WL{\isacharunderscore}{\kern0pt}ex\isactrlsub {\isadigit{2}}\ {\isasymequiv}\ {\isacharparenleft}{\kern0pt}{\isacharprime}{\kern0pt}{\isacharprime}{\kern0pt}x{\isacharprime}{\kern0pt}{\isacharprime}{\kern0pt}\ {\isacharcolon}{\kern0pt}{\isacharequal}{\kern0pt}\ Num\ {\isadigit{6}}{\isacharsemicolon}{\kern0pt}{\isacharsemicolon}{\kern0pt}\ {\isacharprime}{\kern0pt}{\isacharprime}{\kern0pt}y{\isacharprime}{\kern0pt}{\isacharprime}{\kern0pt}\ {\isacharcolon}{\kern0pt}{\isacharequal}{\kern0pt}\ Num\ {\isadigit{1}}{\isacharparenright}{\kern0pt}{\isacharsemicolon}{\kern0pt}{\isacharsemicolon}{\kern0pt}\ \isanewline
\ \ \ \ \ \ \ \ \ \ \ \ \ \ \ \ \ \ \ \ WHILE\ {\isacharparenleft}{\kern0pt}{\isacharparenleft}{\kern0pt}Var\ {\isacharprime}{\kern0pt}{\isacharprime}{\kern0pt}x{\isacharprime}{\kern0pt}{\isacharprime}{\kern0pt}{\isacharparenright}{\kern0pt}\ \isactrlsub Rgeq\ {\isacharparenleft}{\kern0pt}Num\ {\isadigit{2}}{\isacharparenright}{\kern0pt}{\isacharparenright}{\kern0pt}\ DO\ \isanewline
\ \ \ \ \ \ \ \ \ \ \ \ \ \ \ \ \ \ \ \ \ \ \ \ \ \ {\isacharprime}{\kern0pt}{\isacharprime}{\kern0pt}y{\isacharprime}{\kern0pt}{\isacharprime}{\kern0pt}\ {\isacharcolon}{\kern0pt}{\isacharequal}{\kern0pt}\ {\isacharparenleft}{\kern0pt}Var\ {\isacharprime}{\kern0pt}{\isacharprime}{\kern0pt}y{\isacharprime}{\kern0pt}{\isacharprime}{\kern0pt}{\isacharparenright}{\kern0pt}\ \isactrlsub Amul\ {\isacharparenleft}{\kern0pt}Var\ {\isacharprime}{\kern0pt}{\isacharprime}{\kern0pt}x{\isacharprime}{\kern0pt}{\isacharprime}{\kern0pt}{\isacharparenright}{\kern0pt}{\isacharsemicolon}{\kern0pt}{\isacharsemicolon}{\kern0pt}\ \isanewline
\ \ \ \ \ \ \ \ \ \ \ \ \ \ \ \ \ \ \ \ \ \ \ \ \ \ {\isacharprime}{\kern0pt}{\isacharprime}{\kern0pt}x{\isacharprime}{\kern0pt}{\isacharprime}{\kern0pt}\ {\isacharcolon}{\kern0pt}{\isacharequal}{\kern0pt}\ {\isacharparenleft}{\kern0pt}Var\ {\isacharprime}{\kern0pt}{\isacharprime}{\kern0pt}x{\isacharprime}{\kern0pt}{\isacharprime}{\kern0pt}{\isacharparenright}{\kern0pt}\ \isactrlsub Asub\ {\isacharparenleft}{\kern0pt}Num\ {\isadigit{1}}{\isacharparenright}{\kern0pt}\ \isanewline
\ \ \ \ \ \ \ \ \ \ \ \ \ \ \ \ \ \ \ \ OD{\isachardoublequoteclose}%
\isadelimdocument
\endisadelimdocument
\isatagdocument
\isamarkupsubsubsection{Variable Mappings%
}
\isamarkuptrue%
\endisatagdocument
{\isafolddocument}%
\isadelimdocument
\endisadelimdocument
\begin{isamarkuptext}%
We now introduce variable mappings for WL programs mapping specific
  statements to a set of their enclosed free variables. The definition of the 
  function is~straightforward.%
\end{isamarkuptext}\isamarkuptrue%
\ \ \isacommand{fun}\isamarkupfalse%
\isanewline
\ \ \ \ vars\ {\isacharcolon}{\kern0pt}{\isacharcolon}{\kern0pt}\ {\isachardoublequoteopen}stmt\ {\isasymRightarrow}\ var\ set{\isachardoublequoteclose}\ \isakeyword{where}\isanewline
\ \ \ \ {\isachardoublequoteopen}vars\ SKIP\ {\isacharequal}{\kern0pt}\ {\isacharbraceleft}{\kern0pt}{\isacharbraceright}{\kern0pt}{\isachardoublequoteclose}\ {\isacharbar}{\kern0pt}\isanewline
\ \ \ \ {\isachardoublequoteopen}vars\ {\isacharparenleft}{\kern0pt}Assign\ x\ a{\isacharparenright}{\kern0pt}\ {\isacharequal}{\kern0pt}\ {\isacharbraceleft}{\kern0pt}x{\isacharbraceright}{\kern0pt}\ {\isasymunion}\ vars\isactrlsub A{\isacharparenleft}{\kern0pt}a{\isacharparenright}{\kern0pt}{\isachardoublequoteclose}\ {\isacharbar}{\kern0pt}\isanewline
\ \ \ \ {\isachardoublequoteopen}vars\ {\isacharparenleft}{\kern0pt}If\ b\ S{\isacharparenright}{\kern0pt}\ {\isacharequal}{\kern0pt}\ vars\isactrlsub B{\isacharparenleft}{\kern0pt}b{\isacharparenright}{\kern0pt}\ {\isasymunion}\ vars{\isacharparenleft}{\kern0pt}S{\isacharparenright}{\kern0pt}{\isachardoublequoteclose}\ {\isacharbar}{\kern0pt}\isanewline
\ \ \ \ {\isachardoublequoteopen}vars\ {\isacharparenleft}{\kern0pt}While\ b\ S{\isacharparenright}{\kern0pt}\ {\isacharequal}{\kern0pt}\ vars\isactrlsub B{\isacharparenleft}{\kern0pt}b{\isacharparenright}{\kern0pt}\ {\isasymunion}\ vars{\isacharparenleft}{\kern0pt}S{\isacharparenright}{\kern0pt}{\isachardoublequoteclose}\ {\isacharbar}{\kern0pt}\isanewline
\ \ \ \ {\isachardoublequoteopen}vars\ {\isacharparenleft}{\kern0pt}Seq\ S\isactrlsub {\isadigit{1}}\ S\isactrlsub {\isadigit{2}}{\isacharparenright}{\kern0pt}\ {\isacharequal}{\kern0pt}\ vars{\isacharparenleft}{\kern0pt}S\isactrlsub {\isadigit{1}}{\isacharparenright}{\kern0pt}\ {\isasymunion}\ vars{\isacharparenleft}{\kern0pt}S\isactrlsub {\isadigit{2}}{\isacharparenright}{\kern0pt}{\isachardoublequoteclose}%
\begin{isamarkuptext}%
Similar to expressions, we also provide a variable occurrence function
  that maps specific statements onto a list of all variables occurring in the
  statement. Mapping onto a finite list instead of a (theoretically) infinite set 
  ensures that all variables of programs can be systematically traversed, thereby 
  allowing us to later establish a notion of initial program states. Note that the 
  returned list can consist of duplicate variables, depending on the number of 
  variable occurrences in the actual~program.%
\end{isamarkuptext}\isamarkuptrue%
\ \ \isacommand{fun}\isamarkupfalse%
\isanewline
\ \ \ \ occ\ {\isacharcolon}{\kern0pt}{\isacharcolon}{\kern0pt}\ {\isachardoublequoteopen}stmt\ {\isasymRightarrow}\ var\ list{\isachardoublequoteclose}\ \isakeyword{where}\isanewline
\ \ \ \ {\isachardoublequoteopen}occ\ SKIP\ {\isacharequal}{\kern0pt}\ {\isacharbrackleft}{\kern0pt}{\isacharbrackright}{\kern0pt}{\isachardoublequoteclose}\ {\isacharbar}{\kern0pt}\isanewline
\ \ \ \ {\isachardoublequoteopen}occ\ {\isacharparenleft}{\kern0pt}Assign\ x\ a{\isacharparenright}{\kern0pt}\ {\isacharequal}{\kern0pt}\ x\ {\isacharhash}{\kern0pt}\ occ\isactrlsub A{\isacharparenleft}{\kern0pt}a{\isacharparenright}{\kern0pt}{\isachardoublequoteclose}\ {\isacharbar}{\kern0pt}\isanewline
\ \ \ \ {\isachardoublequoteopen}occ\ {\isacharparenleft}{\kern0pt}If\ b\ S{\isacharparenright}{\kern0pt}\ {\isacharequal}{\kern0pt}\ occ\isactrlsub B{\isacharparenleft}{\kern0pt}b{\isacharparenright}{\kern0pt}\ {\isacharat}{\kern0pt}\ occ{\isacharparenleft}{\kern0pt}S{\isacharparenright}{\kern0pt}{\isachardoublequoteclose}\ {\isacharbar}{\kern0pt}\isanewline
\ \ \ \ {\isachardoublequoteopen}occ\ {\isacharparenleft}{\kern0pt}While\ b\ S{\isacharparenright}{\kern0pt}\ {\isacharequal}{\kern0pt}\ occ\isactrlsub B{\isacharparenleft}{\kern0pt}b{\isacharparenright}{\kern0pt}\ {\isacharat}{\kern0pt}\ occ{\isacharparenleft}{\kern0pt}S{\isacharparenright}{\kern0pt}{\isachardoublequoteclose}\ {\isacharbar}{\kern0pt}\isanewline
\ \ \ \ {\isachardoublequoteopen}occ\ {\isacharparenleft}{\kern0pt}Seq\ S\isactrlsub {\isadigit{1}}\ S\isactrlsub {\isadigit{2}}{\isacharparenright}{\kern0pt}\ {\isacharequal}{\kern0pt}\ occ{\isacharparenleft}{\kern0pt}S\isactrlsub {\isadigit{1}}{\isacharparenright}{\kern0pt}\ {\isacharat}{\kern0pt}\ occ{\isacharparenleft}{\kern0pt}S\isactrlsub {\isadigit{2}}{\isacharparenright}{\kern0pt}{\isachardoublequoteclose}%
\begin{isamarkuptext}%
We can now take another look at our earlier program examples and analyze
  the result of applying a variable mapping and variable occurrence~function.%
\end{isamarkuptext}\isamarkuptrue%
\ \ \isacommand{lemma}\isamarkupfalse%
\ {\isachardoublequoteopen}vars\ WL{\isacharunderscore}{\kern0pt}ex\isactrlsub {\isadigit{1}}\ {\isacharequal}{\kern0pt}\ {\isacharbraceleft}{\kern0pt}{\isacharprime}{\kern0pt}{\isacharprime}{\kern0pt}x{\isacharprime}{\kern0pt}{\isacharprime}{\kern0pt}{\isacharcomma}{\kern0pt}\ {\isacharprime}{\kern0pt}{\isacharprime}{\kern0pt}y{\isacharprime}{\kern0pt}{\isacharprime}{\kern0pt}{\isacharcomma}{\kern0pt}\ {\isacharprime}{\kern0pt}{\isacharprime}{\kern0pt}z{\isacharprime}{\kern0pt}{\isacharprime}{\kern0pt}{\isacharbraceright}{\kern0pt}{\isachardoublequoteclose}\isanewline
\isadelimproof
\ \ \ \ %
\endisadelimproof
\isatagproof
\isacommand{by}\isamarkupfalse%
\ {\isacharparenleft}{\kern0pt}auto\ simp\ add{\isacharcolon}{\kern0pt}\ WL{\isacharunderscore}{\kern0pt}ex\isactrlsub {\isadigit{1}}{\isacharunderscore}{\kern0pt}def{\isacharparenright}{\kern0pt}%
\endisatagproof
{\isafoldproof}%
\isadelimproof
\isanewline
\endisadelimproof
\isanewline
\ \ \isacommand{lemma}\isamarkupfalse%
\ {\isachardoublequoteopen}vars\ WL{\isacharunderscore}{\kern0pt}ex\isactrlsub {\isadigit{2}}\ {\isacharequal}{\kern0pt}\ {\isacharbraceleft}{\kern0pt}{\isacharprime}{\kern0pt}{\isacharprime}{\kern0pt}x{\isacharprime}{\kern0pt}{\isacharprime}{\kern0pt}{\isacharcomma}{\kern0pt}\ {\isacharprime}{\kern0pt}{\isacharprime}{\kern0pt}y{\isacharprime}{\kern0pt}{\isacharprime}{\kern0pt}{\isacharbraceright}{\kern0pt}{\isachardoublequoteclose}\isanewline
\isadelimproof
\ \ \ \ %
\endisadelimproof
\isatagproof
\isacommand{by}\isamarkupfalse%
\ {\isacharparenleft}{\kern0pt}auto\ simp\ add{\isacharcolon}{\kern0pt}\ WL{\isacharunderscore}{\kern0pt}ex\isactrlsub {\isadigit{2}}{\isacharunderscore}{\kern0pt}def{\isacharparenright}{\kern0pt}%
\endisatagproof
{\isafoldproof}%
\isadelimproof
\isanewline
\endisadelimproof
\isanewline
\ \ \isacommand{lemma}\isamarkupfalse%
\ {\isachardoublequoteopen}occ\ WL{\isacharunderscore}{\kern0pt}ex\isactrlsub {\isadigit{1}}\ {\isacharequal}{\kern0pt}\ {\isacharbrackleft}{\kern0pt}{\isacharprime}{\kern0pt}{\isacharprime}{\kern0pt}x{\isacharprime}{\kern0pt}{\isacharprime}{\kern0pt}{\isacharcomma}{\kern0pt}\ {\isacharprime}{\kern0pt}{\isacharprime}{\kern0pt}y{\isacharprime}{\kern0pt}{\isacharprime}{\kern0pt}{\isacharcomma}{\kern0pt}\ {\isacharprime}{\kern0pt}{\isacharprime}{\kern0pt}z{\isacharprime}{\kern0pt}{\isacharprime}{\kern0pt}{\isacharcomma}{\kern0pt}\ {\isacharprime}{\kern0pt}{\isacharprime}{\kern0pt}y{\isacharprime}{\kern0pt}{\isacharprime}{\kern0pt}{\isacharcomma}{\kern0pt}\ {\isacharprime}{\kern0pt}{\isacharprime}{\kern0pt}y{\isacharprime}{\kern0pt}{\isacharprime}{\kern0pt}{\isacharcomma}{\kern0pt}\ {\isacharprime}{\kern0pt}{\isacharprime}{\kern0pt}x{\isacharprime}{\kern0pt}{\isacharprime}{\kern0pt}{\isacharcomma}{\kern0pt}\ {\isacharprime}{\kern0pt}{\isacharprime}{\kern0pt}x{\isacharprime}{\kern0pt}{\isacharprime}{\kern0pt}{\isacharcomma}{\kern0pt}\ {\isacharprime}{\kern0pt}{\isacharprime}{\kern0pt}z{\isacharprime}{\kern0pt}{\isacharprime}{\kern0pt}{\isacharbrackright}{\kern0pt}{\isachardoublequoteclose}\isanewline
\isadelimproof
\ \ \ \ %
\endisadelimproof
\isatagproof
\isacommand{by}\isamarkupfalse%
\ {\isacharparenleft}{\kern0pt}auto\ simp\ add{\isacharcolon}{\kern0pt}\ WL{\isacharunderscore}{\kern0pt}ex\isactrlsub {\isadigit{1}}{\isacharunderscore}{\kern0pt}def{\isacharparenright}{\kern0pt}%
\endisatagproof
{\isafoldproof}%
\isadelimproof
\isanewline
\endisadelimproof
\isanewline
\ \ \isacommand{lemma}\isamarkupfalse%
\ {\isachardoublequoteopen}occ\ WL{\isacharunderscore}{\kern0pt}ex\isactrlsub {\isadigit{2}}\ {\isacharequal}{\kern0pt}\ {\isacharbrackleft}{\kern0pt}{\isacharprime}{\kern0pt}{\isacharprime}{\kern0pt}x{\isacharprime}{\kern0pt}{\isacharprime}{\kern0pt}{\isacharcomma}{\kern0pt}\ {\isacharprime}{\kern0pt}{\isacharprime}{\kern0pt}y{\isacharprime}{\kern0pt}{\isacharprime}{\kern0pt}{\isacharcomma}{\kern0pt}\ {\isacharprime}{\kern0pt}{\isacharprime}{\kern0pt}x{\isacharprime}{\kern0pt}{\isacharprime}{\kern0pt}{\isacharcomma}{\kern0pt}\ {\isacharprime}{\kern0pt}{\isacharprime}{\kern0pt}y{\isacharprime}{\kern0pt}{\isacharprime}{\kern0pt}{\isacharcomma}{\kern0pt}\ {\isacharprime}{\kern0pt}{\isacharprime}{\kern0pt}y{\isacharprime}{\kern0pt}{\isacharprime}{\kern0pt}{\isacharcomma}{\kern0pt}\ {\isacharprime}{\kern0pt}{\isacharprime}{\kern0pt}x{\isacharprime}{\kern0pt}{\isacharprime}{\kern0pt}{\isacharcomma}{\kern0pt}\ {\isacharprime}{\kern0pt}{\isacharprime}{\kern0pt}x{\isacharprime}{\kern0pt}{\isacharprime}{\kern0pt}{\isacharcomma}{\kern0pt}\ {\isacharprime}{\kern0pt}{\isacharprime}{\kern0pt}x{\isacharprime}{\kern0pt}{\isacharprime}{\kern0pt}{\isacharbrackright}{\kern0pt}{\isachardoublequoteclose}\isanewline
\isadelimproof
\ \ \ \ %
\endisadelimproof
\isatagproof
\isacommand{by}\isamarkupfalse%
\ {\isacharparenleft}{\kern0pt}auto\ simp\ add{\isacharcolon}{\kern0pt}\ WL{\isacharunderscore}{\kern0pt}ex\isactrlsub {\isadigit{2}}{\isacharunderscore}{\kern0pt}def{\isacharparenright}{\kern0pt}%
\endisatagproof
{\isafoldproof}%
\isadelimproof
\endisadelimproof
\isadelimdocument
\endisadelimdocument
\isatagdocument
\isamarkupsubsubsection{Initial States%
}
\isamarkuptrue%
\endisatagdocument
{\isafolddocument}%
\isadelimdocument
\endisadelimdocument
\begin{isamarkuptext}%
An initial state for a program is a state that maps all variables occurring
  in the program to the arithmetic numeral 0. Contrary to the paper, we decide to 
  formalize the construction of initial program states in an explicit manner, so as to 
  ensure that we can automatically create them instead of having to provide 
  them manually for each program. We abbreviate the initial state of a  
  program~with~\isa{{\isasymsigma}\isactrlsub I}. \par
  Initial program states are constructed as follows: We first apply the variable
  occurrence function on the given program in order to receive a list of all 
  its program variables. We can then traverse this variable list due to its finite 
  nature and construct the desired initial state. Note that this is the 
  exact purpose of the previously defined \isa{get{\isacharunderscore}{\kern0pt}initial\isactrlsub {\isasymSigma}}~function.%
\end{isamarkuptext}\isamarkuptrue%
\ \ \isacommand{fun}\isamarkupfalse%
\isanewline
\ \ \ \ initial\ {\isacharcolon}{\kern0pt}{\isacharcolon}{\kern0pt}\ {\isachardoublequoteopen}stmt\ {\isasymRightarrow}\ {\isasymSigma}{\isachardoublequoteclose}\ {\isacharparenleft}{\kern0pt}{\isachardoublequoteopen}{\isasymsigma}\isactrlsub I{\isachardoublequoteclose}{\isacharparenright}{\kern0pt}\ \isakeyword{where}\isanewline
\ \ \ \ {\isachardoublequoteopen}initial\ S\ {\isacharequal}{\kern0pt}\ get{\isacharunderscore}{\kern0pt}initial\isactrlsub {\isasymSigma}\ {\isacharparenleft}{\kern0pt}occ\ S{\isacharparenright}{\kern0pt}{\isachardoublequoteclose}%
\begin{isamarkuptext}%
We now provide an example for the construction of an initial program state 
  using one of our earlier~programs.%
\end{isamarkuptext}\isamarkuptrue%
\ \ \isacommand{lemma}\isamarkupfalse%
\ {\isachardoublequoteopen}{\isasymsigma}\isactrlsub I\ WL{\isacharunderscore}{\kern0pt}ex\isactrlsub {\isadigit{1}}\ {\isacharequal}{\kern0pt}\ fm{\isacharbrackleft}{\kern0pt}{\isacharparenleft}{\kern0pt}{\isacharprime}{\kern0pt}{\isacharprime}{\kern0pt}x{\isacharprime}{\kern0pt}{\isacharprime}{\kern0pt}{\isacharcomma}{\kern0pt}\ Exp\ {\isacharparenleft}{\kern0pt}Num\ {\isadigit{0}}{\isacharparenright}{\kern0pt}{\isacharparenright}{\kern0pt}{\isacharcomma}{\kern0pt}\ {\isacharparenleft}{\kern0pt}{\isacharprime}{\kern0pt}{\isacharprime}{\kern0pt}y{\isacharprime}{\kern0pt}{\isacharprime}{\kern0pt}{\isacharcomma}{\kern0pt}\ Exp\ {\isacharparenleft}{\kern0pt}Num\ {\isadigit{0}}{\isacharparenright}{\kern0pt}{\isacharparenright}{\kern0pt}{\isacharcomma}{\kern0pt}\ {\isacharparenleft}{\kern0pt}{\isacharprime}{\kern0pt}{\isacharprime}{\kern0pt}z{\isacharprime}{\kern0pt}{\isacharprime}{\kern0pt}{\isacharcomma}{\kern0pt}\ Exp\ {\isacharparenleft}{\kern0pt}Num\ {\isadigit{0}}{\isacharparenright}{\kern0pt}{\isacharparenright}{\kern0pt}{\isacharbrackright}{\kern0pt}{\isachardoublequoteclose}\isanewline
\isadelimproof
\ \ \ \ %
\endisadelimproof
\isatagproof
\isacommand{by}\isamarkupfalse%
\ {\isacharparenleft}{\kern0pt}simp\ add{\isacharcolon}{\kern0pt}\ WL{\isacharunderscore}{\kern0pt}ex\isactrlsub {\isadigit{1}}{\isacharunderscore}{\kern0pt}def\ fmupd{\isacharunderscore}{\kern0pt}reorder{\isacharunderscore}{\kern0pt}neq{\isacharparenright}{\kern0pt}%
\endisatagproof
{\isafoldproof}%
\isadelimproof
\endisadelimproof
\isadelimdocument
\endisadelimdocument
\isatagdocument
\isamarkupsubsection{Continuations%
}
\isamarkuptrue%
\isamarkupsubsubsection{Continuation Markers%
}
\isamarkuptrue%
\endisatagdocument
{\isafolddocument}%
\isadelimdocument
\endisadelimdocument
\begin{isamarkuptext}%
In order to later faithfully capture the local evaluation of our While Language, 
  we will need to define a valuation function that evaluates a statement in a given 
  state and returns the set of all conditioned symbolic traces that can be constructed
  until the next scheduling point is reached. However, composite statements have 
  multiple scheduling points, as they consist of several constituent parts. We 
  therefore need to keep track of the statements that are still left to be evaluated 
  (i.e. all statements after the next scheduling point). This is the exact purpose 
  of a continuation marker. A notion of continuation markers is therefore a crucial 
  prerequisite for setting up the local evaluation of our LAGC~semantics. \par
  A continuation marker has one of the following two forms: \isa{{\isasymlambda}{\isacharbrackleft}{\kern0pt}S{\isacharbrackright}{\kern0pt}} denotes that 
  statement S still needs to be evaluated under the valuation function. \isa{{\isasymlambda}{\isacharbrackleft}{\kern0pt}{\isasymnabla}{\isacharbrackright}{\kern0pt}} 
  refers to the empty continuation, implying that the computation of the program
  has already been fully~completed.%
\end{isamarkuptext}\isamarkuptrue%
\ \ \isacommand{datatype}\isamarkupfalse%
\ cont{\isacharunderscore}{\kern0pt}marker\ {\isacharequal}{\kern0pt}\ \isanewline
\ \ \ \ \ \ Lambda\ stmt\ {\isacharparenleft}{\kern0pt}{\isachardoublequoteopen}{\isasymlambda}{\isacharbrackleft}{\kern0pt}{\isacharunderscore}{\kern0pt}{\isacharbrackright}{\kern0pt}{\isachardoublequoteclose}{\isacharparenright}{\kern0pt}\ %
\isamarkupcmt{Non-Empty Continuation Marker%
}\isanewline
\ \ \ \ {\isacharbar}{\kern0pt}\ Empty\ {\isacharparenleft}{\kern0pt}{\isachardoublequoteopen}{\isasymlambda}{\isacharbrackleft}{\kern0pt}{\isasymnabla}{\isacharbrackright}{\kern0pt}{\isachardoublequoteclose}{\isacharparenright}{\kern0pt}\ %
\isamarkupcmt{Empty Continuation Marker%
}%
\begin{isamarkuptext}%
In contrast to the original paper, we propose an additional variable mapping for 
  continuation markers, mapping a non-empty continuation marker to all free variables
  occurring in its encased statement. Note that the empty continuation marker contains 
  no free~variables.%
\end{isamarkuptext}\isamarkuptrue%
\ \ \isacommand{fun}\isamarkupfalse%
\isanewline
\ \ \ \ mvars\ {\isacharcolon}{\kern0pt}{\isacharcolon}{\kern0pt}\ {\isachardoublequoteopen}cont{\isacharunderscore}{\kern0pt}marker\ {\isasymRightarrow}\ var\ set{\isachardoublequoteclose}\ \isakeyword{where}\isanewline
\ \ \ \ {\isachardoublequoteopen}mvars\ {\isasymlambda}{\isacharbrackleft}{\kern0pt}{\isasymnabla}{\isacharbrackright}{\kern0pt}\ {\isacharequal}{\kern0pt}\ {\isacharbraceleft}{\kern0pt}{\isacharbraceright}{\kern0pt}{\isachardoublequoteclose}\ {\isacharbar}{\kern0pt}\isanewline
\ \ \ \ {\isachardoublequoteopen}mvars\ {\isasymlambda}{\isacharbrackleft}{\kern0pt}S{\isacharbrackright}{\kern0pt}\ {\isacharequal}{\kern0pt}\ vars{\isacharparenleft}{\kern0pt}S{\isacharparenright}{\kern0pt}{\isachardoublequoteclose}%
\isadelimdocument
\endisadelimdocument
\isatagdocument
\isamarkupsubsubsection{Continuation Traces%
}
\isamarkuptrue%
\endisatagdocument
{\isafolddocument}%
\isadelimdocument
\endisadelimdocument
\begin{isamarkuptext}%
Analogous to the original paper, we can now define a continuation trace as a 
  conditioned symbolic trace with an additional appended continuation~marker.%
\end{isamarkuptext}\isamarkuptrue%
\ \ \isacommand{datatype}\isamarkupfalse%
\ cont{\isacharunderscore}{\kern0pt}trace\ {\isacharequal}{\kern0pt}\ \isanewline
\ \ \ \ \ \ Cont\ {\isasymCC}{\isasymT}\ cont{\isacharunderscore}{\kern0pt}marker\ {\isacharparenleft}{\kern0pt}\isakeyword{infix}\ {\isachardoublequoteopen}\isactrlitem {\isachardoublequoteclose}\ {\isadigit{5}}{\isadigit{5}}{\isacharparenright}{\kern0pt}%
\begin{isamarkuptext}%
In order to ease the handling of continuation traces, we propose additional
  projections which map a continuation trace onto its encased~components.%
\end{isamarkuptext}\isamarkuptrue%
\ \ \isacommand{fun}\isamarkupfalse%
\isanewline
\ \ \ \ proj{\isacharunderscore}{\kern0pt}pc\ {\isacharcolon}{\kern0pt}{\isacharcolon}{\kern0pt}\ {\isachardoublequoteopen}cont{\isacharunderscore}{\kern0pt}trace\ {\isasymRightarrow}\ path{\isacharunderscore}{\kern0pt}condition{\isachardoublequoteclose}\ {\isacharparenleft}{\kern0pt}{\isachardoublequoteopen}{\isasymdown}\isactrlsub p{\isachardoublequoteclose}{\isacharparenright}{\kern0pt}\ \isakeyword{where}\isanewline
\ \ \ \ {\isachardoublequoteopen}{\isasymdown}\isactrlsub p\ {\isacharparenleft}{\kern0pt}pc\ {\isasymtriangleright}\ {\isasymtau}\ \isactrlitem \ cm{\isacharparenright}{\kern0pt}\ {\isacharequal}{\kern0pt}\ pc{\isachardoublequoteclose}\ \isanewline
\isanewline
\ \ \isacommand{fun}\isamarkupfalse%
\isanewline
\ \ \ \ proj{\isacharunderscore}{\kern0pt}{\isasymtau}\ {\isacharcolon}{\kern0pt}{\isacharcolon}{\kern0pt}\ {\isachardoublequoteopen}cont{\isacharunderscore}{\kern0pt}trace\ {\isasymRightarrow}\ {\isasymT}{\isachardoublequoteclose}\ {\isacharparenleft}{\kern0pt}{\isachardoublequoteopen}{\isasymdown}\isactrlsub {\isasymtau}{\isachardoublequoteclose}{\isacharparenright}{\kern0pt}\ \isakeyword{where}\isanewline
\ \ \ \ {\isachardoublequoteopen}{\isasymdown}\isactrlsub {\isasymtau}\ {\isacharparenleft}{\kern0pt}pc\ {\isasymtriangleright}\ {\isasymtau}\ \isactrlitem \ cm{\isacharparenright}{\kern0pt}\ {\isacharequal}{\kern0pt}\ {\isasymtau}{\isachardoublequoteclose}\ \isanewline
\isanewline
\ \ \isacommand{fun}\isamarkupfalse%
\isanewline
\ \ \ \ proj{\isacharunderscore}{\kern0pt}cont\ {\isacharcolon}{\kern0pt}{\isacharcolon}{\kern0pt}\ {\isachardoublequoteopen}cont{\isacharunderscore}{\kern0pt}trace\ {\isasymRightarrow}\ cont{\isacharunderscore}{\kern0pt}marker{\isachardoublequoteclose}\ {\isacharparenleft}{\kern0pt}{\isachardoublequoteopen}{\isasymdown}\isactrlsub {\isasymlambda}{\isachardoublequoteclose}{\isacharparenright}{\kern0pt}\ \isakeyword{where}\isanewline
\ \ \ \ {\isachardoublequoteopen}{\isasymdown}\isactrlsub {\isasymlambda}\ {\isacharparenleft}{\kern0pt}cont\ \isactrlitem \ cm{\isacharparenright}{\kern0pt}\ {\isacharequal}{\kern0pt}\ cm{\isachardoublequoteclose}%
\isadelimdocument
\endisadelimdocument
\isatagdocument
\isamarkupsubsection{Local Evaluation%
}
\isamarkuptrue%
\endisatagdocument
{\isafolddocument}%
\isadelimdocument
\endisadelimdocument
\begin{isamarkuptext}%
We can now start with formalizing the local evaluation of our semantics,
  referring to the construction of traces for arbitrary programs in a local
  environment. Our evaluation rules will take a program and a possibly symbolic state 
  as arguments, subsequently returning the set of all possible continuation traces that 
  can be constructed until the next scheduling point~is~reached. \par
  Aiming to setup the local evaluation, we first introduce a helper function, which 
  modifies a given continuation marker by sequentially appending another statement 
  onto the command inside the continuation marker. If the continuation marker is 
  empty, we simply insert the provided~statement.%
\end{isamarkuptext}\isamarkuptrue%
\ \ \isacommand{fun}\isamarkupfalse%
\isanewline
\ \ \ \ cont{\isacharunderscore}{\kern0pt}append\ {\isacharcolon}{\kern0pt}{\isacharcolon}{\kern0pt}\ {\isachardoublequoteopen}cont{\isacharunderscore}{\kern0pt}marker\ {\isasymRightarrow}\ stmt\ {\isasymRightarrow}\ cont{\isacharunderscore}{\kern0pt}marker{\isachardoublequoteclose}\ \isakeyword{where}\isanewline
\ \ \ \ {\isachardoublequoteopen}cont{\isacharunderscore}{\kern0pt}append\ {\isasymlambda}{\isacharbrackleft}{\kern0pt}S\isactrlsub {\isadigit{1}}{\isacharprime}{\kern0pt}{\isacharbrackright}{\kern0pt}\ S\isactrlsub {\isadigit{2}}\ {\isacharequal}{\kern0pt}\ {\isasymlambda}{\isacharbrackleft}{\kern0pt}S\isactrlsub {\isadigit{1}}{\isacharprime}{\kern0pt}{\isacharsemicolon}{\kern0pt}{\isacharsemicolon}{\kern0pt}S\isactrlsub {\isadigit{2}}{\isacharbrackright}{\kern0pt}{\isachardoublequoteclose}\ {\isacharbar}{\kern0pt}\isanewline
\ \ \ \ {\isachardoublequoteopen}cont{\isacharunderscore}{\kern0pt}append\ {\isasymlambda}{\isacharbrackleft}{\kern0pt}{\isasymnabla}{\isacharbrackright}{\kern0pt}\ S\isactrlsub {\isadigit{2}}\ {\isacharequal}{\kern0pt}\ {\isasymlambda}{\isacharbrackleft}{\kern0pt}S\isactrlsub {\isadigit{2}}{\isacharbrackright}{\kern0pt}{\isachardoublequoteclose}%
\begin{isamarkuptext}%
We now have sufficient means to establish the valuation function. Our
  formalization adheres to the following core concepts: \begin{description}
  \item[Skip Statement] The Skip statement called in state \isa{{\isasymsigma}} only generates 
    a singular continuation trace. Its path condition is empty, because the Skip
    statement can be executed in any situation. Its symbolic trace only contains \isa{{\isasymsigma}},
    as no state changes occur. Considering that Skip is an atomic statement, the 
    continuation marker will also~be~empty. 
  \item[Assignment] An assignment called in state \isa{{\isasymsigma}} generates exactly one
    continuation trace, consisting of an empty path condition and a symbolic trace 
    transiting from \isa{{\isasymsigma}} into an updated version of \isa{{\isasymsigma}}. Note that its continuation 
    marker is again empty, as the whole assignment is evaluated in one singular 
    evaluation~step.
  \item[If-Branch] The conditional statement called in state \isa{{\isasymsigma}} generates two 
    distinct continuation traces. The first one (true-case) can only be taken if 
    the Boolean guard evaluates to true, indicated by the path condition. While its 
    symbolic trace contains only the original state \isa{{\isasymsigma}}, its continuation marker 
    encases the statement S, suggesting that S is still left to be evaluated. Note 
    that this implies a scheduling point right after the evaluation of the Boolean 
    expression. Guard and statement are therefore never evaluated in the same 
    evaluation step. The second continuation trace (false-case) can only be chosen 
    if the Boolean guard evaluates to false, subsequently skipping the 
    statement body S. This causes the symbolic trace to only contain \isa{{\isasymsigma}}, and the 
    continuation marker~to~be~empty. \par
    The formalization of the path conditions slightly deviates from the original
    paper. Whilst the original paper compares the evaluation of the guard with
    Boolean truth values, our approach simply evaluates the guard in its 
    normal/negated form. This design choice ensures that our formalization is 
    easier to read, as too long expressions would bloat up our trace~specifications.
  \item[While-Loop] The While-Loop called in state \isa{{\isasymsigma}} also generates two continuation 
    traces, which closely resemble the traces generated by the conditional statement. 
    Their only difference lies in the continuation marker of the true-case. Whilst the 
    If-Branch ensures the subsequent evaluation of its encased statement, the 
    While-Loop additionally enforces the execution of another loop~repetition. \par
    Contrary to the paper, we have decided against rewriting the While-Loop
    as a conditional statement, because Isabelle, in this case, fails at establishing 
    a corresponding termination~argument.
  \item[Sequential Statement] The rule for the sequential statement \isa{S\isactrlsub {\isadigit{1}}{\isacharsemicolon}{\kern0pt}{\isacharsemicolon}{\kern0pt}S\isactrlsub {\isadigit{2}}} is 
    simple. We collect all possible continuation traces generated by \isa{S\isactrlsub {\isadigit{1}}}, and
    append statement \isa{S\isactrlsub {\isadigit{2}}} onto all their continuation~markers. \par
    Considering that we have to adhere to Isabelle syntax, our formalization gets 
    slightly more complex than the definition of the original paper. We first apply 
    \isa{{\isacharparenleft}{\kern0pt}val\ S\isactrlsub {\isadigit{1}}\ {\isasymsigma}{\isacharparenright}{\kern0pt}} to compute the set of all continuation traces generated by \isa{S\isactrlsub {\isadigit{1}}} called 
    in \isa{{\isasymsigma}}. We then apply the predefined operator \isa{{\isacharbackquote}{\kern0pt}} in order to compute the image of 
    this set under a function, which appends \isa{S\isactrlsub {\isadigit{2}}} onto all their continuation markers. 
    For this purpose, we utilize our earlier defined helper~function. \par
    Note that it is not possible to generate code for Isabelle functions if 
    quantifications over infinite types (e.g. states, traces) take place. We circumvent 
    this problem in our formalization by using the element-wise operation \isa{{\isacharbackquote}{\kern0pt}} on a
    finite set of continuation traces instead of using explicit trace quantifications, 
    thus ensuring that Isabelle can automatically generate corresponding~code.
  \end{description} Due to its construction, each application of the valuation 
  function results in only finitely many continuation~traces.%
\end{isamarkuptext}\isamarkuptrue%
\ \ \isacommand{primrec}\isamarkupfalse%
\isanewline
\ \ \ \ val\ {\isacharcolon}{\kern0pt}{\isacharcolon}{\kern0pt}\ {\isachardoublequoteopen}stmt\ {\isasymRightarrow}\ {\isasymSigma}\ {\isasymRightarrow}\ cont{\isacharunderscore}{\kern0pt}trace\ set{\isachardoublequoteclose}\ \isakeyword{where}\isanewline
\ \ \ \ {\isachardoublequoteopen}val\ SKIP\ {\isasymsigma}\ {\isacharequal}{\kern0pt}\ {\isacharbraceleft}{\kern0pt}\ {\isacharbraceleft}{\kern0pt}{\isacharbraceright}{\kern0pt}\ {\isasymtriangleright}\ {\isasymlangle}{\isasymsigma}{\isasymrangle}\ \isactrlitem \ {\isasymlambda}{\isacharbrackleft}{\kern0pt}{\isasymnabla}{\isacharbrackright}{\kern0pt}\ {\isacharbraceright}{\kern0pt}{\isachardoublequoteclose}\ {\isacharbar}{\kern0pt}\isanewline
\ \ \ \ {\isachardoublequoteopen}val\ {\isacharparenleft}{\kern0pt}x\ {\isacharcolon}{\kern0pt}{\isacharequal}{\kern0pt}\ a{\isacharparenright}{\kern0pt}\ {\isasymsigma}\ {\isacharequal}{\kern0pt}\ {\isacharbraceleft}{\kern0pt}\ {\isacharbraceleft}{\kern0pt}{\isacharbraceright}{\kern0pt}\ {\isasymtriangleright}\ {\isasymlangle}{\isasymsigma}{\isasymrangle}\ {\isasymleadsto}\ State{\isasymllangle}{\isacharbrackleft}{\kern0pt}x\ {\isasymlongmapsto}\ Exp\ {\isacharparenleft}{\kern0pt}val\isactrlsub A\ a\ {\isasymsigma}{\isacharparenright}{\kern0pt}{\isacharbrackright}{\kern0pt}\ {\isasymsigma}{\isasymrrangle}\ \isactrlitem \ {\isasymlambda}{\isacharbrackleft}{\kern0pt}{\isasymnabla}{\isacharbrackright}{\kern0pt}\ {\isacharbraceright}{\kern0pt}{\isachardoublequoteclose}\ {\isacharbar}{\kern0pt}\isanewline
\ \ \ \ {\isachardoublequoteopen}val\ {\isacharparenleft}{\kern0pt}IF\ b\ THEN\ S\ FI{\isacharparenright}{\kern0pt}\ {\isasymsigma}\ {\isacharequal}{\kern0pt}\ {\isacharbraceleft}{\kern0pt}\isanewline
\ \ \ \ \ \ \ \ {\isacharbraceleft}{\kern0pt}val\isactrlsub B\ b\ {\isasymsigma}{\isacharbraceright}{\kern0pt}\ {\isasymtriangleright}\ {\isasymlangle}{\isasymsigma}{\isasymrangle}\ \isactrlitem \ {\isasymlambda}{\isacharbrackleft}{\kern0pt}S{\isacharbrackright}{\kern0pt}{\isacharcomma}{\kern0pt}\isanewline
\ \ \ \ \ \ \ \ {\isacharbraceleft}{\kern0pt}val\isactrlsub B\ {\isacharparenleft}{\kern0pt}Not\ b{\isacharparenright}{\kern0pt}\ {\isasymsigma}{\isacharbraceright}{\kern0pt}\ {\isasymtriangleright}\ {\isasymlangle}{\isasymsigma}{\isasymrangle}\ \isactrlitem \ {\isasymlambda}{\isacharbrackleft}{\kern0pt}{\isasymnabla}{\isacharbrackright}{\kern0pt}\isanewline
\ \ \ \ \ \ {\isacharbraceright}{\kern0pt}{\isachardoublequoteclose}\ {\isacharbar}{\kern0pt}\isanewline
\ \ \ \ {\isachardoublequoteopen}val\ {\isacharparenleft}{\kern0pt}WHILE\ b\ DO\ S\ OD{\isacharparenright}{\kern0pt}\ {\isasymsigma}\ {\isacharequal}{\kern0pt}\ {\isacharbraceleft}{\kern0pt}\isanewline
\ \ \ \ \ \ \ \ {\isacharbraceleft}{\kern0pt}val\isactrlsub B\ b\ {\isasymsigma}{\isacharbraceright}{\kern0pt}\ {\isasymtriangleright}\ {\isasymlangle}{\isasymsigma}{\isasymrangle}\ \isactrlitem \ {\isasymlambda}{\isacharbrackleft}{\kern0pt}S{\isacharsemicolon}{\kern0pt}{\isacharsemicolon}{\kern0pt}WHILE\ b\ DO\ S\ OD{\isacharbrackright}{\kern0pt}{\isacharcomma}{\kern0pt}\isanewline
\ \ \ \ \ \ \ \ {\isacharbraceleft}{\kern0pt}val\isactrlsub B\ {\isacharparenleft}{\kern0pt}Not\ b{\isacharparenright}{\kern0pt}\ {\isasymsigma}{\isacharbraceright}{\kern0pt}\ {\isasymtriangleright}\ {\isasymlangle}{\isasymsigma}{\isasymrangle}\ \isactrlitem \ {\isasymlambda}{\isacharbrackleft}{\kern0pt}{\isasymnabla}{\isacharbrackright}{\kern0pt}\isanewline
\ \ \ \ \ \ {\isacharbraceright}{\kern0pt}{\isachardoublequoteclose}\ {\isacharbar}{\kern0pt}\isanewline
\ \ \ \ {\isachardoublequoteopen}val\ {\isacharparenleft}{\kern0pt}S\isactrlsub {\isadigit{1}}{\isacharsemicolon}{\kern0pt}{\isacharsemicolon}{\kern0pt}S\isactrlsub {\isadigit{2}}{\isacharparenright}{\kern0pt}\ {\isasymsigma}\ {\isacharequal}{\kern0pt}\ {\isacharparenleft}{\kern0pt}{\isacharpercent}{\kern0pt}c{\isachardot}{\kern0pt}\ {\isacharparenleft}{\kern0pt}{\isasymdown}\isactrlsub p\ c{\isacharparenright}{\kern0pt}\ {\isasymtriangleright}\ {\isacharparenleft}{\kern0pt}{\isasymdown}\isactrlsub {\isasymtau}\ c{\isacharparenright}{\kern0pt}\ \isactrlitem \ cont{\isacharunderscore}{\kern0pt}append\ {\isacharparenleft}{\kern0pt}{\isasymdown}\isactrlsub {\isasymlambda}\ c{\isacharparenright}{\kern0pt}\ S\isactrlsub {\isadigit{2}}{\isacharparenright}{\kern0pt}\ {\isacharbackquote}{\kern0pt}\ {\isacharparenleft}{\kern0pt}val\ S\isactrlsub {\isadigit{1}}\ {\isasymsigma}{\isacharparenright}{\kern0pt}{\isachardoublequoteclose}%
\begin{isamarkuptext}%
Considering that we have established the local evaluation of our semantics, 
  we can now take a look at several examples of valuation function~applications.%
\end{isamarkuptext}\isamarkuptrue%
\ \ \isacommand{lemma}\isamarkupfalse%
\ {\isachardoublequoteopen}val\ {\isacharparenleft}{\kern0pt}{\isacharprime}{\kern0pt}{\isacharprime}{\kern0pt}x{\isacharprime}{\kern0pt}{\isacharprime}{\kern0pt}\ {\isacharcolon}{\kern0pt}{\isacharequal}{\kern0pt}\ Num\ {\isadigit{2}}{\isacharparenright}{\kern0pt}\ {\isasymsigma}\isactrlsub {\isadigit{1}}\ {\isacharequal}{\kern0pt}\ {\isacharbraceleft}{\kern0pt}\ \isanewline
\ \ \ \ \ \ \ \ \ \ \ \ \ \ \ \ \ \ {\isacharbraceleft}{\kern0pt}{\isacharbraceright}{\kern0pt}\ {\isasymtriangleright}\ {\isasymlangle}{\isasymsigma}\isactrlsub {\isadigit{1}}{\isasymrangle}\ {\isasymleadsto}\ State{\isasymllangle}{\isacharbrackleft}{\kern0pt}{\isacharprime}{\kern0pt}{\isacharprime}{\kern0pt}x{\isacharprime}{\kern0pt}{\isacharprime}{\kern0pt}\ {\isasymlongmapsto}\ Exp\ {\isacharparenleft}{\kern0pt}Num\ {\isadigit{2}}{\isacharparenright}{\kern0pt}{\isacharbrackright}{\kern0pt}\ {\isasymsigma}\isactrlsub {\isadigit{1}}{\isasymrrangle}\ \isactrlitem \ {\isasymlambda}{\isacharbrackleft}{\kern0pt}{\isasymnabla}{\isacharbrackright}{\kern0pt}\ \isanewline
\ \ \ \ \ \ \ \ \ \ \ \ \ \ {\isacharbraceright}{\kern0pt}{\isachardoublequoteclose}\ \isanewline
\isadelimproof
\ \ \ \ %
\endisadelimproof
\isatagproof
\isacommand{by}\isamarkupfalse%
\ simp%
\endisatagproof
{\isafoldproof}%
\isadelimproof
\isanewline
\endisadelimproof
\isanewline
\ \ \isacommand{lemma}\isamarkupfalse%
\ {\isachardoublequoteopen}val\ WL{\isacharunderscore}{\kern0pt}ex\isactrlsub {\isadigit{1}}\ {\isasymsigma}\isactrlsub {\isadigit{2}}\ {\isacharequal}{\kern0pt}\ {\isacharbraceleft}{\kern0pt}\ \isanewline
\ \ \ \ \ \ \ \ \ \ \ \ \ \ \ \ \ \ {\isacharbraceleft}{\kern0pt}Bool\ True{\isacharbraceright}{\kern0pt}\ {\isasymtriangleright}\ {\isasymlangle}{\isasymsigma}\isactrlsub {\isadigit{2}}{\isasymrangle}\ \isactrlitem \ {\isasymlambda}{\isacharbrackleft}{\kern0pt}{\isacharparenleft}{\kern0pt}{\isacharprime}{\kern0pt}{\isacharprime}{\kern0pt}z{\isacharprime}{\kern0pt}{\isacharprime}{\kern0pt}\ {\isacharcolon}{\kern0pt}{\isacharequal}{\kern0pt}\ Var\ {\isacharprime}{\kern0pt}{\isacharprime}{\kern0pt}y{\isacharprime}{\kern0pt}{\isacharprime}{\kern0pt}{\isacharsemicolon}{\kern0pt}{\isacharsemicolon}{\kern0pt}\ {\isacharprime}{\kern0pt}{\isacharprime}{\kern0pt}y{\isacharprime}{\kern0pt}{\isacharprime}{\kern0pt}\ {\isacharcolon}{\kern0pt}{\isacharequal}{\kern0pt}\ Var\ {\isacharprime}{\kern0pt}{\isacharprime}{\kern0pt}x{\isacharprime}{\kern0pt}{\isacharprime}{\kern0pt}{\isacharparenright}{\kern0pt}{\isacharsemicolon}{\kern0pt}{\isacharsemicolon}{\kern0pt}\ {\isacharprime}{\kern0pt}{\isacharprime}{\kern0pt}x{\isacharprime}{\kern0pt}{\isacharprime}{\kern0pt}\ {\isacharcolon}{\kern0pt}{\isacharequal}{\kern0pt}\ Var\ {\isacharprime}{\kern0pt}{\isacharprime}{\kern0pt}z{\isacharprime}{\kern0pt}{\isacharprime}{\kern0pt}{\isacharbrackright}{\kern0pt}{\isacharcomma}{\kern0pt}\isanewline
\ \ \ \ \ \ \ \ \ \ \ \ \ \ \ \ \ \ {\isacharbraceleft}{\kern0pt}Bool\ False{\isacharbraceright}{\kern0pt}\ {\isasymtriangleright}\ {\isasymlangle}{\isasymsigma}\isactrlsub {\isadigit{2}}{\isasymrangle}\ \isactrlitem \ {\isasymlambda}{\isacharbrackleft}{\kern0pt}{\isasymnabla}{\isacharbrackright}{\kern0pt}\isanewline
\ \ \ \ \ \ \ \ \ \ \ \ \ \ {\isacharbraceright}{\kern0pt}{\isachardoublequoteclose}\isanewline
\isadelimproof
\ \ \ \ %
\endisadelimproof
\isatagproof
\isacommand{by}\isamarkupfalse%
\ {\isacharparenleft}{\kern0pt}simp\ add{\isacharcolon}{\kern0pt}\ {\isasymsigma}\isactrlsub {\isadigit{2}}{\isacharunderscore}{\kern0pt}def\ WL{\isacharunderscore}{\kern0pt}ex\isactrlsub {\isadigit{1}}{\isacharunderscore}{\kern0pt}def{\isacharparenright}{\kern0pt}%
\endisatagproof
{\isafoldproof}%
\isadelimproof
\endisadelimproof
\isadelimdocument
\endisadelimdocument
\isatagdocument
\isamarkupsubsection{Trace Composition%
}
\isamarkuptrue%
\isamarkupsubsubsection{Configurations%
}
\isamarkuptrue%
\endisatagdocument
{\isafolddocument}%
\isadelimdocument
\endisadelimdocument
\begin{isamarkuptext}%
Similar to the original paper, we introduce program configurations as 
  tuples of symbolic traces and continuation markers. We call a configuration
  terminal iff the corresponding continuation marker is~empty.%
\end{isamarkuptext}\isamarkuptrue%
\ \ \isacommand{type{\isacharunderscore}{\kern0pt}synonym}\isamarkupfalse%
\ config\ {\isacharequal}{\kern0pt}\ {\isachardoublequoteopen}{\isasymT}\ {\isacharasterisk}{\kern0pt}\ cont{\isacharunderscore}{\kern0pt}marker{\isachardoublequoteclose}%
\isadelimdocument
\endisadelimdocument
\isatagdocument
\isamarkupsubsubsection{\isa{{\isasymdelta}}-function%
}
\isamarkuptrue%
\endisatagdocument
{\isafolddocument}%
\isadelimdocument
\endisadelimdocument
\begin{isamarkuptext}%
The previously defined valuation function already enables us to
construct traces in a local environment. However, these local traces still
need to be composed into concrete, global traces. This motivates the notion of 
trace~compositions. \par
During a trace composition, the valuation function will be repeatedly applied
on a given program, until it has been fully evaluated. After each application,
the constructed traces will be stitched together, and the scheduler will choose 
which process to execute next. Considering that WL supports no method calls, 
the scheduler will only be able to select the main process. As WL is even strictly 
deterministic, we will not be able to generate more than one global trace for any 
arbitrary~program. \par
In the original paper, the composition rule is formalized using an inductive
definition. The rule describes that \isa{{\isacharparenleft}{\kern0pt}sh\ \isactrlemph \isactrlemph \ {\isasymtau}{\isacharcomma}{\kern0pt}\ {\isasymlambda}{\isacharbrackleft}{\kern0pt}S{\isacharprime}{\kern0pt}{\isacharbrackright}{\kern0pt}{\isacharparenright}{\kern0pt}} is a successor configuration
of \isa{{\isacharparenleft}{\kern0pt}sh{\isacharcomma}{\kern0pt}\ {\isasymlambda}{\isacharbrackleft}{\kern0pt}S{\isacharbrackright}{\kern0pt}{\isacharparenright}{\kern0pt}} iff the continuation trace \isa{pc\ {\isasymtriangleright}\ {\isasymtau}\ \isactrlitem \ {\isasymlambda}{\isacharbrackleft}{\kern0pt}S{\isacharprime}{\kern0pt}{\isacharbrackright}{\kern0pt}} with consistent path 
condition \isa{pc} can be generated from \isa{S} called in the last state of \isa{sh}. Such an 
inductive definition could also be provided in Isabelle. However, its
transitive closure would later need to have the following~form: \par
\begin{center} \isa{{\isacharbraceleft}{\kern0pt}{\isasymtau}{\isachardot}{\kern0pt}\ {\isacharparenleft}{\kern0pt}{\isasymlangle}{\isasymsigma}\isactrlsub I{\isasymrangle}{\isacharcomma}{\kern0pt}\ {\isasymlambda}{\isacharbrackleft}{\kern0pt}S{\isacharbrackright}{\kern0pt}{\isacharparenright}{\kern0pt}\ {\isasymlongrightarrow}\isactrlsup {\isasymstar}\ {\isacharparenleft}{\kern0pt}{\isasymtau}{\isacharcomma}{\kern0pt}\ {\isasymlambda}{\isacharbrackleft}{\kern0pt}{\isasymnabla}{\isacharbrackright}{\kern0pt}{\isacharparenright}{\kern0pt}{\isacharbraceright}{\kern0pt}} \end{center} \par
Note that this greatly impedes the automatic code generation in Isabelle, as
we are quantifying over infinitely many traces \isa{{\isasymtau}}. We therefore need to provide an 
alternative formalization, which does not entail explicit trace quantifications, 
thus greatly deviating from the rule denoted in the original~paper. \par
Instead of an inductively defined relation between successor configurations, we 
decide to formalize a deterministic successor function (\isa{{\isasymdelta}}-function) that maps 
a configuration onto all possible successor configurations reachable in 
one evaluation step. Using this approach, we can later denote the transitive
closure with a recursive function, mapping a configuration onto all 
reachable terminal configurations. This in turn can be formalized without 
using trace quantifications, thereby solving the earlier~problem. \par
We formalize the \isa{{\isasymdelta}}-function as follows: We first collect all continuation
traces with consistent path conditions that can be generated by S called in \isa{{\isasymsigma}}. 
Note that the cardinality of this set is finite. We then use the 
predefined image operator \isa{{\isacharbackquote}{\kern0pt}} in order to translate all these continuation traces 
into corresponding configurations, whilst concatenating the previous trace \isa{sh}
with the newly generated trace \isa{{\isasymtau}}. Note that this is a partial function,
as it is undefined if the symbolic trace in the configuration ends
with an event. However, we will not encounter this situation, as long as we ensure
that the composition preserves the wellformedness of symbolic~traces.%
\end{isamarkuptext}\isamarkuptrue%
\ \ \isacommand{fun}\isamarkupfalse%
\isanewline
\ \ \ \ successors\ {\isacharcolon}{\kern0pt}{\isacharcolon}{\kern0pt}\ {\isachardoublequoteopen}config\ {\isasymRightarrow}\ config\ set{\isachardoublequoteclose}\ {\isacharparenleft}{\kern0pt}{\isachardoublequoteopen}{\isasymdelta}{\isachardoublequoteclose}{\isacharparenright}{\kern0pt}\ \ \isakeyword{where}\isanewline
\ \ \ \ {\isachardoublequoteopen}{\isasymdelta}\ {\isacharparenleft}{\kern0pt}sh\ {\isasymleadsto}\ State{\isasymllangle}{\isasymsigma}{\isasymrrangle}{\isacharcomma}{\kern0pt}\ {\isasymlambda}{\isacharbrackleft}{\kern0pt}S{\isacharbrackright}{\kern0pt}{\isacharparenright}{\kern0pt}\ {\isacharequal}{\kern0pt}\ \isanewline
\ \ \ \ \ \ \ \ {\isacharparenleft}{\kern0pt}{\isacharpercent}{\kern0pt}c{\isachardot}{\kern0pt}\ {\isacharparenleft}{\kern0pt}sh\ {\isasymcdot}\ {\isacharparenleft}{\kern0pt}{\isasymdown}\isactrlsub {\isasymtau}\ c{\isacharparenright}{\kern0pt}{\isacharcomma}{\kern0pt}\ {\isasymdown}\isactrlsub {\isasymlambda}\ c{\isacharparenright}{\kern0pt}{\isacharparenright}{\kern0pt}\ {\isacharbackquote}{\kern0pt}\ {\isacharbraceleft}{\kern0pt}cont\ {\isasymin}\ {\isacharparenleft}{\kern0pt}val\ S\ {\isasymsigma}{\isacharparenright}{\kern0pt}{\isachardot}{\kern0pt}\ consistent{\isacharparenleft}{\kern0pt}{\isasymdown}\isactrlsub p\ cont{\isacharparenright}{\kern0pt}{\isacharbraceright}{\kern0pt}{\isachardoublequoteclose}\ {\isacharbar}{\kern0pt}\isanewline
\ \ \ \ {\isachardoublequoteopen}{\isasymdelta}\ {\isacharunderscore}{\kern0pt}\ {\isacharequal}{\kern0pt}\ undefined{\isachardoublequoteclose}%
\isadelimdocument
\endisadelimdocument
\isatagdocument
\isamarkupsubsubsection{Proof Automation%
}
\isamarkuptrue%
\endisatagdocument
{\isafolddocument}%
\isadelimdocument
\endisadelimdocument
\begin{isamarkuptext}%
We strongly desire to establish an automated proof system for the
  construction of global traces in our LAGC semantics. The \isa{{\isasymdelta}}-function plays
  a crucial role in this construction. However, its definition is slightly complex, 
  considering that it builds on top of the valuation function. This greatly impedes
  automatic proofs, causing a need for additional simplification~lemmas. \par
  We will therefore provide a general simplification lemma for each statement of WL,
  thereby covering all application scenarios of the \isa{{\isasymdelta}}-function. This will lay the 
  groundwork for efficient proof derivations, as we can later simply utilize our 
  simplification lemmas when deriving global traces, hence avoiding having to deal 
  with the underlying valuation function. Note that all applications of the \isa{{\isasymdelta}}-function 
  will return singleton configuration sets, considering the determinism~of~WL.%
\end{isamarkuptext}\isamarkuptrue%
\ \ \isacommand{context}\isamarkupfalse%
\ \isakeyword{notes}\ {\isacharbrackleft}{\kern0pt}simp{\isacharbrackright}{\kern0pt}\ {\isacharequal}{\kern0pt}\ consistent{\isacharunderscore}{\kern0pt}def\ \isakeyword{begin}\isanewline
\isanewline
\ \ \isacommand{lemma}\isamarkupfalse%
\ {\isasymdelta}{\isacharunderscore}{\kern0pt}Skip{\isacharcolon}{\kern0pt}\isanewline
\ \ \ \ {\isachardoublequoteopen}{\isasymdelta}\ {\isacharparenleft}{\kern0pt}sh\ {\isasymleadsto}\ State{\isasymllangle}{\isasymsigma}{\isasymrrangle}{\isacharcomma}{\kern0pt}\ {\isasymlambda}{\isacharbrackleft}{\kern0pt}SKIP{\isacharbrackright}{\kern0pt}{\isacharparenright}{\kern0pt}\ {\isacharequal}{\kern0pt}\ {\isacharbraceleft}{\kern0pt}{\isacharparenleft}{\kern0pt}sh\ {\isasymleadsto}\ State{\isasymllangle}{\isasymsigma}{\isasymrrangle}{\isacharcomma}{\kern0pt}\ {\isasymlambda}{\isacharbrackleft}{\kern0pt}{\isasymnabla}{\isacharbrackright}{\kern0pt}{\isacharparenright}{\kern0pt}{\isacharbraceright}{\kern0pt}{\isachardoublequoteclose}\isanewline
\isadelimproof
\ \ \ \ %
\endisadelimproof
\isatagproof
\isacommand{by}\isamarkupfalse%
\ {\isacharparenleft}{\kern0pt}simp\ add{\isacharcolon}{\kern0pt}\ image{\isacharunderscore}{\kern0pt}constant{\isacharunderscore}{\kern0pt}conv{\isacharparenright}{\kern0pt}%
\endisatagproof
{\isafoldproof}%
\isadelimproof
\isanewline
\endisadelimproof
\isanewline
\ \ \isacommand{lemma}\isamarkupfalse%
\ {\isasymdelta}{\isacharunderscore}{\kern0pt}Assign{\isacharcolon}{\kern0pt}\isanewline
\ \ \ \ {\isachardoublequoteopen}{\isasymdelta}\ {\isacharparenleft}{\kern0pt}sh\ {\isasymleadsto}\ State{\isasymllangle}{\isasymsigma}{\isasymrrangle}{\isacharcomma}{\kern0pt}\ {\isasymlambda}{\isacharbrackleft}{\kern0pt}x\ {\isacharcolon}{\kern0pt}{\isacharequal}{\kern0pt}\ a{\isacharbrackright}{\kern0pt}{\isacharparenright}{\kern0pt}\ {\isacharequal}{\kern0pt}\ {\isacharbraceleft}{\kern0pt}{\isacharparenleft}{\kern0pt}{\isacharparenleft}{\kern0pt}sh\ {\isasymleadsto}\ State{\isasymllangle}{\isasymsigma}{\isasymrrangle}{\isacharparenright}{\kern0pt}\ {\isasymleadsto}\ State{\isasymllangle}{\isacharbrackleft}{\kern0pt}x\ {\isasymlongmapsto}\ Exp\ {\isacharparenleft}{\kern0pt}val\isactrlsub A\ a\ {\isasymsigma}{\isacharparenright}{\kern0pt}{\isacharbrackright}{\kern0pt}\ {\isasymsigma}{\isasymrrangle}{\isacharcomma}{\kern0pt}\ {\isasymlambda}{\isacharbrackleft}{\kern0pt}{\isasymnabla}{\isacharbrackright}{\kern0pt}{\isacharparenright}{\kern0pt}{\isacharbraceright}{\kern0pt}{\isachardoublequoteclose}\isanewline
\isadelimproof
\ \ \ \ %
\endisadelimproof
\isatagproof
\isacommand{by}\isamarkupfalse%
\ {\isacharparenleft}{\kern0pt}simp\ add{\isacharcolon}{\kern0pt}\ image{\isacharunderscore}{\kern0pt}constant{\isacharunderscore}{\kern0pt}conv{\isacharparenright}{\kern0pt}%
\endisatagproof
{\isafoldproof}%
\isadelimproof
\endisadelimproof
\begin{isamarkuptext}%
Applying the \isa{{\isasymdelta}}-function on the If-Branch and the While-Loop can cause
  three distinct results. If the path condition containing the evaluated Boolean
  guard is consistent, the statement enters the true-case. If the path condition
  consisting of the evaluated negated Boolean guard is consistent, the statement will
  enter the false-case. The results of both of these cases are~straightforward. \par
  However, it is also possible that the Boolean guard cannot be fully evaluated
  (e.g. due to symbolic variables in the guard expression), thereby differing from 
  the previous cases. Note that the \isa{{\isasymdelta}}-function will return the empty set in this
  case, implying that no successor configuration exists. Such a situation is not 
  supposed to transpire during a trace composition, and will not occur, so long as 
  we start our composition in a concrete~state.%
\end{isamarkuptext}\isamarkuptrue%
\ \ \isacommand{lemma}\isamarkupfalse%
\ {\isasymdelta}{\isacharunderscore}{\kern0pt}If\isactrlsub T{\isacharcolon}{\kern0pt}\isanewline
\ \ \ \ \isakeyword{assumes}\ {\isachardoublequoteopen}consistent\ {\isacharbraceleft}{\kern0pt}{\isacharparenleft}{\kern0pt}val\isactrlsub B\ b\ {\isasymsigma}{\isacharparenright}{\kern0pt}{\isacharbraceright}{\kern0pt}{\isachardoublequoteclose}\isanewline
\ \ \ \ \isakeyword{shows}\ {\isachardoublequoteopen}{\isasymdelta}\ {\isacharparenleft}{\kern0pt}sh\ {\isasymleadsto}\ State{\isasymllangle}{\isasymsigma}{\isasymrrangle}{\isacharcomma}{\kern0pt}\ {\isasymlambda}{\isacharbrackleft}{\kern0pt}IF\ b\ THEN\ S\ FI{\isacharbrackright}{\kern0pt}{\isacharparenright}{\kern0pt}\ {\isacharequal}{\kern0pt}\ {\isacharbraceleft}{\kern0pt}{\isacharparenleft}{\kern0pt}sh\ {\isasymleadsto}\ State{\isasymllangle}{\isasymsigma}{\isasymrrangle}{\isacharcomma}{\kern0pt}\ {\isasymlambda}{\isacharbrackleft}{\kern0pt}S{\isacharbrackright}{\kern0pt}{\isacharparenright}{\kern0pt}{\isacharbraceright}{\kern0pt}{\isachardoublequoteclose}\isanewline
\isadelimproof
\ \ \ \ %
\endisadelimproof
\isatagproof
\isacommand{using}\isamarkupfalse%
\ assms\ \isacommand{by}\isamarkupfalse%
\ force%
\endisatagproof
{\isafoldproof}%
\isadelimproof
\isanewline
\endisadelimproof
\isanewline
\ \ \isacommand{lemma}\isamarkupfalse%
\ {\isasymdelta}{\isacharunderscore}{\kern0pt}If\isactrlsub F{\isacharcolon}{\kern0pt}\isanewline
\ \ \ \ \isakeyword{assumes}\ {\isachardoublequoteopen}consistent\ {\isacharbraceleft}{\kern0pt}{\isacharparenleft}{\kern0pt}val\isactrlsub B\ {\isacharparenleft}{\kern0pt}Not\ b{\isacharparenright}{\kern0pt}\ {\isasymsigma}{\isacharparenright}{\kern0pt}{\isacharbraceright}{\kern0pt}{\isachardoublequoteclose}\isanewline
\ \ \ \ \isakeyword{shows}\ {\isachardoublequoteopen}{\isasymdelta}\ {\isacharparenleft}{\kern0pt}sh\ {\isasymleadsto}\ State{\isasymllangle}{\isasymsigma}{\isasymrrangle}{\isacharcomma}{\kern0pt}\ {\isasymlambda}{\isacharbrackleft}{\kern0pt}IF\ b\ THEN\ S\ FI{\isacharbrackright}{\kern0pt}{\isacharparenright}{\kern0pt}\ {\isacharequal}{\kern0pt}\ {\isacharbraceleft}{\kern0pt}{\isacharparenleft}{\kern0pt}sh\ {\isasymleadsto}\ State{\isasymllangle}{\isasymsigma}{\isasymrrangle}{\isacharcomma}{\kern0pt}\ {\isasymlambda}{\isacharbrackleft}{\kern0pt}{\isasymnabla}{\isacharbrackright}{\kern0pt}{\isacharparenright}{\kern0pt}{\isacharbraceright}{\kern0pt}{\isachardoublequoteclose}\isanewline
\isadelimproof
\ \ \ \ %
\endisadelimproof
\isatagproof
\isacommand{using}\isamarkupfalse%
\ assms\ \isacommand{by}\isamarkupfalse%
\ force%
\endisatagproof
{\isafoldproof}%
\isadelimproof
\isanewline
\endisadelimproof
\isanewline
\ \ \isacommand{lemma}\isamarkupfalse%
\ {\isasymdelta}{\isacharunderscore}{\kern0pt}If\isactrlsub E{\isacharcolon}{\kern0pt}\isanewline
\ \ \ \ \isakeyword{assumes}\ {\isachardoublequoteopen}{\isasymnot}{\isacharparenleft}{\kern0pt}consistent\ {\isacharbraceleft}{\kern0pt}{\isacharparenleft}{\kern0pt}val\isactrlsub B\ b\ {\isasymsigma}{\isacharparenright}{\kern0pt}{\isacharbraceright}{\kern0pt}\ {\isasymor}\ consistent\ {\isacharbraceleft}{\kern0pt}{\isacharparenleft}{\kern0pt}val\isactrlsub B\ {\isacharparenleft}{\kern0pt}Not\ b{\isacharparenright}{\kern0pt}\ {\isasymsigma}{\isacharparenright}{\kern0pt}{\isacharbraceright}{\kern0pt}{\isacharparenright}{\kern0pt}{\isachardoublequoteclose}\isanewline
\ \ \ \ \isakeyword{shows}\ {\isachardoublequoteopen}{\isasymdelta}\ {\isacharparenleft}{\kern0pt}sh\ {\isasymleadsto}\ State{\isasymllangle}{\isasymsigma}{\isasymrrangle}{\isacharcomma}{\kern0pt}\ {\isasymlambda}{\isacharbrackleft}{\kern0pt}IF\ b\ THEN\ S\ FI{\isacharbrackright}{\kern0pt}{\isacharparenright}{\kern0pt}\ {\isacharequal}{\kern0pt}\ {\isacharbraceleft}{\kern0pt}{\isacharbraceright}{\kern0pt}{\isachardoublequoteclose}\isanewline
\isadelimproof
\ \ \ \ %
\endisadelimproof
\isatagproof
\isacommand{using}\isamarkupfalse%
\ assms\ \isacommand{by}\isamarkupfalse%
\ auto%
\endisatagproof
{\isafoldproof}%
\isadelimproof
\isanewline
\endisadelimproof
\isanewline
\ \ \isacommand{lemma}\isamarkupfalse%
\ {\isasymdelta}{\isacharunderscore}{\kern0pt}While\isactrlsub T{\isacharcolon}{\kern0pt}\isanewline
\ \ \ \ \isakeyword{assumes}\ {\isachardoublequoteopen}consistent\ {\isacharbraceleft}{\kern0pt}{\isacharparenleft}{\kern0pt}val\isactrlsub B\ b\ {\isasymsigma}{\isacharparenright}{\kern0pt}{\isacharbraceright}{\kern0pt}{\isachardoublequoteclose}\isanewline
\ \ \ \ \isakeyword{shows}\ {\isachardoublequoteopen}{\isasymdelta}\ {\isacharparenleft}{\kern0pt}sh\ {\isasymleadsto}\ State{\isasymllangle}{\isasymsigma}{\isasymrrangle}{\isacharcomma}{\kern0pt}\ {\isasymlambda}{\isacharbrackleft}{\kern0pt}WHILE\ b\ DO\ S\ OD{\isacharbrackright}{\kern0pt}{\isacharparenright}{\kern0pt}\ {\isacharequal}{\kern0pt}\ {\isacharbraceleft}{\kern0pt}{\isacharparenleft}{\kern0pt}sh\ {\isasymleadsto}\ State{\isasymllangle}{\isasymsigma}{\isasymrrangle}{\isacharcomma}{\kern0pt}\ {\isasymlambda}{\isacharbrackleft}{\kern0pt}S{\isacharsemicolon}{\kern0pt}{\isacharsemicolon}{\kern0pt}WHILE\ b\ DO\ S\ OD{\isacharbrackright}{\kern0pt}{\isacharparenright}{\kern0pt}{\isacharbraceright}{\kern0pt}{\isachardoublequoteclose}\isanewline
\isadelimproof
\ \ \ \ %
\endisadelimproof
\isatagproof
\isacommand{using}\isamarkupfalse%
\ assms\ \isacommand{by}\isamarkupfalse%
\ force%
\endisatagproof
{\isafoldproof}%
\isadelimproof
\isanewline
\endisadelimproof
\isanewline
\ \ \isacommand{lemma}\isamarkupfalse%
\ {\isasymdelta}{\isacharunderscore}{\kern0pt}While\isactrlsub F{\isacharcolon}{\kern0pt}\isanewline
\ \ \ \ \isakeyword{assumes}\ {\isachardoublequoteopen}consistent\ {\isacharbraceleft}{\kern0pt}{\isacharparenleft}{\kern0pt}val\isactrlsub B\ {\isacharparenleft}{\kern0pt}Not\ b{\isacharparenright}{\kern0pt}\ {\isasymsigma}{\isacharparenright}{\kern0pt}{\isacharbraceright}{\kern0pt}{\isachardoublequoteclose}\isanewline
\ \ \ \ \isakeyword{shows}\ {\isachardoublequoteopen}{\isasymdelta}\ {\isacharparenleft}{\kern0pt}sh\ {\isasymleadsto}\ State{\isasymllangle}{\isasymsigma}{\isasymrrangle}{\isacharcomma}{\kern0pt}\ {\isasymlambda}{\isacharbrackleft}{\kern0pt}WHILE\ b\ DO\ S\ OD{\isacharbrackright}{\kern0pt}{\isacharparenright}{\kern0pt}\ {\isacharequal}{\kern0pt}\ {\isacharbraceleft}{\kern0pt}{\isacharparenleft}{\kern0pt}sh\ {\isasymleadsto}\ State{\isasymllangle}{\isasymsigma}{\isasymrrangle}{\isacharcomma}{\kern0pt}\ {\isasymlambda}{\isacharbrackleft}{\kern0pt}{\isasymnabla}{\isacharbrackright}{\kern0pt}{\isacharparenright}{\kern0pt}{\isacharbraceright}{\kern0pt}{\isachardoublequoteclose}\isanewline
\isadelimproof
\ \ \ \ %
\endisadelimproof
\isatagproof
\isacommand{using}\isamarkupfalse%
\ assms\ \isacommand{by}\isamarkupfalse%
\ force%
\endisatagproof
{\isafoldproof}%
\isadelimproof
\isanewline
\endisadelimproof
\isanewline
\ \ \isacommand{lemma}\isamarkupfalse%
\ {\isasymdelta}{\isacharunderscore}{\kern0pt}While\isactrlsub E{\isacharcolon}{\kern0pt}\isanewline
\ \ \ \ \isakeyword{assumes}\ {\isachardoublequoteopen}{\isasymnot}{\isacharparenleft}{\kern0pt}consistent\ {\isacharbraceleft}{\kern0pt}{\isacharparenleft}{\kern0pt}val\isactrlsub B\ b\ {\isasymsigma}{\isacharparenright}{\kern0pt}{\isacharbraceright}{\kern0pt}\ {\isasymor}\ consistent\ {\isacharbraceleft}{\kern0pt}{\isacharparenleft}{\kern0pt}val\isactrlsub B\ {\isacharparenleft}{\kern0pt}Not\ b{\isacharparenright}{\kern0pt}\ {\isasymsigma}{\isacharparenright}{\kern0pt}{\isacharbraceright}{\kern0pt}{\isacharparenright}{\kern0pt}{\isachardoublequoteclose}\isanewline
\ \ \ \ \isakeyword{shows}\ {\isachardoublequoteopen}{\isasymdelta}\ {\isacharparenleft}{\kern0pt}sh\ {\isasymleadsto}\ State{\isasymllangle}{\isasymsigma}{\isasymrrangle}{\isacharcomma}{\kern0pt}\ {\isasymlambda}{\isacharbrackleft}{\kern0pt}WHILE\ b\ DO\ S\ OD{\isacharbrackright}{\kern0pt}{\isacharparenright}{\kern0pt}\ {\isacharequal}{\kern0pt}\ {\isacharbraceleft}{\kern0pt}{\isacharbraceright}{\kern0pt}{\isachardoublequoteclose}\isanewline
\isadelimproof
\ \ \ \ %
\endisadelimproof
\isatagproof
\isacommand{using}\isamarkupfalse%
\ assms\ \isacommand{by}\isamarkupfalse%
\ auto%
\endisatagproof
{\isafoldproof}%
\isadelimproof
\isanewline
\endisadelimproof
\isanewline
\ \ \isacommand{end}\isamarkupfalse%
\begin{isamarkuptext}%
The simplification lemma for the sequential command is slightly more complex. 
  We establish that the successor configurations of any sequential statement \isa{S\isactrlsub {\isadigit{1}}{\isacharsemicolon}{\kern0pt}{\isacharsemicolon}{\kern0pt}S\isactrlsub {\isadigit{2}}} 
  match the successor configurations of \isa{S\isactrlsub {\isadigit{1}}} with \isa{S\isactrlsub {\isadigit{2}}} appended on all their 
  continuation markers. This allows us to simplify \isa{{\isasymdelta}}-function applications on 
  sequential statements to applications on only its first constituent. We infer 
  this equality in Isabelle by proving both subset~relations.%
\end{isamarkuptext}\isamarkuptrue%
\ \ \isacommand{lemma}\isamarkupfalse%
\ {\isasymdelta}{\isacharunderscore}{\kern0pt}Seq\isactrlsub {\isadigit{1}}{\isacharcolon}{\kern0pt}\isanewline
\ \ \ \ {\isachardoublequoteopen}{\isasymdelta}\ {\isacharparenleft}{\kern0pt}sh\ {\isasymleadsto}\ State{\isasymllangle}{\isasymsigma}{\isasymrrangle}{\isacharcomma}{\kern0pt}\ {\isasymlambda}{\isacharbrackleft}{\kern0pt}S\isactrlsub {\isadigit{1}}{\isacharsemicolon}{\kern0pt}{\isacharsemicolon}{\kern0pt}S\isactrlsub {\isadigit{2}}{\isacharbrackright}{\kern0pt}{\isacharparenright}{\kern0pt}\ {\isasymsubseteq}\ {\isacharparenleft}{\kern0pt}{\isacharpercent}{\kern0pt}c{\isachardot}{\kern0pt}\ {\isacharparenleft}{\kern0pt}{\isacharparenleft}{\kern0pt}fst\ c{\isacharparenright}{\kern0pt}{\isacharcomma}{\kern0pt}\ cont{\isacharunderscore}{\kern0pt}append\ {\isacharparenleft}{\kern0pt}snd\ c{\isacharparenright}{\kern0pt}\ S\isactrlsub {\isadigit{2}}{\isacharparenright}{\kern0pt}{\isacharparenright}{\kern0pt}\ {\isacharbackquote}{\kern0pt}\ {\isasymdelta}\ {\isacharparenleft}{\kern0pt}sh\ {\isasymleadsto}\ State{\isasymllangle}{\isasymsigma}{\isasymrrangle}{\isacharcomma}{\kern0pt}\ {\isasymlambda}{\isacharbrackleft}{\kern0pt}S\isactrlsub {\isadigit{1}}{\isacharbrackright}{\kern0pt}{\isacharparenright}{\kern0pt}{\isachardoublequoteclose}\isanewline
\isadelimproof
\ \ %
\endisadelimproof
\isatagproof
\isacommand{proof}\isamarkupfalse%
\ {\isacharparenleft}{\kern0pt}subst\ subset{\isacharunderscore}{\kern0pt}iff{\isacharparenright}{\kern0pt}\isanewline
\ \ \ \ \isacommand{show}\isamarkupfalse%
\ {\isachardoublequoteopen}{\isasymforall}c{\isachardot}{\kern0pt}\ c\ {\isasymin}\ {\isasymdelta}\ {\isacharparenleft}{\kern0pt}sh\ {\isasymleadsto}\ State{\isasymllangle}{\isasymsigma}{\isasymrrangle}{\isacharcomma}{\kern0pt}\ {\isasymlambda}{\isacharbrackleft}{\kern0pt}S\isactrlsub {\isadigit{1}}{\isacharsemicolon}{\kern0pt}{\isacharsemicolon}{\kern0pt}S\isactrlsub {\isadigit{2}}{\isacharbrackright}{\kern0pt}{\isacharparenright}{\kern0pt}\ \isanewline
\ \ \ \ \ \ \ \ \ \ \ \ \ \ {\isasymlongrightarrow}\ c\ {\isasymin}\ {\isacharparenleft}{\kern0pt}{\isacharpercent}{\kern0pt}c{\isachardot}{\kern0pt}\ {\isacharparenleft}{\kern0pt}{\isacharparenleft}{\kern0pt}fst\ c{\isacharparenright}{\kern0pt}{\isacharcomma}{\kern0pt}\ cont{\isacharunderscore}{\kern0pt}append\ {\isacharparenleft}{\kern0pt}snd\ c{\isacharparenright}{\kern0pt}\ S\isactrlsub {\isadigit{2}}{\isacharparenright}{\kern0pt}{\isacharparenright}{\kern0pt}\ {\isacharbackquote}{\kern0pt}\ {\isasymdelta}\ {\isacharparenleft}{\kern0pt}sh\ {\isasymleadsto}\ State{\isasymllangle}{\isasymsigma}{\isasymrrangle}{\isacharcomma}{\kern0pt}\ {\isasymlambda}{\isacharbrackleft}{\kern0pt}S\isactrlsub {\isadigit{1}}{\isacharbrackright}{\kern0pt}{\isacharparenright}{\kern0pt}{\isachardoublequoteclose}\isanewline
\ \ \ \ %
\isamarkupcmt{We first use the subset-iff rule in order to rewrite the subset relation into 
       a semantically equivalent~implication.%
}\isanewline
\ \ \ \ \isacommand{proof}\isamarkupfalse%
\ {\isacharparenleft}{\kern0pt}rule\ allI{\isacharcomma}{\kern0pt}\ rule\ impI{\isacharparenright}{\kern0pt}\isanewline
\ \ \ \ \ \ \isacommand{fix}\isamarkupfalse%
\ c\isanewline
\ \ \ \ \ \ %
\isamarkupcmt{We assume \isa{c} to be an arbitrary, but fixed configuration.%
}\isanewline
\ \ \ \ \ \ \isacommand{assume}\isamarkupfalse%
\ {\isachardoublequoteopen}c\ {\isasymin}\ {\isasymdelta}\ {\isacharparenleft}{\kern0pt}sh\ {\isasymleadsto}\ State{\isasymllangle}{\isasymsigma}{\isasymrrangle}{\isacharcomma}{\kern0pt}\ {\isasymlambda}{\isacharbrackleft}{\kern0pt}S\isactrlsub {\isadigit{1}}{\isacharsemicolon}{\kern0pt}{\isacharsemicolon}{\kern0pt}S\isactrlsub {\isadigit{2}}{\isacharbrackright}{\kern0pt}{\isacharparenright}{\kern0pt}{\isachardoublequoteclose}\isanewline
\ \ \ \ \ \ %
\isamarkupcmt{We assume that the premise holds, implying that \isa{c} is a successor 
         configuration~of~\isa{S\isactrlsub {\isadigit{1}}{\isacharsemicolon}{\kern0pt}{\isacharsemicolon}{\kern0pt}S\isactrlsub {\isadigit{2}}}.%
}\isanewline
\ \ \ \ \ \ \isacommand{then}\isamarkupfalse%
\ \isacommand{obtain}\isamarkupfalse%
\ {\isasympi}\ \isakeyword{where}\ assm\isactrlsub {\isasympi}{\isacharcolon}{\kern0pt}\ \isanewline
\ \ \ \ \ \ \ \ {\isachardoublequoteopen}c\ {\isacharequal}{\kern0pt}\ {\isacharparenleft}{\kern0pt}sh\ {\isasymcdot}\ {\isacharparenleft}{\kern0pt}{\isasymdown}\isactrlsub {\isasymtau}\ {\isasympi}{\isacharparenright}{\kern0pt}{\isacharcomma}{\kern0pt}\ {\isasymdown}\isactrlsub {\isasymlambda}\ {\isasympi}{\isacharparenright}{\kern0pt}\ {\isasymand}\ {\isasympi}\ {\isasymin}\ val\ {\isacharparenleft}{\kern0pt}S\isactrlsub {\isadigit{1}}{\isacharsemicolon}{\kern0pt}{\isacharsemicolon}{\kern0pt}S\isactrlsub {\isadigit{2}}{\isacharparenright}{\kern0pt}\ {\isasymsigma}\ {\isasymand}\ consistent{\isacharparenleft}{\kern0pt}{\isasymdown}\isactrlsub p\ {\isasympi}{\isacharparenright}{\kern0pt}{\isachardoublequoteclose}\ \isacommand{by}\isamarkupfalse%
\ auto\isanewline
\ \ \ \ \ \ %
\isamarkupcmt{Due to the definition of the \isa{{\isasymdelta}}-function, there must exist a continuation 
         trace \isa{{\isasympi}} with a consistent path condition generated from \isa{S\isactrlsub {\isadigit{1}}{\isacharsemicolon}{\kern0pt}{\isacharsemicolon}{\kern0pt}S\isactrlsub {\isadigit{2}}} that can 
         be translated into \isa{c}. We obtain this continuation trace~\isa{{\isasympi}}.%
}\ \isanewline
\ \ \ \ \ \ \isacommand{moreover}\isamarkupfalse%
\ \isacommand{then}\isamarkupfalse%
\ \isacommand{obtain}\isamarkupfalse%
\ {\isasympi}{\isacharprime}{\kern0pt}\ \isakeyword{where}\ assm\isactrlsub {\isasympi}{\isacharprime}{\kern0pt}{\isacharcolon}{\kern0pt}\ \isanewline
\ \ \ \ \ \ \ \ {\isachardoublequoteopen}{\isasympi}\ {\isacharequal}{\kern0pt}\ {\isacharparenleft}{\kern0pt}{\isasymdown}\isactrlsub p\ {\isasympi}{\isacharprime}{\kern0pt}{\isacharparenright}{\kern0pt}\ {\isasymtriangleright}\ {\isacharparenleft}{\kern0pt}{\isasymdown}\isactrlsub {\isasymtau}\ {\isasympi}{\isacharprime}{\kern0pt}{\isacharparenright}{\kern0pt}\ \isactrlitem \ cont{\isacharunderscore}{\kern0pt}append\ {\isacharparenleft}{\kern0pt}{\isasymdown}\isactrlsub {\isasymlambda}\ {\isasympi}{\isacharprime}{\kern0pt}{\isacharparenright}{\kern0pt}\ S\isactrlsub {\isadigit{2}}\ {\isasymand}\ {\isasympi}{\isacharprime}{\kern0pt}\ {\isasymin}\ {\isacharparenleft}{\kern0pt}val\ S\isactrlsub {\isadigit{1}}\ {\isasymsigma}{\isacharparenright}{\kern0pt}{\isachardoublequoteclose}\ \isacommand{by}\isamarkupfalse%
\ auto\isanewline
\ \ \ \ \ \ %
\isamarkupcmt{Aligning with the definition of the valuation function, there must also
         exist a continuation trace \isa{{\isasympi}{\isacharprime}{\kern0pt}} generated from \isa{S\isactrlsub {\isadigit{1}}}, which matches
         \isa{{\isasympi}}, if we appended \isa{S\isactrlsub {\isadigit{2}}} onto its continuation~marker.%
}\isanewline
\ \ \ \ \ \ \isacommand{ultimately}\isamarkupfalse%
\ \isacommand{have}\isamarkupfalse%
\ connect{\isacharcolon}{\kern0pt}\ \isanewline
\ \ \ \ \ \ \ \ {\isachardoublequoteopen}consistent\ {\isacharparenleft}{\kern0pt}{\isasymdown}\isactrlsub p\ {\isasympi}{\isacharprime}{\kern0pt}{\isacharparenright}{\kern0pt}\ {\isasymand}\ {\isacharparenleft}{\kern0pt}{\isasymdown}\isactrlsub {\isasymtau}\ {\isasympi}{\isacharprime}{\kern0pt}{\isacharparenright}{\kern0pt}\ {\isacharequal}{\kern0pt}\ {\isacharparenleft}{\kern0pt}{\isasymdown}\isactrlsub {\isasymtau}\ {\isasympi}{\isacharparenright}{\kern0pt}\ {\isasymand}\ {\isacharparenleft}{\kern0pt}{\isasymdown}\isactrlsub {\isasymlambda}\ {\isasympi}{\isacharparenright}{\kern0pt}\ {\isacharequal}{\kern0pt}\ cont{\isacharunderscore}{\kern0pt}append\ {\isacharparenleft}{\kern0pt}{\isasymdown}\isactrlsub {\isasymlambda}\ {\isasympi}{\isacharprime}{\kern0pt}{\isacharparenright}{\kern0pt}\ S\isactrlsub {\isadigit{2}}{\isachardoublequoteclose}\ \isacommand{by}\isamarkupfalse%
\ simp\isanewline
\ \ \ \ \ \ %
\isamarkupcmt{Considering that the path conditions of \isa{{\isasympi}} and \isa{{\isasympi}{\isacharprime}{\kern0pt}} match, both must be 
         consistent. Their symbolic traces also match. The only difference lies
         in the modified continuation~marker.%
}\isanewline
\ \ \ \ \ \ \isacommand{moreover}\isamarkupfalse%
\ \isacommand{then}\isamarkupfalse%
\ \isacommand{obtain}\isamarkupfalse%
\ c{\isacharprime}{\kern0pt}\ \isakeyword{where}\ {\isachardoublequoteopen}c{\isacharprime}{\kern0pt}\ {\isacharequal}{\kern0pt}\ {\isacharparenleft}{\kern0pt}sh\ {\isasymcdot}\ {\isacharparenleft}{\kern0pt}{\isasymdown}\isactrlsub {\isasymtau}\ {\isasympi}{\isacharprime}{\kern0pt}{\isacharparenright}{\kern0pt}{\isacharcomma}{\kern0pt}\ {\isasymdown}\isactrlsub {\isasymlambda}\ {\isasympi}{\isacharprime}{\kern0pt}{\isacharparenright}{\kern0pt}{\isachardoublequoteclose}\ \isacommand{by}\isamarkupfalse%
\ auto\isanewline
\ \ \ \ \ \ %
\isamarkupcmt{We can then obtain the configuration \isa{c{\isacharprime}{\kern0pt}} translated from the 
         continuation trace~\isa{{\isasympi}{\isacharprime}{\kern0pt}}.%
}\isanewline
\ \ \ \ \ \ \isacommand{moreover}\isamarkupfalse%
\ \isacommand{then}\isamarkupfalse%
\ \isacommand{have}\isamarkupfalse%
\ {\isachardoublequoteopen}fst{\isacharparenleft}{\kern0pt}c{\isacharparenright}{\kern0pt}\ {\isacharequal}{\kern0pt}\ fst{\isacharparenleft}{\kern0pt}c{\isacharprime}{\kern0pt}{\isacharparenright}{\kern0pt}\ {\isasymand}\ snd{\isacharparenleft}{\kern0pt}c{\isacharparenright}{\kern0pt}\ {\isacharequal}{\kern0pt}\ cont{\isacharunderscore}{\kern0pt}append\ {\isacharparenleft}{\kern0pt}snd\ c{\isacharprime}{\kern0pt}{\isacharparenright}{\kern0pt}\ S\isactrlsub {\isadigit{2}}{\isachardoublequoteclose}\ \isanewline
\ \ \ \ \ \ \ \ \isacommand{by}\isamarkupfalse%
\ {\isacharparenleft}{\kern0pt}simp\ add{\isacharcolon}{\kern0pt}\ assm\isactrlsub {\isasympi}\ connect{\isacharparenright}{\kern0pt}\isanewline
\ \ \ \ \ %
\isamarkupcmt{Given this information, both \isa{c} and \isa{c{\isacharprime}{\kern0pt}} must match in their symbolic 
         traces. However, \isa{c} additionally appends \isa{S\isactrlsub {\isadigit{2}}} onto its continuation~marker.%
}\isanewline
\ \ \ \ \ \ \isacommand{ultimately}\isamarkupfalse%
\ \isacommand{show}\isamarkupfalse%
\ {\isachardoublequoteopen}c\ {\isasymin}\ {\isacharparenleft}{\kern0pt}{\isacharpercent}{\kern0pt}c{\isachardot}{\kern0pt}\ {\isacharparenleft}{\kern0pt}{\isacharparenleft}{\kern0pt}fst\ c{\isacharparenright}{\kern0pt}{\isacharcomma}{\kern0pt}\ cont{\isacharunderscore}{\kern0pt}append\ {\isacharparenleft}{\kern0pt}snd\ c{\isacharparenright}{\kern0pt}\ S\isactrlsub {\isadigit{2}}{\isacharparenright}{\kern0pt}{\isacharparenright}{\kern0pt}\ {\isacharbackquote}{\kern0pt}\ {\isasymdelta}\ {\isacharparenleft}{\kern0pt}sh\ {\isasymleadsto}\ State{\isasymllangle}{\isasymsigma}{\isasymrrangle}{\isacharcomma}{\kern0pt}\ {\isasymlambda}{\isacharbrackleft}{\kern0pt}S\isactrlsub {\isadigit{1}}{\isacharbrackright}{\kern0pt}{\isacharparenright}{\kern0pt}{\isachardoublequoteclose}\isanewline
\ \ \ \ \ \ \ \ \isacommand{using}\isamarkupfalse%
\ assm\isactrlsub {\isasympi}\ assm\isactrlsub {\isasympi}{\isacharprime}{\kern0pt}\ image{\isacharunderscore}{\kern0pt}iff\ \isacommand{by}\isamarkupfalse%
\ fastforce\isanewline
\ \ \ \ \ \ %
\isamarkupcmt{Thus \isa{c} must match \isa{c{\isacharprime}{\kern0pt}} with an appended \isa{S\isactrlsub {\isadigit{2}}} in its continuation marker,
         closing the proof.%
}\isanewline
\ \ \ \ \isacommand{qed}\isamarkupfalse%
\isanewline
\ \ \isacommand{qed}\isamarkupfalse%
\endisatagproof
{\isafoldproof}%
\isadelimproof
\isanewline
\endisadelimproof
\isanewline
\ \ \isacommand{lemma}\isamarkupfalse%
\ {\isasymdelta}{\isacharunderscore}{\kern0pt}Seq\isactrlsub {\isadigit{2}}{\isacharcolon}{\kern0pt}\isanewline
\ \ \ \ {\isachardoublequoteopen}{\isacharparenleft}{\kern0pt}{\isacharpercent}{\kern0pt}c{\isachardot}{\kern0pt}\ {\isacharparenleft}{\kern0pt}{\isacharparenleft}{\kern0pt}fst\ c{\isacharparenright}{\kern0pt}{\isacharcomma}{\kern0pt}\ cont{\isacharunderscore}{\kern0pt}append\ {\isacharparenleft}{\kern0pt}snd\ c{\isacharparenright}{\kern0pt}\ S\isactrlsub {\isadigit{2}}{\isacharparenright}{\kern0pt}{\isacharparenright}{\kern0pt}\ {\isacharbackquote}{\kern0pt}\ {\isasymdelta}\ {\isacharparenleft}{\kern0pt}sh\ {\isasymleadsto}\ State{\isasymllangle}{\isasymsigma}{\isasymrrangle}{\isacharcomma}{\kern0pt}\ {\isasymlambda}{\isacharbrackleft}{\kern0pt}S\isactrlsub {\isadigit{1}}{\isacharbrackright}{\kern0pt}{\isacharparenright}{\kern0pt}\ {\isasymsubseteq}\ {\isasymdelta}\ {\isacharparenleft}{\kern0pt}sh\ {\isasymleadsto}\ State{\isasymllangle}{\isasymsigma}{\isasymrrangle}{\isacharcomma}{\kern0pt}\ {\isasymlambda}{\isacharbrackleft}{\kern0pt}S\isactrlsub {\isadigit{1}}{\isacharsemicolon}{\kern0pt}{\isacharsemicolon}{\kern0pt}S\isactrlsub {\isadigit{2}}{\isacharbrackright}{\kern0pt}{\isacharparenright}{\kern0pt}\ {\isachardoublequoteclose}\isanewline
\isadelimproof
\ \ %
\endisadelimproof
\isatagproof
\isacommand{proof}\isamarkupfalse%
\ {\isacharparenleft}{\kern0pt}subst\ subset{\isacharunderscore}{\kern0pt}iff{\isacharparenright}{\kern0pt}\isanewline
\ \ \ \ \isacommand{show}\isamarkupfalse%
\ {\isachardoublequoteopen}{\isasymforall}c{\isachardot}{\kern0pt}\ c\ {\isasymin}\ {\isacharparenleft}{\kern0pt}{\isacharpercent}{\kern0pt}c{\isachardot}{\kern0pt}\ {\isacharparenleft}{\kern0pt}{\isacharparenleft}{\kern0pt}fst\ c{\isacharparenright}{\kern0pt}{\isacharcomma}{\kern0pt}\ cont{\isacharunderscore}{\kern0pt}append\ {\isacharparenleft}{\kern0pt}snd\ c{\isacharparenright}{\kern0pt}\ S\isactrlsub {\isadigit{2}}{\isacharparenright}{\kern0pt}{\isacharparenright}{\kern0pt}\ {\isacharbackquote}{\kern0pt}\ {\isasymdelta}\ {\isacharparenleft}{\kern0pt}sh\ {\isasymleadsto}\ State{\isasymllangle}{\isasymsigma}{\isasymrrangle}{\isacharcomma}{\kern0pt}\ {\isasymlambda}{\isacharbrackleft}{\kern0pt}S\isactrlsub {\isadigit{1}}{\isacharbrackright}{\kern0pt}{\isacharparenright}{\kern0pt}\ \isanewline
\ \ \ \ \ \ \ \ \ \ \ \ \ \ {\isasymlongrightarrow}\ c\ {\isasymin}\ {\isasymdelta}\ {\isacharparenleft}{\kern0pt}sh\ {\isasymleadsto}\ State{\isasymllangle}{\isasymsigma}{\isasymrrangle}{\isacharcomma}{\kern0pt}\ {\isasymlambda}{\isacharbrackleft}{\kern0pt}S\isactrlsub {\isadigit{1}}{\isacharsemicolon}{\kern0pt}{\isacharsemicolon}{\kern0pt}S\isactrlsub {\isadigit{2}}{\isacharbrackright}{\kern0pt}{\isacharparenright}{\kern0pt}{\isachardoublequoteclose}\isanewline
\ \ \ \ %
\isamarkupcmt{We first use the subset-iff rule in order to rewrite the subset relation into 
       a semantically equivalent~implication.%
}\ \isanewline
\ \ \ \ \isacommand{proof}\isamarkupfalse%
\ {\isacharparenleft}{\kern0pt}rule\ allI{\isacharcomma}{\kern0pt}\ rule\ impI{\isacharparenright}{\kern0pt}\isanewline
\ \ \ \ \ \ \isacommand{fix}\isamarkupfalse%
\ c\isanewline
\ \ \ \ \ \ %
\isamarkupcmt{We assume \isa{c} to be an arbitrary, but fixed configuration.%
}\isanewline
\ \ \ \ \ \ \isacommand{assume}\isamarkupfalse%
\ {\isachardoublequoteopen}c\ {\isasymin}\ {\isacharparenleft}{\kern0pt}{\isacharpercent}{\kern0pt}c{\isachardot}{\kern0pt}\ {\isacharparenleft}{\kern0pt}{\isacharparenleft}{\kern0pt}fst\ c{\isacharparenright}{\kern0pt}{\isacharcomma}{\kern0pt}\ cont{\isacharunderscore}{\kern0pt}append\ {\isacharparenleft}{\kern0pt}snd\ c{\isacharparenright}{\kern0pt}\ S\isactrlsub {\isadigit{2}}{\isacharparenright}{\kern0pt}{\isacharparenright}{\kern0pt}\ {\isacharbackquote}{\kern0pt}\ {\isasymdelta}\ {\isacharparenleft}{\kern0pt}sh\ {\isasymleadsto}\ State{\isasymllangle}{\isasymsigma}{\isasymrrangle}{\isacharcomma}{\kern0pt}\ {\isasymlambda}{\isacharbrackleft}{\kern0pt}S\isactrlsub {\isadigit{1}}{\isacharbrackright}{\kern0pt}{\isacharparenright}{\kern0pt}{\isachardoublequoteclose}\isanewline
\ \ \ \ \ \ %
\isamarkupcmt{We assume that the premise holds, implying that \isa{c} is a successor 
         configuration of \isa{S\isactrlsub {\isadigit{1}}} with \isa{S\isactrlsub {\isadigit{2}}} appended onto its continuation~marker.%
}\isanewline
\ \ \ \ \ \ \isacommand{then}\isamarkupfalse%
\ \isacommand{obtain}\isamarkupfalse%
\ c{\isacharprime}{\kern0pt}\ \isakeyword{where}\ assm\isactrlsub c{\isacharprime}{\kern0pt}{\isacharcolon}{\kern0pt}\ \isanewline
\ \ \ \ \ \ \ \ {\isachardoublequoteopen}c{\isacharprime}{\kern0pt}\ {\isasymin}\ {\isasymdelta}\ {\isacharparenleft}{\kern0pt}sh\ {\isasymleadsto}\ State{\isasymllangle}{\isasymsigma}{\isasymrrangle}{\isacharcomma}{\kern0pt}\ {\isasymlambda}{\isacharbrackleft}{\kern0pt}S\isactrlsub {\isadigit{1}}{\isacharbrackright}{\kern0pt}{\isacharparenright}{\kern0pt}\ {\isasymand}\ fst{\isacharparenleft}{\kern0pt}c{\isacharprime}{\kern0pt}{\isacharparenright}{\kern0pt}\ {\isacharequal}{\kern0pt}\ fst{\isacharparenleft}{\kern0pt}c{\isacharparenright}{\kern0pt}\ {\isasymand}\ cont{\isacharunderscore}{\kern0pt}append\ {\isacharparenleft}{\kern0pt}snd\ c{\isacharprime}{\kern0pt}{\isacharparenright}{\kern0pt}\ S\isactrlsub {\isadigit{2}}\ {\isacharequal}{\kern0pt}\ snd{\isacharparenleft}{\kern0pt}c{\isacharparenright}{\kern0pt}{\isachardoublequoteclose}\ \isacommand{by}\isamarkupfalse%
\ force\isanewline
\ \ \ \ \ \ %
\isamarkupcmt{Then there must exist a configuration \isa{c{\isacharprime}{\kern0pt}} that is also a successor 
         configuration of \isa{S\isactrlsub {\isadigit{1}}}, matching with \isa{c} in everything except its
         continuation marker. \isa{c{\isacharprime}{\kern0pt}} with \isa{S\isactrlsub {\isadigit{2}}} appended on its continuation marker
         matches~configuration~\isa{c}.%
}\isanewline
\ \ \ \ \ \ \isacommand{moreover}\isamarkupfalse%
\ \isacommand{then}\isamarkupfalse%
\ \isacommand{obtain}\isamarkupfalse%
\ {\isasympi}\ \isakeyword{where}\ assm\isactrlsub {\isasympi}{\isacharcolon}{\kern0pt}\ \isanewline
\ \ \ \ \ \ \ \ {\isachardoublequoteopen}{\isasympi}\ {\isasymin}\ val\ S\isactrlsub {\isadigit{1}}\ {\isasymsigma}\ {\isasymand}\ consistent{\isacharparenleft}{\kern0pt}{\isasymdown}\isactrlsub p\ {\isasympi}{\isacharparenright}{\kern0pt}\ {\isasymand}\ c{\isacharprime}{\kern0pt}\ {\isacharequal}{\kern0pt}\ {\isacharparenleft}{\kern0pt}sh\ {\isasymcdot}\ {\isacharparenleft}{\kern0pt}{\isasymdown}\isactrlsub {\isasymtau}\ {\isasympi}{\isacharparenright}{\kern0pt}{\isacharcomma}{\kern0pt}\ {\isacharparenleft}{\kern0pt}{\isasymdown}\isactrlsub {\isasymlambda}\ {\isasympi}{\isacharparenright}{\kern0pt}{\isacharparenright}{\kern0pt}{\isachardoublequoteclose}\ \isacommand{by}\isamarkupfalse%
\ auto\isanewline
\ \ \ \ \ \ %
\isamarkupcmt{Hence there must exist a continuation trace \isa{{\isasympi}} with a consistent path 
         condition generated from \isa{S\isactrlsub {\isadigit{1}}}, which translates to configuration~\isa{c{\isacharprime}{\kern0pt}}.%
}\isanewline
\ \ \ \ \ \ \isacommand{moreover}\isamarkupfalse%
\ \isacommand{then}\isamarkupfalse%
\ \isacommand{obtain}\isamarkupfalse%
\ {\isasympi}{\isacharprime}{\kern0pt}\ \isakeyword{where}\ \isanewline
\ \ \ \ \ \ \ \ {\isachardoublequoteopen}{\isasympi}{\isacharprime}{\kern0pt}\ {\isacharequal}{\kern0pt}\ {\isacharparenleft}{\kern0pt}{\isasymdown}\isactrlsub p\ {\isasympi}{\isacharparenright}{\kern0pt}\ {\isasymtriangleright}\ {\isacharparenleft}{\kern0pt}{\isasymdown}\isactrlsub {\isasymtau}\ {\isasympi}{\isacharparenright}{\kern0pt}\ \isactrlitem \ cont{\isacharunderscore}{\kern0pt}append\ {\isacharparenleft}{\kern0pt}{\isasymdown}\isactrlsub {\isasymlambda}\ {\isasympi}{\isacharparenright}{\kern0pt}\ S\isactrlsub {\isadigit{2}}{\isachardoublequoteclose}\ \isacommand{by}\isamarkupfalse%
\ auto\isanewline
\ \ \ \ \ \ %
\isamarkupcmt{We then obtain the continuation trace \isa{{\isasympi}{\isacharprime}{\kern0pt}} which matches \isa{{\isasympi}} except
         having \isa{S\isactrlsub {\isadigit{2}}} additionally appended onto its continuation~marker.%
}\isanewline
\ \ \ \ \ \ \isacommand{ultimately}\isamarkupfalse%
\ \isacommand{have}\isamarkupfalse%
\ connect{\isacharcolon}{\kern0pt}\ {\isachardoublequoteopen}{\isasympi}{\isacharprime}{\kern0pt}\ {\isasymin}\ {\isacharbraceleft}{\kern0pt}c\ {\isasymin}\ val\ {\isacharparenleft}{\kern0pt}S\isactrlsub {\isadigit{1}}{\isacharsemicolon}{\kern0pt}{\isacharsemicolon}{\kern0pt}S\isactrlsub {\isadigit{2}}{\isacharparenright}{\kern0pt}\ {\isasymsigma}{\isachardot}{\kern0pt}\ consistent\ {\isacharparenleft}{\kern0pt}{\isasymdown}\isactrlsub p\ c{\isacharparenright}{\kern0pt}{\isacharbraceright}{\kern0pt}\ {\isasymand}\ {\isacharparenleft}{\kern0pt}{\isasymdown}\isactrlsub {\isasymtau}\ {\isasympi}{\isacharprime}{\kern0pt}{\isacharparenright}{\kern0pt}\ {\isacharequal}{\kern0pt}\ {\isacharparenleft}{\kern0pt}{\isasymdown}\isactrlsub {\isasymtau}\ {\isasympi}{\isacharparenright}{\kern0pt}\ {\isasymand}\ {\isacharparenleft}{\kern0pt}{\isasymdown}\isactrlsub {\isasymlambda}\ {\isasympi}{\isacharprime}{\kern0pt}{\isacharparenright}{\kern0pt}\ {\isacharequal}{\kern0pt}\ cont{\isacharunderscore}{\kern0pt}append\ {\isacharparenleft}{\kern0pt}{\isasymdown}\isactrlsub {\isasymlambda}\ {\isasympi}{\isacharparenright}{\kern0pt}\ S\isactrlsub {\isadigit{2}}{\isachardoublequoteclose}\ \isacommand{by}\isamarkupfalse%
\ auto\ \isanewline
\ \ \ \ \ \ %
\isamarkupcmt{This implies that \isa{{\isasympi}{\isacharprime}{\kern0pt}} must be a continuation trace with a consistent
         path condition generated from \isa{S\isactrlsub {\isadigit{1}}{\isacharsemicolon}{\kern0pt}{\isacharsemicolon}{\kern0pt}S\isactrlsub {\isadigit{2}}}. Note that \isa{{\isasympi}} matches with \isa{{\isasympi}{\isacharprime}{\kern0pt}}
         in its symbolic trace, but not in its continuation marker.%
}\isanewline
\ \ \ \ \ \ \isacommand{then}\isamarkupfalse%
\ \isacommand{obtain}\isamarkupfalse%
\ c{\isacharprime}{\kern0pt}{\isacharprime}{\kern0pt}\ \isakeyword{where}\ assm\isactrlsub c{\isacharprime}{\kern0pt}{\isacharprime}{\kern0pt}{\isacharcolon}{\kern0pt}\ {\isachardoublequoteopen}c{\isacharprime}{\kern0pt}{\isacharprime}{\kern0pt}\ {\isacharequal}{\kern0pt}\ {\isacharparenleft}{\kern0pt}sh\ {\isasymcdot}\ {\isacharparenleft}{\kern0pt}{\isasymdown}\isactrlsub {\isasymtau}\ {\isasympi}{\isacharprime}{\kern0pt}{\isacharparenright}{\kern0pt}{\isacharcomma}{\kern0pt}\ {\isasymdown}\isactrlsub {\isasymlambda}\ {\isasympi}{\isacharprime}{\kern0pt}{\isacharparenright}{\kern0pt}{\isachardoublequoteclose}\ \isacommand{by}\isamarkupfalse%
\ auto\isanewline
\ \ \ \ \ \ %
\isamarkupcmt{We then obtain the configuration \isa{c{\isacharprime}{\kern0pt}{\isacharprime}{\kern0pt}} translated from \isa{{\isasympi}{\isacharprime}{\kern0pt}}.%
}\isanewline
\ \ \ \ \ \ \isacommand{hence}\isamarkupfalse%
\ {\isachardoublequoteopen}c\ {\isacharequal}{\kern0pt}\ c{\isacharprime}{\kern0pt}{\isacharprime}{\kern0pt}{\isachardoublequoteclose}\ \isacommand{using}\isamarkupfalse%
\ assm\isactrlsub c{\isacharprime}{\kern0pt}\ assm\isactrlsub {\isasympi}\ connect\ \isacommand{by}\isamarkupfalse%
\ auto\isanewline
\ \ \ \ \ \ %
\isamarkupcmt{We can now derive that \isa{c} and \isa{c{\isacharprime}{\kern0pt}{\isacharprime}{\kern0pt}} match in both of~their~elements.%
}\isanewline
\ \ \ \ \ \ \isacommand{thus}\isamarkupfalse%
\ {\isachardoublequoteopen}c\ {\isasymin}\ {\isasymdelta}\ {\isacharparenleft}{\kern0pt}sh\ {\isasymleadsto}\ State{\isasymllangle}{\isasymsigma}{\isasymrrangle}{\isacharcomma}{\kern0pt}\ {\isasymlambda}{\isacharbrackleft}{\kern0pt}S\isactrlsub {\isadigit{1}}{\isacharsemicolon}{\kern0pt}{\isacharsemicolon}{\kern0pt}S\isactrlsub {\isadigit{2}}{\isacharbrackright}{\kern0pt}{\isacharparenright}{\kern0pt}{\isachardoublequoteclose}\ \isanewline
\ \ \ \ \ \ \ \ \isacommand{using}\isamarkupfalse%
\ assm\isactrlsub c{\isacharprime}{\kern0pt}{\isacharprime}{\kern0pt}\ connect\ image{\isacharunderscore}{\kern0pt}iff\ \isacommand{by}\isamarkupfalse%
\ fastforce\isanewline
\ \ \ \ \ \ %
\isamarkupcmt{We know that \isa{c{\isacharprime}{\kern0pt}{\isacharprime}{\kern0pt}} is a successor configuration of \isa{S\isactrlsub {\isadigit{1}}{\isacharsemicolon}{\kern0pt}{\isacharsemicolon}{\kern0pt}S\isactrlsub {\isadigit{2}}}. 
         Considering that \isa{c} matches \isa{c{\isacharprime}{\kern0pt}{\isacharprime}{\kern0pt}}, we can finally conclude that 
         \isa{c} must also be a successor configuration of \isa{S\isactrlsub {\isadigit{1}}{\isacharsemicolon}{\kern0pt}{\isacharsemicolon}{\kern0pt}S\isactrlsub {\isadigit{2}}}, which needed to be 
         proven in the first~place.%
}\isanewline
\ \ \ \ \isacommand{qed}\isamarkupfalse%
\isanewline
\ \ \isacommand{qed}\isamarkupfalse%
\endisatagproof
{\isafoldproof}%
\isadelimproof
\isanewline
\endisadelimproof
\ \ \ \ \ \ \ \ \isanewline
\ \ \isacommand{lemma}\isamarkupfalse%
\ {\isasymdelta}{\isacharunderscore}{\kern0pt}Seq{\isacharcolon}{\kern0pt}\isanewline
\ \ \ \ {\isachardoublequoteopen}{\isasymdelta}\ {\isacharparenleft}{\kern0pt}sh\ {\isasymleadsto}\ State{\isasymllangle}{\isasymsigma}{\isasymrrangle}{\isacharcomma}{\kern0pt}\ {\isasymlambda}{\isacharbrackleft}{\kern0pt}S\isactrlsub {\isadigit{1}}{\isacharsemicolon}{\kern0pt}{\isacharsemicolon}{\kern0pt}S\isactrlsub {\isadigit{2}}{\isacharbrackright}{\kern0pt}{\isacharparenright}{\kern0pt}\ {\isacharequal}{\kern0pt}\ {\isacharparenleft}{\kern0pt}{\isacharpercent}{\kern0pt}c{\isachardot}{\kern0pt}\ {\isacharparenleft}{\kern0pt}{\isacharparenleft}{\kern0pt}fst\ c{\isacharparenright}{\kern0pt}{\isacharcomma}{\kern0pt}\ cont{\isacharunderscore}{\kern0pt}append\ {\isacharparenleft}{\kern0pt}snd\ c{\isacharparenright}{\kern0pt}\ S\isactrlsub {\isadigit{2}}{\isacharparenright}{\kern0pt}{\isacharparenright}{\kern0pt}\ {\isacharbackquote}{\kern0pt}\ {\isasymdelta}\ {\isacharparenleft}{\kern0pt}sh\ {\isasymleadsto}\ State{\isasymllangle}{\isasymsigma}{\isasymrrangle}{\isacharcomma}{\kern0pt}\ {\isasymlambda}{\isacharbrackleft}{\kern0pt}S\isactrlsub {\isadigit{1}}{\isacharbrackright}{\kern0pt}{\isacharparenright}{\kern0pt}{\isachardoublequoteclose}\isanewline
\isadelimproof
\ \ \ \ %
\endisadelimproof
\isatagproof
\isacommand{apply}\isamarkupfalse%
\ {\isacharparenleft}{\kern0pt}subst\ set{\isacharunderscore}{\kern0pt}eq{\isacharunderscore}{\kern0pt}subset{\isacharparenright}{\kern0pt}\isanewline
\ \ \ \ \isacommand{using}\isamarkupfalse%
\ {\isasymdelta}{\isacharunderscore}{\kern0pt}Seq\isactrlsub {\isadigit{1}}\ {\isasymdelta}{\isacharunderscore}{\kern0pt}Seq\isactrlsub {\isadigit{2}}\ \isacommand{by}\isamarkupfalse%
\ simp\isanewline
\ \ \ \ %
\isamarkupcmt{We can now use the proof of both subset directions to infer the 
       desired~equality.%
}%
\endisatagproof
{\isafoldproof}%
\isadelimproof
\endisadelimproof
\begin{isamarkuptext}%
In order to finalize the automated proof system of the \isa{{\isasymdelta}}-function, we 
  collect all previously proven lemmas in a set, and call it the \isa{{\isasymdelta}}-system.
  We additionally remove the normal simplifications of the \isa{{\isasymdelta}}-function, thereby 
  ensuring that the Isabelle simplifier will later select the simplification
  lemmas of our system when deriving successor~configurations.%
\end{isamarkuptext}\isamarkuptrue%
\ \ \isacommand{lemmas}\isamarkupfalse%
\ {\isasymdelta}{\isacharunderscore}{\kern0pt}system\ {\isacharequal}{\kern0pt}\ \isanewline
\ \ \ \ {\isasymdelta}{\isacharunderscore}{\kern0pt}Skip\ {\isasymdelta}{\isacharunderscore}{\kern0pt}Assign\ {\isasymdelta}{\isacharunderscore}{\kern0pt}If\isactrlsub T\ {\isasymdelta}{\isacharunderscore}{\kern0pt}If\isactrlsub F\ {\isasymdelta}{\isacharunderscore}{\kern0pt}If\isactrlsub E\ {\isasymdelta}{\isacharunderscore}{\kern0pt}While\isactrlsub T\ {\isasymdelta}{\isacharunderscore}{\kern0pt}While\isactrlsub F\ {\isasymdelta}{\isacharunderscore}{\kern0pt}While\isactrlsub E\ {\isasymdelta}{\isacharunderscore}{\kern0pt}Seq\ consistent{\isacharunderscore}{\kern0pt}def\isanewline
\isanewline
\ \ \isacommand{declare}\isamarkupfalse%
\ successors{\isachardot}{\kern0pt}simps{\isacharbrackleft}{\kern0pt}simp\ del{\isacharbrackright}{\kern0pt}%
\begin{isamarkuptext}%
We can now use the program \isa{WL{\isacharminus}{\kern0pt}ex\isactrlsub {\isadigit{1}}} in order to provide an example for the
  derivation of successor configurations applying our established~\isa{{\isasymdelta}}-system. Note 
  that the Boolean guard of the program evaluates to false in its initial state,  
  which implies that it terminates in one singular evaluation step, also indicated 
  by the derivation~below.%
\end{isamarkuptext}\isamarkuptrue%
\ \ \isacommand{lemma}\isamarkupfalse%
\ {\isachardoublequoteopen}{\isasymdelta}\ {\isacharparenleft}{\kern0pt}{\isasymlangle}{\isacharparenleft}{\kern0pt}{\isasymsigma}\isactrlsub I\ WL{\isacharunderscore}{\kern0pt}ex\isactrlsub {\isadigit{1}}{\isacharparenright}{\kern0pt}{\isasymrangle}{\isacharcomma}{\kern0pt}\ {\isasymlambda}{\isacharbrackleft}{\kern0pt}WL{\isacharunderscore}{\kern0pt}ex\isactrlsub {\isadigit{1}}{\isacharbrackright}{\kern0pt}{\isacharparenright}{\kern0pt}\ {\isacharequal}{\kern0pt}\ {\isacharbraceleft}{\kern0pt}{\isacharparenleft}{\kern0pt}{\isasymlangle}{\isacharparenleft}{\kern0pt}{\isasymsigma}\isactrlsub I\ WL{\isacharunderscore}{\kern0pt}ex\isactrlsub {\isadigit{1}}{\isacharparenright}{\kern0pt}{\isasymrangle}{\isacharcomma}{\kern0pt}\ {\isasymlambda}{\isacharbrackleft}{\kern0pt}{\isasymnabla}{\isacharbrackright}{\kern0pt}{\isacharparenright}{\kern0pt}{\isacharbraceright}{\kern0pt}{\isachardoublequoteclose}\isanewline
\isadelimproof
\ \ \ \ %
\endisadelimproof
\isatagproof
\isacommand{apply}\isamarkupfalse%
\ {\isacharparenleft}{\kern0pt}simp\ add{\isacharcolon}{\kern0pt}\ WL{\isacharunderscore}{\kern0pt}ex\isactrlsub {\isadigit{1}}{\isacharunderscore}{\kern0pt}def{\isacharparenright}{\kern0pt}\isanewline
\ \ \ \ \isacommand{using}\isamarkupfalse%
\ {\isasymdelta}{\isacharunderscore}{\kern0pt}system\ \isacommand{by}\isamarkupfalse%
\ simp%
\endisatagproof
{\isafoldproof}%
\isadelimproof
\endisadelimproof
\isadelimdocument
\endisadelimdocument
\isatagdocument
\isamarkupsubsection{Global Trace Semantics%
}
\isamarkuptrue%
\isamarkupsubsubsection{Bounded Global Traces%
}
\isamarkuptrue%
\endisatagdocument
{\isafolddocument}%
\isadelimdocument
\endisadelimdocument
\begin{isamarkuptext}%
By utilizing the notion of trace compositions, we can finally construct global 
  system traces. A global trace of a given program \isa{S} called in initial state \isa{{\isasymsigma}}
  is a symbolic trace of a terminal configuration reachable from initial 
  configuration \isa{{\isacharparenleft}{\kern0pt}{\isasymlangle}{\isasymsigma}{\isasymrangle}{\isacharcomma}{\kern0pt}\ {\isasymlambda}{\isacharbrackleft}{\kern0pt}S{\isacharbrackright}{\kern0pt}{\isacharparenright}{\kern0pt}}. In the original paper, this construction was 
  realized by computing the transitive closure of the inductive composition rule. 
  However, note that we have formalized the trace composition using a deterministic 
  function instead of an inductive relation. We will therefore need to provide an 
  additional function for modeling the transitive closure, hence strongly deviating 
  from the original paper. This function should map an initial configuration 
  onto all symbolic traces of reachable terminal~configurations. \par
  The concept of this transitive closure gives rise to a crucial predicament.
  Note that a standard function in Isabelle requires a corresponding termination
  argument. However, it is possible to construct diverging programs in WL 
  due to the While-Loop command (e.g.~\isa{WHILE\ {\isacharparenleft}{\kern0pt}Bool\ True{\isacharparenright}{\kern0pt}\ DO\ SKIP\ OD}).
  This implies that a function modeling the transitive closure of the trace
  composition might never terminate, considering that we cannot ensure that we 
  eventually reach a terminal configuration. A solution to this problem must be found, 
  so as to formalize this concept in~Isabelle. \par
  We therefore first compute n-bounded global traces (\isa{{\isasymDelta}\isactrlsub N}-function), halting the 
  transitive closure after a certain bound is reached. The function is provided a 
  bound \isa{n} and an initial configuration \isa{c}. It then returns the set of all 
  terminal configurations reachable from \isa{c} in at most n-steps, as well as all
  configurations reachable in exactly n-steps (regardless of their terminal character). 
  This design choice ensures that we stop the evaluation after a finite number
  of steps, thereby providing the missing termination~argument. \par
  We use Isabelle to formalize this intuitive concept as follows: If a terminal 
  configuration is reached, or the bound is exceeded, then the corresponding 
  configuration is returned in a singleton set. If the bound for a non-terminal 
  configuration is not yet exceeded, we first compute all successor configurations 
  using the \isa{{\isasymdelta}}-function. We can then recursively apply the \isa{{\isasymDelta}\isactrlsub N}-function with 
  bound \isa{{\isacharparenleft}{\kern0pt}n{\isacharminus}{\kern0pt}{\isadigit{1}}{\isacharparenright}{\kern0pt}} on each of those successor configurations, and then merge the results. 
  Note that this formalization avoids trace quantifications, thereby ensuring that 
  Isabelle can automatically generate corresponding~code.%
\end{isamarkuptext}\isamarkuptrue%
\ \ \isacommand{fun}\isamarkupfalse%
\isanewline
\ \ \ \ composition\isactrlsub N\ {\isacharcolon}{\kern0pt}{\isacharcolon}{\kern0pt}\ {\isachardoublequoteopen}nat\ {\isasymRightarrow}\ config\ {\isasymRightarrow}\ config\ set{\isachardoublequoteclose}\ {\isacharparenleft}{\kern0pt}{\isachardoublequoteopen}{\isasymDelta}\isactrlsub N{\isachardoublequoteclose}{\isacharparenright}{\kern0pt}\ \isakeyword{where}\isanewline
\ \ \ \ {\isachardoublequoteopen}{\isasymDelta}\isactrlsub N\ {\isadigit{0}}\ c\ {\isacharequal}{\kern0pt}\ {\isacharbraceleft}{\kern0pt}c{\isacharbraceright}{\kern0pt}{\isachardoublequoteclose}\ {\isacharbar}{\kern0pt}\isanewline
\ \ \ \ {\isachardoublequoteopen}{\isasymDelta}\isactrlsub N\ {\isacharparenleft}{\kern0pt}Suc\ n{\isacharparenright}{\kern0pt}\ {\isacharparenleft}{\kern0pt}{\isasymtau}{\isacharcomma}{\kern0pt}\ {\isasymlambda}{\isacharbrackleft}{\kern0pt}{\isasymnabla}{\isacharbrackright}{\kern0pt}{\isacharparenright}{\kern0pt}\ {\isacharequal}{\kern0pt}\ {\isacharbraceleft}{\kern0pt}{\isacharparenleft}{\kern0pt}{\isasymtau}{\isacharcomma}{\kern0pt}\ {\isasymlambda}{\isacharbrackleft}{\kern0pt}{\isasymnabla}{\isacharbrackright}{\kern0pt}{\isacharparenright}{\kern0pt}{\isacharbraceright}{\kern0pt}{\isachardoublequoteclose}\ {\isacharbar}{\kern0pt}\isanewline
\ \ \ \ {\isachardoublequoteopen}{\isasymDelta}\isactrlsub N\ {\isacharparenleft}{\kern0pt}Suc\ n{\isacharparenright}{\kern0pt}\ {\isacharparenleft}{\kern0pt}{\isasymtau}{\isacharcomma}{\kern0pt}\ {\isasymlambda}{\isacharbrackleft}{\kern0pt}S{\isacharbrackright}{\kern0pt}{\isacharparenright}{\kern0pt}\ {\isacharequal}{\kern0pt}\ {\isasymUnion}{\isacharparenleft}{\kern0pt}{\isacharparenleft}{\kern0pt}{\isacharpercent}{\kern0pt}c{\isachardot}{\kern0pt}\ {\isasymDelta}\isactrlsub N\ n\ c{\isacharparenright}{\kern0pt}\ {\isacharbackquote}{\kern0pt}\ {\isasymdelta}{\isacharparenleft}{\kern0pt}{\isasymtau}{\isacharcomma}{\kern0pt}\ {\isasymlambda}{\isacharbrackleft}{\kern0pt}S{\isacharbrackright}{\kern0pt}{\isacharparenright}{\kern0pt}{\isacharparenright}{\kern0pt}{\isachardoublequoteclose}%
\begin{isamarkuptext}%
The definition for n-bounded global traces generated from program \isa{S} called 
  in \isa{{\isasymsigma}} is now straightforward. We simply return all symbolic traces of the 
  configurations received by calling the \isa{{\isasymDelta}\isactrlsub N}-function in initial configuration 
  \isa{{\isacharparenleft}{\kern0pt}{\isasymlangle}{\isasymsigma}{\isasymrangle}{\isacharcomma}{\kern0pt}\ {\isasymlambda}{\isacharbrackleft}{\kern0pt}S{\isacharbrackright}{\kern0pt}{\isacharparenright}{\kern0pt}} with bound n. Note that this implies that the empty continuation 
  markers of the configurations are~discarded.%
\end{isamarkuptext}\isamarkuptrue%
\ \ \isacommand{definition}\isamarkupfalse%
\isanewline
\ \ \ \ Traces\isactrlsub N\ {\isacharcolon}{\kern0pt}{\isacharcolon}{\kern0pt}\ {\isachardoublequoteopen}stmt\ {\isasymRightarrow}\ {\isasymSigma}\ {\isasymRightarrow}\ nat\ {\isasymRightarrow}\ {\isasymT}\ set{\isachardoublequoteclose}\ {\isacharparenleft}{\kern0pt}{\isachardoublequoteopen}Tr\isactrlsub N{\isachardoublequoteclose}{\isacharparenright}{\kern0pt}\ \isakeyword{where}\isanewline
\ \ \ \ {\isachardoublequoteopen}Tr\isactrlsub N\ S\ {\isasymsigma}\ n\ {\isasymequiv}\ fst\ {\isacharbackquote}{\kern0pt}\ {\isasymDelta}\isactrlsub N\ n\ {\isacharparenleft}{\kern0pt}{\isasymlangle}{\isasymsigma}{\isasymrangle}{\isacharcomma}{\kern0pt}\ {\isasymlambda}{\isacharbrackleft}{\kern0pt}S{\isacharbrackright}{\kern0pt}{\isacharparenright}{\kern0pt}{\isachardoublequoteclose}%
\begin{isamarkuptext}%
Note that bounded global traces can also be used for debugging purposes, as
  it is possible to return intermediate trace results.%
\end{isamarkuptext}\isamarkuptrue%
\isadelimdocument
\endisadelimdocument
\isatagdocument
\isamarkupsubsubsection{Unbounded Global Traces%
}
\isamarkuptrue%
\endisatagdocument
{\isafolddocument}%
\isadelimdocument
\endisadelimdocument
\begin{isamarkuptext}%
We are now interested in constructing global traces without having to 
  explicitly provide a bound as an argument. As we have previously established
  however, While-Loops can be used to write diverging programs, which 
  never reach a terminal configuration. This crushes possible termination 
  arguments for our boundless transitive closure. We therefore have to 
  examine alternative formalizations, so as to circumvent this~predicament. \par
  We first desire to construct a boundless function that maps an initial configuration 
  onto all reachable terminal configurations (\isa{{\isasymDelta}}-function). One idea about the
  bound elimination could center around modeling the function explicitly as 
  partial (\isa{{\isasymrightharpoonup}}), thereby ensuring termination by mapping all diverging programs onto 
  \isa{None} (i.e. they have no function image). However, such a formalization is not
  feasible, as we would have to explicitly formalize which programs map onto \isa{None},
  thereby axiomatizing divergence in WL. This is conceptually not possible, thus 
  eliminating this formalization~option. \par
  We instead decide to construct our transitive closure using the predefined 
  \isa{Partial{\isacharminus}{\kern0pt}Function} theory of the HOL-Library. Using this theory, we remove the 
  need for a corresponding termination argument. Instead, it becomes a necessity 
  to prove the monotonicity of the given function, so as to ensure the existence 
  of a fixpoint that the function may converge~against. \par
  The function is provided a step amount \isa{n} and an initial configuration \isa{c}
  as arguments. It then utilizes the previously defined bounded transitive closure
  (\isa{{\isasymDelta}\isactrlsub N{\isacharminus}{\kern0pt}function}) to compute all configurations reachable from \isa{c} in \isa{n}-steps.
  If all returned configurations are terminal, then the evaluation has already 
  finished, implying that we reached a fixpoint. Otherwise, we simply recursively
  call the \isa{{\isasymDelta}}-function with a higher step amount. A continuous increase of
  the evaluation steps will eventually converge in a result, as long as the
  program does not diverge. The step incrementation amount is independent of the
  function result, and can hence be arbitrarily chosen. However, too low numbers will 
  cause an overhead during the computation due to the high amount of recursive 
  calls, thus lowering the performance. We have arbitrarily decided for an increase 
  of 100 steps, taking into account that most example programs will not exceed that 
  boundary. Note that this should be adapted if large case studies were 
  to~be~analyzed. \par
  We formalize our function in a tailrecursive (\isa{tailrec}) fashion, which is directly
  supported by the \isa{Partial{\isacharminus}{\kern0pt}Function} theory. This ensures that we can use
  the predefined theory for the automatic code generation of our construction.
  Note that the compiled code will only terminate if a corresponding fixpoint
  is reached during the execution. In all other scenarios, the execution of the generated 
  code will diverge. The necessary mononoticity proof is automatically 
  conducted by Isabelle without any need of user~interaction. \par%
\end{isamarkuptext}\isamarkuptrue%
\ \ \isacommand{partial{\isacharunderscore}{\kern0pt}function}\isamarkupfalse%
\ {\isacharparenleft}{\kern0pt}tailrec{\isacharparenright}{\kern0pt}\ composition\ {\isacharcolon}{\kern0pt}{\isacharcolon}{\kern0pt}\ {\isachardoublequoteopen}nat\ {\isasymRightarrow}\ config\ {\isasymRightarrow}\ config\ set{\isachardoublequoteclose}\ {\isacharparenleft}{\kern0pt}{\isachardoublequoteopen}{\isasymDelta}{\isachardoublequoteclose}{\isacharparenright}{\kern0pt}\ \isakeyword{where}\isanewline
\ \ \ \ {\isacharbrackleft}{\kern0pt}code{\isacharbrackright}{\kern0pt}{\isacharcolon}{\kern0pt}\ {\isachardoublequoteopen}{\isasymDelta}\ n\ c\ {\isacharequal}{\kern0pt}\ {\isacharparenleft}{\kern0pt}if\ {\isasymforall}c\ {\isasymin}\ {\isacharparenleft}{\kern0pt}{\isasymDelta}\isactrlsub N\ n\ c{\isacharparenright}{\kern0pt}{\isachardot}{\kern0pt}\ snd{\isacharparenleft}{\kern0pt}c{\isacharparenright}{\kern0pt}\ {\isacharequal}{\kern0pt}\ {\isasymlambda}{\isacharbrackleft}{\kern0pt}{\isasymnabla}{\isacharbrackright}{\kern0pt}\ then\ {\isasymDelta}\isactrlsub N\ n\ c\ else\ {\isasymDelta}\ {\isacharparenleft}{\kern0pt}n\ {\isacharplus}{\kern0pt}\ {\isadigit{1}}{\isadigit{0}}{\isadigit{0}}{\isacharparenright}{\kern0pt}\ c{\isacharparenright}{\kern0pt}{\isachardoublequoteclose}%
\begin{isamarkuptext}%
The construction of unbounded global traces generated from program \isa{S} called 
  in \isa{{\isasymsigma}} is now straightforward. We simply return all symbolic traces of the 
  configurations received by calling the \isa{{\isasymDelta}}-function in initial configuration 
  \isa{{\isacharparenleft}{\kern0pt}{\isasymlangle}{\isasymsigma}{\isasymrangle}{\isacharcomma}{\kern0pt}\ {\isasymlambda}{\isacharbrackleft}{\kern0pt}S{\isacharbrackright}{\kern0pt}{\isacharparenright}{\kern0pt}}. Note that this requires the \isa{{\isasymDelta}}-function to converge in a 
  corresponding~fixpoint.%
\end{isamarkuptext}\isamarkuptrue%
\ \ \isacommand{definition}\isamarkupfalse%
\ \ \isanewline
\ \ \ \ Traces\ {\isacharcolon}{\kern0pt}{\isacharcolon}{\kern0pt}\ {\isachardoublequoteopen}stmt\ {\isasymRightarrow}\ {\isasymSigma}\ {\isasymRightarrow}\ {\isasymT}\ set{\isachardoublequoteclose}\ {\isacharparenleft}{\kern0pt}{\isachardoublequoteopen}Tr{\isachardoublequoteclose}{\isacharparenright}{\kern0pt}\ \isakeyword{where}\isanewline
\ \ \ \ {\isachardoublequoteopen}Tr\ S\ {\isasymsigma}\ {\isasymequiv}\ fst\ {\isacharbackquote}{\kern0pt}\ {\isasymDelta}\ {\isadigit{0}}\ {\isacharparenleft}{\kern0pt}{\isasymlangle}{\isasymsigma}{\isasymrangle}{\isacharcomma}{\kern0pt}\ {\isasymlambda}{\isacharbrackleft}{\kern0pt}S{\isacharbrackright}{\kern0pt}{\isacharparenright}{\kern0pt}{\isachardoublequoteclose}%
\isadelimdocument
\endisadelimdocument
\isatagdocument
\isamarkupsubsubsection{Proof Automation%
}
\isamarkuptrue%
\endisatagdocument
{\isafolddocument}%
\isadelimdocument
\endisadelimdocument
\begin{isamarkuptext}%
Similar to the simplification lemmas for the \isa{{\isasymdelta}}-function, we now choose to 
  introduce simplification lemmas for the \isa{{\isasymDelta}\isactrlsub N}-function. This decision will later 
  ensure a major speedup when deriving bounded global traces in our proof~system. \par 
  We first provide simplifications for situations, in which either the provided bound 
  is exceeded, or a terminal configuration is reached. In these cases, the singular 
  returned configuration matches the configuration provided as the argument. The 
  corresponding proofs are~straightforward.%
\end{isamarkuptext}\isamarkuptrue%
\ \ \isacommand{lemma}\isamarkupfalse%
\ {\isasymDelta}\isactrlsub N{\isacharunderscore}{\kern0pt}Bounded{\isacharcolon}{\kern0pt}\isanewline
\ \ \ \ \isakeyword{shows}\ {\isachardoublequoteopen}{\isasymDelta}\isactrlsub N\ {\isadigit{0}}\ {\isacharparenleft}{\kern0pt}sh\ {\isasymleadsto}\ State{\isasymllangle}{\isasymsigma}{\isasymrrangle}{\isacharcomma}{\kern0pt}\ cm{\isacharparenright}{\kern0pt}\ {\isacharequal}{\kern0pt}\ {\isacharbraceleft}{\kern0pt}{\isacharparenleft}{\kern0pt}sh\ {\isasymleadsto}\ State{\isasymllangle}{\isasymsigma}{\isasymrrangle}{\isacharcomma}{\kern0pt}\ cm{\isacharparenright}{\kern0pt}{\isacharbraceright}{\kern0pt}{\isachardoublequoteclose}\isanewline
\isadelimproof
\ \ \ \ %
\endisadelimproof
\isatagproof
\isacommand{by}\isamarkupfalse%
\ {\isacharparenleft}{\kern0pt}metis\ composition\isactrlsub N{\isachardot}{\kern0pt}simps{\isacharparenleft}{\kern0pt}{\isadigit{1}}{\isacharparenright}{\kern0pt}{\isacharparenright}{\kern0pt}%
\endisatagproof
{\isafoldproof}%
\isadelimproof
\isanewline
\endisadelimproof
\isanewline
\ \ \isacommand{lemma}\isamarkupfalse%
\ {\isasymDelta}\isactrlsub N{\isacharunderscore}{\kern0pt}End{\isacharcolon}{\kern0pt}\isanewline
\ \ \ \ \isakeyword{shows}\ {\isachardoublequoteopen}{\isasymDelta}\isactrlsub N\ n\ {\isacharparenleft}{\kern0pt}sh\ {\isasymleadsto}\ State{\isasymllangle}{\isasymsigma}{\isasymrrangle}{\isacharcomma}{\kern0pt}\ {\isasymlambda}{\isacharbrackleft}{\kern0pt}{\isasymnabla}{\isacharbrackright}{\kern0pt}{\isacharparenright}{\kern0pt}\ {\isacharequal}{\kern0pt}\ {\isacharbraceleft}{\kern0pt}{\isacharparenleft}{\kern0pt}sh\ {\isasymleadsto}\ State{\isasymllangle}{\isasymsigma}{\isasymrrangle}{\isacharcomma}{\kern0pt}\ {\isasymlambda}{\isacharbrackleft}{\kern0pt}{\isasymnabla}{\isacharbrackright}{\kern0pt}{\isacharparenright}{\kern0pt}{\isacharbraceright}{\kern0pt}{\isachardoublequoteclose}\isanewline
\isadelimproof
\ \ \ \ %
\endisadelimproof
\isatagproof
\isacommand{by}\isamarkupfalse%
\ {\isacharparenleft}{\kern0pt}metis\ composition\isactrlsub N{\isachardot}{\kern0pt}simps{\isacharparenleft}{\kern0pt}{\isadigit{2}}{\isacharparenright}{\kern0pt}\ composition\isactrlsub N{\isachardot}{\kern0pt}elims{\isacharparenright}{\kern0pt}%
\endisatagproof
{\isafoldproof}%
\isadelimproof
\endisadelimproof
\begin{isamarkuptext}%
We next provide generic simplification lemmas for each statement of WL, 
  allowing us to repeatedly compute successor configurations using the \isa{{\isasymDelta}\isactrlsub N}-function. 
  All denoted simplifications apply the \isa{{\isasymdelta}}-function on the configuration corresponding 
  to the given statement, whilst also decrementing the step amount for the next 
  recursive call. Similar to the \isa{{\isasymdelta}}-function, we also provide simplifications for 
  situations, in which the guard of an If-Branch or While-Loop evaluates to neither 
  True nor False (due to symbolic variables). The program evaluation terminates 
  in this case, returning the empty configuration~set.%
\end{isamarkuptext}\isamarkuptrue%
\ \ \isacommand{lemma}\isamarkupfalse%
\ {\isasymDelta}\isactrlsub N{\isacharunderscore}{\kern0pt}Skip{\isacharcolon}{\kern0pt}\isanewline
\ \ \ \ \isakeyword{shows}\ {\isachardoublequoteopen}{\isasymDelta}\isactrlsub N\ {\isacharparenleft}{\kern0pt}Suc\ n{\isacharparenright}{\kern0pt}\ {\isacharparenleft}{\kern0pt}sh\ {\isasymleadsto}\ State{\isasymllangle}{\isasymsigma}{\isasymrrangle}{\isacharcomma}{\kern0pt}\ {\isasymlambda}{\isacharbrackleft}{\kern0pt}SKIP{\isacharbrackright}{\kern0pt}{\isacharparenright}{\kern0pt}\ {\isacharequal}{\kern0pt}\ {\isasymDelta}\isactrlsub N\ n\ {\isacharparenleft}{\kern0pt}sh\ {\isasymleadsto}\ State{\isasymllangle}{\isasymsigma}{\isasymrrangle}{\isacharcomma}{\kern0pt}\ {\isasymlambda}{\isacharbrackleft}{\kern0pt}{\isasymnabla}{\isacharbrackright}{\kern0pt}{\isacharparenright}{\kern0pt}{\isachardoublequoteclose}\isanewline
\isadelimproof
\ \ \ \ %
\endisadelimproof
\isatagproof
\isacommand{using}\isamarkupfalse%
\ {\isasymdelta}{\isacharunderscore}{\kern0pt}Skip\ \isacommand{by}\isamarkupfalse%
\ simp%
\endisatagproof
{\isafoldproof}%
\isadelimproof
\isanewline
\endisadelimproof
\isanewline
\ \ \isacommand{lemma}\isamarkupfalse%
\ {\isasymDelta}\isactrlsub N{\isacharunderscore}{\kern0pt}Assign{\isacharcolon}{\kern0pt}\isanewline
\ \ \ \ \isakeyword{shows}\ {\isachardoublequoteopen}{\isasymDelta}\isactrlsub N\ {\isacharparenleft}{\kern0pt}Suc\ n{\isacharparenright}{\kern0pt}\ {\isacharparenleft}{\kern0pt}sh\ {\isasymleadsto}\ State{\isasymllangle}{\isasymsigma}{\isasymrrangle}{\isacharcomma}{\kern0pt}\ {\isasymlambda}{\isacharbrackleft}{\kern0pt}x\ {\isacharcolon}{\kern0pt}{\isacharequal}{\kern0pt}\ a{\isacharbrackright}{\kern0pt}{\isacharparenright}{\kern0pt}\ {\isacharequal}{\kern0pt}\ {\isasymDelta}\isactrlsub N\ n\ {\isacharparenleft}{\kern0pt}{\isacharparenleft}{\kern0pt}sh\ {\isasymleadsto}\ State{\isasymllangle}{\isasymsigma}{\isasymrrangle}{\isacharparenright}{\kern0pt}\ {\isasymleadsto}\ State{\isasymllangle}{\isacharbrackleft}{\kern0pt}x\ {\isasymlongmapsto}\ Exp\ {\isacharparenleft}{\kern0pt}val\isactrlsub A\ a\ {\isasymsigma}{\isacharparenright}{\kern0pt}{\isacharbrackright}{\kern0pt}\ {\isasymsigma}{\isasymrrangle}{\isacharcomma}{\kern0pt}\ {\isasymlambda}{\isacharbrackleft}{\kern0pt}{\isasymnabla}{\isacharbrackright}{\kern0pt}{\isacharparenright}{\kern0pt}{\isachardoublequoteclose}\isanewline
\isadelimproof
\ \ \ \ %
\endisadelimproof
\isatagproof
\isacommand{using}\isamarkupfalse%
\ {\isasymdelta}{\isacharunderscore}{\kern0pt}Assign\ \isacommand{by}\isamarkupfalse%
\ simp%
\endisatagproof
{\isafoldproof}%
\isadelimproof
\isanewline
\endisadelimproof
\isanewline
\ \ \isacommand{lemma}\isamarkupfalse%
\ {\isasymDelta}\isactrlsub N{\isacharunderscore}{\kern0pt}If\isactrlsub T{\isacharcolon}{\kern0pt}\isanewline
\ \ \ \ \isakeyword{assumes}\ {\isachardoublequoteopen}consistent\ {\isacharbraceleft}{\kern0pt}{\isacharparenleft}{\kern0pt}val\isactrlsub B\ b\ {\isasymsigma}{\isacharparenright}{\kern0pt}{\isacharbraceright}{\kern0pt}{\isachardoublequoteclose}\isanewline
\ \ \ \ \isakeyword{shows}\ {\isachardoublequoteopen}{\isasymDelta}\isactrlsub N\ {\isacharparenleft}{\kern0pt}Suc\ n{\isacharparenright}{\kern0pt}\ {\isacharparenleft}{\kern0pt}sh\ {\isasymleadsto}\ State{\isasymllangle}{\isasymsigma}{\isasymrrangle}{\isacharcomma}{\kern0pt}\ {\isasymlambda}{\isacharbrackleft}{\kern0pt}IF\ b\ THEN\ S\ FI{\isacharbrackright}{\kern0pt}{\isacharparenright}{\kern0pt}\ {\isacharequal}{\kern0pt}\ {\isasymDelta}\isactrlsub N\ n\ {\isacharparenleft}{\kern0pt}sh\ {\isasymleadsto}\ State{\isasymllangle}{\isasymsigma}{\isasymrrangle}{\isacharcomma}{\kern0pt}\ {\isasymlambda}{\isacharbrackleft}{\kern0pt}S{\isacharbrackright}{\kern0pt}{\isacharparenright}{\kern0pt}{\isachardoublequoteclose}\isanewline
\isadelimproof
\ \ \ \ %
\endisadelimproof
\isatagproof
\isacommand{using}\isamarkupfalse%
\ assms\ {\isasymdelta}{\isacharunderscore}{\kern0pt}If\isactrlsub T\ \isacommand{by}\isamarkupfalse%
\ simp%
\endisatagproof
{\isafoldproof}%
\isadelimproof
\isanewline
\endisadelimproof
\isanewline
\ \ \isacommand{lemma}\isamarkupfalse%
\ {\isasymDelta}\isactrlsub N{\isacharunderscore}{\kern0pt}If\isactrlsub F{\isacharcolon}{\kern0pt}\isanewline
\ \ \ \ \isakeyword{assumes}\ {\isachardoublequoteopen}consistent\ {\isacharbraceleft}{\kern0pt}{\isacharparenleft}{\kern0pt}val\isactrlsub B\ {\isacharparenleft}{\kern0pt}Not\ b{\isacharparenright}{\kern0pt}\ {\isasymsigma}{\isacharparenright}{\kern0pt}{\isacharbraceright}{\kern0pt}{\isachardoublequoteclose}\isanewline
\ \ \ \ \isakeyword{shows}\ {\isachardoublequoteopen}{\isasymDelta}\isactrlsub N\ {\isacharparenleft}{\kern0pt}Suc\ n{\isacharparenright}{\kern0pt}\ {\isacharparenleft}{\kern0pt}sh\ {\isasymleadsto}\ State{\isasymllangle}{\isasymsigma}{\isasymrrangle}{\isacharcomma}{\kern0pt}\ {\isasymlambda}{\isacharbrackleft}{\kern0pt}IF\ b\ THEN\ S\ FI{\isacharbrackright}{\kern0pt}{\isacharparenright}{\kern0pt}\ {\isacharequal}{\kern0pt}\ {\isasymDelta}\isactrlsub N\ n\ {\isacharparenleft}{\kern0pt}sh\ {\isasymleadsto}\ State{\isasymllangle}{\isasymsigma}{\isasymrrangle}{\isacharcomma}{\kern0pt}\ {\isasymlambda}{\isacharbrackleft}{\kern0pt}{\isasymnabla}{\isacharbrackright}{\kern0pt}{\isacharparenright}{\kern0pt}{\isachardoublequoteclose}\isanewline
\isadelimproof
\ \ \ \ %
\endisadelimproof
\isatagproof
\isacommand{using}\isamarkupfalse%
\ assms\ {\isasymdelta}{\isacharunderscore}{\kern0pt}If\isactrlsub F\ \isacommand{by}\isamarkupfalse%
\ simp%
\endisatagproof
{\isafoldproof}%
\isadelimproof
\isanewline
\endisadelimproof
\isanewline
\ \ \isacommand{lemma}\isamarkupfalse%
\ {\isasymDelta}\isactrlsub N{\isacharunderscore}{\kern0pt}If\isactrlsub E{\isacharcolon}{\kern0pt}\isanewline
\ \ \ \ \isakeyword{assumes}\ {\isachardoublequoteopen}{\isasymnot}consistent\ {\isacharbraceleft}{\kern0pt}{\isacharparenleft}{\kern0pt}val\isactrlsub B\ b\ {\isasymsigma}{\isacharparenright}{\kern0pt}{\isacharbraceright}{\kern0pt}\ {\isasymand}\ {\isasymnot}consistent\ {\isacharbraceleft}{\kern0pt}{\isacharparenleft}{\kern0pt}val\isactrlsub B\ {\isacharparenleft}{\kern0pt}Not\ b{\isacharparenright}{\kern0pt}\ {\isasymsigma}{\isacharparenright}{\kern0pt}{\isacharbraceright}{\kern0pt}{\isachardoublequoteclose}\isanewline
\ \ \ \ \isakeyword{shows}\ {\isachardoublequoteopen}{\isasymDelta}\isactrlsub N\ {\isacharparenleft}{\kern0pt}Suc\ n{\isacharparenright}{\kern0pt}\ {\isacharparenleft}{\kern0pt}sh\ {\isasymleadsto}\ State{\isasymllangle}{\isasymsigma}{\isasymrrangle}{\isacharcomma}{\kern0pt}\ {\isasymlambda}{\isacharbrackleft}{\kern0pt}IF\ b\ THEN\ S\ FI{\isacharbrackright}{\kern0pt}{\isacharparenright}{\kern0pt}\ {\isacharequal}{\kern0pt}\ {\isacharbraceleft}{\kern0pt}{\isacharbraceright}{\kern0pt}{\isachardoublequoteclose}\isanewline
\isadelimproof
\ \ %
\endisadelimproof
\isatagproof
\isacommand{proof}\isamarkupfalse%
\ {\isacharminus}{\kern0pt}\isanewline
\ \ \ \ \isacommand{have}\isamarkupfalse%
\ {\isachardoublequoteopen}{\isasymdelta}\ {\isacharparenleft}{\kern0pt}sh\ {\isasymleadsto}\ State{\isasymllangle}{\isasymsigma}{\isasymrrangle}{\isacharcomma}{\kern0pt}\ {\isasymlambda}{\isacharbrackleft}{\kern0pt}IF\ b\ THEN\ S\ FI{\isacharbrackright}{\kern0pt}{\isacharparenright}{\kern0pt}\ {\isacharequal}{\kern0pt}\ {\isacharbraceleft}{\kern0pt}{\isacharbraceright}{\kern0pt}{\isachardoublequoteclose}\ \isacommand{using}\isamarkupfalse%
\ assms\ {\isasymdelta}{\isacharunderscore}{\kern0pt}If\isactrlsub E\ \isacommand{by}\isamarkupfalse%
\ simp\isanewline
\ \ \ \ \isacommand{thus}\isamarkupfalse%
\ {\isacharquery}{\kern0pt}thesis\ \isacommand{using}\isamarkupfalse%
\ composition\isactrlsub N{\isachardot}{\kern0pt}simps\ \isacommand{by}\isamarkupfalse%
\ fastforce\isanewline
\ \ \isacommand{qed}\isamarkupfalse%
\endisatagproof
{\isafoldproof}%
\isadelimproof
\isanewline
\endisadelimproof
\isanewline
\ \ \isacommand{lemma}\isamarkupfalse%
\ {\isasymDelta}\isactrlsub N{\isacharunderscore}{\kern0pt}While\isactrlsub T{\isacharcolon}{\kern0pt}\isanewline
\ \ \ \ \isakeyword{assumes}\ {\isachardoublequoteopen}consistent\ {\isacharbraceleft}{\kern0pt}{\isacharparenleft}{\kern0pt}val\isactrlsub B\ b\ {\isasymsigma}{\isacharparenright}{\kern0pt}{\isacharbraceright}{\kern0pt}{\isachardoublequoteclose}\isanewline
\ \ \ \ \isakeyword{shows}\ {\isachardoublequoteopen}{\isasymDelta}\isactrlsub N\ {\isacharparenleft}{\kern0pt}Suc\ n{\isacharparenright}{\kern0pt}\ {\isacharparenleft}{\kern0pt}sh\ {\isasymleadsto}\ State{\isasymllangle}{\isasymsigma}{\isasymrrangle}{\isacharcomma}{\kern0pt}\ {\isasymlambda}{\isacharbrackleft}{\kern0pt}WHILE\ b\ DO\ S\ OD{\isacharbrackright}{\kern0pt}{\isacharparenright}{\kern0pt}\ {\isacharequal}{\kern0pt}\ {\isasymDelta}\isactrlsub N\ n\ {\isacharparenleft}{\kern0pt}sh\ {\isasymleadsto}\ State{\isasymllangle}{\isasymsigma}{\isasymrrangle}{\isacharcomma}{\kern0pt}\ {\isasymlambda}{\isacharbrackleft}{\kern0pt}S{\isacharsemicolon}{\kern0pt}{\isacharsemicolon}{\kern0pt}WHILE\ b\ DO\ S\ OD{\isacharbrackright}{\kern0pt}{\isacharparenright}{\kern0pt}{\isachardoublequoteclose}\isanewline
\isadelimproof
\ \ \ \ %
\endisadelimproof
\isatagproof
\isacommand{using}\isamarkupfalse%
\ assms\ {\isasymdelta}{\isacharunderscore}{\kern0pt}While\isactrlsub T\ \isacommand{by}\isamarkupfalse%
\ simp%
\endisatagproof
{\isafoldproof}%
\isadelimproof
\isanewline
\endisadelimproof
\isanewline
\ \ \isacommand{lemma}\isamarkupfalse%
\ {\isasymDelta}\isactrlsub N{\isacharunderscore}{\kern0pt}While\isactrlsub F{\isacharcolon}{\kern0pt}\isanewline
\ \ \ \ \isakeyword{assumes}\ {\isachardoublequoteopen}consistent\ {\isacharbraceleft}{\kern0pt}{\isacharparenleft}{\kern0pt}val\isactrlsub B\ {\isacharparenleft}{\kern0pt}Not\ b{\isacharparenright}{\kern0pt}\ {\isasymsigma}{\isacharparenright}{\kern0pt}{\isacharbraceright}{\kern0pt}{\isachardoublequoteclose}\isanewline
\ \ \ \ \isakeyword{shows}\ {\isachardoublequoteopen}{\isasymDelta}\isactrlsub N\ {\isacharparenleft}{\kern0pt}Suc\ n{\isacharparenright}{\kern0pt}\ {\isacharparenleft}{\kern0pt}sh\ {\isasymleadsto}\ State{\isasymllangle}{\isasymsigma}{\isasymrrangle}{\isacharcomma}{\kern0pt}\ {\isasymlambda}{\isacharbrackleft}{\kern0pt}WHILE\ b\ DO\ S\ OD{\isacharbrackright}{\kern0pt}{\isacharparenright}{\kern0pt}\ {\isacharequal}{\kern0pt}\ {\isasymDelta}\isactrlsub N\ n\ {\isacharparenleft}{\kern0pt}sh\ {\isasymleadsto}\ State{\isasymllangle}{\isasymsigma}{\isasymrrangle}{\isacharcomma}{\kern0pt}\ {\isasymlambda}{\isacharbrackleft}{\kern0pt}{\isasymnabla}{\isacharbrackright}{\kern0pt}{\isacharparenright}{\kern0pt}{\isachardoublequoteclose}\isanewline
\isadelimproof
\ \ \ \ %
\endisadelimproof
\isatagproof
\isacommand{using}\isamarkupfalse%
\ assms\ {\isasymdelta}{\isacharunderscore}{\kern0pt}While\isactrlsub F\ \isacommand{by}\isamarkupfalse%
\ simp%
\endisatagproof
{\isafoldproof}%
\isadelimproof
\isanewline
\endisadelimproof
\isanewline
\ \ \isacommand{lemma}\isamarkupfalse%
\ {\isasymDelta}\isactrlsub N{\isacharunderscore}{\kern0pt}While\isactrlsub E{\isacharcolon}{\kern0pt}\isanewline
\ \ \ \ \isakeyword{assumes}\ {\isachardoublequoteopen}{\isasymnot}consistent\ {\isacharbraceleft}{\kern0pt}{\isacharparenleft}{\kern0pt}val\isactrlsub B\ b\ {\isasymsigma}{\isacharparenright}{\kern0pt}{\isacharbraceright}{\kern0pt}\ {\isasymand}\ {\isasymnot}consistent\ {\isacharbraceleft}{\kern0pt}{\isacharparenleft}{\kern0pt}val\isactrlsub B\ {\isacharparenleft}{\kern0pt}Not\ b{\isacharparenright}{\kern0pt}\ {\isasymsigma}{\isacharparenright}{\kern0pt}{\isacharbraceright}{\kern0pt}{\isachardoublequoteclose}\isanewline
\ \ \ \ \isakeyword{shows}\ {\isachardoublequoteopen}{\isasymDelta}\isactrlsub N\ {\isacharparenleft}{\kern0pt}Suc\ n{\isacharparenright}{\kern0pt}\ {\isacharparenleft}{\kern0pt}sh\ {\isasymleadsto}\ State{\isasymllangle}{\isasymsigma}{\isasymrrangle}{\isacharcomma}{\kern0pt}\ {\isasymlambda}{\isacharbrackleft}{\kern0pt}WHILE\ b\ DO\ S\ OD{\isacharbrackright}{\kern0pt}{\isacharparenright}{\kern0pt}\ {\isacharequal}{\kern0pt}\ {\isacharbraceleft}{\kern0pt}{\isacharbraceright}{\kern0pt}{\isachardoublequoteclose}\isanewline
\isadelimproof
\ \ %
\endisadelimproof
\isatagproof
\isacommand{proof}\isamarkupfalse%
\ {\isacharminus}{\kern0pt}\isanewline
\ \ \ \ \isacommand{have}\isamarkupfalse%
\ {\isachardoublequoteopen}{\isasymdelta}\ {\isacharparenleft}{\kern0pt}sh\ {\isasymleadsto}\ State{\isasymllangle}{\isasymsigma}{\isasymrrangle}{\isacharcomma}{\kern0pt}\ {\isasymlambda}{\isacharbrackleft}{\kern0pt}WHILE\ b\ DO\ S\ OD{\isacharbrackright}{\kern0pt}{\isacharparenright}{\kern0pt}\ {\isacharequal}{\kern0pt}\ {\isacharbraceleft}{\kern0pt}{\isacharbraceright}{\kern0pt}{\isachardoublequoteclose}\ \isacommand{using}\isamarkupfalse%
\ assms\ {\isasymdelta}{\isacharunderscore}{\kern0pt}While\isactrlsub E\ \isacommand{by}\isamarkupfalse%
\ simp\isanewline
\ \ \ \ \isacommand{thus}\isamarkupfalse%
\ {\isacharquery}{\kern0pt}thesis\ \isacommand{using}\isamarkupfalse%
\ composition\isactrlsub N{\isachardot}{\kern0pt}simps\ \isacommand{by}\isamarkupfalse%
\ fastforce\isanewline
\ \ \isacommand{qed}\isamarkupfalse%
\endisatagproof
{\isafoldproof}%
\isadelimproof
\isanewline
\endisadelimproof
\isanewline
\ \ \isacommand{lemma}\isamarkupfalse%
\ {\isasymDelta}\isactrlsub N{\isacharunderscore}{\kern0pt}Seq{\isacharcolon}{\kern0pt}\isanewline
\ \ \ \ \isakeyword{shows}\ {\isachardoublequoteopen}{\isasymDelta}\isactrlsub N\ {\isacharparenleft}{\kern0pt}Suc\ n{\isacharparenright}{\kern0pt}\ {\isacharparenleft}{\kern0pt}sh\ {\isasymleadsto}\ State{\isasymllangle}{\isasymsigma}{\isasymrrangle}{\isacharcomma}{\kern0pt}\ {\isasymlambda}{\isacharbrackleft}{\kern0pt}S\isactrlsub {\isadigit{1}}{\isacharsemicolon}{\kern0pt}{\isacharsemicolon}{\kern0pt}S\isactrlsub {\isadigit{2}}{\isacharbrackright}{\kern0pt}{\isacharparenright}{\kern0pt}\ {\isacharequal}{\kern0pt}\ {\isasymUnion}{\isacharparenleft}{\kern0pt}{\isacharparenleft}{\kern0pt}{\isacharpercent}{\kern0pt}c{\isachardot}{\kern0pt}\ {\isasymDelta}\isactrlsub N\ n\ c{\isacharparenright}{\kern0pt}\ {\isacharbackquote}{\kern0pt}\ {\isasymdelta}\ {\isacharparenleft}{\kern0pt}sh\ {\isasymleadsto}\ State{\isasymllangle}{\isasymsigma}{\isasymrrangle}{\isacharcomma}{\kern0pt}\ {\isasymlambda}{\isacharbrackleft}{\kern0pt}S\isactrlsub {\isadigit{1}}{\isacharsemicolon}{\kern0pt}{\isacharsemicolon}{\kern0pt}S\isactrlsub {\isadigit{2}}{\isacharbrackright}{\kern0pt}{\isacharparenright}{\kern0pt}{\isacharparenright}{\kern0pt}{\isachardoublequoteclose}\isanewline
\isadelimproof
\ \ \ \ %
\endisadelimproof
\isatagproof
\isacommand{by}\isamarkupfalse%
\ simp%
\endisatagproof
{\isafoldproof}%
\isadelimproof
\endisadelimproof
\begin{isamarkuptext}%
Subsequently, we also provide simplification lemmas for the \isa{{\isasymDelta}}-function. The
  first lemma establishes that the result of the \isa{{\isasymDelta}}-function matches the result of 
  the n-bounded \isa{{\isasymDelta}\isactrlsub N}-function iff the program always terminates in at most n 
  evaluation steps, implying that the \isa{{\isasymDelta}}-function has reached a corresponding 
  fixpoint at step amount n. The second lemma infers that the \isa{{\isasymDelta}}-function executes 
  a recursive call with an increased step amount iff the fixpoint has not yet 
  been~reached.%
\end{isamarkuptext}\isamarkuptrue%
\ \ \isacommand{lemma}\isamarkupfalse%
\ {\isasymDelta}{\isacharunderscore}{\kern0pt}fixpoint{\isacharunderscore}{\kern0pt}reached{\isacharcolon}{\kern0pt}\isanewline
\ \ \ \ \isakeyword{assumes}\ {\isachardoublequoteopen}{\isasymforall}c\ {\isasymin}\ {\isacharparenleft}{\kern0pt}{\isasymDelta}\isactrlsub N\ n\ c{\isacharparenright}{\kern0pt}{\isachardot}{\kern0pt}\ snd{\isacharparenleft}{\kern0pt}c{\isacharparenright}{\kern0pt}\ {\isacharequal}{\kern0pt}\ {\isasymlambda}{\isacharbrackleft}{\kern0pt}{\isasymnabla}{\isacharbrackright}{\kern0pt}{\isachardoublequoteclose}\isanewline
\ \ \ \ \isakeyword{shows}\ {\isachardoublequoteopen}{\isasymDelta}\ n\ c\ {\isacharequal}{\kern0pt}\ {\isasymDelta}\isactrlsub N\ n\ c{\isachardoublequoteclose}\isanewline
\isadelimproof
\ \ \ \ %
\endisadelimproof
\isatagproof
\isacommand{using}\isamarkupfalse%
\ composition{\isachardot}{\kern0pt}simps\ \isacommand{by}\isamarkupfalse%
\ {\isacharparenleft}{\kern0pt}simp\ add{\isacharcolon}{\kern0pt}\ assms{\isacharparenright}{\kern0pt}%
\endisatagproof
{\isafoldproof}%
\isadelimproof
\isanewline
\endisadelimproof
\isanewline
\ \ \isacommand{lemma}\isamarkupfalse%
\ {\isasymDelta}{\isacharunderscore}{\kern0pt}iteration{\isacharcolon}{\kern0pt}\isanewline
\ \ \ \ \isakeyword{assumes}\ {\isachardoublequoteopen}{\isasymnot}{\isacharparenleft}{\kern0pt}{\isasymforall}c\ {\isasymin}\ {\isacharparenleft}{\kern0pt}{\isasymDelta}\isactrlsub N\ n\ c{\isacharparenright}{\kern0pt}{\isachardot}{\kern0pt}\ snd{\isacharparenleft}{\kern0pt}c{\isacharparenright}{\kern0pt}\ {\isacharequal}{\kern0pt}\ {\isasymlambda}{\isacharbrackleft}{\kern0pt}{\isasymnabla}{\isacharbrackright}{\kern0pt}{\isacharparenright}{\kern0pt}{\isachardoublequoteclose}\isanewline
\ \ \ \ \isakeyword{shows}\ {\isachardoublequoteopen}{\isasymDelta}\ n\ c\ {\isacharequal}{\kern0pt}\ {\isasymDelta}\ {\isacharparenleft}{\kern0pt}n\ {\isacharplus}{\kern0pt}\ {\isadigit{1}}{\isadigit{0}}{\isadigit{0}}{\isacharparenright}{\kern0pt}\ c{\isachardoublequoteclose}\isanewline
\isadelimproof
\ \ \ \ %
\endisadelimproof
\isatagproof
\isacommand{using}\isamarkupfalse%
\ composition{\isachardot}{\kern0pt}simps\ \isacommand{by}\isamarkupfalse%
\ {\isacharparenleft}{\kern0pt}simp\ add{\isacharcolon}{\kern0pt}\ assms{\isacharparenright}{\kern0pt}%
\endisatagproof
{\isafoldproof}%
\isadelimproof
\endisadelimproof
\begin{isamarkuptext}%
In order to finalize the automated proof system of the 
  \isa{{\isasymDelta}\isactrlsub N}-function/\isa{{\isasymDelta}}-function, we collect all previously proven lemmas in a set, 
  and call it the \isa{{\isasymDelta}\isactrlsub N}-system/\isa{{\isasymDelta}}-system. We additionally add the
  \isa{fmupd{\isacharminus}{\kern0pt}reorder{\isacharminus}{\kern0pt}neq} simplification to our \isa{{\isasymDelta}\isactrlsub N}-system, thereby ensuring that
  Isabelle can switch the order of updates when proving the equality of
  finite maps. The \isa{eval{\isacharminus}{\kern0pt}nat{\isacharminus}{\kern0pt}numeral} lemma is also included, such that 
  the simplifier can freely swap between the Suc/Zero notation and the numeral 
  representation when inferring results from the \isa{{\isasymDelta}\isactrlsub N}-function. This is necessary, 
  as we use the \isa{Suc} constructor in the function definition, but numerals when
  decrementing~the~bound. \par
  Note that we additionally remove the normal function simplifications of the
  \isa{{\isasymDelta}\isactrlsub N}-function, such that the Isabelle simplifier will later select our \isa{{\isasymDelta}\isactrlsub N}-system
  when deriving global traces. An explicitly defined \isa{partial{\isacharminus}{\kern0pt}function} does not 
  automatically add its simplifications to the simplifier, hence there are no 
  equations to remove for~the~\isa{{\isasymDelta}}-function.%
\end{isamarkuptext}\isamarkuptrue%
\ \ \isacommand{lemmas}\isamarkupfalse%
\ {\isasymDelta}\isactrlsub N{\isacharunderscore}{\kern0pt}system\ {\isacharequal}{\kern0pt}\ \isanewline
\ \ \ \ {\isasymDelta}\isactrlsub N{\isacharunderscore}{\kern0pt}Bounded\ {\isasymDelta}\isactrlsub N{\isacharunderscore}{\kern0pt}End\ {\isasymDelta}\isactrlsub N{\isacharunderscore}{\kern0pt}Skip\ {\isasymDelta}\isactrlsub N{\isacharunderscore}{\kern0pt}Assign\ {\isasymDelta}\isactrlsub N{\isacharunderscore}{\kern0pt}If\isactrlsub T\ {\isasymDelta}\isactrlsub N{\isacharunderscore}{\kern0pt}If\isactrlsub F\ {\isasymDelta}\isactrlsub N{\isacharunderscore}{\kern0pt}If\isactrlsub E\ \isanewline
\ \ \ \ {\isasymDelta}\isactrlsub N{\isacharunderscore}{\kern0pt}While\isactrlsub T\ {\isasymDelta}\isactrlsub N{\isacharunderscore}{\kern0pt}While\isactrlsub F\ {\isasymDelta}\isactrlsub N{\isacharunderscore}{\kern0pt}While\isactrlsub E\ {\isasymDelta}\isactrlsub N{\isacharunderscore}{\kern0pt}Seq\ fmupd{\isacharunderscore}{\kern0pt}reorder{\isacharunderscore}{\kern0pt}neq\ eval{\isacharunderscore}{\kern0pt}nat{\isacharunderscore}{\kern0pt}numeral\ \isanewline
\isanewline
\ \ \isacommand{lemmas}\isamarkupfalse%
\ {\isasymDelta}{\isacharunderscore}{\kern0pt}system\ {\isacharequal}{\kern0pt}\ \isanewline
\ \ \ \ {\isasymDelta}{\isacharunderscore}{\kern0pt}fixpoint{\isacharunderscore}{\kern0pt}reached\ {\isasymDelta}{\isacharunderscore}{\kern0pt}iteration\isanewline
\isanewline
\ \ \isacommand{declare}\isamarkupfalse%
\ composition\isactrlsub N{\isachardot}{\kern0pt}simps{\isacharbrackleft}{\kern0pt}simp\ del{\isacharbrackright}{\kern0pt}%
\begin{isamarkuptext}%
We now collect all our proof systems in one set, and name it the 
  WL-derivation-system. Note that we will later be able to use this system for all 
  global trace derivations~in~WL.%
\end{isamarkuptext}\isamarkuptrue%
\ \ \isacommand{lemmas}\isamarkupfalse%
\ WL{\isacharunderscore}{\kern0pt}derivation{\isacharunderscore}{\kern0pt}system\ {\isacharequal}{\kern0pt}\ \isanewline
\ \ \ \ {\isasymdelta}{\isacharunderscore}{\kern0pt}system\ {\isasymDelta}\isactrlsub N{\isacharunderscore}{\kern0pt}system\ {\isasymDelta}{\isacharunderscore}{\kern0pt}system%
\isadelimdocument
\endisadelimdocument
\isatagdocument
\isamarkupsubsubsection{Trace Derivation Examples%
}
\isamarkuptrue%
\endisatagdocument
{\isafolddocument}%
\isadelimdocument
\endisadelimdocument
\begin{isamarkuptext}%
We can now use our automated proof system to derive global traces for several
  example programs. Note that the trace derivation speed directly depends on the length 
  of the corresponding program evaluation. The derivations are already performant enough 
  for small-scale programs. However, longer programs (e.g. for case studies) may need 
  additional optimizations, which we propose as an idea for further extensions 
  of~the~model. \par
  Program \isa{WL{\isacharunderscore}{\kern0pt}ex\isactrlsub {\isadigit{1}}} corresponds to an If-Branch that switches the contents
  of program variables \isa{x} and \isa{y}. The implementation works as follows:
  If the values of \isa{x} and \isa{y} are different, the conditional statement enters the 
  then-case, which switches the contents of both variables by using an intermediate 
  variable \isa{z}. However, if the value of \isa{x} and \isa{y} already match at the start of the 
  program, no switch needs to occur, thus causing the program to 
  immediately~terminate. \par 
  Note that our program evaluation starts in the initial state induced by \isa{WL{\isacharunderscore}{\kern0pt}ex\isactrlsub {\isadigit{1}}}. This
  implies that all occurring variables are initialized with 0, indicating that the 
  contents of \isa{x} and \isa{y} trivially match. Hence, the program terminates after the 
  evaluation of the Boolean guard, thus resulting in exactly one global trace containing 
  only the initial state \isa{{\isacharparenleft}{\kern0pt}{\isasymsigma}\isactrlsub I\ WL{\isacharunderscore}{\kern0pt}ex\isactrlsub {\isadigit{1}}{\isacharparenright}{\kern0pt}}. We infer this conclusion in~Isabelle.%
\end{isamarkuptext}\isamarkuptrue%
\ \ \isacommand{lemma}\isamarkupfalse%
\ {\isachardoublequoteopen}Tr\ WL{\isacharunderscore}{\kern0pt}ex\isactrlsub {\isadigit{1}}\ {\isacharparenleft}{\kern0pt}{\isasymsigma}\isactrlsub I\ WL{\isacharunderscore}{\kern0pt}ex\isactrlsub {\isadigit{1}}{\isacharparenright}{\kern0pt}\ \isanewline
\ \ \ \ \ \ \ \ {\isacharequal}{\kern0pt}\ {\isacharbraceleft}{\kern0pt}{\isasymlangle}fm\ {\isacharbrackleft}{\kern0pt}{\isacharparenleft}{\kern0pt}{\isacharprime}{\kern0pt}{\isacharprime}{\kern0pt}y{\isacharprime}{\kern0pt}{\isacharprime}{\kern0pt}{\isacharcomma}{\kern0pt}\ Exp\ {\isacharparenleft}{\kern0pt}Num\ {\isadigit{0}}{\isacharparenright}{\kern0pt}{\isacharparenright}{\kern0pt}{\isacharcomma}{\kern0pt}\ {\isacharparenleft}{\kern0pt}{\isacharprime}{\kern0pt}{\isacharprime}{\kern0pt}x{\isacharprime}{\kern0pt}{\isacharprime}{\kern0pt}{\isacharcomma}{\kern0pt}\ Exp\ {\isacharparenleft}{\kern0pt}Num\ {\isadigit{0}}{\isacharparenright}{\kern0pt}{\isacharparenright}{\kern0pt}{\isacharcomma}{\kern0pt}\ {\isacharparenleft}{\kern0pt}{\isacharprime}{\kern0pt}{\isacharprime}{\kern0pt}z{\isacharprime}{\kern0pt}{\isacharprime}{\kern0pt}{\isacharcomma}{\kern0pt}\ Exp\ {\isacharparenleft}{\kern0pt}Num\ {\isadigit{0}}{\isacharparenright}{\kern0pt}{\isacharparenright}{\kern0pt}{\isacharbrackright}{\kern0pt}{\isasymrangle}{\isacharbraceright}{\kern0pt}{\isachardoublequoteclose}\isanewline
\isadelimproof
\ \ \ \ %
\endisadelimproof
\isatagproof
\isacommand{by}\isamarkupfalse%
\ {\isacharparenleft}{\kern0pt}simp\ add{\isacharcolon}{\kern0pt}\ WL{\isacharunderscore}{\kern0pt}ex\isactrlsub {\isadigit{1}}{\isacharunderscore}{\kern0pt}def\ Traces{\isacharunderscore}{\kern0pt}def\ WL{\isacharunderscore}{\kern0pt}derivation{\isacharunderscore}{\kern0pt}system{\isacharparenright}{\kern0pt}%
\endisatagproof
{\isafoldproof}%
\isadelimproof
\endisadelimproof
\begin{isamarkuptext}%
Program \isa{WL{\isacharunderscore}{\kern0pt}ex\isactrlsub {\isadigit{2}}} computes the factorial of 6 using a While-Loop. The implementation
  works as follows: In the beginning, variable \isa{x} is assigned to 6, whilst variable 
  \isa{y} is assigned to 1. We then utilize a While-Loop in order to faithfully capture
  the intended factorial computation. In every iteration of the While-Loop, \isa{y} is updated 
  with the result of the multiplication \isa{x}~*~\isa{y}, and variable \isa{x} is decremented. 
  When the value stored in \isa{x} drops below 2, the Boolean guard of the While-Loop
  evaluates to false, thus terminating the~program. \par
  Note that this exact program behaviour can also be observed in the singular global 
  trace inferred below. The value of variable \isa{y} matches 720 in the final state
  of the trace, thereby directly corresponding to the desired evaluation of \isa{{\isadigit{6}}{\isacharbang}{\kern0pt}}. 
  Due to the determinism of the program, only one global trace can exist.%
\end{isamarkuptext}\isamarkuptrue%
\ \ \isacommand{lemma}\isamarkupfalse%
\ {\isachardoublequoteopen}Tr\ WL{\isacharunderscore}{\kern0pt}ex\isactrlsub {\isadigit{2}}\ {\isacharparenleft}{\kern0pt}{\isasymsigma}\isactrlsub I\ WL{\isacharunderscore}{\kern0pt}ex\isactrlsub {\isadigit{2}}{\isacharparenright}{\kern0pt}\isanewline
\ \ \ \ \ \ \ \ {\isacharequal}{\kern0pt}\ {\isacharbraceleft}{\kern0pt}{\isacharparenleft}{\kern0pt}{\isacharparenleft}{\kern0pt}{\isacharparenleft}{\kern0pt}{\isacharparenleft}{\kern0pt}{\isacharparenleft}{\kern0pt}{\isacharparenleft}{\kern0pt}{\isacharparenleft}{\kern0pt}{\isacharparenleft}{\kern0pt}{\isacharparenleft}{\kern0pt}{\isacharparenleft}{\kern0pt}{\isacharparenleft}{\kern0pt}{\isasymlangle}fm\ {\isacharbrackleft}{\kern0pt}{\isacharparenleft}{\kern0pt}{\isacharprime}{\kern0pt}{\isacharprime}{\kern0pt}y{\isacharprime}{\kern0pt}{\isacharprime}{\kern0pt}{\isacharcomma}{\kern0pt}\ Exp\ {\isacharparenleft}{\kern0pt}Num\ {\isadigit{0}}{\isacharparenright}{\kern0pt}{\isacharparenright}{\kern0pt}{\isacharcomma}{\kern0pt}\ {\isacharparenleft}{\kern0pt}{\isacharprime}{\kern0pt}{\isacharprime}{\kern0pt}x{\isacharprime}{\kern0pt}{\isacharprime}{\kern0pt}{\isacharcomma}{\kern0pt}\ Exp\ {\isacharparenleft}{\kern0pt}Num\ {\isadigit{0}}{\isacharparenright}{\kern0pt}{\isacharparenright}{\kern0pt}{\isacharbrackright}{\kern0pt}{\isasymrangle}\ {\isasymleadsto}\isanewline
\ \ \ \ \ \ \ \ \ \ \ \ \ \ \ \ State\ {\isasymllangle}fm\ {\isacharbrackleft}{\kern0pt}{\isacharparenleft}{\kern0pt}{\isacharprime}{\kern0pt}{\isacharprime}{\kern0pt}y{\isacharprime}{\kern0pt}{\isacharprime}{\kern0pt}{\isacharcomma}{\kern0pt}\ Exp\ {\isacharparenleft}{\kern0pt}Num\ {\isadigit{0}}{\isacharparenright}{\kern0pt}{\isacharparenright}{\kern0pt}{\isacharcomma}{\kern0pt}\ {\isacharparenleft}{\kern0pt}{\isacharprime}{\kern0pt}{\isacharprime}{\kern0pt}x{\isacharprime}{\kern0pt}{\isacharprime}{\kern0pt}{\isacharcomma}{\kern0pt}\ Exp\ {\isacharparenleft}{\kern0pt}Num\ {\isadigit{6}}{\isacharparenright}{\kern0pt}{\isacharparenright}{\kern0pt}{\isacharbrackright}{\kern0pt}{\isasymrrangle}{\isacharparenright}{\kern0pt}\ {\isasymleadsto}\isanewline
\ \ \ \ \ \ \ \ \ \ \ \ \ \ \ \ State\ {\isasymllangle}fm\ {\isacharbrackleft}{\kern0pt}{\isacharparenleft}{\kern0pt}{\isacharprime}{\kern0pt}{\isacharprime}{\kern0pt}y{\isacharprime}{\kern0pt}{\isacharprime}{\kern0pt}{\isacharcomma}{\kern0pt}\ Exp\ {\isacharparenleft}{\kern0pt}Num\ {\isadigit{1}}{\isacharparenright}{\kern0pt}{\isacharparenright}{\kern0pt}{\isacharcomma}{\kern0pt}\ {\isacharparenleft}{\kern0pt}{\isacharprime}{\kern0pt}{\isacharprime}{\kern0pt}x{\isacharprime}{\kern0pt}{\isacharprime}{\kern0pt}{\isacharcomma}{\kern0pt}\ Exp\ {\isacharparenleft}{\kern0pt}Num\ {\isadigit{6}}{\isacharparenright}{\kern0pt}{\isacharparenright}{\kern0pt}{\isacharbrackright}{\kern0pt}{\isasymrrangle}{\isacharparenright}{\kern0pt}\ {\isasymleadsto}\isanewline
\ \ \ \ \ \ \ \ \ \ \ \ \ \ State\ {\isasymllangle}fm\ {\isacharbrackleft}{\kern0pt}{\isacharparenleft}{\kern0pt}{\isacharprime}{\kern0pt}{\isacharprime}{\kern0pt}y{\isacharprime}{\kern0pt}{\isacharprime}{\kern0pt}{\isacharcomma}{\kern0pt}\ Exp\ {\isacharparenleft}{\kern0pt}Num\ {\isadigit{6}}{\isacharparenright}{\kern0pt}{\isacharparenright}{\kern0pt}{\isacharcomma}{\kern0pt}\ {\isacharparenleft}{\kern0pt}{\isacharprime}{\kern0pt}{\isacharprime}{\kern0pt}x{\isacharprime}{\kern0pt}{\isacharprime}{\kern0pt}{\isacharcomma}{\kern0pt}\ Exp\ {\isacharparenleft}{\kern0pt}Num\ {\isadigit{6}}{\isacharparenright}{\kern0pt}{\isacharparenright}{\kern0pt}{\isacharbrackright}{\kern0pt}{\isasymrrangle}{\isacharparenright}{\kern0pt}\ {\isasymleadsto}\isanewline
\ \ \ \ \ \ \ \ \ \ \ \ \ \ State\ {\isasymllangle}fm\ {\isacharbrackleft}{\kern0pt}{\isacharparenleft}{\kern0pt}{\isacharprime}{\kern0pt}{\isacharprime}{\kern0pt}y{\isacharprime}{\kern0pt}{\isacharprime}{\kern0pt}{\isacharcomma}{\kern0pt}\ Exp\ {\isacharparenleft}{\kern0pt}Num\ {\isadigit{6}}{\isacharparenright}{\kern0pt}{\isacharparenright}{\kern0pt}{\isacharcomma}{\kern0pt}\ {\isacharparenleft}{\kern0pt}{\isacharprime}{\kern0pt}{\isacharprime}{\kern0pt}x{\isacharprime}{\kern0pt}{\isacharprime}{\kern0pt}{\isacharcomma}{\kern0pt}\ Exp\ {\isacharparenleft}{\kern0pt}Num\ {\isadigit{5}}{\isacharparenright}{\kern0pt}{\isacharparenright}{\kern0pt}{\isacharbrackright}{\kern0pt}{\isasymrrangle}{\isacharparenright}{\kern0pt}\ {\isasymleadsto}\isanewline
\ \ \ \ \ \ \ \ \ \ \ \ State\ {\isasymllangle}fm\ {\isacharbrackleft}{\kern0pt}{\isacharparenleft}{\kern0pt}{\isacharprime}{\kern0pt}{\isacharprime}{\kern0pt}y{\isacharprime}{\kern0pt}{\isacharprime}{\kern0pt}{\isacharcomma}{\kern0pt}\ Exp\ {\isacharparenleft}{\kern0pt}Num\ {\isadigit{3}}{\isadigit{0}}{\isacharparenright}{\kern0pt}{\isacharparenright}{\kern0pt}{\isacharcomma}{\kern0pt}\ {\isacharparenleft}{\kern0pt}{\isacharprime}{\kern0pt}{\isacharprime}{\kern0pt}x{\isacharprime}{\kern0pt}{\isacharprime}{\kern0pt}{\isacharcomma}{\kern0pt}\ Exp\ {\isacharparenleft}{\kern0pt}Num\ {\isadigit{5}}{\isacharparenright}{\kern0pt}{\isacharparenright}{\kern0pt}{\isacharbrackright}{\kern0pt}{\isasymrrangle}{\isacharparenright}{\kern0pt}\ {\isasymleadsto}\isanewline
\ \ \ \ \ \ \ \ \ \ \ \ State\ {\isasymllangle}fm\ {\isacharbrackleft}{\kern0pt}{\isacharparenleft}{\kern0pt}{\isacharprime}{\kern0pt}{\isacharprime}{\kern0pt}y{\isacharprime}{\kern0pt}{\isacharprime}{\kern0pt}{\isacharcomma}{\kern0pt}\ Exp\ {\isacharparenleft}{\kern0pt}Num\ {\isadigit{3}}{\isadigit{0}}{\isacharparenright}{\kern0pt}{\isacharparenright}{\kern0pt}{\isacharcomma}{\kern0pt}\ {\isacharparenleft}{\kern0pt}{\isacharprime}{\kern0pt}{\isacharprime}{\kern0pt}x{\isacharprime}{\kern0pt}{\isacharprime}{\kern0pt}{\isacharcomma}{\kern0pt}\ Exp\ {\isacharparenleft}{\kern0pt}Num\ {\isadigit{4}}{\isacharparenright}{\kern0pt}{\isacharparenright}{\kern0pt}{\isacharbrackright}{\kern0pt}{\isasymrrangle}{\isacharparenright}{\kern0pt}\ {\isasymleadsto}\isanewline
\ \ \ \ \ \ \ \ \ \ State\ {\isasymllangle}fm\ {\isacharbrackleft}{\kern0pt}{\isacharparenleft}{\kern0pt}{\isacharprime}{\kern0pt}{\isacharprime}{\kern0pt}y{\isacharprime}{\kern0pt}{\isacharprime}{\kern0pt}{\isacharcomma}{\kern0pt}\ Exp\ {\isacharparenleft}{\kern0pt}Num\ {\isadigit{1}}{\isadigit{2}}{\isadigit{0}}{\isacharparenright}{\kern0pt}{\isacharparenright}{\kern0pt}{\isacharcomma}{\kern0pt}\ {\isacharparenleft}{\kern0pt}{\isacharprime}{\kern0pt}{\isacharprime}{\kern0pt}x{\isacharprime}{\kern0pt}{\isacharprime}{\kern0pt}{\isacharcomma}{\kern0pt}\ Exp\ {\isacharparenleft}{\kern0pt}Num\ {\isadigit{4}}{\isacharparenright}{\kern0pt}{\isacharparenright}{\kern0pt}{\isacharbrackright}{\kern0pt}{\isasymrrangle}{\isacharparenright}{\kern0pt}\ {\isasymleadsto}\isanewline
\ \ \ \ \ \ \ \ \ \ State\ {\isasymllangle}fm\ {\isacharbrackleft}{\kern0pt}{\isacharparenleft}{\kern0pt}{\isacharprime}{\kern0pt}{\isacharprime}{\kern0pt}y{\isacharprime}{\kern0pt}{\isacharprime}{\kern0pt}{\isacharcomma}{\kern0pt}\ Exp\ {\isacharparenleft}{\kern0pt}Num\ {\isadigit{1}}{\isadigit{2}}{\isadigit{0}}{\isacharparenright}{\kern0pt}{\isacharparenright}{\kern0pt}{\isacharcomma}{\kern0pt}\ {\isacharparenleft}{\kern0pt}{\isacharprime}{\kern0pt}{\isacharprime}{\kern0pt}x{\isacharprime}{\kern0pt}{\isacharprime}{\kern0pt}{\isacharcomma}{\kern0pt}\ Exp\ {\isacharparenleft}{\kern0pt}Num\ {\isadigit{3}}{\isacharparenright}{\kern0pt}{\isacharparenright}{\kern0pt}{\isacharbrackright}{\kern0pt}{\isasymrrangle}{\isacharparenright}{\kern0pt}\ {\isasymleadsto}\isanewline
\ \ \ \ \ \ \ \ State\ {\isasymllangle}fm\ {\isacharbrackleft}{\kern0pt}{\isacharparenleft}{\kern0pt}{\isacharprime}{\kern0pt}{\isacharprime}{\kern0pt}y{\isacharprime}{\kern0pt}{\isacharprime}{\kern0pt}{\isacharcomma}{\kern0pt}\ Exp\ {\isacharparenleft}{\kern0pt}Num\ {\isadigit{3}}{\isadigit{6}}{\isadigit{0}}{\isacharparenright}{\kern0pt}{\isacharparenright}{\kern0pt}{\isacharcomma}{\kern0pt}\ {\isacharparenleft}{\kern0pt}{\isacharprime}{\kern0pt}{\isacharprime}{\kern0pt}x{\isacharprime}{\kern0pt}{\isacharprime}{\kern0pt}{\isacharcomma}{\kern0pt}\ Exp\ {\isacharparenleft}{\kern0pt}Num\ {\isadigit{3}}{\isacharparenright}{\kern0pt}{\isacharparenright}{\kern0pt}{\isacharbrackright}{\kern0pt}{\isasymrrangle}{\isacharparenright}{\kern0pt}\ {\isasymleadsto}\isanewline
\ \ \ \ \ \ \ \ State\ {\isasymllangle}fm\ {\isacharbrackleft}{\kern0pt}{\isacharparenleft}{\kern0pt}{\isacharprime}{\kern0pt}{\isacharprime}{\kern0pt}y{\isacharprime}{\kern0pt}{\isacharprime}{\kern0pt}{\isacharcomma}{\kern0pt}\ Exp\ {\isacharparenleft}{\kern0pt}Num\ {\isadigit{3}}{\isadigit{6}}{\isadigit{0}}{\isacharparenright}{\kern0pt}{\isacharparenright}{\kern0pt}{\isacharcomma}{\kern0pt}\ {\isacharparenleft}{\kern0pt}{\isacharprime}{\kern0pt}{\isacharprime}{\kern0pt}x{\isacharprime}{\kern0pt}{\isacharprime}{\kern0pt}{\isacharcomma}{\kern0pt}\ Exp\ {\isacharparenleft}{\kern0pt}Num\ {\isadigit{2}}{\isacharparenright}{\kern0pt}{\isacharparenright}{\kern0pt}{\isacharbrackright}{\kern0pt}{\isasymrrangle}{\isacharparenright}{\kern0pt}\ {\isasymleadsto}\isanewline
\ \ \ \ \ \ State\ {\isasymllangle}fm\ {\isacharbrackleft}{\kern0pt}{\isacharparenleft}{\kern0pt}{\isacharprime}{\kern0pt}{\isacharprime}{\kern0pt}y{\isacharprime}{\kern0pt}{\isacharprime}{\kern0pt}{\isacharcomma}{\kern0pt}\ Exp\ {\isacharparenleft}{\kern0pt}Num\ {\isadigit{7}}{\isadigit{2}}{\isadigit{0}}{\isacharparenright}{\kern0pt}{\isacharparenright}{\kern0pt}{\isacharcomma}{\kern0pt}\ {\isacharparenleft}{\kern0pt}{\isacharprime}{\kern0pt}{\isacharprime}{\kern0pt}x{\isacharprime}{\kern0pt}{\isacharprime}{\kern0pt}{\isacharcomma}{\kern0pt}\ Exp\ {\isacharparenleft}{\kern0pt}Num\ {\isadigit{2}}{\isacharparenright}{\kern0pt}{\isacharparenright}{\kern0pt}{\isacharbrackright}{\kern0pt}{\isasymrrangle}{\isacharparenright}{\kern0pt}\ {\isasymleadsto}\isanewline
\ \ \ \ \ \ State\ {\isasymllangle}fm\ {\isacharbrackleft}{\kern0pt}{\isacharparenleft}{\kern0pt}{\isacharprime}{\kern0pt}{\isacharprime}{\kern0pt}y{\isacharprime}{\kern0pt}{\isacharprime}{\kern0pt}{\isacharcomma}{\kern0pt}\ Exp\ {\isacharparenleft}{\kern0pt}Num\ {\isadigit{7}}{\isadigit{2}}{\isadigit{0}}{\isacharparenright}{\kern0pt}{\isacharparenright}{\kern0pt}{\isacharcomma}{\kern0pt}\ {\isacharparenleft}{\kern0pt}{\isacharprime}{\kern0pt}{\isacharprime}{\kern0pt}x{\isacharprime}{\kern0pt}{\isacharprime}{\kern0pt}{\isacharcomma}{\kern0pt}\ Exp\ {\isacharparenleft}{\kern0pt}Num\ {\isadigit{1}}{\isacharparenright}{\kern0pt}{\isacharparenright}{\kern0pt}{\isacharbrackright}{\kern0pt}{\isasymrrangle}{\isacharbraceright}{\kern0pt}{\isachardoublequoteclose}\isanewline
\isadelimproof
\ \ \ \ %
\endisadelimproof
\isatagproof
\isacommand{by}\isamarkupfalse%
\ {\isacharparenleft}{\kern0pt}simp\ add{\isacharcolon}{\kern0pt}\ WL{\isacharunderscore}{\kern0pt}ex\isactrlsub {\isadigit{2}}{\isacharunderscore}{\kern0pt}def\ Traces{\isacharunderscore}{\kern0pt}def\ WL{\isacharunderscore}{\kern0pt}derivation{\isacharunderscore}{\kern0pt}system{\isacharparenright}{\kern0pt}%
\endisatagproof
{\isafoldproof}%
\isadelimproof
\endisadelimproof
\isadelimdocument
\endisadelimdocument
\isatagdocument
\isamarkupsubsubsection{Code Generation%
}
\isamarkuptrue%
\endisatagdocument
{\isafolddocument}%
\isadelimdocument
\endisadelimdocument
\begin{isamarkuptext}%
The idea of making all function definitions executable finally pays off, 
  as we can now generate code for the construction of global system traces by using 
  the \isa{value} keyword. Note that the code execution itself is very performant, 
  and can therefore be used to compute the global traces for any arbitrary program 
  in any arbitrary initial state. We additionally propose further work on exports of
  this code to several other programming languages supported by Isabelle (e.g. Haskell, 
  Scala) as an idea for extending the work of this~thesis.%
\end{isamarkuptext}\isamarkuptrue%
\ \ \isacommand{value}\isamarkupfalse%
\ {\isachardoublequoteopen}Tr\ WL{\isacharunderscore}{\kern0pt}ex\isactrlsub {\isadigit{1}}\ {\isacharparenleft}{\kern0pt}{\isasymsigma}\isactrlsub I\ WL{\isacharunderscore}{\kern0pt}ex\isactrlsub {\isadigit{1}}{\isacharparenright}{\kern0pt}{\isachardoublequoteclose}\isanewline
\ \ \isacommand{value}\isamarkupfalse%
\ {\isachardoublequoteopen}Tr\ WL{\isacharunderscore}{\kern0pt}ex\isactrlsub {\isadigit{2}}\ {\isacharparenleft}{\kern0pt}{\isasymsigma}\isactrlsub I\ WL{\isacharunderscore}{\kern0pt}ex\isactrlsub {\isadigit{2}}{\isacharparenright}{\kern0pt}{\isachardoublequoteclose}%
\isadelimdocument
\endisadelimdocument
\isatagdocument
\isamarkupsubsection{Trace Equivalence%
}
\isamarkuptrue%
\endisatagdocument
{\isafolddocument}%
\isadelimdocument
\endisadelimdocument
\begin{isamarkuptext}%
In contrast to the paper, we additionally propose a notion of equivalence 
  between programs. We call two programs \isa{S} and \isa{S{\isacharprime}{\kern0pt}} of \isa{WL} trace equivalent under a 
  given initial state \isa{{\isasymsigma}} iff \isa{S} and \isa{S{\isacharprime}{\kern0pt}} called in \isa{{\isasymsigma}} generate the exact same 
  set of global traces upon termination. This equivalence property can later be 
  utilized to quickly prove that two programs match in their 
  trace~behaviour. \par
  We formalize this equivalence notion using an inductive predicate. The
  inductive formalization in Isabelle is straightforward, as no step case 
  needs to be considered. Note that we introduce the notation \isa{{\isacharparenleft}{\kern0pt}S\ {\isasymsim}\ S{\isacharprime}{\kern0pt}{\isacharparenright}{\kern0pt}\ {\isacharbrackleft}{\kern0pt}{\isasymsigma}{\isacharbrackright}{\kern0pt}} to 
  denote the trace equivalence of \isa{S} and \isa{S{\isacharprime}{\kern0pt}} under initial~state~\isa{{\isasymsigma}}.%
\end{isamarkuptext}\isamarkuptrue%
\ \ \isacommand{inductive}\isamarkupfalse%
\isanewline
\ \ \ \ tequivalent\ {\isacharcolon}{\kern0pt}{\isacharcolon}{\kern0pt}\ {\isachardoublequoteopen}stmt\ {\isasymRightarrow}\ stmt\ {\isasymRightarrow}\ {\isasymSigma}\ {\isasymRightarrow}\ bool{\isachardoublequoteclose}\ {\isacharparenleft}{\kern0pt}{\isachardoublequoteopen}{\isacharunderscore}{\kern0pt}\ {\isasymsim}\ {\isacharunderscore}{\kern0pt}\ {\isacharbrackleft}{\kern0pt}{\isacharunderscore}{\kern0pt}{\isacharbrackright}{\kern0pt}{\isachardoublequoteclose}\ {\isadigit{8}}{\isadigit{0}}{\isacharparenright}{\kern0pt}\ \isakeyword{where}\isanewline
\ \ \ \ {\isachardoublequoteopen}{\isasymlbrakk}\ Tr\ S\ {\isasymsigma}\ {\isacharequal}{\kern0pt}\ Tr\ S{\isacharprime}{\kern0pt}\ {\isasymsigma}\ {\isasymrbrakk}\ {\isasymLongrightarrow}\ S\ {\isasymsim}\ S{\isacharprime}{\kern0pt}\ {\isacharbrackleft}{\kern0pt}{\isasymsigma}{\isacharbrackright}{\kern0pt}{\isachardoublequoteclose}%
\begin{isamarkuptext}%
We can now use the \isa{code{\isacharminus}{\kern0pt}pred} keyword in order to automatically
  generate code for the inductive definition above. This will later ensure
  that we can simply output the result of the inductive predicate in the
  console using the \isa{value}~keyword.%
\end{isamarkuptext}\isamarkuptrue%
\ \ \isacommand{code{\isacharunderscore}{\kern0pt}pred}\isamarkupfalse%
\ tequivalent%
\isadelimproof
\ %
\endisadelimproof
\isatagproof
\isacommand{{\isachardot}{\kern0pt}}\isamarkupfalse%
\endisatagproof
{\isafoldproof}%
\isadelimproof
\endisadelimproof
\begin{isamarkuptext}%
We furthermore automatically generate inductive simplifications for our trace
  equivalence notion using the \isa{inductive{\isacharminus}{\kern0pt}simps} keyword, thereby adding them to the 
  Isabelle simplifier equations. This guarantees that we can later resolve the
  trace equivalence notion \isa{{\isasymsim}}, allowing us to conduct proofs inferring trace
  equivalence.%
\end{isamarkuptext}\isamarkuptrue%
\ \ \isacommand{inductive{\isacharunderscore}{\kern0pt}simps}\isamarkupfalse%
\ tequivalence{\isacharcolon}{\kern0pt}\ {\isachardoublequoteopen}S\ {\isasymsim}\ S{\isacharprime}{\kern0pt}\ {\isacharbrackleft}{\kern0pt}{\isasymsigma}{\isacharbrackright}{\kern0pt}{\isachardoublequoteclose}%
\begin{isamarkuptext}%
Finally, we derive several trace equivalence conclusions in Isabelle using 
  the following practical~examples.%
\end{isamarkuptext}\isamarkuptrue%
\ \ \isacommand{lemma}\isamarkupfalse%
\ {\isachardoublequoteopen}SKIP\ {\isasymsim}\ {\isacharparenleft}{\kern0pt}SKIP{\isacharsemicolon}{\kern0pt}{\isacharsemicolon}{\kern0pt}SKIP{\isacharparenright}{\kern0pt}\ {\isacharbrackleft}{\kern0pt}{\isasymsigma}{\isacharbrackright}{\kern0pt}{\isachardoublequoteclose}\isanewline
\isadelimproof
\ \ \ \ %
\endisadelimproof
\isatagproof
\isacommand{using}\isamarkupfalse%
\ tequivalence\ \isacommand{by}\isamarkupfalse%
\ {\isacharparenleft}{\kern0pt}simp\ add{\isacharcolon}{\kern0pt}\ Traces{\isacharunderscore}{\kern0pt}def\ WL{\isacharunderscore}{\kern0pt}derivation{\isacharunderscore}{\kern0pt}system{\isacharparenright}{\kern0pt}%
\endisatagproof
{\isafoldproof}%
\isadelimproof
\isanewline
\endisadelimproof
\isanewline
\ \ \isacommand{lemma}\isamarkupfalse%
\ {\isachardoublequoteopen}{\isacharparenleft}{\kern0pt}IF\ {\isacharparenleft}{\kern0pt}{\isacharparenleft}{\kern0pt}Var\ {\isacharprime}{\kern0pt}{\isacharprime}{\kern0pt}x{\isacharprime}{\kern0pt}{\isacharprime}{\kern0pt}{\isacharparenright}{\kern0pt}\ \isactrlsub Req\ {\isacharparenleft}{\kern0pt}Num\ {\isadigit{1}}{\isacharparenright}{\kern0pt}{\isacharparenright}{\kern0pt}\ THEN\ {\isacharprime}{\kern0pt}{\isacharprime}{\kern0pt}x{\isacharprime}{\kern0pt}{\isacharprime}{\kern0pt}\ {\isacharcolon}{\kern0pt}{\isacharequal}{\kern0pt}\ Num\ {\isadigit{0}}\ FI{\isacharparenright}{\kern0pt}\ \isanewline
\ \ \ \ \ \ \ \ \ \ \ \ \ \ \ \ {\isasymsim}\ {\isacharparenleft}{\kern0pt}{\isacharprime}{\kern0pt}{\isacharprime}{\kern0pt}x{\isacharprime}{\kern0pt}{\isacharprime}{\kern0pt}\ {\isacharcolon}{\kern0pt}{\isacharequal}{\kern0pt}\ Num\ {\isadigit{0}}{\isacharparenright}{\kern0pt}\ {\isacharbrackleft}{\kern0pt}{\isacharbrackleft}{\kern0pt}{\isacharprime}{\kern0pt}{\isacharprime}{\kern0pt}x{\isacharprime}{\kern0pt}{\isacharprime}{\kern0pt}\ {\isasymlongmapsto}\ Exp\ {\isacharparenleft}{\kern0pt}Num\ {\isadigit{1}}{\isacharparenright}{\kern0pt}{\isacharbrackright}{\kern0pt}\ {\isasymcircle}{\isacharbrackright}{\kern0pt}{\isachardoublequoteclose}\isanewline
\isadelimproof
\ \ \ \ %
\endisadelimproof
\isatagproof
\isacommand{using}\isamarkupfalse%
\ tequivalence\ \isacommand{by}\isamarkupfalse%
\ {\isacharparenleft}{\kern0pt}simp\ add{\isacharcolon}{\kern0pt}\ Traces{\isacharunderscore}{\kern0pt}def\ WL{\isacharunderscore}{\kern0pt}derivation{\isacharunderscore}{\kern0pt}system{\isacharparenright}{\kern0pt}%
\endisatagproof
{\isafoldproof}%
\isadelimproof
\isanewline
\endisadelimproof
\isadelimtheory
\isanewline
\endisadelimtheory
\isatagtheory
\isacommand{end}\isamarkupfalse%
\endisatagtheory
{\isafoldtheory}%
\isadelimtheory
\endisadelimtheory
\end{isabellebody}%

%% file: LAGC_WL_Extended.tex
\begin{isabellebody}%
\setisabellecontext{LAGC{\isacharunderscore}{\kern0pt}WL{\isacharunderscore}{\kern0pt}Extended}%
\isadelimdocument
\endisadelimdocument
\isatagdocument
\isamarkupsection{LAGC Semantics for $WL_{EXT}$%
}
\isamarkuptrue%
\endisatagdocument
{\isafolddocument}%
\isadelimdocument
\endisadelimdocument
\isadelimtheory
\endisadelimtheory
\isatagtheory
\isacommand{theory}\isamarkupfalse%
\ LAGC{\isacharunderscore}{\kern0pt}WL{\isacharunderscore}{\kern0pt}Extended\isanewline
\ \ \isakeyword{imports}\ {\isachardoublequoteopen}{\isachardot}{\kern0pt}{\isachardot}{\kern0pt}{\isacharslash}{\kern0pt}basics{\isacharslash}{\kern0pt}LAGC{\isacharunderscore}{\kern0pt}Base{\isachardoublequoteclose}\ {\isachardoublequoteopen}{\isachardot}{\kern0pt}{\isachardot}{\kern0pt}{\isacharslash}{\kern0pt}supplementary{\isacharslash}{\kern0pt}LAGC{\isacharunderscore}{\kern0pt}Base{\isacharunderscore}{\kern0pt}Supplementary{\isachardoublequoteclose}\ {\isachardoublequoteopen}HOL{\isacharminus}{\kern0pt}Library{\isachardot}{\kern0pt}Multiset{\isachardoublequoteclose}\ \isanewline
\isakeyword{begin}%
\endisatagtheory
{\isafoldtheory}%
\isadelimtheory
\endisadelimtheory
\begin{isamarkuptext}%
In this chapter we formalize the LAGC semantics for the extended 
  While Language $WL_{EXT}$ as described in section~5 of the original paper.
  For this purpose, we faithfully expand on the LAGC semantics for \isa{WL} in 
  order to accommodate new programming language concepts (e.g. concurrency, scopes).
  We show that this adaption can take place without endangering the automatic
  generation of corresponding code in Isabelle. We furthermore enhance our
  proof automation system in order to preserve our ability to systematically
  derive global traces for any specific program. However, note that our
  derivation system will get slightly more complex due to the introduction
  of trace concretizations, which will be applied in every step of the
  trace composition.%
\end{isamarkuptext}\isamarkuptrue%
\isadelimdocument
\endisadelimdocument
\isatagdocument
\isamarkupsubsection{Extended While Language ($WL_{EXT}$)%
}
\isamarkuptrue%
\isamarkupsubsubsection{Syntax%
}
\isamarkuptrue%
\endisatagdocument
{\isafolddocument}%
\isadelimdocument
\endisadelimdocument
\begin{isamarkuptext}%
We begin by introducing the syntax of the extended While Language $WL_{EXT}$,
  which is an expansion of the standard While Language \isa{WL}. \par 
  However, before defining the statements of $WL_{EXT}$, we first introduce a new 
  datatype that models the declaration of fresh variables upon opening new scopes.
  These variables can then only be used inside the newly opened scope of the 
  corresponding program, implying that they are only alive when the scope is open. We 
  decide to model these variable declarations as finite sequences of variables, which 
  are separated by semicolons. Note that \isa{{\isasymnu}} represents the empty variable declaration, 
  referring to the case in which no variable is declared upon opening a new~scope.%
\end{isamarkuptext}\isamarkuptrue%
\ \ \isacommand{datatype}\isamarkupfalse%
\ varDecl\ {\isacharequal}{\kern0pt}\isanewline
\ \ \ \ \ \ \ \ Nu\ {\isacharparenleft}{\kern0pt}{\isachardoublequoteopen}{\isasymnu}{\isachardoublequoteclose}{\isacharparenright}{\kern0pt}\ \ %
\isamarkupcmt{Empty Declaration%
}\isanewline
\ \ \ \ \ \ {\isacharbar}{\kern0pt}\ Declaration\ var\ varDecl\ {\isacharparenleft}{\kern0pt}\isakeyword{infix}\ {\isachardoublequoteopen}{\isacharsemicolon}{\kern0pt}{\isachardoublequoteclose}\ {\isadigit{5}}{\isadigit{8}}{\isacharparenright}{\kern0pt}\ \ %
\isamarkupcmt{Variable Declaration%
}%
\begin{isamarkuptext}%
Statements of $WL_{EXT}$ expand the statements of \isa{WL} with the following 
      new concepts:
      \begin{description}
  \item[Local Parallelism] This command executes two statements concurrently,
      until both statements have been fully~executed.
  \item[Local Memory] The extended While Language also supports a hierarchical block 
      structure. When opening a new scope with the local memory command, an arbitrary 
      amount of fresh variables can be declared, which are then usable inside the body of 
      the~scope.
  \item[Input] In this command the system receives an unknown input, which is then 
      stored inside a provided variable. This later motivates the usage of 
      symbolic~variables.
  \item[Guarded Statement] This command hides a statement behind a Boolean
      guard. When the guard evaluates to true, the statement body is executed.
      Otherwise, the command blocks, halting the execution of the~statement.
  \item[Call Statement] This command calls a new method associated with
     a provided method name, passing it an arithmetic expression as an actual 
     parameter. Note that this implies that all method arguments must be of arithmetic 
     nature. Additionally, only one singular parameter can be passed as an argument. 
     Both of these design choices simplify our model without any loss of generality.
     The call statement will furthermore introduce implicit concurrency, meaning that
     caller and callee can both be scheduled after the~call. 
  \end{description} 
  Note that our model does not include a formalization of atomic statements, as
  this would later violate the termination argument of the valuation 
  function. Although it is possible to circumvent this predicament (e.g. by using the
  \isa{partial{\isacharunderscore}{\kern0pt}function} theory), the corresponding formalization would add another 
  layer of complexity to our model, thereby later complicating our proof 
  automation. We therefore propose the addition of an atomic statement as an idea for
  an extension of this~model.%
\end{isamarkuptext}\isamarkuptrue%
\ \ \isacommand{datatype}\isamarkupfalse%
\ stmt\ {\isacharequal}{\kern0pt}\ \isanewline
\ \ \ \ \ \ \ \ SKIP\ \ %
\isamarkupcmt{No-Op%
}\isanewline
\ \ \ \ \ \ {\isacharbar}{\kern0pt}\ Assign\ var\ aexp\ %
\isamarkupcmt{Assignment of a variable%
}\isanewline
\ \ \ \ \ \ {\isacharbar}{\kern0pt}\ If\ bexp\ stmt\ \ %
\isamarkupcmt{Conditional Branch%
}\isanewline
\ \ \ \ \ \ {\isacharbar}{\kern0pt}\ While\ bexp\ stmt\ %
\isamarkupcmt{While Loop%
}\isanewline
\ \ \ \ \ \ {\isacharbar}{\kern0pt}\ Seq\ stmt\ stmt\ %
\isamarkupcmt{Sequential Statement%
}\isanewline
\isanewline
\ \ \ \ \ \ %
\isamarkupcmt{Extension of the standard While Language:%
}\isanewline
\isanewline
\ \ \ \ \ \ {\isacharbar}{\kern0pt}\ LocPar\ stmt\ stmt\ \ %
\isamarkupcmt{Local Parallelism%
}\isanewline
\ \ \ \ \ \ {\isacharbar}{\kern0pt}\ LocMem\ varDecl\ stmt\ %
\isamarkupcmt{Local Memory%
}\isanewline
\ \ \ \ \ \ {\isacharbar}{\kern0pt}\ Input\ var\ %
\isamarkupcmt{Input%
}\ \ \isanewline
\ \ \ \ \ \ {\isacharbar}{\kern0pt}\ Guard\ bexp\ stmt\ %
\isamarkupcmt{Guard Statement%
}\isanewline
\ \ \ \ \ \ {\isacharbar}{\kern0pt}\ Call\ method{\isacharunderscore}{\kern0pt}name\ aexp\ %
\isamarkupcmt{Call Statement%
}%
\begin{isamarkuptext}%
We can now build on top of the previously defined statements in order to 
  introduce a notion of methods. A method consists of a method  name (for 
  identification purposes), a formal parameter, as well as a corresponding 
  method~body.%
\end{isamarkuptext}\isamarkuptrue%
\ \ \isacommand{datatype}\isamarkupfalse%
\ method\ {\isacharequal}{\kern0pt}\ \isanewline
\ \ \ \ \ \ \ \ Method\ method{\isacharunderscore}{\kern0pt}name\ var\ stmt%
\begin{isamarkuptext}%
A program consists of a finite list of methods, as well as a main statement 
  body, which is executed upon starting the program. Contrary to the original paper, 
  we enforce the list of methods to be finite, such that we can later traverse them 
  in finite time. Note that this design choice does not reduce the expressivity of 
  the programming~language.%
\end{isamarkuptext}\isamarkuptrue%
\ \ \isacommand{datatype}\isamarkupfalse%
\ program\ {\isacharequal}{\kern0pt}\isanewline
\ \ \ \ \ \ \ \ Program\ {\isachardoublequoteopen}method\ list{\isachardoublequoteclose}\ stmt%
\begin{isamarkuptext}%
We also add a minimal concrete syntax for our programming language, thereby 
  greatly improving the readability of~programs.%
\end{isamarkuptext}\isamarkuptrue%
\ \ \isacommand{notation}\isamarkupfalse%
\ Assign\ {\isacharparenleft}{\kern0pt}\isakeyword{infix}\ {\isachardoublequoteopen}{\isacharcolon}{\kern0pt}{\isacharequal}{\kern0pt}{\isachardoublequoteclose}\ {\isadigit{6}}{\isadigit{1}}{\isacharparenright}{\kern0pt}\isanewline
\ \ \isacommand{notation}\isamarkupfalse%
\ If\ {\isacharparenleft}{\kern0pt}{\isachardoublequoteopen}{\isacharparenleft}{\kern0pt}IF\ {\isacharunderscore}{\kern0pt}{\isacharslash}{\kern0pt}\ THEN\ {\isacharunderscore}{\kern0pt}{\isacharslash}{\kern0pt}\ FI{\isacharparenright}{\kern0pt}{\isachardoublequoteclose}\ {\isacharbrackleft}{\kern0pt}{\isadigit{1}}{\isadigit{0}}{\isadigit{0}}{\isadigit{0}}{\isacharcomma}{\kern0pt}\ {\isadigit{0}}{\isacharbrackright}{\kern0pt}\ {\isadigit{6}}{\isadigit{1}}{\isacharparenright}{\kern0pt}\isanewline
\ \ \isacommand{notation}\isamarkupfalse%
\ While\ {\isacharparenleft}{\kern0pt}{\isachardoublequoteopen}{\isacharparenleft}{\kern0pt}WHILE\ {\isacharunderscore}{\kern0pt}{\isacharslash}{\kern0pt}\ DO\ {\isacharunderscore}{\kern0pt}{\isacharslash}{\kern0pt}\ OD{\isacharparenright}{\kern0pt}{\isachardoublequoteclose}\ {\isacharbrackleft}{\kern0pt}{\isadigit{1}}{\isadigit{0}}{\isadigit{0}}{\isadigit{0}}{\isacharcomma}{\kern0pt}\ {\isadigit{0}}{\isacharbrackright}{\kern0pt}\ {\isadigit{6}}{\isadigit{1}}{\isacharparenright}{\kern0pt}\isanewline
\ \ \isacommand{notation}\isamarkupfalse%
\ Seq\ {\isacharparenleft}{\kern0pt}\isakeyword{infix}\ {\isachardoublequoteopen}{\isacharsemicolon}{\kern0pt}{\isacharsemicolon}{\kern0pt}{\isachardoublequoteclose}\ {\isadigit{6}}{\isadigit{0}}{\isacharparenright}{\kern0pt}\isanewline
\ \ \isacommand{notation}\isamarkupfalse%
\ LocPar\ {\isacharparenleft}{\kern0pt}{\isachardoublequoteopen}{\isacharparenleft}{\kern0pt}CO\ {\isacharunderscore}{\kern0pt}{\isacharslash}{\kern0pt}\ {\isasymparallel}\ {\isacharunderscore}{\kern0pt}{\isacharslash}{\kern0pt}\ OC{\isacharparenright}{\kern0pt}{\isachardoublequoteclose}\ {\isacharbrackleft}{\kern0pt}{\isadigit{0}}{\isacharcomma}{\kern0pt}\ {\isadigit{0}}{\isacharbrackright}{\kern0pt}\ {\isadigit{6}}{\isadigit{1}}{\isacharparenright}{\kern0pt}\isanewline
\ \ \isacommand{notation}\isamarkupfalse%
\ LocMem\ {\isacharparenleft}{\kern0pt}{\isachardoublequoteopen}{\isasymlbrace}{\isacharunderscore}{\kern0pt}{\isacharslash}{\kern0pt}\ {\isacharunderscore}{\kern0pt}{\isacharslash}{\kern0pt}{\isasymrbrace}{\isachardoublequoteclose}\ {\isacharbrackleft}{\kern0pt}{\isadigit{1}}{\isadigit{0}}{\isadigit{0}}{\isadigit{0}}{\isacharcomma}{\kern0pt}\ {\isadigit{0}}{\isacharbrackright}{\kern0pt}\ {\isadigit{6}}{\isadigit{1}}{\isacharparenright}{\kern0pt}\isanewline
\ \ \isacommand{notation}\isamarkupfalse%
\ Input\ {\isacharparenleft}{\kern0pt}{\isachardoublequoteopen}{\isacharparenleft}{\kern0pt}INPUT\ {\isacharunderscore}{\kern0pt}{\isacharslash}{\kern0pt}{\isacharparenright}{\kern0pt}{\isachardoublequoteclose}\ {\isacharbrackleft}{\kern0pt}{\isadigit{1}}{\isadigit{0}}{\isadigit{0}}{\isadigit{0}}{\isacharbrackright}{\kern0pt}\ {\isadigit{6}}{\isadigit{1}}{\isacharparenright}{\kern0pt}\ \isanewline
\ \ \isacommand{notation}\isamarkupfalse%
\ Guard\ {\isacharparenleft}{\kern0pt}{\isachardoublequoteopen}{\isacharparenleft}{\kern0pt}{\isacharcolon}{\kern0pt}{\isacharcolon}{\kern0pt}\ {\isacharunderscore}{\kern0pt}{\isacharslash}{\kern0pt}\ {\isacharsemicolon}{\kern0pt}{\isacharsemicolon}{\kern0pt}\ {\isacharunderscore}{\kern0pt}{\isacharslash}{\kern0pt}\ END{\isacharparenright}{\kern0pt}{\isachardoublequoteclose}\ {\isacharbrackleft}{\kern0pt}{\isadigit{1}}{\isadigit{0}}{\isadigit{0}}{\isadigit{0}}{\isacharcomma}{\kern0pt}\ {\isadigit{0}}{\isacharbrackright}{\kern0pt}\ {\isadigit{6}}{\isadigit{1}}{\isacharparenright}{\kern0pt}\isanewline
\ \ \isacommand{notation}\isamarkupfalse%
\ Call\ {\isacharparenleft}{\kern0pt}{\isachardoublequoteopen}{\isacharparenleft}{\kern0pt}CALL\ {\isacharunderscore}{\kern0pt}{\isacharslash}{\kern0pt}\ {\isacharunderscore}{\kern0pt}{\isacharslash}{\kern0pt}{\isacharparenright}{\kern0pt}{\isachardoublequoteclose}\ {\isacharbrackleft}{\kern0pt}{\isadigit{1}}{\isadigit{0}}{\isadigit{0}}{\isadigit{0}}{\isacharcomma}{\kern0pt}\ {\isadigit{1}}{\isadigit{0}}{\isadigit{0}}{\isadigit{0}}{\isacharbrackright}{\kern0pt}\ {\isadigit{6}}{\isadigit{1}}{\isacharparenright}{\kern0pt}\isanewline
\ \ \isacommand{notation}\isamarkupfalse%
\ Method\ {\isacharparenleft}{\kern0pt}{\isachardoublequoteopen}Method\ {\isacharunderscore}{\kern0pt}\ {\isacharunderscore}{\kern0pt}\ {\isasymlbrace}{\isacharunderscore}{\kern0pt}{\isasymrbrace}{\isachardoublequoteclose}\ {\isacharbrackleft}{\kern0pt}{\isadigit{0}}{\isacharcomma}{\kern0pt}\ {\isadigit{0}}{\isacharcomma}{\kern0pt}\ {\isadigit{0}}{\isacharbrackright}{\kern0pt}\ {\isadigit{6}}{\isadigit{2}}{\isacharparenright}{\kern0pt}\isanewline
\ \ \isacommand{notation}\isamarkupfalse%
\ Program\ {\isacharparenleft}{\kern0pt}{\isachardoublequoteopen}Program\ {\isacharunderscore}{\kern0pt}\ {\isasymlbrace}{\isacharunderscore}{\kern0pt}{\isasymrbrace}{\isachardoublequoteclose}\ {\isacharbrackleft}{\kern0pt}{\isadigit{0}}{\isacharcomma}{\kern0pt}\ {\isadigit{0}}{\isacharbrackright}{\kern0pt}\ {\isadigit{6}}{\isadigit{2}}{\isacharparenright}{\kern0pt}%
\begin{isamarkuptext}%
In order to ease the handling of methods, we propose additional
  straightforward projections, which map a method onto its encased~components.%
\end{isamarkuptext}\isamarkuptrue%
\ \ \isacommand{fun}\isamarkupfalse%
\isanewline
\ \ \ \ method{\isacharunderscore}{\kern0pt}name{\isacharunderscore}{\kern0pt}proj\ {\isacharcolon}{\kern0pt}{\isacharcolon}{\kern0pt}\ {\isachardoublequoteopen}method\ {\isasymRightarrow}\ method{\isacharunderscore}{\kern0pt}name{\isachardoublequoteclose}\ {\isacharparenleft}{\kern0pt}{\isachardoublequoteopen}{\isasymUp}\isactrlsub n{\isachardoublequoteclose}{\isacharparenright}{\kern0pt}\ \isakeyword{where}\isanewline
\ \ \ \ {\isachardoublequoteopen}{\isasymUp}\isactrlsub n\ {\isacharparenleft}{\kern0pt}Method\ m\ x\ {\isasymlbrace}\ S\ {\isasymrbrace}{\isacharparenright}{\kern0pt}\ {\isacharequal}{\kern0pt}\ m{\isachardoublequoteclose}\isanewline
\isanewline
\ \ \isacommand{fun}\isamarkupfalse%
\isanewline
\ \ \ \ method{\isacharunderscore}{\kern0pt}var{\isacharunderscore}{\kern0pt}proj\ {\isacharcolon}{\kern0pt}{\isacharcolon}{\kern0pt}\ {\isachardoublequoteopen}method\ {\isasymRightarrow}\ var{\isachardoublequoteclose}\ {\isacharparenleft}{\kern0pt}{\isachardoublequoteopen}{\isasymUp}\isactrlsub v{\isachardoublequoteclose}{\isacharparenright}{\kern0pt}\ \isakeyword{where}\isanewline
\ \ \ \ {\isachardoublequoteopen}{\isasymUp}\isactrlsub v\ {\isacharparenleft}{\kern0pt}Method\ m\ x\ {\isasymlbrace}\ S\ {\isasymrbrace}{\isacharparenright}{\kern0pt}\ {\isacharequal}{\kern0pt}\ x{\isachardoublequoteclose}\isanewline
\isanewline
\ \ \isacommand{fun}\isamarkupfalse%
\isanewline
\ \ \ \ method{\isacharunderscore}{\kern0pt}stmt{\isacharunderscore}{\kern0pt}proj\ {\isacharcolon}{\kern0pt}{\isacharcolon}{\kern0pt}\ {\isachardoublequoteopen}method\ {\isasymRightarrow}\ stmt{\isachardoublequoteclose}\ {\isacharparenleft}{\kern0pt}{\isachardoublequoteopen}{\isasymUp}\isactrlsub s{\isachardoublequoteclose}{\isacharparenright}{\kern0pt}\ \isakeyword{where}\isanewline
\ \ \ \ {\isachardoublequoteopen}{\isasymUp}\isactrlsub s\ {\isacharparenleft}{\kern0pt}Method\ m\ x\ {\isasymlbrace}\ S\ {\isasymrbrace}{\isacharparenright}{\kern0pt}\ {\isacharequal}{\kern0pt}\ S{\isachardoublequoteclose}%
\begin{isamarkuptext}%
Using our previously defined grammar, it is now possible to derive 
  syntactically correct $WL_{EXT}$ programs. We demonstrate this by presenting short 
  examples for programs utilizing our minimal concrete syntax. The first program 
  non-deterministically assigns a variable of an inner scope the value 1 or 2, the 
  second program demonstrates the usage of method calls, and the third program
  presents a computation on an input variable.%
\end{isamarkuptext}\isamarkuptrue%
\ \ \isacommand{definition}\isamarkupfalse%
\isanewline
\ \ \ \ WL{\isacharunderscore}{\kern0pt}ex\isactrlsub {\isadigit{1}}\ {\isacharcolon}{\kern0pt}{\isacharcolon}{\kern0pt}\ program\ \isakeyword{where}\isanewline
\ \ \ \ {\isachardoublequoteopen}WL{\isacharunderscore}{\kern0pt}ex\isactrlsub {\isadigit{1}}\ {\isasymequiv}\ Program\ {\isacharbrackleft}{\kern0pt}{\isacharbrackright}{\kern0pt}\ {\isasymlbrace}\ {\isasymlbrace}\ {\isacharparenleft}{\kern0pt}{\isacharprime}{\kern0pt}{\isacharprime}{\kern0pt}x{\isacharprime}{\kern0pt}{\isacharprime}{\kern0pt}{\isacharsemicolon}{\kern0pt}{\isasymnu}{\isacharparenright}{\kern0pt}\ CO\ {\isacharprime}{\kern0pt}{\isacharprime}{\kern0pt}x{\isacharprime}{\kern0pt}{\isacharprime}{\kern0pt}\ {\isacharcolon}{\kern0pt}{\isacharequal}{\kern0pt}\ Num\ {\isadigit{1}}\ {\isasymparallel}\ {\isacharprime}{\kern0pt}{\isacharprime}{\kern0pt}x{\isacharprime}{\kern0pt}{\isacharprime}{\kern0pt}\ {\isacharcolon}{\kern0pt}{\isacharequal}{\kern0pt}\ Num\ {\isadigit{2}}\ OC\ {\isasymrbrace}\ {\isasymrbrace}{\isachardoublequoteclose}\isanewline
\isanewline
\ \ \isacommand{definition}\isamarkupfalse%
\isanewline
\ \ \ \ WL{\isacharunderscore}{\kern0pt}ex\isactrlsub {\isadigit{2}}\ {\isacharcolon}{\kern0pt}{\isacharcolon}{\kern0pt}\ program\ \isakeyword{where}\isanewline
\ \ \ \ {\isachardoublequoteopen}WL{\isacharunderscore}{\kern0pt}ex\isactrlsub {\isadigit{2}}\ {\isasymequiv}\ Program\ {\isacharbrackleft}{\kern0pt}\isanewline
\ \ \ \ \ \ \ \ \ \ \ \ \ \ \ \ \ \ \ \ \ {\isacharparenleft}{\kern0pt}Method\ {\isacharprime}{\kern0pt}{\isacharprime}{\kern0pt}foo{\isacharprime}{\kern0pt}{\isacharprime}{\kern0pt}\ {\isacharprime}{\kern0pt}{\isacharprime}{\kern0pt}x{\isacharprime}{\kern0pt}{\isacharprime}{\kern0pt}\ {\isasymlbrace}\ {\isacharprime}{\kern0pt}{\isacharprime}{\kern0pt}x{\isacharprime}{\kern0pt}{\isacharprime}{\kern0pt}\ {\isacharcolon}{\kern0pt}{\isacharequal}{\kern0pt}\ Num\ {\isadigit{2}}\ {\isasymrbrace}{\isacharparenright}{\kern0pt}\isanewline
\ \ \ \ \ \ \ \ \ \ \ \ \ \ \ \ \ \ \ \ {\isacharbrackright}{\kern0pt}\ {\isasymlbrace}\ {\isacharparenleft}{\kern0pt}{\isacharprime}{\kern0pt}{\isacharprime}{\kern0pt}x{\isacharprime}{\kern0pt}{\isacharprime}{\kern0pt}\ {\isacharcolon}{\kern0pt}{\isacharequal}{\kern0pt}\ Num\ {\isadigit{0}}{\isacharsemicolon}{\kern0pt}{\isacharsemicolon}{\kern0pt}\ CALL\ {\isacharprime}{\kern0pt}{\isacharprime}{\kern0pt}foo{\isacharprime}{\kern0pt}{\isacharprime}{\kern0pt}\ {\isacharparenleft}{\kern0pt}Var\ {\isacharprime}{\kern0pt}{\isacharprime}{\kern0pt}x{\isacharprime}{\kern0pt}{\isacharprime}{\kern0pt}{\isacharparenright}{\kern0pt}{\isacharparenright}{\kern0pt}{\isacharsemicolon}{\kern0pt}{\isacharsemicolon}{\kern0pt}\ {\isacharprime}{\kern0pt}{\isacharprime}{\kern0pt}x{\isacharprime}{\kern0pt}{\isacharprime}{\kern0pt}\ {\isacharcolon}{\kern0pt}{\isacharequal}{\kern0pt}\ Num\ {\isadigit{1}}\ {\isasymrbrace}{\isachardoublequoteclose}\isanewline
\isanewline
\ \ \isacommand{definition}\isamarkupfalse%
\isanewline
\ \ \ \ WL{\isacharunderscore}{\kern0pt}ex\isactrlsub {\isadigit{3}}\ {\isacharcolon}{\kern0pt}{\isacharcolon}{\kern0pt}\ program\ \isakeyword{where}\isanewline
\ \ \ \ {\isachardoublequoteopen}WL{\isacharunderscore}{\kern0pt}ex\isactrlsub {\isadigit{3}}\ {\isasymequiv}\ Program\ {\isacharbrackleft}{\kern0pt}{\isacharbrackright}{\kern0pt}\ {\isasymlbrace}\ INPUT\ {\isacharprime}{\kern0pt}{\isacharprime}{\kern0pt}x{\isacharprime}{\kern0pt}{\isacharprime}{\kern0pt}{\isacharsemicolon}{\kern0pt}{\isacharsemicolon}{\kern0pt}\ {\isacharprime}{\kern0pt}{\isacharprime}{\kern0pt}x{\isacharprime}{\kern0pt}{\isacharprime}{\kern0pt}\ {\isacharcolon}{\kern0pt}{\isacharequal}{\kern0pt}\ {\isacharparenleft}{\kern0pt}{\isacharparenleft}{\kern0pt}Var\ {\isacharprime}{\kern0pt}{\isacharprime}{\kern0pt}x{\isacharprime}{\kern0pt}{\isacharprime}{\kern0pt}{\isacharparenright}{\kern0pt}\ \isactrlsub Aadd\ {\isacharparenleft}{\kern0pt}Num\ {\isadigit{1}}{\isacharparenright}{\kern0pt}{\isacharparenright}{\kern0pt}\ {\isasymrbrace}{\isachardoublequoteclose}%
\isadelimdocument
\endisadelimdocument
\isatagdocument
\isamarkupsubsubsection{Variable Mappings%
}
\isamarkuptrue%
\endisatagdocument
{\isafolddocument}%
\isadelimdocument
\endisadelimdocument
\begin{isamarkuptext}%
We introduce variable mappings for $WL_{EXT}$ programs, mapping specific
  programs to a set of their enclosed free variables. Note that the declared scope 
  variables and formal method parameters are not considered free, hence 
  we have to explicitly exclude them. The definition of the corresponding recursive 
  functions are~straightforward.%
\end{isamarkuptext}\isamarkuptrue%
\ \ \isacommand{fun}\isamarkupfalse%
\isanewline
\ \ \ \ vars\isactrlsub d\ {\isacharcolon}{\kern0pt}{\isacharcolon}{\kern0pt}\ {\isachardoublequoteopen}varDecl\ {\isasymRightarrow}\ var\ set{\isachardoublequoteclose}\ \isakeyword{where}\isanewline
\ \ \ \ {\isachardoublequoteopen}vars\isactrlsub d\ {\isasymnu}\ {\isacharequal}{\kern0pt}\ {\isacharbraceleft}{\kern0pt}{\isacharbraceright}{\kern0pt}{\isachardoublequoteclose}\ {\isacharbar}{\kern0pt}\isanewline
\ \ \ \ {\isachardoublequoteopen}vars\isactrlsub d\ {\isacharparenleft}{\kern0pt}x{\isacharsemicolon}{\kern0pt}d{\isacharparenright}{\kern0pt}\ {\isacharequal}{\kern0pt}\ {\isacharbraceleft}{\kern0pt}x{\isacharbraceright}{\kern0pt}\ {\isasymunion}\ vars\isactrlsub d{\isacharparenleft}{\kern0pt}d{\isacharparenright}{\kern0pt}{\isachardoublequoteclose}\ \isanewline
\isanewline
\ \ \isacommand{fun}\isamarkupfalse%
\isanewline
\ \ \ \ vars\isactrlsub s\ {\isacharcolon}{\kern0pt}{\isacharcolon}{\kern0pt}\ {\isachardoublequoteopen}stmt\ {\isasymRightarrow}\ var\ set{\isachardoublequoteclose}\ \isakeyword{where}\isanewline
\ \ \ \ {\isachardoublequoteopen}vars\isactrlsub s\ SKIP\ {\isacharequal}{\kern0pt}\ {\isacharbraceleft}{\kern0pt}{\isacharbraceright}{\kern0pt}{\isachardoublequoteclose}\ {\isacharbar}{\kern0pt}\isanewline
\ \ \ \ {\isachardoublequoteopen}vars\isactrlsub s\ {\isacharparenleft}{\kern0pt}Assign\ x\ a{\isacharparenright}{\kern0pt}\ {\isacharequal}{\kern0pt}\ {\isacharbraceleft}{\kern0pt}x{\isacharbraceright}{\kern0pt}\ {\isasymunion}\ vars\isactrlsub A{\isacharparenleft}{\kern0pt}a{\isacharparenright}{\kern0pt}{\isachardoublequoteclose}\ {\isacharbar}{\kern0pt}\isanewline
\ \ \ \ {\isachardoublequoteopen}vars\isactrlsub s\ {\isacharparenleft}{\kern0pt}If\ b\ S{\isacharparenright}{\kern0pt}\ {\isacharequal}{\kern0pt}\ vars\isactrlsub B{\isacharparenleft}{\kern0pt}b{\isacharparenright}{\kern0pt}\ {\isasymunion}\ vars\isactrlsub s{\isacharparenleft}{\kern0pt}S{\isacharparenright}{\kern0pt}{\isachardoublequoteclose}\ {\isacharbar}{\kern0pt}\isanewline
\ \ \ \ {\isachardoublequoteopen}vars\isactrlsub s\ {\isacharparenleft}{\kern0pt}While\ b\ S{\isacharparenright}{\kern0pt}\ {\isacharequal}{\kern0pt}\ vars\isactrlsub B{\isacharparenleft}{\kern0pt}b{\isacharparenright}{\kern0pt}\ {\isasymunion}\ vars\isactrlsub s{\isacharparenleft}{\kern0pt}S{\isacharparenright}{\kern0pt}{\isachardoublequoteclose}\ {\isacharbar}{\kern0pt}\isanewline
\ \ \ \ {\isachardoublequoteopen}vars\isactrlsub s\ {\isacharparenleft}{\kern0pt}Seq\ S\isactrlsub {\isadigit{1}}\ S\isactrlsub {\isadigit{2}}{\isacharparenright}{\kern0pt}\ {\isacharequal}{\kern0pt}\ vars\isactrlsub s{\isacharparenleft}{\kern0pt}S\isactrlsub {\isadigit{1}}{\isacharparenright}{\kern0pt}\ {\isasymunion}\ vars\isactrlsub s{\isacharparenleft}{\kern0pt}S\isactrlsub {\isadigit{2}}{\isacharparenright}{\kern0pt}{\isachardoublequoteclose}\ {\isacharbar}{\kern0pt}\isanewline
\ \ \ \ {\isachardoublequoteopen}vars\isactrlsub s\ {\isacharparenleft}{\kern0pt}LocPar\ S\isactrlsub {\isadigit{1}}\ S\isactrlsub {\isadigit{2}}{\isacharparenright}{\kern0pt}\ {\isacharequal}{\kern0pt}\ vars\isactrlsub s{\isacharparenleft}{\kern0pt}S\isactrlsub {\isadigit{1}}{\isacharparenright}{\kern0pt}\ {\isasymunion}\ vars\isactrlsub s{\isacharparenleft}{\kern0pt}S\isactrlsub {\isadigit{2}}{\isacharparenright}{\kern0pt}{\isachardoublequoteclose}\ {\isacharbar}{\kern0pt}\isanewline
\ \ \ \ {\isachardoublequoteopen}vars\isactrlsub s\ {\isacharparenleft}{\kern0pt}LocMem\ D\ S{\isacharparenright}{\kern0pt}\ {\isacharequal}{\kern0pt}\ vars\isactrlsub s{\isacharparenleft}{\kern0pt}S{\isacharparenright}{\kern0pt}\ {\isacharminus}{\kern0pt}\ vars\isactrlsub d{\isacharparenleft}{\kern0pt}D{\isacharparenright}{\kern0pt}{\isachardoublequoteclose}\ {\isacharbar}{\kern0pt}\isanewline
\ \ \ \ {\isachardoublequoteopen}vars\isactrlsub s\ {\isacharparenleft}{\kern0pt}Input\ x{\isacharparenright}{\kern0pt}\ {\isacharequal}{\kern0pt}\ {\isacharbraceleft}{\kern0pt}x{\isacharbraceright}{\kern0pt}{\isachardoublequoteclose}\ {\isacharbar}{\kern0pt}\isanewline
\ \ \ \ {\isachardoublequoteopen}vars\isactrlsub s\ {\isacharparenleft}{\kern0pt}Guard\ g\ S{\isacharparenright}{\kern0pt}\ {\isacharequal}{\kern0pt}\ vars\isactrlsub B{\isacharparenleft}{\kern0pt}g{\isacharparenright}{\kern0pt}\ {\isasymunion}\ vars\isactrlsub s{\isacharparenleft}{\kern0pt}S{\isacharparenright}{\kern0pt}{\isachardoublequoteclose}\ {\isacharbar}{\kern0pt}\isanewline
\ \ \ \ {\isachardoublequoteopen}vars\isactrlsub s\ {\isacharparenleft}{\kern0pt}Call\ m\ a{\isacharparenright}{\kern0pt}\ {\isacharequal}{\kern0pt}\ vars\isactrlsub A{\isacharparenleft}{\kern0pt}a{\isacharparenright}{\kern0pt}{\isachardoublequoteclose}\isanewline
\isanewline
\ \ \isacommand{fun}\isamarkupfalse%
\isanewline
\ \ \ \ vars\isactrlsub m\ {\isacharcolon}{\kern0pt}{\isacharcolon}{\kern0pt}\ {\isachardoublequoteopen}method\ {\isasymRightarrow}\ var\ set{\isachardoublequoteclose}\ \isakeyword{where}\isanewline
\ \ \ \ {\isachardoublequoteopen}vars\isactrlsub m\ {\isacharparenleft}{\kern0pt}Method\ m\ x\ {\isasymlbrace}\ S\ {\isasymrbrace}{\isacharparenright}{\kern0pt}\ {\isacharequal}{\kern0pt}\ vars\isactrlsub s{\isacharparenleft}{\kern0pt}S{\isacharparenright}{\kern0pt}\ {\isacharminus}{\kern0pt}\ {\isacharbraceleft}{\kern0pt}x{\isacharbraceright}{\kern0pt}{\isachardoublequoteclose}\isanewline
\isanewline
\ \ \isacommand{fun}\isamarkupfalse%
\isanewline
\ \ \ \ lvars\isactrlsub m\ {\isacharcolon}{\kern0pt}{\isacharcolon}{\kern0pt}\ {\isachardoublequoteopen}method\ list\ {\isasymRightarrow}\ var\ set{\isachardoublequoteclose}\ \isakeyword{where}\isanewline
\ \ \ \ {\isachardoublequoteopen}lvars\isactrlsub m\ {\isacharbrackleft}{\kern0pt}{\isacharbrackright}{\kern0pt}\ {\isacharequal}{\kern0pt}\ {\isacharbraceleft}{\kern0pt}{\isacharbraceright}{\kern0pt}{\isachardoublequoteclose}\ {\isacharbar}{\kern0pt}\isanewline
\ \ \ \ {\isachardoublequoteopen}lvars\isactrlsub m\ {\isacharparenleft}{\kern0pt}m\ {\isacharhash}{\kern0pt}\ rest{\isacharparenright}{\kern0pt}\ {\isacharequal}{\kern0pt}\ vars\isactrlsub m{\isacharparenleft}{\kern0pt}m{\isacharparenright}{\kern0pt}\ {\isasymunion}\ lvars\isactrlsub m{\isacharparenleft}{\kern0pt}rest{\isacharparenright}{\kern0pt}{\isachardoublequoteclose}\isanewline
\isanewline
\ \ \isacommand{fun}\isamarkupfalse%
\isanewline
\ \ \ \ vars\isactrlsub p\ {\isacharcolon}{\kern0pt}{\isacharcolon}{\kern0pt}\ {\isachardoublequoteopen}program\ {\isasymRightarrow}\ var\ set{\isachardoublequoteclose}\ \isakeyword{where}\isanewline
\ \ \ \ {\isachardoublequoteopen}vars\isactrlsub p\ {\isacharparenleft}{\kern0pt}Program\ M\ {\isasymlbrace}\ S\ {\isasymrbrace}{\isacharparenright}{\kern0pt}\ {\isacharequal}{\kern0pt}\ lvars\isactrlsub m{\isacharparenleft}{\kern0pt}M{\isacharparenright}{\kern0pt}\ {\isasymunion}\ vars\isactrlsub s{\isacharparenleft}{\kern0pt}S{\isacharparenright}{\kern0pt}{\isachardoublequoteclose}%
\begin{isamarkuptext}%
Similar to \isa{WL}, we again provide variable occurrence functions,
  mapping a program onto a list of all free variables occurring in it. This 
  again ensures that all free program variables can be systematically traversed,
  thereby later allowing us to establish a notion of initial program states.%
\end{isamarkuptext}\isamarkuptrue%
\ \ \isacommand{fun}\isamarkupfalse%
\isanewline
\ \ \ \ occ\isactrlsub d\ {\isacharcolon}{\kern0pt}{\isacharcolon}{\kern0pt}\ {\isachardoublequoteopen}varDecl\ {\isasymRightarrow}\ var\ list{\isachardoublequoteclose}\ \isakeyword{where}\isanewline
\ \ \ \ {\isachardoublequoteopen}occ\isactrlsub d\ {\isasymnu}\ {\isacharequal}{\kern0pt}\ {\isacharbrackleft}{\kern0pt}{\isacharbrackright}{\kern0pt}{\isachardoublequoteclose}\ {\isacharbar}{\kern0pt}\isanewline
\ \ \ \ {\isachardoublequoteopen}occ\isactrlsub d\ {\isacharparenleft}{\kern0pt}x{\isacharsemicolon}{\kern0pt}d{\isacharparenright}{\kern0pt}\ {\isacharequal}{\kern0pt}\ x\ {\isacharhash}{\kern0pt}\ occ\isactrlsub d{\isacharparenleft}{\kern0pt}d{\isacharparenright}{\kern0pt}{\isachardoublequoteclose}\isanewline
\isanewline
\ \ \isacommand{fun}\isamarkupfalse%
\isanewline
\ \ \ \ occ\isactrlsub s\ {\isacharcolon}{\kern0pt}{\isacharcolon}{\kern0pt}\ {\isachardoublequoteopen}stmt\ {\isasymRightarrow}\ var\ list{\isachardoublequoteclose}\ \isakeyword{where}\isanewline
\ \ \ \ {\isachardoublequoteopen}occ\isactrlsub s\ SKIP\ {\isacharequal}{\kern0pt}\ {\isacharbrackleft}{\kern0pt}{\isacharbrackright}{\kern0pt}{\isachardoublequoteclose}\ {\isacharbar}{\kern0pt}\isanewline
\ \ \ \ {\isachardoublequoteopen}occ\isactrlsub s\ {\isacharparenleft}{\kern0pt}Assign\ x\ a{\isacharparenright}{\kern0pt}\ {\isacharequal}{\kern0pt}\ x\ {\isacharhash}{\kern0pt}\ occ\isactrlsub A{\isacharparenleft}{\kern0pt}a{\isacharparenright}{\kern0pt}{\isachardoublequoteclose}\ {\isacharbar}{\kern0pt}\isanewline
\ \ \ \ {\isachardoublequoteopen}occ\isactrlsub s\ {\isacharparenleft}{\kern0pt}If\ b\ S{\isacharparenright}{\kern0pt}\ {\isacharequal}{\kern0pt}\ occ\isactrlsub B{\isacharparenleft}{\kern0pt}b{\isacharparenright}{\kern0pt}\ {\isacharat}{\kern0pt}\ occ\isactrlsub s{\isacharparenleft}{\kern0pt}S{\isacharparenright}{\kern0pt}{\isachardoublequoteclose}\ {\isacharbar}{\kern0pt}\isanewline
\ \ \ \ {\isachardoublequoteopen}occ\isactrlsub s\ {\isacharparenleft}{\kern0pt}While\ b\ S{\isacharparenright}{\kern0pt}\ {\isacharequal}{\kern0pt}\ occ\isactrlsub B{\isacharparenleft}{\kern0pt}b{\isacharparenright}{\kern0pt}\ {\isacharat}{\kern0pt}\ occ\isactrlsub s{\isacharparenleft}{\kern0pt}S{\isacharparenright}{\kern0pt}{\isachardoublequoteclose}\ {\isacharbar}{\kern0pt}\isanewline
\ \ \ \ {\isachardoublequoteopen}occ\isactrlsub s\ {\isacharparenleft}{\kern0pt}Seq\ S\isactrlsub {\isadigit{1}}\ S\isactrlsub {\isadigit{2}}{\isacharparenright}{\kern0pt}\ {\isacharequal}{\kern0pt}\ occ\isactrlsub s{\isacharparenleft}{\kern0pt}S\isactrlsub {\isadigit{1}}{\isacharparenright}{\kern0pt}\ {\isacharat}{\kern0pt}\ occ\isactrlsub s{\isacharparenleft}{\kern0pt}S\isactrlsub {\isadigit{2}}{\isacharparenright}{\kern0pt}{\isachardoublequoteclose}\ {\isacharbar}{\kern0pt}\isanewline
\ \ \ \ {\isachardoublequoteopen}occ\isactrlsub s\ {\isacharparenleft}{\kern0pt}LocPar\ S\isactrlsub {\isadigit{1}}\ S\isactrlsub {\isadigit{2}}{\isacharparenright}{\kern0pt}\ {\isacharequal}{\kern0pt}\ occ\isactrlsub s{\isacharparenleft}{\kern0pt}S\isactrlsub {\isadigit{1}}{\isacharparenright}{\kern0pt}\ {\isacharat}{\kern0pt}\ occ\isactrlsub s{\isacharparenleft}{\kern0pt}S\isactrlsub {\isadigit{2}}{\isacharparenright}{\kern0pt}{\isachardoublequoteclose}\ {\isacharbar}{\kern0pt}\isanewline
\ \ \ \ {\isachardoublequoteopen}occ\isactrlsub s\ {\isacharparenleft}{\kern0pt}LocMem\ D\ S{\isacharparenright}{\kern0pt}\ {\isacharequal}{\kern0pt}\ filter\ {\isacharparenleft}{\kern0pt}{\isasymlambda}x{\isachardot}{\kern0pt}\ {\isasymnot}List{\isachardot}{\kern0pt}member\ {\isacharparenleft}{\kern0pt}occ\isactrlsub d\ D{\isacharparenright}{\kern0pt}\ x{\isacharparenright}{\kern0pt}\ {\isacharparenleft}{\kern0pt}occ\isactrlsub s\ S{\isacharparenright}{\kern0pt}{\isachardoublequoteclose}\ {\isacharbar}{\kern0pt}\isanewline
\ \ \ \ {\isachardoublequoteopen}occ\isactrlsub s\ {\isacharparenleft}{\kern0pt}Input\ x{\isacharparenright}{\kern0pt}\ {\isacharequal}{\kern0pt}\ {\isacharbrackleft}{\kern0pt}x{\isacharbrackright}{\kern0pt}{\isachardoublequoteclose}\ {\isacharbar}{\kern0pt}\isanewline
\ \ \ \ {\isachardoublequoteopen}occ\isactrlsub s\ {\isacharparenleft}{\kern0pt}Guard\ g\ S{\isacharparenright}{\kern0pt}\ {\isacharequal}{\kern0pt}\ occ\isactrlsub B{\isacharparenleft}{\kern0pt}g{\isacharparenright}{\kern0pt}\ {\isacharat}{\kern0pt}\ occ\isactrlsub s{\isacharparenleft}{\kern0pt}S{\isacharparenright}{\kern0pt}{\isachardoublequoteclose}\ {\isacharbar}{\kern0pt}\isanewline
\ \ \ \ {\isachardoublequoteopen}occ\isactrlsub s\ {\isacharparenleft}{\kern0pt}Call\ m\ a{\isacharparenright}{\kern0pt}\ {\isacharequal}{\kern0pt}\ occ\isactrlsub A{\isacharparenleft}{\kern0pt}a{\isacharparenright}{\kern0pt}{\isachardoublequoteclose}\isanewline
\isanewline
\ \ \isacommand{fun}\isamarkupfalse%
\isanewline
\ \ \ \ occ\isactrlsub m\ {\isacharcolon}{\kern0pt}{\isacharcolon}{\kern0pt}\ {\isachardoublequoteopen}method\ {\isasymRightarrow}\ var\ list{\isachardoublequoteclose}\ \isakeyword{where}\isanewline
\ \ \ \ {\isachardoublequoteopen}occ\isactrlsub m\ {\isacharparenleft}{\kern0pt}Method\ m\ x\ {\isasymlbrace}\ S\ {\isasymrbrace}{\isacharparenright}{\kern0pt}\ {\isacharequal}{\kern0pt}\ removeAll\ x\ {\isacharparenleft}{\kern0pt}occ\isactrlsub s\ S{\isacharparenright}{\kern0pt}{\isachardoublequoteclose}\isanewline
\isanewline
\ \ \isacommand{fun}\isamarkupfalse%
\isanewline
\ \ \ \ locc\isactrlsub m\ {\isacharcolon}{\kern0pt}{\isacharcolon}{\kern0pt}\ {\isachardoublequoteopen}method\ list\ {\isasymRightarrow}\ var\ list{\isachardoublequoteclose}\ \isakeyword{where}\isanewline
\ \ \ \ {\isachardoublequoteopen}locc\isactrlsub m\ {\isacharbrackleft}{\kern0pt}{\isacharbrackright}{\kern0pt}\ {\isacharequal}{\kern0pt}\ {\isacharbrackleft}{\kern0pt}{\isacharbrackright}{\kern0pt}{\isachardoublequoteclose}\ {\isacharbar}{\kern0pt}\isanewline
\ \ \ \ {\isachardoublequoteopen}locc\isactrlsub m\ {\isacharparenleft}{\kern0pt}m\ {\isacharhash}{\kern0pt}\ rest{\isacharparenright}{\kern0pt}\ {\isacharequal}{\kern0pt}\ occ\isactrlsub m{\isacharparenleft}{\kern0pt}m{\isacharparenright}{\kern0pt}\ {\isacharat}{\kern0pt}\ locc\isactrlsub m{\isacharparenleft}{\kern0pt}rest{\isacharparenright}{\kern0pt}{\isachardoublequoteclose}\isanewline
\isanewline
\ \ \isacommand{fun}\isamarkupfalse%
\isanewline
\ \ \ \ occ\isactrlsub p\ {\isacharcolon}{\kern0pt}{\isacharcolon}{\kern0pt}\ {\isachardoublequoteopen}program\ {\isasymRightarrow}\ var\ list{\isachardoublequoteclose}\ \isakeyword{where}\isanewline
\ \ \ \ {\isachardoublequoteopen}occ\isactrlsub p\ {\isacharparenleft}{\kern0pt}Program\ M\ {\isasymlbrace}\ S\ {\isasymrbrace}{\isacharparenright}{\kern0pt}\ {\isacharequal}{\kern0pt}\ locc\isactrlsub m{\isacharparenleft}{\kern0pt}M{\isacharparenright}{\kern0pt}\ {\isacharat}{\kern0pt}\ occ\isactrlsub s{\isacharparenleft}{\kern0pt}S{\isacharparenright}{\kern0pt}{\isachardoublequoteclose}%
\begin{isamarkuptext}%
We can now take another look at our earlier program examples and analyze
  the result of applying a variable mapping and variable occurrence~function.%
\end{isamarkuptext}\isamarkuptrue%
\ \ \isacommand{lemma}\isamarkupfalse%
\ {\isachardoublequoteopen}vars\isactrlsub p\ WL{\isacharunderscore}{\kern0pt}ex\isactrlsub {\isadigit{1}}\ {\isacharequal}{\kern0pt}\ {\isacharbraceleft}{\kern0pt}{\isacharbraceright}{\kern0pt}{\isachardoublequoteclose}\isanewline
\isadelimproof
\ \ \ \ %
\endisadelimproof
\isatagproof
\isacommand{by}\isamarkupfalse%
\ {\isacharparenleft}{\kern0pt}auto\ simp\ add{\isacharcolon}{\kern0pt}\ WL{\isacharunderscore}{\kern0pt}ex\isactrlsub {\isadigit{1}}{\isacharunderscore}{\kern0pt}def{\isacharparenright}{\kern0pt}%
\endisatagproof
{\isafoldproof}%
\isadelimproof
\isanewline
\endisadelimproof
\isanewline
\ \ \isacommand{lemma}\isamarkupfalse%
\ {\isachardoublequoteopen}vars\isactrlsub p\ WL{\isacharunderscore}{\kern0pt}ex\isactrlsub {\isadigit{2}}\ {\isacharequal}{\kern0pt}\ {\isacharbraceleft}{\kern0pt}{\isacharprime}{\kern0pt}{\isacharprime}{\kern0pt}x{\isacharprime}{\kern0pt}{\isacharprime}{\kern0pt}{\isacharbraceright}{\kern0pt}{\isachardoublequoteclose}\isanewline
\isadelimproof
\ \ \ \ %
\endisadelimproof
\isatagproof
\isacommand{by}\isamarkupfalse%
\ {\isacharparenleft}{\kern0pt}auto\ simp\ add{\isacharcolon}{\kern0pt}\ WL{\isacharunderscore}{\kern0pt}ex\isactrlsub {\isadigit{2}}{\isacharunderscore}{\kern0pt}def{\isacharparenright}{\kern0pt}%
\endisatagproof
{\isafoldproof}%
\isadelimproof
\isanewline
\endisadelimproof
\isanewline
\ \ \isacommand{lemma}\isamarkupfalse%
\ {\isachardoublequoteopen}occ\isactrlsub p\ WL{\isacharunderscore}{\kern0pt}ex\isactrlsub {\isadigit{1}}\ {\isacharequal}{\kern0pt}\ {\isacharbrackleft}{\kern0pt}{\isacharbrackright}{\kern0pt}{\isachardoublequoteclose}\isanewline
\isadelimproof
\ \ \ \ %
\endisadelimproof
\isatagproof
\isacommand{by}\isamarkupfalse%
\ {\isacharparenleft}{\kern0pt}auto\ simp\ add{\isacharcolon}{\kern0pt}\ WL{\isacharunderscore}{\kern0pt}ex\isactrlsub {\isadigit{1}}{\isacharunderscore}{\kern0pt}def\ member{\isacharunderscore}{\kern0pt}rec{\isacharparenleft}{\kern0pt}{\isadigit{1}}{\isacharparenright}{\kern0pt}{\isacharparenright}{\kern0pt}%
\endisatagproof
{\isafoldproof}%
\isadelimproof
\isanewline
\endisadelimproof
\isanewline
\ \ \isacommand{lemma}\isamarkupfalse%
\ {\isachardoublequoteopen}occ\isactrlsub p\ WL{\isacharunderscore}{\kern0pt}ex\isactrlsub {\isadigit{2}}\ {\isacharequal}{\kern0pt}\ {\isacharbrackleft}{\kern0pt}{\isacharprime}{\kern0pt}{\isacharprime}{\kern0pt}x{\isacharprime}{\kern0pt}{\isacharprime}{\kern0pt}{\isacharcomma}{\kern0pt}\ {\isacharprime}{\kern0pt}{\isacharprime}{\kern0pt}x{\isacharprime}{\kern0pt}{\isacharprime}{\kern0pt}{\isacharcomma}{\kern0pt}\ {\isacharprime}{\kern0pt}{\isacharprime}{\kern0pt}x{\isacharprime}{\kern0pt}{\isacharprime}{\kern0pt}{\isacharbrackright}{\kern0pt}{\isachardoublequoteclose}\isanewline
\isadelimproof
\ \ \ \ %
\endisadelimproof
\isatagproof
\isacommand{by}\isamarkupfalse%
\ {\isacharparenleft}{\kern0pt}auto\ simp\ add{\isacharcolon}{\kern0pt}\ WL{\isacharunderscore}{\kern0pt}ex\isactrlsub {\isadigit{2}}{\isacharunderscore}{\kern0pt}def{\isacharparenright}{\kern0pt}%
\endisatagproof
{\isafoldproof}%
\isadelimproof
\endisadelimproof
\isadelimdocument
\endisadelimdocument
\isatagdocument
\isamarkupsubsubsection{Variable Substitutions%
}
\isamarkuptrue%
\endisatagdocument
{\isafolddocument}%
\isadelimdocument
\endisadelimdocument
\begin{isamarkuptext}%
In order to later handle variable conflicts, it becomes necessary to replace
  conflicting variables with fresh variables, thus motivating a notion of variable 
  substitutions. We therefore introduce recursive variable substitution functions, 
  which substitute every occurrence of a variable in a program with another variable. 
  Note that we will from now on use the abbreviation \isa{c\ {\isacharbrackleft}{\kern0pt}v\ {\isasymleftarrow}\ z{\isacharbrackright}{\kern0pt}} when referring to 
  the substitution of \isa{v} with \isa{z} in~command~\isa{c}.%
\end{isamarkuptext}\isamarkuptrue%
\ \ \isacommand{primrec}\isamarkupfalse%
\isanewline
\ \ \ \ substitute\isactrlsub d\ {\isacharcolon}{\kern0pt}{\isacharcolon}{\kern0pt}\ {\isachardoublequoteopen}varDecl\ {\isasymRightarrow}\ var\ {\isasymRightarrow}\ var\ {\isasymRightarrow}\ varDecl{\isachardoublequoteclose}\ {\isacharparenleft}{\kern0pt}{\isachardoublequoteopen}{\isacharunderscore}{\kern0pt}\ {\isacharbrackleft}{\kern0pt}{\isacharunderscore}{\kern0pt}\ {\isasymleftarrow}\isactrlsub d\ {\isacharunderscore}{\kern0pt}{\isacharbrackright}{\kern0pt}{\isachardoublequoteclose}\ {\isadigit{7}}{\isadigit{0}}{\isacharparenright}{\kern0pt}\ \isakeyword{where}\isanewline
\ \ \ \ {\isachardoublequoteopen}{\isasymnu}\ {\isacharbrackleft}{\kern0pt}v\ {\isasymleftarrow}\isactrlsub d\ z{\isacharbrackright}{\kern0pt}\ {\isacharequal}{\kern0pt}\ {\isasymnu}{\isachardoublequoteclose}\ {\isacharbar}{\kern0pt}\isanewline
\ \ \ \ {\isachardoublequoteopen}{\isacharparenleft}{\kern0pt}x{\isacharsemicolon}{\kern0pt}d{\isacharparenright}{\kern0pt}\ {\isacharbrackleft}{\kern0pt}v\ {\isasymleftarrow}\isactrlsub d\ z{\isacharbrackright}{\kern0pt}\ {\isacharequal}{\kern0pt}\ {\isacharparenleft}{\kern0pt}if\ x\ {\isacharequal}{\kern0pt}\ v\ then\ z{\isacharsemicolon}{\kern0pt}\ {\isacharparenleft}{\kern0pt}d\ {\isacharbrackleft}{\kern0pt}v\ {\isasymleftarrow}\isactrlsub d\ z{\isacharbrackright}{\kern0pt}{\isacharparenright}{\kern0pt}\ else\ x{\isacharsemicolon}{\kern0pt}\ {\isacharparenleft}{\kern0pt}d\ {\isacharbrackleft}{\kern0pt}v\ {\isasymleftarrow}\isactrlsub d\ z{\isacharbrackright}{\kern0pt}{\isacharparenright}{\kern0pt}{\isacharparenright}{\kern0pt}{\isachardoublequoteclose}\isanewline
\isanewline
\ \ \isacommand{primrec}\isamarkupfalse%
\isanewline
\ \ \ \ substitute\isactrlsub s\ {\isacharcolon}{\kern0pt}{\isacharcolon}{\kern0pt}\ {\isachardoublequoteopen}stmt\ {\isasymRightarrow}\ var\ {\isasymRightarrow}\ var\ {\isasymRightarrow}\ stmt{\isachardoublequoteclose}\ {\isacharparenleft}{\kern0pt}{\isachardoublequoteopen}{\isacharunderscore}{\kern0pt}\ {\isacharbrackleft}{\kern0pt}{\isacharunderscore}{\kern0pt}\ {\isasymleftarrow}\isactrlsub s\ {\isacharunderscore}{\kern0pt}{\isacharbrackright}{\kern0pt}{\isachardoublequoteclose}\ {\isadigit{7}}{\isadigit{0}}{\isacharparenright}{\kern0pt}\ \isakeyword{where}\isanewline
\ \ \ \ {\isachardoublequoteopen}SKIP\ {\isacharbrackleft}{\kern0pt}v\ {\isasymleftarrow}\isactrlsub s\ z{\isacharbrackright}{\kern0pt}\ {\isacharequal}{\kern0pt}\ SKIP{\isachardoublequoteclose}\ {\isacharbar}{\kern0pt}\isanewline
\ \ \ \ {\isachardoublequoteopen}{\isacharparenleft}{\kern0pt}Assign\ x\ a{\isacharparenright}{\kern0pt}\ {\isacharbrackleft}{\kern0pt}v\ {\isasymleftarrow}\isactrlsub s\ z{\isacharbrackright}{\kern0pt}\ {\isacharequal}{\kern0pt}\ {\isacharparenleft}{\kern0pt}if\ x\ {\isacharequal}{\kern0pt}\ v\ then\ {\isacharparenleft}{\kern0pt}z\ {\isacharcolon}{\kern0pt}{\isacharequal}{\kern0pt}\ substitute\isactrlsub A\ a\ v\ z{\isacharparenright}{\kern0pt}\ else\ {\isacharparenleft}{\kern0pt}x\ {\isacharcolon}{\kern0pt}{\isacharequal}{\kern0pt}\ substitute\isactrlsub A\ a\ v\ z{\isacharparenright}{\kern0pt}{\isacharparenright}{\kern0pt}{\isachardoublequoteclose}\ {\isacharbar}{\kern0pt}\isanewline
\ \ \ \ {\isachardoublequoteopen}{\isacharparenleft}{\kern0pt}If\ b\ S{\isacharparenright}{\kern0pt}\ {\isacharbrackleft}{\kern0pt}v\ {\isasymleftarrow}\isactrlsub s\ z{\isacharbrackright}{\kern0pt}\ {\isacharequal}{\kern0pt}\ IF\ {\isacharparenleft}{\kern0pt}substitute\isactrlsub B\ b\ v\ z{\isacharparenright}{\kern0pt}\ THEN\ {\isacharparenleft}{\kern0pt}S\ {\isacharbrackleft}{\kern0pt}v\ {\isasymleftarrow}\isactrlsub s\ z{\isacharbrackright}{\kern0pt}{\isacharparenright}{\kern0pt}\ FI{\isachardoublequoteclose}\ {\isacharbar}{\kern0pt}\isanewline
\ \ \ \ {\isachardoublequoteopen}{\isacharparenleft}{\kern0pt}While\ b\ S{\isacharparenright}{\kern0pt}\ {\isacharbrackleft}{\kern0pt}v\ {\isasymleftarrow}\isactrlsub s\ z{\isacharbrackright}{\kern0pt}\ {\isacharequal}{\kern0pt}\ WHILE\ {\isacharparenleft}{\kern0pt}substitute\isactrlsub B\ b\ v\ z{\isacharparenright}{\kern0pt}\ DO\ {\isacharparenleft}{\kern0pt}S\ {\isacharbrackleft}{\kern0pt}v\ {\isasymleftarrow}\isactrlsub s\ z{\isacharbrackright}{\kern0pt}{\isacharparenright}{\kern0pt}\ OD{\isachardoublequoteclose}\ {\isacharbar}{\kern0pt}\isanewline
\ \ \ \ {\isachardoublequoteopen}{\isacharparenleft}{\kern0pt}Seq\ S\isactrlsub {\isadigit{1}}\ S\isactrlsub {\isadigit{2}}{\isacharparenright}{\kern0pt}\ {\isacharbrackleft}{\kern0pt}v\ {\isasymleftarrow}\isactrlsub s\ z{\isacharbrackright}{\kern0pt}\ {\isacharequal}{\kern0pt}\ {\isacharparenleft}{\kern0pt}S\isactrlsub {\isadigit{1}}\ {\isacharbrackleft}{\kern0pt}v\ {\isasymleftarrow}\isactrlsub s\ z{\isacharbrackright}{\kern0pt}{\isacharparenright}{\kern0pt}{\isacharsemicolon}{\kern0pt}{\isacharsemicolon}{\kern0pt}\ {\isacharparenleft}{\kern0pt}S\isactrlsub {\isadigit{2}}\ {\isacharbrackleft}{\kern0pt}v\ {\isasymleftarrow}\isactrlsub s\ z{\isacharbrackright}{\kern0pt}{\isacharparenright}{\kern0pt}{\isachardoublequoteclose}\ {\isacharbar}{\kern0pt}\isanewline
\ \ \ \ {\isachardoublequoteopen}{\isacharparenleft}{\kern0pt}LocPar\ S\isactrlsub {\isadigit{1}}\ S\isactrlsub {\isadigit{2}}{\isacharparenright}{\kern0pt}\ {\isacharbrackleft}{\kern0pt}v\ {\isasymleftarrow}\isactrlsub s\ z{\isacharbrackright}{\kern0pt}\ {\isacharequal}{\kern0pt}\ CO\ {\isacharparenleft}{\kern0pt}S\isactrlsub {\isadigit{1}}\ {\isacharbrackleft}{\kern0pt}v\ {\isasymleftarrow}\isactrlsub s\ z{\isacharbrackright}{\kern0pt}{\isacharparenright}{\kern0pt}\ {\isasymparallel}\ {\isacharparenleft}{\kern0pt}S\isactrlsub {\isadigit{2}}\ {\isacharbrackleft}{\kern0pt}v\ {\isasymleftarrow}\isactrlsub s\ z{\isacharbrackright}{\kern0pt}{\isacharparenright}{\kern0pt}\ OC{\isachardoublequoteclose}\ {\isacharbar}{\kern0pt}\isanewline
\ \ \ \ {\isachardoublequoteopen}{\isacharparenleft}{\kern0pt}LocMem\ D\ S{\isacharparenright}{\kern0pt}\ {\isacharbrackleft}{\kern0pt}v\ {\isasymleftarrow}\isactrlsub s\ z{\isacharbrackright}{\kern0pt}\ {\isacharequal}{\kern0pt}\ {\isasymlbrace}\ {\isacharparenleft}{\kern0pt}substitute\isactrlsub d\ D\ v\ z{\isacharparenright}{\kern0pt}\ {\isacharparenleft}{\kern0pt}S\ {\isacharbrackleft}{\kern0pt}v\ {\isasymleftarrow}\isactrlsub s\ z{\isacharbrackright}{\kern0pt}{\isacharparenright}{\kern0pt}\ {\isasymrbrace}{\isachardoublequoteclose}\ {\isacharbar}{\kern0pt}\isanewline
\ \ \ \ {\isachardoublequoteopen}{\isacharparenleft}{\kern0pt}Input\ x{\isacharparenright}{\kern0pt}\ {\isacharbrackleft}{\kern0pt}v\ {\isasymleftarrow}\isactrlsub s\ z{\isacharbrackright}{\kern0pt}\ {\isacharequal}{\kern0pt}\ {\isacharparenleft}{\kern0pt}if\ x\ {\isacharequal}{\kern0pt}\ v\ then\ INPUT\ z\ else\ INPUT\ x{\isacharparenright}{\kern0pt}{\isachardoublequoteclose}\ {\isacharbar}{\kern0pt}\isanewline
\ \ \ \ {\isachardoublequoteopen}{\isacharparenleft}{\kern0pt}Guard\ g\ S{\isacharparenright}{\kern0pt}\ {\isacharbrackleft}{\kern0pt}v\ {\isasymleftarrow}\isactrlsub s\ z{\isacharbrackright}{\kern0pt}\ {\isacharequal}{\kern0pt}\ {\isacharcolon}{\kern0pt}{\isacharcolon}{\kern0pt}\ {\isacharparenleft}{\kern0pt}substitute\isactrlsub B\ g\ v\ z{\isacharparenright}{\kern0pt}{\isacharsemicolon}{\kern0pt}{\isacharsemicolon}{\kern0pt}\ {\isacharparenleft}{\kern0pt}S\ {\isacharbrackleft}{\kern0pt}v\ {\isasymleftarrow}\isactrlsub s\ z{\isacharbrackright}{\kern0pt}{\isacharparenright}{\kern0pt}\ END{\isachardoublequoteclose}\ {\isacharbar}{\kern0pt}\isanewline
\ \ \ \ {\isachardoublequoteopen}{\isacharparenleft}{\kern0pt}Call\ m\ a{\isacharparenright}{\kern0pt}\ {\isacharbrackleft}{\kern0pt}v\ {\isasymleftarrow}\isactrlsub s\ z{\isacharbrackright}{\kern0pt}\ {\isacharequal}{\kern0pt}\ CALL\ m\ {\isacharparenleft}{\kern0pt}substitute\isactrlsub A\ a\ v\ z{\isacharparenright}{\kern0pt}{\isachardoublequoteclose}\ \isanewline
\isanewline
\ \ \isacommand{primrec}\isamarkupfalse%
\isanewline
\ \ \ \ substitute\isactrlsub m\ {\isacharcolon}{\kern0pt}{\isacharcolon}{\kern0pt}\ {\isachardoublequoteopen}method\ {\isasymRightarrow}\ var\ {\isasymRightarrow}\ var\ {\isasymRightarrow}\ method{\isachardoublequoteclose}\ {\isacharparenleft}{\kern0pt}{\isachardoublequoteopen}{\isacharunderscore}{\kern0pt}\ {\isacharbrackleft}{\kern0pt}{\isacharunderscore}{\kern0pt}\ {\isasymleftarrow}\isactrlsub m\ {\isacharunderscore}{\kern0pt}{\isacharbrackright}{\kern0pt}{\isachardoublequoteclose}\ {\isadigit{7}}{\isadigit{0}}{\isacharparenright}{\kern0pt}\ \isakeyword{where}\isanewline
\ \ \ \ {\isachardoublequoteopen}{\isacharparenleft}{\kern0pt}Method\ m\ x\ {\isasymlbrace}\ S\ {\isasymrbrace}{\isacharparenright}{\kern0pt}\ {\isacharbrackleft}{\kern0pt}v\ {\isasymleftarrow}\isactrlsub m\ z{\isacharbrackright}{\kern0pt}\ {\isacharequal}{\kern0pt}\ {\isacharparenleft}{\kern0pt}if\ x\ {\isacharequal}{\kern0pt}\ v\ then\ {\isacharparenleft}{\kern0pt}Method\ m\ z\ {\isasymlbrace}\ S\ {\isacharbrackleft}{\kern0pt}v\ {\isasymleftarrow}\isactrlsub s\ z{\isacharbrackright}{\kern0pt}\ {\isasymrbrace}{\isacharparenright}{\kern0pt}\ else\ {\isacharparenleft}{\kern0pt}Method\ m\ x\ {\isasymlbrace}\ S\ {\isacharbrackleft}{\kern0pt}v\ {\isasymleftarrow}\isactrlsub s\ z{\isacharbrackright}{\kern0pt}\ {\isasymrbrace}{\isacharparenright}{\kern0pt}{\isacharparenright}{\kern0pt}{\isachardoublequoteclose}\isanewline
\isanewline
\ \ \isacommand{primrec}\isamarkupfalse%
\isanewline
\ \ \ \ lsubstitute\isactrlsub m\ {\isacharcolon}{\kern0pt}{\isacharcolon}{\kern0pt}\ {\isachardoublequoteopen}method\ list\ {\isasymRightarrow}\ var\ {\isasymRightarrow}\ var\ {\isasymRightarrow}\ method\ list{\isachardoublequoteclose}\ \isakeyword{where}\isanewline
\ \ \ \ {\isachardoublequoteopen}lsubstitute\isactrlsub m\ {\isacharbrackleft}{\kern0pt}{\isacharbrackright}{\kern0pt}\ v\ z\ {\isacharequal}{\kern0pt}\ {\isacharbrackleft}{\kern0pt}{\isacharbrackright}{\kern0pt}{\isachardoublequoteclose}\ {\isacharbar}{\kern0pt}\isanewline
\ \ \ \ {\isachardoublequoteopen}lsubstitute\isactrlsub m\ {\isacharparenleft}{\kern0pt}m\ {\isacharhash}{\kern0pt}\ rest{\isacharparenright}{\kern0pt}\ v\ z\ {\isacharequal}{\kern0pt}\ {\isacharparenleft}{\kern0pt}m\ {\isacharbrackleft}{\kern0pt}v\ {\isasymleftarrow}\isactrlsub m\ z{\isacharbrackright}{\kern0pt}{\isacharparenright}{\kern0pt}\ {\isacharhash}{\kern0pt}\ {\isacharparenleft}{\kern0pt}lsubstitute\isactrlsub m\ rest\ v\ z{\isacharparenright}{\kern0pt}{\isachardoublequoteclose}\isanewline
\isanewline
\ \ \isacommand{primrec}\isamarkupfalse%
\isanewline
\ \ \ \ substitute\isactrlsub p\ {\isacharcolon}{\kern0pt}{\isacharcolon}{\kern0pt}\ {\isachardoublequoteopen}program\ {\isasymRightarrow}\ var\ {\isasymRightarrow}\ var\ {\isasymRightarrow}\ program{\isachardoublequoteclose}\ {\isacharparenleft}{\kern0pt}{\isachardoublequoteopen}{\isacharunderscore}{\kern0pt}\ {\isacharbrackleft}{\kern0pt}{\isacharunderscore}{\kern0pt}\ {\isasymleftarrow}\isactrlsub p\ {\isacharunderscore}{\kern0pt}{\isacharbrackright}{\kern0pt}{\isachardoublequoteclose}\ {\isadigit{7}}{\isadigit{0}}{\isacharparenright}{\kern0pt}\ \isakeyword{where}\isanewline
\ \ \ \ {\isachardoublequoteopen}{\isacharparenleft}{\kern0pt}Program\ M\ {\isasymlbrace}\ S\ {\isasymrbrace}{\isacharparenright}{\kern0pt}\ {\isacharbrackleft}{\kern0pt}v\ {\isasymleftarrow}\isactrlsub p\ z{\isacharbrackright}{\kern0pt}\ {\isacharequal}{\kern0pt}\ {\isacharparenleft}{\kern0pt}Program\ {\isacharparenleft}{\kern0pt}lsubstitute\isactrlsub m\ M\ v\ z{\isacharparenright}{\kern0pt}\ {\isasymlbrace}\ S\ {\isacharbrackleft}{\kern0pt}v\ {\isasymleftarrow}\isactrlsub s\ z{\isacharbrackright}{\kern0pt}\ {\isasymrbrace}{\isacharparenright}{\kern0pt}{\isachardoublequoteclose}%
\begin{isamarkuptext}%
We can now look at an example application of the variable substitution functions 
  by utilizing one of our earlier defined example~programs.%
\end{isamarkuptext}\isamarkuptrue%
\ \ \isacommand{lemma}\isamarkupfalse%
\ {\isachardoublequoteopen}WL{\isacharunderscore}{\kern0pt}ex\isactrlsub {\isadigit{1}}\ {\isacharbrackleft}{\kern0pt}{\isacharprime}{\kern0pt}{\isacharprime}{\kern0pt}x{\isacharprime}{\kern0pt}{\isacharprime}{\kern0pt}\ {\isasymleftarrow}\isactrlsub p\ {\isacharprime}{\kern0pt}{\isacharprime}{\kern0pt}y{\isacharprime}{\kern0pt}{\isacharprime}{\kern0pt}{\isacharbrackright}{\kern0pt}\ {\isacharequal}{\kern0pt}\ Program\ {\isacharbrackleft}{\kern0pt}{\isacharbrackright}{\kern0pt}\ {\isasymlbrace}\ {\isasymlbrace}\ {\isacharparenleft}{\kern0pt}{\isacharprime}{\kern0pt}{\isacharprime}{\kern0pt}y{\isacharprime}{\kern0pt}{\isacharprime}{\kern0pt}{\isacharsemicolon}{\kern0pt}{\isasymnu}{\isacharparenright}{\kern0pt}\ CO\ {\isacharprime}{\kern0pt}{\isacharprime}{\kern0pt}y{\isacharprime}{\kern0pt}{\isacharprime}{\kern0pt}\ {\isacharcolon}{\kern0pt}{\isacharequal}{\kern0pt}\ Num\ {\isadigit{1}}\ {\isasymparallel}\ {\isacharprime}{\kern0pt}{\isacharprime}{\kern0pt}y{\isacharprime}{\kern0pt}{\isacharprime}{\kern0pt}\ {\isacharcolon}{\kern0pt}{\isacharequal}{\kern0pt}\ Num\ {\isadigit{2}}\ OC\ {\isasymrbrace}\ {\isasymrbrace}{\isachardoublequoteclose}\isanewline
\isadelimproof
\ \ \ \ %
\endisadelimproof
\isatagproof
\isacommand{by}\isamarkupfalse%
\ {\isacharparenleft}{\kern0pt}simp\ add{\isacharcolon}{\kern0pt}\ WL{\isacharunderscore}{\kern0pt}ex\isactrlsub {\isadigit{1}}{\isacharunderscore}{\kern0pt}def{\isacharparenright}{\kern0pt}%
\endisatagproof
{\isafoldproof}%
\isadelimproof
\endisadelimproof
\isadelimdocument
\endisadelimdocument
\isatagdocument
\isamarkupsubsubsection{Initial States%
}
\isamarkuptrue%
\endisatagdocument
{\isafolddocument}%
\isadelimdocument
\endisadelimdocument
\begin{isamarkuptext}%
Using our variable occurrence functions, we now introduce initial 
  program states in the same manner as we did for the standard While Language \isa{WL}.%
\end{isamarkuptext}\isamarkuptrue%
\ \ \isacommand{fun}\isamarkupfalse%
\isanewline
\ \ \ \ initial\ {\isacharcolon}{\kern0pt}{\isacharcolon}{\kern0pt}\ {\isachardoublequoteopen}program\ {\isasymRightarrow}\ {\isasymSigma}{\isachardoublequoteclose}\ {\isacharparenleft}{\kern0pt}{\isachardoublequoteopen}{\isasymsigma}\isactrlsub I{\isachardoublequoteclose}{\isacharparenright}{\kern0pt}\ \isakeyword{where}\isanewline
\ \ \ \ {\isachardoublequoteopen}initial\ prog\ {\isacharequal}{\kern0pt}\ get{\isacharunderscore}{\kern0pt}initial\isactrlsub {\isasymSigma}\ {\isacharparenleft}{\kern0pt}occ\isactrlsub p\ prog{\isacharparenright}{\kern0pt}{\isachardoublequoteclose}%
\begin{isamarkuptext}%
Using Isabelle, we can now infer that every initial state, constructed using
  the function above, must be of concrete nature. This trivially holds due to
  the definition of the \isa{get{\isacharunderscore}{\kern0pt}initial\isactrlsub {\isasymSigma}} function, which assigns every variable occurring
  in the provided program the concrete arithmetic~expression~0.%
\end{isamarkuptext}\isamarkuptrue%
\ \ \isacommand{lemma}\isamarkupfalse%
\ initial{\isacharunderscore}{\kern0pt}concrete{\isacharcolon}{\kern0pt}\ {\isachardoublequoteopen}concrete\isactrlsub {\isasymSigma}\ {\isacharparenleft}{\kern0pt}get{\isacharunderscore}{\kern0pt}initial\isactrlsub {\isasymSigma}\ l{\isacharparenright}{\kern0pt}{\isachardoublequoteclose}\isanewline
\isadelimproof
\ \ \ \ %
\endisadelimproof
\isatagproof
\isacommand{by}\isamarkupfalse%
\ {\isacharparenleft}{\kern0pt}induct\ l{\isacharsemicolon}{\kern0pt}\ simp\ add{\isacharcolon}{\kern0pt}\ concrete\isactrlsub {\isasymSigma}{\isacharunderscore}{\kern0pt}def{\isacharparenright}{\kern0pt}%
\endisatagproof
{\isafoldproof}%
\isadelimproof
\isanewline
\endisadelimproof
\isanewline
\ \ \isacommand{lemma}\isamarkupfalse%
\ {\isasymsigma}\isactrlsub I{\isacharunderscore}{\kern0pt}concrete{\isacharcolon}{\kern0pt}\ {\isachardoublequoteopen}concrete\isactrlsub {\isasymSigma}\ {\isacharparenleft}{\kern0pt}{\isasymsigma}\isactrlsub I\ S{\isacharparenright}{\kern0pt}{\isachardoublequoteclose}\isanewline
\isadelimproof
\ \ \ \ %
\endisadelimproof
\isatagproof
\isacommand{using}\isamarkupfalse%
\ initial{\isacharunderscore}{\kern0pt}concrete\ \isacommand{by}\isamarkupfalse%
\ simp%
\endisatagproof
{\isafoldproof}%
\isadelimproof
\endisadelimproof
\begin{isamarkuptext}%
We provide an example for the construction of an initial program state 
  using one of our earlier~programs.%
\end{isamarkuptext}\isamarkuptrue%
\ \ \isacommand{lemma}\isamarkupfalse%
\ {\isachardoublequoteopen}{\isasymsigma}\isactrlsub I\ WL{\isacharunderscore}{\kern0pt}ex\isactrlsub {\isadigit{2}}\ {\isacharequal}{\kern0pt}\ fm{\isacharbrackleft}{\kern0pt}{\isacharparenleft}{\kern0pt}{\isacharprime}{\kern0pt}{\isacharprime}{\kern0pt}x{\isacharprime}{\kern0pt}{\isacharprime}{\kern0pt}{\isacharcomma}{\kern0pt}\ Exp\ {\isacharparenleft}{\kern0pt}Num\ {\isadigit{0}}{\isacharparenright}{\kern0pt}{\isacharparenright}{\kern0pt}{\isacharbrackright}{\kern0pt}{\isachardoublequoteclose}\isanewline
\isadelimproof
\ \ \ \ %
\endisadelimproof
\isatagproof
\isacommand{by}\isamarkupfalse%
\ {\isacharparenleft}{\kern0pt}simp\ add{\isacharcolon}{\kern0pt}\ WL{\isacharunderscore}{\kern0pt}ex\isactrlsub {\isadigit{2}}{\isacharunderscore}{\kern0pt}def\ fmupd{\isacharunderscore}{\kern0pt}reorder{\isacharunderscore}{\kern0pt}neq{\isacharparenright}{\kern0pt}%
\endisatagproof
{\isafoldproof}%
\isadelimproof
\endisadelimproof
\isadelimdocument
\endisadelimdocument
\isatagdocument
\isamarkupsubsection{Continuations%
}
\isamarkuptrue%
\isamarkupsubsubsection{Continuation Markers%
}
\isamarkuptrue%
\endisatagdocument
{\isafolddocument}%
\isadelimdocument
\endisadelimdocument
\begin{isamarkuptext}%
The notion of continuation markers of the standard While Language \isa{WL} 
  does not need to be changed in order to handle $WL_{EXT}$.%
\end{isamarkuptext}\isamarkuptrue%
\ \ \isacommand{datatype}\isamarkupfalse%
\ cont{\isacharunderscore}{\kern0pt}marker\ {\isacharequal}{\kern0pt}\ \isanewline
\ \ \ \ \ \ Lambda\ stmt\ {\isacharparenleft}{\kern0pt}{\isachardoublequoteopen}{\isasymlambda}{\isacharbrackleft}{\kern0pt}{\isacharunderscore}{\kern0pt}{\isacharbrackright}{\kern0pt}{\isachardoublequoteclose}{\isacharparenright}{\kern0pt}\ \ %
\isamarkupcmt{Non-Empty Continuation Marker%
}\isanewline
\ \ \ \ \ \ {\isacharbar}{\kern0pt}\ Empty\ {\isacharparenleft}{\kern0pt}{\isachardoublequoteopen}{\isasymlambda}{\isacharbrackleft}{\kern0pt}{\isasymnabla}{\isacharbrackright}{\kern0pt}{\isachardoublequoteclose}{\isacharparenright}{\kern0pt}\ \ %
\isamarkupcmt{Empty Continuation Marker%
}\isanewline
\isanewline
\ \ \isacommand{fun}\isamarkupfalse%
\isanewline
\ \ \ \ mvars\ {\isacharcolon}{\kern0pt}{\isacharcolon}{\kern0pt}\ {\isachardoublequoteopen}cont{\isacharunderscore}{\kern0pt}marker\ {\isasymRightarrow}\ var\ set{\isachardoublequoteclose}\ \isakeyword{where}\isanewline
\ \ \ \ {\isachardoublequoteopen}mvars\ {\isasymlambda}{\isacharbrackleft}{\kern0pt}{\isasymnabla}{\isacharbrackright}{\kern0pt}\ {\isacharequal}{\kern0pt}\ {\isacharbraceleft}{\kern0pt}{\isacharbraceright}{\kern0pt}{\isachardoublequoteclose}\ {\isacharbar}{\kern0pt}\isanewline
\ \ \ \ {\isachardoublequoteopen}mvars\ {\isasymlambda}{\isacharbrackleft}{\kern0pt}S{\isacharbrackright}{\kern0pt}\ {\isacharequal}{\kern0pt}\ vars\isactrlsub s{\isacharparenleft}{\kern0pt}S{\isacharparenright}{\kern0pt}{\isachardoublequoteclose}%
\isadelimdocument
\endisadelimdocument
\isatagdocument
\isamarkupsubsubsection{Continuation Traces%
}
\isamarkuptrue%
\endisatagdocument
{\isafolddocument}%
\isadelimdocument
\endisadelimdocument
\begin{isamarkuptext}%
Analogous to \isa{WL}, we again define a continuation trace as a 
  conditioned symbolic trace with an additional appended continuation~marker.%
\end{isamarkuptext}\isamarkuptrue%
\ \ \isacommand{datatype}\isamarkupfalse%
\ cont{\isacharunderscore}{\kern0pt}trace\ {\isacharequal}{\kern0pt}\ \isanewline
\ \ \ \ \ \ Cont\ {\isasymCC}{\isasymT}\ cont{\isacharunderscore}{\kern0pt}marker\ {\isacharparenleft}{\kern0pt}\isakeyword{infix}\ {\isachardoublequoteopen}\isactrlitem {\isachardoublequoteclose}\ {\isadigit{5}}{\isadigit{5}}{\isacharparenright}{\kern0pt}%
\begin{isamarkuptext}%
In order to ease the handling of continuation traces, we propose additional
  projections, which map a continuation trace onto its encased~components.%
\end{isamarkuptext}\isamarkuptrue%
\ \ \isacommand{fun}\isamarkupfalse%
\isanewline
\ \ \ \ proj{\isacharunderscore}{\kern0pt}pc\ {\isacharcolon}{\kern0pt}{\isacharcolon}{\kern0pt}\ {\isachardoublequoteopen}cont{\isacharunderscore}{\kern0pt}trace\ {\isasymRightarrow}\ path{\isacharunderscore}{\kern0pt}condition{\isachardoublequoteclose}\ {\isacharparenleft}{\kern0pt}{\isachardoublequoteopen}{\isasymdown}\isactrlsub p{\isachardoublequoteclose}{\isacharparenright}{\kern0pt}\ \isakeyword{where}\isanewline
\ \ \ \ {\isachardoublequoteopen}{\isasymdown}\isactrlsub p\ {\isacharparenleft}{\kern0pt}pc\ {\isasymtriangleright}\ {\isasymtau}\ \isactrlitem \ cm{\isacharparenright}{\kern0pt}\ {\isacharequal}{\kern0pt}\ pc{\isachardoublequoteclose}\ \isanewline
\isanewline
\ \ \isacommand{fun}\isamarkupfalse%
\isanewline
\ \ \ \ proj{\isacharunderscore}{\kern0pt}{\isasymtau}\ {\isacharcolon}{\kern0pt}{\isacharcolon}{\kern0pt}\ {\isachardoublequoteopen}cont{\isacharunderscore}{\kern0pt}trace\ {\isasymRightarrow}\ {\isasymT}{\isachardoublequoteclose}\ {\isacharparenleft}{\kern0pt}{\isachardoublequoteopen}{\isasymdown}\isactrlsub {\isasymtau}{\isachardoublequoteclose}{\isacharparenright}{\kern0pt}\ \isakeyword{where}\isanewline
\ \ \ \ {\isachardoublequoteopen}{\isasymdown}\isactrlsub {\isasymtau}\ {\isacharparenleft}{\kern0pt}pc\ {\isasymtriangleright}\ {\isasymtau}\ \isactrlitem \ cm{\isacharparenright}{\kern0pt}\ {\isacharequal}{\kern0pt}\ {\isasymtau}{\isachardoublequoteclose}\ \isanewline
\isanewline
\ \ \isacommand{fun}\isamarkupfalse%
\isanewline
\ \ \ \ proj{\isacharunderscore}{\kern0pt}cont\ {\isacharcolon}{\kern0pt}{\isacharcolon}{\kern0pt}\ {\isachardoublequoteopen}cont{\isacharunderscore}{\kern0pt}trace\ {\isasymRightarrow}\ cont{\isacharunderscore}{\kern0pt}marker{\isachardoublequoteclose}\ {\isacharparenleft}{\kern0pt}{\isachardoublequoteopen}{\isasymdown}\isactrlsub {\isasymlambda}{\isachardoublequoteclose}{\isacharparenright}{\kern0pt}\ \isakeyword{where}\isanewline
\ \ \ \ {\isachardoublequoteopen}{\isasymdown}\isactrlsub {\isasymlambda}\ {\isacharparenleft}{\kern0pt}cont\ \isactrlitem \ cm{\isacharparenright}{\kern0pt}\ {\isacharequal}{\kern0pt}\ cm{\isachardoublequoteclose}%
\isadelimdocument
\endisadelimdocument
\isatagdocument
\isamarkupsubsection{Local Evaluation%
}
\isamarkuptrue%
\endisatagdocument
{\isafolddocument}%
\isadelimdocument
\endisadelimdocument
\begin{isamarkuptext}%
In order to faithfully handle the new language concepts of $WL_{EXT}$, adapting
  the local evaluation of statements is crucial. However, note that the 
  local semantics of the original \isa{WL} commands do not have to be changed, as their
  local semantics are identical to the standard While Language. This directly implies 
  that we only have to add support for the newly added programming 
  language~concepts. \par
  For this purpose, we first provide a helper function, which modifies a given 
  continuation marker by sequentially appending another statement onto the command 
  inside the continuation marker. If the continuation marker is empty, we simply 
  insert the provided~statement.%
\end{isamarkuptext}\isamarkuptrue%
\ \ \isacommand{fun}\isamarkupfalse%
\isanewline
\ \ \ \ cont{\isacharunderscore}{\kern0pt}append\ {\isacharcolon}{\kern0pt}{\isacharcolon}{\kern0pt}\ {\isachardoublequoteopen}cont{\isacharunderscore}{\kern0pt}marker\ {\isasymRightarrow}\ stmt\ {\isasymRightarrow}\ cont{\isacharunderscore}{\kern0pt}marker{\isachardoublequoteclose}\ \isakeyword{where}\isanewline
\ \ \ \ {\isachardoublequoteopen}cont{\isacharunderscore}{\kern0pt}append\ {\isasymlambda}{\isacharbrackleft}{\kern0pt}S\isactrlsub {\isadigit{1}}{\isacharprime}{\kern0pt}{\isacharbrackright}{\kern0pt}\ S\isactrlsub {\isadigit{2}}\ {\isacharequal}{\kern0pt}\ {\isasymlambda}{\isacharbrackleft}{\kern0pt}S\isactrlsub {\isadigit{1}}{\isacharprime}{\kern0pt}{\isacharsemicolon}{\kern0pt}{\isacharsemicolon}{\kern0pt}S\isactrlsub {\isadigit{2}}{\isacharbrackright}{\kern0pt}{\isachardoublequoteclose}\ {\isacharbar}{\kern0pt}\isanewline
\ \ \ \ {\isachardoublequoteopen}cont{\isacharunderscore}{\kern0pt}append\ {\isasymlambda}{\isacharbrackleft}{\kern0pt}{\isasymnabla}{\isacharbrackright}{\kern0pt}\ S\isactrlsub {\isadigit{2}}\ {\isacharequal}{\kern0pt}\ {\isasymlambda}{\isacharbrackleft}{\kern0pt}S\isactrlsub {\isadigit{2}}{\isacharbrackright}{\kern0pt}{\isachardoublequoteclose}%
\begin{isamarkuptext}%
We also define another helper function, which will later assist us when
  trying to reconstruct the local parallelism construct from two commands. The 
  function receives two continuation markers as arguments. If both continuation 
  markers are non-empty, we return a continuation marker containing the local 
  parallelism construct made up of both argument continuation marker contents. 
  If at least one continuation marker is empty, we simply return the other
  continuation~marker.%
\end{isamarkuptext}\isamarkuptrue%
\ \ \isacommand{fun}\isamarkupfalse%
\isanewline
\ \ \ \ parallel\ {\isacharcolon}{\kern0pt}{\isacharcolon}{\kern0pt}\ {\isachardoublequoteopen}cont{\isacharunderscore}{\kern0pt}marker\ {\isasymRightarrow}\ cont{\isacharunderscore}{\kern0pt}marker\ {\isasymRightarrow}\ cont{\isacharunderscore}{\kern0pt}marker{\isachardoublequoteclose}\ \isakeyword{where}\isanewline
\ \ \ \ {\isachardoublequoteopen}parallel\ {\isasymlambda}{\isacharbrackleft}{\kern0pt}S\isactrlsub {\isadigit{1}}{\isacharbrackright}{\kern0pt}\ {\isasymlambda}{\isacharbrackleft}{\kern0pt}S\isactrlsub {\isadigit{2}}{\isacharbrackright}{\kern0pt}\ {\isacharequal}{\kern0pt}\ {\isasymlambda}{\isacharbrackleft}{\kern0pt}CO\ S\isactrlsub {\isadigit{1}}\ {\isasymparallel}\ S\isactrlsub {\isadigit{2}}\ OC{\isacharbrackright}{\kern0pt}{\isachardoublequoteclose}\ {\isacharbar}{\kern0pt}\isanewline
\ \ \ \ {\isachardoublequoteopen}parallel\ {\isasymlambda}{\isacharbrackleft}{\kern0pt}{\isasymnabla}{\isacharbrackright}{\kern0pt}\ {\isasymlambda}{\isacharbrackleft}{\kern0pt}S{\isacharbrackright}{\kern0pt}\ {\isacharequal}{\kern0pt}\ {\isasymlambda}{\isacharbrackleft}{\kern0pt}S{\isacharbrackright}{\kern0pt}{\isachardoublequoteclose}\ {\isacharbar}{\kern0pt}\isanewline
\ \ \ \ {\isachardoublequoteopen}parallel\ {\isasymlambda}{\isacharbrackleft}{\kern0pt}S{\isacharbrackright}{\kern0pt}\ {\isasymlambda}{\isacharbrackleft}{\kern0pt}{\isasymnabla}{\isacharbrackright}{\kern0pt}\ {\isacharequal}{\kern0pt}\ {\isasymlambda}{\isacharbrackleft}{\kern0pt}S{\isacharbrackright}{\kern0pt}{\isachardoublequoteclose}\ {\isacharbar}{\kern0pt}\isanewline
\ \ \ \ {\isachardoublequoteopen}parallel\ {\isasymlambda}{\isacharbrackleft}{\kern0pt}{\isasymnabla}{\isacharbrackright}{\kern0pt}\ {\isasymlambda}{\isacharbrackleft}{\kern0pt}{\isasymnabla}{\isacharbrackright}{\kern0pt}\ {\isacharequal}{\kern0pt}\ {\isasymlambda}{\isacharbrackleft}{\kern0pt}{\isasymnabla}{\isacharbrackright}{\kern0pt}{\isachardoublequoteclose}%
\begin{isamarkuptext}%
We now have sufficient means to establish the valuation function, which
  maps statements in a given state onto a corresponding set of possible
  continuation traces. Note that the continuation traces for the standard \isa{WL} 
  commands do not change, thus we simply have to add function values for the
  newly added statements. For this purpose, we adhere to the following~core~ideas: 
  \begin{description}
  \item[Local Parallelism] The rule for evaluating the local parallelism construct 
    \isa{CO\ S\isactrlsub {\isadigit{1}}{\isacharsemicolon}{\kern0pt}{\isacharsemicolon}{\kern0pt}S\isactrlsub {\isadigit{2}}\ OC} called in \isa{{\isasymsigma}} is simple. Considering that the choice of 
    evaluating \isa{S\isactrlsub {\isadigit{1}}} or \isa{S\isactrlsub {\isadigit{2}}} for one step is non-deterministic, two potential scenarios 
    have to be taken into consideration. If \isa{S\isactrlsub {\isadigit{1}}} is evaluated to \isa{S\isactrlsub {\isadigit{1}}{\isacharprime}{\kern0pt}}, the subsequent 
    continuation marker should have the form \isa{{\isasymlambda}{\isacharbrackleft}{\kern0pt}CO\ S\isactrlsub {\isadigit{1}}{\isacharprime}{\kern0pt}{\isacharsemicolon}{\kern0pt}{\isacharsemicolon}{\kern0pt}S\isactrlsub {\isadigit{2}}\ OC{\isacharbrackright}{\kern0pt}}. If \isa{S\isactrlsub {\isadigit{2}}} is evaluated 
    to \isa{S\isactrlsub {\isadigit{2}}{\isacharprime}{\kern0pt}}, the subsequent continuation marker should have the form 
    \isa{{\isasymlambda}{\isacharbrackleft}{\kern0pt}CO\ S\isactrlsub {\isadigit{1}}{\isacharsemicolon}{\kern0pt}{\isacharsemicolon}{\kern0pt}S\isactrlsub {\isadigit{2}}{\isacharprime}{\kern0pt}\ OC{\isacharbrackright}{\kern0pt}}. However, if the respective evaluated statement terminates in 
    one step, the local parallelism construct is abolished, whilst the other command 
    still remains to be evaluated. \par
    Considering that we have to adhere to Isabelle syntax, our formalization gets 
    slightly more complex than the definition of the original paper. We first compute 
    \isa{{\isacharparenleft}{\kern0pt}val}~\isa{S\isactrlsub {\isadigit{1}}}~\isa{{\isasymsigma}{\isacharparenright}{\kern0pt}} and \isa{{\isacharparenleft}{\kern0pt}val\ S\isactrlsub {\isadigit{2}}\ {\isasymsigma}{\isacharparenright}{\kern0pt}} in order to figure out all continuation 
    traces generated by \isa{S\isactrlsub {\isadigit{1}}} and \isa{S\isactrlsub {\isadigit{2}}} called in \isa{{\isasymsigma}}. We can then apply the 
    predefined operator \isa{{\isacharbackquote}{\kern0pt}} in order to compute the image of both sets under 
    a function, which reconstructs the local parallelism construct with the other 
    command in all their continuation markers. For this purpose, we utilize one of our 
    earlier defined helper functions. Finally, we merge both sets of continuation 
    traces, thus faithfully capturing the semantics of local~parallelism. \par
    Note that we again circumvent quantifications over infinite types (e.g. traces)
    in order to ensure that Isabelle can automatically generate efficient code for
    this~function.
  \item[Local Memory] The local memory command corresponds to two distinct rules,
    covering both empty and non-empty variable declarations. If no variable is 
    declared upon opening the scope in state \isa{{\isasymsigma}}, the continuation traces generated 
    from the local memory command match the continuation traces generated from the 
    scope body. If a variable \isa{x} is declared when opening the scope in state \isa{{\isasymsigma}}, 
    one single continuation trace can be generated. The path condition of this 
    continuation trace is empty, considering that there are no constraints for
    variable declarations in our programming language. Its symbolic trace transits 
    from \isa{{\isasymsigma}} into an updated version of \isa{{\isasymsigma}}, in which a freshly generated variable 
    \isa{x{\isacharprime}{\kern0pt}} maps to initial value 0. This is done to ensure that every declared variable
    is automatically initialized with 0. The continuation marker also needs to be 
    adapted, such that all occurrences of \isa{x} in the scope body are substituted by 
    \isa{x{\isacharprime}{\kern0pt}}. Note that we rename the declared variable in order to avoid variable 
    conflicts, as declarations in different scopes could theoretically introduce the 
    same variable name. The fresh variable is then only usable inside the opened
    scope, thus cleanly aligning with the desired scope~semantics. \par
    Note that the formalization of this valuation rule has been a major obstacle
    during the modeling procedure. In the original paper, the constraint \isa{x{\isacharprime}{\kern0pt}\ {\isasymnotin}\ dom{\isacharparenleft}{\kern0pt}{\isasymsigma}{\isacharparenright}{\kern0pt}}
    is used to ensure that \isa{x{\isacharprime}{\kern0pt}} is fresh, implying that \isa{x{\isacharprime}{\kern0pt}} is arbitrarily chosen
    out of the variables not occurring in the domain of \isa{{\isasymsigma}}. However, this causes the 
    choice of \isa{x{\isacharprime}{\kern0pt}} to be non-deterministic, thereby indicating an infinite
    set of possible continuation traces. This endangers our code generation 
    objective. In order to circumvent this predicament, we have therefore decided 
    to model the generation of fresh variables in a deterministic manner using a 
    self-defined variable generation function, thereby greatly deviating
    from the definition of the original paper. This design choice also simplifies
    the proof automation, as only one continuation trace needs to be~considered. \par
    We additionally propose the name convention \isa{{\isachardollar}{\kern0pt}x{\isacharcolon}{\kern0pt}{\isacharcolon}{\kern0pt}Scope} for new scope
    variables, which ensures that we can easier associate the variables
    occurring in traces with their corresponding~scopes.  
  \item[Input] Evaluating the input command in state \isa{{\isasymsigma}} generates exactly one
    continuation trace. Its path condition and continuation marker are empty,
    implying that an input statement always terminates in one singular 
    evaluation step. However, the construction of its symbolic trace is slightly more 
    complicated. The variable, which should store the input (i.e. the unknown value), 
    will be updated, such that it maps onto a freshly generated variable \isa{x{\isacharprime}{\kern0pt}}. This 
    freshly generated variable must in turn be initialized with the symbolic
    value \isa{\isactrlemph }, thereby modeling the lack of knowledge about the input. Rerouting the
    input variable via \isa{x{\isacharprime}{\kern0pt}} to \isa{\isactrlemph } (instead of directly mapping the input variable
    to \isa{\isactrlemph }) ensures the possibility of further symbolic computations on the received
    input. We furthermore insert an event capturing the introduction 
    of the new variable \isa{x{\isacharprime}{\kern0pt}}, thus putting forth a possible interaction point for 
    the trace composition. \par
    Note that we again slightly deviate from the original paper by utilizing the 
    self-defined variable generation function, so as to deterministically generate a 
    fresh symbolic~variable. 
  \item[Guarded Statement] The guarded statement called in state \isa{{\isasymsigma}} generates exactly 
    one continuation trace. This continuation trace can only be taken iff 
    the Boolean guard evaluates to true, indicated by its path condition. While its 
    symbolic trace contains only the original state \isa{{\isasymsigma}}, its continuation marker 
    encases the statement S, suggesting that the statement body is still left to be 
    evaluated. Note that this implies a scheduling point right after the evaluation 
    of the Boolean guard expression. Guard and statement are therefore never evaluated 
    in the same evaluation step. Considering that there is no second possible
    continuation trace, this statement can only be evaluated iff the guard holds,
    implying that it blocks~otherwise. \par
    The semantics of this statement greatly deviate from the original paper, as the
    original paper additionally suggests a second continuation trace that preserves
    the guarded statement in its continuation marker iff the guard evaluates to false.
    However, this kind of model would later cause the transitive closure of our
    trace composition to unfavourably diverge when dealing with blocked guarded 
    statements, as they could just be continuously evaluated without making any 
    progress (i.e. stutter). This is a result of a missing fairness notion in our 
    semantics. In order to circumvent this problem, we therefore completely exclude 
    this continuation trace, thus completely eliminating possible stuttering. \par
    Note that the guarded statement introduces deadlocks in our semantics, as it
    could possibly block continuously. Due to our deviations, the method of handling 
    these deadlocks has drastically changed. In our formalization, reaching a deadlock 
    terminates the program, as there cannot exist a possible continuation trace with a 
    consistent path condition. However, the definition of this function in the original 
    paper would indicate infinite stuttering, thus always implying the construction
    of an infinite trace in the case of a~deadlock. 
  \item[Call Statement] Evaluating the call statement in state \isa{{\isasymsigma}} results in
    one singular continuation trace, which consists of an empty path condition, as 
    well as an empty continuation marker. This again indicates that a call statement
    terminates after exactly one evaluation step. The symbolic trace is appended with
    a method invocation event, which contains the method name of the callee and
    an arithmetic expression as its argument. The method name is later used
    to identify the callee, whilst the arithmetic expression models the actual
    parameter. \par
    Note that the formalization of this continuation trace was slightly tricky,
    as the original paper just assumes that method names can be passed alongside
    the argument via the event. However, considering that our formalization
    enforces strict adherence to type constraints, this cannot pass. Two 
    alternative solutions to this predicament come into mind: Firstly, it would
    be an option to model a method name using reserved variable names. However, this
    would later conflict with our concreteness notion, as method names should
    not be simplified. The second option involves modeling a method name 
    as a new type of expression, thereby greatly deviating from the expression 
    syntax of the original paper. We select this second alternative, as this design 
    choice cleanly decouples method names from other kinds of expressions, hence 
    avoiding possible~conflicts.  
  \end{description} Due to its construction, each application of the valuation 
  function results in only finitely many continuation~traces, thereby simplifying
  our proof automation and code~generation.%
\end{isamarkuptext}\isamarkuptrue%
\ \ \isacommand{fun}\isamarkupfalse%
\ \isanewline
\ \ \ \ val\isactrlsub s\ {\isacharcolon}{\kern0pt}{\isacharcolon}{\kern0pt}\ {\isachardoublequoteopen}stmt\ {\isasymRightarrow}\ {\isasymSigma}\ {\isasymRightarrow}\ cont{\isacharunderscore}{\kern0pt}trace\ set{\isachardoublequoteclose}\ \isakeyword{where}\isanewline
\ \ \ \ {\isachardoublequoteopen}val\isactrlsub s\ SKIP\ {\isasymsigma}\ {\isacharequal}{\kern0pt}\ {\isacharbraceleft}{\kern0pt}\ {\isacharbraceleft}{\kern0pt}{\isacharbraceright}{\kern0pt}\ {\isasymtriangleright}\ {\isasymlangle}{\isasymsigma}{\isasymrangle}\ \isactrlitem \ {\isasymlambda}{\isacharbrackleft}{\kern0pt}{\isasymnabla}{\isacharbrackright}{\kern0pt}\ {\isacharbraceright}{\kern0pt}{\isachardoublequoteclose}\ {\isacharbar}{\kern0pt}\isanewline
\ \ \ \ {\isachardoublequoteopen}val\isactrlsub s\ {\isacharparenleft}{\kern0pt}x\ {\isacharcolon}{\kern0pt}{\isacharequal}{\kern0pt}\ a{\isacharparenright}{\kern0pt}\ {\isasymsigma}\ {\isacharequal}{\kern0pt}\ {\isacharbraceleft}{\kern0pt}\ {\isacharbraceleft}{\kern0pt}{\isacharbraceright}{\kern0pt}\ {\isasymtriangleright}\ {\isasymlangle}{\isasymsigma}{\isasymrangle}\ {\isasymleadsto}\ State{\isasymllangle}{\isacharbrackleft}{\kern0pt}x\ {\isasymlongmapsto}\ Exp\ {\isacharparenleft}{\kern0pt}val\isactrlsub A\ a\ {\isasymsigma}{\isacharparenright}{\kern0pt}{\isacharbrackright}{\kern0pt}\ {\isasymsigma}{\isasymrrangle}\ \isactrlitem \ {\isasymlambda}{\isacharbrackleft}{\kern0pt}{\isasymnabla}{\isacharbrackright}{\kern0pt}\ {\isacharbraceright}{\kern0pt}{\isachardoublequoteclose}\ {\isacharbar}{\kern0pt}\isanewline
\ \ \ \ {\isachardoublequoteopen}val\isactrlsub s\ {\isacharparenleft}{\kern0pt}IF\ b\ THEN\ S\ FI{\isacharparenright}{\kern0pt}\ {\isasymsigma}\ {\isacharequal}{\kern0pt}\ {\isacharbraceleft}{\kern0pt}\isanewline
\ \ \ \ \ \ \ \ {\isacharbraceleft}{\kern0pt}val\isactrlsub B\ b\ {\isasymsigma}{\isacharbraceright}{\kern0pt}\ {\isasymtriangleright}\ {\isasymlangle}{\isasymsigma}{\isasymrangle}\ \isactrlitem \ {\isasymlambda}{\isacharbrackleft}{\kern0pt}S{\isacharbrackright}{\kern0pt}{\isacharcomma}{\kern0pt}\isanewline
\ \ \ \ \ \ \ \ {\isacharbraceleft}{\kern0pt}val\isactrlsub B\ {\isacharparenleft}{\kern0pt}Not\ b{\isacharparenright}{\kern0pt}\ {\isasymsigma}{\isacharbraceright}{\kern0pt}\ {\isasymtriangleright}\ {\isasymlangle}{\isasymsigma}{\isasymrangle}\ \isactrlitem \ {\isasymlambda}{\isacharbrackleft}{\kern0pt}{\isasymnabla}{\isacharbrackright}{\kern0pt}\isanewline
\ \ \ \ \ \ {\isacharbraceright}{\kern0pt}{\isachardoublequoteclose}\ {\isacharbar}{\kern0pt}\isanewline
\ \ \ \ {\isachardoublequoteopen}val\isactrlsub s\ {\isacharparenleft}{\kern0pt}WHILE\ b\ DO\ S\ OD{\isacharparenright}{\kern0pt}\ {\isasymsigma}\ {\isacharequal}{\kern0pt}\ {\isacharbraceleft}{\kern0pt}\isanewline
\ \ \ \ \ \ \ \ {\isacharbraceleft}{\kern0pt}val\isactrlsub B\ b\ {\isasymsigma}{\isacharbraceright}{\kern0pt}\ {\isasymtriangleright}\ {\isasymlangle}{\isasymsigma}{\isasymrangle}\ \isactrlitem \ {\isasymlambda}{\isacharbrackleft}{\kern0pt}S{\isacharsemicolon}{\kern0pt}{\isacharsemicolon}{\kern0pt}WHILE\ b\ DO\ S\ OD{\isacharbrackright}{\kern0pt}{\isacharcomma}{\kern0pt}\isanewline
\ \ \ \ \ \ \ \ {\isacharbraceleft}{\kern0pt}val\isactrlsub B\ {\isacharparenleft}{\kern0pt}Not\ b{\isacharparenright}{\kern0pt}\ {\isasymsigma}{\isacharbraceright}{\kern0pt}\ {\isasymtriangleright}\ {\isasymlangle}{\isasymsigma}{\isasymrangle}\ \isactrlitem \ {\isasymlambda}{\isacharbrackleft}{\kern0pt}{\isasymnabla}{\isacharbrackright}{\kern0pt}\isanewline
\ \ \ \ \ \ {\isacharbraceright}{\kern0pt}{\isachardoublequoteclose}\ {\isacharbar}{\kern0pt}\isanewline
\ \ \ \ {\isachardoublequoteopen}val\isactrlsub s\ {\isacharparenleft}{\kern0pt}S\isactrlsub {\isadigit{1}}{\isacharsemicolon}{\kern0pt}{\isacharsemicolon}{\kern0pt}S\isactrlsub {\isadigit{2}}{\isacharparenright}{\kern0pt}\ {\isasymsigma}\ {\isacharequal}{\kern0pt}\ {\isacharparenleft}{\kern0pt}{\isacharpercent}{\kern0pt}c{\isachardot}{\kern0pt}\ {\isacharparenleft}{\kern0pt}{\isasymdown}\isactrlsub p\ c{\isacharparenright}{\kern0pt}\ {\isasymtriangleright}\ {\isacharparenleft}{\kern0pt}{\isasymdown}\isactrlsub {\isasymtau}\ c{\isacharparenright}{\kern0pt}\ \isactrlitem \ cont{\isacharunderscore}{\kern0pt}append\ {\isacharparenleft}{\kern0pt}{\isasymdown}\isactrlsub {\isasymlambda}\ c{\isacharparenright}{\kern0pt}\ S\isactrlsub {\isadigit{2}}{\isacharparenright}{\kern0pt}\ {\isacharbackquote}{\kern0pt}\ {\isacharparenleft}{\kern0pt}val\isactrlsub s\ S\isactrlsub {\isadigit{1}}\ {\isasymsigma}{\isacharparenright}{\kern0pt}{\isachardoublequoteclose}\ {\isacharbar}{\kern0pt}\isanewline
\ \ \ \ {\isachardoublequoteopen}val\isactrlsub s\ {\isacharparenleft}{\kern0pt}CO\ S\isactrlsub {\isadigit{1}}\ {\isasymparallel}\ S\isactrlsub {\isadigit{2}}\ OC{\isacharparenright}{\kern0pt}\ {\isasymsigma}\ {\isacharequal}{\kern0pt}\ \isanewline
\ \ \ \ \ \ \ \ \ \ {\isacharparenleft}{\kern0pt}{\isacharpercent}{\kern0pt}c{\isachardot}{\kern0pt}\ {\isacharparenleft}{\kern0pt}{\isasymdown}\isactrlsub p\ c{\isacharparenright}{\kern0pt}\ {\isasymtriangleright}\ {\isacharparenleft}{\kern0pt}{\isasymdown}\isactrlsub {\isasymtau}\ c{\isacharparenright}{\kern0pt}\ \isactrlitem \ parallel\ {\isacharparenleft}{\kern0pt}{\isasymdown}\isactrlsub {\isasymlambda}\ c{\isacharparenright}{\kern0pt}\ {\isasymlambda}{\isacharbrackleft}{\kern0pt}S\isactrlsub {\isadigit{2}}{\isacharbrackright}{\kern0pt}{\isacharparenright}{\kern0pt}\ {\isacharbackquote}{\kern0pt}\ {\isacharparenleft}{\kern0pt}val\isactrlsub s\ S\isactrlsub {\isadigit{1}}\ {\isasymsigma}{\isacharparenright}{\kern0pt}\ \isanewline
\ \ \ \ \ \ \ \ \ \ {\isasymunion}\ {\isacharparenleft}{\kern0pt}{\isacharpercent}{\kern0pt}c{\isachardot}{\kern0pt}\ {\isacharparenleft}{\kern0pt}{\isasymdown}\isactrlsub p\ c{\isacharparenright}{\kern0pt}\ {\isasymtriangleright}\ {\isacharparenleft}{\kern0pt}{\isasymdown}\isactrlsub {\isasymtau}\ c{\isacharparenright}{\kern0pt}\ \isactrlitem \ parallel\ {\isasymlambda}{\isacharbrackleft}{\kern0pt}S\isactrlsub {\isadigit{1}}{\isacharbrackright}{\kern0pt}\ {\isacharparenleft}{\kern0pt}{\isasymdown}\isactrlsub {\isasymlambda}\ c{\isacharparenright}{\kern0pt}{\isacharparenright}{\kern0pt}\ {\isacharbackquote}{\kern0pt}\ {\isacharparenleft}{\kern0pt}val\isactrlsub s\ S\isactrlsub {\isadigit{2}}\ {\isasymsigma}{\isacharparenright}{\kern0pt}{\isachardoublequoteclose}\ {\isacharbar}{\kern0pt}\isanewline
\ \ \ \ {\isachardoublequoteopen}val\isactrlsub s\ {\isacharparenleft}{\kern0pt}{\isasymlbrace}\ {\isasymnu}\ S\ {\isasymrbrace}{\isacharparenright}{\kern0pt}\ {\isasymsigma}\ {\isacharequal}{\kern0pt}\ val\isactrlsub s\ S\ {\isasymsigma}{\isachardoublequoteclose}\ {\isacharbar}{\kern0pt}\isanewline
\ \ \ \ {\isachardoublequoteopen}val\isactrlsub s\ {\isacharparenleft}{\kern0pt}{\isasymlbrace}\ {\isacharparenleft}{\kern0pt}x{\isacharsemicolon}{\kern0pt}d{\isacharparenright}{\kern0pt}\ S\ {\isasymrbrace}{\isacharparenright}{\kern0pt}\ {\isasymsigma}\ {\isacharequal}{\kern0pt}\ {\isacharbraceleft}{\kern0pt}\ \isanewline
\ \ \ \ \ \ \ \ {\isacharbraceleft}{\kern0pt}{\isacharbraceright}{\kern0pt}\ {\isasymtriangleright}\ {\isasymlangle}{\isasymsigma}{\isasymrangle}\ {\isasymleadsto}\ State{\isasymllangle}{\isacharbrackleft}{\kern0pt}{\isacharparenleft}{\kern0pt}vargen\ {\isasymsigma}\ {\isadigit{0}}\ {\isadigit{1}}{\isadigit{0}}{\isadigit{0}}\ {\isacharparenleft}{\kern0pt}{\isacharprime}{\kern0pt}{\isacharprime}{\kern0pt}{\isachardollar}{\kern0pt}{\isacharprime}{\kern0pt}{\isacharprime}{\kern0pt}\ {\isacharat}{\kern0pt}\ x\ {\isacharat}{\kern0pt}\ {\isacharprime}{\kern0pt}{\isacharprime}{\kern0pt}{\isacharcolon}{\kern0pt}{\isacharcolon}{\kern0pt}Scope{\isacharprime}{\kern0pt}{\isacharprime}{\kern0pt}{\isacharparenright}{\kern0pt}{\isacharparenright}{\kern0pt}\ {\isasymlongmapsto}\ Exp\ {\isacharparenleft}{\kern0pt}Num\ {\isadigit{0}}{\isacharparenright}{\kern0pt}{\isacharbrackright}{\kern0pt}\ {\isasymsigma}{\isasymrrangle}\ \isanewline
\ \ \ \ \ \ \ \ \ \ \ \ \ \ \ \ \ \ \ \isactrlitem \ {\isasymlambda}{\isacharbrackleft}{\kern0pt}{\isasymlbrace}\ d\ S\ {\isasymrbrace}\ {\isacharbrackleft}{\kern0pt}x\ {\isasymleftarrow}\isactrlsub s\ {\isacharparenleft}{\kern0pt}vargen\ {\isasymsigma}\ {\isadigit{0}}\ {\isadigit{1}}{\isadigit{0}}{\isadigit{0}}\ {\isacharparenleft}{\kern0pt}{\isacharprime}{\kern0pt}{\isacharprime}{\kern0pt}{\isachardollar}{\kern0pt}{\isacharprime}{\kern0pt}{\isacharprime}{\kern0pt}\ {\isacharat}{\kern0pt}\ x\ {\isacharat}{\kern0pt}\ {\isacharprime}{\kern0pt}{\isacharprime}{\kern0pt}{\isacharcolon}{\kern0pt}{\isacharcolon}{\kern0pt}Scope{\isacharprime}{\kern0pt}{\isacharprime}{\kern0pt}{\isacharparenright}{\kern0pt}{\isacharparenright}{\kern0pt}{\isacharbrackright}{\kern0pt}\ {\isacharbrackright}{\kern0pt}\ {\isacharbraceright}{\kern0pt}{\isachardoublequoteclose}\ {\isacharbar}{\kern0pt}\isanewline
\ \ \ \ {\isachardoublequoteopen}val\isactrlsub s\ {\isacharparenleft}{\kern0pt}INPUT\ x{\isacharparenright}{\kern0pt}\ {\isasymsigma}\ {\isacharequal}{\kern0pt}\ {\isacharbraceleft}{\kern0pt}\ \isanewline
\ \ \ \ \ \ \ \ {\isacharbraceleft}{\kern0pt}{\isacharbraceright}{\kern0pt}\ {\isasymtriangleright}\ {\isacharparenleft}{\kern0pt}{\isasymlangle}{\isasymsigma}{\isasymrangle}\ {\isasymcdot}\ {\isacharparenleft}{\kern0pt}gen{\isacharunderscore}{\kern0pt}event\ inpEv\ {\isacharparenleft}{\kern0pt}{\isacharbrackleft}{\kern0pt}x\ {\isasymlongmapsto}\ Exp\ {\isacharparenleft}{\kern0pt}Var\ {\isacharparenleft}{\kern0pt}vargen\ {\isasymsigma}\ {\isadigit{0}}\ {\isadigit{1}}{\isadigit{0}}{\isadigit{0}}\ {\isacharparenleft}{\kern0pt}{\isacharprime}{\kern0pt}{\isacharprime}{\kern0pt}{\isachardollar}{\kern0pt}{\isacharprime}{\kern0pt}{\isacharprime}{\kern0pt}\ {\isacharat}{\kern0pt}\ x\ {\isacharat}{\kern0pt}\ {\isacharprime}{\kern0pt}{\isacharprime}{\kern0pt}{\isacharcolon}{\kern0pt}{\isacharcolon}{\kern0pt}Input{\isacharprime}{\kern0pt}{\isacharprime}{\kern0pt}{\isacharparenright}{\kern0pt}{\isacharparenright}{\kern0pt}{\isacharparenright}{\kern0pt}{\isacharbrackright}{\kern0pt}\ \isanewline
\ \ \ \ \ \ \ \ \ \ \ \ \ \ \ \ \ \ \ {\isacharbrackleft}{\kern0pt}{\isacharparenleft}{\kern0pt}vargen\ {\isasymsigma}\ {\isadigit{0}}\ {\isadigit{1}}{\isadigit{0}}{\isadigit{0}}\ {\isacharparenleft}{\kern0pt}{\isacharprime}{\kern0pt}{\isacharprime}{\kern0pt}{\isachardollar}{\kern0pt}{\isacharprime}{\kern0pt}{\isacharprime}{\kern0pt}\ {\isacharat}{\kern0pt}\ x\ {\isacharat}{\kern0pt}\ {\isacharprime}{\kern0pt}{\isacharprime}{\kern0pt}{\isacharcolon}{\kern0pt}{\isacharcolon}{\kern0pt}Input{\isacharprime}{\kern0pt}{\isacharprime}{\kern0pt}{\isacharparenright}{\kern0pt}{\isacharparenright}{\kern0pt}\ {\isasymlongmapsto}\ \isactrlemph {\isacharbrackright}{\kern0pt}\ {\isasymsigma}{\isacharparenright}{\kern0pt}\ \isanewline
\ \ \ \ \ \ \ \ \ \ \ \ \ \ \ \ \ \ \ {\isacharbrackleft}{\kern0pt}A\ {\isacharparenleft}{\kern0pt}Var\ {\isacharparenleft}{\kern0pt}vargen\ {\isasymsigma}\ {\isadigit{0}}\ {\isadigit{1}}{\isadigit{0}}{\isadigit{0}}\ {\isacharparenleft}{\kern0pt}{\isacharprime}{\kern0pt}{\isacharprime}{\kern0pt}{\isachardollar}{\kern0pt}{\isacharprime}{\kern0pt}{\isacharprime}{\kern0pt}\ {\isacharat}{\kern0pt}\ x\ {\isacharat}{\kern0pt}\ {\isacharprime}{\kern0pt}{\isacharprime}{\kern0pt}{\isacharcolon}{\kern0pt}{\isacharcolon}{\kern0pt}Input{\isacharprime}{\kern0pt}{\isacharprime}{\kern0pt}{\isacharparenright}{\kern0pt}{\isacharparenright}{\kern0pt}{\isacharparenright}{\kern0pt}{\isacharbrackright}{\kern0pt}{\isacharparenright}{\kern0pt}{\isacharparenright}{\kern0pt}\ \isactrlitem \ {\isasymlambda}{\isacharbrackleft}{\kern0pt}{\isasymnabla}{\isacharbrackright}{\kern0pt}\ \isanewline
\ \ \ \ \ \ {\isacharbraceright}{\kern0pt}{\isachardoublequoteclose}\ {\isacharbar}{\kern0pt}\isanewline
\ \ \ \ {\isachardoublequoteopen}val\isactrlsub s\ {\isacharparenleft}{\kern0pt}{\isacharcolon}{\kern0pt}{\isacharcolon}{\kern0pt}\ g{\isacharsemicolon}{\kern0pt}{\isacharsemicolon}{\kern0pt}\ S\ END{\isacharparenright}{\kern0pt}\ {\isasymsigma}\ {\isacharequal}{\kern0pt}\ {\isacharbraceleft}{\kern0pt}\isanewline
\ \ \ \ \ \ \ \ {\isacharbraceleft}{\kern0pt}val\isactrlsub B\ g\ {\isasymsigma}{\isacharbraceright}{\kern0pt}\ {\isasymtriangleright}\ {\isasymlangle}{\isasymsigma}{\isasymrangle}\ \isactrlitem \ {\isasymlambda}{\isacharbrackleft}{\kern0pt}S{\isacharbrackright}{\kern0pt}\isanewline
\ \ \ \ \ \ {\isacharbraceright}{\kern0pt}{\isachardoublequoteclose}\ {\isacharbar}{\kern0pt}\isanewline
\ \ \ \ {\isachardoublequoteopen}val\isactrlsub s\ {\isacharparenleft}{\kern0pt}CALL\ m\ a{\isacharparenright}{\kern0pt}\ {\isasymsigma}\ {\isacharequal}{\kern0pt}\ {\isacharbraceleft}{\kern0pt}\ {\isacharbraceleft}{\kern0pt}{\isacharbraceright}{\kern0pt}\ {\isasymtriangleright}\ {\isacharparenleft}{\kern0pt}gen{\isacharunderscore}{\kern0pt}event\ invEv\ {\isasymsigma}\ {\isacharbrackleft}{\kern0pt}P\ m{\isacharcomma}{\kern0pt}\ A\ a{\isacharbrackright}{\kern0pt}{\isacharparenright}{\kern0pt}\ \isactrlitem \ {\isasymlambda}{\isacharbrackleft}{\kern0pt}{\isasymnabla}{\isacharbrackright}{\kern0pt}\ {\isacharbraceright}{\kern0pt}{\isachardoublequoteclose}%
\begin{isamarkuptext}%
As we have extended the local evaluation of our semantics, we can now analyze 
  several examples of valuation function applications including our new 
  language~concepts.%
\end{isamarkuptext}\isamarkuptrue%
\ \ \isacommand{lemma}\isamarkupfalse%
\ {\isachardoublequoteopen}val\isactrlsub s\ {\isacharparenleft}{\kern0pt}CO\ {\isacharprime}{\kern0pt}{\isacharprime}{\kern0pt}x{\isacharprime}{\kern0pt}{\isacharprime}{\kern0pt}\ {\isacharcolon}{\kern0pt}{\isacharequal}{\kern0pt}\ Num\ {\isadigit{1}}\ {\isasymparallel}\ {\isacharprime}{\kern0pt}{\isacharprime}{\kern0pt}x{\isacharprime}{\kern0pt}{\isacharprime}{\kern0pt}\ {\isacharcolon}{\kern0pt}{\isacharequal}{\kern0pt}\ Num\ {\isadigit{2}}{\isacharsemicolon}{\kern0pt}{\isacharsemicolon}{\kern0pt}\ SKIP\ OC{\isacharparenright}{\kern0pt}\ {\isasymsigma}\isactrlsub {\isadigit{1}}\ {\isacharequal}{\kern0pt}\ {\isacharbraceleft}{\kern0pt}\ \isanewline
\ \ \ \ \ \ \ \ \ \ \ \ \ \ \ \ \ \ {\isacharbraceleft}{\kern0pt}{\isacharbraceright}{\kern0pt}\ {\isasymtriangleright}\ {\isasymlangle}{\isasymsigma}\isactrlsub {\isadigit{1}}{\isasymrangle}\ {\isasymleadsto}\ State{\isasymllangle}{\isacharbrackleft}{\kern0pt}{\isacharprime}{\kern0pt}{\isacharprime}{\kern0pt}x{\isacharprime}{\kern0pt}{\isacharprime}{\kern0pt}\ {\isasymlongmapsto}\ Exp\ {\isacharparenleft}{\kern0pt}Num\ {\isadigit{1}}{\isacharparenright}{\kern0pt}{\isacharbrackright}{\kern0pt}\ {\isasymsigma}\isactrlsub {\isadigit{1}}{\isasymrrangle}\ \isactrlitem \ {\isasymlambda}{\isacharbrackleft}{\kern0pt}{\isacharprime}{\kern0pt}{\isacharprime}{\kern0pt}x{\isacharprime}{\kern0pt}{\isacharprime}{\kern0pt}\ {\isacharcolon}{\kern0pt}{\isacharequal}{\kern0pt}\ Num\ {\isadigit{2}}{\isacharsemicolon}{\kern0pt}{\isacharsemicolon}{\kern0pt}\ SKIP{\isacharbrackright}{\kern0pt}{\isacharcomma}{\kern0pt}\ \isanewline
\ \ \ \ \ \ \ \ \ \ \ \ \ \ \ \ \ \ {\isacharbraceleft}{\kern0pt}{\isacharbraceright}{\kern0pt}\ {\isasymtriangleright}\ {\isasymlangle}{\isasymsigma}\isactrlsub {\isadigit{1}}{\isasymrangle}\ {\isasymleadsto}\ State{\isasymllangle}{\isacharbrackleft}{\kern0pt}{\isacharprime}{\kern0pt}{\isacharprime}{\kern0pt}x{\isacharprime}{\kern0pt}{\isacharprime}{\kern0pt}\ {\isasymlongmapsto}\ Exp\ {\isacharparenleft}{\kern0pt}Num\ {\isadigit{2}}{\isacharparenright}{\kern0pt}{\isacharbrackright}{\kern0pt}\ {\isasymsigma}\isactrlsub {\isadigit{1}}{\isasymrrangle}\ \isactrlitem \ {\isasymlambda}{\isacharbrackleft}{\kern0pt}CO\ {\isacharprime}{\kern0pt}{\isacharprime}{\kern0pt}x{\isacharprime}{\kern0pt}{\isacharprime}{\kern0pt}\ {\isacharcolon}{\kern0pt}{\isacharequal}{\kern0pt}\ Num\ {\isadigit{1}}\ {\isasymparallel}\ SKIP\ OC{\isacharbrackright}{\kern0pt}\isanewline
\ \ \ \ \ \ \ \ \ \ \ \ \ \ {\isacharbraceright}{\kern0pt}{\isachardoublequoteclose}\ \isanewline
\isadelimproof
\ \ \ \ %
\endisadelimproof
\isatagproof
\isacommand{by}\isamarkupfalse%
\ {\isacharparenleft}{\kern0pt}simp\ add{\isacharcolon}{\kern0pt}\ {\isasymsigma}\isactrlsub {\isadigit{1}}{\isacharunderscore}{\kern0pt}def\ insert{\isacharunderscore}{\kern0pt}commute{\isacharparenright}{\kern0pt}%
\endisatagproof
{\isafoldproof}%
\isadelimproof
\isanewline
\endisadelimproof
\isanewline
\ \ \isacommand{lemma}\isamarkupfalse%
\ {\isachardoublequoteopen}val\isactrlsub s\ {\isacharparenleft}{\kern0pt}{\isasymlbrace}\ {\isacharparenleft}{\kern0pt}{\isacharprime}{\kern0pt}{\isacharprime}{\kern0pt}x{\isacharprime}{\kern0pt}{\isacharprime}{\kern0pt}{\isacharsemicolon}{\kern0pt}{\isasymnu}{\isacharparenright}{\kern0pt}\ {\isacharprime}{\kern0pt}{\isacharprime}{\kern0pt}x{\isacharprime}{\kern0pt}{\isacharprime}{\kern0pt}\ {\isacharcolon}{\kern0pt}{\isacharequal}{\kern0pt}\ Num\ {\isadigit{5}}\ {\isasymrbrace}{\isacharparenright}{\kern0pt}\ {\isasymsigma}\isactrlsub {\isadigit{1}}\ {\isacharequal}{\kern0pt}\ {\isacharbraceleft}{\kern0pt}\ \isanewline
\ \ \ \ \ \ \ \ \ \ \ \ \ \ \ \ \ \ {\isacharbraceleft}{\kern0pt}{\isacharbraceright}{\kern0pt}\ {\isasymtriangleright}\ {\isasymlangle}{\isasymsigma}\isactrlsub {\isadigit{1}}{\isasymrangle}\ {\isasymleadsto}\ State{\isasymllangle}{\isacharbrackleft}{\kern0pt}{\isacharprime}{\kern0pt}{\isacharprime}{\kern0pt}{\isachardollar}{\kern0pt}x{\isacharcolon}{\kern0pt}{\isacharcolon}{\kern0pt}Scope{\isacharprime}{\kern0pt}{\isacharprime}{\kern0pt}\ {\isasymlongmapsto}\ Exp\ {\isacharparenleft}{\kern0pt}Num\ {\isadigit{0}}{\isacharparenright}{\kern0pt}{\isacharbrackright}{\kern0pt}\ {\isasymsigma}\isactrlsub {\isadigit{1}}{\isasymrrangle}\ \isactrlitem \ {\isasymlambda}{\isacharbrackleft}{\kern0pt}{\isasymlbrace}{\isasymnu}\ {\isacharprime}{\kern0pt}{\isacharprime}{\kern0pt}{\isachardollar}{\kern0pt}x{\isacharcolon}{\kern0pt}{\isacharcolon}{\kern0pt}Scope{\isacharprime}{\kern0pt}{\isacharprime}{\kern0pt}\ {\isacharcolon}{\kern0pt}{\isacharequal}{\kern0pt}\ Num\ {\isadigit{5}}\ {\isasymrbrace}{\isacharbrackright}{\kern0pt}\ \isanewline
\ \ \ \ \ \ \ \ \ \ \ \ \ \ {\isacharbraceright}{\kern0pt}{\isachardoublequoteclose}\ \isanewline
\isadelimproof
\ \ \ \ %
\endisadelimproof
\isatagproof
\isacommand{by}\isamarkupfalse%
\ {\isacharparenleft}{\kern0pt}simp\ add{\isacharcolon}{\kern0pt}\ {\isasymsigma}\isactrlsub {\isadigit{1}}{\isacharunderscore}{\kern0pt}def\ eval{\isacharunderscore}{\kern0pt}nat{\isacharunderscore}{\kern0pt}numeral{\isacharparenright}{\kern0pt}%
\endisatagproof
{\isafoldproof}%
\isadelimproof
\isanewline
\endisadelimproof
\isanewline
\ \ \isacommand{lemma}\isamarkupfalse%
\ {\isachardoublequoteopen}val\isactrlsub s\ {\isacharparenleft}{\kern0pt}CALL\ {\isacharprime}{\kern0pt}{\isacharprime}{\kern0pt}foo{\isacharprime}{\kern0pt}{\isacharprime}{\kern0pt}\ {\isacharparenleft}{\kern0pt}Var\ {\isacharprime}{\kern0pt}{\isacharprime}{\kern0pt}x{\isacharprime}{\kern0pt}{\isacharprime}{\kern0pt}{\isacharparenright}{\kern0pt}{\isacharparenright}{\kern0pt}\ {\isasymsigma}\isactrlsub {\isadigit{2}}\ {\isacharequal}{\kern0pt}\ {\isacharbraceleft}{\kern0pt}\ \isanewline
\ \ \ \ \ \ \ \ \ \ \ \ \ \ \ \ \ \ {\isacharbraceleft}{\kern0pt}{\isacharbraceright}{\kern0pt}\ {\isasymtriangleright}\ {\isacharparenleft}{\kern0pt}{\isasymlangle}{\isasymsigma}\isactrlsub {\isadigit{2}}{\isasymrangle}\ {\isasymleadsto}\ Event{\isasymllangle}invEv{\isacharcomma}{\kern0pt}\ {\isacharbrackleft}{\kern0pt}P\ {\isacharprime}{\kern0pt}{\isacharprime}{\kern0pt}foo{\isacharprime}{\kern0pt}{\isacharprime}{\kern0pt}{\isacharcomma}{\kern0pt}\ A\ {\isacharparenleft}{\kern0pt}Num\ {\isadigit{8}}{\isacharparenright}{\kern0pt}{\isacharbrackright}{\kern0pt}{\isasymrrangle}{\isacharparenright}{\kern0pt}\ {\isasymleadsto}\ State{\isasymllangle}{\isasymsigma}\isactrlsub {\isadigit{2}}{\isasymrrangle}\ \isactrlitem \ {\isasymlambda}{\isacharbrackleft}{\kern0pt}{\isasymnabla}{\isacharbrackright}{\kern0pt}\isanewline
\ \ \ \ \ \ \ \ \ \ \ \ \ \ {\isacharbraceright}{\kern0pt}{\isachardoublequoteclose}\isanewline
\isadelimproof
\ \ \ \ %
\endisadelimproof
\isatagproof
\isacommand{by}\isamarkupfalse%
\ {\isacharparenleft}{\kern0pt}simp\ add{\isacharcolon}{\kern0pt}\ {\isasymsigma}\isactrlsub {\isadigit{2}}{\isacharunderscore}{\kern0pt}def\ eval{\isacharunderscore}{\kern0pt}nat{\isacharunderscore}{\kern0pt}numeral{\isacharparenright}{\kern0pt}%
\endisatagproof
{\isafoldproof}%
\isadelimproof
\endisadelimproof
\isadelimdocument
\endisadelimdocument
\isatagdocument
\isamarkupsubsection{Trace Composition%
}
\isamarkuptrue%
\isamarkupsubsubsection{Configurations%
}
\isamarkuptrue%
\endisatagdocument
{\isafolddocument}%
\isadelimdocument
\endisadelimdocument
\begin{isamarkuptext}%
In order to setup the trace composition for $WL_{EXT}$, a notion of program
  configurations becomes indispensable. In \isa{WL}, we introduced program
  configurations as tuples of symbolic traces and continuation markers. However,
  note that $WL_{EXT}$ expands the standard While Language with the call command. 
  The semantics of the call command enforce the caller and callee process to 
  execute in a concurrent manner, specifically depending on the underlying scheduling 
  algorithm. Considering that each process is associated with exactly one continuation 
  marker, the previous notion of program configurations (i.e. configurations only 
  containing one single continuation marker) is simply not strong~enough. \par
  We first introduce \isa{basic} program configurations as tuples consisting of a 
  symbolic trace and one single continuation marker. Note that this notion directly 
  corresponds to the notion of program configurations for \isa{WL}. We then
  introduce program configurations for $WL_{EXT}$ as tuples consisting of a symbolic 
  trace and a continuation marker multiset. This multiset will later contain the 
  continuation markers of all concurrently executing~processes. \par
  Note that the distinction between basic and non-basic program configurations 
  is not made in the original paper. However, both notions will later turn out
  to be useful for the purpose of establishing a high level of modularity in our 
  trace~composition.%
\end{isamarkuptext}\isamarkuptrue%
\ \ \isacommand{type{\isacharunderscore}{\kern0pt}synonym}\isamarkupfalse%
\ basic{\isacharunderscore}{\kern0pt}config\ {\isacharequal}{\kern0pt}\ {\isachardoublequoteopen}{\isasymT}\ {\isacharasterisk}{\kern0pt}\ cont{\isacharunderscore}{\kern0pt}marker{\isachardoublequoteclose}\ \isanewline
\ \ \isacommand{type{\isacharunderscore}{\kern0pt}synonym}\isamarkupfalse%
\ config\ {\isacharequal}{\kern0pt}\ {\isachardoublequoteopen}{\isasymT}\ {\isacharasterisk}{\kern0pt}\ {\isacharparenleft}{\kern0pt}cont{\isacharunderscore}{\kern0pt}marker\ multiset{\isacharparenright}{\kern0pt}{\isachardoublequoteclose}%
\isadelimdocument
\endisadelimdocument
\isatagdocument
\isamarkupsubsubsection{$\isa{{\isasymdelta}}_1$-function%
}
\isamarkuptrue%
\endisatagdocument
{\isafolddocument}%
\isadelimdocument
\endisadelimdocument
\begin{isamarkuptext}%
Our next objective is to faithfully adjust the successor function (\isa{{\isasymdelta}}-function),
  which maps each configuration onto the set of all possible successor configurations
  (i.e. all configurations reachable in one evaluation step). Considering that
  $WL_{EXT}$ is more complex than \isa{WL} (due to its higher expressivity), we 
  decide to split up the \isa{{\isasymdelta}}-function into two separate mappings, which correspond 
  to the two inductive rules given in the original~paper. \par
  We first aim to formalize the composition rule, which selects and removes
  one continuation marker from the corresponding multiset (i.e. schedules the
  associated process), evaluates it to the next scheduling point, and then inserts the 
  updated continuation marker back into the multiset. Also note that the 
  composed trace needs to be concretized after every evaluation step in order to 
  turn the local trace into a global~trace. \par 
  In order to ensure the modularity of our formalization, we choose to first introduce 
  a basic successor function~($\isa{{\isasymdelta}}_s$-function), which maps a basic configuration onto 
  all reachable basic successor configurations. This design choice later reduces the 
  complexity of the non-basic successor function ($\isa{{\isasymdelta}}_1$-function), whilst also 
  guaranteeing a higher~readability. \par
  The basic successor function is defined as follows: Let us assume we have a 
  trace (\isa{sh}~\isa{{\isasymleadsto}}~\isa{{\isasymsigma}}), whilst statement \isa{S} is still left to be evaluated. 
  \begin{description}
  \item[Step 1] We first collect all continuation traces \isa{{\isasymPi}} generated from \isa{S} in \isa{{\isasymsigma}}, 
  which have a consistent path condition. Note that we beforehand, similar to the 
  original paper, simplify each path condition under the minimal concretization mapping 
  of the corresponding continuation trace \isa{{\isasympi}\ {\isasymin}\ {\isasymPi}}, as the path condition could contain 
  symbolic variables (e.g. due to the input~statement). 
  \item[Step 2] The symbolic traces of all consistent continuation traces \isa{{\isasympi}\ {\isasymin}\ {\isasymPi}} 
  are then concretized, thus ensuring the concreteness of all composed traces.
  We can then translate the concretized symbolic traces and continuation markers into 
  corresponding concrete basic successor~configurations. 
  \end{description} 
  Note that our formalization again greatly deviates from the original paper, as we 
  do not model the composition using an inductive rule, but with a deterministic 
  function. Remember that this is done in order to circumvent quantifications over 
  infinitely many~traces. \par 
  In the original paper, there are no constraints set up for the
  applied concretization mapping, implying that any valid trace concretization
  mapping could be used during the trace composition. Considering that there are 
  infinitely many concretization mappings, this would entail that the set of all basic 
  successor configurations is infinite. Hence, this kind of model would strongly 
  interfere with our objective of providing an efficient code generation. We therefore 
  decide to deviate from the paper by requiring the corresponding trace concretization 
  mapping to be minimal. We furthermore also enforce that each concretization mapping 
  concretizes all symbolic variables with 0. Although this guarantees the finiteness 
  of all basic successor configurations, it also greatly restricts the reachable 
  configurations of our model. We therefore propose a more faithful 
  representation of the trace concretization as an idea for an extension of this~work.%
\end{isamarkuptext}\isamarkuptrue%
\ \ \isacommand{fun}\isamarkupfalse%
\isanewline
\ \ \ \ basic{\isacharunderscore}{\kern0pt}successors\ {\isacharcolon}{\kern0pt}{\isacharcolon}{\kern0pt}\ {\isachardoublequoteopen}basic{\isacharunderscore}{\kern0pt}config\ {\isasymRightarrow}\ basic{\isacharunderscore}{\kern0pt}config\ set{\isachardoublequoteclose}\ {\isacharparenleft}{\kern0pt}{\isachardoublequoteopen}{\isasymdelta}\isactrlsub s{\isachardoublequoteclose}{\isacharparenright}{\kern0pt}\ \ \isakeyword{where}\isanewline
\ \ \ \ {\isachardoublequoteopen}{\isasymdelta}\isactrlsub s\ {\isacharparenleft}{\kern0pt}sh\ {\isasymleadsto}\ State{\isasymllangle}{\isasymsigma}{\isasymrrangle}{\isacharcomma}{\kern0pt}\ {\isasymlambda}{\isacharbrackleft}{\kern0pt}S{\isacharbrackright}{\kern0pt}{\isacharparenright}{\kern0pt}\ {\isacharequal}{\kern0pt}\ \isanewline
\ \ \ \ \ \ \ \ {\isacharparenleft}{\kern0pt}{\isacharpercent}{\kern0pt}c{\isachardot}{\kern0pt}\ {\isacharparenleft}{\kern0pt}trace{\isacharunderscore}{\kern0pt}conc\ {\isacharparenleft}{\kern0pt}min{\isacharunderscore}{\kern0pt}conc{\isacharunderscore}{\kern0pt}map\isactrlsub {\isasymT}\ {\isacharparenleft}{\kern0pt}sh\ {\isasymcdot}\ {\isacharparenleft}{\kern0pt}{\isasymdown}\isactrlsub {\isasymtau}\ c{\isacharparenright}{\kern0pt}{\isacharparenright}{\kern0pt}\ {\isadigit{0}}{\isacharparenright}{\kern0pt}\ {\isacharparenleft}{\kern0pt}sh\ {\isasymcdot}\ {\isacharparenleft}{\kern0pt}{\isasymdown}\isactrlsub {\isasymtau}\ c{\isacharparenright}{\kern0pt}{\isacharparenright}{\kern0pt}{\isacharcomma}{\kern0pt}\ {\isasymdown}\isactrlsub {\isasymlambda}\ c{\isacharparenright}{\kern0pt}{\isacharparenright}{\kern0pt}\ \ \isanewline
\ \ \ \ \ \ \ \ {\isacharbackquote}{\kern0pt}\ {\isacharbraceleft}{\kern0pt}cont\ {\isasymin}\ {\isacharparenleft}{\kern0pt}val\isactrlsub s\ S\ {\isasymsigma}{\isacharparenright}{\kern0pt}{\isachardot}{\kern0pt}\ consistent{\isacharparenleft}{\kern0pt}sval\isactrlsub B\ {\isacharparenleft}{\kern0pt}{\isasymdown}\isactrlsub p\ cont{\isacharparenright}{\kern0pt}\ {\isacharparenleft}{\kern0pt}min{\isacharunderscore}{\kern0pt}conc{\isacharunderscore}{\kern0pt}map\isactrlsub {\isasymT}\ {\isacharparenleft}{\kern0pt}{\isasymdown}\isactrlsub {\isasymtau}\ cont{\isacharparenright}{\kern0pt}\ {\isadigit{0}}{\isacharparenright}{\kern0pt}{\isacharparenright}{\kern0pt}{\isacharbraceright}{\kern0pt}{\isachardoublequoteclose}\ {\isacharbar}{\kern0pt}\isanewline
\ \ \ \ {\isachardoublequoteopen}{\isasymdelta}\isactrlsub s\ {\isacharunderscore}{\kern0pt}\ {\isacharequal}{\kern0pt}\ undefined{\isachardoublequoteclose}%
\begin{isamarkuptext}%
By utilizing the basic successor function, we can now define the normal
  successor function ($\isa{{\isasymdelta}}_1$-function) as follows: Let us assume we have a
  trace (\isa{sh\ {\isasymleadsto}\ {\isasymsigma}}), while the statements of the continuation markers in multiset 
  \isa{q} still need to be evaluated. 
  \begin{description}
  \item[Step 1] We begin by translating the multiset \isa{q} into a normal set \isa{M}, such 
  that we can use the element-wise operator \isa{{\isacharbackquote}{\kern0pt}} in order to apply a function on every 
  continuation marker contained in \isa{q}. 
  \item[Step 2a] If the continuation marker \isa{cm\ {\isasymin}\ M} is empty, it returns the empty set. 
  Knowing that the process has already terminated, no successor configurations can be 
  generated. 
  \item[Step 2b] If the continuation marker \isa{cm\ {\isasymin}\ M} still contains a statement, we 
  utilize the basic successor function in order to map \isa{cm} onto the set of all 
  reachable basic successor configurations \isa{C}. We then translate all these basic 
  successor configurations \isa{cm{\isacharprime}{\kern0pt}\ {\isasymin}\ C} into non-basic successor configurations by 
  adding \isa{cm{\isacharprime}{\kern0pt}} to \isa{q{\isacharbackslash}{\kern0pt}{\isacharbraceleft}{\kern0pt}cm{\isacharbraceright}{\kern0pt}}. 
  \end{description} Note that this makes up the set of all successor configurations,
  which can be constructed by the continuation markers in \isa{q}. This smoothly aligns
  with what the $\isa{{\isasymdelta}}_1$-function is supposed~to~compute.%
\end{isamarkuptext}\isamarkuptrue%
\ \ \isacommand{fun}\isamarkupfalse%
\isanewline
\ \ \ \ successors\isactrlsub {\isadigit{1}}\ {\isacharcolon}{\kern0pt}{\isacharcolon}{\kern0pt}\ {\isachardoublequoteopen}config\ {\isasymRightarrow}\ config\ set{\isachardoublequoteclose}\ {\isacharparenleft}{\kern0pt}{\isachardoublequoteopen}{\isasymdelta}\isactrlsub {\isadigit{1}}{\isachardoublequoteclose}{\isacharparenright}{\kern0pt}\ \ \isakeyword{where}\isanewline
\ \ \ \ {\isachardoublequoteopen}{\isasymdelta}\isactrlsub {\isadigit{1}}\ {\isacharparenleft}{\kern0pt}sh\ {\isasymleadsto}\ State{\isasymllangle}{\isasymsigma}{\isasymrrangle}{\isacharcomma}{\kern0pt}\ q{\isacharparenright}{\kern0pt}\ {\isacharequal}{\kern0pt}\ \isanewline
\ \ \ \ \ \ \ \ {\isasymUnion}{\isacharparenleft}{\kern0pt}{\isacharparenleft}{\kern0pt}{\isacharpercent}{\kern0pt}cm{\isachardot}{\kern0pt}\ {\isacharparenleft}{\kern0pt}if\ cm\ {\isacharequal}{\kern0pt}\ {\isasymlambda}{\isacharbrackleft}{\kern0pt}{\isasymnabla}{\isacharbrackright}{\kern0pt}\ then\ {\isacharbraceleft}{\kern0pt}{\isacharbraceright}{\kern0pt}\ else\ {\isacharparenleft}{\kern0pt}{\isacharpercent}{\kern0pt}c{\isachardot}{\kern0pt}\ {\isacharparenleft}{\kern0pt}fst{\isacharparenleft}{\kern0pt}c{\isacharparenright}{\kern0pt}{\isacharcomma}{\kern0pt}\ {\isacharparenleft}{\kern0pt}q\ {\isacharminus}{\kern0pt}\ {\isacharbraceleft}{\kern0pt}{\isacharhash}{\kern0pt}\ cm\ {\isacharhash}{\kern0pt}{\isacharbraceright}{\kern0pt}{\isacharparenright}{\kern0pt}\ {\isacharplus}{\kern0pt}\ {\isacharbraceleft}{\kern0pt}{\isacharhash}{\kern0pt}\ snd{\isacharparenleft}{\kern0pt}c{\isacharparenright}{\kern0pt}\ {\isacharhash}{\kern0pt}{\isacharbraceright}{\kern0pt}{\isacharparenright}{\kern0pt}{\isacharparenright}{\kern0pt}\ \isanewline
\ \ \ \ \ \ \ \ \ \ \ \ {\isacharbackquote}{\kern0pt}\ {\isacharparenleft}{\kern0pt}{\isasymdelta}\isactrlsub s\ {\isacharparenleft}{\kern0pt}sh\ {\isasymleadsto}\ State{\isasymllangle}{\isasymsigma}{\isasymrrangle}{\isacharcomma}{\kern0pt}\ cm{\isacharparenright}{\kern0pt}{\isacharparenright}{\kern0pt}{\isacharparenright}{\kern0pt}{\isacharparenright}{\kern0pt}\ \isanewline
\ \ \ \ \ \ \ \ \ \ {\isacharbackquote}{\kern0pt}\ set{\isacharunderscore}{\kern0pt}mset\ q{\isacharparenright}{\kern0pt}{\isachardoublequoteclose}\ {\isacharbar}{\kern0pt}\isanewline
\ \ \ \ {\isachardoublequoteopen}{\isasymdelta}\isactrlsub {\isadigit{1}}\ {\isacharunderscore}{\kern0pt}\ {\isacharequal}{\kern0pt}\ undefined{\isachardoublequoteclose}%
\isadelimdocument
\endisadelimdocument
\isatagdocument
\isamarkupsubsubsection{$\isa{{\isasymdelta}}_2$-function%
}
\isamarkuptrue%
\endisatagdocument
{\isafolddocument}%
\isadelimdocument
\endisadelimdocument
\begin{isamarkuptext}%
The $\isa{{\isasymdelta}}_1$-function allows the evaluation of call statements, which insert
  correlating invocation events into the symbolic trace during the composition
  procedure. However, we have not yet modeled the corresponding reaction of the called
  methods (i.e. the process creation). This motivates the definition of a separate 
  deterministic function ($\isa{{\isasymdelta}}_2$-function), which maps a given program configuration 
  onto all successor configurations containing a newly created process. Note that
  we will only allow a reaction to occur iff a corresponding method invocation 
  took place~beforehand. \par
  Before we can begin with the actual formalization of this function, we first 
  introduce a helper function, which counts the occurrences of a specific trace atom 
  in a provided symbolic trace. The definition of this function is straightforward
  due to the use of~recursion.%
\end{isamarkuptext}\isamarkuptrue%
\ \ \isacommand{fun}\isamarkupfalse%
\isanewline
\ \ \ \ counter\ {\isacharcolon}{\kern0pt}{\isacharcolon}{\kern0pt}\ {\isachardoublequoteopen}{\isasymT}\ {\isasymRightarrow}\ trace{\isacharunderscore}{\kern0pt}atom\ {\isasymRightarrow}\ nat{\isachardoublequoteclose}\ {\isacharparenleft}{\kern0pt}{\isachardoublequoteopen}{\isacharhash}{\kern0pt}\isactrlsub {\isasymT}{\isachardoublequoteclose}\ {\isadigit{6}}{\isadigit{5}}{\isacharparenright}{\kern0pt}\ \isakeyword{where}\isanewline
\ \ \ \ {\isachardoublequoteopen}{\isacharparenleft}{\kern0pt}{\isacharhash}{\kern0pt}\isactrlsub {\isasymT}{\isacharparenright}{\kern0pt}\ {\isasymepsilon}\ ta\ {\isacharequal}{\kern0pt}\ {\isadigit{0}}{\isachardoublequoteclose}\ {\isacharbar}{\kern0pt}\isanewline
\ \ \ \ {\isachardoublequoteopen}{\isacharparenleft}{\kern0pt}{\isacharhash}{\kern0pt}\isactrlsub {\isasymT}{\isacharparenright}{\kern0pt}\ {\isacharparenleft}{\kern0pt}{\isasymtau}\ {\isasymleadsto}\ Event{\isasymllangle}ev{\isacharcomma}{\kern0pt}\ e{\isasymrrangle}{\isacharparenright}{\kern0pt}\ ta\ {\isacharequal}{\kern0pt}\ {\isacharparenleft}{\kern0pt}if\ ta\ {\isacharequal}{\kern0pt}\ Event{\isasymllangle}ev{\isacharcomma}{\kern0pt}\ e{\isasymrrangle}\ then\ {\isadigit{1}}\ {\isacharplus}{\kern0pt}\ {\isacharparenleft}{\kern0pt}{\isacharhash}{\kern0pt}\isactrlsub {\isasymT}{\isacharparenright}{\kern0pt}\ {\isasymtau}\ ta\ else\ {\isacharparenleft}{\kern0pt}{\isacharhash}{\kern0pt}\isactrlsub {\isasymT}{\isacharparenright}{\kern0pt}\ {\isasymtau}\ ta{\isacharparenright}{\kern0pt}{\isachardoublequoteclose}\ {\isacharbar}{\kern0pt}\isanewline
\ \ \ \ {\isachardoublequoteopen}{\isacharparenleft}{\kern0pt}{\isacharhash}{\kern0pt}\isactrlsub {\isasymT}{\isacharparenright}{\kern0pt}\ {\isacharparenleft}{\kern0pt}{\isasymtau}\ {\isasymleadsto}\ State{\isasymllangle}{\isasymsigma}{\isasymrrangle}{\isacharparenright}{\kern0pt}\ ta\ {\isacharequal}{\kern0pt}\ {\isacharparenleft}{\kern0pt}if\ ta\ {\isacharequal}{\kern0pt}\ State{\isasymllangle}{\isasymsigma}{\isasymrrangle}\ then\ {\isadigit{1}}\ {\isacharplus}{\kern0pt}\ {\isacharparenleft}{\kern0pt}{\isacharhash}{\kern0pt}\isactrlsub {\isasymT}{\isacharparenright}{\kern0pt}\ {\isasymtau}\ ta\ else\ {\isacharparenleft}{\kern0pt}{\isacharhash}{\kern0pt}\isactrlsub {\isasymT}{\isacharparenright}{\kern0pt}\ {\isasymtau}\ ta{\isacharparenright}{\kern0pt}{\isachardoublequoteclose}%
\begin{isamarkuptext}%
We can now setup a separate wellformedness condition on traces,
  which will later ensure that processes can only be created iff they have been 
  invocated at an earlier point of the program. Whilst an invocation event 
  represents that a method is called, an invocation reaction event models the 
  corresponding reaction of the callee (i.e. the process creation). We call a 
  trace during the composition procedure wellformed iff there is an injective
  function that maps every invocation reaction event ocurring in the trace onto 
  a matching (preceding) invocation event. Note that we can easily formalize this 
  predicate by recursively traversing the provided trace, whilst checking that every 
  invocation reaction~event has been preceded by a matching incovation event
  that has not yet been~reacted~to.%
\end{isamarkuptext}\isamarkuptrue%
\ \ \isacommand{fun}\isamarkupfalse%
\isanewline
\ \ \ \ wellformed\ {\isacharcolon}{\kern0pt}{\isacharcolon}{\kern0pt}\ {\isachardoublequoteopen}{\isasymT}\ {\isasymRightarrow}\ bool{\isachardoublequoteclose}\ \isakeyword{where}\isanewline
\ \ \ \ {\isachardoublequoteopen}wellformed\ {\isasymepsilon}\ {\isacharequal}{\kern0pt}\ True{\isachardoublequoteclose}\ {\isacharbar}{\kern0pt}\isanewline
\ \ \ \ {\isachardoublequoteopen}wellformed\ {\isacharparenleft}{\kern0pt}{\isasymtau}\ {\isasymleadsto}\ Event{\isasymllangle}invREv{\isacharcomma}{\kern0pt}\ e{\isasymrrangle}{\isacharparenright}{\kern0pt}\ {\isacharequal}{\kern0pt}\ {\isacharparenleft}{\kern0pt}wellformed{\isacharparenleft}{\kern0pt}{\isasymtau}{\isacharparenright}{\kern0pt}\ {\isasymand}\ {\isacharparenleft}{\kern0pt}{\isacharhash}{\kern0pt}\isactrlsub {\isasymT}{\isacharparenright}{\kern0pt}\ {\isasymtau}\ {\isacharparenleft}{\kern0pt}Event{\isasymllangle}invEv{\isacharcomma}{\kern0pt}\ e{\isasymrrangle}{\isacharparenright}{\kern0pt}\ {\isachargreater}{\kern0pt}\ {\isacharparenleft}{\kern0pt}{\isacharhash}{\kern0pt}\isactrlsub {\isasymT}{\isacharparenright}{\kern0pt}\ {\isasymtau}\ {\isacharparenleft}{\kern0pt}Event{\isasymllangle}invREv{\isacharcomma}{\kern0pt}\ e{\isasymrrangle}{\isacharparenright}{\kern0pt}{\isacharparenright}{\kern0pt}{\isachardoublequoteclose}\ {\isacharbar}{\kern0pt}\isanewline
\ \ \ \ {\isachardoublequoteopen}wellformed\ {\isacharparenleft}{\kern0pt}{\isasymtau}\ {\isasymleadsto}\ t{\isacharparenright}{\kern0pt}\ {\isacharequal}{\kern0pt}\ wellformed{\isacharparenleft}{\kern0pt}{\isasymtau}{\isacharparenright}{\kern0pt}{\isachardoublequoteclose}%
\begin{isamarkuptext}%
We also provide another helper function, which projects a symbolic trace \isa{{\isasymtau}}
  onto all method arguments that were passed in invocation events occurring
  in \isa{{\isasymtau}}. Every invocation reaction event occurring directly after \isa{last{\isacharparenleft}{\kern0pt}{\isasymtau}{\isacharparenright}{\kern0pt}} 
  will only be allowed to receive an argument from a method call that 
  has already been executed at an earlier point. Hence, this definition restricts 
  the infinite set of possible actual parameters received in invocation reaction 
  events onto a finite set. Note that this will turn out to be a major advantage 
  when trying to setup the code generation for the~$\isa{{\isasymdelta}}_2$-function.%
\end{isamarkuptext}\isamarkuptrue%
\ \ \isacommand{fun}\isamarkupfalse%
\isanewline
\ \ \ \ params\ {\isacharcolon}{\kern0pt}{\isacharcolon}{\kern0pt}\ {\isachardoublequoteopen}{\isasymT}\ {\isasymRightarrow}\ exp\ set{\isachardoublequoteclose}\ \isakeyword{where}\isanewline
\ \ \ \ {\isachardoublequoteopen}params\ {\isasymepsilon}\ {\isacharequal}{\kern0pt}\ {\isacharbraceleft}{\kern0pt}{\isacharbraceright}{\kern0pt}{\isachardoublequoteclose}\ {\isacharbar}{\kern0pt}\isanewline
\ \ \ \ {\isachardoublequoteopen}params\ {\isacharparenleft}{\kern0pt}{\isasymtau}\ {\isasymleadsto}\ Event{\isasymllangle}invEv{\isacharcomma}{\kern0pt}\ {\isacharbrackleft}{\kern0pt}P\ m{\isacharcomma}{\kern0pt}\ A\ a{\isacharbrackright}{\kern0pt}{\isasymrrangle}{\isacharparenright}{\kern0pt}\ {\isacharequal}{\kern0pt}\ params{\isacharparenleft}{\kern0pt}{\isasymtau}{\isacharparenright}{\kern0pt}\ {\isasymunion}\ {\isacharbraceleft}{\kern0pt}A\ a{\isacharbraceright}{\kern0pt}{\isachardoublequoteclose}\ {\isacharbar}{\kern0pt}\isanewline
\ \ \ \ {\isachardoublequoteopen}params\ {\isacharparenleft}{\kern0pt}{\isasymtau}\ {\isasymleadsto}\ t{\isacharparenright}{\kern0pt}\ {\isacharequal}{\kern0pt}\ params{\isacharparenleft}{\kern0pt}{\isasymtau}{\isacharparenright}{\kern0pt}{\isachardoublequoteclose}%
\begin{isamarkuptext}%
Using the helper functions above, we can finally provide a definition
  for the $\isa{{\isasymdelta}}_2$-function. We formalize the function as follows: Let us assume
  we have a trace (\isa{sh\ {\isasymleadsto}\ {\isasymsigma}}), while the statements of the continuation markers 
  in multiset \isa{q} still need to be evaluated.
  \begin{description}
  \item[Step 1] We begin by computing all tuples of methods and arithmetic 
    method arguments \isa{M\ {\isasymtimes}\ params{\isacharparenleft}{\kern0pt}sh{\isacharparenright}{\kern0pt}}, which could be used as parameters for invocation 
    reaction events directly after state~\isa{{\isasymsigma}}. Note that we use the \isa{filter}-function of 
    the \isa{Set}-theory in order to filter out all tuples, which would violate the 
    previously established notion of wellformedness. This results in the set of all 
    allowed tuples \isa{T}. \par
    Note that it would be more intuitive to directly filter out the wellformed
    tuples out of the \isa{M\ {\isasymtimes}\ aexp} tuple set, as the result would be the same. 
    However, this model would interfere with our code generation, considering that
    the type of arithmetic expressions is infinitely big. Hence, we decide to 
    circumvent this by utilizing the previously defined helper function \isa{params},
    thus ensuring that we only have to consider finitely many (concrete) 
    arithmetic~arguments.
  \item[Step 2] We can now use the element-wise operator \isa{{\isacharbackquote}{\kern0pt}} in order to map
    each of the previously established tuples \isa{{\isacharparenleft}{\kern0pt}m{\isacharcomma}{\kern0pt}\ v{\isacharparenright}{\kern0pt}\ {\isasymin}\ T} onto a corresponding 
    successor configuration. For this purpose, \isa{sh} is expanded by appending
    the correlating invocation reaction event, which consists of \isa{m{\isacharprime}{\kern0pt}s} method name
    and argument \isa{v} as parameters. The trace afterwards transits into an updated
    version of \isa{{\isasymsigma}}, in which a freshly generated variable \isa{x{\isacharprime}{\kern0pt}} maps onto \isa{v}. Note
    that \isa{x{\isacharprime}{\kern0pt}} represents the disambiguated call parameter (i.e. formal parameter) 
    of the~callee. \par
    We then merge \isa{q} with the continuation marker consisting of \isa{m{\isacharprime}{\kern0pt}s} method 
    body in order to result in the multiset of the desired successor configuration. 
    This represents the creation of a new process. Note that we additionally have 
    to substitute every occurrence of the formal parameter \isa{x} in \isa{m} with the 
    disambiguated call parameter \isa{x{\isacharprime}{\kern0pt}}, so as to avoid possible variable~conflicts. 
  \end{description} In contrast to the paper, we again use our self-defined 
  deterministic variable generation function in order to generate the fresh
  disambiguated call parameter \isa{x{\isacharprime}{\kern0pt}}. This ensures the existence of only finitely 
  many successor configurations, hence not endangering our code~generation. \par
  Note that the formalization of this function is based on the draft of the original
  paper from June 2021, thus we are slightly deviating from its final version. We
  therefore propose a faithful adaption of this function as an idea for further
  work on this~model.%
\end{isamarkuptext}\isamarkuptrue%
\ \ \isacommand{fun}\isamarkupfalse%
\isanewline
\ \ \ \ successors\isactrlsub {\isadigit{2}}\ {\isacharcolon}{\kern0pt}{\isacharcolon}{\kern0pt}\ {\isachardoublequoteopen}method\ set\ {\isasymRightarrow}\ config\ {\isasymRightarrow}\ config\ set{\isachardoublequoteclose}\ {\isacharparenleft}{\kern0pt}{\isachardoublequoteopen}{\isasymdelta}\isactrlsub {\isadigit{2}}{\isachardoublequoteclose}{\isacharparenright}{\kern0pt}\ \ \isakeyword{where}\isanewline
\ \ \ \ {\isachardoublequoteopen}{\isasymdelta}\isactrlsub {\isadigit{2}}\ M\ {\isacharparenleft}{\kern0pt}sh\ {\isasymleadsto}\ State{\isasymllangle}{\isasymsigma}{\isasymrrangle}{\isacharcomma}{\kern0pt}\ q{\isacharparenright}{\kern0pt}\ {\isacharequal}{\kern0pt}\ \isanewline
\ \ \ \ \ \ \ \ {\isacharparenleft}{\kern0pt}{\isacharpercent}{\kern0pt}{\isacharparenleft}{\kern0pt}m{\isacharcomma}{\kern0pt}\ v{\isacharparenright}{\kern0pt}{\isachardot}{\kern0pt}\ {\isacharparenleft}{\kern0pt}{\isacharparenleft}{\kern0pt}{\isacharparenleft}{\kern0pt}sh\ {\isasymcdot}\ {\isacharparenleft}{\kern0pt}gen{\isacharunderscore}{\kern0pt}event\ invREv\ {\isasymsigma}\ {\isacharbrackleft}{\kern0pt}P\ {\isacharparenleft}{\kern0pt}{\isasymUp}\isactrlsub n\ m{\isacharparenright}{\kern0pt}{\isacharcomma}{\kern0pt}\ v{\isacharbrackright}{\kern0pt}{\isacharparenright}{\kern0pt}{\isacharparenright}{\kern0pt}\ \isanewline
\ \ \ \ \ \ \ \ \ \ \ \ \ \ \ \ \ \ \ \ \ \ {\isasymleadsto}\ State{\isasymllangle}{\isacharbrackleft}{\kern0pt}{\isacharparenleft}{\kern0pt}vargen\ {\isasymsigma}\ {\isadigit{0}}\ {\isadigit{1}}{\isadigit{0}}{\isadigit{0}}\ {\isacharparenleft}{\kern0pt}{\isacharprime}{\kern0pt}{\isacharprime}{\kern0pt}{\isachardollar}{\kern0pt}{\isacharprime}{\kern0pt}{\isacharprime}{\kern0pt}\ {\isacharat}{\kern0pt}\ {\isacharparenleft}{\kern0pt}{\isasymUp}\isactrlsub n\ m{\isacharparenright}{\kern0pt}\ {\isacharat}{\kern0pt}\ {\isacharprime}{\kern0pt}{\isacharprime}{\kern0pt}{\isacharcolon}{\kern0pt}{\isacharcolon}{\kern0pt}Param{\isacharprime}{\kern0pt}{\isacharprime}{\kern0pt}{\isacharparenright}{\kern0pt}{\isacharparenright}{\kern0pt}\ {\isasymlongmapsto}\ Exp\ {\isacharparenleft}{\kern0pt}proj\isactrlsub A\ v{\isacharparenright}{\kern0pt}{\isacharbrackright}{\kern0pt}\ {\isasymsigma}{\isasymrrangle}{\isacharparenright}{\kern0pt}{\isacharcomma}{\kern0pt}\ \isanewline
\ \ \ \ \ \ \ \ \ \ \ \ \ \ \ \ \ \ q\ {\isacharplus}{\kern0pt}\ {\isacharbraceleft}{\kern0pt}{\isacharhash}{\kern0pt}\ {\isasymlambda}{\isacharbrackleft}{\kern0pt}{\isacharparenleft}{\kern0pt}{\isasymUp}\isactrlsub s\ m{\isacharparenright}{\kern0pt}\ {\isacharbrackleft}{\kern0pt}{\isacharparenleft}{\kern0pt}{\isasymUp}\isactrlsub v\ m{\isacharparenright}{\kern0pt}\ {\isasymleftarrow}\isactrlsub s\ {\isacharparenleft}{\kern0pt}vargen\ {\isasymsigma}\ {\isadigit{0}}\ {\isadigit{1}}{\isadigit{0}}{\isadigit{0}}\ {\isacharparenleft}{\kern0pt}{\isacharprime}{\kern0pt}{\isacharprime}{\kern0pt}{\isachardollar}{\kern0pt}{\isacharprime}{\kern0pt}{\isacharprime}{\kern0pt}\ {\isacharat}{\kern0pt}\ {\isacharparenleft}{\kern0pt}{\isasymUp}\isactrlsub n\ m{\isacharparenright}{\kern0pt}\ {\isacharat}{\kern0pt}\ {\isacharprime}{\kern0pt}{\isacharprime}{\kern0pt}{\isacharcolon}{\kern0pt}{\isacharcolon}{\kern0pt}Param{\isacharprime}{\kern0pt}{\isacharprime}{\kern0pt}{\isacharparenright}{\kern0pt}{\isacharparenright}{\kern0pt}{\isacharbrackright}{\kern0pt}\ {\isacharbrackright}{\kern0pt}\ {\isacharhash}{\kern0pt}{\isacharbraceright}{\kern0pt}{\isacharparenright}{\kern0pt}{\isacharparenright}{\kern0pt}\ \isanewline
\ \ \ \ \ \ \ \ {\isacharbackquote}{\kern0pt}\ Set{\isachardot}{\kern0pt}filter\ {\isacharparenleft}{\kern0pt}{\isacharpercent}{\kern0pt}{\isacharparenleft}{\kern0pt}m{\isacharcomma}{\kern0pt}\ v{\isacharparenright}{\kern0pt}{\isachardot}{\kern0pt}\ wellformed\ {\isacharparenleft}{\kern0pt}sh\ {\isasymcdot}\ {\isacharparenleft}{\kern0pt}gen{\isacharunderscore}{\kern0pt}event\ invREv\ {\isasymsigma}\ {\isacharbrackleft}{\kern0pt}P\ {\isacharparenleft}{\kern0pt}{\isasymUp}\isactrlsub n\ m{\isacharparenright}{\kern0pt}{\isacharcomma}{\kern0pt}\ v{\isacharbrackright}{\kern0pt}{\isacharparenright}{\kern0pt}{\isacharparenright}{\kern0pt}{\isacharparenright}{\kern0pt}\ {\isacharparenleft}{\kern0pt}M\ {\isasymtimes}\ params{\isacharparenleft}{\kern0pt}sh{\isacharparenright}{\kern0pt}{\isacharparenright}{\kern0pt}{\isachardoublequoteclose}\ {\isacharbar}{\kern0pt}\isanewline
\ \ \ \ {\isachardoublequoteopen}{\isasymdelta}\isactrlsub {\isadigit{2}}\ M\ {\isacharunderscore}{\kern0pt}\ {\isacharequal}{\kern0pt}\ undefined{\isachardoublequoteclose}%
\isadelimdocument
\endisadelimdocument
\isatagdocument
\isamarkupsubsubsection{\isa{{\isasymdelta}}-function%
}
\isamarkuptrue%
\endisatagdocument
{\isafolddocument}%
\isadelimdocument
\endisadelimdocument
\begin{isamarkuptext}%
In order to map a program configuration onto all possible successor 
  configurations, we can now merge the results of the $\isa{{\isasymdelta}}_1$ and 
  $\isa{{\isasymdelta}}_2$-function, thereby establishing the definition of the \isa{{\isasymdelta}}-function.
  Note that each \isa{{\isasymdelta}}-application can only result in finitely many successor 
  configurations due to the earlier established properties of the $\isa{{\isasymdelta}}_1$ and 
  $\isa{{\isasymdelta}}_2$-function.%
\end{isamarkuptext}\isamarkuptrue%
\ \ \isacommand{fun}\isamarkupfalse%
\isanewline
\ \ \ \ successors\ {\isacharcolon}{\kern0pt}{\isacharcolon}{\kern0pt}\ {\isachardoublequoteopen}method\ set\ {\isasymRightarrow}\ config\ {\isasymRightarrow}\ config\ set{\isachardoublequoteclose}\ {\isacharparenleft}{\kern0pt}{\isachardoublequoteopen}{\isasymdelta}{\isachardoublequoteclose}{\isacharparenright}{\kern0pt}\ \isakeyword{where}\isanewline
\ \ \ \ {\isachardoublequoteopen}{\isasymdelta}\ M\ c\ {\isacharequal}{\kern0pt}\ {\isasymdelta}\isactrlsub {\isadigit{1}}\ c\ {\isasymunion}\ {\isasymdelta}\isactrlsub {\isadigit{2}}\ M\ c{\isachardoublequoteclose}%
\isadelimdocument
\endisadelimdocument
\isatagdocument
\isamarkupsubsubsection{Proof Automation%
}
\isamarkuptrue%
\endisatagdocument
{\isafolddocument}%
\isadelimdocument
\endisadelimdocument
\begin{isamarkuptext}%
In order to establish an automated proof system for the construction of 
  global traces, we now desire to provide simplification lemmas for the $\isa{{\isasymdelta}}_s$
  and $\isa{{\isasymdelta}}_2$-function. \par
  We start by providing corresponding simplifications for the $\isa{{\isasymdelta}}_s$-function. For
  this purpose, we aim to define a general simplification lemma for each statement of 
  $WL_{EXT}$. This will lay the groundwork for efficient proof derivations, as 
  we can later simply utilize our simplification lemmas when deriving successor 
  configurations, hence avoiding having to deal with the underlying valuation function. 
  Note that $WL_{EXT}$ is non-deterministic due to its notion of concurrency, thus 
  implying that an application of the $\isa{{\isasymdelta}}_s$-function may return multiple 
  successor~configurations. \par
  Our simplification lemmas will furthermore assume the concreteness of the trace
  provided as the function argument. This ensures that we only have to deal with 
  trace concretizations after evaluations of input statements, as it is the only
  statement that introduces symbolic variables. The concretizations for 
  any other statement do not change the composed trace, as we have already
  proven that the concretization of a concrete trace under its minimal
  concretization mapping preserves the original trace. \par
  This concreteness assumption is reasonable, and not a problem, as we always concretize 
  all composed traces after every evaluation step of the trace composition, which 
  in turn implies that the resulting trace will always be concrete. Hence, our proof 
  derivation can always be applied, as long as we start our trace composition in a 
  concrete state. Note that we make this assumption in order to greatly reduce the 
  complexity of the following simplification~lemmas.%
\end{isamarkuptext}\isamarkuptrue%
\ \ \isacommand{context}\isamarkupfalse%
\ \isakeyword{notes}\ {\isacharbrackleft}{\kern0pt}simp{\isacharbrackright}{\kern0pt}\ {\isacharequal}{\kern0pt}\ consistent{\isacharunderscore}{\kern0pt}def\ concrete\isactrlsub {\isasymSigma}{\isacharunderscore}{\kern0pt}def\ \isakeyword{begin}\isanewline
\isanewline
\ \ \isacommand{lemma}\isamarkupfalse%
\ {\isasymdelta}\isactrlsub s{\isacharunderscore}{\kern0pt}Skip{\isacharcolon}{\kern0pt}\isanewline
\ \ \ \ \isakeyword{assumes}\ {\isachardoublequoteopen}concrete\isactrlsub {\isasymT}{\isacharparenleft}{\kern0pt}sh\ {\isasymleadsto}\ State{\isasymllangle}{\isasymsigma}{\isasymrrangle}{\isacharparenright}{\kern0pt}{\isachardoublequoteclose}\ \isanewline
\ \ \ \ \isakeyword{shows}\ {\isachardoublequoteopen}{\isasymdelta}\isactrlsub s\ {\isacharparenleft}{\kern0pt}sh\ {\isasymleadsto}\ State{\isasymllangle}{\isasymsigma}{\isasymrrangle}{\isacharcomma}{\kern0pt}\ {\isasymlambda}{\isacharbrackleft}{\kern0pt}SKIP{\isacharbrackright}{\kern0pt}{\isacharparenright}{\kern0pt}\ {\isacharequal}{\kern0pt}\ {\isacharbraceleft}{\kern0pt}{\isacharparenleft}{\kern0pt}sh\ {\isasymleadsto}\ State{\isasymllangle}{\isasymsigma}{\isasymrrangle}{\isacharcomma}{\kern0pt}\ {\isasymlambda}{\isacharbrackleft}{\kern0pt}{\isasymnabla}{\isacharbrackright}{\kern0pt}{\isacharparenright}{\kern0pt}{\isacharbraceright}{\kern0pt}{\isachardoublequoteclose}\isanewline
\isadelimproof
\ \ %
\endisadelimproof
\isatagproof
\isacommand{proof}\isamarkupfalse%
\ {\isacharminus}{\kern0pt}\isanewline
\ \ \ \ \isacommand{have}\isamarkupfalse%
\ {\isachardoublequoteopen}min{\isacharunderscore}{\kern0pt}conc{\isacharunderscore}{\kern0pt}map\isactrlsub {\isasymT}\ {\isacharparenleft}{\kern0pt}sh\ {\isasymleadsto}\ State{\isasymllangle}{\isasymsigma}{\isasymrrangle}{\isacharparenright}{\kern0pt}\ {\isadigit{0}}\ {\isacharequal}{\kern0pt}\ {\isasymcircle}{\isachardoublequoteclose}\isanewline
\ \ \ \ \ \ \isacommand{using}\isamarkupfalse%
\ assms\ min{\isacharunderscore}{\kern0pt}conc{\isacharunderscore}{\kern0pt}map{\isacharunderscore}{\kern0pt}of{\isacharunderscore}{\kern0pt}concrete\isactrlsub {\isasymT}\ \isacommand{by}\isamarkupfalse%
\ presburger\isanewline
\ \ \ \ %
\isamarkupcmt{Considering that \isa{{\isacharparenleft}{\kern0pt}sh\ {\isasymleadsto}\ {\isasymsigma}{\isacharparenright}{\kern0pt}} is concrete, it contains no symbolic variables.
       Using the \isa{min{\isacharunderscore}{\kern0pt}conc{\isacharunderscore}{\kern0pt}map{\isacharunderscore}{\kern0pt}of{\isacharunderscore}{\kern0pt}concrete\isactrlsub {\isasymT}} theorem, we can then infer that the minimal 
       concretization mapping of \isa{{\isacharparenleft}{\kern0pt}sh\ {\isasymleadsto}\ {\isasymsigma}{\isacharparenright}{\kern0pt}} matches the empty state.%
}\isanewline
\ \ \ \ \isacommand{hence}\isamarkupfalse%
\ {\isachardoublequoteopen}trace{\isacharunderscore}{\kern0pt}conc\ {\isacharparenleft}{\kern0pt}min{\isacharunderscore}{\kern0pt}conc{\isacharunderscore}{\kern0pt}map\isactrlsub {\isasymT}\ {\isacharparenleft}{\kern0pt}sh\ {\isasymleadsto}\ State{\isasymllangle}{\isasymsigma}{\isasymrrangle}{\isacharparenright}{\kern0pt}\ {\isadigit{0}}{\isacharparenright}{\kern0pt}\ {\isacharparenleft}{\kern0pt}sh\ {\isasymleadsto}\ State{\isasymllangle}{\isasymsigma}{\isasymrrangle}{\isacharparenright}{\kern0pt}\ {\isacharequal}{\kern0pt}\ sh\ {\isasymleadsto}\ State{\isasymllangle}{\isasymsigma}{\isasymrrangle}{\isachardoublequoteclose}\isanewline
\ \ \ \ \ \ \isacommand{using}\isamarkupfalse%
\ assms\ trace{\isacharunderscore}{\kern0pt}conc{\isacharunderscore}{\kern0pt}pr\ \isacommand{by}\isamarkupfalse%
\ fastforce\isanewline
\ \ \ \ %
\isamarkupcmt{Utilizing a supplementary theorem, we conclude that the trace concretization of 
       \isa{{\isacharparenleft}{\kern0pt}sh\ {\isasymleadsto}\ {\isasymsigma}{\isacharparenright}{\kern0pt}} under its minimal concretization mapping must be \isa{{\isacharparenleft}{\kern0pt}sh\ {\isasymleadsto}\ {\isasymsigma}{\isacharparenright}{\kern0pt}} itself.%
}\isanewline
\ \ \ \ \isacommand{thus}\isamarkupfalse%
\ {\isacharquery}{\kern0pt}thesis\ \isacommand{by}\isamarkupfalse%
\ auto\isanewline
\ \ \ \ %
\isamarkupcmt{We can now use Isabelle in order to infer the conclusion of the lemma.%
}\isanewline
\ \ \isacommand{qed}\isamarkupfalse%
\endisatagproof
{\isafoldproof}%
\isadelimproof
\isanewline
\endisadelimproof
\isanewline
\ \ \isacommand{lemma}\isamarkupfalse%
\ {\isasymdelta}\isactrlsub s{\isacharunderscore}{\kern0pt}Assign{\isacharcolon}{\kern0pt}\isanewline
\ \ \ \ \isakeyword{assumes}\ {\isachardoublequoteopen}concrete\isactrlsub {\isasymT}{\isacharparenleft}{\kern0pt}sh\ {\isasymleadsto}\ State{\isasymllangle}{\isasymsigma}{\isasymrrangle}{\isacharparenright}{\kern0pt}\ {\isasymand}\ vars\isactrlsub A{\isacharparenleft}{\kern0pt}a{\isacharparenright}{\kern0pt}\ {\isasymsubseteq}\ fmdom{\isacharprime}{\kern0pt}{\isacharparenleft}{\kern0pt}{\isasymsigma}{\isacharparenright}{\kern0pt}{\isachardoublequoteclose}\ \isanewline
\ \ \ \ \isakeyword{shows}\ {\isachardoublequoteopen}{\isasymdelta}\isactrlsub s\ {\isacharparenleft}{\kern0pt}sh\ {\isasymleadsto}\ State{\isasymllangle}{\isasymsigma}{\isasymrrangle}{\isacharcomma}{\kern0pt}\ {\isasymlambda}{\isacharbrackleft}{\kern0pt}x\ {\isacharcolon}{\kern0pt}{\isacharequal}{\kern0pt}\ a{\isacharbrackright}{\kern0pt}{\isacharparenright}{\kern0pt}\ {\isacharequal}{\kern0pt}\ {\isacharbraceleft}{\kern0pt}{\isacharparenleft}{\kern0pt}{\isacharparenleft}{\kern0pt}sh\ {\isasymleadsto}\ State{\isasymllangle}{\isasymsigma}{\isasymrrangle}{\isacharparenright}{\kern0pt}\ {\isasymleadsto}\ State{\isasymllangle}{\isacharbrackleft}{\kern0pt}x\ {\isasymlongmapsto}\ Exp\ {\isacharparenleft}{\kern0pt}val\isactrlsub A\ a\ {\isasymsigma}{\isacharparenright}{\kern0pt}{\isacharbrackright}{\kern0pt}\ {\isasymsigma}{\isasymrrangle}{\isacharcomma}{\kern0pt}\ {\isasymlambda}{\isacharbrackleft}{\kern0pt}{\isasymnabla}{\isacharbrackright}{\kern0pt}{\isacharparenright}{\kern0pt}{\isacharbraceright}{\kern0pt}{\isachardoublequoteclose}\isanewline
\isadelimproof
\ \ %
\endisadelimproof
\isatagproof
\isacommand{proof}\isamarkupfalse%
\ {\isacharminus}{\kern0pt}\isanewline
\ \ \ \ \isacommand{have}\isamarkupfalse%
\ {\isachardoublequoteopen}concrete\isactrlsub S\ {\isacharparenleft}{\kern0pt}Exp\ {\isacharparenleft}{\kern0pt}val\isactrlsub A\ a\ {\isasymsigma}{\isacharparenright}{\kern0pt}{\isacharparenright}{\kern0pt}{\isachardoublequoteclose}\isanewline
\ \ \ \ \ \ \isacommand{using}\isamarkupfalse%
\ assms\ concrete{\isacharunderscore}{\kern0pt}imp\isactrlsub A\ \isacommand{by}\isamarkupfalse%
\ {\isacharparenleft}{\kern0pt}metis\ concrete\isactrlsub {\isasymT}{\isachardot}{\kern0pt}simps{\isacharparenleft}{\kern0pt}{\isadigit{2}}{\isacharparenright}{\kern0pt}\ sexp{\isachardot}{\kern0pt}simps{\isacharparenleft}{\kern0pt}{\isadigit{4}}{\isacharparenright}{\kern0pt}{\isacharparenright}{\kern0pt}\isanewline
\ \ \ \ %
\isamarkupcmt{Due to the concreteness of \isa{{\isasymsigma}}, the evaluation of a under \isa{{\isasymsigma}} must be of 
       concrete nature. This can be inferred using the \isa{concrete{\isacharunderscore}{\kern0pt}imp\isactrlsub A} theorem
       of the base theory.%
}\isanewline
\ \ \ \ \isacommand{moreover}\isamarkupfalse%
\ \isacommand{hence}\isamarkupfalse%
\ {\isachardoublequoteopen}concrete\isactrlsub {\isasymT}\ {\isacharparenleft}{\kern0pt}{\isacharparenleft}{\kern0pt}sh\ {\isasymleadsto}\ State{\isasymllangle}{\isasymsigma}{\isasymrrangle}{\isacharparenright}{\kern0pt}\ {\isasymleadsto}\ State{\isasymllangle}{\isacharbrackleft}{\kern0pt}x\ {\isasymlongmapsto}\ Exp\ {\isacharparenleft}{\kern0pt}val\isactrlsub A\ a\ {\isasymsigma}{\isacharparenright}{\kern0pt}{\isacharbrackright}{\kern0pt}\ {\isasymsigma}{\isasymrrangle}{\isacharparenright}{\kern0pt}{\isachardoublequoteclose}\isanewline
\ \ \ \ \ \ \isacommand{using}\isamarkupfalse%
\ assms\ \isacommand{by}\isamarkupfalse%
\ simp\isanewline
\ \ \ \ %
\isamarkupcmt{Hence the symbolic trace generated by the the assign statement must also 
       be concrete.%
}\isanewline
\ \ \ \ \isacommand{moreover}\isamarkupfalse%
\ \isacommand{hence}\isamarkupfalse%
\ {\isachardoublequoteopen}min{\isacharunderscore}{\kern0pt}conc{\isacharunderscore}{\kern0pt}map\isactrlsub {\isasymT}\ {\isacharparenleft}{\kern0pt}{\isacharparenleft}{\kern0pt}sh\ {\isasymleadsto}\ State{\isasymllangle}{\isasymsigma}{\isasymrrangle}{\isacharparenright}{\kern0pt}\ {\isasymleadsto}\ State{\isasymllangle}{\isacharbrackleft}{\kern0pt}x\ {\isasymlongmapsto}\ Exp\ {\isacharparenleft}{\kern0pt}val\isactrlsub A\ a\ {\isasymsigma}{\isacharparenright}{\kern0pt}{\isacharbrackright}{\kern0pt}\ {\isasymsigma}{\isasymrrangle}{\isacharparenright}{\kern0pt}\ {\isadigit{0}}\ {\isacharequal}{\kern0pt}\ {\isasymcircle}{\isachardoublequoteclose}\isanewline
\ \ \ \ \ \ \isacommand{using}\isamarkupfalse%
\ assms\ min{\isacharunderscore}{\kern0pt}conc{\isacharunderscore}{\kern0pt}map{\isacharunderscore}{\kern0pt}of{\isacharunderscore}{\kern0pt}concrete\isactrlsub {\isasymT}\ \isacommand{by}\isamarkupfalse%
\ presburger\isanewline
\ \ \ \ %
\isamarkupcmt{Considering that the computed trace is concrete, it must not contain any 
       symbolic variables. Using the \isa{min{\isacharunderscore}{\kern0pt}conc{\isacharunderscore}{\kern0pt}map{\isacharunderscore}{\kern0pt}of{\isacharunderscore}{\kern0pt}concrete\isactrlsub {\isasymT}} theorem, we can then 
       infer that its minimal concretization must be the empty state.%
}\isanewline
\ \ \ \ \isacommand{ultimately}\isamarkupfalse%
\ \isacommand{have}\isamarkupfalse%
\ {\isachardoublequoteopen}trace{\isacharunderscore}{\kern0pt}conc\ {\isacharparenleft}{\kern0pt}min{\isacharunderscore}{\kern0pt}conc{\isacharunderscore}{\kern0pt}map\isactrlsub {\isasymT}\ {\isacharparenleft}{\kern0pt}{\isacharparenleft}{\kern0pt}sh\ {\isasymleadsto}\ State{\isasymllangle}{\isasymsigma}{\isasymrrangle}{\isacharparenright}{\kern0pt}\ {\isasymleadsto}\ State{\isasymllangle}{\isacharbrackleft}{\kern0pt}x\ {\isasymlongmapsto}\ Exp\ {\isacharparenleft}{\kern0pt}val\isactrlsub A\ a\ {\isasymsigma}{\isacharparenright}{\kern0pt}{\isacharbrackright}{\kern0pt}\ {\isasymsigma}{\isasymrrangle}{\isacharparenright}{\kern0pt}\ {\isadigit{0}}{\isacharparenright}{\kern0pt}\ {\isacharparenleft}{\kern0pt}{\isacharparenleft}{\kern0pt}sh\ {\isasymleadsto}\ State{\isasymllangle}{\isasymsigma}{\isasymrrangle}{\isacharparenright}{\kern0pt}\ {\isasymleadsto}\ State{\isasymllangle}{\isacharbrackleft}{\kern0pt}x\ {\isasymlongmapsto}\ Exp\ {\isacharparenleft}{\kern0pt}val\isactrlsub A\ a\ {\isasymsigma}{\isacharparenright}{\kern0pt}{\isacharbrackright}{\kern0pt}\ {\isasymsigma}{\isasymrrangle}{\isacharparenright}{\kern0pt}\ {\isacharequal}{\kern0pt}\ {\isacharparenleft}{\kern0pt}{\isacharparenleft}{\kern0pt}sh\ {\isasymleadsto}\ State{\isasymllangle}{\isasymsigma}{\isasymrrangle}{\isacharparenright}{\kern0pt}\ {\isasymleadsto}\ State{\isasymllangle}{\isacharbrackleft}{\kern0pt}x\ {\isasymlongmapsto}\ Exp\ {\isacharparenleft}{\kern0pt}val\isactrlsub A\ a\ {\isasymsigma}{\isacharparenright}{\kern0pt}{\isacharbrackright}{\kern0pt}\ {\isasymsigma}{\isasymrrangle}{\isacharparenright}{\kern0pt}{\isachardoublequoteclose}\isanewline
\ \ \ \ \ \ \isacommand{using}\isamarkupfalse%
\ assms\ trace{\isacharunderscore}{\kern0pt}conc{\isacharunderscore}{\kern0pt}pr\ \isacommand{by}\isamarkupfalse%
\ presburger\isanewline
\ \ \ \ %
\isamarkupcmt{Utilizing a supplementary theorem, we conclude that the trace concretization of 
       the computed trace under its minimal concretization mapping must be 
       the computed trace itself.%
}\isanewline
\ \ \ \ \isacommand{thus}\isamarkupfalse%
\ {\isacharquery}{\kern0pt}thesis\ \isacommand{by}\isamarkupfalse%
\ auto\isanewline
\ \ \ \ %
\isamarkupcmt{We can now use Isabelle in order to infer the conclusion of the lemma.%
}\isanewline
\ \ \isacommand{qed}\isamarkupfalse%
\endisatagproof
{\isafoldproof}%
\isadelimproof
\endisadelimproof
\begin{isamarkuptext}%
Applying the $\isa{{\isasymdelta}}_s$-function on the If-Branch and the While-Loop can cause
  two distinct results. If the path condition containing the evaluated Boolean
  guard is consistent, the statement enters the true-case. If the path condition
  consisting of the evaluated negated Boolean guard is consistent, the statement will
  enter the false-case. We formalize a simplification lemma for each of these
  cases. \par
  Our proof system ensures that path conditions turn concrete when simplified, 
  as they are evaluated under state \isa{{\isasymsigma}}, which we guarantee to be of
  concrete~nature.%
\end{isamarkuptext}\isamarkuptrue%
\ \ \isacommand{lemma}\isamarkupfalse%
\ {\isasymdelta}\isactrlsub s{\isacharunderscore}{\kern0pt}If\isactrlsub T{\isacharcolon}{\kern0pt}\isanewline
\ \ \ \ \isakeyword{assumes}\ {\isachardoublequoteopen}consistent\ {\isacharparenleft}{\kern0pt}{\isacharbraceleft}{\kern0pt}val\isactrlsub B\ b\ {\isasymsigma}{\isacharbraceright}{\kern0pt}{\isacharparenright}{\kern0pt}\ {\isasymand}\ concrete\isactrlsub {\isasymT}{\isacharparenleft}{\kern0pt}sh\ {\isasymleadsto}\ State{\isasymllangle}{\isasymsigma}{\isasymrrangle}{\isacharparenright}{\kern0pt}{\isachardoublequoteclose}\isanewline
\ \ \ \ \isakeyword{shows}\ {\isachardoublequoteopen}{\isasymdelta}\isactrlsub s\ {\isacharparenleft}{\kern0pt}sh\ {\isasymleadsto}\ State{\isasymllangle}{\isasymsigma}{\isasymrrangle}{\isacharcomma}{\kern0pt}\ {\isasymlambda}{\isacharbrackleft}{\kern0pt}IF\ b\ THEN\ S\ FI{\isacharbrackright}{\kern0pt}{\isacharparenright}{\kern0pt}\ {\isacharequal}{\kern0pt}\ {\isacharbraceleft}{\kern0pt}{\isacharparenleft}{\kern0pt}sh\ {\isasymleadsto}\ State{\isasymllangle}{\isasymsigma}{\isasymrrangle}{\isacharcomma}{\kern0pt}\ {\isasymlambda}{\isacharbrackleft}{\kern0pt}S{\isacharbrackright}{\kern0pt}{\isacharparenright}{\kern0pt}{\isacharbraceright}{\kern0pt}{\isachardoublequoteclose}\isanewline
\isadelimproof
\ \ %
\endisadelimproof
\isatagproof
\isacommand{proof}\isamarkupfalse%
\ {\isacharminus}{\kern0pt}\isanewline
\ \ \ \ \isacommand{have}\isamarkupfalse%
\ {\isachardoublequoteopen}consistent{\isacharparenleft}{\kern0pt}sval\isactrlsub B\ {\isacharbraceleft}{\kern0pt}val\isactrlsub B\ b\ {\isasymsigma}{\isacharbraceright}{\kern0pt}\ {\isacharparenleft}{\kern0pt}min{\isacharunderscore}{\kern0pt}conc{\isacharunderscore}{\kern0pt}map\isactrlsub {\isasymT}\ {\isacharparenleft}{\kern0pt}{\isasymlangle}{\isasymsigma}{\isasymrangle}{\isacharparenright}{\kern0pt}\ {\isadigit{0}}{\isacharparenright}{\kern0pt}{\isacharparenright}{\kern0pt}{\isachardoublequoteclose}\isanewline
\ \ \ \ \ \ \isacommand{using}\isamarkupfalse%
\ assms\ consistent{\isacharunderscore}{\kern0pt}pr\ \isacommand{by}\isamarkupfalse%
\ blast\isanewline
\ \ \ \ %
\isamarkupcmt{As we have already proven in the base theory, the concretization of a 
       concrete expression always preserves the original expression. Note that
       this theorem also holds for consistent path conditions. Considering that we 
       have already assumed the consistency of the path condition containing b   
       (i.e. the true-case), the consistency of the path condition must then be 
       preserved if further simplified under a minimal concretization mapping.%
}\isanewline
\ \ \ \ \isacommand{moreover}\isamarkupfalse%
\ \isacommand{have}\isamarkupfalse%
\ {\isachardoublequoteopen}{\isasymnot}consistent{\isacharparenleft}{\kern0pt}sval\isactrlsub B\ {\isacharbraceleft}{\kern0pt}val\isactrlsub B\ {\isacharparenleft}{\kern0pt}not\ b{\isacharparenright}{\kern0pt}\ {\isasymsigma}{\isacharbraceright}{\kern0pt}\ {\isacharparenleft}{\kern0pt}min{\isacharunderscore}{\kern0pt}conc{\isacharunderscore}{\kern0pt}map\isactrlsub {\isasymT}\ {\isacharparenleft}{\kern0pt}{\isasymlangle}{\isasymsigma}{\isasymrangle}{\isacharparenright}{\kern0pt}\ {\isadigit{0}}{\isacharparenright}{\kern0pt}{\isacharparenright}{\kern0pt}{\isachardoublequoteclose}\isanewline
\ \ \ \ \ \ \isacommand{using}\isamarkupfalse%
\ assms\ \isacommand{by}\isamarkupfalse%
\ simp\isanewline
\ \ \ \ %
\isamarkupcmt{Hence, the path condition of the continuation trace corresponding to
       the false-case cannot~be~consistent.%
}\isanewline
\ \ \ \ \isacommand{ultimately}\isamarkupfalse%
\ \isacommand{have}\isamarkupfalse%
\ {\isachardoublequoteopen}{\isacharbraceleft}{\kern0pt}cont\ {\isasymin}\ {\isacharparenleft}{\kern0pt}val\isactrlsub s\ {\isacharparenleft}{\kern0pt}IF\ b\ THEN\ S\ FI{\isacharparenright}{\kern0pt}\ {\isasymsigma}{\isacharparenright}{\kern0pt}{\isachardot}{\kern0pt}\ consistent{\isacharparenleft}{\kern0pt}sval\isactrlsub B\ {\isacharparenleft}{\kern0pt}{\isasymdown}\isactrlsub p\ cont{\isacharparenright}{\kern0pt}\ {\isacharparenleft}{\kern0pt}min{\isacharunderscore}{\kern0pt}conc{\isacharunderscore}{\kern0pt}map\isactrlsub {\isasymT}\ {\isacharparenleft}{\kern0pt}{\isasymdown}\isactrlsub {\isasymtau}\ cont{\isacharparenright}{\kern0pt}\ {\isadigit{0}}{\isacharparenright}{\kern0pt}{\isacharparenright}{\kern0pt}{\isacharbraceright}{\kern0pt}\ {\isacharequal}{\kern0pt}\ {\isacharbraceleft}{\kern0pt}\ {\isacharbraceleft}{\kern0pt}val\isactrlsub B\ b\ {\isasymsigma}{\isacharbraceright}{\kern0pt}\ {\isasymtriangleright}\ {\isasymlangle}{\isasymsigma}{\isasymrangle}\ \isactrlitem \ {\isasymlambda}{\isacharbrackleft}{\kern0pt}S{\isacharbrackright}{\kern0pt}\ {\isacharbraceright}{\kern0pt}{\isachardoublequoteclose}\isanewline
\ \ \ \ \ \ \isacommand{using}\isamarkupfalse%
\ assms\ \isacommand{by}\isamarkupfalse%
\ auto\isanewline
\ \ \ \ %
\isamarkupcmt{Thus, the set of consistent continuation traces can only contain the
       continuation trace corresponding to the true-case.%
}\isanewline
\ \ \ \ \isacommand{moreover}\isamarkupfalse%
\ \isacommand{have}\isamarkupfalse%
\ {\isachardoublequoteopen}min{\isacharunderscore}{\kern0pt}conc{\isacharunderscore}{\kern0pt}map\isactrlsub {\isasymT}\ {\isacharparenleft}{\kern0pt}sh\ {\isasymleadsto}\ State{\isasymllangle}{\isasymsigma}{\isasymrrangle}{\isacharparenright}{\kern0pt}\ {\isadigit{0}}\ {\isacharequal}{\kern0pt}\ {\isasymcircle}{\isachardoublequoteclose}\isanewline
\ \ \ \ \ \ \isacommand{using}\isamarkupfalse%
\ assms\ min{\isacharunderscore}{\kern0pt}conc{\isacharunderscore}{\kern0pt}map{\isacharunderscore}{\kern0pt}of{\isacharunderscore}{\kern0pt}concrete\isactrlsub {\isasymT}\ \isacommand{by}\isamarkupfalse%
\ presburger\isanewline
\ \ \ \ %
\isamarkupcmt{Considering that \isa{{\isacharparenleft}{\kern0pt}sh\ {\isasymleadsto}\ {\isasymsigma}{\isacharparenright}{\kern0pt}} is concrete, it contains no symbolic variables.
       Using the \isa{min{\isacharunderscore}{\kern0pt}conc{\isacharunderscore}{\kern0pt}map{\isacharunderscore}{\kern0pt}of{\isacharunderscore}{\kern0pt}concrete\isactrlsub {\isasymT}} theorem, we can then infer that the minimal 
       concretization mapping of \isa{{\isacharparenleft}{\kern0pt}sh\ {\isasymleadsto}\ {\isasymsigma}{\isacharparenright}{\kern0pt}} matches the empty state.%
}\isanewline
\ \ \ \ \isacommand{hence}\isamarkupfalse%
\ {\isachardoublequoteopen}trace{\isacharunderscore}{\kern0pt}conc\ {\isacharparenleft}{\kern0pt}min{\isacharunderscore}{\kern0pt}conc{\isacharunderscore}{\kern0pt}map\isactrlsub {\isasymT}\ {\isacharparenleft}{\kern0pt}sh\ {\isasymleadsto}\ State{\isasymllangle}{\isasymsigma}{\isasymrrangle}{\isacharparenright}{\kern0pt}\ {\isadigit{0}}{\isacharparenright}{\kern0pt}\ {\isacharparenleft}{\kern0pt}sh\ {\isasymleadsto}\ State{\isasymllangle}{\isasymsigma}{\isasymrrangle}{\isacharparenright}{\kern0pt}\ {\isacharequal}{\kern0pt}\ sh\ {\isasymleadsto}\ State{\isasymllangle}{\isasymsigma}{\isasymrrangle}{\isachardoublequoteclose}\isanewline
\ \ \ \ \ \ \isacommand{using}\isamarkupfalse%
\ assms\ trace{\isacharunderscore}{\kern0pt}conc{\isacharunderscore}{\kern0pt}pr\ LAGC{\isacharunderscore}{\kern0pt}Base{\isachardot}{\kern0pt}concat{\isachardot}{\kern0pt}simps{\isacharparenleft}{\kern0pt}{\isadigit{1}}{\isacharparenright}{\kern0pt}\ \isacommand{by}\isamarkupfalse%
\ presburger\isanewline
\ \ \ \ %
\isamarkupcmt{Utilizing a supplementary theorem, we conclude that the trace concretization of 
       \isa{{\isacharparenleft}{\kern0pt}sh\ {\isasymleadsto}\ {\isasymsigma}{\isacharparenright}{\kern0pt}} under its minimal concretization mapping must be \isa{{\isacharparenleft}{\kern0pt}sh\ {\isasymleadsto}\ {\isasymsigma}{\isacharparenright}{\kern0pt}} itself.%
}\isanewline
\ \ \ \ \isacommand{ultimately}\isamarkupfalse%
\ \isacommand{show}\isamarkupfalse%
\ {\isacharquery}{\kern0pt}thesis\ \isacommand{by}\isamarkupfalse%
\ auto\isanewline
\ \ \ \ %
\isamarkupcmt{We can now use Isabelle in order to infer the conclusion of the lemma.%
}\isanewline
\ \ \isacommand{qed}\isamarkupfalse%
\endisatagproof
{\isafoldproof}%
\isadelimproof
\isanewline
\endisadelimproof
\isanewline
\ \ \isacommand{lemma}\isamarkupfalse%
\ {\isasymdelta}\isactrlsub s{\isacharunderscore}{\kern0pt}If\isactrlsub F{\isacharcolon}{\kern0pt}\isanewline
\ \ \ \ \isakeyword{assumes}\ {\isachardoublequoteopen}consistent\ {\isacharbraceleft}{\kern0pt}{\isacharparenleft}{\kern0pt}val\isactrlsub B\ {\isacharparenleft}{\kern0pt}not\ b{\isacharparenright}{\kern0pt}\ {\isasymsigma}{\isacharparenright}{\kern0pt}{\isacharbraceright}{\kern0pt}\ {\isasymand}\ concrete\isactrlsub {\isasymT}{\isacharparenleft}{\kern0pt}sh\ {\isasymleadsto}\ State{\isasymllangle}{\isasymsigma}{\isasymrrangle}{\isacharparenright}{\kern0pt}{\isachardoublequoteclose}\isanewline
\ \ \ \ \isakeyword{shows}\ {\isachardoublequoteopen}{\isasymdelta}\isactrlsub s\ {\isacharparenleft}{\kern0pt}sh\ {\isasymleadsto}\ State{\isasymllangle}{\isasymsigma}{\isasymrrangle}{\isacharcomma}{\kern0pt}\ {\isasymlambda}{\isacharbrackleft}{\kern0pt}IF\ b\ THEN\ S\ FI{\isacharbrackright}{\kern0pt}{\isacharparenright}{\kern0pt}\ {\isacharequal}{\kern0pt}\ {\isacharbraceleft}{\kern0pt}{\isacharparenleft}{\kern0pt}sh\ {\isasymleadsto}\ State{\isasymllangle}{\isasymsigma}{\isasymrrangle}{\isacharcomma}{\kern0pt}\ {\isasymlambda}{\isacharbrackleft}{\kern0pt}{\isasymnabla}{\isacharbrackright}{\kern0pt}{\isacharparenright}{\kern0pt}{\isacharbraceright}{\kern0pt}{\isachardoublequoteclose}\isanewline
\isadelimproof
\ \ %
\endisadelimproof
\isatagproof
\isacommand{proof}\isamarkupfalse%
\ {\isacharminus}{\kern0pt}\isanewline
\ \ \ \ \isacommand{have}\isamarkupfalse%
\ {\isachardoublequoteopen}consistent{\isacharparenleft}{\kern0pt}sval\isactrlsub B\ {\isacharbraceleft}{\kern0pt}val\isactrlsub B\ {\isacharparenleft}{\kern0pt}not\ b{\isacharparenright}{\kern0pt}\ {\isasymsigma}{\isacharbraceright}{\kern0pt}\ {\isacharparenleft}{\kern0pt}min{\isacharunderscore}{\kern0pt}conc{\isacharunderscore}{\kern0pt}map\isactrlsub {\isasymT}\ {\isacharparenleft}{\kern0pt}{\isasymlangle}{\isasymsigma}{\isasymrangle}{\isacharparenright}{\kern0pt}\ {\isadigit{0}}{\isacharparenright}{\kern0pt}{\isacharparenright}{\kern0pt}{\isachardoublequoteclose}\isanewline
\ \ \ \ \ \ \isacommand{using}\isamarkupfalse%
\ assms\ consistent{\isacharunderscore}{\kern0pt}pr\ \isacommand{by}\isamarkupfalse%
\ blast\isanewline
\ \ \ \ %
\isamarkupcmt{As we have already assumed that the path condition containing \isa{not\ b} 
       (i.e. the false-case) is consistent, the consistency of the path condition must
       then be preserved if further simplified under a minimal concretization mapping.%
}\isanewline
\ \ \ \ \isacommand{moreover}\isamarkupfalse%
\ \isacommand{have}\isamarkupfalse%
\ {\isachardoublequoteopen}{\isasymnot}consistent{\isacharparenleft}{\kern0pt}sval\isactrlsub B\ {\isacharbraceleft}{\kern0pt}val\isactrlsub B\ b\ {\isasymsigma}{\isacharbraceright}{\kern0pt}\ {\isacharparenleft}{\kern0pt}min{\isacharunderscore}{\kern0pt}conc{\isacharunderscore}{\kern0pt}map\isactrlsub {\isasymT}\ {\isacharparenleft}{\kern0pt}{\isasymlangle}{\isasymsigma}{\isasymrangle}{\isacharparenright}{\kern0pt}\ {\isadigit{0}}{\isacharparenright}{\kern0pt}{\isacharparenright}{\kern0pt}{\isachardoublequoteclose}\isanewline
\ \ \ \ \ \ \isacommand{using}\isamarkupfalse%
\ assms\ conc{\isacharunderscore}{\kern0pt}pc{\isacharunderscore}{\kern0pt}pr\isactrlsub B\isactrlsub N\ consistent{\isacharunderscore}{\kern0pt}def\ s{\isacharunderscore}{\kern0pt}value{\isacharunderscore}{\kern0pt}pr\isactrlsub B\ \isanewline
\ \ \ \ \ \ \isacommand{by}\isamarkupfalse%
\ {\isacharparenleft}{\kern0pt}metis\ bexp{\isachardot}{\kern0pt}simps{\isacharparenleft}{\kern0pt}{\isadigit{1}}{\isadigit{7}}{\isacharparenright}{\kern0pt}\ singletonI\ val\isactrlsub B{\isachardot}{\kern0pt}simps{\isacharparenleft}{\kern0pt}{\isadigit{2}}{\isacharparenright}{\kern0pt}{\isacharparenright}{\kern0pt}\isanewline
\ \ \ \ %
\isamarkupcmt{Hence, the path condition of the continuation trace corresponding to the
       true-case cannot be~consistent.%
}\isanewline
\ \ \ \ \isacommand{ultimately}\isamarkupfalse%
\ \isacommand{have}\isamarkupfalse%
\ {\isachardoublequoteopen}{\isacharbraceleft}{\kern0pt}cont\ {\isasymin}\ {\isacharparenleft}{\kern0pt}val\isactrlsub s\ {\isacharparenleft}{\kern0pt}IF\ b\ THEN\ S\ FI{\isacharparenright}{\kern0pt}\ {\isasymsigma}{\isacharparenright}{\kern0pt}{\isachardot}{\kern0pt}\ consistent{\isacharparenleft}{\kern0pt}sval\isactrlsub B\ {\isacharparenleft}{\kern0pt}{\isasymdown}\isactrlsub p\ cont{\isacharparenright}{\kern0pt}\ {\isacharparenleft}{\kern0pt}min{\isacharunderscore}{\kern0pt}conc{\isacharunderscore}{\kern0pt}map\isactrlsub {\isasymT}\ {\isacharparenleft}{\kern0pt}{\isasymdown}\isactrlsub {\isasymtau}\ cont{\isacharparenright}{\kern0pt}\ {\isadigit{0}}{\isacharparenright}{\kern0pt}{\isacharparenright}{\kern0pt}{\isacharbraceright}{\kern0pt}\ {\isacharequal}{\kern0pt}\ {\isacharbraceleft}{\kern0pt}\ {\isacharbraceleft}{\kern0pt}val\isactrlsub B\ {\isacharparenleft}{\kern0pt}not\ b{\isacharparenright}{\kern0pt}\ {\isasymsigma}{\isacharbraceright}{\kern0pt}\ {\isasymtriangleright}\ {\isasymlangle}{\isasymsigma}{\isasymrangle}\ \isactrlitem \ {\isasymlambda}{\isacharbrackleft}{\kern0pt}{\isasymnabla}{\isacharbrackright}{\kern0pt}\ {\isacharbraceright}{\kern0pt}{\isachardoublequoteclose}\isanewline
\ \ \ \ \ \ \isacommand{using}\isamarkupfalse%
\ assms\ \isacommand{by}\isamarkupfalse%
\ auto\isanewline
\ \ \ \ %
\isamarkupcmt{Thus, the set of consistent continuation traces can only contain the
       continuation trace corresponding to the false-case.%
}\isanewline
\ \ \ \ \isacommand{moreover}\isamarkupfalse%
\ \isacommand{have}\isamarkupfalse%
\ {\isachardoublequoteopen}min{\isacharunderscore}{\kern0pt}conc{\isacharunderscore}{\kern0pt}map\isactrlsub {\isasymT}\ {\isacharparenleft}{\kern0pt}sh\ {\isasymleadsto}\ State{\isasymllangle}{\isasymsigma}{\isasymrrangle}{\isacharparenright}{\kern0pt}\ {\isadigit{0}}\ {\isacharequal}{\kern0pt}\ {\isasymcircle}{\isachardoublequoteclose}\isanewline
\ \ \ \ \ \ \isacommand{using}\isamarkupfalse%
\ assms\ min{\isacharunderscore}{\kern0pt}conc{\isacharunderscore}{\kern0pt}map{\isacharunderscore}{\kern0pt}of{\isacharunderscore}{\kern0pt}concrete\isactrlsub {\isasymT}\ \isacommand{by}\isamarkupfalse%
\ presburger\isanewline
\ \ \ \ %
\isamarkupcmt{Considering that \isa{{\isacharparenleft}{\kern0pt}sh\ {\isasymleadsto}\ {\isasymsigma}{\isacharparenright}{\kern0pt}} is concrete, it contains no symbolic variables.
       Using the \isa{min{\isacharunderscore}{\kern0pt}conc{\isacharunderscore}{\kern0pt}map{\isacharunderscore}{\kern0pt}of{\isacharunderscore}{\kern0pt}concrete\isactrlsub {\isasymT}} theorem, we can then infer that the minimal 
       concretization mapping of \isa{{\isacharparenleft}{\kern0pt}sh\ {\isasymleadsto}\ {\isasymsigma}{\isacharparenright}{\kern0pt}} matches the empty state.%
}\isanewline
\ \ \ \ \isacommand{hence}\isamarkupfalse%
\ {\isachardoublequoteopen}trace{\isacharunderscore}{\kern0pt}conc\ {\isacharparenleft}{\kern0pt}min{\isacharunderscore}{\kern0pt}conc{\isacharunderscore}{\kern0pt}map\isactrlsub {\isasymT}\ {\isacharparenleft}{\kern0pt}sh\ {\isasymleadsto}\ State{\isasymllangle}{\isasymsigma}{\isasymrrangle}{\isacharparenright}{\kern0pt}\ {\isadigit{0}}{\isacharparenright}{\kern0pt}\ {\isacharparenleft}{\kern0pt}sh\ {\isasymleadsto}\ State{\isasymllangle}{\isasymsigma}{\isasymrrangle}{\isacharparenright}{\kern0pt}\ {\isacharequal}{\kern0pt}\ sh\ {\isasymleadsto}\ State{\isasymllangle}{\isasymsigma}{\isasymrrangle}{\isachardoublequoteclose}\isanewline
\ \ \ \ \ \ \isacommand{using}\isamarkupfalse%
\ assms\ trace{\isacharunderscore}{\kern0pt}conc{\isacharunderscore}{\kern0pt}pr\ LAGC{\isacharunderscore}{\kern0pt}Base{\isachardot}{\kern0pt}concat{\isachardot}{\kern0pt}simps{\isacharparenleft}{\kern0pt}{\isadigit{1}}{\isacharparenright}{\kern0pt}\ \isacommand{by}\isamarkupfalse%
\ presburger\ \isanewline
\ \ \ \ %
\isamarkupcmt{Utilizing a supplementary theorem, we conclude that the trace concretization of 
       \isa{{\isacharparenleft}{\kern0pt}sh\ {\isasymleadsto}\ {\isasymsigma}{\isacharparenright}{\kern0pt}} under its minimal concretization mapping must be \isa{{\isacharparenleft}{\kern0pt}sh\ {\isasymleadsto}\ {\isasymsigma}{\isacharparenright}{\kern0pt}} itself.%
}\isanewline
\ \ \ \ \isacommand{ultimately}\isamarkupfalse%
\ \isacommand{show}\isamarkupfalse%
\ {\isacharquery}{\kern0pt}thesis\ \isacommand{by}\isamarkupfalse%
\ auto\isanewline
\ \ \ \ %
\isamarkupcmt{We can now use Isabelle in order to infer the conclusion of the lemma.%
}\isanewline
\ \ \isacommand{qed}\isamarkupfalse%
\endisatagproof
{\isafoldproof}%
\isadelimproof
\isanewline
\endisadelimproof
\isanewline
\ \ \isacommand{lemma}\isamarkupfalse%
\ {\isasymdelta}\isactrlsub s{\isacharunderscore}{\kern0pt}While\isactrlsub T{\isacharcolon}{\kern0pt}\isanewline
\ \ \ \ \isakeyword{assumes}\ {\isachardoublequoteopen}consistent\ {\isacharbraceleft}{\kern0pt}{\isacharparenleft}{\kern0pt}val\isactrlsub B\ b\ {\isasymsigma}{\isacharparenright}{\kern0pt}{\isacharbraceright}{\kern0pt}\ {\isasymand}\ concrete\isactrlsub {\isasymT}{\isacharparenleft}{\kern0pt}sh\ {\isasymleadsto}\ State{\isasymllangle}{\isasymsigma}{\isasymrrangle}{\isacharparenright}{\kern0pt}{\isachardoublequoteclose}\isanewline
\ \ \ \ \isakeyword{shows}\ {\isachardoublequoteopen}{\isasymdelta}\isactrlsub s\ {\isacharparenleft}{\kern0pt}sh\ {\isasymleadsto}\ State{\isasymllangle}{\isasymsigma}{\isasymrrangle}{\isacharcomma}{\kern0pt}\ {\isasymlambda}{\isacharbrackleft}{\kern0pt}WHILE\ b\ DO\ S\ OD{\isacharbrackright}{\kern0pt}{\isacharparenright}{\kern0pt}\ {\isacharequal}{\kern0pt}\ {\isacharbraceleft}{\kern0pt}{\isacharparenleft}{\kern0pt}sh\ {\isasymleadsto}\ State{\isasymllangle}{\isasymsigma}{\isasymrrangle}{\isacharcomma}{\kern0pt}\ {\isasymlambda}{\isacharbrackleft}{\kern0pt}S{\isacharsemicolon}{\kern0pt}{\isacharsemicolon}{\kern0pt}WHILE\ b\ DO\ S\ OD{\isacharbrackright}{\kern0pt}{\isacharparenright}{\kern0pt}{\isacharbraceright}{\kern0pt}{\isachardoublequoteclose}\isanewline
\isadelimproof
\ \ %
\endisadelimproof
\isatagproof
\isacommand{proof}\isamarkupfalse%
\ {\isacharminus}{\kern0pt}\isanewline
\ \ \ \ \isacommand{have}\isamarkupfalse%
\ {\isachardoublequoteopen}consistent{\isacharparenleft}{\kern0pt}sval\isactrlsub B\ {\isacharbraceleft}{\kern0pt}val\isactrlsub B\ b\ {\isasymsigma}{\isacharbraceright}{\kern0pt}\ {\isacharparenleft}{\kern0pt}min{\isacharunderscore}{\kern0pt}conc{\isacharunderscore}{\kern0pt}map\isactrlsub {\isasymT}\ {\isacharparenleft}{\kern0pt}{\isasymlangle}{\isasymsigma}{\isasymrangle}{\isacharparenright}{\kern0pt}\ {\isadigit{0}}{\isacharparenright}{\kern0pt}{\isacharparenright}{\kern0pt}{\isachardoublequoteclose}\isanewline
\ \ \ \ \ \ \isacommand{using}\isamarkupfalse%
\ assms\ consistent{\isacharunderscore}{\kern0pt}pr\ \isacommand{by}\isamarkupfalse%
\ blast\isanewline
\ \ \ \ %
\isamarkupcmt{As we have already assumed that the path condition containing b 
       (i.e. the true-case) is consistent, the consistency of the path condition must
       then be preserved if further simplified under a minimal concretization mapping.%
}\isanewline
\ \ \ \ \isacommand{moreover}\isamarkupfalse%
\ \isacommand{have}\isamarkupfalse%
\ {\isachardoublequoteopen}{\isasymnot}consistent{\isacharparenleft}{\kern0pt}sval\isactrlsub B\ {\isacharbraceleft}{\kern0pt}val\isactrlsub B\ {\isacharparenleft}{\kern0pt}not\ b{\isacharparenright}{\kern0pt}\ {\isasymsigma}{\isacharbraceright}{\kern0pt}\ {\isacharparenleft}{\kern0pt}min{\isacharunderscore}{\kern0pt}conc{\isacharunderscore}{\kern0pt}map\isactrlsub {\isasymT}\ {\isacharparenleft}{\kern0pt}{\isasymlangle}{\isasymsigma}{\isasymrangle}{\isacharparenright}{\kern0pt}\ {\isadigit{0}}{\isacharparenright}{\kern0pt}{\isacharparenright}{\kern0pt}{\isachardoublequoteclose}\isanewline
\ \ \ \ \ \ \isacommand{using}\isamarkupfalse%
\ assms\ \isacommand{by}\isamarkupfalse%
\ simp\isanewline
\ \ \ \ %
\isamarkupcmt{Hence, the path condition of the continuation trace corresponding to the
       false-case cannot be~consistent.%
}\isanewline
\ \ \ \ \isacommand{ultimately}\isamarkupfalse%
\ \isacommand{have}\isamarkupfalse%
\ {\isachardoublequoteopen}{\isacharbraceleft}{\kern0pt}cont\ {\isasymin}\ {\isacharparenleft}{\kern0pt}val\isactrlsub s\ {\isacharparenleft}{\kern0pt}WHILE\ b\ DO\ S\ OD{\isacharparenright}{\kern0pt}\ {\isasymsigma}{\isacharparenright}{\kern0pt}{\isachardot}{\kern0pt}\ consistent{\isacharparenleft}{\kern0pt}sval\isactrlsub B\ {\isacharparenleft}{\kern0pt}{\isasymdown}\isactrlsub p\ cont{\isacharparenright}{\kern0pt}\ {\isacharparenleft}{\kern0pt}min{\isacharunderscore}{\kern0pt}conc{\isacharunderscore}{\kern0pt}map\isactrlsub {\isasymT}\ {\isacharparenleft}{\kern0pt}{\isasymdown}\isactrlsub {\isasymtau}\ cont{\isacharparenright}{\kern0pt}\ {\isadigit{0}}{\isacharparenright}{\kern0pt}{\isacharparenright}{\kern0pt}{\isacharbraceright}{\kern0pt}\ {\isacharequal}{\kern0pt}\ {\isacharbraceleft}{\kern0pt}\ {\isacharbraceleft}{\kern0pt}val\isactrlsub B\ b\ {\isasymsigma}{\isacharbraceright}{\kern0pt}\ {\isasymtriangleright}\ {\isasymlangle}{\isasymsigma}{\isasymrangle}\ \isactrlitem \ {\isasymlambda}{\isacharbrackleft}{\kern0pt}S{\isacharsemicolon}{\kern0pt}{\isacharsemicolon}{\kern0pt}WHILE\ b\ DO\ S\ OD{\isacharbrackright}{\kern0pt}\ {\isacharbraceright}{\kern0pt}{\isachardoublequoteclose}\isanewline
\ \ \ \ \ \ \isacommand{using}\isamarkupfalse%
\ assms\ \isacommand{by}\isamarkupfalse%
\ auto\isanewline
\ \ \ \ %
\isamarkupcmt{Thus, the set of consistent continuation traces can only contain the
       continuation trace corresponding to the true-case.%
}\isanewline
\ \ \ \ \isacommand{moreover}\isamarkupfalse%
\ \isacommand{have}\isamarkupfalse%
\ {\isachardoublequoteopen}min{\isacharunderscore}{\kern0pt}conc{\isacharunderscore}{\kern0pt}map\isactrlsub {\isasymT}\ {\isacharparenleft}{\kern0pt}sh\ {\isasymleadsto}\ State{\isasymllangle}{\isasymsigma}{\isasymrrangle}{\isacharparenright}{\kern0pt}\ {\isadigit{0}}\ {\isacharequal}{\kern0pt}\ {\isasymcircle}{\isachardoublequoteclose}\isanewline
\ \ \ \ \ \ \isacommand{using}\isamarkupfalse%
\ assms\ min{\isacharunderscore}{\kern0pt}conc{\isacharunderscore}{\kern0pt}map{\isacharunderscore}{\kern0pt}of{\isacharunderscore}{\kern0pt}concrete\isactrlsub {\isasymT}\ \isacommand{by}\isamarkupfalse%
\ presburger\isanewline
\ \ \ \ %
\isamarkupcmt{Considering that \isa{{\isacharparenleft}{\kern0pt}sh\ {\isasymleadsto}\ {\isasymsigma}{\isacharparenright}{\kern0pt}} is concrete, it contains no symbolic variables.
       Using the \isa{min{\isacharunderscore}{\kern0pt}conc{\isacharunderscore}{\kern0pt}map{\isacharunderscore}{\kern0pt}of{\isacharunderscore}{\kern0pt}concrete\isactrlsub {\isasymT}} theorem, we can then infer that the minimal 
       concretization mapping of \isa{{\isacharparenleft}{\kern0pt}sh\ {\isasymleadsto}\ {\isasymsigma}{\isacharparenright}{\kern0pt}} matches the empty state.%
}\isanewline
\ \ \ \ \isacommand{hence}\isamarkupfalse%
\ {\isachardoublequoteopen}trace{\isacharunderscore}{\kern0pt}conc\ {\isacharparenleft}{\kern0pt}min{\isacharunderscore}{\kern0pt}conc{\isacharunderscore}{\kern0pt}map\isactrlsub {\isasymT}\ {\isacharparenleft}{\kern0pt}sh\ {\isasymleadsto}\ State{\isasymllangle}{\isasymsigma}{\isasymrrangle}{\isacharparenright}{\kern0pt}\ {\isadigit{0}}{\isacharparenright}{\kern0pt}\ {\isacharparenleft}{\kern0pt}sh\ {\isasymleadsto}\ State{\isasymllangle}{\isasymsigma}{\isasymrrangle}{\isacharparenright}{\kern0pt}\ {\isacharequal}{\kern0pt}\ sh\ {\isasymleadsto}\ State{\isasymllangle}{\isasymsigma}{\isasymrrangle}{\isachardoublequoteclose}\isanewline
\ \ \ \ \ \ \isacommand{using}\isamarkupfalse%
\ assms\ trace{\isacharunderscore}{\kern0pt}conc{\isacharunderscore}{\kern0pt}pr\ \isacommand{by}\isamarkupfalse%
\ presburger\isanewline
\ \ \ \ %
\isamarkupcmt{Utilizing a supplementary theorem, we conclude that the trace concretization of 
       \isa{{\isacharparenleft}{\kern0pt}sh\ {\isasymleadsto}\ {\isasymsigma}{\isacharparenright}{\kern0pt}} under its minimal concretization mapping must be \isa{{\isacharparenleft}{\kern0pt}sh\ {\isasymleadsto}\ {\isasymsigma}{\isacharparenright}{\kern0pt}} itself.%
}\isanewline
\ \ \ \ \isacommand{ultimately}\isamarkupfalse%
\ \isacommand{show}\isamarkupfalse%
\ {\isacharquery}{\kern0pt}thesis\ \isacommand{by}\isamarkupfalse%
\ auto\isanewline
\ \ \ \ %
\isamarkupcmt{We can now use Isabelle in order to infer the conclusion of the lemma.%
}\isanewline
\ \ \isacommand{qed}\isamarkupfalse%
\endisatagproof
{\isafoldproof}%
\isadelimproof
\isanewline
\endisadelimproof
\isanewline
\ \ \isacommand{lemma}\isamarkupfalse%
\ {\isasymdelta}\isactrlsub s{\isacharunderscore}{\kern0pt}While\isactrlsub F{\isacharcolon}{\kern0pt}\isanewline
\ \ \ \ \isakeyword{assumes}\ {\isachardoublequoteopen}consistent\ {\isacharbraceleft}{\kern0pt}{\isacharparenleft}{\kern0pt}val\isactrlsub B\ {\isacharparenleft}{\kern0pt}not\ b{\isacharparenright}{\kern0pt}\ {\isasymsigma}{\isacharparenright}{\kern0pt}{\isacharbraceright}{\kern0pt}\ {\isasymand}\ concrete\isactrlsub {\isasymT}{\isacharparenleft}{\kern0pt}sh\ {\isasymleadsto}\ State{\isasymllangle}{\isasymsigma}{\isasymrrangle}{\isacharparenright}{\kern0pt}{\isachardoublequoteclose}\isanewline
\ \ \ \ \isakeyword{shows}\ {\isachardoublequoteopen}{\isasymdelta}\isactrlsub s\ {\isacharparenleft}{\kern0pt}sh\ {\isasymleadsto}\ State{\isasymllangle}{\isasymsigma}{\isasymrrangle}{\isacharcomma}{\kern0pt}\ {\isasymlambda}{\isacharbrackleft}{\kern0pt}WHILE\ b\ DO\ S\ OD{\isacharbrackright}{\kern0pt}{\isacharparenright}{\kern0pt}\ {\isacharequal}{\kern0pt}\ {\isacharbraceleft}{\kern0pt}{\isacharparenleft}{\kern0pt}sh\ {\isasymleadsto}\ State{\isasymllangle}{\isasymsigma}{\isasymrrangle}{\isacharcomma}{\kern0pt}\ {\isasymlambda}{\isacharbrackleft}{\kern0pt}{\isasymnabla}{\isacharbrackright}{\kern0pt}{\isacharparenright}{\kern0pt}{\isacharbraceright}{\kern0pt}{\isachardoublequoteclose}\isanewline
\isadelimproof
\ \ %
\endisadelimproof
\isatagproof
\isacommand{proof}\isamarkupfalse%
\ {\isacharminus}{\kern0pt}\isanewline
\ \ \ \ \isacommand{have}\isamarkupfalse%
\ {\isachardoublequoteopen}consistent{\isacharparenleft}{\kern0pt}sval\isactrlsub B\ {\isacharbraceleft}{\kern0pt}val\isactrlsub B\ {\isacharparenleft}{\kern0pt}not\ b{\isacharparenright}{\kern0pt}\ {\isasymsigma}{\isacharbraceright}{\kern0pt}\ {\isacharparenleft}{\kern0pt}min{\isacharunderscore}{\kern0pt}conc{\isacharunderscore}{\kern0pt}map\isactrlsub {\isasymT}\ {\isacharparenleft}{\kern0pt}{\isasymlangle}{\isasymsigma}{\isasymrangle}{\isacharparenright}{\kern0pt}\ {\isadigit{0}}{\isacharparenright}{\kern0pt}{\isacharparenright}{\kern0pt}{\isachardoublequoteclose}\isanewline
\ \ \ \ \ \ \isacommand{using}\isamarkupfalse%
\ assms\ consistent{\isacharunderscore}{\kern0pt}pr\ \isacommand{by}\isamarkupfalse%
\ blast\isanewline
\ \ \ \ %
\isamarkupcmt{As we have already assumed that the path condition containing \isa{not\ b} 
       (i.e. the false-case) is consistent, the consistency of the path condition must
       then be preserved if further simplified under a minimal concretization mapping.%
}\isanewline
\ \ \ \ \isacommand{moreover}\isamarkupfalse%
\ \isacommand{have}\isamarkupfalse%
\ {\isachardoublequoteopen}{\isasymnot}consistent{\isacharparenleft}{\kern0pt}sval\isactrlsub B\ {\isacharbraceleft}{\kern0pt}val\isactrlsub B\ b\ {\isasymsigma}{\isacharbraceright}{\kern0pt}\ {\isacharparenleft}{\kern0pt}min{\isacharunderscore}{\kern0pt}conc{\isacharunderscore}{\kern0pt}map\isactrlsub {\isasymT}\ {\isacharparenleft}{\kern0pt}{\isasymlangle}{\isasymsigma}{\isasymrangle}{\isacharparenright}{\kern0pt}\ {\isadigit{0}}{\isacharparenright}{\kern0pt}{\isacharparenright}{\kern0pt}{\isachardoublequoteclose}\isanewline
\ \ \ \ \ \ \isacommand{using}\isamarkupfalse%
\ assms\ conc{\isacharunderscore}{\kern0pt}pc{\isacharunderscore}{\kern0pt}pr\isactrlsub B\isactrlsub N\ consistent{\isacharunderscore}{\kern0pt}def\ s{\isacharunderscore}{\kern0pt}value{\isacharunderscore}{\kern0pt}pr\isactrlsub B\ \isanewline
\ \ \ \ \ \ \isacommand{by}\isamarkupfalse%
\ {\isacharparenleft}{\kern0pt}metis\ bexp{\isachardot}{\kern0pt}simps{\isacharparenleft}{\kern0pt}{\isadigit{1}}{\isadigit{7}}{\isacharparenright}{\kern0pt}\ singletonI\ val\isactrlsub B{\isachardot}{\kern0pt}simps{\isacharparenleft}{\kern0pt}{\isadigit{2}}{\isacharparenright}{\kern0pt}{\isacharparenright}{\kern0pt}\isanewline
\ \ \ \ %
\isamarkupcmt{Hence, the path condition of the true-case is not consistent.%
}\isanewline
\ \ \ \ \isacommand{ultimately}\isamarkupfalse%
\ \isacommand{have}\isamarkupfalse%
\ {\isachardoublequoteopen}{\isacharbraceleft}{\kern0pt}cont\ {\isasymin}\ {\isacharparenleft}{\kern0pt}val\isactrlsub s\ {\isacharparenleft}{\kern0pt}WHILE\ b\ DO\ S\ OD{\isacharparenright}{\kern0pt}\ {\isasymsigma}{\isacharparenright}{\kern0pt}{\isachardot}{\kern0pt}\ consistent{\isacharparenleft}{\kern0pt}sval\isactrlsub B\ {\isacharparenleft}{\kern0pt}{\isasymdown}\isactrlsub p\ cont{\isacharparenright}{\kern0pt}\ {\isacharparenleft}{\kern0pt}min{\isacharunderscore}{\kern0pt}conc{\isacharunderscore}{\kern0pt}map\isactrlsub {\isasymT}\ {\isacharparenleft}{\kern0pt}{\isasymdown}\isactrlsub {\isasymtau}\ cont{\isacharparenright}{\kern0pt}\ {\isadigit{0}}{\isacharparenright}{\kern0pt}{\isacharparenright}{\kern0pt}{\isacharbraceright}{\kern0pt}\ {\isacharequal}{\kern0pt}\ {\isacharbraceleft}{\kern0pt}\ {\isacharbraceleft}{\kern0pt}val\isactrlsub B\ {\isacharparenleft}{\kern0pt}not\ b{\isacharparenright}{\kern0pt}\ {\isasymsigma}{\isacharbraceright}{\kern0pt}\ {\isasymtriangleright}\ {\isasymlangle}{\isasymsigma}{\isasymrangle}\ \isactrlitem \ {\isasymlambda}{\isacharbrackleft}{\kern0pt}{\isasymnabla}{\isacharbrackright}{\kern0pt}\ {\isacharbraceright}{\kern0pt}{\isachardoublequoteclose}\isanewline
\ \ \ \ \ \ \isacommand{using}\isamarkupfalse%
\ assms\ \isacommand{by}\isamarkupfalse%
\ auto\isanewline
\ \ \ \ %
\isamarkupcmt{Thus, the set of consistent continuation traces can only contain the
       continuation trace corresponding to the false-case.%
}\isanewline
\ \ \ \ \isacommand{moreover}\isamarkupfalse%
\ \isacommand{have}\isamarkupfalse%
\ {\isachardoublequoteopen}min{\isacharunderscore}{\kern0pt}conc{\isacharunderscore}{\kern0pt}map\isactrlsub {\isasymT}\ {\isacharparenleft}{\kern0pt}sh\ {\isasymleadsto}\ State{\isasymllangle}{\isasymsigma}{\isasymrrangle}{\isacharparenright}{\kern0pt}\ {\isadigit{0}}\ {\isacharequal}{\kern0pt}\ {\isasymcircle}{\isachardoublequoteclose}\isanewline
\ \ \ \ \ \ \isacommand{using}\isamarkupfalse%
\ assms\ min{\isacharunderscore}{\kern0pt}conc{\isacharunderscore}{\kern0pt}map{\isacharunderscore}{\kern0pt}of{\isacharunderscore}{\kern0pt}concrete\isactrlsub {\isasymT}\ \isacommand{by}\isamarkupfalse%
\ presburger\isanewline
\ \ \ \ %
\isamarkupcmt{Considering that \isa{{\isacharparenleft}{\kern0pt}sh\ {\isasymleadsto}\ {\isasymsigma}{\isacharparenright}{\kern0pt}} is concrete, it contains no symbolic variables.
       Using the \isa{min{\isacharunderscore}{\kern0pt}conc{\isacharunderscore}{\kern0pt}map{\isacharunderscore}{\kern0pt}of{\isacharunderscore}{\kern0pt}concrete\isactrlsub {\isasymT}} theorem, we can then infer that the minimal 
       concretization mapping of \isa{{\isacharparenleft}{\kern0pt}sh\ {\isasymleadsto}\ {\isasymsigma}{\isacharparenright}{\kern0pt}} matches the empty state.%
}\isanewline
\ \ \ \ \isacommand{hence}\isamarkupfalse%
\ {\isachardoublequoteopen}trace{\isacharunderscore}{\kern0pt}conc\ {\isacharparenleft}{\kern0pt}min{\isacharunderscore}{\kern0pt}conc{\isacharunderscore}{\kern0pt}map\isactrlsub {\isasymT}\ {\isacharparenleft}{\kern0pt}sh\ {\isasymleadsto}\ State{\isasymllangle}{\isasymsigma}{\isasymrrangle}{\isacharparenright}{\kern0pt}\ {\isadigit{0}}{\isacharparenright}{\kern0pt}\ {\isacharparenleft}{\kern0pt}sh\ {\isasymleadsto}\ State{\isasymllangle}{\isasymsigma}{\isasymrrangle}{\isacharparenright}{\kern0pt}\ {\isacharequal}{\kern0pt}\ sh\ {\isasymleadsto}\ State{\isasymllangle}{\isasymsigma}{\isasymrrangle}{\isachardoublequoteclose}\isanewline
\ \ \ \ \ \ \isacommand{using}\isamarkupfalse%
\ assms\ trace{\isacharunderscore}{\kern0pt}conc{\isacharunderscore}{\kern0pt}pr\ \isacommand{by}\isamarkupfalse%
\ presburger\isanewline
\ \ \ \ %
\isamarkupcmt{Utilizing a supplementary theorem, we conclude that the trace concretization of 
       \isa{{\isacharparenleft}{\kern0pt}sh\ {\isasymleadsto}\ {\isasymsigma}{\isacharparenright}{\kern0pt}} under its minimal concretization mapping must be \isa{{\isacharparenleft}{\kern0pt}sh\ {\isasymleadsto}\ {\isasymsigma}{\isacharparenright}{\kern0pt}} itself.%
}\isanewline
\ \ \ \ \isacommand{ultimately}\isamarkupfalse%
\ \isacommand{show}\isamarkupfalse%
\ {\isacharquery}{\kern0pt}thesis\ \isacommand{by}\isamarkupfalse%
\ auto\isanewline
\ \ \ \ %
\isamarkupcmt{We can now use Isabelle in order to infer the conclusion of the lemma.%
}\isanewline
\ \ \isacommand{qed}\isamarkupfalse%
\endisatagproof
{\isafoldproof}%
\isadelimproof
\endisadelimproof
\begin{isamarkuptext}%
The simplification lemma for the sequential command is slightly more complex. 
  We establish that the successor configurations of any sequential statement \isa{S\isactrlsub {\isadigit{1}}{\isacharsemicolon}{\kern0pt}{\isacharsemicolon}{\kern0pt}S\isactrlsub {\isadigit{2}}} 
  match the successor configurations of \isa{S\isactrlsub {\isadigit{1}}} with \isa{S\isactrlsub {\isadigit{2}}} appended on all their 
  continuation markers. This allows us to simplify $\isa{{\isasymdelta}}_s$-function applications on 
  sequential statements to applications on only its first constituent. We infer 
  this equality in Isabelle by proving both subset~relations.%
\end{isamarkuptext}\isamarkuptrue%
\ \ \isacommand{lemma}\isamarkupfalse%
\ {\isasymdelta}\isactrlsub s{\isacharunderscore}{\kern0pt}Seq\isactrlsub {\isadigit{1}}{\isacharcolon}{\kern0pt}\isanewline
\ \ \ \ \isakeyword{assumes}\ {\isachardoublequoteopen}concrete\isactrlsub {\isasymT}{\isacharparenleft}{\kern0pt}sh\ {\isasymleadsto}\ State{\isasymllangle}{\isasymsigma}{\isasymrrangle}{\isacharparenright}{\kern0pt}{\isachardoublequoteclose}\isanewline
\ \ \ \ \isakeyword{shows}\ {\isachardoublequoteopen}{\isasymdelta}\isactrlsub s\ {\isacharparenleft}{\kern0pt}sh\ {\isasymleadsto}\ State{\isasymllangle}{\isasymsigma}{\isasymrrangle}{\isacharcomma}{\kern0pt}\ {\isasymlambda}{\isacharbrackleft}{\kern0pt}S\isactrlsub {\isadigit{1}}{\isacharsemicolon}{\kern0pt}{\isacharsemicolon}{\kern0pt}S\isactrlsub {\isadigit{2}}{\isacharbrackright}{\kern0pt}{\isacharparenright}{\kern0pt}\ \isanewline
\ \ \ \ \ \ \ \ \ \ \ \ \ \ \ \ {\isasymsubseteq}\ {\isacharparenleft}{\kern0pt}{\isacharpercent}{\kern0pt}c{\isachardot}{\kern0pt}\ {\isacharparenleft}{\kern0pt}{\isacharparenleft}{\kern0pt}fst\ c{\isacharparenright}{\kern0pt}{\isacharcomma}{\kern0pt}\ cont{\isacharunderscore}{\kern0pt}append\ {\isacharparenleft}{\kern0pt}snd\ c{\isacharparenright}{\kern0pt}\ S\isactrlsub {\isadigit{2}}{\isacharparenright}{\kern0pt}{\isacharparenright}{\kern0pt}\ {\isacharbackquote}{\kern0pt}\ {\isasymdelta}\isactrlsub s\ {\isacharparenleft}{\kern0pt}sh\ {\isasymleadsto}\ State{\isasymllangle}{\isasymsigma}{\isasymrrangle}{\isacharcomma}{\kern0pt}\ {\isasymlambda}{\isacharbrackleft}{\kern0pt}S\isactrlsub {\isadigit{1}}{\isacharbrackright}{\kern0pt}{\isacharparenright}{\kern0pt}{\isachardoublequoteclose}\isanewline
\isadelimproof
\ \ %
\endisadelimproof
\isatagproof
\isacommand{proof}\isamarkupfalse%
\ {\isacharparenleft}{\kern0pt}subst\ subset{\isacharunderscore}{\kern0pt}iff{\isacharparenright}{\kern0pt}\isanewline
\ \ \ \ \isacommand{show}\isamarkupfalse%
\ {\isachardoublequoteopen}{\isasymforall}c{\isachardot}{\kern0pt}\ c\ {\isasymin}\ {\isasymdelta}\isactrlsub s\ {\isacharparenleft}{\kern0pt}sh\ {\isasymleadsto}\ State{\isasymllangle}{\isasymsigma}{\isasymrrangle}{\isacharcomma}{\kern0pt}\ {\isasymlambda}{\isacharbrackleft}{\kern0pt}S\isactrlsub {\isadigit{1}}{\isacharsemicolon}{\kern0pt}{\isacharsemicolon}{\kern0pt}S\isactrlsub {\isadigit{2}}{\isacharbrackright}{\kern0pt}{\isacharparenright}{\kern0pt}\ \isanewline
\ \ \ \ \ \ \ \ \ \ \ \ \ \ {\isasymlongrightarrow}\ c\ {\isasymin}\ {\isacharparenleft}{\kern0pt}{\isacharpercent}{\kern0pt}c{\isachardot}{\kern0pt}\ {\isacharparenleft}{\kern0pt}{\isacharparenleft}{\kern0pt}fst\ c{\isacharparenright}{\kern0pt}{\isacharcomma}{\kern0pt}\ cont{\isacharunderscore}{\kern0pt}append\ {\isacharparenleft}{\kern0pt}snd\ c{\isacharparenright}{\kern0pt}\ S\isactrlsub {\isadigit{2}}{\isacharparenright}{\kern0pt}{\isacharparenright}{\kern0pt}\ {\isacharbackquote}{\kern0pt}\ {\isasymdelta}\isactrlsub s\ {\isacharparenleft}{\kern0pt}sh\ {\isasymleadsto}\ State{\isasymllangle}{\isasymsigma}{\isasymrrangle}{\isacharcomma}{\kern0pt}\ {\isasymlambda}{\isacharbrackleft}{\kern0pt}S\isactrlsub {\isadigit{1}}{\isacharbrackright}{\kern0pt}{\isacharparenright}{\kern0pt}{\isachardoublequoteclose}\isanewline
\ \ \ \ %
\isamarkupcmt{We first use the subset-iff rule in order to rewrite the subset relation into 
       a semantically equivalent~implication.%
}\isanewline
\ \ \ \ \isacommand{proof}\isamarkupfalse%
\ {\isacharparenleft}{\kern0pt}rule\ allI{\isacharcomma}{\kern0pt}\ rule\ impI{\isacharparenright}{\kern0pt}\isanewline
\ \ \ \ \ \ \isacommand{fix}\isamarkupfalse%
\ c\isanewline
\ \ \ \ \ \ %
\isamarkupcmt{We assume \isa{c} to be an arbitrary, but fixed configuration.%
}\isanewline
\ \ \ \ \ \ \isacommand{assume}\isamarkupfalse%
\ {\isachardoublequoteopen}c\ {\isasymin}\ {\isasymdelta}\isactrlsub s\ {\isacharparenleft}{\kern0pt}sh\ {\isasymleadsto}\ State{\isasymllangle}{\isasymsigma}{\isasymrrangle}{\isacharcomma}{\kern0pt}\ {\isasymlambda}{\isacharbrackleft}{\kern0pt}S\isactrlsub {\isadigit{1}}{\isacharsemicolon}{\kern0pt}{\isacharsemicolon}{\kern0pt}S\isactrlsub {\isadigit{2}}{\isacharbrackright}{\kern0pt}{\isacharparenright}{\kern0pt}{\isachardoublequoteclose}\isanewline
\ \ \ \ \ \ %
\isamarkupcmt{We assume that the premise holds, implying that \isa{c} is a successor 
         configuration~of~\isa{S\isactrlsub {\isadigit{1}}{\isacharsemicolon}{\kern0pt}{\isacharsemicolon}{\kern0pt}S\isactrlsub {\isadigit{2}}}.%
}\isanewline
\ \ \ \ \ \ \isacommand{then}\isamarkupfalse%
\ \isacommand{obtain}\isamarkupfalse%
\ {\isasympi}\ \isakeyword{where}\ assm\isactrlsub {\isasympi}{\isacharcolon}{\kern0pt}\ \isanewline
\ \ \ \ \ \ \ \ {\isachardoublequoteopen}c\ {\isacharequal}{\kern0pt}\ {\isacharparenleft}{\kern0pt}trace{\isacharunderscore}{\kern0pt}conc\ {\isacharparenleft}{\kern0pt}min{\isacharunderscore}{\kern0pt}conc{\isacharunderscore}{\kern0pt}map\isactrlsub {\isasymT}\ {\isacharparenleft}{\kern0pt}sh\ {\isasymcdot}\ {\isacharparenleft}{\kern0pt}{\isasymdown}\isactrlsub {\isasymtau}\ {\isasympi}{\isacharparenright}{\kern0pt}{\isacharparenright}{\kern0pt}\ {\isadigit{0}}{\isacharparenright}{\kern0pt}\ {\isacharparenleft}{\kern0pt}sh\ {\isasymcdot}\ {\isacharparenleft}{\kern0pt}{\isasymdown}\isactrlsub {\isasymtau}\ {\isasympi}{\isacharparenright}{\kern0pt}{\isacharparenright}{\kern0pt}{\isacharcomma}{\kern0pt}\ {\isasymdown}\isactrlsub {\isasymlambda}\ {\isasympi}{\isacharparenright}{\kern0pt}\ \isanewline
\ \ \ \ \ \ \ \ \ {\isasymand}\ {\isasympi}\ {\isasymin}\ val\isactrlsub s\ {\isacharparenleft}{\kern0pt}S\isactrlsub {\isadigit{1}}{\isacharsemicolon}{\kern0pt}{\isacharsemicolon}{\kern0pt}S\isactrlsub {\isadigit{2}}{\isacharparenright}{\kern0pt}\ {\isasymsigma}\ {\isasymand}\ consistent{\isacharparenleft}{\kern0pt}sval\isactrlsub B\ {\isacharparenleft}{\kern0pt}{\isasymdown}\isactrlsub p\ {\isasympi}{\isacharparenright}{\kern0pt}\ {\isacharparenleft}{\kern0pt}min{\isacharunderscore}{\kern0pt}conc{\isacharunderscore}{\kern0pt}map\isactrlsub {\isasymT}\ {\isacharparenleft}{\kern0pt}{\isasymdown}\isactrlsub {\isasymtau}\ {\isasympi}{\isacharparenright}{\kern0pt}\ {\isadigit{0}}{\isacharparenright}{\kern0pt}{\isacharparenright}{\kern0pt}{\isachardoublequoteclose}\ \isacommand{by}\isamarkupfalse%
\ auto\isanewline
\ \ \ \ \ \ %
\isamarkupcmt{Due to the definition of the $\isa{{\isasymdelta}}_s$-function, there must exist a continuation 
         trace \isa{{\isasympi}} with a consistent path condition generated from \isa{S\isactrlsub {\isadigit{1}}{\isacharsemicolon}{\kern0pt}{\isacharsemicolon}{\kern0pt}S\isactrlsub {\isadigit{2}}} that can 
         be translated into \isa{c}. We obtain this continuation trace~\isa{{\isasympi}}.%
}\ \isanewline
\ \ \ \ \ \ \isacommand{moreover}\isamarkupfalse%
\ \isacommand{then}\isamarkupfalse%
\ \isacommand{obtain}\isamarkupfalse%
\ {\isasympi}{\isacharprime}{\kern0pt}\ \isakeyword{where}\ assm\isactrlsub {\isasympi}{\isacharprime}{\kern0pt}{\isacharcolon}{\kern0pt}\ \isanewline
\ \ \ \ \ \ \ \ {\isachardoublequoteopen}{\isasympi}\ {\isacharequal}{\kern0pt}\ {\isacharparenleft}{\kern0pt}{\isasymdown}\isactrlsub p\ {\isasympi}{\isacharprime}{\kern0pt}{\isacharparenright}{\kern0pt}\ {\isasymtriangleright}\ {\isacharparenleft}{\kern0pt}{\isasymdown}\isactrlsub {\isasymtau}\ {\isasympi}{\isacharprime}{\kern0pt}{\isacharparenright}{\kern0pt}\ \isactrlitem \ cont{\isacharunderscore}{\kern0pt}append\ {\isacharparenleft}{\kern0pt}{\isasymdown}\isactrlsub {\isasymlambda}\ {\isasympi}{\isacharprime}{\kern0pt}{\isacharparenright}{\kern0pt}\ S\isactrlsub {\isadigit{2}}\ {\isasymand}\ {\isasympi}{\isacharprime}{\kern0pt}\ {\isasymin}\ {\isacharparenleft}{\kern0pt}val\isactrlsub s\ S\isactrlsub {\isadigit{1}}\ {\isasymsigma}{\isacharparenright}{\kern0pt}{\isachardoublequoteclose}\ \isacommand{by}\isamarkupfalse%
\ auto\isanewline
\ \ \ \ \ \ %
\isamarkupcmt{Aligning with the definition of the valuation function, there must also
         exist a continuation trace \isa{{\isasympi}{\isacharprime}{\kern0pt}} generated from \isa{S\isactrlsub {\isadigit{1}}}, which matches
         \isa{{\isasympi}}, if we appended \isa{S\isactrlsub {\isadigit{2}}} onto its continuation~marker.%
}\isanewline
\ \ \ \ \ \ \isacommand{ultimately}\isamarkupfalse%
\ \isacommand{have}\isamarkupfalse%
\ connect{\isacharcolon}{\kern0pt}\ \isanewline
\ \ \ \ \ \ \ \ {\isachardoublequoteopen}consistent\ {\isacharparenleft}{\kern0pt}sval\isactrlsub B\ {\isacharparenleft}{\kern0pt}{\isasymdown}\isactrlsub p\ {\isasympi}{\isacharprime}{\kern0pt}{\isacharparenright}{\kern0pt}\ {\isacharparenleft}{\kern0pt}min{\isacharunderscore}{\kern0pt}conc{\isacharunderscore}{\kern0pt}map\isactrlsub {\isasymT}\ {\isacharparenleft}{\kern0pt}{\isasymdown}\isactrlsub {\isasymtau}\ {\isasympi}{\isacharprime}{\kern0pt}{\isacharparenright}{\kern0pt}\ {\isadigit{0}}{\isacharparenright}{\kern0pt}{\isacharparenright}{\kern0pt}\ \isanewline
\ \ \ \ \ \ \ \ \ {\isasymand}\ {\isacharparenleft}{\kern0pt}{\isasymdown}\isactrlsub {\isasymtau}\ {\isasympi}{\isacharprime}{\kern0pt}{\isacharparenright}{\kern0pt}\ {\isacharequal}{\kern0pt}\ {\isacharparenleft}{\kern0pt}{\isasymdown}\isactrlsub {\isasymtau}\ {\isasympi}{\isacharparenright}{\kern0pt}\ {\isasymand}\ {\isacharparenleft}{\kern0pt}{\isasymdown}\isactrlsub {\isasymlambda}\ {\isasympi}{\isacharparenright}{\kern0pt}\ {\isacharequal}{\kern0pt}\ cont{\isacharunderscore}{\kern0pt}append\ {\isacharparenleft}{\kern0pt}{\isasymdown}\isactrlsub {\isasymlambda}\ {\isasympi}{\isacharprime}{\kern0pt}{\isacharparenright}{\kern0pt}\ S\isactrlsub {\isadigit{2}}{\isachardoublequoteclose}\ \isacommand{by}\isamarkupfalse%
\ simp\isanewline
\ \ \ \ \ \ %
\isamarkupcmt{Considering that the path conditions of \isa{{\isasympi}} and \isa{{\isasympi}{\isacharprime}{\kern0pt}} match, both must be 
         consistent. Their symbolic traces also match. The only difference lies
         in the modified continuation~marker.%
}\isanewline
\ \ \ \ \ \ \isacommand{moreover}\isamarkupfalse%
\ \isacommand{then}\isamarkupfalse%
\ \isacommand{obtain}\isamarkupfalse%
\ c{\isacharprime}{\kern0pt}\ \isakeyword{where}\ \isanewline
\ \ \ \ \ \ \ \ {\isachardoublequoteopen}c{\isacharprime}{\kern0pt}\ {\isacharequal}{\kern0pt}\ {\isacharparenleft}{\kern0pt}trace{\isacharunderscore}{\kern0pt}conc\ {\isacharparenleft}{\kern0pt}min{\isacharunderscore}{\kern0pt}conc{\isacharunderscore}{\kern0pt}map\isactrlsub {\isasymT}\ {\isacharparenleft}{\kern0pt}sh\ {\isasymcdot}\ {\isacharparenleft}{\kern0pt}{\isasymdown}\isactrlsub {\isasymtau}\ {\isasympi}{\isacharprime}{\kern0pt}{\isacharparenright}{\kern0pt}{\isacharparenright}{\kern0pt}\ {\isadigit{0}}{\isacharparenright}{\kern0pt}\ {\isacharparenleft}{\kern0pt}sh\ {\isasymcdot}\ {\isacharparenleft}{\kern0pt}{\isasymdown}\isactrlsub {\isasymtau}\ {\isasympi}{\isacharprime}{\kern0pt}{\isacharparenright}{\kern0pt}{\isacharparenright}{\kern0pt}{\isacharcomma}{\kern0pt}\ {\isasymdown}\isactrlsub {\isasymlambda}\ {\isasympi}{\isacharprime}{\kern0pt}{\isacharparenright}{\kern0pt}{\isachardoublequoteclose}\ \isacommand{by}\isamarkupfalse%
\ auto\isanewline
\ \ \ \ \ \ %
\isamarkupcmt{We can then obtain the configuration \isa{c{\isacharprime}{\kern0pt}} translated from the 
         continuation trace~\isa{{\isasympi}{\isacharprime}{\kern0pt}}.%
}\isanewline
\ \ \ \ \ \ \isacommand{moreover}\isamarkupfalse%
\ \isacommand{then}\isamarkupfalse%
\ \isacommand{have}\isamarkupfalse%
\ {\isachardoublequoteopen}fst{\isacharparenleft}{\kern0pt}c{\isacharparenright}{\kern0pt}\ {\isacharequal}{\kern0pt}\ fst{\isacharparenleft}{\kern0pt}c{\isacharprime}{\kern0pt}{\isacharparenright}{\kern0pt}\ {\isasymand}\ snd{\isacharparenleft}{\kern0pt}c{\isacharparenright}{\kern0pt}\ {\isacharequal}{\kern0pt}\ cont{\isacharunderscore}{\kern0pt}append\ {\isacharparenleft}{\kern0pt}snd\ c{\isacharprime}{\kern0pt}{\isacharparenright}{\kern0pt}\ S\isactrlsub {\isadigit{2}}{\isachardoublequoteclose}\isanewline
\ \ \ \ \ \ \ \ \isacommand{by}\isamarkupfalse%
\ {\isacharparenleft}{\kern0pt}simp\ add{\isacharcolon}{\kern0pt}\ assm\isactrlsub {\isasympi}\ connect{\isacharparenright}{\kern0pt}\isanewline
\ \ \ \ \ \ %
\isamarkupcmt{Given this information, both \isa{c} and \isa{c{\isacharprime}{\kern0pt}} must match in their symbolic 
         traces. However, \isa{c} additionally appends \isa{S\isactrlsub {\isadigit{2}}} onto its continuation~marker.%
}\isanewline
\ \ \ \ \ \ \isacommand{ultimately}\isamarkupfalse%
\ \isacommand{show}\isamarkupfalse%
\ {\isachardoublequoteopen}c\ {\isasymin}\ {\isacharparenleft}{\kern0pt}{\isacharpercent}{\kern0pt}c{\isachardot}{\kern0pt}\ {\isacharparenleft}{\kern0pt}{\isacharparenleft}{\kern0pt}fst\ c{\isacharparenright}{\kern0pt}{\isacharcomma}{\kern0pt}\ cont{\isacharunderscore}{\kern0pt}append\ {\isacharparenleft}{\kern0pt}snd\ c{\isacharparenright}{\kern0pt}\ S\isactrlsub {\isadigit{2}}{\isacharparenright}{\kern0pt}{\isacharparenright}{\kern0pt}\ {\isacharbackquote}{\kern0pt}\ {\isasymdelta}\isactrlsub s\ {\isacharparenleft}{\kern0pt}sh\ {\isasymleadsto}\ State{\isasymllangle}{\isasymsigma}{\isasymrrangle}{\isacharcomma}{\kern0pt}\ {\isasymlambda}{\isacharbrackleft}{\kern0pt}S\isactrlsub {\isadigit{1}}{\isacharbrackright}{\kern0pt}{\isacharparenright}{\kern0pt}{\isachardoublequoteclose}\isanewline
\ \ \ \ \ \ \ \ \isacommand{using}\isamarkupfalse%
\ assm\isactrlsub {\isasympi}{\isacharprime}{\kern0pt}\ assm\isactrlsub {\isasympi}\ image{\isacharunderscore}{\kern0pt}iff\ \isacommand{by}\isamarkupfalse%
\ fastforce\isanewline
\ \ \ \ \ \ %
\isamarkupcmt{Thus \isa{c} must match \isa{c{\isacharprime}{\kern0pt}} with an appended \isa{S\isactrlsub {\isadigit{2}}} in its continuation marker,
         closing the proof.%
}\isanewline
\ \ \ \ \isacommand{qed}\isamarkupfalse%
\isanewline
\ \ \isacommand{qed}\isamarkupfalse%
\endisatagproof
{\isafoldproof}%
\isadelimproof
\isanewline
\endisadelimproof
\isanewline
\ \ \isacommand{lemma}\isamarkupfalse%
\ {\isasymdelta}\isactrlsub s{\isacharunderscore}{\kern0pt}Seq\isactrlsub {\isadigit{2}}{\isacharcolon}{\kern0pt}\isanewline
\ \ \ \ \isakeyword{assumes}\ {\isachardoublequoteopen}concrete\isactrlsub {\isasymT}{\isacharparenleft}{\kern0pt}sh\ {\isasymleadsto}\ State{\isasymllangle}{\isasymsigma}{\isasymrrangle}{\isacharparenright}{\kern0pt}{\isachardoublequoteclose}\isanewline
\ \ \ \ \isakeyword{shows}\ {\isachardoublequoteopen}{\isacharparenleft}{\kern0pt}{\isacharpercent}{\kern0pt}c{\isachardot}{\kern0pt}\ {\isacharparenleft}{\kern0pt}{\isacharparenleft}{\kern0pt}fst\ c{\isacharparenright}{\kern0pt}{\isacharcomma}{\kern0pt}\ cont{\isacharunderscore}{\kern0pt}append\ {\isacharparenleft}{\kern0pt}snd\ c{\isacharparenright}{\kern0pt}\ S\isactrlsub {\isadigit{2}}{\isacharparenright}{\kern0pt}{\isacharparenright}{\kern0pt}\ {\isacharbackquote}{\kern0pt}\ {\isasymdelta}\isactrlsub s\ {\isacharparenleft}{\kern0pt}sh\ {\isasymleadsto}\ State{\isasymllangle}{\isasymsigma}{\isasymrrangle}{\isacharcomma}{\kern0pt}\ {\isasymlambda}{\isacharbrackleft}{\kern0pt}S\isactrlsub {\isadigit{1}}{\isacharbrackright}{\kern0pt}{\isacharparenright}{\kern0pt}\ \isanewline
\ \ \ \ \ \ \ \ \ \ \ \ \ \ \ \ {\isasymsubseteq}\ {\isasymdelta}\isactrlsub s\ {\isacharparenleft}{\kern0pt}sh\ {\isasymleadsto}\ State{\isasymllangle}{\isasymsigma}{\isasymrrangle}{\isacharcomma}{\kern0pt}\ {\isasymlambda}{\isacharbrackleft}{\kern0pt}S\isactrlsub {\isadigit{1}}{\isacharsemicolon}{\kern0pt}{\isacharsemicolon}{\kern0pt}S\isactrlsub {\isadigit{2}}{\isacharbrackright}{\kern0pt}{\isacharparenright}{\kern0pt}{\isachardoublequoteclose}\isanewline
\isadelimproof
\ \ %
\endisadelimproof
\isatagproof
\isacommand{proof}\isamarkupfalse%
\ {\isacharparenleft}{\kern0pt}subst\ subset{\isacharunderscore}{\kern0pt}iff{\isacharparenright}{\kern0pt}\isanewline
\ \ \ \ \isacommand{show}\isamarkupfalse%
\ {\isachardoublequoteopen}{\isasymforall}c{\isachardot}{\kern0pt}\ c\ {\isasymin}\ {\isacharparenleft}{\kern0pt}{\isacharpercent}{\kern0pt}c{\isachardot}{\kern0pt}\ {\isacharparenleft}{\kern0pt}{\isacharparenleft}{\kern0pt}fst\ c{\isacharparenright}{\kern0pt}{\isacharcomma}{\kern0pt}\ cont{\isacharunderscore}{\kern0pt}append\ {\isacharparenleft}{\kern0pt}snd\ c{\isacharparenright}{\kern0pt}\ S\isactrlsub {\isadigit{2}}{\isacharparenright}{\kern0pt}{\isacharparenright}{\kern0pt}\ {\isacharbackquote}{\kern0pt}\ {\isasymdelta}\isactrlsub s\ {\isacharparenleft}{\kern0pt}sh\ {\isasymleadsto}\ State{\isasymllangle}{\isasymsigma}{\isasymrrangle}{\isacharcomma}{\kern0pt}\ {\isasymlambda}{\isacharbrackleft}{\kern0pt}S\isactrlsub {\isadigit{1}}{\isacharbrackright}{\kern0pt}{\isacharparenright}{\kern0pt}\ \isanewline
\ \ \ \ \ \ \ \ \ \ \ \ \ \ {\isasymlongrightarrow}\ c\ {\isasymin}\ {\isasymdelta}\isactrlsub s\ {\isacharparenleft}{\kern0pt}sh\ {\isasymleadsto}\ State{\isasymllangle}{\isasymsigma}{\isasymrrangle}{\isacharcomma}{\kern0pt}\ {\isasymlambda}{\isacharbrackleft}{\kern0pt}S\isactrlsub {\isadigit{1}}{\isacharsemicolon}{\kern0pt}{\isacharsemicolon}{\kern0pt}S\isactrlsub {\isadigit{2}}{\isacharbrackright}{\kern0pt}{\isacharparenright}{\kern0pt}{\isachardoublequoteclose}\isanewline
\ \ \ \ %
\isamarkupcmt{We first use the subset-iff rule in order to rewrite the subset relation into 
       a semantically equivalent~implication.%
}\ \isanewline
\ \ \ \ \isacommand{proof}\isamarkupfalse%
\ {\isacharparenleft}{\kern0pt}rule\ allI{\isacharcomma}{\kern0pt}\ rule\ impI{\isacharparenright}{\kern0pt}\isanewline
\ \ \ \ \ \ \isacommand{fix}\isamarkupfalse%
\ c\isanewline
\ \ \ \ \ \ %
\isamarkupcmt{We assume \isa{c} to be an arbitrary, but fixed configuration.%
}\isanewline
\ \ \ \ \ \ \isacommand{assume}\isamarkupfalse%
\ {\isachardoublequoteopen}c\ {\isasymin}\ {\isacharparenleft}{\kern0pt}{\isacharpercent}{\kern0pt}c{\isachardot}{\kern0pt}\ {\isacharparenleft}{\kern0pt}{\isacharparenleft}{\kern0pt}fst\ c{\isacharparenright}{\kern0pt}{\isacharcomma}{\kern0pt}\ cont{\isacharunderscore}{\kern0pt}append\ {\isacharparenleft}{\kern0pt}snd\ c{\isacharparenright}{\kern0pt}\ S\isactrlsub {\isadigit{2}}{\isacharparenright}{\kern0pt}{\isacharparenright}{\kern0pt}\ {\isacharbackquote}{\kern0pt}\ {\isasymdelta}\isactrlsub s\ {\isacharparenleft}{\kern0pt}sh\ {\isasymleadsto}\ State{\isasymllangle}{\isasymsigma}{\isasymrrangle}{\isacharcomma}{\kern0pt}\ {\isasymlambda}{\isacharbrackleft}{\kern0pt}S\isactrlsub {\isadigit{1}}{\isacharbrackright}{\kern0pt}{\isacharparenright}{\kern0pt}{\isachardoublequoteclose}\isanewline
\ \ \ \ \ \ %
\isamarkupcmt{We assume that the premise holds, implying that \isa{c} is a successor 
         configuration of \isa{S\isactrlsub {\isadigit{1}}} with \isa{S\isactrlsub {\isadigit{2}}} appended onto its continuation~marker.%
}\isanewline
\ \ \ \ \ \ \isacommand{then}\isamarkupfalse%
\ \isacommand{obtain}\isamarkupfalse%
\ c{\isacharprime}{\kern0pt}\ \isakeyword{where}\ assm\isactrlsub c{\isacharprime}{\kern0pt}{\isacharcolon}{\kern0pt}\isanewline
\ \ \ \ \ \ \ \ {\isachardoublequoteopen}c{\isacharprime}{\kern0pt}\ {\isasymin}\ {\isasymdelta}\isactrlsub s\ {\isacharparenleft}{\kern0pt}sh\ {\isasymleadsto}\ State{\isasymllangle}{\isasymsigma}{\isasymrrangle}{\isacharcomma}{\kern0pt}\ {\isasymlambda}{\isacharbrackleft}{\kern0pt}S\isactrlsub {\isadigit{1}}{\isacharbrackright}{\kern0pt}{\isacharparenright}{\kern0pt}\ {\isasymand}\ fst{\isacharparenleft}{\kern0pt}c{\isacharprime}{\kern0pt}{\isacharparenright}{\kern0pt}\ {\isacharequal}{\kern0pt}\ fst{\isacharparenleft}{\kern0pt}c{\isacharparenright}{\kern0pt}\ {\isasymand}\ cont{\isacharunderscore}{\kern0pt}append\ {\isacharparenleft}{\kern0pt}snd\ c{\isacharprime}{\kern0pt}{\isacharparenright}{\kern0pt}\ S\isactrlsub {\isadigit{2}}\ {\isacharequal}{\kern0pt}\ snd{\isacharparenleft}{\kern0pt}c{\isacharparenright}{\kern0pt}{\isachardoublequoteclose}\ \isacommand{by}\isamarkupfalse%
\ force\isanewline
\ \ \ \ \ \ %
\isamarkupcmt{Then there must exist a configuration \isa{c{\isacharprime}{\kern0pt}} that is also a successor 
         configuration of \isa{S\isactrlsub {\isadigit{1}}}, matching with \isa{c} in everything except its
         continuation marker. \isa{c{\isacharprime}{\kern0pt}} with \isa{S\isactrlsub {\isadigit{2}}} appended on its continuation marker
         matches~configuration~\isa{c}.%
}\isanewline
\ \ \ \ \ \ \isacommand{moreover}\isamarkupfalse%
\ \isacommand{then}\isamarkupfalse%
\ \isacommand{obtain}\isamarkupfalse%
\ {\isasympi}\ \isakeyword{where}\ assm\isactrlsub {\isasympi}{\isacharcolon}{\kern0pt}\isanewline
\ \ \ \ \ \ \ \ {\isachardoublequoteopen}{\isasympi}\ {\isasymin}\ val\isactrlsub s\ S\isactrlsub {\isadigit{1}}\ {\isasymsigma}\ {\isasymand}\ consistent{\isacharparenleft}{\kern0pt}sval\isactrlsub B\ {\isacharparenleft}{\kern0pt}{\isasymdown}\isactrlsub p\ {\isasympi}{\isacharparenright}{\kern0pt}\ {\isacharparenleft}{\kern0pt}min{\isacharunderscore}{\kern0pt}conc{\isacharunderscore}{\kern0pt}map\isactrlsub {\isasymT}\ {\isacharparenleft}{\kern0pt}{\isasymdown}\isactrlsub {\isasymtau}\ {\isasympi}{\isacharparenright}{\kern0pt}\ {\isadigit{0}}{\isacharparenright}{\kern0pt}{\isacharparenright}{\kern0pt}\ \isanewline
\ \ \ \ \ \ \ \ \ {\isasymand}\ c{\isacharprime}{\kern0pt}\ {\isacharequal}{\kern0pt}\ {\isacharparenleft}{\kern0pt}trace{\isacharunderscore}{\kern0pt}conc\ {\isacharparenleft}{\kern0pt}min{\isacharunderscore}{\kern0pt}conc{\isacharunderscore}{\kern0pt}map\isactrlsub {\isasymT}\ {\isacharparenleft}{\kern0pt}sh\ {\isasymcdot}\ {\isacharparenleft}{\kern0pt}{\isasymdown}\isactrlsub {\isasymtau}\ {\isasympi}{\isacharparenright}{\kern0pt}{\isacharparenright}{\kern0pt}\ {\isadigit{0}}{\isacharparenright}{\kern0pt}\ {\isacharparenleft}{\kern0pt}sh\ {\isasymcdot}\ {\isacharparenleft}{\kern0pt}{\isasymdown}\isactrlsub {\isasymtau}\ {\isasympi}{\isacharparenright}{\kern0pt}{\isacharparenright}{\kern0pt}{\isacharcomma}{\kern0pt}\ {\isacharparenleft}{\kern0pt}{\isasymdown}\isactrlsub {\isasymlambda}\ {\isasympi}{\isacharparenright}{\kern0pt}{\isacharparenright}{\kern0pt}{\isachardoublequoteclose}\ \isacommand{by}\isamarkupfalse%
\ auto\isanewline
\ \ \ \ \ \ %
\isamarkupcmt{Hence there must exist a continuation trace \isa{{\isasympi}} with a consistent path 
         condition generated from \isa{S\isactrlsub {\isadigit{1}}}, which translates to configuration~\isa{c{\isacharprime}{\kern0pt}}.%
}\isanewline
\ \ \ \ \ \ \isacommand{moreover}\isamarkupfalse%
\ \isacommand{then}\isamarkupfalse%
\ \isacommand{obtain}\isamarkupfalse%
\ {\isasympi}{\isacharprime}{\kern0pt}\ \isakeyword{where}\isanewline
\ \ \ \ \ \ \ \ {\isachardoublequoteopen}{\isasympi}{\isacharprime}{\kern0pt}\ {\isacharequal}{\kern0pt}\ {\isacharparenleft}{\kern0pt}{\isasymdown}\isactrlsub p\ {\isasympi}{\isacharparenright}{\kern0pt}\ {\isasymtriangleright}\ {\isacharparenleft}{\kern0pt}{\isasymdown}\isactrlsub {\isasymtau}\ {\isasympi}{\isacharparenright}{\kern0pt}\ \isactrlitem \ cont{\isacharunderscore}{\kern0pt}append\ {\isacharparenleft}{\kern0pt}{\isasymdown}\isactrlsub {\isasymlambda}\ {\isasympi}{\isacharparenright}{\kern0pt}\ S\isactrlsub {\isadigit{2}}{\isachardoublequoteclose}\ \isacommand{by}\isamarkupfalse%
\ auto\isanewline
\ \ \ \ \ \ %
\isamarkupcmt{We then obtain the continuation trace \isa{{\isasympi}{\isacharprime}{\kern0pt}} which matches \isa{{\isasympi}} except
         having \isa{S\isactrlsub {\isadigit{2}}} additionally appended onto its continuation~marker.%
}\isanewline
\ \ \ \ \ \ \isacommand{ultimately}\isamarkupfalse%
\ \isacommand{have}\isamarkupfalse%
\ connect{\isacharcolon}{\kern0pt}\ \isanewline
\ \ \ \ \ \ \ \ {\isachardoublequoteopen}{\isasympi}{\isacharprime}{\kern0pt}\ {\isasymin}\ val\isactrlsub s\ {\isacharparenleft}{\kern0pt}S\isactrlsub {\isadigit{1}}{\isacharsemicolon}{\kern0pt}{\isacharsemicolon}{\kern0pt}S\isactrlsub {\isadigit{2}}{\isacharparenright}{\kern0pt}\ {\isasymsigma}\ {\isasymand}\ consistent{\isacharparenleft}{\kern0pt}sval\isactrlsub B\ {\isacharparenleft}{\kern0pt}{\isasymdown}\isactrlsub p\ {\isasympi}{\isacharprime}{\kern0pt}{\isacharparenright}{\kern0pt}\ {\isacharparenleft}{\kern0pt}min{\isacharunderscore}{\kern0pt}conc{\isacharunderscore}{\kern0pt}map\isactrlsub {\isasymT}\ {\isacharparenleft}{\kern0pt}{\isasymdown}\isactrlsub {\isasymtau}\ {\isasympi}{\isacharprime}{\kern0pt}{\isacharparenright}{\kern0pt}\ {\isadigit{0}}{\isacharparenright}{\kern0pt}{\isacharparenright}{\kern0pt}\ \isanewline
\ \ \ \ \ \ \ \ \ {\isasymand}\ {\isacharparenleft}{\kern0pt}{\isasymdown}\isactrlsub {\isasymtau}\ {\isasympi}{\isacharprime}{\kern0pt}{\isacharparenright}{\kern0pt}\ {\isacharequal}{\kern0pt}\ {\isacharparenleft}{\kern0pt}{\isasymdown}\isactrlsub {\isasymtau}\ {\isasympi}{\isacharparenright}{\kern0pt}\ {\isasymand}\ {\isacharparenleft}{\kern0pt}{\isasymdown}\isactrlsub {\isasymlambda}\ {\isasympi}{\isacharprime}{\kern0pt}{\isacharparenright}{\kern0pt}\ {\isacharequal}{\kern0pt}\ cont{\isacharunderscore}{\kern0pt}append\ {\isacharparenleft}{\kern0pt}{\isasymdown}\isactrlsub {\isasymlambda}\ {\isasympi}{\isacharparenright}{\kern0pt}\ S\isactrlsub {\isadigit{2}}{\isachardoublequoteclose}\ \isacommand{by}\isamarkupfalse%
\ auto\ \isanewline
\ \ \ \ \ \ %
\isamarkupcmt{This implies that \isa{{\isasympi}{\isacharprime}{\kern0pt}} must be a continuation trace with a consistent
         path condition generated from \isa{S\isactrlsub {\isadigit{1}}{\isacharsemicolon}{\kern0pt}{\isacharsemicolon}{\kern0pt}S\isactrlsub {\isadigit{2}}}. Note that \isa{{\isasympi}} matches with \isa{{\isasympi}{\isacharprime}{\kern0pt}}
         in its symbolic trace, but not in its continuation marker.%
}\isanewline
\ \ \ \ \ \ \isacommand{then}\isamarkupfalse%
\ \isacommand{obtain}\isamarkupfalse%
\ c{\isacharprime}{\kern0pt}{\isacharprime}{\kern0pt}\ \isakeyword{where}\ assm\isactrlsub c{\isacharprime}{\kern0pt}{\isacharprime}{\kern0pt}{\isacharcolon}{\kern0pt}\isanewline
\ \ \ \ \ \ \ \ {\isachardoublequoteopen}c{\isacharprime}{\kern0pt}{\isacharprime}{\kern0pt}\ {\isacharequal}{\kern0pt}\ {\isacharparenleft}{\kern0pt}trace{\isacharunderscore}{\kern0pt}conc\ {\isacharparenleft}{\kern0pt}min{\isacharunderscore}{\kern0pt}conc{\isacharunderscore}{\kern0pt}map\isactrlsub {\isasymT}\ {\isacharparenleft}{\kern0pt}sh\ {\isasymcdot}\ {\isacharparenleft}{\kern0pt}{\isasymdown}\isactrlsub {\isasymtau}\ {\isasympi}{\isacharprime}{\kern0pt}{\isacharparenright}{\kern0pt}{\isacharparenright}{\kern0pt}\ {\isadigit{0}}{\isacharparenright}{\kern0pt}\ {\isacharparenleft}{\kern0pt}sh\ {\isasymcdot}\ {\isacharparenleft}{\kern0pt}{\isasymdown}\isactrlsub {\isasymtau}\ {\isasympi}{\isacharprime}{\kern0pt}{\isacharparenright}{\kern0pt}{\isacharparenright}{\kern0pt}{\isacharcomma}{\kern0pt}\ {\isasymdown}\isactrlsub {\isasymlambda}\ {\isasympi}{\isacharprime}{\kern0pt}{\isacharparenright}{\kern0pt}{\isachardoublequoteclose}\ \isacommand{by}\isamarkupfalse%
\ auto\isanewline
\ \ \ \ \ \ %
\isamarkupcmt{We then obtain the configuration \isa{c{\isacharprime}{\kern0pt}{\isacharprime}{\kern0pt}} translated from \isa{{\isasympi}{\isacharprime}{\kern0pt}}.%
}\isanewline
\ \ \ \ \ \ \isacommand{hence}\isamarkupfalse%
\ {\isachardoublequoteopen}c\ {\isacharequal}{\kern0pt}\ c{\isacharprime}{\kern0pt}{\isacharprime}{\kern0pt}{\isachardoublequoteclose}\ \isacommand{using}\isamarkupfalse%
\ assm\isactrlsub c{\isacharprime}{\kern0pt}\ assm\isactrlsub {\isasympi}\ connect\ \isacommand{by}\isamarkupfalse%
\ auto\isanewline
\ \ \ \ \ \ %
\isamarkupcmt{We can now derive that \isa{c} and \isa{c{\isacharprime}{\kern0pt}{\isacharprime}{\kern0pt}} match in both of~their~elements.%
}\isanewline
\ \ \ \ \ \ \isacommand{thus}\isamarkupfalse%
\ {\isachardoublequoteopen}c\ {\isasymin}\ {\isasymdelta}\isactrlsub s\ {\isacharparenleft}{\kern0pt}sh\ {\isasymleadsto}\ State{\isasymllangle}{\isasymsigma}{\isasymrrangle}{\isacharcomma}{\kern0pt}\ {\isasymlambda}{\isacharbrackleft}{\kern0pt}S\isactrlsub {\isadigit{1}}{\isacharsemicolon}{\kern0pt}{\isacharsemicolon}{\kern0pt}S\isactrlsub {\isadigit{2}}{\isacharbrackright}{\kern0pt}{\isacharparenright}{\kern0pt}{\isachardoublequoteclose}\ \isanewline
\ \ \ \ \ \ \ \ \isacommand{using}\isamarkupfalse%
\ assm\isactrlsub c{\isacharprime}{\kern0pt}{\isacharprime}{\kern0pt}\ connect\ image{\isacharunderscore}{\kern0pt}iff\ \isacommand{by}\isamarkupfalse%
\ fastforce\isanewline
\ \ \ \ \ \ %
\isamarkupcmt{We know that \isa{c{\isacharprime}{\kern0pt}{\isacharprime}{\kern0pt}} is a successor configuration of \isa{S\isactrlsub {\isadigit{1}}{\isacharsemicolon}{\kern0pt}{\isacharsemicolon}{\kern0pt}S\isactrlsub {\isadigit{2}}}. 
         Considering that \isa{c} matches \isa{c{\isacharprime}{\kern0pt}{\isacharprime}{\kern0pt}}, we can finally conclude that 
         \isa{c} must also be a successor configuration of \isa{S\isactrlsub {\isadigit{1}}{\isacharsemicolon}{\kern0pt}{\isacharsemicolon}{\kern0pt}S\isactrlsub {\isadigit{2}}}, which needed to be 
         proven in the first~place.%
}\isanewline
\ \ \ \ \isacommand{qed}\isamarkupfalse%
\isanewline
\ \ \isacommand{qed}\isamarkupfalse%
\endisatagproof
{\isafoldproof}%
\isadelimproof
\isanewline
\endisadelimproof
\ \ \ \ \ \ \ \ \isanewline
\ \ \isacommand{lemma}\isamarkupfalse%
\ {\isasymdelta}\isactrlsub s{\isacharunderscore}{\kern0pt}Seq{\isacharcolon}{\kern0pt}\isanewline
\ \ \ \ \isakeyword{assumes}\ {\isachardoublequoteopen}concrete\isactrlsub {\isasymT}{\isacharparenleft}{\kern0pt}sh\ {\isasymleadsto}\ State{\isasymllangle}{\isasymsigma}{\isasymrrangle}{\isacharparenright}{\kern0pt}{\isachardoublequoteclose}\isanewline
\ \ \ \ \isakeyword{shows}\ {\isachardoublequoteopen}{\isasymdelta}\isactrlsub s\ {\isacharparenleft}{\kern0pt}sh\ {\isasymleadsto}\ State{\isasymllangle}{\isasymsigma}{\isasymrrangle}{\isacharcomma}{\kern0pt}\ {\isasymlambda}{\isacharbrackleft}{\kern0pt}S\isactrlsub {\isadigit{1}}{\isacharsemicolon}{\kern0pt}{\isacharsemicolon}{\kern0pt}S\isactrlsub {\isadigit{2}}{\isacharbrackright}{\kern0pt}{\isacharparenright}{\kern0pt}\ \isanewline
\ \ \ \ \ \ \ \ \ \ \ \ \ \ \ \ {\isacharequal}{\kern0pt}\ {\isacharparenleft}{\kern0pt}{\isacharpercent}{\kern0pt}c{\isachardot}{\kern0pt}\ {\isacharparenleft}{\kern0pt}{\isacharparenleft}{\kern0pt}fst\ c{\isacharparenright}{\kern0pt}{\isacharcomma}{\kern0pt}\ cont{\isacharunderscore}{\kern0pt}append\ {\isacharparenleft}{\kern0pt}snd\ c{\isacharparenright}{\kern0pt}\ S\isactrlsub {\isadigit{2}}{\isacharparenright}{\kern0pt}{\isacharparenright}{\kern0pt}\ {\isacharbackquote}{\kern0pt}\ {\isasymdelta}\isactrlsub s\ {\isacharparenleft}{\kern0pt}sh\ {\isasymleadsto}\ State{\isasymllangle}{\isasymsigma}{\isasymrrangle}{\isacharcomma}{\kern0pt}\ {\isasymlambda}{\isacharbrackleft}{\kern0pt}S\isactrlsub {\isadigit{1}}{\isacharbrackright}{\kern0pt}{\isacharparenright}{\kern0pt}{\isachardoublequoteclose}\isanewline
\isadelimproof
\ \ \ \ %
\endisadelimproof
\isatagproof
\isacommand{apply}\isamarkupfalse%
\ {\isacharparenleft}{\kern0pt}subst\ set{\isacharunderscore}{\kern0pt}eq{\isacharunderscore}{\kern0pt}subset{\isacharparenright}{\kern0pt}\isanewline
\ \ \ \ \isacommand{using}\isamarkupfalse%
\ assms\ {\isasymdelta}\isactrlsub s{\isacharunderscore}{\kern0pt}Seq\isactrlsub {\isadigit{1}}\ {\isasymdelta}\isactrlsub s{\isacharunderscore}{\kern0pt}Seq\isactrlsub {\isadigit{2}}\ \isacommand{by}\isamarkupfalse%
\ auto\isanewline
\ \ \ \ %
\isamarkupcmt{We can now use the proof of both subset directions to infer the 
       desired~equality.%
}%
\endisatagproof
{\isafoldproof}%
\isadelimproof
\endisadelimproof
\begin{isamarkuptext}%
The simplification lemma for the local parallelism command can be formalized
  in a similar manner. We establish that the successor configurations of any
  local parallelism command \isa{CO\ S\isactrlsub {\isadigit{1}}\ {\isasymparallel}\ S\isactrlsub {\isadigit{2}}\ OC} match the successor configurations of 
  \isa{S\isactrlsub {\isadigit{1}}} with the local parallelism under \isa{S\isactrlsub {\isadigit{2}}} reconstructed in all their continuation
  markers, merged with the successor configurations of \isa{S\isactrlsub {\isadigit{2}}} with the local parallelism 
  under \isa{S\isactrlsub {\isadigit{1}}} reconstructed in all their continuation markers. This allows us to 
  simplify $\isa{{\isasymdelta}}_s$-function applications on local parallelism commands to applications 
  on both of its constituents. We again infer this equality in Isabelle by proving 
  both corresponding subset~relations.%
\end{isamarkuptext}\isamarkuptrue%
\ \ \isacommand{lemma}\isamarkupfalse%
\ {\isasymdelta}\isactrlsub s{\isacharunderscore}{\kern0pt}LocPar\isactrlsub {\isadigit{1}}{\isacharcolon}{\kern0pt}\isanewline
\ \ \ \ \isakeyword{assumes}\ {\isachardoublequoteopen}concrete\isactrlsub {\isasymT}{\isacharparenleft}{\kern0pt}sh\ {\isasymleadsto}\ State{\isasymllangle}{\isasymsigma}{\isasymrrangle}{\isacharparenright}{\kern0pt}{\isachardoublequoteclose}\isanewline
\ \ \ \ \isakeyword{shows}\ {\isachardoublequoteopen}{\isasymdelta}\isactrlsub s\ {\isacharparenleft}{\kern0pt}sh\ {\isasymleadsto}\ State{\isasymllangle}{\isasymsigma}{\isasymrrangle}{\isacharcomma}{\kern0pt}\ {\isasymlambda}{\isacharbrackleft}{\kern0pt}CO\ S\isactrlsub {\isadigit{1}}\ {\isasymparallel}\ S\isactrlsub {\isadigit{2}}\ OC{\isacharbrackright}{\kern0pt}{\isacharparenright}{\kern0pt}\ \isanewline
\ \ \ \ \ \ \ \ \ \ \ \ \ \ \ \ {\isasymsubseteq}\ {\isacharparenleft}{\kern0pt}{\isacharpercent}{\kern0pt}c{\isachardot}{\kern0pt}\ {\isacharparenleft}{\kern0pt}fst{\isacharparenleft}{\kern0pt}c{\isacharparenright}{\kern0pt}{\isacharcomma}{\kern0pt}\ parallel\ {\isacharparenleft}{\kern0pt}snd\ c{\isacharparenright}{\kern0pt}\ {\isasymlambda}{\isacharbrackleft}{\kern0pt}S\isactrlsub {\isadigit{2}}{\isacharbrackright}{\kern0pt}{\isacharparenright}{\kern0pt}{\isacharparenright}{\kern0pt}\ {\isacharbackquote}{\kern0pt}\ {\isasymdelta}\isactrlsub s\ {\isacharparenleft}{\kern0pt}sh\ {\isasymleadsto}\ State{\isasymllangle}{\isasymsigma}{\isasymrrangle}{\isacharcomma}{\kern0pt}\ {\isasymlambda}{\isacharbrackleft}{\kern0pt}S\isactrlsub {\isadigit{1}}{\isacharbrackright}{\kern0pt}{\isacharparenright}{\kern0pt}\ {\isasymunion}\isanewline
\ \ \ \ \ \ \ \ \ \ \ \ \ \ \ \ \ \ \ \ {\isacharparenleft}{\kern0pt}{\isacharpercent}{\kern0pt}c{\isachardot}{\kern0pt}\ {\isacharparenleft}{\kern0pt}fst{\isacharparenleft}{\kern0pt}c{\isacharparenright}{\kern0pt}{\isacharcomma}{\kern0pt}\ parallel\ {\isasymlambda}{\isacharbrackleft}{\kern0pt}S\isactrlsub {\isadigit{1}}{\isacharbrackright}{\kern0pt}\ {\isacharparenleft}{\kern0pt}snd\ c{\isacharparenright}{\kern0pt}{\isacharparenright}{\kern0pt}{\isacharparenright}{\kern0pt}\ {\isacharbackquote}{\kern0pt}\ {\isasymdelta}\isactrlsub s\ {\isacharparenleft}{\kern0pt}sh\ {\isasymleadsto}\ State{\isasymllangle}{\isasymsigma}{\isasymrrangle}{\isacharcomma}{\kern0pt}\ {\isasymlambda}{\isacharbrackleft}{\kern0pt}S\isactrlsub {\isadigit{2}}{\isacharbrackright}{\kern0pt}{\isacharparenright}{\kern0pt}{\isachardoublequoteclose}\isanewline
\isadelimproof
\ \ %
\endisadelimproof
\isatagproof
\isacommand{proof}\isamarkupfalse%
\ {\isacharparenleft}{\kern0pt}subst\ subset{\isacharunderscore}{\kern0pt}iff{\isacharparenright}{\kern0pt}\isanewline
\ \ \ \ \isacommand{show}\isamarkupfalse%
\ {\isachardoublequoteopen}{\isasymforall}c{\isachardot}{\kern0pt}\ c\ {\isasymin}\ {\isasymdelta}\isactrlsub s\ {\isacharparenleft}{\kern0pt}sh\ {\isasymleadsto}\ State{\isasymllangle}{\isasymsigma}{\isasymrrangle}{\isacharcomma}{\kern0pt}\ {\isasymlambda}{\isacharbrackleft}{\kern0pt}CO\ S\isactrlsub {\isadigit{1}}\ {\isasymparallel}\ S\isactrlsub {\isadigit{2}}\ OC{\isacharbrackright}{\kern0pt}{\isacharparenright}{\kern0pt}\ \isanewline
\ \ \ \ \ \ \ \ \ \ \ \ \ \ {\isasymlongrightarrow}\ c\ {\isasymin}\ {\isacharparenleft}{\kern0pt}{\isacharpercent}{\kern0pt}c{\isachardot}{\kern0pt}\ {\isacharparenleft}{\kern0pt}fst{\isacharparenleft}{\kern0pt}c{\isacharparenright}{\kern0pt}{\isacharcomma}{\kern0pt}\ parallel\ {\isacharparenleft}{\kern0pt}snd\ c{\isacharparenright}{\kern0pt}\ {\isasymlambda}{\isacharbrackleft}{\kern0pt}S\isactrlsub {\isadigit{2}}{\isacharbrackright}{\kern0pt}{\isacharparenright}{\kern0pt}{\isacharparenright}{\kern0pt}\ {\isacharbackquote}{\kern0pt}\ {\isasymdelta}\isactrlsub s\ {\isacharparenleft}{\kern0pt}sh\ {\isasymleadsto}\ State{\isasymllangle}{\isasymsigma}{\isasymrrangle}{\isacharcomma}{\kern0pt}\ {\isasymlambda}{\isacharbrackleft}{\kern0pt}S\isactrlsub {\isadigit{1}}{\isacharbrackright}{\kern0pt}{\isacharparenright}{\kern0pt}\ {\isasymunion}\ \isanewline
\ \ \ \ \ \ \ \ \ \ \ \ \ \ \ \ \ \ \ \ \ \ \ \ \ \ \ {\isacharparenleft}{\kern0pt}{\isacharpercent}{\kern0pt}c{\isachardot}{\kern0pt}\ {\isacharparenleft}{\kern0pt}fst{\isacharparenleft}{\kern0pt}c{\isacharparenright}{\kern0pt}{\isacharcomma}{\kern0pt}\ parallel\ {\isasymlambda}{\isacharbrackleft}{\kern0pt}S\isactrlsub {\isadigit{1}}{\isacharbrackright}{\kern0pt}\ {\isacharparenleft}{\kern0pt}snd\ c{\isacharparenright}{\kern0pt}{\isacharparenright}{\kern0pt}{\isacharparenright}{\kern0pt}\ {\isacharbackquote}{\kern0pt}\ {\isasymdelta}\isactrlsub s\ {\isacharparenleft}{\kern0pt}sh\ {\isasymleadsto}\ State{\isasymllangle}{\isasymsigma}{\isasymrrangle}{\isacharcomma}{\kern0pt}\ {\isasymlambda}{\isacharbrackleft}{\kern0pt}S\isactrlsub {\isadigit{2}}{\isacharbrackright}{\kern0pt}{\isacharparenright}{\kern0pt}{\isachardoublequoteclose}\isanewline
\ \ \ \ %
\isamarkupcmt{We first use the subset-iff rule in order to rewrite the subset relation into 
       a semantically equivalent~implication.%
}\ \ \isanewline
\ \ \ \ \isacommand{proof}\isamarkupfalse%
\ {\isacharparenleft}{\kern0pt}rule\ allI{\isacharcomma}{\kern0pt}\ rule\ impI{\isacharparenright}{\kern0pt}\isanewline
\ \ \ \ \ \ \isacommand{fix}\isamarkupfalse%
\ c\isanewline
\ \ \ \ \ \ %
\isamarkupcmt{We assume \isa{c} to be an arbitrary, but fixed configuration.%
}\isanewline
\ \ \ \ \ \ \isacommand{assume}\isamarkupfalse%
\ {\isachardoublequoteopen}c\ {\isasymin}\ {\isasymdelta}\isactrlsub s\ {\isacharparenleft}{\kern0pt}sh\ {\isasymleadsto}\ State{\isasymllangle}{\isasymsigma}{\isasymrrangle}{\isacharcomma}{\kern0pt}\ {\isasymlambda}{\isacharbrackleft}{\kern0pt}CO\ S\isactrlsub {\isadigit{1}}\ {\isasymparallel}\ S\isactrlsub {\isadigit{2}}\ OC{\isacharbrackright}{\kern0pt}{\isacharparenright}{\kern0pt}{\isachardoublequoteclose}\isanewline
\ \ \ \ \ \ %
\isamarkupcmt{We assume that the premise holds, implying that \isa{c} is a successor 
         configuration of the local parallelism command \isa{CO\ S\isactrlsub {\isadigit{1}}\ {\isasymparallel}\ S\isactrlsub {\isadigit{2}}\ OC}.%
}\isanewline
\ \ \ \ \ \ \isacommand{then}\isamarkupfalse%
\ \isacommand{obtain}\isamarkupfalse%
\ {\isasympi}\ \isakeyword{where}\ assm\isactrlsub {\isasympi}{\isacharcolon}{\kern0pt}\ \isanewline
\ \ \ \ \ \ \ \ {\isachardoublequoteopen}c\ {\isacharequal}{\kern0pt}\ {\isacharparenleft}{\kern0pt}trace{\isacharunderscore}{\kern0pt}conc\ {\isacharparenleft}{\kern0pt}min{\isacharunderscore}{\kern0pt}conc{\isacharunderscore}{\kern0pt}map\isactrlsub {\isasymT}\ {\isacharparenleft}{\kern0pt}sh\ {\isasymcdot}\ {\isacharparenleft}{\kern0pt}{\isasymdown}\isactrlsub {\isasymtau}\ {\isasympi}{\isacharparenright}{\kern0pt}{\isacharparenright}{\kern0pt}\ {\isadigit{0}}{\isacharparenright}{\kern0pt}\ {\isacharparenleft}{\kern0pt}sh\ {\isasymcdot}\ {\isacharparenleft}{\kern0pt}{\isasymdown}\isactrlsub {\isasymtau}\ {\isasympi}{\isacharparenright}{\kern0pt}{\isacharparenright}{\kern0pt}{\isacharcomma}{\kern0pt}\ {\isasymdown}\isactrlsub {\isasymlambda}\ {\isasympi}{\isacharparenright}{\kern0pt}\ \isanewline
\ \ \ \ \ \ \ \ \ {\isasymand}\ {\isasympi}\ {\isasymin}\ val\isactrlsub s\ {\isacharparenleft}{\kern0pt}CO\ S\isactrlsub {\isadigit{1}}\ {\isasymparallel}\ S\isactrlsub {\isadigit{2}}\ OC{\isacharparenright}{\kern0pt}\ {\isasymsigma}\ {\isasymand}\ consistent{\isacharparenleft}{\kern0pt}sval\isactrlsub B\ {\isacharparenleft}{\kern0pt}{\isasymdown}\isactrlsub p\ {\isasympi}{\isacharparenright}{\kern0pt}\ {\isacharparenleft}{\kern0pt}min{\isacharunderscore}{\kern0pt}conc{\isacharunderscore}{\kern0pt}map\isactrlsub {\isasymT}\ {\isacharparenleft}{\kern0pt}{\isasymdown}\isactrlsub {\isasymtau}\ {\isasympi}{\isacharparenright}{\kern0pt}\ {\isadigit{0}}{\isacharparenright}{\kern0pt}{\isacharparenright}{\kern0pt}{\isachardoublequoteclose}\ \isacommand{by}\isamarkupfalse%
\ auto\isanewline
\ \ \ \ \ \ %
\isamarkupcmt{Due to the definition of the $\isa{{\isasymdelta}}_s$-function, there must exist a continuation 
         trace \isa{{\isasympi}} with a consistent path condition generated from \isa{CO\ S\isactrlsub {\isadigit{1}}\ {\isasymparallel}\ S\isactrlsub {\isadigit{2}}\ OC} 
         that can be translated into \isa{c}. We obtain this continuation trace~\isa{{\isasympi}}.%
}\ \isanewline
\ \ \ \ \ \ \isacommand{moreover}\isamarkupfalse%
\ \isacommand{then}\isamarkupfalse%
\ \isacommand{have}\isamarkupfalse%
\ {\isachardoublequoteopen}{\isacharparenleft}{\kern0pt}{\isasymexists}{\isasympi}{\isacharprime}{\kern0pt}{\isachardot}{\kern0pt}\ {\isasympi}\ {\isacharequal}{\kern0pt}\ {\isacharparenleft}{\kern0pt}{\isasymdown}\isactrlsub p\ {\isasympi}{\isacharprime}{\kern0pt}{\isacharparenright}{\kern0pt}\ {\isasymtriangleright}\ {\isacharparenleft}{\kern0pt}{\isasymdown}\isactrlsub {\isasymtau}\ {\isasympi}{\isacharprime}{\kern0pt}{\isacharparenright}{\kern0pt}\ \isactrlitem \ parallel\ {\isacharparenleft}{\kern0pt}{\isasymdown}\isactrlsub {\isasymlambda}\ {\isasympi}{\isacharprime}{\kern0pt}{\isacharparenright}{\kern0pt}\ {\isasymlambda}{\isacharbrackleft}{\kern0pt}S\isactrlsub {\isadigit{2}}{\isacharbrackright}{\kern0pt}\ {\isasymand}\ {\isasympi}{\isacharprime}{\kern0pt}\ {\isasymin}\ {\isacharparenleft}{\kern0pt}val\isactrlsub s\ S\isactrlsub {\isadigit{1}}\ {\isasymsigma}{\isacharparenright}{\kern0pt}{\isacharparenright}{\kern0pt}\ \isanewline
\ \ \ \ \ \ \ \ \ {\isasymor}\ {\isacharparenleft}{\kern0pt}{\isasymexists}{\isasympi}{\isacharprime}{\kern0pt}{\isachardot}{\kern0pt}\ {\isasympi}\ {\isacharequal}{\kern0pt}\ {\isacharparenleft}{\kern0pt}{\isasymdown}\isactrlsub p\ {\isasympi}{\isacharprime}{\kern0pt}{\isacharparenright}{\kern0pt}\ {\isasymtriangleright}\ {\isacharparenleft}{\kern0pt}{\isasymdown}\isactrlsub {\isasymtau}\ {\isasympi}{\isacharprime}{\kern0pt}{\isacharparenright}{\kern0pt}\ \isactrlitem \ parallel\ {\isasymlambda}{\isacharbrackleft}{\kern0pt}S\isactrlsub {\isadigit{1}}{\isacharbrackright}{\kern0pt}\ {\isacharparenleft}{\kern0pt}{\isasymdown}\isactrlsub {\isasymlambda}\ {\isasympi}{\isacharprime}{\kern0pt}{\isacharparenright}{\kern0pt}\ {\isasymand}\ {\isasympi}{\isacharprime}{\kern0pt}\ {\isasymin}\ {\isacharparenleft}{\kern0pt}val\isactrlsub s\ S\isactrlsub {\isadigit{2}}\ {\isasymsigma}{\isacharparenright}{\kern0pt}{\isacharparenright}{\kern0pt}{\isachardoublequoteclose}\ \isacommand{by}\isamarkupfalse%
\ auto\isanewline
\ \ \ \ \ \ %
\isamarkupcmt{Aligning with the definition of the valuation function, there must exist
         a continuation trace \isa{{\isasympi}{\isacharprime}{\kern0pt}}, which satisifies one of the following cases: In 
         the first case, \isa{{\isasympi}{\isacharprime}{\kern0pt}} is generated from \isa{S\isactrlsub {\isadigit{1}}}, and matches \isa{{\isasympi}} when
         reconstructing the local parallelism construct with \isa{S\isactrlsub {\isadigit{2}}} in its 
         continuation marker. In the second case, \isa{{\isasympi}{\isacharprime}{\kern0pt}} is generated from \isa{S\isactrlsub {\isadigit{2}}} and 
         matches \isa{{\isasympi}} when reconstructing the local parallelism construct with \isa{S\isactrlsub {\isadigit{1}}} 
         in its continuation~marker.%
}\isanewline
\ \ \ \ \ \ \isacommand{thus}\isamarkupfalse%
\ {\isachardoublequoteopen}c\ {\isasymin}\ {\isacharparenleft}{\kern0pt}{\isacharparenleft}{\kern0pt}{\isacharpercent}{\kern0pt}c{\isachardot}{\kern0pt}\ {\isacharparenleft}{\kern0pt}fst{\isacharparenleft}{\kern0pt}c{\isacharparenright}{\kern0pt}{\isacharcomma}{\kern0pt}\ parallel\ {\isacharparenleft}{\kern0pt}snd\ c{\isacharparenright}{\kern0pt}\ {\isasymlambda}{\isacharbrackleft}{\kern0pt}S\isactrlsub {\isadigit{2}}{\isacharbrackright}{\kern0pt}{\isacharparenright}{\kern0pt}{\isacharparenright}{\kern0pt}\ {\isacharbackquote}{\kern0pt}\ {\isasymdelta}\isactrlsub s\ {\isacharparenleft}{\kern0pt}sh\ {\isasymleadsto}\ State{\isasymllangle}{\isasymsigma}{\isasymrrangle}{\isacharcomma}{\kern0pt}\ {\isasymlambda}{\isacharbrackleft}{\kern0pt}S\isactrlsub {\isadigit{1}}{\isacharbrackright}{\kern0pt}{\isacharparenright}{\kern0pt}{\isacharparenright}{\kern0pt}\ {\isasymunion}\ {\isacharparenleft}{\kern0pt}{\isacharparenleft}{\kern0pt}{\isacharpercent}{\kern0pt}c{\isachardot}{\kern0pt}\ {\isacharparenleft}{\kern0pt}fst{\isacharparenleft}{\kern0pt}c{\isacharparenright}{\kern0pt}{\isacharcomma}{\kern0pt}\ parallel\ {\isasymlambda}{\isacharbrackleft}{\kern0pt}S\isactrlsub {\isadigit{1}}{\isacharbrackright}{\kern0pt}\ {\isacharparenleft}{\kern0pt}snd\ c{\isacharparenright}{\kern0pt}{\isacharparenright}{\kern0pt}{\isacharparenright}{\kern0pt}\ {\isacharbackquote}{\kern0pt}\ {\isasymdelta}\isactrlsub s\ {\isacharparenleft}{\kern0pt}sh\ {\isasymleadsto}\ State{\isasymllangle}{\isasymsigma}{\isasymrrangle}{\isacharcomma}{\kern0pt}\ {\isasymlambda}{\isacharbrackleft}{\kern0pt}S\isactrlsub {\isadigit{2}}{\isacharbrackright}{\kern0pt}{\isacharparenright}{\kern0pt}{\isacharparenright}{\kern0pt}{\isachardoublequoteclose}\isanewline
\ \ \ \ \ \ \isacommand{proof}\isamarkupfalse%
\ {\isacharparenleft}{\kern0pt}rule\ disjE{\isacharparenright}{\kern0pt}\isanewline
\ \ \ \ \ \ %
\isamarkupcmt{We perform a case distinction over those two cases.%
}\isanewline
\ \ \ \ \ \ \ \ \isacommand{assume}\isamarkupfalse%
\ {\isachardoublequoteopen}{\isasymexists}{\isasympi}{\isacharprime}{\kern0pt}{\isachardot}{\kern0pt}\ {\isasympi}\ {\isacharequal}{\kern0pt}\ {\isacharparenleft}{\kern0pt}{\isasymdown}\isactrlsub p\ {\isasympi}{\isacharprime}{\kern0pt}{\isacharparenright}{\kern0pt}\ {\isasymtriangleright}\ {\isacharparenleft}{\kern0pt}{\isasymdown}\isactrlsub {\isasymtau}\ {\isasympi}{\isacharprime}{\kern0pt}{\isacharparenright}{\kern0pt}\ \isactrlitem \ parallel\ {\isacharparenleft}{\kern0pt}{\isasymdown}\isactrlsub {\isasymlambda}\ {\isasympi}{\isacharprime}{\kern0pt}{\isacharparenright}{\kern0pt}\ {\isasymlambda}{\isacharbrackleft}{\kern0pt}S\isactrlsub {\isadigit{2}}{\isacharbrackright}{\kern0pt}\ {\isasymand}\ {\isasympi}{\isacharprime}{\kern0pt}\ {\isasymin}\ {\isacharparenleft}{\kern0pt}val\isactrlsub s\ S\isactrlsub {\isadigit{1}}\ {\isasymsigma}{\isacharparenright}{\kern0pt}{\isachardoublequoteclose}\isanewline
\ \ \ \ \ \ \ \ %
\isamarkupcmt{Let us assume that the first case holds.%
}\isanewline
\ \ \ \ \ \ \ \ \isacommand{moreover}\isamarkupfalse%
\ \isacommand{then}\isamarkupfalse%
\ \isacommand{obtain}\isamarkupfalse%
\ {\isasympi}{\isacharprime}{\kern0pt}\ \isakeyword{where}\ assm\isactrlsub {\isasympi}{\isacharprime}{\kern0pt}{\isacharcolon}{\kern0pt}\isanewline
\ \ \ \ \ \ \ \ \ \ {\isachardoublequoteopen}{\isasympi}\ {\isacharequal}{\kern0pt}\ {\isacharparenleft}{\kern0pt}{\isasymdown}\isactrlsub p\ {\isasympi}{\isacharprime}{\kern0pt}{\isacharparenright}{\kern0pt}\ {\isasymtriangleright}\ {\isacharparenleft}{\kern0pt}{\isasymdown}\isactrlsub {\isasymtau}\ {\isasympi}{\isacharprime}{\kern0pt}{\isacharparenright}{\kern0pt}\ \isactrlitem \ parallel\ {\isacharparenleft}{\kern0pt}{\isasymdown}\isactrlsub {\isasymlambda}\ {\isasympi}{\isacharprime}{\kern0pt}{\isacharparenright}{\kern0pt}\ {\isasymlambda}{\isacharbrackleft}{\kern0pt}S\isactrlsub {\isadigit{2}}{\isacharbrackright}{\kern0pt}\ {\isasymand}\ {\isasympi}{\isacharprime}{\kern0pt}\ {\isasymin}\ {\isacharparenleft}{\kern0pt}val\isactrlsub s\ S\isactrlsub {\isadigit{1}}\ {\isasymsigma}{\isacharparenright}{\kern0pt}{\isachardoublequoteclose}\ \isacommand{by}\isamarkupfalse%
\ auto\isanewline
\ \ \ \ \ \ \ \ %
\isamarkupcmt{We then obtain this continuation trace \isa{{\isasympi}{\isacharprime}{\kern0pt}}.%
}\isanewline
\ \ \ \ \ \ \ \ \isacommand{ultimately}\isamarkupfalse%
\ \isacommand{have}\isamarkupfalse%
\ connect{\isacharcolon}{\kern0pt}\ \isanewline
\ \ \ \ \ \ \ \ \ \ {\isachardoublequoteopen}consistent\ {\isacharparenleft}{\kern0pt}sval\isactrlsub B\ {\isacharparenleft}{\kern0pt}{\isasymdown}\isactrlsub p\ {\isasympi}{\isacharprime}{\kern0pt}{\isacharparenright}{\kern0pt}\ {\isacharparenleft}{\kern0pt}min{\isacharunderscore}{\kern0pt}conc{\isacharunderscore}{\kern0pt}map\isactrlsub {\isasymT}\ {\isacharparenleft}{\kern0pt}{\isasymdown}\isactrlsub {\isasymtau}\ {\isasympi}{\isacharprime}{\kern0pt}{\isacharparenright}{\kern0pt}\ {\isadigit{0}}{\isacharparenright}{\kern0pt}{\isacharparenright}{\kern0pt}\ \isanewline
\ \ \ \ \ \ \ \ \ \ {\isasymand}\ {\isacharparenleft}{\kern0pt}{\isasymdown}\isactrlsub {\isasymtau}\ {\isasympi}{\isacharprime}{\kern0pt}{\isacharparenright}{\kern0pt}\ {\isacharequal}{\kern0pt}\ {\isacharparenleft}{\kern0pt}{\isasymdown}\isactrlsub {\isasymtau}\ {\isasympi}{\isacharparenright}{\kern0pt}\ {\isasymand}\ {\isacharparenleft}{\kern0pt}{\isasymdown}\isactrlsub {\isasymlambda}\ {\isasympi}{\isacharparenright}{\kern0pt}\ {\isacharequal}{\kern0pt}\ parallel\ {\isacharparenleft}{\kern0pt}{\isasymdown}\isactrlsub {\isasymlambda}\ {\isasympi}{\isacharprime}{\kern0pt}{\isacharparenright}{\kern0pt}\ {\isasymlambda}{\isacharbrackleft}{\kern0pt}S\isactrlsub {\isadigit{2}}{\isacharbrackright}{\kern0pt}{\isachardoublequoteclose}\ \isanewline
\ \ \ \ \ \ \ \ \ \ \isacommand{using}\isamarkupfalse%
\ assm\isactrlsub {\isasympi}\ \isacommand{by}\isamarkupfalse%
\ auto\isanewline
\ \ \ \ \ \ \ \ %
\isamarkupcmt{Considering that the path conditions of \isa{{\isasympi}} and \isa{{\isasympi}{\isacharprime}{\kern0pt}} match, both must be 
           consistent. Their symbolic traces also match. The only difference lies
           in the modified continuation~marker.%
}\isanewline
\ \ \ \ \ \ \ \ \isacommand{moreover}\isamarkupfalse%
\ \isacommand{then}\isamarkupfalse%
\ \isacommand{obtain}\isamarkupfalse%
\ c{\isacharprime}{\kern0pt}\ \isakeyword{where}\isanewline
\ \ \ \ \ \ \ \ \ \ {\isachardoublequoteopen}c{\isacharprime}{\kern0pt}\ {\isacharequal}{\kern0pt}\ {\isacharparenleft}{\kern0pt}trace{\isacharunderscore}{\kern0pt}conc\ {\isacharparenleft}{\kern0pt}min{\isacharunderscore}{\kern0pt}conc{\isacharunderscore}{\kern0pt}map\isactrlsub {\isasymT}\ {\isacharparenleft}{\kern0pt}sh\ {\isasymcdot}\ {\isacharparenleft}{\kern0pt}{\isasymdown}\isactrlsub {\isasymtau}\ {\isasympi}{\isacharprime}{\kern0pt}{\isacharparenright}{\kern0pt}{\isacharparenright}{\kern0pt}\ {\isadigit{0}}{\isacharparenright}{\kern0pt}\ {\isacharparenleft}{\kern0pt}sh\ {\isasymcdot}\ {\isacharparenleft}{\kern0pt}{\isasymdown}\isactrlsub {\isasymtau}\ {\isasympi}{\isacharprime}{\kern0pt}{\isacharparenright}{\kern0pt}{\isacharparenright}{\kern0pt}{\isacharcomma}{\kern0pt}\ {\isasymdown}\isactrlsub {\isasymlambda}\ {\isasympi}{\isacharprime}{\kern0pt}{\isacharparenright}{\kern0pt}{\isachardoublequoteclose}\ \isacommand{by}\isamarkupfalse%
\ auto\isanewline
\ \ \ \ \ \ \ \ %
\isamarkupcmt{We can then obtain the configuration \isa{c{\isacharprime}{\kern0pt}} translated from the 
           continuation trace~\isa{{\isasympi}{\isacharprime}{\kern0pt}}.%
}\isanewline
\ \ \ \ \ \ \ \ \isacommand{moreover}\isamarkupfalse%
\ \isacommand{then}\isamarkupfalse%
\ \isacommand{have}\isamarkupfalse%
\ {\isachardoublequoteopen}fst{\isacharparenleft}{\kern0pt}c{\isacharparenright}{\kern0pt}\ {\isacharequal}{\kern0pt}\ fst{\isacharparenleft}{\kern0pt}c{\isacharprime}{\kern0pt}{\isacharparenright}{\kern0pt}\ {\isasymand}\ snd{\isacharparenleft}{\kern0pt}c{\isacharparenright}{\kern0pt}\ {\isacharequal}{\kern0pt}\ parallel\ {\isacharparenleft}{\kern0pt}snd\ c{\isacharprime}{\kern0pt}{\isacharparenright}{\kern0pt}\ {\isasymlambda}{\isacharbrackleft}{\kern0pt}S\isactrlsub {\isadigit{2}}{\isacharbrackright}{\kern0pt}{\isachardoublequoteclose}\isanewline
\ \ \ \ \ \ \ \ \ \ \isacommand{by}\isamarkupfalse%
\ {\isacharparenleft}{\kern0pt}simp\ add{\isacharcolon}{\kern0pt}\ assm\isactrlsub {\isasympi}\ connect{\isacharparenright}{\kern0pt}\isanewline
\ \ \ \ \ \ \ \ %
\isamarkupcmt{Given this information, both \isa{c} and \isa{c{\isacharprime}{\kern0pt}} must match in their symbolic 
           trace. However, \isa{c} additionally reconstructs the local parallelism
           with \isa{S\isactrlsub {\isadigit{2}}} in its continuation~marker.%
}\isanewline
\ \ \ \ \ \ \ \ \isacommand{ultimately}\isamarkupfalse%
\ \isacommand{have}\isamarkupfalse%
\ {\isachardoublequoteopen}c\ {\isasymin}\ {\isacharparenleft}{\kern0pt}{\isacharpercent}{\kern0pt}c{\isachardot}{\kern0pt}\ {\isacharparenleft}{\kern0pt}fst{\isacharparenleft}{\kern0pt}c{\isacharparenright}{\kern0pt}{\isacharcomma}{\kern0pt}\ parallel\ {\isacharparenleft}{\kern0pt}snd\ c{\isacharparenright}{\kern0pt}\ {\isasymlambda}{\isacharbrackleft}{\kern0pt}S\isactrlsub {\isadigit{2}}{\isacharbrackright}{\kern0pt}{\isacharparenright}{\kern0pt}{\isacharparenright}{\kern0pt}\ {\isacharbackquote}{\kern0pt}\ {\isasymdelta}\isactrlsub s\ {\isacharparenleft}{\kern0pt}sh\ {\isasymleadsto}\ State{\isasymllangle}{\isasymsigma}{\isasymrrangle}{\isacharcomma}{\kern0pt}\ {\isasymlambda}{\isacharbrackleft}{\kern0pt}S\isactrlsub {\isadigit{1}}{\isacharbrackright}{\kern0pt}{\isacharparenright}{\kern0pt}{\isachardoublequoteclose}\isanewline
\ \ \ \ \ \ \ \ \ \ \isacommand{using}\isamarkupfalse%
\ assm\isactrlsub {\isasympi}{\isacharprime}{\kern0pt}\ assm\isactrlsub {\isasympi}\ image{\isacharunderscore}{\kern0pt}iff\ \isacommand{by}\isamarkupfalse%
\ fastforce\isanewline
\ \ \ \ \ \ \ \ %
\isamarkupcmt{Thus, configuration \isa{c{\isacharprime}{\kern0pt}} with the local parallelism under \isa{S\isactrlsub {\isadigit{2}}} reconstructed
           in its continuation marker must match with \isa{c}.%
}\isanewline
\ \ \ \ \ \ \ \ \isacommand{thus}\isamarkupfalse%
\ {\isachardoublequoteopen}c\ {\isasymin}\ {\isacharparenleft}{\kern0pt}{\isacharpercent}{\kern0pt}c{\isachardot}{\kern0pt}\ {\isacharparenleft}{\kern0pt}fst{\isacharparenleft}{\kern0pt}c{\isacharparenright}{\kern0pt}{\isacharcomma}{\kern0pt}\ parallel\ {\isacharparenleft}{\kern0pt}snd\ c{\isacharparenright}{\kern0pt}\ {\isasymlambda}{\isacharbrackleft}{\kern0pt}S\isactrlsub {\isadigit{2}}{\isacharbrackright}{\kern0pt}{\isacharparenright}{\kern0pt}{\isacharparenright}{\kern0pt}\ {\isacharbackquote}{\kern0pt}\ {\isasymdelta}\isactrlsub s\ {\isacharparenleft}{\kern0pt}sh\ {\isasymleadsto}\ State{\isasymllangle}{\isasymsigma}{\isasymrrangle}{\isacharcomma}{\kern0pt}\ {\isasymlambda}{\isacharbrackleft}{\kern0pt}S\isactrlsub {\isadigit{1}}{\isacharbrackright}{\kern0pt}{\isacharparenright}{\kern0pt}\ {\isasymunion}\ \isanewline
\ \ \ \ \ \ \ \ \ \ \ \ \ \ \ \ \ \ \ \ \ {\isacharparenleft}{\kern0pt}{\isacharpercent}{\kern0pt}c{\isachardot}{\kern0pt}\ {\isacharparenleft}{\kern0pt}fst{\isacharparenleft}{\kern0pt}c{\isacharparenright}{\kern0pt}{\isacharcomma}{\kern0pt}\ parallel\ {\isasymlambda}{\isacharbrackleft}{\kern0pt}S\isactrlsub {\isadigit{1}}{\isacharbrackright}{\kern0pt}\ {\isacharparenleft}{\kern0pt}snd\ c{\isacharparenright}{\kern0pt}{\isacharparenright}{\kern0pt}{\isacharparenright}{\kern0pt}\ {\isacharbackquote}{\kern0pt}\ {\isasymdelta}\isactrlsub s\ {\isacharparenleft}{\kern0pt}sh\ {\isasymleadsto}\ State{\isasymllangle}{\isasymsigma}{\isasymrrangle}{\isacharcomma}{\kern0pt}\ {\isasymlambda}{\isacharbrackleft}{\kern0pt}S\isactrlsub {\isadigit{2}}{\isacharbrackright}{\kern0pt}{\isacharparenright}{\kern0pt}{\isachardoublequoteclose}\ \isacommand{by}\isamarkupfalse%
\ simp\isanewline
\ \ \ \ \ \ \ \ %
\isamarkupcmt{Hence it must also be in the merged set, thus closing this case.%
}\isanewline
\ \ \ \ \ \ \isacommand{next}\isamarkupfalse%
\isanewline
\ \ \ \ \ \ \ \ \isacommand{assume}\isamarkupfalse%
\ {\isachardoublequoteopen}{\isasymexists}{\isasympi}{\isacharprime}{\kern0pt}{\isachardot}{\kern0pt}\ {\isasympi}\ {\isacharequal}{\kern0pt}\ {\isacharparenleft}{\kern0pt}{\isasymdown}\isactrlsub p\ {\isasympi}{\isacharprime}{\kern0pt}{\isacharparenright}{\kern0pt}\ {\isasymtriangleright}\ {\isacharparenleft}{\kern0pt}{\isasymdown}\isactrlsub {\isasymtau}\ {\isasympi}{\isacharprime}{\kern0pt}{\isacharparenright}{\kern0pt}\ \isactrlitem \ parallel\ {\isasymlambda}{\isacharbrackleft}{\kern0pt}S\isactrlsub {\isadigit{1}}{\isacharbrackright}{\kern0pt}\ {\isacharparenleft}{\kern0pt}{\isasymdown}\isactrlsub {\isasymlambda}\ {\isasympi}{\isacharprime}{\kern0pt}{\isacharparenright}{\kern0pt}\ {\isasymand}\ {\isasympi}{\isacharprime}{\kern0pt}\ {\isasymin}\ {\isacharparenleft}{\kern0pt}val\isactrlsub s\ S\isactrlsub {\isadigit{2}}\ {\isasymsigma}{\isacharparenright}{\kern0pt}{\isachardoublequoteclose}\isanewline
\ \ \ \ \ \ \ \ %
\isamarkupcmt{Let us assume that the second case holds.%
}\isanewline
\ \ \ \ \ \ \ \ \isacommand{moreover}\isamarkupfalse%
\ \isacommand{then}\isamarkupfalse%
\ \isacommand{obtain}\isamarkupfalse%
\ {\isasympi}{\isacharprime}{\kern0pt}\ \isakeyword{where}\ assm\isactrlsub {\isasympi}{\isacharprime}{\kern0pt}{\isacharcolon}{\kern0pt}\isanewline
\ \ \ \ \ \ \ \ \ \ {\isachardoublequoteopen}{\isasympi}\ {\isacharequal}{\kern0pt}\ {\isacharparenleft}{\kern0pt}{\isasymdown}\isactrlsub p\ {\isasympi}{\isacharprime}{\kern0pt}{\isacharparenright}{\kern0pt}\ {\isasymtriangleright}\ {\isacharparenleft}{\kern0pt}{\isasymdown}\isactrlsub {\isasymtau}\ {\isasympi}{\isacharprime}{\kern0pt}{\isacharparenright}{\kern0pt}\ \isactrlitem \ parallel\ {\isasymlambda}{\isacharbrackleft}{\kern0pt}S\isactrlsub {\isadigit{1}}{\isacharbrackright}{\kern0pt}\ {\isacharparenleft}{\kern0pt}{\isasymdown}\isactrlsub {\isasymlambda}\ {\isasympi}{\isacharprime}{\kern0pt}{\isacharparenright}{\kern0pt}\ {\isasymand}\ {\isasympi}{\isacharprime}{\kern0pt}\ {\isasymin}\ {\isacharparenleft}{\kern0pt}val\isactrlsub s\ S\isactrlsub {\isadigit{2}}\ {\isasymsigma}{\isacharparenright}{\kern0pt}{\isachardoublequoteclose}\ \isacommand{by}\isamarkupfalse%
\ auto\isanewline
\ \ \ \ \ \ \ \ %
\isamarkupcmt{We then obtain this continuation trace \isa{{\isasympi}{\isacharprime}{\kern0pt}}.%
}\isanewline
\ \ \ \ \ \ \ \ \isacommand{ultimately}\isamarkupfalse%
\ \isacommand{have}\isamarkupfalse%
\ connect{\isacharcolon}{\kern0pt}\isanewline
\ \ \ \ \ \ \ \ \ \ {\isachardoublequoteopen}consistent\ {\isacharparenleft}{\kern0pt}sval\isactrlsub B\ {\isacharparenleft}{\kern0pt}{\isasymdown}\isactrlsub p\ {\isasympi}{\isacharprime}{\kern0pt}{\isacharparenright}{\kern0pt}\ {\isacharparenleft}{\kern0pt}min{\isacharunderscore}{\kern0pt}conc{\isacharunderscore}{\kern0pt}map\isactrlsub {\isasymT}\ {\isacharparenleft}{\kern0pt}{\isasymdown}\isactrlsub {\isasymtau}\ {\isasympi}{\isacharprime}{\kern0pt}{\isacharparenright}{\kern0pt}\ {\isadigit{0}}{\isacharparenright}{\kern0pt}{\isacharparenright}{\kern0pt}\ \isanewline
\ \ \ \ \ \ \ \ \ \ {\isasymand}\ {\isacharparenleft}{\kern0pt}{\isasymdown}\isactrlsub {\isasymtau}\ {\isasympi}{\isacharprime}{\kern0pt}{\isacharparenright}{\kern0pt}\ {\isacharequal}{\kern0pt}\ {\isacharparenleft}{\kern0pt}{\isasymdown}\isactrlsub {\isasymtau}\ {\isasympi}{\isacharparenright}{\kern0pt}\ {\isasymand}\ {\isacharparenleft}{\kern0pt}{\isasymdown}\isactrlsub {\isasymlambda}\ {\isasympi}{\isacharparenright}{\kern0pt}\ {\isacharequal}{\kern0pt}\ parallel\ {\isasymlambda}{\isacharbrackleft}{\kern0pt}S\isactrlsub {\isadigit{1}}{\isacharbrackright}{\kern0pt}\ {\isacharparenleft}{\kern0pt}{\isasymdown}\isactrlsub {\isasymlambda}\ {\isasympi}{\isacharprime}{\kern0pt}{\isacharparenright}{\kern0pt}{\isachardoublequoteclose}\ \isanewline
\ \ \ \ \ \ \ \ \ \ \isacommand{using}\isamarkupfalse%
\ assm\isactrlsub {\isasympi}\ \isacommand{by}\isamarkupfalse%
\ auto\isanewline
\ \ \ \ \ \ \ \ %
\isamarkupcmt{Considering that the path conditions of \isa{{\isasympi}} and \isa{{\isasympi}{\isacharprime}{\kern0pt}} match, both must be 
           consistent. Their symbolic traces also match. The only difference lies
           in the modified continuation~marker.%
}\isanewline
\ \ \ \ \ \ \ \ \isacommand{moreover}\isamarkupfalse%
\ \isacommand{then}\isamarkupfalse%
\ \isacommand{obtain}\isamarkupfalse%
\ c{\isacharprime}{\kern0pt}\ \isakeyword{where}\isanewline
\ \ \ \ \ \ \ \ \ \ {\isachardoublequoteopen}c{\isacharprime}{\kern0pt}\ {\isacharequal}{\kern0pt}\ {\isacharparenleft}{\kern0pt}trace{\isacharunderscore}{\kern0pt}conc\ {\isacharparenleft}{\kern0pt}min{\isacharunderscore}{\kern0pt}conc{\isacharunderscore}{\kern0pt}map\isactrlsub {\isasymT}\ {\isacharparenleft}{\kern0pt}sh\ {\isasymcdot}\ {\isacharparenleft}{\kern0pt}{\isasymdown}\isactrlsub {\isasymtau}\ {\isasympi}{\isacharprime}{\kern0pt}{\isacharparenright}{\kern0pt}{\isacharparenright}{\kern0pt}\ {\isadigit{0}}{\isacharparenright}{\kern0pt}\ {\isacharparenleft}{\kern0pt}sh\ {\isasymcdot}\ {\isacharparenleft}{\kern0pt}{\isasymdown}\isactrlsub {\isasymtau}\ {\isasympi}{\isacharprime}{\kern0pt}{\isacharparenright}{\kern0pt}{\isacharparenright}{\kern0pt}{\isacharcomma}{\kern0pt}\ {\isasymdown}\isactrlsub {\isasymlambda}\ {\isasympi}{\isacharprime}{\kern0pt}{\isacharparenright}{\kern0pt}{\isachardoublequoteclose}\ \isacommand{by}\isamarkupfalse%
\ auto\isanewline
\ \ \ \ \ \ \ \ %
\isamarkupcmt{We can then obtain the configuration \isa{c{\isacharprime}{\kern0pt}} translated from the 
           continuation trace~\isa{{\isasympi}{\isacharprime}{\kern0pt}}.%
}\isanewline
\ \ \ \ \ \ \ \ \isacommand{moreover}\isamarkupfalse%
\ \isacommand{then}\isamarkupfalse%
\ \isacommand{have}\isamarkupfalse%
\ {\isachardoublequoteopen}fst{\isacharparenleft}{\kern0pt}c{\isacharparenright}{\kern0pt}\ {\isacharequal}{\kern0pt}\ fst{\isacharparenleft}{\kern0pt}c{\isacharprime}{\kern0pt}{\isacharparenright}{\kern0pt}\ {\isasymand}\ snd{\isacharparenleft}{\kern0pt}c{\isacharparenright}{\kern0pt}\ {\isacharequal}{\kern0pt}\ parallel\ {\isasymlambda}{\isacharbrackleft}{\kern0pt}S\isactrlsub {\isadigit{1}}{\isacharbrackright}{\kern0pt}\ {\isacharparenleft}{\kern0pt}snd\ c{\isacharprime}{\kern0pt}{\isacharparenright}{\kern0pt}{\isachardoublequoteclose}\isanewline
\ \ \ \ \ \ \ \ \ \ \isacommand{by}\isamarkupfalse%
\ {\isacharparenleft}{\kern0pt}simp\ add{\isacharcolon}{\kern0pt}\ assm\isactrlsub {\isasympi}\ connect{\isacharparenright}{\kern0pt}\isanewline
\ \ \ \ \ \ \ \ %
\isamarkupcmt{Given this information, both \isa{c} and \isa{c{\isacharprime}{\kern0pt}} must match in their symbolic 
           trace. However, \isa{c} additionally reconstructs the local parallelism
           with \isa{S\isactrlsub {\isadigit{1}}} in its continuation~marker.%
}\isanewline
\ \ \ \ \ \ \ \ \isacommand{ultimately}\isamarkupfalse%
\ \isacommand{have}\isamarkupfalse%
\ {\isachardoublequoteopen}c\ {\isasymin}\ {\isacharparenleft}{\kern0pt}{\isacharpercent}{\kern0pt}c{\isachardot}{\kern0pt}\ {\isacharparenleft}{\kern0pt}fst{\isacharparenleft}{\kern0pt}c{\isacharparenright}{\kern0pt}{\isacharcomma}{\kern0pt}\ parallel\ {\isasymlambda}{\isacharbrackleft}{\kern0pt}S\isactrlsub {\isadigit{1}}{\isacharbrackright}{\kern0pt}\ {\isacharparenleft}{\kern0pt}snd\ c{\isacharparenright}{\kern0pt}{\isacharparenright}{\kern0pt}{\isacharparenright}{\kern0pt}\ {\isacharbackquote}{\kern0pt}\ {\isasymdelta}\isactrlsub s\ {\isacharparenleft}{\kern0pt}sh\ {\isasymleadsto}\ State{\isasymllangle}{\isasymsigma}{\isasymrrangle}{\isacharcomma}{\kern0pt}\ {\isasymlambda}{\isacharbrackleft}{\kern0pt}S\isactrlsub {\isadigit{2}}{\isacharbrackright}{\kern0pt}{\isacharparenright}{\kern0pt}{\isachardoublequoteclose}\isanewline
\ \ \ \ \ \ \ \ \ \ \isacommand{using}\isamarkupfalse%
\ assm\isactrlsub {\isasympi}{\isacharprime}{\kern0pt}\ assm\isactrlsub {\isasympi}\ image{\isacharunderscore}{\kern0pt}iff\ \isacommand{by}\isamarkupfalse%
\ fastforce\isanewline
\ \ \ \ \ \ \ \ %
\isamarkupcmt{Thus, configuration \isa{c{\isacharprime}{\kern0pt}} with the local parallelism under \isa{S\isactrlsub {\isadigit{1}}} reconstructed
           in its continuation marker must match with \isa{c}.%
}\isanewline
\ \ \ \ \ \ \ \ \isacommand{thus}\isamarkupfalse%
\ {\isachardoublequoteopen}c\ {\isasymin}\ {\isacharparenleft}{\kern0pt}{\isacharpercent}{\kern0pt}c{\isachardot}{\kern0pt}\ {\isacharparenleft}{\kern0pt}fst{\isacharparenleft}{\kern0pt}c{\isacharparenright}{\kern0pt}{\isacharcomma}{\kern0pt}\ parallel\ {\isacharparenleft}{\kern0pt}snd\ c{\isacharparenright}{\kern0pt}\ {\isasymlambda}{\isacharbrackleft}{\kern0pt}S\isactrlsub {\isadigit{2}}{\isacharbrackright}{\kern0pt}{\isacharparenright}{\kern0pt}{\isacharparenright}{\kern0pt}\ {\isacharbackquote}{\kern0pt}\ {\isasymdelta}\isactrlsub s\ {\isacharparenleft}{\kern0pt}sh\ {\isasymleadsto}\ State{\isasymllangle}{\isasymsigma}{\isasymrrangle}{\isacharcomma}{\kern0pt}\ {\isasymlambda}{\isacharbrackleft}{\kern0pt}S\isactrlsub {\isadigit{1}}{\isacharbrackright}{\kern0pt}{\isacharparenright}{\kern0pt}\ {\isasymunion}\ \isanewline
\ \ \ \ \ \ \ \ \ \ \ \ \ \ \ \ \ \ \ \ \ \ {\isacharparenleft}{\kern0pt}{\isacharpercent}{\kern0pt}c{\isachardot}{\kern0pt}\ {\isacharparenleft}{\kern0pt}fst{\isacharparenleft}{\kern0pt}c{\isacharparenright}{\kern0pt}{\isacharcomma}{\kern0pt}\ parallel\ {\isasymlambda}{\isacharbrackleft}{\kern0pt}S\isactrlsub {\isadigit{1}}{\isacharbrackright}{\kern0pt}\ {\isacharparenleft}{\kern0pt}snd\ c{\isacharparenright}{\kern0pt}{\isacharparenright}{\kern0pt}{\isacharparenright}{\kern0pt}\ {\isacharbackquote}{\kern0pt}\ {\isasymdelta}\isactrlsub s\ {\isacharparenleft}{\kern0pt}sh\ {\isasymleadsto}\ State{\isasymllangle}{\isasymsigma}{\isasymrrangle}{\isacharcomma}{\kern0pt}\ {\isasymlambda}{\isacharbrackleft}{\kern0pt}S\isactrlsub {\isadigit{2}}{\isacharbrackright}{\kern0pt}{\isacharparenright}{\kern0pt}{\isachardoublequoteclose}\ \isacommand{by}\isamarkupfalse%
\ simp\isanewline
\ \ \ \ \ \ \ \ %
\isamarkupcmt{Hence it must also be in the merged set, thus closing the second case.%
}\isanewline
\ \ \ \ \ \ \isacommand{qed}\isamarkupfalse%
\isanewline
\ \ \ \ \isacommand{qed}\isamarkupfalse%
\isanewline
\ \ \isacommand{qed}\isamarkupfalse%
\endisatagproof
{\isafoldproof}%
\isadelimproof
\isanewline
\endisadelimproof
\isanewline
\ \ \isacommand{lemma}\isamarkupfalse%
\ {\isasymdelta}\isactrlsub s{\isacharunderscore}{\kern0pt}LocPar\isactrlsub {\isadigit{2}}{\isacharcolon}{\kern0pt}\isanewline
\ \ \ \ \isakeyword{assumes}\ {\isachardoublequoteopen}concrete\isactrlsub {\isasymT}{\isacharparenleft}{\kern0pt}sh\ {\isasymleadsto}\ State{\isasymllangle}{\isasymsigma}{\isasymrrangle}{\isacharparenright}{\kern0pt}{\isachardoublequoteclose}\isanewline
\ \ \ \ \isakeyword{shows}\ {\isachardoublequoteopen}{\isacharparenleft}{\kern0pt}{\isacharpercent}{\kern0pt}c{\isachardot}{\kern0pt}\ {\isacharparenleft}{\kern0pt}fst{\isacharparenleft}{\kern0pt}c{\isacharparenright}{\kern0pt}{\isacharcomma}{\kern0pt}\ parallel\ {\isacharparenleft}{\kern0pt}snd\ c{\isacharparenright}{\kern0pt}\ {\isasymlambda}{\isacharbrackleft}{\kern0pt}S\isactrlsub {\isadigit{2}}{\isacharbrackright}{\kern0pt}{\isacharparenright}{\kern0pt}{\isacharparenright}{\kern0pt}\ {\isacharbackquote}{\kern0pt}\ {\isasymdelta}\isactrlsub s\ {\isacharparenleft}{\kern0pt}sh\ {\isasymleadsto}\ State{\isasymllangle}{\isasymsigma}{\isasymrrangle}{\isacharcomma}{\kern0pt}\ {\isasymlambda}{\isacharbrackleft}{\kern0pt}S\isactrlsub {\isadigit{1}}{\isacharbrackright}{\kern0pt}{\isacharparenright}{\kern0pt}\ {\isasymunion}\ \isanewline
\ \ \ \ \ \ \ \ \ \ \ \ \ \ \ {\isacharparenleft}{\kern0pt}{\isacharpercent}{\kern0pt}c{\isachardot}{\kern0pt}\ {\isacharparenleft}{\kern0pt}fst{\isacharparenleft}{\kern0pt}c{\isacharparenright}{\kern0pt}{\isacharcomma}{\kern0pt}\ parallel\ {\isasymlambda}{\isacharbrackleft}{\kern0pt}S\isactrlsub {\isadigit{1}}{\isacharbrackright}{\kern0pt}\ {\isacharparenleft}{\kern0pt}snd\ c{\isacharparenright}{\kern0pt}{\isacharparenright}{\kern0pt}{\isacharparenright}{\kern0pt}\ {\isacharbackquote}{\kern0pt}\ {\isasymdelta}\isactrlsub s\ {\isacharparenleft}{\kern0pt}sh\ {\isasymleadsto}\ State{\isasymllangle}{\isasymsigma}{\isasymrrangle}{\isacharcomma}{\kern0pt}\ {\isasymlambda}{\isacharbrackleft}{\kern0pt}S\isactrlsub {\isadigit{2}}{\isacharbrackright}{\kern0pt}{\isacharparenright}{\kern0pt}\ \isanewline
\ \ \ \ \ \ \ \ \ \ \ \ \ \ \ \ {\isasymsubseteq}\ {\isasymdelta}\isactrlsub s\ {\isacharparenleft}{\kern0pt}sh\ {\isasymleadsto}\ State{\isasymllangle}{\isasymsigma}{\isasymrrangle}{\isacharcomma}{\kern0pt}\ {\isasymlambda}{\isacharbrackleft}{\kern0pt}CO\ S\isactrlsub {\isadigit{1}}\ {\isasymparallel}\ S\isactrlsub {\isadigit{2}}\ OC{\isacharbrackright}{\kern0pt}{\isacharparenright}{\kern0pt}{\isachardoublequoteclose}\isanewline
\isadelimproof
\ \ %
\endisadelimproof
\isatagproof
\isacommand{proof}\isamarkupfalse%
\ {\isacharparenleft}{\kern0pt}subst\ subset{\isacharunderscore}{\kern0pt}iff{\isacharparenright}{\kern0pt}\isanewline
\ \ \ \ \isacommand{show}\isamarkupfalse%
\ {\isachardoublequoteopen}{\isasymforall}c{\isachardot}{\kern0pt}\ c\ {\isasymin}\ {\isacharparenleft}{\kern0pt}{\isacharpercent}{\kern0pt}c{\isachardot}{\kern0pt}\ {\isacharparenleft}{\kern0pt}fst{\isacharparenleft}{\kern0pt}c{\isacharparenright}{\kern0pt}{\isacharcomma}{\kern0pt}\ parallel\ {\isacharparenleft}{\kern0pt}snd\ c{\isacharparenright}{\kern0pt}\ {\isasymlambda}{\isacharbrackleft}{\kern0pt}S\isactrlsub {\isadigit{2}}{\isacharbrackright}{\kern0pt}{\isacharparenright}{\kern0pt}{\isacharparenright}{\kern0pt}\ {\isacharbackquote}{\kern0pt}\ {\isasymdelta}\isactrlsub s\ {\isacharparenleft}{\kern0pt}sh\ {\isasymleadsto}\ State{\isasymllangle}{\isasymsigma}{\isasymrrangle}{\isacharcomma}{\kern0pt}\ {\isasymlambda}{\isacharbrackleft}{\kern0pt}S\isactrlsub {\isadigit{1}}{\isacharbrackright}{\kern0pt}{\isacharparenright}{\kern0pt}\ {\isasymunion}\ \isanewline
\ \ \ \ \ \ \ \ \ \ \ \ \ \ \ \ \ \ \ \ \ \ \ \ \ \ {\isacharparenleft}{\kern0pt}{\isacharpercent}{\kern0pt}c{\isachardot}{\kern0pt}\ {\isacharparenleft}{\kern0pt}fst{\isacharparenleft}{\kern0pt}c{\isacharparenright}{\kern0pt}{\isacharcomma}{\kern0pt}\ parallel\ {\isasymlambda}{\isacharbrackleft}{\kern0pt}S\isactrlsub {\isadigit{1}}{\isacharbrackright}{\kern0pt}\ {\isacharparenleft}{\kern0pt}snd\ c{\isacharparenright}{\kern0pt}{\isacharparenright}{\kern0pt}{\isacharparenright}{\kern0pt}\ {\isacharbackquote}{\kern0pt}\ {\isasymdelta}\isactrlsub s\ {\isacharparenleft}{\kern0pt}sh\ {\isasymleadsto}\ State{\isasymllangle}{\isasymsigma}{\isasymrrangle}{\isacharcomma}{\kern0pt}\ {\isasymlambda}{\isacharbrackleft}{\kern0pt}S\isactrlsub {\isadigit{2}}{\isacharbrackright}{\kern0pt}{\isacharparenright}{\kern0pt}\ \isanewline
\ \ \ \ \ \ \ \ \ \ \ \ \ \ {\isasymlongrightarrow}\ c\ {\isasymin}\ {\isasymdelta}\isactrlsub s\ {\isacharparenleft}{\kern0pt}sh\ {\isasymleadsto}\ State{\isasymllangle}{\isasymsigma}{\isasymrrangle}{\isacharcomma}{\kern0pt}\ {\isasymlambda}{\isacharbrackleft}{\kern0pt}CO\ S\isactrlsub {\isadigit{1}}\ {\isasymparallel}\ S\isactrlsub {\isadigit{2}}\ OC{\isacharbrackright}{\kern0pt}{\isacharparenright}{\kern0pt}{\isachardoublequoteclose}\isanewline
\ \ \ \ %
\isamarkupcmt{We first use the subset-iff rule in order to rewrite the subset relation into 
       a semantically equivalent~implication.%
}\ \ \isanewline
\ \ \ \ \isacommand{proof}\isamarkupfalse%
\ {\isacharparenleft}{\kern0pt}rule\ allI{\isacharcomma}{\kern0pt}\ rule\ impI{\isacharparenright}{\kern0pt}\isanewline
\ \ \ \ \ \ \isacommand{fix}\isamarkupfalse%
\ c\isanewline
\ \ \ \ \ \ %
\isamarkupcmt{We assume \isa{c} to be an arbitrary, but fixed configuration.%
}\isanewline
\ \ \ \ \ \ \isacommand{assume}\isamarkupfalse%
\ {\isachardoublequoteopen}c\ {\isasymin}\ {\isacharparenleft}{\kern0pt}{\isacharpercent}{\kern0pt}c{\isachardot}{\kern0pt}\ {\isacharparenleft}{\kern0pt}fst{\isacharparenleft}{\kern0pt}c{\isacharparenright}{\kern0pt}{\isacharcomma}{\kern0pt}\ parallel\ {\isacharparenleft}{\kern0pt}snd\ c{\isacharparenright}{\kern0pt}\ {\isasymlambda}{\isacharbrackleft}{\kern0pt}S\isactrlsub {\isadigit{2}}{\isacharbrackright}{\kern0pt}{\isacharparenright}{\kern0pt}{\isacharparenright}{\kern0pt}\ {\isacharbackquote}{\kern0pt}\ {\isasymdelta}\isactrlsub s\ {\isacharparenleft}{\kern0pt}sh\ {\isasymleadsto}\ State{\isasymllangle}{\isasymsigma}{\isasymrrangle}{\isacharcomma}{\kern0pt}\ {\isasymlambda}{\isacharbrackleft}{\kern0pt}S\isactrlsub {\isadigit{1}}{\isacharbrackright}{\kern0pt}{\isacharparenright}{\kern0pt}\ {\isasymunion}\ \isanewline
\ \ \ \ \ \ \ \ \ \ \ \ \ \ \ \ \ \ \ \ \ \ \ \ \ \ {\isacharparenleft}{\kern0pt}{\isacharpercent}{\kern0pt}c{\isachardot}{\kern0pt}\ {\isacharparenleft}{\kern0pt}fst{\isacharparenleft}{\kern0pt}c{\isacharparenright}{\kern0pt}{\isacharcomma}{\kern0pt}\ parallel\ {\isasymlambda}{\isacharbrackleft}{\kern0pt}S\isactrlsub {\isadigit{1}}{\isacharbrackright}{\kern0pt}\ {\isacharparenleft}{\kern0pt}snd\ c{\isacharparenright}{\kern0pt}{\isacharparenright}{\kern0pt}{\isacharparenright}{\kern0pt}\ {\isacharbackquote}{\kern0pt}\ {\isasymdelta}\isactrlsub s\ {\isacharparenleft}{\kern0pt}sh\ {\isasymleadsto}\ State{\isasymllangle}{\isasymsigma}{\isasymrrangle}{\isacharcomma}{\kern0pt}\ {\isasymlambda}{\isacharbrackleft}{\kern0pt}S\isactrlsub {\isadigit{2}}{\isacharbrackright}{\kern0pt}{\isacharparenright}{\kern0pt}{\isachardoublequoteclose}\isanewline
\ \ \ \ \ \ %
\isamarkupcmt{We assume that the premise holds, implying that \isa{c} is either a successor 
         configuration of \isa{S\isactrlsub {\isadigit{1}}} with the local parallelism under \isa{S\isactrlsub {\isadigit{2}}} reconstructed in 
         its continuation marker, or a successor configuration of \isa{S\isactrlsub {\isadigit{2}}} with the local 
         parallelism under \isa{S\isactrlsub {\isadigit{1}}} reconstructed in its continuation marker.%
}\isanewline
\ \ \ \ \ \ \isacommand{then}\isamarkupfalse%
\ \isacommand{obtain}\isamarkupfalse%
\ c{\isacharprime}{\kern0pt}\ \isakeyword{where}\ assm\isactrlsub c{\isacharprime}{\kern0pt}{\isacharcolon}{\kern0pt}\isanewline
\ \ \ \ \ \ \ \ {\isachardoublequoteopen}{\isacharparenleft}{\kern0pt}c{\isacharprime}{\kern0pt}\ {\isasymin}\ {\isasymdelta}\isactrlsub s\ {\isacharparenleft}{\kern0pt}sh\ {\isasymleadsto}\ State{\isasymllangle}{\isasymsigma}{\isasymrrangle}{\isacharcomma}{\kern0pt}\ {\isasymlambda}{\isacharbrackleft}{\kern0pt}S\isactrlsub {\isadigit{1}}{\isacharbrackright}{\kern0pt}{\isacharparenright}{\kern0pt}\ {\isasymand}\ fst{\isacharparenleft}{\kern0pt}c{\isacharprime}{\kern0pt}{\isacharparenright}{\kern0pt}\ {\isacharequal}{\kern0pt}\ fst{\isacharparenleft}{\kern0pt}c{\isacharparenright}{\kern0pt}\ {\isasymand}\ parallel\ {\isacharparenleft}{\kern0pt}snd\ c{\isacharprime}{\kern0pt}{\isacharparenright}{\kern0pt}\ {\isasymlambda}{\isacharbrackleft}{\kern0pt}S\isactrlsub {\isadigit{2}}{\isacharbrackright}{\kern0pt}\ {\isacharequal}{\kern0pt}\ snd{\isacharparenleft}{\kern0pt}c{\isacharparenright}{\kern0pt}{\isacharparenright}{\kern0pt}\ {\isasymor}\ \isanewline
\ \ \ \ \ \ \ \ {\isacharparenleft}{\kern0pt}c{\isacharprime}{\kern0pt}\ {\isasymin}\ {\isasymdelta}\isactrlsub s\ {\isacharparenleft}{\kern0pt}sh\ {\isasymleadsto}\ State{\isasymllangle}{\isasymsigma}{\isasymrrangle}{\isacharcomma}{\kern0pt}\ {\isasymlambda}{\isacharbrackleft}{\kern0pt}S\isactrlsub {\isadigit{2}}{\isacharbrackright}{\kern0pt}{\isacharparenright}{\kern0pt}\ {\isasymand}\ fst{\isacharparenleft}{\kern0pt}c{\isacharprime}{\kern0pt}{\isacharparenright}{\kern0pt}\ {\isacharequal}{\kern0pt}\ fst{\isacharparenleft}{\kern0pt}c{\isacharparenright}{\kern0pt}\ {\isasymand}\ parallel\ {\isasymlambda}{\isacharbrackleft}{\kern0pt}S\isactrlsub {\isadigit{1}}{\isacharbrackright}{\kern0pt}\ {\isacharparenleft}{\kern0pt}snd\ c{\isacharprime}{\kern0pt}{\isacharparenright}{\kern0pt}\ {\isacharequal}{\kern0pt}\ snd{\isacharparenleft}{\kern0pt}c{\isacharparenright}{\kern0pt}{\isacharparenright}{\kern0pt}{\isachardoublequoteclose}\ \isanewline
\ \ \ \ \ \ \ \ \isacommand{by}\isamarkupfalse%
\ fastforce\isanewline
\ \ \ \ \ \ %
\isamarkupcmt{Then there must exist a configuration \isa{c{\isacharprime}{\kern0pt}} that satisfies one of the two
         cases: In the first case, \isa{c{\isacharprime}{\kern0pt}} is a successor configuration of \isa{S\isactrlsub {\isadigit{1}}}, whilst
         \isa{c} matches \isa{c{\isacharprime}{\kern0pt}} with the local parallelism under \isa{S\isactrlsub {\isadigit{2}}} reconstructed in its 
         continuation marker. In the second case, \isa{c{\isacharprime}{\kern0pt}} is a successor configuration 
         of \isa{S\isactrlsub {\isadigit{2}}}, whilst \isa{c} matches \isa{c{\isacharprime}{\kern0pt}} with the local parallelism under \isa{S\isactrlsub {\isadigit{1}}} 
         reconstructed in its continuation~marker.%
}\isanewline
\ \ \ \ \ \ \isacommand{thus}\isamarkupfalse%
\ {\isachardoublequoteopen}c\ {\isasymin}\ {\isasymdelta}\isactrlsub s\ {\isacharparenleft}{\kern0pt}sh\ {\isasymleadsto}\ State{\isasymllangle}{\isasymsigma}{\isasymrrangle}{\isacharcomma}{\kern0pt}\ {\isasymlambda}{\isacharbrackleft}{\kern0pt}CO\ S\isactrlsub {\isadigit{1}}\ {\isasymparallel}\ S\isactrlsub {\isadigit{2}}\ OC{\isacharbrackright}{\kern0pt}{\isacharparenright}{\kern0pt}{\isachardoublequoteclose}\isanewline
\ \ \ \ \ \ \isacommand{proof}\isamarkupfalse%
\ {\isacharparenleft}{\kern0pt}cases{\isacharparenright}{\kern0pt}\isanewline
\ \ \ \ \ \ %
\isamarkupcmt{We perform a case distinction over those two cases.%
}\isanewline
\ \ \ \ \ \ \ \ \isacommand{assume}\isamarkupfalse%
\ p{\isacharcolon}{\kern0pt}\ {\isachardoublequoteopen}c{\isacharprime}{\kern0pt}\ {\isasymin}\ {\isasymdelta}\isactrlsub s\ {\isacharparenleft}{\kern0pt}sh\ {\isasymleadsto}\ State{\isasymllangle}{\isasymsigma}{\isasymrrangle}{\isacharcomma}{\kern0pt}\ {\isasymlambda}{\isacharbrackleft}{\kern0pt}S\isactrlsub {\isadigit{1}}{\isacharbrackright}{\kern0pt}{\isacharparenright}{\kern0pt}\ {\isasymand}\ fst{\isacharparenleft}{\kern0pt}c{\isacharprime}{\kern0pt}{\isacharparenright}{\kern0pt}\ {\isacharequal}{\kern0pt}\ fst{\isacharparenleft}{\kern0pt}c{\isacharparenright}{\kern0pt}\ {\isasymand}\ parallel\ {\isacharparenleft}{\kern0pt}snd\ c{\isacharprime}{\kern0pt}{\isacharparenright}{\kern0pt}\ {\isasymlambda}{\isacharbrackleft}{\kern0pt}S\isactrlsub {\isadigit{2}}{\isacharbrackright}{\kern0pt}\ {\isacharequal}{\kern0pt}\ snd{\isacharparenleft}{\kern0pt}c{\isacharparenright}{\kern0pt}{\isachardoublequoteclose}\isanewline
\ \ \ \ \ \ \ \ %
\isamarkupcmt{Let us assume that the first case holds.%
}\isanewline
\ \ \ \ \ \ \ \ \isacommand{moreover}\isamarkupfalse%
\ \isacommand{then}\isamarkupfalse%
\ \isacommand{obtain}\isamarkupfalse%
\ {\isasympi}\ \isakeyword{where}\ assm\isactrlsub {\isasympi}{\isacharcolon}{\kern0pt}\isanewline
\ \ \ \ \ \ \ \ \ \ {\isachardoublequoteopen}{\isasympi}\ {\isasymin}\ val\isactrlsub s\ S\isactrlsub {\isadigit{1}}\ {\isasymsigma}\ {\isasymand}\ consistent{\isacharparenleft}{\kern0pt}sval\isactrlsub B\ {\isacharparenleft}{\kern0pt}{\isasymdown}\isactrlsub p\ {\isasympi}{\isacharparenright}{\kern0pt}\ {\isacharparenleft}{\kern0pt}min{\isacharunderscore}{\kern0pt}conc{\isacharunderscore}{\kern0pt}map\isactrlsub {\isasymT}\ {\isacharparenleft}{\kern0pt}{\isasymdown}\isactrlsub {\isasymtau}\ {\isasympi}{\isacharparenright}{\kern0pt}\ {\isadigit{0}}{\isacharparenright}{\kern0pt}{\isacharparenright}{\kern0pt}\ \isanewline
\ \ \ \ \ \ \ \ \ \ {\isasymand}\ c{\isacharprime}{\kern0pt}\ {\isacharequal}{\kern0pt}\ {\isacharparenleft}{\kern0pt}trace{\isacharunderscore}{\kern0pt}conc\ {\isacharparenleft}{\kern0pt}min{\isacharunderscore}{\kern0pt}conc{\isacharunderscore}{\kern0pt}map\isactrlsub {\isasymT}\ {\isacharparenleft}{\kern0pt}sh\ {\isasymcdot}\ {\isacharparenleft}{\kern0pt}{\isasymdown}\isactrlsub {\isasymtau}\ {\isasympi}{\isacharparenright}{\kern0pt}{\isacharparenright}{\kern0pt}\ {\isadigit{0}}{\isacharparenright}{\kern0pt}\ {\isacharparenleft}{\kern0pt}sh\ {\isasymcdot}\ {\isacharparenleft}{\kern0pt}{\isasymdown}\isactrlsub {\isasymtau}\ {\isasympi}{\isacharparenright}{\kern0pt}{\isacharparenright}{\kern0pt}{\isacharcomma}{\kern0pt}\ {\isacharparenleft}{\kern0pt}{\isasymdown}\isactrlsub {\isasymlambda}\ {\isasympi}{\isacharparenright}{\kern0pt}{\isacharparenright}{\kern0pt}{\isachardoublequoteclose}\ \isacommand{by}\isamarkupfalse%
\ auto\isanewline
\ \ \ \ \ \ \ \ %
\isamarkupcmt{Hence there must exist a continuation trace \isa{{\isasympi}} with a consistent path 
           condition generated from \isa{S\isactrlsub {\isadigit{1}}}, which translates to configuration~\isa{c{\isacharprime}{\kern0pt}}.%
}\isanewline
\ \ \ \ \ \ \ \ \isacommand{moreover}\isamarkupfalse%
\ \isacommand{then}\isamarkupfalse%
\ \isacommand{obtain}\isamarkupfalse%
\ {\isasympi}{\isacharprime}{\kern0pt}\ \isakeyword{where}\isanewline
\ \ \ \ \ \ \ \ \ \ {\isachardoublequoteopen}{\isasympi}{\isacharprime}{\kern0pt}\ {\isacharequal}{\kern0pt}\ {\isacharparenleft}{\kern0pt}{\isasymdown}\isactrlsub p\ {\isasympi}{\isacharparenright}{\kern0pt}\ {\isasymtriangleright}\ {\isacharparenleft}{\kern0pt}{\isasymdown}\isactrlsub {\isasymtau}\ {\isasympi}{\isacharparenright}{\kern0pt}\ \isactrlitem \ parallel\ {\isacharparenleft}{\kern0pt}{\isasymdown}\isactrlsub {\isasymlambda}\ {\isasympi}{\isacharparenright}{\kern0pt}\ {\isasymlambda}{\isacharbrackleft}{\kern0pt}S\isactrlsub {\isadigit{2}}{\isacharbrackright}{\kern0pt}{\isachardoublequoteclose}\ \isacommand{by}\isamarkupfalse%
\ auto\isanewline
\ \ \ \ \ \ \ \ %
\isamarkupcmt{We then obtain the continuation trace \isa{{\isasympi}{\isacharprime}{\kern0pt}} which matches with \isa{{\isasympi}} except
           having the local parallelism under \isa{S\isactrlsub {\isadigit{2}}} reconstructed in its 
           continuation~marker.%
}\isanewline
\ \ \ \ \ \ \ \ \isacommand{ultimately}\isamarkupfalse%
\ \isacommand{have}\isamarkupfalse%
\ connect{\isacharcolon}{\kern0pt}\ \isanewline
\ \ \ \ \ \ \ \ \ \ {\isachardoublequoteopen}{\isasympi}{\isacharprime}{\kern0pt}\ {\isasymin}\ val\isactrlsub s\ {\isacharparenleft}{\kern0pt}CO\ S\isactrlsub {\isadigit{1}}\ {\isasymparallel}\ S\isactrlsub {\isadigit{2}}\ OC{\isacharparenright}{\kern0pt}\ {\isasymsigma}\ {\isasymand}\ consistent{\isacharparenleft}{\kern0pt}sval\isactrlsub B\ {\isacharparenleft}{\kern0pt}{\isasymdown}\isactrlsub p\ {\isasympi}{\isacharprime}{\kern0pt}{\isacharparenright}{\kern0pt}\ {\isacharparenleft}{\kern0pt}min{\isacharunderscore}{\kern0pt}conc{\isacharunderscore}{\kern0pt}map\isactrlsub {\isasymT}\ {\isacharparenleft}{\kern0pt}{\isasymdown}\isactrlsub {\isasymtau}\ {\isasympi}{\isacharprime}{\kern0pt}{\isacharparenright}{\kern0pt}\ {\isadigit{0}}{\isacharparenright}{\kern0pt}{\isacharparenright}{\kern0pt}\ \isanewline
\ \ \ \ \ \ \ \ \ \ {\isasymand}\ {\isacharparenleft}{\kern0pt}{\isasymdown}\isactrlsub {\isasymtau}\ {\isasympi}{\isacharprime}{\kern0pt}{\isacharparenright}{\kern0pt}\ {\isacharequal}{\kern0pt}\ {\isacharparenleft}{\kern0pt}{\isasymdown}\isactrlsub {\isasymtau}\ {\isasympi}{\isacharparenright}{\kern0pt}\ {\isasymand}\ {\isacharparenleft}{\kern0pt}{\isasymdown}\isactrlsub {\isasymlambda}\ {\isasympi}{\isacharprime}{\kern0pt}{\isacharparenright}{\kern0pt}\ {\isacharequal}{\kern0pt}\ parallel\ {\isacharparenleft}{\kern0pt}{\isasymdown}\isactrlsub {\isasymlambda}\ {\isasympi}{\isacharparenright}{\kern0pt}\ {\isasymlambda}{\isacharbrackleft}{\kern0pt}S\isactrlsub {\isadigit{2}}{\isacharbrackright}{\kern0pt}{\isachardoublequoteclose}\ \isacommand{by}\isamarkupfalse%
\ auto\ \isanewline
\ \ \ \ \ \ \ \ %
\isamarkupcmt{This implies that \isa{{\isasympi}{\isacharprime}{\kern0pt}} must be a continuation trace with a consistent
           path condition generated from \isa{CO\ S\isactrlsub {\isadigit{1}}\ {\isasymparallel}\ S\isactrlsub {\isadigit{2}}\ OC}. Note that \isa{{\isasympi}} matches with 
           \isa{{\isasympi}{\isacharprime}{\kern0pt}} in its symbolic trace, but not in its continuation~marker.%
}\isanewline
\ \ \ \ \ \ \ \ \isacommand{then}\isamarkupfalse%
\ \isacommand{obtain}\isamarkupfalse%
\ c{\isacharprime}{\kern0pt}{\isacharprime}{\kern0pt}\ \isakeyword{where}\ assm\isactrlsub c{\isacharprime}{\kern0pt}{\isacharprime}{\kern0pt}{\isacharcolon}{\kern0pt}\isanewline
\ \ \ \ \ \ \ \ \ \ {\isachardoublequoteopen}c{\isacharprime}{\kern0pt}{\isacharprime}{\kern0pt}\ {\isacharequal}{\kern0pt}\ {\isacharparenleft}{\kern0pt}trace{\isacharunderscore}{\kern0pt}conc\ {\isacharparenleft}{\kern0pt}min{\isacharunderscore}{\kern0pt}conc{\isacharunderscore}{\kern0pt}map\isactrlsub {\isasymT}\ {\isacharparenleft}{\kern0pt}sh\ {\isasymcdot}\ {\isacharparenleft}{\kern0pt}{\isasymdown}\isactrlsub {\isasymtau}\ {\isasympi}{\isacharprime}{\kern0pt}{\isacharparenright}{\kern0pt}{\isacharparenright}{\kern0pt}\ {\isadigit{0}}{\isacharparenright}{\kern0pt}\ {\isacharparenleft}{\kern0pt}sh\ {\isasymcdot}\ {\isacharparenleft}{\kern0pt}{\isasymdown}\isactrlsub {\isasymtau}\ {\isasympi}{\isacharprime}{\kern0pt}{\isacharparenright}{\kern0pt}{\isacharparenright}{\kern0pt}{\isacharcomma}{\kern0pt}\ {\isasymdown}\isactrlsub {\isasymlambda}\ {\isasympi}{\isacharprime}{\kern0pt}{\isacharparenright}{\kern0pt}{\isachardoublequoteclose}\ \isacommand{by}\isamarkupfalse%
\ auto\isanewline
\ \ \ \ \ \ \ \ %
\isamarkupcmt{We then obtain the configuration \isa{c{\isacharprime}{\kern0pt}{\isacharprime}{\kern0pt}} translated from \isa{{\isasympi}{\isacharprime}{\kern0pt}}.%
}\isanewline
\ \ \ \ \ \ \ \ \isacommand{hence}\isamarkupfalse%
\ {\isachardoublequoteopen}c\ {\isacharequal}{\kern0pt}\ c{\isacharprime}{\kern0pt}{\isacharprime}{\kern0pt}{\isachardoublequoteclose}\ \isacommand{by}\isamarkupfalse%
\ {\isacharparenleft}{\kern0pt}metis\ p\ assm\isactrlsub c{\isacharprime}{\kern0pt}\ assm\isactrlsub {\isasympi}\ connect\ fst{\isacharunderscore}{\kern0pt}eqD\ prod{\isachardot}{\kern0pt}collapse\ snd{\isacharunderscore}{\kern0pt}eqD{\isacharparenright}{\kern0pt}\isanewline
\ \ \ \ \ \ \ \ %
\isamarkupcmt{We can now derive that \isa{c} and \isa{c{\isacharprime}{\kern0pt}{\isacharprime}{\kern0pt}} match in both of~their~elements.%
}\isanewline
\ \ \ \ \ \ \ \ \isacommand{moreover}\isamarkupfalse%
\ \isacommand{have}\isamarkupfalse%
\ {\isachardoublequoteopen}c{\isacharprime}{\kern0pt}{\isacharprime}{\kern0pt}\ {\isasymin}\ {\isacharparenleft}{\kern0pt}{\isacharpercent}{\kern0pt}c{\isachardot}{\kern0pt}\ {\isacharparenleft}{\kern0pt}trace{\isacharunderscore}{\kern0pt}conc\ {\isacharparenleft}{\kern0pt}min{\isacharunderscore}{\kern0pt}conc{\isacharunderscore}{\kern0pt}map\isactrlsub {\isasymT}\ {\isacharparenleft}{\kern0pt}sh\ {\isasymcdot}\ {\isacharparenleft}{\kern0pt}{\isasymdown}\isactrlsub {\isasymtau}\ c{\isacharparenright}{\kern0pt}{\isacharparenright}{\kern0pt}\ {\isadigit{0}}{\isacharparenright}{\kern0pt}\ {\isacharparenleft}{\kern0pt}sh\ {\isasymcdot}\ {\isacharparenleft}{\kern0pt}{\isasymdown}\isactrlsub {\isasymtau}\ c{\isacharparenright}{\kern0pt}{\isacharparenright}{\kern0pt}{\isacharcomma}{\kern0pt}\ {\isasymdown}\isactrlsub {\isasymlambda}\ c{\isacharparenright}{\kern0pt}{\isacharparenright}{\kern0pt}\ \isanewline
\ \ \ \ \ \ \ \ \ \ {\isacharbackquote}{\kern0pt}\ {\isacharbraceleft}{\kern0pt}c\ {\isasymin}\ val\isactrlsub s\ {\isacharparenleft}{\kern0pt}CO\ S\isactrlsub {\isadigit{1}}\ {\isasymparallel}\ S\isactrlsub {\isadigit{2}}\ OC{\isacharparenright}{\kern0pt}\ {\isasymsigma}{\isachardot}{\kern0pt}\ consistent{\isacharparenleft}{\kern0pt}sval\isactrlsub B\ {\isacharparenleft}{\kern0pt}{\isasymdown}\isactrlsub p\ c{\isacharparenright}{\kern0pt}\ {\isacharparenleft}{\kern0pt}min{\isacharunderscore}{\kern0pt}conc{\isacharunderscore}{\kern0pt}map\isactrlsub {\isasymT}\ {\isacharparenleft}{\kern0pt}{\isasymdown}\isactrlsub {\isasymtau}\ c{\isacharparenright}{\kern0pt}\ {\isadigit{0}}{\isacharparenright}{\kern0pt}{\isacharparenright}{\kern0pt}{\isacharbraceright}{\kern0pt}{\isachardoublequoteclose}\isanewline
\ \ \ \ \ \ \ \ \ \ \isacommand{using}\isamarkupfalse%
\ assm\isactrlsub c{\isacharprime}{\kern0pt}{\isacharprime}{\kern0pt}\ connect\ \isacommand{by}\isamarkupfalse%
\ blast\isanewline
\ \ \ \ \ \ \ \ %
\isamarkupcmt{Using the setup of \isa{c{\isacharprime}{\kern0pt}{\isacharprime}{\kern0pt}}, we can now infer that \isa{c{\isacharprime}{\kern0pt}{\isacharprime}{\kern0pt}} must be a successor 
           configuration of the local parallelism statement \isa{CO\ S\isactrlsub {\isadigit{1}}\ {\isasymparallel}\ S\isactrlsub {\isadigit{2}}\ OC}.%
}\isanewline
\ \ \ \ \ \ \ \ \isacommand{ultimately}\isamarkupfalse%
\ \isacommand{show}\isamarkupfalse%
\ {\isachardoublequoteopen}c\ {\isasymin}\ {\isasymdelta}\isactrlsub s\ {\isacharparenleft}{\kern0pt}sh\ {\isasymleadsto}\ State{\isasymllangle}{\isasymsigma}{\isasymrrangle}{\isacharcomma}{\kern0pt}\ {\isasymlambda}{\isacharbrackleft}{\kern0pt}CO\ S\isactrlsub {\isadigit{1}}\ {\isasymparallel}\ S\isactrlsub {\isadigit{2}}\ OC{\isacharbrackright}{\kern0pt}{\isacharparenright}{\kern0pt}{\isachardoublequoteclose}\isanewline
\ \ \ \ \ \ \ \ \ \ \isacommand{by}\isamarkupfalse%
\ auto\isanewline
\ \ \ \ \ \ \ \ %
\isamarkupcmt{Taking into consideration that \isa{c} matches with \isa{c{\isacharprime}{\kern0pt}{\isacharprime}{\kern0pt}}, we can finally 
           conclude that \isa{c} must also be a successor configuration of the local
           parallelism statement \isa{CO\ S\isactrlsub {\isadigit{1}}\ {\isasymparallel}\ S\isactrlsub {\isadigit{2}}\ OC}, which needed to be proven in the 
           first~place.%
}\isanewline
\ \ \ \ \ \ \isacommand{next}\isamarkupfalse%
\isanewline
\ \ \ \ \ \ \ \ \isacommand{assume}\isamarkupfalse%
\ {\isachardoublequoteopen}{\isasymnot}{\isacharparenleft}{\kern0pt}c{\isacharprime}{\kern0pt}\ {\isasymin}\ {\isasymdelta}\isactrlsub s\ {\isacharparenleft}{\kern0pt}sh\ {\isasymleadsto}\ State{\isasymllangle}{\isasymsigma}{\isasymrrangle}{\isacharcomma}{\kern0pt}\ {\isasymlambda}{\isacharbrackleft}{\kern0pt}S\isactrlsub {\isadigit{1}}{\isacharbrackright}{\kern0pt}{\isacharparenright}{\kern0pt}\ {\isasymand}\ fst{\isacharparenleft}{\kern0pt}c{\isacharprime}{\kern0pt}{\isacharparenright}{\kern0pt}\ {\isacharequal}{\kern0pt}\ fst{\isacharparenleft}{\kern0pt}c{\isacharparenright}{\kern0pt}\ {\isasymand}\ parallel\ {\isacharparenleft}{\kern0pt}snd\ c{\isacharprime}{\kern0pt}{\isacharparenright}{\kern0pt}\ {\isasymlambda}{\isacharbrackleft}{\kern0pt}S\isactrlsub {\isadigit{2}}{\isacharbrackright}{\kern0pt}\ {\isacharequal}{\kern0pt}\ snd{\isacharparenleft}{\kern0pt}c{\isacharparenright}{\kern0pt}{\isacharparenright}{\kern0pt}{\isachardoublequoteclose}\isanewline
\ \ \ \ \ \ \ \ %
\isamarkupcmt{Let us assume the first case does not hold.%
}\isanewline
\ \ \ \ \ \ \ \ \isacommand{hence}\isamarkupfalse%
\ p{\isacharcolon}{\kern0pt}\ {\isachardoublequoteopen}c{\isacharprime}{\kern0pt}\ {\isasymin}\ {\isasymdelta}\isactrlsub s\ {\isacharparenleft}{\kern0pt}sh\ {\isasymleadsto}\ State{\isasymllangle}{\isasymsigma}{\isasymrrangle}{\isacharcomma}{\kern0pt}\ {\isasymlambda}{\isacharbrackleft}{\kern0pt}S\isactrlsub {\isadigit{2}}{\isacharbrackright}{\kern0pt}{\isacharparenright}{\kern0pt}\ {\isasymand}\ fst{\isacharparenleft}{\kern0pt}c{\isacharprime}{\kern0pt}{\isacharparenright}{\kern0pt}\ {\isacharequal}{\kern0pt}\ fst{\isacharparenleft}{\kern0pt}c{\isacharparenright}{\kern0pt}\ {\isasymand}\ parallel\ {\isasymlambda}{\isacharbrackleft}{\kern0pt}S\isactrlsub {\isadigit{1}}{\isacharbrackright}{\kern0pt}\ {\isacharparenleft}{\kern0pt}snd\ c{\isacharprime}{\kern0pt}{\isacharparenright}{\kern0pt}\ {\isacharequal}{\kern0pt}\ snd{\isacharparenleft}{\kern0pt}c{\isacharparenright}{\kern0pt}{\isachardoublequoteclose}\isanewline
\ \ \ \ \ \ \ \ \ \ \isacommand{using}\isamarkupfalse%
\ assm\isactrlsub c{\isacharprime}{\kern0pt}\ \isacommand{by}\isamarkupfalse%
\ auto\isanewline
\ \ \ \ \ \ \ \ %
\isamarkupcmt{Then the second case must hold.%
}\isanewline
\ \ \ \ \ \ \ \ \isacommand{moreover}\isamarkupfalse%
\ \isacommand{then}\isamarkupfalse%
\ \isacommand{obtain}\isamarkupfalse%
\ {\isasympi}\ \isakeyword{where}\ assm\isactrlsub {\isasympi}{\isacharcolon}{\kern0pt}\isanewline
\ \ \ \ \ \ \ \ \ \ {\isachardoublequoteopen}{\isasympi}\ {\isasymin}\ val\isactrlsub s\ S\isactrlsub {\isadigit{2}}\ {\isasymsigma}\ {\isasymand}\ consistent{\isacharparenleft}{\kern0pt}sval\isactrlsub B\ {\isacharparenleft}{\kern0pt}{\isasymdown}\isactrlsub p\ {\isasympi}{\isacharparenright}{\kern0pt}\ {\isacharparenleft}{\kern0pt}min{\isacharunderscore}{\kern0pt}conc{\isacharunderscore}{\kern0pt}map\isactrlsub {\isasymT}\ {\isacharparenleft}{\kern0pt}{\isasymdown}\isactrlsub {\isasymtau}\ {\isasympi}{\isacharparenright}{\kern0pt}\ {\isadigit{0}}{\isacharparenright}{\kern0pt}{\isacharparenright}{\kern0pt}\ \isanewline
\ \ \ \ \ \ \ \ \ \ {\isasymand}\ c{\isacharprime}{\kern0pt}\ {\isacharequal}{\kern0pt}\ {\isacharparenleft}{\kern0pt}trace{\isacharunderscore}{\kern0pt}conc\ {\isacharparenleft}{\kern0pt}min{\isacharunderscore}{\kern0pt}conc{\isacharunderscore}{\kern0pt}map\isactrlsub {\isasymT}\ {\isacharparenleft}{\kern0pt}sh\ {\isasymcdot}\ {\isacharparenleft}{\kern0pt}{\isasymdown}\isactrlsub {\isasymtau}\ {\isasympi}{\isacharparenright}{\kern0pt}{\isacharparenright}{\kern0pt}\ {\isadigit{0}}{\isacharparenright}{\kern0pt}\ {\isacharparenleft}{\kern0pt}sh\ {\isasymcdot}\ {\isacharparenleft}{\kern0pt}{\isasymdown}\isactrlsub {\isasymtau}\ {\isasympi}{\isacharparenright}{\kern0pt}{\isacharparenright}{\kern0pt}{\isacharcomma}{\kern0pt}\ {\isacharparenleft}{\kern0pt}{\isasymdown}\isactrlsub {\isasymlambda}\ {\isasympi}{\isacharparenright}{\kern0pt}{\isacharparenright}{\kern0pt}{\isachardoublequoteclose}\ \isacommand{by}\isamarkupfalse%
\ auto\isanewline
\ \ \ \ \ \ \ \ %
\isamarkupcmt{Hence there must exist a continuation trace \isa{{\isasympi}} with a consistent path 
           condition generated from \isa{S\isactrlsub {\isadigit{2}}}, which translates to configuration~\isa{c{\isacharprime}{\kern0pt}}.%
}\isanewline
\ \ \ \ \ \ \ \ \isacommand{moreover}\isamarkupfalse%
\ \isacommand{then}\isamarkupfalse%
\ \isacommand{obtain}\isamarkupfalse%
\ {\isasympi}{\isacharprime}{\kern0pt}\ \isakeyword{where}\isanewline
\ \ \ \ \ \ \ \ \ \ {\isachardoublequoteopen}{\isasympi}{\isacharprime}{\kern0pt}\ {\isacharequal}{\kern0pt}\ {\isacharparenleft}{\kern0pt}{\isasymdown}\isactrlsub p\ {\isasympi}{\isacharparenright}{\kern0pt}\ {\isasymtriangleright}\ {\isacharparenleft}{\kern0pt}{\isasymdown}\isactrlsub {\isasymtau}\ {\isasympi}{\isacharparenright}{\kern0pt}\ \isactrlitem \ parallel\ {\isasymlambda}{\isacharbrackleft}{\kern0pt}S\isactrlsub {\isadigit{1}}{\isacharbrackright}{\kern0pt}\ {\isacharparenleft}{\kern0pt}{\isasymdown}\isactrlsub {\isasymlambda}\ {\isasympi}{\isacharparenright}{\kern0pt}{\isachardoublequoteclose}\ \isacommand{by}\isamarkupfalse%
\ auto\isanewline
\ \ \ \ \ \ \ \ %
\isamarkupcmt{We then obtain the continuation trace \isa{{\isasympi}{\isacharprime}{\kern0pt}} which matches with \isa{{\isasympi}} except
           having the local parallelism under \isa{S\isactrlsub {\isadigit{1}}} reconstructed in its 
           continuation~marker.%
}\isanewline
\ \ \ \ \ \ \ \ \isacommand{ultimately}\isamarkupfalse%
\ \isacommand{have}\isamarkupfalse%
\ connect{\isacharcolon}{\kern0pt}\isanewline
\ \ \ \ \ \ \ \ \ \ {\isachardoublequoteopen}{\isasympi}{\isacharprime}{\kern0pt}\ {\isasymin}\ val\isactrlsub s\ {\isacharparenleft}{\kern0pt}CO\ S\isactrlsub {\isadigit{1}}\ {\isasymparallel}\ S\isactrlsub {\isadigit{2}}\ OC{\isacharparenright}{\kern0pt}\ {\isasymsigma}\ {\isasymand}\ consistent{\isacharparenleft}{\kern0pt}sval\isactrlsub B\ {\isacharparenleft}{\kern0pt}{\isasymdown}\isactrlsub p\ {\isasympi}{\isacharprime}{\kern0pt}{\isacharparenright}{\kern0pt}\ {\isacharparenleft}{\kern0pt}min{\isacharunderscore}{\kern0pt}conc{\isacharunderscore}{\kern0pt}map\isactrlsub {\isasymT}\ {\isacharparenleft}{\kern0pt}{\isasymdown}\isactrlsub {\isasymtau}\ {\isasympi}{\isacharprime}{\kern0pt}{\isacharparenright}{\kern0pt}\ {\isadigit{0}}{\isacharparenright}{\kern0pt}{\isacharparenright}{\kern0pt}\ \isanewline
\ \ \ \ \ \ \ \ \ \ {\isasymand}\ {\isacharparenleft}{\kern0pt}{\isasymdown}\isactrlsub {\isasymtau}\ {\isasympi}{\isacharprime}{\kern0pt}{\isacharparenright}{\kern0pt}\ {\isacharequal}{\kern0pt}\ {\isacharparenleft}{\kern0pt}{\isasymdown}\isactrlsub {\isasymtau}\ {\isasympi}{\isacharparenright}{\kern0pt}\ {\isasymand}\ {\isacharparenleft}{\kern0pt}{\isasymdown}\isactrlsub {\isasymlambda}\ {\isasympi}{\isacharprime}{\kern0pt}{\isacharparenright}{\kern0pt}\ {\isacharequal}{\kern0pt}\ parallel\ {\isasymlambda}{\isacharbrackleft}{\kern0pt}S\isactrlsub {\isadigit{1}}{\isacharbrackright}{\kern0pt}\ {\isacharparenleft}{\kern0pt}{\isasymdown}\isactrlsub {\isasymlambda}\ {\isasympi}{\isacharparenright}{\kern0pt}{\isachardoublequoteclose}\ \isacommand{by}\isamarkupfalse%
\ auto\ \isanewline
\ \ \ \ \ \ \ \ %
\isamarkupcmt{This implies that \isa{{\isasympi}{\isacharprime}{\kern0pt}} must be a continuation trace with a consistent
           path condition generated from \isa{CO\ S\isactrlsub {\isadigit{1}}\ {\isasymparallel}\ S\isactrlsub {\isadigit{2}}\ OC}. Note that \isa{{\isasympi}} matches with 
           \isa{{\isasympi}{\isacharprime}{\kern0pt}} in its symbolic trace, but not in its continuation~marker.%
}\isanewline
\ \ \ \ \ \ \ \ \isacommand{then}\isamarkupfalse%
\ \isacommand{obtain}\isamarkupfalse%
\ c{\isacharprime}{\kern0pt}{\isacharprime}{\kern0pt}\ \isakeyword{where}\ assm\isactrlsub c{\isacharprime}{\kern0pt}{\isacharprime}{\kern0pt}{\isacharcolon}{\kern0pt}\isanewline
\ \ \ \ \ \ \ \ \ \ {\isachardoublequoteopen}c{\isacharprime}{\kern0pt}{\isacharprime}{\kern0pt}\ {\isacharequal}{\kern0pt}\ {\isacharparenleft}{\kern0pt}trace{\isacharunderscore}{\kern0pt}conc\ {\isacharparenleft}{\kern0pt}min{\isacharunderscore}{\kern0pt}conc{\isacharunderscore}{\kern0pt}map\isactrlsub {\isasymT}\ {\isacharparenleft}{\kern0pt}sh\ {\isasymcdot}\ {\isacharparenleft}{\kern0pt}{\isasymdown}\isactrlsub {\isasymtau}\ {\isasympi}{\isacharprime}{\kern0pt}{\isacharparenright}{\kern0pt}{\isacharparenright}{\kern0pt}\ {\isadigit{0}}{\isacharparenright}{\kern0pt}\ {\isacharparenleft}{\kern0pt}sh\ {\isasymcdot}\ {\isacharparenleft}{\kern0pt}{\isasymdown}\isactrlsub {\isasymtau}\ {\isasympi}{\isacharprime}{\kern0pt}{\isacharparenright}{\kern0pt}{\isacharparenright}{\kern0pt}{\isacharcomma}{\kern0pt}\ {\isasymdown}\isactrlsub {\isasymlambda}\ {\isasympi}{\isacharprime}{\kern0pt}{\isacharparenright}{\kern0pt}{\isachardoublequoteclose}\ \isacommand{by}\isamarkupfalse%
\ auto\isanewline
\ \ \ \ \ \ \ \ %
\isamarkupcmt{We then obtain the configuration \isa{c{\isacharprime}{\kern0pt}{\isacharprime}{\kern0pt}} translated from \isa{{\isasympi}{\isacharprime}{\kern0pt}}.%
}\isanewline
\ \ \ \ \ \ \ \ \isacommand{hence}\isamarkupfalse%
\ {\isachardoublequoteopen}c\ {\isacharequal}{\kern0pt}\ c{\isacharprime}{\kern0pt}{\isacharprime}{\kern0pt}{\isachardoublequoteclose}\ \isacommand{by}\isamarkupfalse%
\ {\isacharparenleft}{\kern0pt}metis\ p\ assm\isactrlsub c{\isacharprime}{\kern0pt}\ assm\isactrlsub {\isasympi}\ connect\ fst{\isacharunderscore}{\kern0pt}eqD\ prod{\isachardot}{\kern0pt}collapse\ snd{\isacharunderscore}{\kern0pt}eqD{\isacharparenright}{\kern0pt}\isanewline
\ \ \ \ \ \ \ \ %
\isamarkupcmt{We can now derive that \isa{c} and \isa{c{\isacharprime}{\kern0pt}{\isacharprime}{\kern0pt}} match in both of~their~elements.%
}\isanewline
\ \ \ \ \ \ \ \ \isacommand{moreover}\isamarkupfalse%
\ \isacommand{have}\isamarkupfalse%
\ {\isachardoublequoteopen}c{\isacharprime}{\kern0pt}{\isacharprime}{\kern0pt}\ {\isasymin}\ {\isacharparenleft}{\kern0pt}{\isacharpercent}{\kern0pt}c{\isachardot}{\kern0pt}\ {\isacharparenleft}{\kern0pt}trace{\isacharunderscore}{\kern0pt}conc\ {\isacharparenleft}{\kern0pt}min{\isacharunderscore}{\kern0pt}conc{\isacharunderscore}{\kern0pt}map\isactrlsub {\isasymT}\ {\isacharparenleft}{\kern0pt}sh\ {\isasymcdot}\ {\isacharparenleft}{\kern0pt}{\isasymdown}\isactrlsub {\isasymtau}\ c{\isacharparenright}{\kern0pt}{\isacharparenright}{\kern0pt}\ {\isadigit{0}}{\isacharparenright}{\kern0pt}\ {\isacharparenleft}{\kern0pt}sh\ {\isasymcdot}\ {\isacharparenleft}{\kern0pt}{\isasymdown}\isactrlsub {\isasymtau}\ c{\isacharparenright}{\kern0pt}{\isacharparenright}{\kern0pt}{\isacharcomma}{\kern0pt}\ {\isasymdown}\isactrlsub {\isasymlambda}\ c{\isacharparenright}{\kern0pt}{\isacharparenright}{\kern0pt}\ \isanewline
\ \ \ \ \ \ \ \ \ \ {\isacharbackquote}{\kern0pt}\ {\isacharbraceleft}{\kern0pt}c\ {\isasymin}\ val\isactrlsub s\ {\isacharparenleft}{\kern0pt}CO\ S\isactrlsub {\isadigit{1}}\ {\isasymparallel}\ S\isactrlsub {\isadigit{2}}\ OC{\isacharparenright}{\kern0pt}\ {\isasymsigma}{\isachardot}{\kern0pt}\ consistent{\isacharparenleft}{\kern0pt}sval\isactrlsub B\ {\isacharparenleft}{\kern0pt}{\isasymdown}\isactrlsub p\ c{\isacharparenright}{\kern0pt}\ {\isacharparenleft}{\kern0pt}min{\isacharunderscore}{\kern0pt}conc{\isacharunderscore}{\kern0pt}map\isactrlsub {\isasymT}\ {\isacharparenleft}{\kern0pt}{\isasymdown}\isactrlsub {\isasymtau}\ c{\isacharparenright}{\kern0pt}\ {\isadigit{0}}{\isacharparenright}{\kern0pt}{\isacharparenright}{\kern0pt}{\isacharbraceright}{\kern0pt}{\isachardoublequoteclose}\isanewline
\ \ \ \ \ \ \ \ \ \ \isacommand{using}\isamarkupfalse%
\ assm\isactrlsub c{\isacharprime}{\kern0pt}{\isacharprime}{\kern0pt}\ connect\ \isacommand{by}\isamarkupfalse%
\ blast\isanewline
\ \ \ \ \ \ \ \ %
\isamarkupcmt{Using the setup of \isa{c{\isacharprime}{\kern0pt}{\isacharprime}{\kern0pt}}, we can now infer that \isa{c{\isacharprime}{\kern0pt}{\isacharprime}{\kern0pt}} must be a successor 
           configuration of the local parallelism command \isa{CO\ S\isactrlsub {\isadigit{1}}\ {\isasymparallel}\ S\isactrlsub {\isadigit{2}}\ OC}.%
}\isanewline
\ \ \ \ \ \ \ \ \isacommand{ultimately}\isamarkupfalse%
\ \isacommand{show}\isamarkupfalse%
\ {\isachardoublequoteopen}c\ {\isasymin}\ {\isasymdelta}\isactrlsub s\ {\isacharparenleft}{\kern0pt}sh\ {\isasymleadsto}\ State{\isasymllangle}{\isasymsigma}{\isasymrrangle}{\isacharcomma}{\kern0pt}\ {\isasymlambda}{\isacharbrackleft}{\kern0pt}CO\ S\isactrlsub {\isadigit{1}}\ {\isasymparallel}\ S\isactrlsub {\isadigit{2}}\ OC{\isacharbrackright}{\kern0pt}{\isacharparenright}{\kern0pt}{\isachardoublequoteclose}\isanewline
\ \ \ \ \ \ \ \ \ \ \isacommand{by}\isamarkupfalse%
\ auto\isanewline
\ \ \ \ \ \ \ \ %
\isamarkupcmt{Taking into consideration that \isa{c} matches with \isa{c{\isacharprime}{\kern0pt}{\isacharprime}{\kern0pt}}, we can finally 
           conclude that \isa{c} must also be a successor configuration of the local
           parallelism statement \isa{CO\ S\isactrlsub {\isadigit{1}}\ {\isasymparallel}\ S\isactrlsub {\isadigit{2}}\ OC}, which needed to be proven in the 
           first~place.%
}\isanewline
\ \ \ \ \ \ \isacommand{qed}\isamarkupfalse%
\isanewline
\ \ \ \ \isacommand{qed}\isamarkupfalse%
\isanewline
\ \ \isacommand{qed}\isamarkupfalse%
\endisatagproof
{\isafoldproof}%
\isadelimproof
\isanewline
\endisadelimproof
\isanewline
\ \ \isacommand{lemma}\isamarkupfalse%
\ {\isasymdelta}\isactrlsub s{\isacharunderscore}{\kern0pt}LocPar{\isacharcolon}{\kern0pt}\isanewline
\ \ \ \ \isakeyword{assumes}\ {\isachardoublequoteopen}concrete\isactrlsub {\isasymT}{\isacharparenleft}{\kern0pt}sh\ {\isasymleadsto}\ State{\isasymllangle}{\isasymsigma}{\isasymrrangle}{\isacharparenright}{\kern0pt}{\isachardoublequoteclose}\isanewline
\ \ \ \ \isakeyword{shows}\ {\isachardoublequoteopen}{\isasymdelta}\isactrlsub s\ {\isacharparenleft}{\kern0pt}sh\ {\isasymleadsto}\ State{\isasymllangle}{\isasymsigma}{\isasymrrangle}{\isacharcomma}{\kern0pt}\ {\isasymlambda}{\isacharbrackleft}{\kern0pt}CO\ S\isactrlsub {\isadigit{1}}\ {\isasymparallel}\ S\isactrlsub {\isadigit{2}}\ OC{\isacharbrackright}{\kern0pt}{\isacharparenright}{\kern0pt}\ {\isacharequal}{\kern0pt}\ {\isacharparenleft}{\kern0pt}{\isacharparenleft}{\kern0pt}{\isacharpercent}{\kern0pt}c{\isachardot}{\kern0pt}\ {\isacharparenleft}{\kern0pt}fst{\isacharparenleft}{\kern0pt}c{\isacharparenright}{\kern0pt}{\isacharcomma}{\kern0pt}\ parallel\ {\isacharparenleft}{\kern0pt}snd\ c{\isacharparenright}{\kern0pt}\ {\isasymlambda}{\isacharbrackleft}{\kern0pt}S\isactrlsub {\isadigit{2}}{\isacharbrackright}{\kern0pt}{\isacharparenright}{\kern0pt}{\isacharparenright}{\kern0pt}\ {\isacharbackquote}{\kern0pt}\ {\isasymdelta}\isactrlsub s\ {\isacharparenleft}{\kern0pt}sh\ {\isasymleadsto}\ State{\isasymllangle}{\isasymsigma}{\isasymrrangle}{\isacharcomma}{\kern0pt}\ {\isasymlambda}{\isacharbrackleft}{\kern0pt}S\isactrlsub {\isadigit{1}}{\isacharbrackright}{\kern0pt}{\isacharparenright}{\kern0pt}{\isacharparenright}{\kern0pt}\ {\isasymunion}\ {\isacharparenleft}{\kern0pt}{\isacharparenleft}{\kern0pt}{\isacharpercent}{\kern0pt}c{\isachardot}{\kern0pt}\ {\isacharparenleft}{\kern0pt}fst{\isacharparenleft}{\kern0pt}c{\isacharparenright}{\kern0pt}{\isacharcomma}{\kern0pt}\ parallel\ {\isasymlambda}{\isacharbrackleft}{\kern0pt}S\isactrlsub {\isadigit{1}}{\isacharbrackright}{\kern0pt}\ {\isacharparenleft}{\kern0pt}snd\ c{\isacharparenright}{\kern0pt}{\isacharparenright}{\kern0pt}{\isacharparenright}{\kern0pt}\ {\isacharbackquote}{\kern0pt}\ {\isasymdelta}\isactrlsub s\ {\isacharparenleft}{\kern0pt}sh\ {\isasymleadsto}\ State{\isasymllangle}{\isasymsigma}{\isasymrrangle}{\isacharcomma}{\kern0pt}\ {\isasymlambda}{\isacharbrackleft}{\kern0pt}S\isactrlsub {\isadigit{2}}{\isacharbrackright}{\kern0pt}{\isacharparenright}{\kern0pt}{\isacharparenright}{\kern0pt}{\isachardoublequoteclose}\isanewline
\isadelimproof
\ \ \ \ %
\endisadelimproof
\isatagproof
\isacommand{apply}\isamarkupfalse%
\ {\isacharparenleft}{\kern0pt}subst\ set{\isacharunderscore}{\kern0pt}eq{\isacharunderscore}{\kern0pt}subset{\isacharparenright}{\kern0pt}\isanewline
\ \ \ \ \isacommand{using}\isamarkupfalse%
\ assms\ {\isasymdelta}\isactrlsub s{\isacharunderscore}{\kern0pt}LocPar\isactrlsub {\isadigit{1}}\ {\isasymdelta}\isactrlsub s{\isacharunderscore}{\kern0pt}LocPar\isactrlsub {\isadigit{2}}\ \isacommand{by}\isamarkupfalse%
\ auto\isanewline
\ \ \ \ %
\isamarkupcmt{We can now use the proof of both subset directions to infer the 
       desired~equality.%
}%
\endisatagproof
{\isafoldproof}%
\isadelimproof
\endisadelimproof
\begin{isamarkuptext}%
We next provide the simplification lemmas for the local memory command
  of $WL_{EXT}$. Similar to the valuation function, we choose to provide two distinct 
  rules in order to manage the scope evaluation. The first rule handles the case, in which 
  no variable declarations occur upon opening the scope, whilst the second rule 
  simplifies the evaluation of the command with existent variable~declarations.%
\end{isamarkuptext}\isamarkuptrue%
\ \ \isacommand{lemma}\isamarkupfalse%
\ {\isasymdelta}\isactrlsub s{\isacharunderscore}{\kern0pt}Scope\isactrlsub A{\isacharcolon}{\kern0pt}\isanewline
\ \ \ \ \isakeyword{assumes}\ {\isachardoublequoteopen}concrete\isactrlsub {\isasymT}{\isacharparenleft}{\kern0pt}sh\ {\isasymleadsto}\ State{\isasymllangle}{\isasymsigma}{\isasymrrangle}{\isacharparenright}{\kern0pt}{\isachardoublequoteclose}\ \isanewline
\ \ \ \ \isakeyword{shows}\ {\isachardoublequoteopen}{\isasymdelta}\isactrlsub s\ {\isacharparenleft}{\kern0pt}sh\ {\isasymleadsto}\ State{\isasymllangle}{\isasymsigma}{\isasymrrangle}{\isacharcomma}{\kern0pt}\ {\isasymlambda}{\isacharbrackleft}{\kern0pt}{\isasymlbrace}\ {\isasymnu}\ S\ {\isasymrbrace}{\isacharbrackright}{\kern0pt}{\isacharparenright}{\kern0pt}\ {\isacharequal}{\kern0pt}\ {\isasymdelta}\isactrlsub s\ {\isacharparenleft}{\kern0pt}sh\ {\isasymleadsto}\ State{\isasymllangle}{\isasymsigma}{\isasymrrangle}{\isacharcomma}{\kern0pt}\ {\isasymlambda}{\isacharbrackleft}{\kern0pt}S{\isacharbrackright}{\kern0pt}{\isacharparenright}{\kern0pt}{\isachardoublequoteclose}\isanewline
\isadelimproof
\ \ %
\endisadelimproof
\isatagproof
\isacommand{proof}\isamarkupfalse%
\ {\isacharminus}{\kern0pt}\isanewline
\ \ \ \ \isacommand{have}\isamarkupfalse%
\ {\isachardoublequoteopen}min{\isacharunderscore}{\kern0pt}conc{\isacharunderscore}{\kern0pt}map\isactrlsub {\isasymT}\ {\isacharparenleft}{\kern0pt}sh\ {\isasymleadsto}\ State{\isasymllangle}{\isasymsigma}{\isasymrrangle}{\isacharparenright}{\kern0pt}\ {\isadigit{0}}\ {\isacharequal}{\kern0pt}\ {\isasymcircle}{\isachardoublequoteclose}\isanewline
\ \ \ \ \ \ \isacommand{using}\isamarkupfalse%
\ assms\ min{\isacharunderscore}{\kern0pt}conc{\isacharunderscore}{\kern0pt}map{\isacharunderscore}{\kern0pt}of{\isacharunderscore}{\kern0pt}concrete\isactrlsub {\isasymT}\ \isacommand{by}\isamarkupfalse%
\ presburger\isanewline
\ \ \ \ %
\isamarkupcmt{Considering that \isa{{\isacharparenleft}{\kern0pt}sh\ {\isasymleadsto}\ {\isasymsigma}{\isacharparenright}{\kern0pt}} is concrete, it contains no symbolic variables.
       Using the \isa{min{\isacharunderscore}{\kern0pt}conc{\isacharunderscore}{\kern0pt}map{\isacharunderscore}{\kern0pt}of{\isacharunderscore}{\kern0pt}concrete\isactrlsub {\isasymT}} theorem, we can then infer that the minimal 
       concretization mapping of \isa{{\isacharparenleft}{\kern0pt}sh\ {\isasymleadsto}\ {\isasymsigma}{\isacharparenright}{\kern0pt}} matches the empty state.%
}\isanewline
\ \ \ \ \isacommand{hence}\isamarkupfalse%
\ {\isachardoublequoteopen}trace{\isacharunderscore}{\kern0pt}conc\ {\isacharparenleft}{\kern0pt}min{\isacharunderscore}{\kern0pt}conc{\isacharunderscore}{\kern0pt}map\isactrlsub {\isasymT}\ {\isacharparenleft}{\kern0pt}sh\ {\isasymleadsto}\ State{\isasymllangle}{\isasymsigma}{\isasymrrangle}{\isacharparenright}{\kern0pt}\ {\isadigit{0}}{\isacharparenright}{\kern0pt}\ {\isacharparenleft}{\kern0pt}sh\ {\isasymleadsto}\ State{\isasymllangle}{\isasymsigma}{\isasymrrangle}{\isacharparenright}{\kern0pt}\ {\isacharequal}{\kern0pt}\ sh\ {\isasymleadsto}\ State{\isasymllangle}{\isasymsigma}{\isasymrrangle}{\isachardoublequoteclose}\isanewline
\ \ \ \ \ \ \isacommand{using}\isamarkupfalse%
\ assms\ trace{\isacharunderscore}{\kern0pt}conc{\isacharunderscore}{\kern0pt}pr\ \isacommand{by}\isamarkupfalse%
\ fastforce\isanewline
\ \ \ \ %
\isamarkupcmt{Utilizing a supplementary theorem, we conclude that the trace concretization of 
       \isa{{\isacharparenleft}{\kern0pt}sh\ {\isasymleadsto}\ {\isasymsigma}{\isacharparenright}{\kern0pt}} under its minimal concretization mapping must be \isa{{\isacharparenleft}{\kern0pt}sh\ {\isasymleadsto}\ {\isasymsigma}{\isacharparenright}{\kern0pt}} itself.%
}\isanewline
\ \ \ \ \isacommand{thus}\isamarkupfalse%
\ {\isacharquery}{\kern0pt}thesis\ \isacommand{by}\isamarkupfalse%
\ auto\isanewline
\ \ \ \ %
\isamarkupcmt{We can now use Isabelle in order to infer the conclusion of this lemma.%
}\isanewline
\ \ \isacommand{qed}\isamarkupfalse%
\endisatagproof
{\isafoldproof}%
\isadelimproof
\isanewline
\endisadelimproof
\isanewline
\ \ \isacommand{lemma}\isamarkupfalse%
\ {\isasymdelta}\isactrlsub s{\isacharunderscore}{\kern0pt}Scope\isactrlsub B{\isacharcolon}{\kern0pt}\isanewline
\ \ \ \ \isakeyword{assumes}\ {\isachardoublequoteopen}concrete\isactrlsub {\isasymT}{\isacharparenleft}{\kern0pt}sh\ {\isasymleadsto}\ State{\isasymllangle}{\isasymsigma}{\isasymrrangle}{\isacharparenright}{\kern0pt}{\isachardoublequoteclose}\ \isanewline
\ \ \ \ \isakeyword{shows}\ {\isachardoublequoteopen}{\isasymdelta}\isactrlsub s\ {\isacharparenleft}{\kern0pt}sh\ {\isasymleadsto}\ State{\isasymllangle}{\isasymsigma}{\isasymrrangle}{\isacharcomma}{\kern0pt}\ {\isasymlambda}{\isacharbrackleft}{\kern0pt}{\isasymlbrace}\ {\isacharparenleft}{\kern0pt}x{\isacharsemicolon}{\kern0pt}d{\isacharparenright}{\kern0pt}\ S\ {\isasymrbrace}{\isacharbrackright}{\kern0pt}{\isacharparenright}{\kern0pt}\ {\isacharequal}{\kern0pt}\ {\isacharbraceleft}{\kern0pt}{\isacharparenleft}{\kern0pt}{\isacharparenleft}{\kern0pt}sh\ {\isasymleadsto}\ State{\isasymllangle}{\isasymsigma}{\isasymrrangle}{\isacharparenright}{\kern0pt}\ {\isasymleadsto}\ State{\isasymllangle}{\isacharbrackleft}{\kern0pt}{\isacharparenleft}{\kern0pt}vargen\ {\isasymsigma}\ {\isadigit{0}}\ {\isadigit{1}}{\isadigit{0}}{\isadigit{0}}\ {\isacharparenleft}{\kern0pt}{\isacharprime}{\kern0pt}{\isacharprime}{\kern0pt}{\isachardollar}{\kern0pt}{\isacharprime}{\kern0pt}{\isacharprime}{\kern0pt}\ {\isacharat}{\kern0pt}\ x\ {\isacharat}{\kern0pt}\ {\isacharprime}{\kern0pt}{\isacharprime}{\kern0pt}{\isacharcolon}{\kern0pt}{\isacharcolon}{\kern0pt}Scope{\isacharprime}{\kern0pt}{\isacharprime}{\kern0pt}{\isacharparenright}{\kern0pt}{\isacharparenright}{\kern0pt}\ {\isasymlongmapsto}\ Exp\ {\isacharparenleft}{\kern0pt}Num\ {\isadigit{0}}{\isacharparenright}{\kern0pt}{\isacharbrackright}{\kern0pt}\ {\isasymsigma}{\isasymrrangle}{\isacharcomma}{\kern0pt}\ {\isasymlambda}{\isacharbrackleft}{\kern0pt}{\isasymlbrace}\ d\ S\ {\isasymrbrace}\ {\isacharbrackleft}{\kern0pt}x\ {\isasymleftarrow}\isactrlsub s\ {\isacharparenleft}{\kern0pt}vargen\ {\isasymsigma}\ {\isadigit{0}}\ {\isadigit{1}}{\isadigit{0}}{\isadigit{0}}\ {\isacharparenleft}{\kern0pt}{\isacharprime}{\kern0pt}{\isacharprime}{\kern0pt}{\isachardollar}{\kern0pt}{\isacharprime}{\kern0pt}{\isacharprime}{\kern0pt}\ {\isacharat}{\kern0pt}\ x\ {\isacharat}{\kern0pt}\ {\isacharprime}{\kern0pt}{\isacharprime}{\kern0pt}{\isacharcolon}{\kern0pt}{\isacharcolon}{\kern0pt}Scope{\isacharprime}{\kern0pt}{\isacharprime}{\kern0pt}{\isacharparenright}{\kern0pt}{\isacharparenright}{\kern0pt}{\isacharbrackright}{\kern0pt}\ {\isacharbrackright}{\kern0pt}{\isacharparenright}{\kern0pt}{\isacharbraceright}{\kern0pt}{\isachardoublequoteclose}\isanewline
\isadelimproof
\ \ %
\endisadelimproof
\isatagproof
\isacommand{proof}\isamarkupfalse%
\ {\isacharminus}{\kern0pt}\isanewline
\ \ \ \ \isacommand{have}\isamarkupfalse%
\ {\isachardoublequoteopen}concrete\isactrlsub {\isasymT}\ {\isacharparenleft}{\kern0pt}{\isacharparenleft}{\kern0pt}sh\ {\isasymleadsto}\ State{\isasymllangle}{\isasymsigma}{\isasymrrangle}{\isacharparenright}{\kern0pt}\ {\isasymleadsto}\ State{\isasymllangle}{\isacharbrackleft}{\kern0pt}{\isacharparenleft}{\kern0pt}vargen\ {\isasymsigma}\ {\isadigit{0}}\ {\isadigit{1}}{\isadigit{0}}{\isadigit{0}}\ {\isacharparenleft}{\kern0pt}{\isacharprime}{\kern0pt}{\isacharprime}{\kern0pt}{\isachardollar}{\kern0pt}{\isacharprime}{\kern0pt}{\isacharprime}{\kern0pt}\ {\isacharat}{\kern0pt}\ x\ {\isacharat}{\kern0pt}\ {\isacharprime}{\kern0pt}{\isacharprime}{\kern0pt}{\isacharcolon}{\kern0pt}{\isacharcolon}{\kern0pt}Scope{\isacharprime}{\kern0pt}{\isacharprime}{\kern0pt}{\isacharparenright}{\kern0pt}{\isacharparenright}{\kern0pt}\ {\isasymlongmapsto}\ Exp\ {\isacharparenleft}{\kern0pt}Num\ {\isadigit{0}}{\isacharparenright}{\kern0pt}{\isacharbrackright}{\kern0pt}\ {\isasymsigma}{\isasymrrangle}{\isacharparenright}{\kern0pt}{\isachardoublequoteclose}\isanewline
\ \ \ \ \ \ \isacommand{using}\isamarkupfalse%
\ assms\ \isacommand{by}\isamarkupfalse%
\ simp\isanewline
\ \ \ \ %
\isamarkupcmt{Due to the concreteness of \isa{{\isacharparenleft}{\kern0pt}sh\ {\isasymleadsto}\ {\isasymsigma}{\isacharparenright}{\kern0pt}}, and the fact that 0 is a concrete 
       arithmetic numeral, we conclude that the symbolic trace generated by the 
       scope statement must be of concrete nature.%
}\isanewline
\ \ \ \ \isacommand{moreover}\isamarkupfalse%
\ \isacommand{hence}\isamarkupfalse%
\ {\isachardoublequoteopen}min{\isacharunderscore}{\kern0pt}conc{\isacharunderscore}{\kern0pt}map\isactrlsub {\isasymT}\ {\isacharparenleft}{\kern0pt}{\isacharparenleft}{\kern0pt}sh\ {\isasymleadsto}\ State{\isasymllangle}{\isasymsigma}{\isasymrrangle}{\isacharparenright}{\kern0pt}\ {\isasymleadsto}\ State{\isasymllangle}{\isacharbrackleft}{\kern0pt}{\isacharparenleft}{\kern0pt}vargen\ {\isasymsigma}\ {\isadigit{0}}\ {\isadigit{1}}{\isadigit{0}}{\isadigit{0}}\ {\isacharparenleft}{\kern0pt}{\isacharprime}{\kern0pt}{\isacharprime}{\kern0pt}{\isachardollar}{\kern0pt}{\isacharprime}{\kern0pt}{\isacharprime}{\kern0pt}\ {\isacharat}{\kern0pt}\ x\ {\isacharat}{\kern0pt}\ {\isacharprime}{\kern0pt}{\isacharprime}{\kern0pt}{\isacharcolon}{\kern0pt}{\isacharcolon}{\kern0pt}Scope{\isacharprime}{\kern0pt}{\isacharprime}{\kern0pt}{\isacharparenright}{\kern0pt}{\isacharparenright}{\kern0pt}\ {\isasymlongmapsto}\ Exp\ {\isacharparenleft}{\kern0pt}Num\ {\isadigit{0}}{\isacharparenright}{\kern0pt}{\isacharbrackright}{\kern0pt}\ {\isasymsigma}{\isasymrrangle}{\isacharparenright}{\kern0pt}\ {\isadigit{0}}\ {\isacharequal}{\kern0pt}\ {\isasymcircle}{\isachardoublequoteclose}\isanewline
\ \ \ \ \ \ \isacommand{using}\isamarkupfalse%
\ assms\ min{\isacharunderscore}{\kern0pt}conc{\isacharunderscore}{\kern0pt}map{\isacharunderscore}{\kern0pt}of{\isacharunderscore}{\kern0pt}concrete\isactrlsub {\isasymT}\ \isacommand{by}\isamarkupfalse%
\ presburger\isanewline
\ \ \ \ %
\isamarkupcmt{Considering that the computed trace is concrete, it contains no symbolic variables.
       Using the \isa{min{\isacharunderscore}{\kern0pt}conc{\isacharunderscore}{\kern0pt}map{\isacharunderscore}{\kern0pt}of{\isacharunderscore}{\kern0pt}concrete\isactrlsub {\isasymT}} theorem, we can hence infer that the minimal 
       concretization mapping must match the empty state.%
}\isanewline
\ \ \ \ \isacommand{ultimately}\isamarkupfalse%
\ \isacommand{have}\isamarkupfalse%
\ {\isachardoublequoteopen}trace{\isacharunderscore}{\kern0pt}conc\ {\isacharparenleft}{\kern0pt}min{\isacharunderscore}{\kern0pt}conc{\isacharunderscore}{\kern0pt}map\isactrlsub {\isasymT}\ {\isacharparenleft}{\kern0pt}{\isacharparenleft}{\kern0pt}sh\ {\isasymleadsto}\ State{\isasymllangle}{\isasymsigma}{\isasymrrangle}{\isacharparenright}{\kern0pt}\ {\isasymleadsto}\ State{\isasymllangle}{\isacharbrackleft}{\kern0pt}{\isacharparenleft}{\kern0pt}vargen\ {\isasymsigma}\ {\isadigit{0}}\ {\isadigit{1}}{\isadigit{0}}{\isadigit{0}}\ {\isacharparenleft}{\kern0pt}{\isacharprime}{\kern0pt}{\isacharprime}{\kern0pt}{\isachardollar}{\kern0pt}{\isacharprime}{\kern0pt}{\isacharprime}{\kern0pt}\ {\isacharat}{\kern0pt}\ x\ {\isacharat}{\kern0pt}\ {\isacharprime}{\kern0pt}{\isacharprime}{\kern0pt}{\isacharcolon}{\kern0pt}{\isacharcolon}{\kern0pt}Scope{\isacharprime}{\kern0pt}{\isacharprime}{\kern0pt}{\isacharparenright}{\kern0pt}{\isacharparenright}{\kern0pt}\ {\isasymlongmapsto}\ Exp\ {\isacharparenleft}{\kern0pt}Num\ {\isadigit{0}}{\isacharparenright}{\kern0pt}{\isacharbrackright}{\kern0pt}\ {\isasymsigma}{\isasymrrangle}{\isacharparenright}{\kern0pt}\ {\isadigit{0}}{\isacharparenright}{\kern0pt}\ {\isacharparenleft}{\kern0pt}{\isacharparenleft}{\kern0pt}sh\ {\isasymleadsto}\ State{\isasymllangle}{\isasymsigma}{\isasymrrangle}{\isacharparenright}{\kern0pt}\ {\isasymleadsto}\ State{\isasymllangle}{\isacharbrackleft}{\kern0pt}{\isacharparenleft}{\kern0pt}vargen\ {\isasymsigma}\ {\isadigit{0}}\ {\isadigit{1}}{\isadigit{0}}{\isadigit{0}}\ {\isacharparenleft}{\kern0pt}{\isacharprime}{\kern0pt}{\isacharprime}{\kern0pt}{\isachardollar}{\kern0pt}{\isacharprime}{\kern0pt}{\isacharprime}{\kern0pt}\ {\isacharat}{\kern0pt}\ x\ {\isacharat}{\kern0pt}\ {\isacharprime}{\kern0pt}{\isacharprime}{\kern0pt}{\isacharcolon}{\kern0pt}{\isacharcolon}{\kern0pt}Scope{\isacharprime}{\kern0pt}{\isacharprime}{\kern0pt}{\isacharparenright}{\kern0pt}{\isacharparenright}{\kern0pt}\ {\isasymlongmapsto}\ Exp\ {\isacharparenleft}{\kern0pt}Num\ {\isadigit{0}}{\isacharparenright}{\kern0pt}{\isacharbrackright}{\kern0pt}\ {\isasymsigma}{\isasymrrangle}{\isacharparenright}{\kern0pt}\ {\isacharequal}{\kern0pt}\ {\isacharparenleft}{\kern0pt}sh\ {\isasymleadsto}\ State{\isasymllangle}{\isasymsigma}{\isasymrrangle}{\isacharparenright}{\kern0pt}\ {\isasymleadsto}\ State{\isasymllangle}{\isacharbrackleft}{\kern0pt}{\isacharparenleft}{\kern0pt}vargen\ {\isasymsigma}\ {\isadigit{0}}\ {\isadigit{1}}{\isadigit{0}}{\isadigit{0}}\ {\isacharparenleft}{\kern0pt}{\isacharprime}{\kern0pt}{\isacharprime}{\kern0pt}{\isachardollar}{\kern0pt}{\isacharprime}{\kern0pt}{\isacharprime}{\kern0pt}\ {\isacharat}{\kern0pt}\ x\ {\isacharat}{\kern0pt}\ {\isacharprime}{\kern0pt}{\isacharprime}{\kern0pt}{\isacharcolon}{\kern0pt}{\isacharcolon}{\kern0pt}Scope{\isacharprime}{\kern0pt}{\isacharprime}{\kern0pt}{\isacharparenright}{\kern0pt}{\isacharparenright}{\kern0pt}\ {\isasymlongmapsto}\ Exp\ {\isacharparenleft}{\kern0pt}Num\ {\isadigit{0}}{\isacharparenright}{\kern0pt}{\isacharbrackright}{\kern0pt}\ {\isasymsigma}{\isasymrrangle}{\isachardoublequoteclose}\isanewline
\ \ \ \ \ \ \isacommand{using}\isamarkupfalse%
\ assms\ trace{\isacharunderscore}{\kern0pt}conc{\isacharunderscore}{\kern0pt}pr\ \isacommand{by}\isamarkupfalse%
\ fastforce\isanewline
\ \ \ \ %
\isamarkupcmt{Utilizing a supplementary theorem, we conclude that the trace concretization of 
       the computed trace under its minimal concretization mapping must be the
       computed trace itself.%
}\isanewline
\ \ \ \ \isacommand{thus}\isamarkupfalse%
\ {\isacharquery}{\kern0pt}thesis\ \isacommand{by}\isamarkupfalse%
\ auto\isanewline
\ \ \ \ %
\isamarkupcmt{We can now use Isabelle in order to infer the conclusion of this lemma.%
}\isanewline
\ \ \isacommand{qed}\isamarkupfalse%
\endisatagproof
{\isafoldproof}%
\isadelimproof
\endisadelimproof
\begin{isamarkuptext}%
We now desire to setup the simplification lemma for the input statement, which
  is the only $WL_{EXT}$ statement that introduces symbolic variables. Hence, it is
  the only statement in our model, for which the trace concretization is not trivially
  computable. However, note that our lemma does not simplify the actual trace 
  concretization procedure in order to keep the model as modular as possible. During
  the proof automization, the concretization will therefore need to be handled by 
  separate~lemmas.%
\end{isamarkuptext}\isamarkuptrue%
\ \ \isacommand{lemma}\isamarkupfalse%
\ {\isasymdelta}\isactrlsub s{\isacharunderscore}{\kern0pt}Input{\isacharcolon}{\kern0pt}\isanewline
\ \ \ \ \isakeyword{assumes}\ {\isachardoublequoteopen}concrete\isactrlsub {\isasymT}{\isacharparenleft}{\kern0pt}sh\ {\isasymleadsto}\ State{\isasymllangle}{\isasymsigma}{\isasymrrangle}{\isacharparenright}{\kern0pt}\ {\isasymand}\ x\ {\isasymnoteq}\ {\isacharparenleft}{\kern0pt}vargen\ {\isasymsigma}\ {\isadigit{0}}\ {\isadigit{1}}{\isadigit{0}}{\isadigit{0}}\ {\isacharparenleft}{\kern0pt}{\isacharprime}{\kern0pt}{\isacharprime}{\kern0pt}{\isachardollar}{\kern0pt}{\isacharprime}{\kern0pt}{\isacharprime}{\kern0pt}\ {\isacharat}{\kern0pt}\ x\ {\isacharat}{\kern0pt}\ {\isacharprime}{\kern0pt}{\isacharprime}{\kern0pt}{\isacharcolon}{\kern0pt}{\isacharcolon}{\kern0pt}Input{\isacharprime}{\kern0pt}{\isacharprime}{\kern0pt}{\isacharparenright}{\kern0pt}{\isacharparenright}{\kern0pt}{\isachardoublequoteclose}\ \isanewline
\ \ \ \ \isakeyword{shows}\ {\isachardoublequoteopen}{\isasymdelta}\isactrlsub s\ {\isacharparenleft}{\kern0pt}sh\ {\isasymleadsto}\ State{\isasymllangle}{\isasymsigma}{\isasymrrangle}{\isacharcomma}{\kern0pt}\ {\isasymlambda}{\isacharbrackleft}{\kern0pt}INPUT\ x{\isacharbrackright}{\kern0pt}{\isacharparenright}{\kern0pt}\ {\isacharequal}{\kern0pt}\ {\isacharbraceleft}{\kern0pt}{\isacharparenleft}{\kern0pt}trace{\isacharunderscore}{\kern0pt}conc\ {\isacharparenleft}{\kern0pt}{\isacharbrackleft}{\kern0pt}{\isacharparenleft}{\kern0pt}vargen\ {\isasymsigma}\ {\isadigit{0}}\ {\isadigit{1}}{\isadigit{0}}{\isadigit{0}}\ {\isacharparenleft}{\kern0pt}{\isacharprime}{\kern0pt}{\isacharprime}{\kern0pt}{\isachardollar}{\kern0pt}{\isacharprime}{\kern0pt}{\isacharprime}{\kern0pt}\ {\isacharat}{\kern0pt}\ x\ {\isacharat}{\kern0pt}\ {\isacharprime}{\kern0pt}{\isacharprime}{\kern0pt}{\isacharcolon}{\kern0pt}{\isacharcolon}{\kern0pt}Input{\isacharprime}{\kern0pt}{\isacharprime}{\kern0pt}{\isacharparenright}{\kern0pt}{\isacharparenright}{\kern0pt}\ {\isasymlongmapsto}\ Exp\ {\isacharparenleft}{\kern0pt}Num\ {\isadigit{0}}{\isacharparenright}{\kern0pt}{\isacharbrackright}{\kern0pt}\ {\isasymcircle}{\isacharparenright}{\kern0pt}\ {\isacharparenleft}{\kern0pt}{\isacharparenleft}{\kern0pt}{\isacharparenleft}{\kern0pt}{\isacharparenleft}{\kern0pt}sh\ {\isasymleadsto}\ State{\isasymllangle}{\isasymsigma}{\isasymrrangle}{\isacharparenright}{\kern0pt}\ {\isasymleadsto}\ State{\isasymllangle}{\isacharbrackleft}{\kern0pt}x\ {\isasymlongmapsto}\ Exp\ {\isacharparenleft}{\kern0pt}Var\ {\isacharparenleft}{\kern0pt}vargen\ {\isasymsigma}\ {\isadigit{0}}\ {\isadigit{1}}{\isadigit{0}}{\isadigit{0}}\ {\isacharparenleft}{\kern0pt}{\isacharprime}{\kern0pt}{\isacharprime}{\kern0pt}{\isachardollar}{\kern0pt}{\isacharprime}{\kern0pt}{\isacharprime}{\kern0pt}\ {\isacharat}{\kern0pt}\ x\ {\isacharat}{\kern0pt}\ {\isacharprime}{\kern0pt}{\isacharprime}{\kern0pt}{\isacharcolon}{\kern0pt}{\isacharcolon}{\kern0pt}Input{\isacharprime}{\kern0pt}{\isacharprime}{\kern0pt}{\isacharparenright}{\kern0pt}{\isacharparenright}{\kern0pt}{\isacharparenright}{\kern0pt}{\isacharbrackright}{\kern0pt}\ {\isacharbrackleft}{\kern0pt}{\isacharparenleft}{\kern0pt}vargen\ {\isasymsigma}\ {\isadigit{0}}\ {\isadigit{1}}{\isadigit{0}}{\isadigit{0}}\ {\isacharparenleft}{\kern0pt}{\isacharprime}{\kern0pt}{\isacharprime}{\kern0pt}{\isachardollar}{\kern0pt}{\isacharprime}{\kern0pt}{\isacharprime}{\kern0pt}\ {\isacharat}{\kern0pt}\ x\ {\isacharat}{\kern0pt}\ {\isacharprime}{\kern0pt}{\isacharprime}{\kern0pt}{\isacharcolon}{\kern0pt}{\isacharcolon}{\kern0pt}Input{\isacharprime}{\kern0pt}{\isacharprime}{\kern0pt}{\isacharparenright}{\kern0pt}{\isacharparenright}{\kern0pt}\ {\isasymlongmapsto}\ \isactrlemph {\isacharbrackright}{\kern0pt}\ {\isasymsigma}{\isasymrrangle}{\isacharparenright}{\kern0pt}\ {\isasymleadsto}\ Event{\isasymllangle}inpEv{\isacharcomma}{\kern0pt}\ {\isacharbrackleft}{\kern0pt}A\ {\isacharparenleft}{\kern0pt}Var\ {\isacharparenleft}{\kern0pt}vargen\ {\isasymsigma}\ {\isadigit{0}}\ {\isadigit{1}}{\isadigit{0}}{\isadigit{0}}\ {\isacharparenleft}{\kern0pt}{\isacharprime}{\kern0pt}{\isacharprime}{\kern0pt}{\isachardollar}{\kern0pt}{\isacharprime}{\kern0pt}{\isacharprime}{\kern0pt}\ {\isacharat}{\kern0pt}\ x\ {\isacharat}{\kern0pt}\ {\isacharprime}{\kern0pt}{\isacharprime}{\kern0pt}{\isacharcolon}{\kern0pt}{\isacharcolon}{\kern0pt}Input{\isacharprime}{\kern0pt}{\isacharprime}{\kern0pt}{\isacharparenright}{\kern0pt}{\isacharparenright}{\kern0pt}{\isacharparenright}{\kern0pt}{\isacharbrackright}{\kern0pt}{\isasymrrangle}{\isacharparenright}{\kern0pt}\ {\isasymleadsto}\ State{\isasymllangle}{\isacharbrackleft}{\kern0pt}x\ {\isasymlongmapsto}\ Exp\ {\isacharparenleft}{\kern0pt}Var\ {\isacharparenleft}{\kern0pt}vargen\ {\isasymsigma}\ {\isadigit{0}}\ {\isadigit{1}}{\isadigit{0}}{\isadigit{0}}\ {\isacharparenleft}{\kern0pt}{\isacharprime}{\kern0pt}{\isacharprime}{\kern0pt}{\isachardollar}{\kern0pt}{\isacharprime}{\kern0pt}{\isacharprime}{\kern0pt}\ {\isacharat}{\kern0pt}\ x\ {\isacharat}{\kern0pt}\ {\isacharprime}{\kern0pt}{\isacharprime}{\kern0pt}{\isacharcolon}{\kern0pt}{\isacharcolon}{\kern0pt}Input{\isacharprime}{\kern0pt}{\isacharprime}{\kern0pt}{\isacharparenright}{\kern0pt}{\isacharparenright}{\kern0pt}{\isacharparenright}{\kern0pt}{\isacharbrackright}{\kern0pt}\ {\isacharbrackleft}{\kern0pt}{\isacharparenleft}{\kern0pt}vargen\ {\isasymsigma}\ {\isadigit{0}}\ {\isadigit{1}}{\isadigit{0}}{\isadigit{0}}\ {\isacharparenleft}{\kern0pt}{\isacharprime}{\kern0pt}{\isacharprime}{\kern0pt}{\isachardollar}{\kern0pt}{\isacharprime}{\kern0pt}{\isacharprime}{\kern0pt}\ {\isacharat}{\kern0pt}\ x\ {\isacharat}{\kern0pt}\ {\isacharprime}{\kern0pt}{\isacharprime}{\kern0pt}{\isacharcolon}{\kern0pt}{\isacharcolon}{\kern0pt}Input{\isacharprime}{\kern0pt}{\isacharprime}{\kern0pt}{\isacharparenright}{\kern0pt}{\isacharparenright}{\kern0pt}\ {\isasymlongmapsto}\ \isactrlemph {\isacharbrackright}{\kern0pt}\ {\isasymsigma}{\isasymrrangle}{\isacharparenright}{\kern0pt}{\isacharcomma}{\kern0pt}\ {\isasymlambda}{\isacharbrackleft}{\kern0pt}{\isasymnabla}{\isacharbrackright}{\kern0pt}{\isacharparenright}{\kern0pt}{\isacharbraceright}{\kern0pt}{\isachardoublequoteclose}\isanewline
\isadelimproof
\ \ %
\endisadelimproof
\isatagproof
\isacommand{proof}\isamarkupfalse%
\ {\isacharminus}{\kern0pt}\isanewline
\ \ \ \ \isacommand{have}\isamarkupfalse%
\ {\isachardoublequoteopen}min{\isacharunderscore}{\kern0pt}conc{\isacharunderscore}{\kern0pt}map\isactrlsub {\isasymSigma}{\isacharparenleft}{\kern0pt}{\isacharbrackleft}{\kern0pt}{\isacharparenleft}{\kern0pt}vargen\ {\isasymsigma}\ {\isadigit{0}}\ {\isadigit{1}}{\isadigit{0}}{\isadigit{0}}\ {\isacharparenleft}{\kern0pt}{\isacharprime}{\kern0pt}{\isacharprime}{\kern0pt}{\isachardollar}{\kern0pt}{\isacharprime}{\kern0pt}{\isacharprime}{\kern0pt}\ {\isacharat}{\kern0pt}\ x\ {\isacharat}{\kern0pt}\ {\isacharprime}{\kern0pt}{\isacharprime}{\kern0pt}{\isacharcolon}{\kern0pt}{\isacharcolon}{\kern0pt}Input{\isacharprime}{\kern0pt}{\isacharprime}{\kern0pt}{\isacharparenright}{\kern0pt}{\isacharparenright}{\kern0pt}\ {\isasymlongmapsto}\ \isactrlemph {\isacharbrackright}{\kern0pt}\ {\isasymsigma}{\isacharparenright}{\kern0pt}\ {\isadigit{0}}\ {\isacharequal}{\kern0pt}\ {\isacharbrackleft}{\kern0pt}{\isacharparenleft}{\kern0pt}vargen\ {\isasymsigma}\ {\isadigit{0}}\ {\isadigit{1}}{\isadigit{0}}{\isadigit{0}}\ {\isacharparenleft}{\kern0pt}{\isacharprime}{\kern0pt}{\isacharprime}{\kern0pt}{\isachardollar}{\kern0pt}{\isacharprime}{\kern0pt}{\isacharprime}{\kern0pt}\ {\isacharat}{\kern0pt}\ x\ {\isacharat}{\kern0pt}\ {\isacharprime}{\kern0pt}{\isacharprime}{\kern0pt}{\isacharcolon}{\kern0pt}{\isacharcolon}{\kern0pt}Input{\isacharprime}{\kern0pt}{\isacharprime}{\kern0pt}{\isacharparenright}{\kern0pt}{\isacharparenright}{\kern0pt}\ {\isasymlongmapsto}\ Exp\ {\isacharparenleft}{\kern0pt}Num\ {\isadigit{0}}{\isacharparenright}{\kern0pt}{\isacharbrackright}{\kern0pt}\ {\isasymcircle}{\isachardoublequoteclose}\isanewline
\ \ \ \ \ \ \isacommand{using}\isamarkupfalse%
\ assms\ min{\isacharunderscore}{\kern0pt}conc{\isacharunderscore}{\kern0pt}map{\isacharunderscore}{\kern0pt}of{\isacharunderscore}{\kern0pt}concrete{\isacharunderscore}{\kern0pt}upd\isactrlsub {\isasymSigma}\ \isacommand{by}\isamarkupfalse%
\ simp\isanewline
\ \ \ \ %
\isamarkupcmt{Let us first investigate state \isa{{\isasymsigma}}, which is updated by mapping a fresh
       disambiguated input variable \isa{x{\isacharprime}{\kern0pt}} onto the symbolic value. Using the 
       \isa{min{\isacharunderscore}{\kern0pt}conc{\isacharunderscore}{\kern0pt}map{\isacharunderscore}{\kern0pt}of{\isacharunderscore}{\kern0pt}concrete{\isacharunderscore}{\kern0pt}upd\isactrlsub {\isasymSigma}} lemma of the \isa{base}-theory, we
       infer that the minimal trace concretization mapping of this updated state 
       only maps \isa{x{\isacharprime}{\kern0pt}} onto the arithmetic numeral 0, implying that it is undefined 
       for any other~variable.%
}\isanewline
\ \ \ \ \isacommand{moreover}\isamarkupfalse%
\ \isacommand{hence}\isamarkupfalse%
\ {\isachardoublequoteopen}symb\isactrlsub {\isasymSigma}{\isacharparenleft}{\kern0pt}{\isacharbrackleft}{\kern0pt}{\isacharparenleft}{\kern0pt}vargen\ {\isasymsigma}\ {\isadigit{0}}\ {\isadigit{1}}{\isadigit{0}}{\isadigit{0}}\ {\isacharparenleft}{\kern0pt}{\isacharprime}{\kern0pt}{\isacharprime}{\kern0pt}{\isachardollar}{\kern0pt}{\isacharprime}{\kern0pt}{\isacharprime}{\kern0pt}\ {\isacharat}{\kern0pt}\ x\ {\isacharat}{\kern0pt}\ {\isacharprime}{\kern0pt}{\isacharprime}{\kern0pt}{\isacharcolon}{\kern0pt}{\isacharcolon}{\kern0pt}Input{\isacharprime}{\kern0pt}{\isacharprime}{\kern0pt}{\isacharparenright}{\kern0pt}{\isacharparenright}{\kern0pt}\ {\isasymlongmapsto}\ \isactrlemph {\isacharbrackright}{\kern0pt}\ {\isasymsigma}{\isacharparenright}{\kern0pt}\ {\isacharequal}{\kern0pt}\ {\isacharbraceleft}{\kern0pt}{\isacharparenleft}{\kern0pt}vargen\ {\isasymsigma}\ {\isadigit{0}}\ {\isadigit{1}}{\isadigit{0}}{\isadigit{0}}\ {\isacharparenleft}{\kern0pt}{\isacharprime}{\kern0pt}{\isacharprime}{\kern0pt}{\isachardollar}{\kern0pt}{\isacharprime}{\kern0pt}{\isacharprime}{\kern0pt}\ {\isacharat}{\kern0pt}\ x\ {\isacharat}{\kern0pt}\ {\isacharprime}{\kern0pt}{\isacharprime}{\kern0pt}{\isacharcolon}{\kern0pt}{\isacharcolon}{\kern0pt}Input{\isacharprime}{\kern0pt}{\isacharprime}{\kern0pt}{\isacharparenright}{\kern0pt}{\isacharparenright}{\kern0pt}{\isacharbraceright}{\kern0pt}{\isachardoublequoteclose}\isanewline
\ \ \ \ \ \ \isacommand{using}\isamarkupfalse%
\ assms\ symb\isactrlsub {\isasymSigma}{\isacharunderscore}{\kern0pt}def\ \isacommand{by}\isamarkupfalse%
\ auto\isanewline
\ \ \ \ %
\isamarkupcmt{Due to the concreteness of \isa{{\isasymsigma}}, the updated version of \isa{{\isasymsigma}} only contains
       one singular symbolic variable, which is \isa{x{\isacharprime}{\kern0pt}}.%
}\isanewline
\ \ \ \ \isacommand{ultimately}\isamarkupfalse%
\ \isacommand{have}\isamarkupfalse%
\ {\isachardoublequoteopen}min{\isacharunderscore}{\kern0pt}conc{\isacharunderscore}{\kern0pt}map\isactrlsub {\isasymSigma}{\isacharparenleft}{\kern0pt}{\isacharbrackleft}{\kern0pt}x\ {\isasymlongmapsto}\ Exp\ {\isacharparenleft}{\kern0pt}Var\ {\isacharparenleft}{\kern0pt}vargen\ {\isasymsigma}\ {\isadigit{0}}\ {\isadigit{1}}{\isadigit{0}}{\isadigit{0}}\ {\isacharparenleft}{\kern0pt}{\isacharprime}{\kern0pt}{\isacharprime}{\kern0pt}{\isachardollar}{\kern0pt}{\isacharprime}{\kern0pt}{\isacharprime}{\kern0pt}\ {\isacharat}{\kern0pt}\ x\ {\isacharat}{\kern0pt}\ {\isacharprime}{\kern0pt}{\isacharprime}{\kern0pt}{\isacharcolon}{\kern0pt}{\isacharcolon}{\kern0pt}Input{\isacharprime}{\kern0pt}{\isacharprime}{\kern0pt}{\isacharparenright}{\kern0pt}{\isacharparenright}{\kern0pt}{\isacharparenright}{\kern0pt}{\isacharbrackright}{\kern0pt}\ {\isacharbrackleft}{\kern0pt}{\isacharparenleft}{\kern0pt}vargen\ {\isasymsigma}\ {\isadigit{0}}\ {\isadigit{1}}{\isadigit{0}}{\isadigit{0}}\ {\isacharparenleft}{\kern0pt}{\isacharprime}{\kern0pt}{\isacharprime}{\kern0pt}{\isachardollar}{\kern0pt}{\isacharprime}{\kern0pt}{\isacharprime}{\kern0pt}\ {\isacharat}{\kern0pt}\ x\ {\isacharat}{\kern0pt}\ {\isacharprime}{\kern0pt}{\isacharprime}{\kern0pt}{\isacharcolon}{\kern0pt}{\isacharcolon}{\kern0pt}Input{\isacharprime}{\kern0pt}{\isacharprime}{\kern0pt}{\isacharparenright}{\kern0pt}{\isacharparenright}{\kern0pt}\ {\isasymlongmapsto}\ \isactrlemph {\isacharbrackright}{\kern0pt}\ {\isasymsigma}{\isacharparenright}{\kern0pt}\ {\isadigit{0}}\ {\isacharequal}{\kern0pt}\ {\isacharbrackleft}{\kern0pt}{\isacharparenleft}{\kern0pt}vargen\ {\isasymsigma}\ {\isadigit{0}}\ {\isadigit{1}}{\isadigit{0}}{\isadigit{0}}\ {\isacharparenleft}{\kern0pt}{\isacharprime}{\kern0pt}{\isacharprime}{\kern0pt}{\isachardollar}{\kern0pt}{\isacharprime}{\kern0pt}{\isacharprime}{\kern0pt}\ {\isacharat}{\kern0pt}\ x\ {\isacharat}{\kern0pt}\ {\isacharprime}{\kern0pt}{\isacharprime}{\kern0pt}{\isacharcolon}{\kern0pt}{\isacharcolon}{\kern0pt}Input{\isacharprime}{\kern0pt}{\isacharprime}{\kern0pt}{\isacharparenright}{\kern0pt}{\isacharparenright}{\kern0pt}\ {\isasymlongmapsto}\ Exp\ {\isacharparenleft}{\kern0pt}Num\ {\isadigit{0}}{\isacharparenright}{\kern0pt}{\isacharbrackright}{\kern0pt}\ {\isasymcircle}{\isachardoublequoteclose}\isanewline
\ \ \ \ \ \ \isacommand{using}\isamarkupfalse%
\ assms\ min{\isacharunderscore}{\kern0pt}conc{\isacharunderscore}{\kern0pt}map{\isacharunderscore}{\kern0pt}of{\isacharunderscore}{\kern0pt}concrete{\isacharunderscore}{\kern0pt}no{\isacharunderscore}{\kern0pt}upd\isactrlsub {\isasymSigma}\ \isacommand{by}\isamarkupfalse%
\ {\isacharparenleft}{\kern0pt}metis\ singletonD{\isacharparenright}{\kern0pt}\ \ \isanewline
\ \ \ \ %
\isamarkupcmt{Let us again update the state, such that \isa{x} maps onto \isa{x{\isacharprime}{\kern0pt}}. Then the earlier
       established minimal concretization mapping does not need to be changed in order 
       to be a valid concretization mapping for the updated state. This holds, because
       the updated variable \isa{x} is not of symbolic nature. Note that we use the
       earlier given \isa{min{\isacharunderscore}{\kern0pt}conc{\isacharunderscore}{\kern0pt}map{\isacharunderscore}{\kern0pt}of{\isacharunderscore}{\kern0pt}concrete{\isacharunderscore}{\kern0pt}no{\isacharunderscore}{\kern0pt}upd\isactrlsub {\isasymSigma}} lemma to infer this
       conclusion.%
}\isanewline
\ \ \ \ \isacommand{hence}\isamarkupfalse%
\ {\isachardoublequoteopen}min{\isacharunderscore}{\kern0pt}conc{\isacharunderscore}{\kern0pt}map\isactrlsub {\isasymT}{\isacharparenleft}{\kern0pt}{\isacharparenleft}{\kern0pt}{\isacharparenleft}{\kern0pt}{\isacharparenleft}{\kern0pt}{\isacharparenleft}{\kern0pt}sh\ {\isasymleadsto}\ State{\isasymllangle}{\isasymsigma}{\isasymrrangle}{\isacharparenright}{\kern0pt}\ {\isasymleadsto}\ State{\isasymllangle}{\isacharbrackleft}{\kern0pt}x\ {\isasymlongmapsto}\ Exp\ {\isacharparenleft}{\kern0pt}Var\ {\isacharparenleft}{\kern0pt}vargen\ {\isasymsigma}\ {\isadigit{0}}\ {\isadigit{1}}{\isadigit{0}}{\isadigit{0}}\ {\isacharparenleft}{\kern0pt}{\isacharprime}{\kern0pt}{\isacharprime}{\kern0pt}{\isachardollar}{\kern0pt}{\isacharprime}{\kern0pt}{\isacharprime}{\kern0pt}\ {\isacharat}{\kern0pt}\ x\ {\isacharat}{\kern0pt}\ {\isacharprime}{\kern0pt}{\isacharprime}{\kern0pt}{\isacharcolon}{\kern0pt}{\isacharcolon}{\kern0pt}Input{\isacharprime}{\kern0pt}{\isacharprime}{\kern0pt}{\isacharparenright}{\kern0pt}{\isacharparenright}{\kern0pt}{\isacharparenright}{\kern0pt}{\isacharbrackright}{\kern0pt}\ {\isacharbrackleft}{\kern0pt}{\isacharparenleft}{\kern0pt}vargen\ {\isasymsigma}\ {\isadigit{0}}\ {\isadigit{1}}{\isadigit{0}}{\isadigit{0}}\ {\isacharparenleft}{\kern0pt}{\isacharprime}{\kern0pt}{\isacharprime}{\kern0pt}{\isachardollar}{\kern0pt}{\isacharprime}{\kern0pt}{\isacharprime}{\kern0pt}\ {\isacharat}{\kern0pt}\ x\ {\isacharat}{\kern0pt}\ {\isacharprime}{\kern0pt}{\isacharprime}{\kern0pt}{\isacharcolon}{\kern0pt}{\isacharcolon}{\kern0pt}Input{\isacharprime}{\kern0pt}{\isacharprime}{\kern0pt}{\isacharparenright}{\kern0pt}{\isacharparenright}{\kern0pt}\ {\isasymlongmapsto}\ \isactrlemph {\isacharbrackright}{\kern0pt}\ {\isasymsigma}{\isasymrrangle}{\isacharparenright}{\kern0pt}\ {\isasymleadsto}\ Event{\isasymllangle}inpEv{\isacharcomma}{\kern0pt}\ {\isacharbrackleft}{\kern0pt}A\ {\isacharparenleft}{\kern0pt}Var\ {\isacharparenleft}{\kern0pt}vargen\ {\isasymsigma}\ {\isadigit{0}}\ {\isadigit{1}}{\isadigit{0}}{\isadigit{0}}\ {\isacharparenleft}{\kern0pt}{\isacharprime}{\kern0pt}{\isacharprime}{\kern0pt}{\isachardollar}{\kern0pt}{\isacharprime}{\kern0pt}{\isacharprime}{\kern0pt}\ {\isacharat}{\kern0pt}\ x\ {\isacharat}{\kern0pt}\ {\isacharprime}{\kern0pt}{\isacharprime}{\kern0pt}{\isacharcolon}{\kern0pt}{\isacharcolon}{\kern0pt}Input{\isacharprime}{\kern0pt}{\isacharprime}{\kern0pt}{\isacharparenright}{\kern0pt}{\isacharparenright}{\kern0pt}{\isacharparenright}{\kern0pt}{\isacharbrackright}{\kern0pt}{\isasymrrangle}{\isacharparenright}{\kern0pt}\ {\isasymleadsto}\ State{\isasymllangle}{\isacharbrackleft}{\kern0pt}x\ {\isasymlongmapsto}\ Exp\ {\isacharparenleft}{\kern0pt}Var\ {\isacharparenleft}{\kern0pt}vargen\ {\isasymsigma}\ {\isadigit{0}}\ {\isadigit{1}}{\isadigit{0}}{\isadigit{0}}\ {\isacharparenleft}{\kern0pt}{\isacharprime}{\kern0pt}{\isacharprime}{\kern0pt}{\isachardollar}{\kern0pt}{\isacharprime}{\kern0pt}{\isacharprime}{\kern0pt}\ {\isacharat}{\kern0pt}\ x\ {\isacharat}{\kern0pt}\ {\isacharprime}{\kern0pt}{\isacharprime}{\kern0pt}{\isacharcolon}{\kern0pt}{\isacharcolon}{\kern0pt}Input{\isacharprime}{\kern0pt}{\isacharprime}{\kern0pt}{\isacharparenright}{\kern0pt}{\isacharparenright}{\kern0pt}{\isacharparenright}{\kern0pt}{\isacharbrackright}{\kern0pt}\ {\isacharbrackleft}{\kern0pt}{\isacharparenleft}{\kern0pt}vargen\ {\isasymsigma}\ {\isadigit{0}}\ {\isadigit{1}}{\isadigit{0}}{\isadigit{0}}\ {\isacharparenleft}{\kern0pt}{\isacharprime}{\kern0pt}{\isacharprime}{\kern0pt}{\isachardollar}{\kern0pt}{\isacharprime}{\kern0pt}{\isacharprime}{\kern0pt}\ {\isacharat}{\kern0pt}\ x\ {\isacharat}{\kern0pt}\ {\isacharprime}{\kern0pt}{\isacharprime}{\kern0pt}{\isacharcolon}{\kern0pt}{\isacharcolon}{\kern0pt}Input{\isacharprime}{\kern0pt}{\isacharprime}{\kern0pt}{\isacharparenright}{\kern0pt}{\isacharparenright}{\kern0pt}\ {\isasymlongmapsto}\ \isactrlemph {\isacharbrackright}{\kern0pt}\ {\isasymsigma}{\isasymrrangle}{\isacharparenright}{\kern0pt}{\isacharparenright}{\kern0pt}\ {\isadigit{0}}\ {\isacharequal}{\kern0pt}\ {\isacharbrackleft}{\kern0pt}{\isacharparenleft}{\kern0pt}vargen\ {\isasymsigma}\ {\isadigit{0}}\ {\isadigit{1}}{\isadigit{0}}{\isadigit{0}}\ {\isacharparenleft}{\kern0pt}{\isacharprime}{\kern0pt}{\isacharprime}{\kern0pt}{\isachardollar}{\kern0pt}{\isacharprime}{\kern0pt}{\isacharprime}{\kern0pt}\ {\isacharat}{\kern0pt}\ x\ {\isacharat}{\kern0pt}\ {\isacharprime}{\kern0pt}{\isacharprime}{\kern0pt}{\isacharcolon}{\kern0pt}{\isacharcolon}{\kern0pt}Input{\isacharprime}{\kern0pt}{\isacharprime}{\kern0pt}{\isacharparenright}{\kern0pt}{\isacharparenright}{\kern0pt}\ {\isasymlongmapsto}\ Exp\ {\isacharparenleft}{\kern0pt}Num\ {\isadigit{0}}{\isacharparenright}{\kern0pt}{\isacharbrackright}{\kern0pt}\ {\isasymcircle}{\isachardoublequoteclose}\isanewline
\ \ \ \ \ \ \isacommand{using}\isamarkupfalse%
\ assms\ min{\isacharunderscore}{\kern0pt}conc{\isacharunderscore}{\kern0pt}map{\isacharunderscore}{\kern0pt}of{\isacharunderscore}{\kern0pt}concrete\isactrlsub {\isasymT}\ \isacommand{by}\isamarkupfalse%
\ {\isacharparenleft}{\kern0pt}smt\ {\isacharparenleft}{\kern0pt}z{\isadigit{3}}{\isacharparenright}{\kern0pt}\ fmadd{\isacharunderscore}{\kern0pt}empty{\isacharparenleft}{\kern0pt}{\isadigit{1}}{\isacharparenright}{\kern0pt}\ fmadd{\isacharunderscore}{\kern0pt}idempotent\ min{\isacharunderscore}{\kern0pt}conc{\isacharunderscore}{\kern0pt}map\isactrlsub {\isasymT}{\isachardot}{\kern0pt}simps{\isacharparenleft}{\kern0pt}{\isadigit{2}}{\isacharparenright}{\kern0pt}\ min{\isacharunderscore}{\kern0pt}conc{\isacharunderscore}{\kern0pt}map\isactrlsub {\isasymT}{\isachardot}{\kern0pt}simps{\isacharparenleft}{\kern0pt}{\isadigit{3}}{\isacharparenright}{\kern0pt}{\isacharparenright}{\kern0pt}\isanewline
\ \ \ \ %
\isamarkupcmt{The established concretization mapping must then be a valid concretization 
       mapping for the whole trace generated by the input statement.%
}\isanewline
\ \ \ \ \isacommand{thus}\isamarkupfalse%
\ {\isacharquery}{\kern0pt}thesis\ \isacommand{by}\isamarkupfalse%
\ fastforce\isanewline
\ \ \ \ %
\isamarkupcmt{We can now use Isabelle in order to infer the conclusion of this lemma.%
}\isanewline
\ \ \isacommand{qed}\isamarkupfalse%
\endisatagproof
{\isafoldproof}%
\isadelimproof
\endisadelimproof
\begin{isamarkuptext}%
Applying the $\isa{{\isasymdelta}}_s$-function on the guarded statement can again cause
  two distinct results. If the path condition containing the evaluated Boolean
  guard is consistent, the statement behind the Boolean guard is executed. 
  If the path condition containing the evaluated negated Boolean guard
  is consistent, the statement blocks, implying that no successor configuration
  exists. We formalize a simplification lemma for both~cases. \par%
\end{isamarkuptext}\isamarkuptrue%
\ \ \isacommand{lemma}\isamarkupfalse%
\ {\isasymdelta}\isactrlsub s{\isacharunderscore}{\kern0pt}Guard\isactrlsub T{\isacharcolon}{\kern0pt}\isanewline
\ \ \ \ \isakeyword{assumes}\ {\isachardoublequoteopen}consistent\ {\isacharparenleft}{\kern0pt}{\isacharbraceleft}{\kern0pt}val\isactrlsub B\ b\ {\isasymsigma}{\isacharbraceright}{\kern0pt}{\isacharparenright}{\kern0pt}\ {\isasymand}\ concrete\isactrlsub {\isasymT}{\isacharparenleft}{\kern0pt}sh\ {\isasymleadsto}\ State{\isasymllangle}{\isasymsigma}{\isasymrrangle}{\isacharparenright}{\kern0pt}{\isachardoublequoteclose}\isanewline
\ \ \ \ \isakeyword{shows}\ {\isachardoublequoteopen}{\isasymdelta}\isactrlsub s\ {\isacharparenleft}{\kern0pt}sh\ {\isasymleadsto}\ State{\isasymllangle}{\isasymsigma}{\isasymrrangle}{\isacharcomma}{\kern0pt}\ {\isasymlambda}{\isacharbrackleft}{\kern0pt}{\isacharcolon}{\kern0pt}{\isacharcolon}{\kern0pt}\ b{\isacharsemicolon}{\kern0pt}{\isacharsemicolon}{\kern0pt}\ S\ END{\isacharbrackright}{\kern0pt}{\isacharparenright}{\kern0pt}\ {\isacharequal}{\kern0pt}\ {\isacharbraceleft}{\kern0pt}{\isacharparenleft}{\kern0pt}sh\ {\isasymleadsto}\ State{\isasymllangle}{\isasymsigma}{\isasymrrangle}{\isacharcomma}{\kern0pt}\ {\isasymlambda}{\isacharbrackleft}{\kern0pt}S{\isacharbrackright}{\kern0pt}{\isacharparenright}{\kern0pt}{\isacharbraceright}{\kern0pt}{\isachardoublequoteclose}\isanewline
\isadelimproof
\ \ %
\endisadelimproof
\isatagproof
\isacommand{proof}\isamarkupfalse%
\ {\isacharminus}{\kern0pt}\isanewline
\ \ \ \ \isacommand{have}\isamarkupfalse%
\ {\isachardoublequoteopen}consistent{\isacharparenleft}{\kern0pt}sval\isactrlsub B\ {\isacharbraceleft}{\kern0pt}val\isactrlsub B\ b\ {\isasymsigma}{\isacharbraceright}{\kern0pt}\ {\isacharparenleft}{\kern0pt}min{\isacharunderscore}{\kern0pt}conc{\isacharunderscore}{\kern0pt}map\isactrlsub {\isasymT}\ {\isacharparenleft}{\kern0pt}{\isasymlangle}{\isasymsigma}{\isasymrangle}{\isacharparenright}{\kern0pt}\ {\isadigit{0}}{\isacharparenright}{\kern0pt}{\isacharparenright}{\kern0pt}{\isachardoublequoteclose}\isanewline
\ \ \ \ \ \ \isacommand{using}\isamarkupfalse%
\ assms\ consistent{\isacharunderscore}{\kern0pt}pr\ \isacommand{by}\isamarkupfalse%
\ blast\isanewline
\ \ \ \ %
\isamarkupcmt{As we have already assumed that the path condition containing b 
       (i.e. the true-case) is consistent, the consistency of the path condition must
       then be preserved if further simplified under a minimal concretization mapping.%
}\isanewline
\ \ \ \ \isacommand{moreover}\isamarkupfalse%
\ \isacommand{have}\isamarkupfalse%
\ {\isachardoublequoteopen}{\isasymnot}consistent{\isacharparenleft}{\kern0pt}sval\isactrlsub B\ {\isacharbraceleft}{\kern0pt}val\isactrlsub B\ {\isacharparenleft}{\kern0pt}not\ b{\isacharparenright}{\kern0pt}\ {\isasymsigma}{\isacharbraceright}{\kern0pt}\ {\isacharparenleft}{\kern0pt}min{\isacharunderscore}{\kern0pt}conc{\isacharunderscore}{\kern0pt}map\isactrlsub {\isasymT}\ {\isacharparenleft}{\kern0pt}{\isasymlangle}{\isasymsigma}{\isasymrangle}{\isacharparenright}{\kern0pt}\ {\isadigit{0}}{\isacharparenright}{\kern0pt}{\isacharparenright}{\kern0pt}{\isachardoublequoteclose}\isanewline
\ \ \ \ \ \ \isacommand{using}\isamarkupfalse%
\ assms\ \isacommand{by}\isamarkupfalse%
\ simp\isanewline
\ \ \ \ %
\isamarkupcmt{Hence, the path condition of the continuation trace corresponding to
       the false-case cannot~be~consistent.%
}\isanewline
\ \ \ \ \isacommand{ultimately}\isamarkupfalse%
\ \isacommand{have}\isamarkupfalse%
\ {\isachardoublequoteopen}{\isacharbraceleft}{\kern0pt}cont\ {\isasymin}\ {\isacharparenleft}{\kern0pt}val\isactrlsub s\ {\isacharparenleft}{\kern0pt}{\isacharcolon}{\kern0pt}{\isacharcolon}{\kern0pt}\ b{\isacharsemicolon}{\kern0pt}{\isacharsemicolon}{\kern0pt}\ S\ END{\isacharparenright}{\kern0pt}\ {\isasymsigma}{\isacharparenright}{\kern0pt}{\isachardot}{\kern0pt}\ consistent{\isacharparenleft}{\kern0pt}sval\isactrlsub B\ {\isacharparenleft}{\kern0pt}{\isasymdown}\isactrlsub p\ cont{\isacharparenright}{\kern0pt}\ {\isacharparenleft}{\kern0pt}min{\isacharunderscore}{\kern0pt}conc{\isacharunderscore}{\kern0pt}map\isactrlsub {\isasymT}\ {\isacharparenleft}{\kern0pt}{\isasymdown}\isactrlsub {\isasymtau}\ cont{\isacharparenright}{\kern0pt}\ {\isadigit{0}}{\isacharparenright}{\kern0pt}{\isacharparenright}{\kern0pt}{\isacharbraceright}{\kern0pt}\ {\isacharequal}{\kern0pt}\ {\isacharbraceleft}{\kern0pt}\ {\isacharbraceleft}{\kern0pt}val\isactrlsub B\ b\ {\isasymsigma}{\isacharbraceright}{\kern0pt}\ {\isasymtriangleright}\ {\isasymlangle}{\isasymsigma}{\isasymrangle}\ \isactrlitem \ {\isasymlambda}{\isacharbrackleft}{\kern0pt}S{\isacharbrackright}{\kern0pt}\ {\isacharbraceright}{\kern0pt}{\isachardoublequoteclose}\isanewline
\ \ \ \ \ \ \isacommand{using}\isamarkupfalse%
\ assms\ \isacommand{by}\isamarkupfalse%
\ auto\isanewline
\ \ \ \ %
\isamarkupcmt{Thus, the set of consistent continuation traces can only contain the
       continuation trace corresponding to the true-case.%
}\isanewline
\ \ \ \ \isacommand{moreover}\isamarkupfalse%
\ \isacommand{have}\isamarkupfalse%
\ {\isachardoublequoteopen}min{\isacharunderscore}{\kern0pt}conc{\isacharunderscore}{\kern0pt}map\isactrlsub {\isasymT}\ {\isacharparenleft}{\kern0pt}sh\ {\isasymleadsto}\ State{\isasymllangle}{\isasymsigma}{\isasymrrangle}{\isacharparenright}{\kern0pt}\ {\isadigit{0}}\ {\isacharequal}{\kern0pt}\ {\isasymcircle}{\isachardoublequoteclose}\isanewline
\ \ \ \ \ \ \isacommand{using}\isamarkupfalse%
\ assms\ min{\isacharunderscore}{\kern0pt}conc{\isacharunderscore}{\kern0pt}map{\isacharunderscore}{\kern0pt}of{\isacharunderscore}{\kern0pt}concrete\isactrlsub {\isasymT}\ \isacommand{by}\isamarkupfalse%
\ presburger\isanewline
\ \ \ \ %
\isamarkupcmt{Considering that \isa{{\isacharparenleft}{\kern0pt}sh\ {\isasymleadsto}\ {\isasymsigma}{\isacharparenright}{\kern0pt}} is concrete, it contains no symbolic variables.
       Using the \isa{min{\isacharunderscore}{\kern0pt}conc{\isacharunderscore}{\kern0pt}map{\isacharunderscore}{\kern0pt}of{\isacharunderscore}{\kern0pt}concrete\isactrlsub {\isasymT}} theorem, we can then infer that the minimal 
       concretization mapping of \isa{{\isacharparenleft}{\kern0pt}sh\ {\isasymleadsto}\ {\isasymsigma}{\isacharparenright}{\kern0pt}} matches the empty state.%
}\isanewline
\ \ \ \ \isacommand{hence}\isamarkupfalse%
\ {\isachardoublequoteopen}trace{\isacharunderscore}{\kern0pt}conc\ {\isacharparenleft}{\kern0pt}min{\isacharunderscore}{\kern0pt}conc{\isacharunderscore}{\kern0pt}map\isactrlsub {\isasymT}\ {\isacharparenleft}{\kern0pt}sh\ {\isasymleadsto}\ State{\isasymllangle}{\isasymsigma}{\isasymrrangle}{\isacharparenright}{\kern0pt}\ {\isadigit{0}}{\isacharparenright}{\kern0pt}\ {\isacharparenleft}{\kern0pt}sh\ {\isasymleadsto}\ State{\isasymllangle}{\isasymsigma}{\isasymrrangle}{\isacharparenright}{\kern0pt}\ {\isacharequal}{\kern0pt}\ sh\ {\isasymleadsto}\ State{\isasymllangle}{\isasymsigma}{\isasymrrangle}{\isachardoublequoteclose}\isanewline
\ \ \ \ \ \ \isacommand{using}\isamarkupfalse%
\ assms\ trace{\isacharunderscore}{\kern0pt}conc{\isacharunderscore}{\kern0pt}pr\ \isacommand{by}\isamarkupfalse%
\ presburger\isanewline
\ \ \ \ %
\isamarkupcmt{Utilizing a supplementary theorem, we conclude that the trace concretization of 
       \isa{{\isacharparenleft}{\kern0pt}sh\ {\isasymleadsto}\ {\isasymsigma}{\isacharparenright}{\kern0pt}} under its minimal concretization mapping must be \isa{{\isacharparenleft}{\kern0pt}sh\ {\isasymleadsto}\ {\isasymsigma}{\isacharparenright}{\kern0pt}} itself.%
}\isanewline
\ \ \ \ \isacommand{ultimately}\isamarkupfalse%
\ \isacommand{show}\isamarkupfalse%
\ {\isacharquery}{\kern0pt}thesis\ \isacommand{by}\isamarkupfalse%
\ auto\isanewline
\ \ \ \ %
\isamarkupcmt{We can now use Isabelle in order to infer the conclusion of this lemma.%
}\isanewline
\ \ \isacommand{qed}\isamarkupfalse%
\endisatagproof
{\isafoldproof}%
\isadelimproof
\isanewline
\endisadelimproof
\isanewline
\ \ \isacommand{lemma}\isamarkupfalse%
\ {\isasymdelta}\isactrlsub s{\isacharunderscore}{\kern0pt}Guard\isactrlsub F{\isacharcolon}{\kern0pt}\isanewline
\ \ \ \ \isakeyword{assumes}\ {\isachardoublequoteopen}consistent\ {\isacharbraceleft}{\kern0pt}{\isacharparenleft}{\kern0pt}val\isactrlsub B\ {\isacharparenleft}{\kern0pt}not\ b{\isacharparenright}{\kern0pt}\ {\isasymsigma}{\isacharparenright}{\kern0pt}{\isacharbraceright}{\kern0pt}\ {\isasymand}\ concrete\isactrlsub {\isasymT}{\isacharparenleft}{\kern0pt}sh\ {\isasymleadsto}\ State{\isasymllangle}{\isasymsigma}{\isasymrrangle}{\isacharparenright}{\kern0pt}{\isachardoublequoteclose}\isanewline
\ \ \ \ \isakeyword{shows}\ {\isachardoublequoteopen}{\isasymdelta}\isactrlsub s\ {\isacharparenleft}{\kern0pt}sh\ {\isasymleadsto}\ State{\isasymllangle}{\isasymsigma}{\isasymrrangle}{\isacharcomma}{\kern0pt}\ {\isasymlambda}{\isacharbrackleft}{\kern0pt}{\isacharcolon}{\kern0pt}{\isacharcolon}{\kern0pt}\ b{\isacharsemicolon}{\kern0pt}{\isacharsemicolon}{\kern0pt}\ S\ END{\isacharbrackright}{\kern0pt}{\isacharparenright}{\kern0pt}\ {\isacharequal}{\kern0pt}\ {\isacharbraceleft}{\kern0pt}{\isacharbraceright}{\kern0pt}{\isachardoublequoteclose}\isanewline
\isadelimproof
\ \ %
\endisadelimproof
\isatagproof
\isacommand{proof}\isamarkupfalse%
\ {\isacharminus}{\kern0pt}\isanewline
\ \ \ \ \isacommand{have}\isamarkupfalse%
\ {\isachardoublequoteopen}consistent{\isacharparenleft}{\kern0pt}sval\isactrlsub B\ {\isacharbraceleft}{\kern0pt}val\isactrlsub B\ {\isacharparenleft}{\kern0pt}not\ b{\isacharparenright}{\kern0pt}\ {\isasymsigma}{\isacharbraceright}{\kern0pt}\ {\isacharparenleft}{\kern0pt}min{\isacharunderscore}{\kern0pt}conc{\isacharunderscore}{\kern0pt}map\isactrlsub {\isasymT}\ {\isacharparenleft}{\kern0pt}{\isasymlangle}{\isasymsigma}{\isasymrangle}{\isacharparenright}{\kern0pt}\ {\isadigit{0}}{\isacharparenright}{\kern0pt}{\isacharparenright}{\kern0pt}{\isachardoublequoteclose}\isanewline
\ \ \ \ \ \ \isacommand{using}\isamarkupfalse%
\ assms\ consistent{\isacharunderscore}{\kern0pt}pr\ \isacommand{by}\isamarkupfalse%
\ blast\isanewline
\ \ \ \ %
\isamarkupcmt{As we have already assumed that the path condition containing \isa{not\ b} 
       (i.e. the false-case) is consistent, the consistency of the path condition must
       then be preserved if further simplified under a minimal concretization mapping.%
}\isanewline
\ \ \ \ \isacommand{moreover}\isamarkupfalse%
\ \isacommand{have}\isamarkupfalse%
\ {\isachardoublequoteopen}{\isasymnot}consistent{\isacharparenleft}{\kern0pt}sval\isactrlsub B\ {\isacharbraceleft}{\kern0pt}val\isactrlsub B\ b\ {\isasymsigma}{\isacharbraceright}{\kern0pt}\ {\isacharparenleft}{\kern0pt}min{\isacharunderscore}{\kern0pt}conc{\isacharunderscore}{\kern0pt}map\isactrlsub {\isasymT}\ {\isacharparenleft}{\kern0pt}{\isasymlangle}{\isasymsigma}{\isasymrangle}{\isacharparenright}{\kern0pt}\ {\isadigit{0}}{\isacharparenright}{\kern0pt}{\isacharparenright}{\kern0pt}{\isachardoublequoteclose}\isanewline
\ \ \ \ \ \ \isacommand{using}\isamarkupfalse%
\ assms\ conc{\isacharunderscore}{\kern0pt}pc{\isacharunderscore}{\kern0pt}pr\isactrlsub B\isactrlsub N\ consistent{\isacharunderscore}{\kern0pt}def\ s{\isacharunderscore}{\kern0pt}value{\isacharunderscore}{\kern0pt}pr\isactrlsub B\ \isanewline
\ \ \ \ \ \ \isacommand{by}\isamarkupfalse%
\ {\isacharparenleft}{\kern0pt}metis\ bexp{\isachardot}{\kern0pt}simps{\isacharparenleft}{\kern0pt}{\isadigit{1}}{\isadigit{7}}{\isacharparenright}{\kern0pt}\ singletonI\ val\isactrlsub B{\isachardot}{\kern0pt}simps{\isacharparenleft}{\kern0pt}{\isadigit{2}}{\isacharparenright}{\kern0pt}{\isacharparenright}{\kern0pt}\isanewline
\ \ \ \ %
\isamarkupcmt{Hence, the path condition corresponding to the true-case cannot be~consistent.%
}\isanewline
\ \ \ \ \isacommand{ultimately}\isamarkupfalse%
\ \isacommand{have}\isamarkupfalse%
\ {\isachardoublequoteopen}{\isacharbraceleft}{\kern0pt}cont\ {\isasymin}\ {\isacharparenleft}{\kern0pt}val\isactrlsub s\ {\isacharparenleft}{\kern0pt}{\isacharcolon}{\kern0pt}{\isacharcolon}{\kern0pt}\ b{\isacharsemicolon}{\kern0pt}{\isacharsemicolon}{\kern0pt}\ S\ END{\isacharparenright}{\kern0pt}\ {\isasymsigma}{\isacharparenright}{\kern0pt}{\isachardot}{\kern0pt}\ consistent{\isacharparenleft}{\kern0pt}sval\isactrlsub B\ {\isacharparenleft}{\kern0pt}{\isasymdown}\isactrlsub p\ cont{\isacharparenright}{\kern0pt}\ {\isacharparenleft}{\kern0pt}min{\isacharunderscore}{\kern0pt}conc{\isacharunderscore}{\kern0pt}map\isactrlsub {\isasymT}\ {\isacharparenleft}{\kern0pt}{\isasymdown}\isactrlsub {\isasymtau}\ cont{\isacharparenright}{\kern0pt}\ {\isadigit{0}}{\isacharparenright}{\kern0pt}{\isacharparenright}{\kern0pt}{\isacharbraceright}{\kern0pt}\ {\isacharequal}{\kern0pt}\ {\isacharbraceleft}{\kern0pt}{\isacharbraceright}{\kern0pt}{\isachardoublequoteclose}\isanewline
\ \ \ \ \ \ \isacommand{using}\isamarkupfalse%
\ assms\ \isacommand{by}\isamarkupfalse%
\ auto\isanewline
\ \ \ \ %
\isamarkupcmt{This implies that the set of consistent continuation traces must be empty.%
}\isanewline
\ \ \ \ \isacommand{thus}\isamarkupfalse%
\ {\isacharquery}{\kern0pt}thesis\ \isacommand{by}\isamarkupfalse%
\ simp\isanewline
\ \ \ \ %
\isamarkupcmt{Thus, no possible successor configuration can exist, concluding this lemma.%
}\isanewline
\ \ \isacommand{qed}\isamarkupfalse%
\endisatagproof
{\isafoldproof}%
\isadelimproof
\endisadelimproof
\begin{isamarkuptext}%
We also provide a straightforward simplification lemma for the call statement.%
\end{isamarkuptext}\isamarkuptrue%
\ \ \isacommand{lemma}\isamarkupfalse%
\ {\isasymdelta}\isactrlsub s{\isacharunderscore}{\kern0pt}Call{\isacharcolon}{\kern0pt}\isanewline
\ \ \ \ \isakeyword{assumes}\ {\isachardoublequoteopen}concrete\isactrlsub {\isasymT}{\isacharparenleft}{\kern0pt}sh\ {\isasymleadsto}\ State{\isasymllangle}{\isasymsigma}{\isasymrrangle}{\isacharparenright}{\kern0pt}\ {\isasymand}\ vars\isactrlsub A{\isacharparenleft}{\kern0pt}a{\isacharparenright}{\kern0pt}\ {\isasymsubseteq}\ fmdom{\isacharprime}{\kern0pt}{\isacharparenleft}{\kern0pt}{\isasymsigma}{\isacharparenright}{\kern0pt}{\isachardoublequoteclose}\ \isanewline
\ \ \ \ \isakeyword{shows}\ {\isachardoublequoteopen}{\isasymdelta}\isactrlsub s\ {\isacharparenleft}{\kern0pt}sh\ {\isasymleadsto}\ State{\isasymllangle}{\isasymsigma}{\isasymrrangle}{\isacharcomma}{\kern0pt}\ {\isasymlambda}{\isacharbrackleft}{\kern0pt}CALL\ m\ a{\isacharbrackright}{\kern0pt}{\isacharparenright}{\kern0pt}\ {\isacharequal}{\kern0pt}\ {\isacharbraceleft}{\kern0pt}{\isacharparenleft}{\kern0pt}{\isacharparenleft}{\kern0pt}{\isacharparenleft}{\kern0pt}sh\ {\isasymleadsto}\ State{\isasymllangle}{\isasymsigma}{\isasymrrangle}{\isacharparenright}{\kern0pt}\ {\isasymleadsto}\ Event{\isasymllangle}invEv{\isacharcomma}{\kern0pt}\ lval\isactrlsub E\ {\isacharbrackleft}{\kern0pt}P\ m{\isacharcomma}{\kern0pt}\ A\ a{\isacharbrackright}{\kern0pt}\ {\isasymsigma}{\isasymrrangle}{\isacharparenright}{\kern0pt}\ {\isasymleadsto}\ State{\isasymllangle}{\isasymsigma}{\isasymrrangle}{\isacharcomma}{\kern0pt}\ {\isasymlambda}{\isacharbrackleft}{\kern0pt}{\isasymnabla}{\isacharbrackright}{\kern0pt}{\isacharparenright}{\kern0pt}{\isacharbraceright}{\kern0pt}{\isachardoublequoteclose}\isanewline
\isadelimproof
\ \ %
\endisadelimproof
\isatagproof
\isacommand{proof}\isamarkupfalse%
\ {\isacharminus}{\kern0pt}\isanewline
\ \ \ \ \isacommand{have}\isamarkupfalse%
\ {\isachardoublequoteopen}concrete\isactrlsub {\isasymT}\ {\isacharparenleft}{\kern0pt}{\isacharparenleft}{\kern0pt}{\isacharparenleft}{\kern0pt}sh\ {\isasymleadsto}\ State{\isasymllangle}{\isasymsigma}{\isasymrrangle}{\isacharparenright}{\kern0pt}\ {\isasymleadsto}\ Event{\isasymllangle}invEv{\isacharcomma}{\kern0pt}\ lval\isactrlsub E\ {\isacharbrackleft}{\kern0pt}P\ m{\isacharcomma}{\kern0pt}\ A\ a{\isacharbrackright}{\kern0pt}\ {\isasymsigma}{\isasymrrangle}{\isacharparenright}{\kern0pt}\ {\isasymleadsto}\ State{\isasymllangle}{\isasymsigma}{\isasymrrangle}{\isacharparenright}{\kern0pt}{\isachardoublequoteclose}\isanewline
\ \ \ \ \ \ \isacommand{using}\isamarkupfalse%
\ assms\ l{\isacharunderscore}{\kern0pt}concrete{\isacharunderscore}{\kern0pt}imp\isactrlsub E\ concrete\isactrlsub {\isasymT}{\isachardot}{\kern0pt}simps\ vars\isactrlsub E{\isachardot}{\kern0pt}simps\ \isanewline
\ \ \ \ \ \ \isacommand{by}\isamarkupfalse%
\ {\isacharparenleft}{\kern0pt}metis\ Un{\isacharunderscore}{\kern0pt}commute\ Un{\isacharunderscore}{\kern0pt}empty{\isacharunderscore}{\kern0pt}right\ lvars\isactrlsub E{\isachardot}{\kern0pt}simps{\isacharparenright}{\kern0pt}\isanewline
\ \ \ \ %
\isamarkupcmt{Due to the premise of this lemma, \isa{{\isasymsigma}} must be of concrete nature. The event
       occurring in the generated trace must also be concrete, as its expression
       turns concrete after the simplification. Hence, we can infer the 
       concreteness of the whole generated trace.%
}\isanewline
\ \ \ \ \isacommand{moreover}\isamarkupfalse%
\ \isacommand{hence}\isamarkupfalse%
\ {\isachardoublequoteopen}min{\isacharunderscore}{\kern0pt}conc{\isacharunderscore}{\kern0pt}map\isactrlsub {\isasymT}\ {\isacharparenleft}{\kern0pt}{\isacharparenleft}{\kern0pt}{\isacharparenleft}{\kern0pt}sh\ {\isasymleadsto}\ State{\isasymllangle}{\isasymsigma}{\isasymrrangle}{\isacharparenright}{\kern0pt}\ {\isasymleadsto}\ Event{\isasymllangle}invEv{\isacharcomma}{\kern0pt}\ lval\isactrlsub E\ {\isacharbrackleft}{\kern0pt}P\ m{\isacharcomma}{\kern0pt}\ A\ a{\isacharbrackright}{\kern0pt}\ {\isasymsigma}{\isasymrrangle}{\isacharparenright}{\kern0pt}\ {\isasymleadsto}\ State{\isasymllangle}{\isasymsigma}{\isasymrrangle}{\isacharparenright}{\kern0pt}\ {\isadigit{0}}\ {\isacharequal}{\kern0pt}\ {\isasymcircle}{\isachardoublequoteclose}\isanewline
\ \ \ \ \ \ \isacommand{using}\isamarkupfalse%
\ assms\ min{\isacharunderscore}{\kern0pt}conc{\isacharunderscore}{\kern0pt}map{\isacharunderscore}{\kern0pt}of{\isacharunderscore}{\kern0pt}concrete\isactrlsub {\isasymT}\ \isacommand{by}\isamarkupfalse%
\ presburger\isanewline
\ \ \ \ %
\isamarkupcmt{Considering that the computed trace is concrete, it contains no symbolic 
       variables. Using the \isa{min{\isacharunderscore}{\kern0pt}conc{\isacharunderscore}{\kern0pt}map{\isacharunderscore}{\kern0pt}of{\isacharunderscore}{\kern0pt}concrete\isactrlsub {\isasymT}} theorem, we can then deduce that 
       the minimal concretization mapping of the computed trace matches the empty state.%
}\isanewline
\ \ \ \ \isacommand{ultimately}\isamarkupfalse%
\ \isacommand{have}\isamarkupfalse%
\ {\isachardoublequoteopen}trace{\isacharunderscore}{\kern0pt}conc\ {\isacharparenleft}{\kern0pt}min{\isacharunderscore}{\kern0pt}conc{\isacharunderscore}{\kern0pt}map\isactrlsub {\isasymT}\ {\isacharparenleft}{\kern0pt}{\isacharparenleft}{\kern0pt}{\isacharparenleft}{\kern0pt}sh\ {\isasymleadsto}\ State{\isasymllangle}{\isasymsigma}{\isasymrrangle}{\isacharparenright}{\kern0pt}\ {\isasymleadsto}\ Event{\isasymllangle}invEv{\isacharcomma}{\kern0pt}\ lval\isactrlsub E\ {\isacharbrackleft}{\kern0pt}P\ m{\isacharcomma}{\kern0pt}\ A\ a{\isacharbrackright}{\kern0pt}\ {\isasymsigma}{\isasymrrangle}{\isacharparenright}{\kern0pt}\ {\isasymleadsto}\ State{\isasymllangle}{\isasymsigma}{\isasymrrangle}{\isacharparenright}{\kern0pt}\ {\isadigit{0}}{\isacharparenright}{\kern0pt}\ {\isacharparenleft}{\kern0pt}{\isacharparenleft}{\kern0pt}{\isacharparenleft}{\kern0pt}sh\ {\isasymleadsto}\ State{\isasymllangle}{\isasymsigma}{\isasymrrangle}{\isacharparenright}{\kern0pt}\ {\isasymleadsto}\ Event{\isasymllangle}invEv{\isacharcomma}{\kern0pt}\ lval\isactrlsub E\ {\isacharbrackleft}{\kern0pt}P\ m{\isacharcomma}{\kern0pt}\ A\ a{\isacharbrackright}{\kern0pt}\ {\isasymsigma}{\isasymrrangle}{\isacharparenright}{\kern0pt}\ {\isasymleadsto}\ State{\isasymllangle}{\isasymsigma}{\isasymrrangle}{\isacharparenright}{\kern0pt}\ {\isacharequal}{\kern0pt}\ {\isacharparenleft}{\kern0pt}{\isacharparenleft}{\kern0pt}sh\ {\isasymleadsto}\ State{\isasymllangle}{\isasymsigma}{\isasymrrangle}{\isacharparenright}{\kern0pt}\ {\isasymleadsto}\ Event{\isasymllangle}invEv{\isacharcomma}{\kern0pt}\ lval\isactrlsub E\ {\isacharbrackleft}{\kern0pt}P\ m{\isacharcomma}{\kern0pt}\ A\ a{\isacharbrackright}{\kern0pt}\ {\isasymsigma}{\isasymrrangle}{\isacharparenright}{\kern0pt}\ {\isasymleadsto}\ State{\isasymllangle}{\isasymsigma}{\isasymrrangle}{\isachardoublequoteclose}\isanewline
\ \ \ \ \ \ \isacommand{using}\isamarkupfalse%
\ assms\ trace{\isacharunderscore}{\kern0pt}conc{\isacharunderscore}{\kern0pt}pr\ \isacommand{by}\isamarkupfalse%
\ presburger\isanewline
\ \ \ \ %
\isamarkupcmt{Utilizing a supplementary theorem, we conclude that the trace concretization of 
       the computed trace under its minimal concretization mapping must be the
       computed trace itself.%
}\isanewline
\ \ \ \ \isacommand{thus}\isamarkupfalse%
\ {\isacharquery}{\kern0pt}thesis\ \isacommand{by}\isamarkupfalse%
\ fastforce\isanewline
\ \ \ \ %
\isamarkupcmt{We can now use Isabelle in order to infer the conclusion of this lemma.%
}\isanewline
\ \ \isacommand{qed}\isamarkupfalse%
\endisatagproof
{\isafoldproof}%
\isadelimproof
\isanewline
\endisadelimproof
\isanewline
\ \ \isacommand{end}\isamarkupfalse%
\begin{isamarkuptext}%
Note that our formalization of the $\isa{{\isasymdelta}}_2$-function was optimized for our 
  code-generation objective. However, this decision causes a weakness in the overlying 
  proof system, as it cannot derive the results of the $\isa{{\isasymdelta}}_2$-function using its 
  automatically generated simplifications alone. The origin of this problem is the 
  \isa{Set{\isachardot}{\kern0pt}filter} function, which we use in order to filter out all tuples that are 
  violating the established notion of wellformedness. However, Isabelle cannot 
  automatically resolve this filtering process using its standard simplifications. \par
  In order to provide a solution for this problem, we therefore propose an alternative
  formalization of the $\isa{{\isasymdelta}}_2$-function, which models the filtering process of the
  tuples in a slightly different manner. Instead of using the \isa{Set{\isachardot}{\kern0pt}filter} function,
  we choose to intersect all tuples of methods and arithmetic method parameters with 
  the set of all wellformed tuples. Note that this is equivalent to the original 
  definition of the $\isa{{\isasymdelta}}_2$-function. \par
  However, we cannot replace the original formalization of the $\isa{{\isasymdelta}}_2$-function with 
  this alternative definition, as the set of all wellformed tuples of methods and 
  arithmetic expressions (i.e. one of the sets we intersect) is infinite. Hence, 
  this kind of formalization would greatly interfere with our code-generation. 
  Thus, we choose to use the standard definition for the purpose of generating 
  executable code, and this supplementary simplification for ensuring
  automated~reasoning.%
\end{isamarkuptext}\isamarkuptrue%
\ \ \isacommand{lemma}\isamarkupfalse%
\ {\isasymdelta}\isactrlsub {\isadigit{2}}{\isacharunderscore}{\kern0pt}simp{\isacharcolon}{\kern0pt}\ {\isachardoublequoteopen}{\isasymdelta}\isactrlsub {\isadigit{2}}\ M\ {\isacharparenleft}{\kern0pt}sh\ {\isasymleadsto}\ State{\isasymllangle}{\isasymsigma}{\isasymrrangle}{\isacharcomma}{\kern0pt}\ q{\isacharparenright}{\kern0pt}\ {\isacharequal}{\kern0pt}\ \isanewline
\ \ \ \ \ \ \ \ \ \ {\isacharparenleft}{\kern0pt}{\isacharpercent}{\kern0pt}{\isacharparenleft}{\kern0pt}m{\isacharcomma}{\kern0pt}\ v{\isacharparenright}{\kern0pt}{\isachardot}{\kern0pt}\ {\isacharparenleft}{\kern0pt}{\isacharparenleft}{\kern0pt}{\isacharparenleft}{\kern0pt}sh\ {\isasymcdot}\ {\isacharparenleft}{\kern0pt}gen{\isacharunderscore}{\kern0pt}event\ invREv\ {\isasymsigma}\ {\isacharbrackleft}{\kern0pt}P\ {\isacharparenleft}{\kern0pt}{\isasymUp}\isactrlsub n\ m{\isacharparenright}{\kern0pt}{\isacharcomma}{\kern0pt}\ v{\isacharbrackright}{\kern0pt}{\isacharparenright}{\kern0pt}{\isacharparenright}{\kern0pt}\ \isanewline
\ \ \ \ \ \ \ \ \ \ \ \ \ \ \ \ \ \ \ \ \ \ \ \ {\isasymleadsto}\ State{\isasymllangle}{\isacharbrackleft}{\kern0pt}{\isacharparenleft}{\kern0pt}vargen\ {\isasymsigma}\ {\isadigit{0}}\ {\isadigit{1}}{\isadigit{0}}{\isadigit{0}}\ {\isacharparenleft}{\kern0pt}{\isacharprime}{\kern0pt}{\isacharprime}{\kern0pt}{\isachardollar}{\kern0pt}{\isacharprime}{\kern0pt}{\isacharprime}{\kern0pt}\ {\isacharat}{\kern0pt}\ {\isacharparenleft}{\kern0pt}{\isasymUp}\isactrlsub n\ m{\isacharparenright}{\kern0pt}\ {\isacharat}{\kern0pt}\ {\isacharprime}{\kern0pt}{\isacharprime}{\kern0pt}{\isacharcolon}{\kern0pt}{\isacharcolon}{\kern0pt}Param{\isacharprime}{\kern0pt}{\isacharprime}{\kern0pt}{\isacharparenright}{\kern0pt}{\isacharparenright}{\kern0pt}\ {\isasymlongmapsto}\ Exp\ {\isacharparenleft}{\kern0pt}proj\isactrlsub A\ v{\isacharparenright}{\kern0pt}{\isacharbrackright}{\kern0pt}\ {\isasymsigma}{\isasymrrangle}{\isacharparenright}{\kern0pt}{\isacharcomma}{\kern0pt}\ \isanewline
\ \ \ \ \ \ \ \ \ \ \ \ \ \ \ \ \ \ \ \ q\ {\isacharplus}{\kern0pt}\ {\isacharbraceleft}{\kern0pt}{\isacharhash}{\kern0pt}\ {\isasymlambda}{\isacharbrackleft}{\kern0pt}{\isacharparenleft}{\kern0pt}{\isasymUp}\isactrlsub s\ m{\isacharparenright}{\kern0pt}\ {\isacharbrackleft}{\kern0pt}{\isacharparenleft}{\kern0pt}{\isasymUp}\isactrlsub v\ m{\isacharparenright}{\kern0pt}\ {\isasymleftarrow}\isactrlsub s\ {\isacharparenleft}{\kern0pt}vargen\ {\isasymsigma}\ {\isadigit{0}}\ {\isadigit{1}}{\isadigit{0}}{\isadigit{0}}\ {\isacharparenleft}{\kern0pt}{\isacharprime}{\kern0pt}{\isacharprime}{\kern0pt}{\isachardollar}{\kern0pt}{\isacharprime}{\kern0pt}{\isacharprime}{\kern0pt}\ {\isacharat}{\kern0pt}\ {\isacharparenleft}{\kern0pt}{\isasymUp}\isactrlsub n\ m{\isacharparenright}{\kern0pt}\ {\isacharat}{\kern0pt}\ {\isacharprime}{\kern0pt}{\isacharprime}{\kern0pt}{\isacharcolon}{\kern0pt}{\isacharcolon}{\kern0pt}Param{\isacharprime}{\kern0pt}{\isacharprime}{\kern0pt}{\isacharparenright}{\kern0pt}{\isacharparenright}{\kern0pt}{\isacharbrackright}{\kern0pt}\ {\isacharbrackright}{\kern0pt}\ {\isacharhash}{\kern0pt}{\isacharbraceright}{\kern0pt}{\isacharparenright}{\kern0pt}{\isacharparenright}{\kern0pt}\ \isanewline
\ \ \ \ \ \ \ \ \ \ {\isacharbackquote}{\kern0pt}\ {\isacharparenleft}{\kern0pt}{\isacharbraceleft}{\kern0pt}{\isacharparenleft}{\kern0pt}m{\isacharcomma}{\kern0pt}\ v{\isacharparenright}{\kern0pt}{\isachardot}{\kern0pt}\ wellformed\ {\isacharparenleft}{\kern0pt}sh\ {\isasymcdot}\ {\isacharparenleft}{\kern0pt}gen{\isacharunderscore}{\kern0pt}event\ invREv\ {\isasymsigma}\ {\isacharbrackleft}{\kern0pt}P\ {\isacharparenleft}{\kern0pt}{\isasymUp}\isactrlsub n\ m{\isacharparenright}{\kern0pt}{\isacharcomma}{\kern0pt}\ v{\isacharbrackright}{\kern0pt}{\isacharparenright}{\kern0pt}{\isacharparenright}{\kern0pt}{\isacharbraceright}{\kern0pt}\ {\isasyminter}\ {\isacharparenleft}{\kern0pt}M\ {\isasymtimes}\ params{\isacharparenleft}{\kern0pt}sh{\isacharparenright}{\kern0pt}{\isacharparenright}{\kern0pt}{\isacharparenright}{\kern0pt}{\isachardoublequoteclose}\isanewline
\isadelimproof
\ \ \ \ %
\endisadelimproof
\isatagproof
\isacommand{by}\isamarkupfalse%
\ fastforce%
\endisatagproof
{\isafoldproof}%
\isadelimproof
\endisadelimproof
\begin{isamarkuptext}%
In order to finalize the automated proof system of the \isa{{\isasymdelta}}-function, we 
  collect all necessary simplifications in a set, and call it the \isa{{\isasymdelta}}-system.
  Our proof derivation system then consists of the following lemmas: \begin{description}
  \item[\isa{{\isasymdelta}\isactrlsub s}-simplifications] These lemmas correspond to our earlier established
    simplifications for the \isa{{\isasymdelta}\isactrlsub s}-function. Remember that these lemmas ensure a major 
    speedup of our proof derivations, as we avoid dealing with the underlying
    valuation function.
  \item[\isa{{\isasymdelta}\isactrlsub {\isadigit{2}}}-simplifications] We also have to add the \isa{{\isasymdelta}\isactrlsub {\isadigit{2}}} simplification lemma
    in order to ensure that we can automatically compute function values
    of the \isa{{\isasymdelta}\isactrlsub {\isadigit{2}}}-function.
  \item[Concreteness lemmas] The concreteness lemmas fulfill two distinct
    purposes: \par
    Remember that the simplification lemmas for the \isa{{\isasymdelta}\isactrlsub s}-function are only
    usable iff we guarantee that the argument trace is of concrete
    nature. Hence, we first need to establish the concreteness of the initial 
    program state (e.g. using the \isa{{\isasymsigma}\isactrlsub I{\isacharunderscore}{\kern0pt}concrete} lemma), so as to ensure that 
    our proof derivation system is even applicable. \par
    Secondly, we provide the \isa{concrete{\isacharunderscore}{\kern0pt}upd{\isacharunderscore}{\kern0pt}pr\isactrlsub A} lemma of the base theory, so as
    to enable the prover to inductively compute the concreteness of states. Note
    that we do not add the standard simplification of the state concreteness
    notion, as this would greatly reduce our derivation speed.
  \item[Trace Concretization lemmas] Remember that the trace concretization
    procedure of the input command is not automatically resolved by the corresponding 
    \isa{{\isasymdelta}\isactrlsub s}-simplification. Hence, it becomes indispensable to add the previously proven 
    trace concretization lemmas of the base theory to our proof automation system. 
  \item[Consistency simplifications] Our proof system also needs to resolve
    the consistency notion in order to verify the consistency of path conditions
    during the trace composition. Hence, we choose to add its standard simplification
    to our \isa{{\isasymdelta}{\isacharunderscore}{\kern0pt}system}.
  \item[Supplementary lemmas] Similar to \isa{WL}, we additionally add the
  \isa{eval{\isacharminus}{\kern0pt}nat{\isacharminus}{\kern0pt}numeral} lemma, so as to ensure that the simplifier can freely swap 
  between the Suc/Zero notation and the numeral representation when inferring results 
  from the variable generation function. This is necessary, as we use the \isa{Suc} 
  constructor in its function definition, but numerals when decrementing~the~bound. \par
  We furthermore add several other simplifications: The \isa{fmap{\isacharunderscore}{\kern0pt}ext} lemma is used in
  order to add a state equality notion to our proof system. The \isa{member{\isacharunderscore}{\kern0pt}rec} 
  simplification ensures that we can derive list memberships when computing the 
  initial state (\isa{{\isasymsigma}\isactrlsub I}~\isa{S}) for a program S. The \isa{insert{\isacharunderscore}{\kern0pt}commute} and
  \isa{fmupd{\isacharunderscore}{\kern0pt}reorder{\isacharunderscore}{\kern0pt}neq} lemmas additionally guarantee that the prover can freely 
  permute state updates and set elements during the proof derivation.
  \end{description} We additionally remove the normal simplifications of the $\isa{{\isasymdelta}}_s$ 
  and $\isa{{\isasymdelta}}_2$-function, thereby ensuring that the Isabelle simplifier will later 
  select the \isa{{\isasymdelta}}-system when deriving successor~configurations. \par
  Note that there is still room for further improvements in our proof automation
  (e.g. speed-related improvements), which we propose as an idea for further work on
  this~model.%
\end{isamarkuptext}\isamarkuptrue%
\ \ \isacommand{lemmas}\isamarkupfalse%
\ {\isasymdelta}{\isacharunderscore}{\kern0pt}system\ {\isacharequal}{\kern0pt}\ \isanewline
\ \ \ \ {\isasymdelta}\isactrlsub s{\isacharunderscore}{\kern0pt}Skip\ {\isasymdelta}\isactrlsub s{\isacharunderscore}{\kern0pt}Assign\ {\isasymdelta}\isactrlsub s{\isacharunderscore}{\kern0pt}If\isactrlsub T\ {\isasymdelta}\isactrlsub s{\isacharunderscore}{\kern0pt}If\isactrlsub F\ {\isasymdelta}\isactrlsub s{\isacharunderscore}{\kern0pt}While\isactrlsub T\ {\isasymdelta}\isactrlsub s{\isacharunderscore}{\kern0pt}While\isactrlsub F\ {\isasymdelta}\isactrlsub s{\isacharunderscore}{\kern0pt}Seq\ \ %
\isamarkupcmt{\isa{{\isasymdelta}\isactrlsub s}-simplifications%
}\isanewline
\ \ \ \ {\isasymdelta}\isactrlsub s{\isacharunderscore}{\kern0pt}LocPar\ {\isasymdelta}\isactrlsub s{\isacharunderscore}{\kern0pt}Scope\isactrlsub A\ {\isasymdelta}\isactrlsub s{\isacharunderscore}{\kern0pt}Scope\isactrlsub B\ {\isasymdelta}\isactrlsub s{\isacharunderscore}{\kern0pt}Input\ {\isasymdelta}\isactrlsub s{\isacharunderscore}{\kern0pt}Guard\isactrlsub T\ {\isasymdelta}\isactrlsub s{\isacharunderscore}{\kern0pt}Guard\isactrlsub F\ {\isasymdelta}\isactrlsub s{\isacharunderscore}{\kern0pt}Call\ %
\isamarkupcmt{\isa{{\isasymdelta}\isactrlsub s}-simplifications%
}\isanewline
\ \ \ \ {\isasymdelta}\isactrlsub {\isadigit{2}}{\isacharunderscore}{\kern0pt}simp\ \ %
\isamarkupcmt{\isa{{\isasymdelta}\isactrlsub {\isadigit{2}}}-simplifications%
}\ \ \isanewline
\ \ \ \ conc{\isacharunderscore}{\kern0pt}concrete\ {\isasymsigma}\isactrlsub I{\isacharunderscore}{\kern0pt}concrete\ concrete{\isacharunderscore}{\kern0pt}upd{\isacharunderscore}{\kern0pt}pr\isactrlsub A\ %
\isamarkupcmt{Concreteness lemmas%
}\ \isanewline
\ \ \ \ fmmap{\isacharunderscore}{\kern0pt}keys{\isacharunderscore}{\kern0pt}empty\ fmmap{\isacharunderscore}{\kern0pt}keys{\isacharunderscore}{\kern0pt}conc\ %
\isamarkupcmt{Trace Concretization lemmas%
}\isanewline
\ \ \ \ fmmap{\isacharunderscore}{\kern0pt}keys{\isacharunderscore}{\kern0pt}input\ fmmap{\isacharunderscore}{\kern0pt}keys{\isacharunderscore}{\kern0pt}dom{\isacharunderscore}{\kern0pt}upd\ %
\isamarkupcmt{Trace Concretization lemmas%
}\isanewline
\ \ \ \ consistent{\isacharunderscore}{\kern0pt}def\ %
\isamarkupcmt{Consistency simplifications%
}\isanewline
\ \ \ \ eval{\isacharunderscore}{\kern0pt}nat{\isacharunderscore}{\kern0pt}numeral\ fmap{\isacharunderscore}{\kern0pt}ext\ member{\isacharunderscore}{\kern0pt}rec\ %
\isamarkupcmt{Supplementary lemmas%
}\isanewline
\ \ \ \ insert{\isacharunderscore}{\kern0pt}commute\ fmupd{\isacharunderscore}{\kern0pt}reorder{\isacharunderscore}{\kern0pt}neq\ %
\isamarkupcmt{Supplementary lemmas%
}\isanewline
\isanewline
\ \ \isacommand{declare}\isamarkupfalse%
\ basic{\isacharunderscore}{\kern0pt}successors{\isachardot}{\kern0pt}simps{\isacharbrackleft}{\kern0pt}simp\ del{\isacharbrackright}{\kern0pt}\isanewline
\ \ \isacommand{declare}\isamarkupfalse%
\ successors\isactrlsub {\isadigit{2}}{\isachardot}{\kern0pt}simps{\isacharbrackleft}{\kern0pt}simp\ del{\isacharbrackright}{\kern0pt}%
\isadelimdocument
\endisadelimdocument
\isatagdocument
\isamarkupsubsection{Global Trace Semantics%
}
\isamarkuptrue%
\isamarkupsubsubsection{Bounded Global Traces%
}
\isamarkuptrue%
\endisatagdocument
{\isafolddocument}%
\isadelimdocument
\endisadelimdocument
\begin{isamarkuptext}%
Similar to \isa{WL}, we now desire to setup the transitive closure of the
  trace composition, thus later allowing us to compute global traces in our 
  semantics. For this purpose, we first adapt the computation of n-bounded global 
  traces (\isa{{\isasymDelta}\isactrlsub N}-function), which halts the transitive closure after a certain bound 
  is reached. \par
  The formalization of this function is almost identical to the definition for \isa{WL}.
  However, note that the notion of a terminal configuration has drastically changed.
  In \isa{WL}, we were simply able to stop the transitive closure when the reached
  configuration contained an empty continuation marker, as this implied its terminal
  character. In $WL_{EXT}$ however, a configuration containing only empty
  continuation markers must not necessary be terminal, as there may still be 
  pending invocations. \par
  This in turn implies that we have to axiomatize terminal configurations of $WL_{EXT}$ 
  in a different manner. We choose to call a $WL_{EXT}$ configuration \isa{c} terminal 
  iff the set of its successor configurations is empty (i.e. \isa{{\isasymdelta}\ M\ c\ {\isacharequal}{\kern0pt}\ {\isacharbraceleft}{\kern0pt}{\isacharbraceright}{\kern0pt}}). This also 
  ensures that the processes corresponding to all called methods have already been 
  created, thus smoothly aligning with our intuition~of~termination.%
\end{isamarkuptext}\isamarkuptrue%
\ \ \isacommand{fun}\isamarkupfalse%
\isanewline
\ \ \ \ composition\isactrlsub N\ {\isacharcolon}{\kern0pt}{\isacharcolon}{\kern0pt}\ {\isachardoublequoteopen}nat\ {\isasymRightarrow}\ method\ set\ {\isasymRightarrow}\ config\ {\isasymRightarrow}\ config\ set{\isachardoublequoteclose}\ {\isacharparenleft}{\kern0pt}{\isachardoublequoteopen}{\isasymDelta}\isactrlsub N{\isachardoublequoteclose}{\isacharparenright}{\kern0pt}\ \isakeyword{where}\isanewline
\ \ \ \ {\isachardoublequoteopen}{\isasymDelta}\isactrlsub N\ {\isadigit{0}}\ M\ c\ {\isacharequal}{\kern0pt}\ {\isacharbraceleft}{\kern0pt}c{\isacharbraceright}{\kern0pt}{\isachardoublequoteclose}\ {\isacharbar}{\kern0pt}\isanewline
\ \ \ \ {\isachardoublequoteopen}{\isasymDelta}\isactrlsub N\ {\isacharparenleft}{\kern0pt}Suc\ n{\isacharparenright}{\kern0pt}\ M\ c\ {\isacharequal}{\kern0pt}\ {\isacharparenleft}{\kern0pt}if\ {\isasymdelta}\ M\ c\ {\isacharequal}{\kern0pt}\ {\isacharbraceleft}{\kern0pt}{\isacharbraceright}{\kern0pt}\ then\ {\isacharbraceleft}{\kern0pt}c{\isacharbraceright}{\kern0pt}\ else\ {\isasymUnion}{\isacharparenleft}{\kern0pt}{\isacharparenleft}{\kern0pt}{\isacharpercent}{\kern0pt}c{\isachardot}{\kern0pt}\ {\isasymDelta}\isactrlsub N\ n\ M\ c{\isacharparenright}{\kern0pt}\ {\isacharbackquote}{\kern0pt}\ {\isasymdelta}\ M\ c{\isacharparenright}{\kern0pt}{\isacharparenright}{\kern0pt}{\isachardoublequoteclose}%
\begin{isamarkuptext}%
The definition for n-bounded global traces generated from program \isa{S} called 
  in \isa{{\isasymsigma}} is now straightforward. We simply return all symbolic traces of the 
  configurations received by calling the \isa{{\isasymDelta}\isactrlsub N}-function in initial configuration 
  \isa{{\isacharparenleft}{\kern0pt}{\isasymlangle}{\isasymsigma}{\isasymrangle}{\isacharcomma}{\kern0pt}\ {\isacharbraceleft}{\kern0pt}{\isacharhash}{\kern0pt}\ {\isasymlambda}{\isacharbrackleft}{\kern0pt}S{\isacharbrackright}{\kern0pt}\ {\isacharhash}{\kern0pt}{\isacharbraceright}{\kern0pt}{\isacharparenright}{\kern0pt}} with bound n.%
\end{isamarkuptext}\isamarkuptrue%
\ \ \isacommand{fun}\isamarkupfalse%
\isanewline
\ \ \ \ Traces\isactrlsub N\ {\isacharcolon}{\kern0pt}{\isacharcolon}{\kern0pt}\ {\isachardoublequoteopen}program\ {\isasymRightarrow}\ {\isasymSigma}\ {\isasymRightarrow}\ nat\ {\isasymRightarrow}\ {\isasymT}\ set{\isachardoublequoteclose}\ {\isacharparenleft}{\kern0pt}{\isachardoublequoteopen}Tr\isactrlsub N{\isachardoublequoteclose}{\isacharparenright}{\kern0pt}\ \isakeyword{where}\isanewline
\ \ \ \ {\isachardoublequoteopen}Tr\isactrlsub N\ {\isacharparenleft}{\kern0pt}Program\ M\ {\isasymlbrace}\ S\ {\isasymrbrace}{\isacharparenright}{\kern0pt}\ {\isasymsigma}\ n\ {\isacharequal}{\kern0pt}\ fst\ {\isacharbackquote}{\kern0pt}\ {\isasymDelta}\isactrlsub N\ n\ {\isacharparenleft}{\kern0pt}set\ {\isacharparenleft}{\kern0pt}remdups\ M{\isacharparenright}{\kern0pt}{\isacharparenright}{\kern0pt}\ {\isacharparenleft}{\kern0pt}{\isasymlangle}{\isasymsigma}{\isasymrangle}{\isacharcomma}{\kern0pt}\ {\isacharbraceleft}{\kern0pt}{\isacharhash}{\kern0pt}\ {\isasymlambda}{\isacharbrackleft}{\kern0pt}S{\isacharbrackright}{\kern0pt}\ {\isacharhash}{\kern0pt}{\isacharbraceright}{\kern0pt}{\isacharparenright}{\kern0pt}{\isachardoublequoteclose}%
\isadelimdocument
\endisadelimdocument
\isatagdocument
\isamarkupsubsubsection{Unbounded Global Traces%
}
\isamarkuptrue%
\endisatagdocument
{\isafolddocument}%
\isadelimdocument
\endisadelimdocument
\begin{isamarkuptext}%
The construction of unbounded global traces (\isa{{\isasymDelta}}-function) is
  straightforward, as it is almost identical to the definition for \isa{WL}. However,
  note that we again have to adapt the formalization of terminal configurations
  in the condition of the If-case.%
\end{isamarkuptext}\isamarkuptrue%
\ \ \isacommand{partial{\isacharunderscore}{\kern0pt}function}\isamarkupfalse%
\ {\isacharparenleft}{\kern0pt}tailrec{\isacharparenright}{\kern0pt}\ composition\ {\isacharcolon}{\kern0pt}{\isacharcolon}{\kern0pt}\ {\isachardoublequoteopen}nat\ {\isasymRightarrow}\ method\ set\ {\isasymRightarrow}\ config\ {\isasymRightarrow}\ config\ set{\isachardoublequoteclose}\ {\isacharparenleft}{\kern0pt}{\isachardoublequoteopen}{\isasymDelta}{\isachardoublequoteclose}{\isacharparenright}{\kern0pt}\ \isakeyword{where}\isanewline
\ \ \ \ {\isacharbrackleft}{\kern0pt}code{\isacharbrackright}{\kern0pt}{\isacharcolon}{\kern0pt}\ {\isachardoublequoteopen}{\isasymDelta}\ n\ M\ c\ {\isacharequal}{\kern0pt}\ {\isacharparenleft}{\kern0pt}if\ {\isasymforall}c\ {\isasymin}\ {\isasymDelta}\isactrlsub N\ n\ M\ c{\isachardot}{\kern0pt}\ {\isasymdelta}\ M\ c\ {\isacharequal}{\kern0pt}\ {\isacharbraceleft}{\kern0pt}{\isacharbraceright}{\kern0pt}\ then\ {\isasymDelta}\isactrlsub N\ n\ M\ c\ else\ {\isasymDelta}\ {\isacharparenleft}{\kern0pt}n\ {\isacharplus}{\kern0pt}\ {\isadigit{1}}{\isadigit{0}}{\isadigit{0}}{\isacharparenright}{\kern0pt}\ M\ c{\isacharparenright}{\kern0pt}{\isachardoublequoteclose}%
\begin{isamarkuptext}%
We can now compute all unbounded global traces for any program \isa{S} called 
  in a given state \isa{{\isasymsigma}}. We simply return all symbolic traces of the configurations 
  received by calling the \isa{{\isasymDelta}}-function in initial configuration \isa{{\isacharparenleft}{\kern0pt}{\isasymlangle}{\isasymsigma}{\isasymrangle}{\isacharcomma}{\kern0pt}\ {\isacharbraceleft}{\kern0pt}{\isacharhash}{\kern0pt}\ {\isasymlambda}{\isacharbrackleft}{\kern0pt}S{\isacharbrackright}{\kern0pt}\ {\isacharhash}{\kern0pt}{\isacharbraceright}{\kern0pt}{\isacharparenright}{\kern0pt}}. 
  Note that this again requires the \isa{{\isasymDelta}}-function to converge in a
  corresponding~fixpoint.%
\end{isamarkuptext}\isamarkuptrue%
\ \ \isacommand{fun}\isamarkupfalse%
\isanewline
\ \ \ \ Traces\ {\isacharcolon}{\kern0pt}{\isacharcolon}{\kern0pt}\ {\isachardoublequoteopen}program\ {\isasymRightarrow}\ {\isasymSigma}\ {\isasymRightarrow}\ {\isasymT}\ set{\isachardoublequoteclose}\ {\isacharparenleft}{\kern0pt}{\isachardoublequoteopen}Tr{\isachardoublequoteclose}{\isacharparenright}{\kern0pt}\ \isakeyword{where}\isanewline
\ \ \ \ {\isachardoublequoteopen}Tr\ {\isacharparenleft}{\kern0pt}Program\ M\ {\isasymlbrace}\ S\ {\isasymrbrace}{\isacharparenright}{\kern0pt}\ {\isasymsigma}\ {\isacharequal}{\kern0pt}\ fst\ {\isacharbackquote}{\kern0pt}\ {\isasymDelta}\ {\isadigit{0}}\ {\isacharparenleft}{\kern0pt}set\ {\isacharparenleft}{\kern0pt}remdups\ M{\isacharparenright}{\kern0pt}{\isacharparenright}{\kern0pt}\ {\isacharparenleft}{\kern0pt}{\isasymlangle}{\isasymsigma}{\isasymrangle}{\isacharcomma}{\kern0pt}\ {\isacharbraceleft}{\kern0pt}{\isacharhash}{\kern0pt}\ {\isasymlambda}{\isacharbrackleft}{\kern0pt}S{\isacharbrackright}{\kern0pt}\ {\isacharhash}{\kern0pt}{\isacharbraceright}{\kern0pt}{\isacharparenright}{\kern0pt}{\isachardoublequoteclose}%
\isadelimdocument
\endisadelimdocument
\isatagdocument
\isamarkupsubsubsection{Proof Automation%
}
\isamarkuptrue%
\endisatagdocument
{\isafolddocument}%
\isadelimdocument
\endisadelimdocument
\begin{isamarkuptext}%
Similar to \isa{WL}, we again choose to introduce simplification lemmas for the \isa{{\isasymDelta}\isactrlsub N} 
  and \isa{{\isasymDelta}}-function. This decision will later ensure a major speedup when deriving 
  global traces in our proof~system.%
\end{isamarkuptext}\isamarkuptrue%
\ \ \isacommand{lemma}\isamarkupfalse%
\ {\isasymDelta}\isactrlsub N{\isacharunderscore}{\kern0pt}Bound{\isacharcolon}{\kern0pt}\ \isanewline
\ \ \ \ \isakeyword{shows}\ {\isachardoublequoteopen}{\isasymDelta}\isactrlsub N\ {\isadigit{0}}\ M\ c\ {\isacharequal}{\kern0pt}\ {\isacharbraceleft}{\kern0pt}c{\isacharbraceright}{\kern0pt}{\isachardoublequoteclose}\isanewline
\isadelimproof
\ \ \ \ %
\endisadelimproof
\isatagproof
\isacommand{by}\isamarkupfalse%
\ simp%
\endisatagproof
{\isafoldproof}%
\isadelimproof
\isanewline
\endisadelimproof
\isanewline
\ \ \isacommand{lemma}\isamarkupfalse%
\ {\isasymDelta}\isactrlsub N{\isacharunderscore}{\kern0pt}Termination{\isacharcolon}{\kern0pt}\ \isanewline
\ \ \ \ \isakeyword{assumes}\ {\isachardoublequoteopen}{\isasymdelta}\ M\ c\ {\isacharequal}{\kern0pt}\ {\isacharbraceleft}{\kern0pt}{\isacharbraceright}{\kern0pt}{\isachardoublequoteclose}\isanewline
\ \ \ \ \isakeyword{shows}{\isachardoublequoteopen}{\isasymDelta}\isactrlsub N\ {\isacharparenleft}{\kern0pt}Suc\ n{\isacharparenright}{\kern0pt}\ M\ c\ {\isacharequal}{\kern0pt}\ {\isacharbraceleft}{\kern0pt}c{\isacharbraceright}{\kern0pt}{\isachardoublequoteclose}\isanewline
\isadelimproof
\ \ \ \ %
\endisadelimproof
\isatagproof
\isacommand{using}\isamarkupfalse%
\ assms\ \isacommand{by}\isamarkupfalse%
\ simp%
\endisatagproof
{\isafoldproof}%
\isadelimproof
\isanewline
\endisadelimproof
\isanewline
\ \ \isacommand{lemma}\isamarkupfalse%
\ {\isasymDelta}\isactrlsub N{\isacharunderscore}{\kern0pt}Step{\isacharcolon}{\kern0pt}\ \isanewline
\ \ \ \ \isakeyword{assumes}\ {\isachardoublequoteopen}{\isasymnot}{\isasymdelta}\ M\ c\ {\isacharequal}{\kern0pt}\ {\isacharbraceleft}{\kern0pt}{\isacharbraceright}{\kern0pt}{\isachardoublequoteclose}\isanewline
\ \ \ \ \isakeyword{shows}{\isachardoublequoteopen}{\isasymDelta}\isactrlsub N\ {\isacharparenleft}{\kern0pt}Suc\ n{\isacharparenright}{\kern0pt}\ M\ c\ {\isacharequal}{\kern0pt}\ {\isasymUnion}{\isacharparenleft}{\kern0pt}{\isacharparenleft}{\kern0pt}{\isacharpercent}{\kern0pt}c{\isachardot}{\kern0pt}\ {\isasymDelta}\isactrlsub N\ n\ M\ c{\isacharparenright}{\kern0pt}\ {\isacharbackquote}{\kern0pt}\ {\isasymdelta}\ M\ c{\isacharparenright}{\kern0pt}{\isachardoublequoteclose}\isanewline
\isadelimproof
\ \ \ \ %
\endisadelimproof
\isatagproof
\isacommand{using}\isamarkupfalse%
\ assms\ \isacommand{by}\isamarkupfalse%
\ simp%
\endisatagproof
{\isafoldproof}%
\isadelimproof
\isanewline
\endisadelimproof
\isanewline
\ \ \isacommand{lemma}\isamarkupfalse%
\ {\isasymDelta}{\isacharunderscore}{\kern0pt}fixpoint{\isacharunderscore}{\kern0pt}reached{\isacharcolon}{\kern0pt}\isanewline
\ \ \ \ \isakeyword{assumes}\ {\isachardoublequoteopen}{\isasymforall}c\ {\isasymin}\ {\isasymDelta}\isactrlsub N\ n\ M\ c{\isachardot}{\kern0pt}\ {\isasymdelta}\ M\ c\ {\isacharequal}{\kern0pt}\ {\isacharbraceleft}{\kern0pt}{\isacharbraceright}{\kern0pt}{\isachardoublequoteclose}\isanewline
\ \ \ \ \isakeyword{shows}\ {\isachardoublequoteopen}{\isasymDelta}\ n\ M\ c\ {\isacharequal}{\kern0pt}\ {\isasymDelta}\isactrlsub N\ n\ M\ c{\isachardoublequoteclose}\isanewline
\isadelimproof
\ \ \ \ %
\endisadelimproof
\isatagproof
\isacommand{using}\isamarkupfalse%
\ composition{\isachardot}{\kern0pt}simps\ assms\ \isacommand{by}\isamarkupfalse%
\ auto%
\endisatagproof
{\isafoldproof}%
\isadelimproof
\isanewline
\endisadelimproof
\isanewline
\ \ \isacommand{lemma}\isamarkupfalse%
\ {\isasymDelta}{\isacharunderscore}{\kern0pt}iteration{\isacharcolon}{\kern0pt}\isanewline
\ \ \ \ \isakeyword{assumes}\ {\isachardoublequoteopen}{\isasymnot}{\isacharparenleft}{\kern0pt}{\isasymforall}c\ {\isasymin}\ {\isasymDelta}\isactrlsub N\ n\ M\ c{\isachardot}{\kern0pt}\ {\isasymdelta}\ M\ c\ {\isacharequal}{\kern0pt}\ {\isacharbraceleft}{\kern0pt}{\isacharbraceright}{\kern0pt}{\isacharparenright}{\kern0pt}{\isachardoublequoteclose}\isanewline
\ \ \ \ \isakeyword{shows}\ {\isachardoublequoteopen}{\isasymDelta}\ n\ M\ c\ {\isacharequal}{\kern0pt}\ {\isasymDelta}\ {\isacharparenleft}{\kern0pt}n\ {\isacharplus}{\kern0pt}\ {\isadigit{1}}{\isadigit{0}}{\isadigit{0}}{\isacharparenright}{\kern0pt}\ M\ c{\isachardoublequoteclose}\isanewline
\isadelimproof
\ \ \ \ %
\endisadelimproof
\isatagproof
\isacommand{using}\isamarkupfalse%
\ composition{\isachardot}{\kern0pt}simps\ assms\ \isacommand{by}\isamarkupfalse%
\ meson%
\endisatagproof
{\isafoldproof}%
\isadelimproof
\endisadelimproof
\begin{isamarkuptext}%
In order to finalize the automated proof system of the 
  \isa{{\isasymDelta}\isactrlsub N}-function/\isa{{\isasymDelta}}-function, we collect all previously proven lemmas in a set, 
  and call it the \isa{{\isasymDelta}\isactrlsub N}-system/\isa{{\isasymDelta}}-system. \par
  Note that we additionally remove the normal function simplifications of the
  \isa{{\isasymDelta}\isactrlsub N}-function, such that the Isabelle simplifier will later select our \isa{{\isasymDelta}\isactrlsub N}-system
  when deriving global traces. An explicitly defined \isa{partial{\isacharminus}{\kern0pt}function} does not 
  automatically add its simplifications to the simplifier, hence there are no 
  equations to remove for~the~\isa{{\isasymDelta}}-function.%
\end{isamarkuptext}\isamarkuptrue%
\ \ \isacommand{lemmas}\isamarkupfalse%
\ {\isasymDelta}\isactrlsub N{\isacharunderscore}{\kern0pt}system\ {\isacharequal}{\kern0pt}\ \isanewline
\ \ \ \ {\isasymDelta}\isactrlsub N{\isacharunderscore}{\kern0pt}Bound\ {\isasymDelta}\isactrlsub N{\isacharunderscore}{\kern0pt}Termination\ {\isasymDelta}\isactrlsub N{\isacharunderscore}{\kern0pt}Step\isanewline
\isanewline
\ \ \isacommand{lemmas}\isamarkupfalse%
\ {\isasymDelta}{\isacharunderscore}{\kern0pt}system\ {\isacharequal}{\kern0pt}\ \isanewline
\ \ \ \ {\isasymDelta}{\isacharunderscore}{\kern0pt}fixpoint{\isacharunderscore}{\kern0pt}reached\ {\isasymDelta}{\isacharunderscore}{\kern0pt}iteration\isanewline
\isanewline
\ \ \isacommand{declare}\isamarkupfalse%
\ composition\isactrlsub N{\isachardot}{\kern0pt}simps{\isacharbrackleft}{\kern0pt}simp\ del{\isacharbrackright}{\kern0pt}%
\begin{isamarkuptext}%
We now collect the \isa{{\isasymdelta}}-system, \isa{{\isasymDelta}\isactrlsub N}-system and \isa{{\isasymDelta}}-system in one set,
  naming it the \isa{WL\isactrlsub E}-derivation-system. Note that we will later be able to use this
  system for all global trace derivations~in~$WL_{EXT}$.%
\end{isamarkuptext}\isamarkuptrue%
\ \ \isacommand{lemmas}\isamarkupfalse%
\ WL\isactrlsub E{\isacharunderscore}{\kern0pt}derivation{\isacharunderscore}{\kern0pt}system\ {\isacharequal}{\kern0pt}\ \isanewline
\ \ \ \ {\isasymdelta}{\isacharunderscore}{\kern0pt}system\ {\isasymDelta}\isactrlsub N{\isacharunderscore}{\kern0pt}system\ {\isasymDelta}{\isacharunderscore}{\kern0pt}system%
\isadelimdocument
\endisadelimdocument
\isatagdocument
\isamarkupsubsubsection{Trace Derivation Examples%
}
\isamarkuptrue%
\endisatagdocument
{\isafolddocument}%
\isadelimdocument
\endisadelimdocument
\begin{isamarkuptext}%
We now take another look at our previously defined example programs for
  $WL_{EXT}$, analyzing their global traces using our proof automation system. \par
  Program \isa{WL{\isacharunderscore}{\kern0pt}ex\isactrlsub {\isadigit{1}}} first opens a new scope that declares a local variable x, which
  is only active inside the corresponding scope. Utilizing the local parallelism
  command, we then non-deterministically assign this variable to either numeral 1 
  or numeral 2. \par
  In order to disambiguate the fresh scope variable, \isa{x} is replaced by the 
  standardized variable \isa{{\isachardollar}{\kern0pt}x{\isacharcolon}{\kern0pt}{\isacharcolon}{\kern0pt}Scope}, which aligns with our convention. Due to the 
  underlying non-determinism of the local parallelism command, two distinct global 
  traces can be generated. The first trace listed below corresponds to the scenario, 
  in which \isa{{\isachardollar}{\kern0pt}x{\isacharcolon}{\kern0pt}{\isacharcolon}{\kern0pt}Scope} is first assigned to 2, before it is assigned to 1. The second
  trace matches the case, in which \isa{{\isachardollar}{\kern0pt}x{\isacharcolon}{\kern0pt}{\isacharcolon}{\kern0pt}Scope} is first assigned to 1, before being
  assigned to numeral 2. Note that this smoothly aligns with our intuition of
  the local parallelism~command.%
\end{isamarkuptext}\isamarkuptrue%
\ \ \isacommand{lemma}\isamarkupfalse%
\ {\isachardoublequoteopen}Tr\ WL{\isacharunderscore}{\kern0pt}ex\isactrlsub {\isadigit{1}}\ {\isacharparenleft}{\kern0pt}{\isasymsigma}\isactrlsub I\ WL{\isacharunderscore}{\kern0pt}ex\isactrlsub {\isadigit{1}}{\isacharparenright}{\kern0pt}\ {\isacharequal}{\kern0pt}\isanewline
\ \ \ \ \ \ \ \ \ \ \ \ {\isacharbraceleft}{\kern0pt}{\isacharparenleft}{\kern0pt}{\isacharparenleft}{\kern0pt}{\isasymlangle}fm\ {\isacharbrackleft}{\kern0pt}{\isacharbrackright}{\kern0pt}{\isasymrangle}\ {\isasymleadsto}\ State\ {\isasymllangle}fm\ {\isacharbrackleft}{\kern0pt}{\isacharparenleft}{\kern0pt}{\isacharprime}{\kern0pt}{\isacharprime}{\kern0pt}{\isachardollar}{\kern0pt}x{\isacharcolon}{\kern0pt}{\isacharcolon}{\kern0pt}Scope{\isacharprime}{\kern0pt}{\isacharprime}{\kern0pt}{\isacharcomma}{\kern0pt}\ Exp\ {\isacharparenleft}{\kern0pt}Num\ {\isadigit{0}}{\isacharparenright}{\kern0pt}{\isacharparenright}{\kern0pt}{\isacharbrackright}{\kern0pt}{\isasymrrangle}{\isacharparenright}{\kern0pt}\ {\isasymleadsto}\isanewline
\ \ \ \ \ \ \ \ \ \ \ \ State\ {\isasymllangle}fm\ {\isacharbrackleft}{\kern0pt}{\isacharparenleft}{\kern0pt}{\isacharprime}{\kern0pt}{\isacharprime}{\kern0pt}{\isachardollar}{\kern0pt}x{\isacharcolon}{\kern0pt}{\isacharcolon}{\kern0pt}Scope{\isacharprime}{\kern0pt}{\isacharprime}{\kern0pt}{\isacharcomma}{\kern0pt}\ Exp\ {\isacharparenleft}{\kern0pt}Num\ {\isadigit{1}}{\isacharparenright}{\kern0pt}{\isacharparenright}{\kern0pt}{\isacharbrackright}{\kern0pt}{\isasymrrangle}{\isacharparenright}{\kern0pt}\ {\isasymleadsto}\isanewline
\ \ \ \ \ \ \ \ \ \ \ \ State\ {\isasymllangle}fm\ {\isacharbrackleft}{\kern0pt}{\isacharparenleft}{\kern0pt}{\isacharprime}{\kern0pt}{\isacharprime}{\kern0pt}{\isachardollar}{\kern0pt}x{\isacharcolon}{\kern0pt}{\isacharcolon}{\kern0pt}Scope{\isacharprime}{\kern0pt}{\isacharprime}{\kern0pt}{\isacharcomma}{\kern0pt}\ Exp\ {\isacharparenleft}{\kern0pt}Num\ {\isadigit{2}}{\isacharparenright}{\kern0pt}{\isacharparenright}{\kern0pt}{\isacharbrackright}{\kern0pt}{\isasymrrangle}{\isacharcomma}{\kern0pt}\isanewline
\ \ \ \ \ \ \ \ \ \ \ \ {\isacharparenleft}{\kern0pt}{\isacharparenleft}{\kern0pt}{\isasymlangle}fm\ {\isacharbrackleft}{\kern0pt}{\isacharbrackright}{\kern0pt}{\isasymrangle}\ {\isasymleadsto}\ State\ {\isasymllangle}fm\ {\isacharbrackleft}{\kern0pt}{\isacharparenleft}{\kern0pt}{\isacharprime}{\kern0pt}{\isacharprime}{\kern0pt}{\isachardollar}{\kern0pt}x{\isacharcolon}{\kern0pt}{\isacharcolon}{\kern0pt}Scope{\isacharprime}{\kern0pt}{\isacharprime}{\kern0pt}{\isacharcomma}{\kern0pt}\ Exp\ {\isacharparenleft}{\kern0pt}Num\ {\isadigit{0}}{\isacharparenright}{\kern0pt}{\isacharparenright}{\kern0pt}{\isacharbrackright}{\kern0pt}{\isasymrrangle}{\isacharparenright}{\kern0pt}\ {\isasymleadsto}\isanewline
\ \ \ \ \ \ \ \ \ \ \ \ State\ {\isasymllangle}fm\ {\isacharbrackleft}{\kern0pt}{\isacharparenleft}{\kern0pt}{\isacharprime}{\kern0pt}{\isacharprime}{\kern0pt}{\isachardollar}{\kern0pt}x{\isacharcolon}{\kern0pt}{\isacharcolon}{\kern0pt}Scope{\isacharprime}{\kern0pt}{\isacharprime}{\kern0pt}{\isacharcomma}{\kern0pt}\ Exp\ {\isacharparenleft}{\kern0pt}Num\ {\isadigit{2}}{\isacharparenright}{\kern0pt}{\isacharparenright}{\kern0pt}{\isacharbrackright}{\kern0pt}{\isasymrrangle}{\isacharparenright}{\kern0pt}\ {\isasymleadsto}\isanewline
\ \ \ \ \ \ \ \ \ \ \ \ State\ {\isasymllangle}fm\ {\isacharbrackleft}{\kern0pt}{\isacharparenleft}{\kern0pt}{\isacharprime}{\kern0pt}{\isacharprime}{\kern0pt}{\isachardollar}{\kern0pt}x{\isacharcolon}{\kern0pt}{\isacharcolon}{\kern0pt}Scope{\isacharprime}{\kern0pt}{\isacharprime}{\kern0pt}{\isacharcomma}{\kern0pt}\ Exp\ {\isacharparenleft}{\kern0pt}Num\ {\isadigit{1}}{\isacharparenright}{\kern0pt}{\isacharparenright}{\kern0pt}{\isacharbrackright}{\kern0pt}{\isasymrrangle}{\isacharbraceright}{\kern0pt}{\isachardoublequoteclose}\isanewline
\isadelimproof
\ \ \ \ %
\endisadelimproof
\isatagproof
\isacommand{by}\isamarkupfalse%
\ {\isacharparenleft}{\kern0pt}simp\ add{\isacharcolon}{\kern0pt}\ WL\isactrlsub E{\isacharunderscore}{\kern0pt}derivation{\isacharunderscore}{\kern0pt}system\ WL{\isacharunderscore}{\kern0pt}ex\isactrlsub {\isadigit{1}}{\isacharunderscore}{\kern0pt}def{\isacharparenright}{\kern0pt}%
\endisatagproof
{\isafoldproof}%
\isadelimproof
\endisadelimproof
\begin{isamarkuptext}%
Program \isa{WL{\isacharunderscore}{\kern0pt}ex\isactrlsub {\isadigit{2}}} first assigns \isa{x} to numeral 0. It then executes a method call on
  method \isa{foo}, passing it the parameter \isa{x}. The main body then finishes execution
  by assigning \isa{x} to numeral 1. Method \isa{foo} in turn contains only a singular
  statement, which assigns its formal parameter \isa{x} to the numeral 2. \par
  Note that the formal parameter of \isa{foo} will again be renamed for disambiguation
  purposes. The new name \isa{{\isachardollar}{\kern0pt}foo{\isacharcolon}{\kern0pt}Param}, as can be seen below, matches our established
  variable convention. Due to the non-deterministic execution of the main body and \isa{foo}, 
  three distinct global traces can be generated. The first and second trace, as inferred 
  below, match the cases, in which the process corresponding to \isa{foo} reacts to the 
  invocation directly after the evaluation of the method call. Afterwards, both the main 
  body and \isa{foo} still have a statement left to execute. Note that both traces differ in 
  which statement is executed first. The third trace matches the scenario, in which the
  process corresponding to \isa{foo} only reacts to the invocation, once the main body
  has finished its execution. As can be seen, the amount of global traces can
  easily blow up in $WL_{EXT}$, which is a direct consequence of the underlying 
  non-determinism.%
\end{isamarkuptext}\isamarkuptrue%
\ \ \isacommand{lemma}\isamarkupfalse%
\ {\isachardoublequoteopen}Tr\ WL{\isacharunderscore}{\kern0pt}ex\isactrlsub {\isadigit{2}}\ {\isacharparenleft}{\kern0pt}{\isasymsigma}\isactrlsub I\ WL{\isacharunderscore}{\kern0pt}ex\isactrlsub {\isadigit{2}}{\isacharparenright}{\kern0pt}\ {\isacharequal}{\kern0pt}\isanewline
\ \ \ \ \ \ \ \ \ \ \ \ {\isacharbraceleft}{\kern0pt}{\isacharparenleft}{\kern0pt}{\isacharparenleft}{\kern0pt}{\isacharparenleft}{\kern0pt}{\isacharparenleft}{\kern0pt}{\isacharparenleft}{\kern0pt}{\isacharparenleft}{\kern0pt}{\isacharparenleft}{\kern0pt}{\isasymlangle}fm\ {\isacharbrackleft}{\kern0pt}{\isacharparenleft}{\kern0pt}{\isacharprime}{\kern0pt}{\isacharprime}{\kern0pt}x{\isacharprime}{\kern0pt}{\isacharprime}{\kern0pt}{\isacharcomma}{\kern0pt}\ Exp\ {\isacharparenleft}{\kern0pt}Num\ {\isadigit{0}}{\isacharparenright}{\kern0pt}{\isacharparenright}{\kern0pt}{\isacharbrackright}{\kern0pt}{\isasymrangle}\ {\isasymleadsto}\ State\ {\isasymllangle}fm\ {\isacharbrackleft}{\kern0pt}{\isacharparenleft}{\kern0pt}{\isacharprime}{\kern0pt}{\isacharprime}{\kern0pt}x{\isacharprime}{\kern0pt}{\isacharprime}{\kern0pt}{\isacharcomma}{\kern0pt}\ Exp\ {\isacharparenleft}{\kern0pt}Num\ {\isadigit{0}}{\isacharparenright}{\kern0pt}{\isacharparenright}{\kern0pt}{\isacharbrackright}{\kern0pt}{\isasymrrangle}{\isacharparenright}{\kern0pt}\ {\isasymleadsto}\isanewline
\ \ \ \ \ \ \ \ Event\ {\isasymllangle}invEv{\isacharcomma}{\kern0pt}{\isacharbrackleft}{\kern0pt}P\ {\isacharprime}{\kern0pt}{\isacharprime}{\kern0pt}foo{\isacharprime}{\kern0pt}{\isacharprime}{\kern0pt}{\isacharcomma}{\kern0pt}\ A\ {\isacharparenleft}{\kern0pt}Num\ {\isadigit{0}}{\isacharparenright}{\kern0pt}{\isacharbrackright}{\kern0pt}{\isasymrrangle}{\isacharparenright}{\kern0pt}\ {\isasymleadsto}\isanewline
\ \ \ \ \ \ \ State\ {\isasymllangle}fm\ {\isacharbrackleft}{\kern0pt}{\isacharparenleft}{\kern0pt}{\isacharprime}{\kern0pt}{\isacharprime}{\kern0pt}x{\isacharprime}{\kern0pt}{\isacharprime}{\kern0pt}{\isacharcomma}{\kern0pt}\ Exp\ {\isacharparenleft}{\kern0pt}Num\ {\isadigit{0}}{\isacharparenright}{\kern0pt}{\isacharparenright}{\kern0pt}{\isacharbrackright}{\kern0pt}{\isasymrrangle}{\isacharparenright}{\kern0pt}\ {\isasymleadsto}\isanewline
\ \ \ \ \ \ Event\ {\isasymllangle}invREv{\isacharcomma}{\kern0pt}{\isacharbrackleft}{\kern0pt}P\ {\isacharprime}{\kern0pt}{\isacharprime}{\kern0pt}foo{\isacharprime}{\kern0pt}{\isacharprime}{\kern0pt}{\isacharcomma}{\kern0pt}\ A\ {\isacharparenleft}{\kern0pt}Num\ {\isadigit{0}}{\isacharparenright}{\kern0pt}{\isacharbrackright}{\kern0pt}{\isasymrrangle}{\isacharparenright}{\kern0pt}\ {\isasymleadsto}\isanewline
\ \ \ \ \ State\ {\isasymllangle}fm\ {\isacharbrackleft}{\kern0pt}{\isacharparenleft}{\kern0pt}{\isacharprime}{\kern0pt}{\isacharprime}{\kern0pt}x{\isacharprime}{\kern0pt}{\isacharprime}{\kern0pt}{\isacharcomma}{\kern0pt}\ Exp\ {\isacharparenleft}{\kern0pt}Num\ {\isadigit{0}}{\isacharparenright}{\kern0pt}{\isacharparenright}{\kern0pt}{\isacharbrackright}{\kern0pt}{\isasymrrangle}{\isacharparenright}{\kern0pt}\ {\isasymleadsto}\isanewline
\ \ \ \ State\ {\isasymllangle}fm\ {\isacharbrackleft}{\kern0pt}{\isacharparenleft}{\kern0pt}{\isacharprime}{\kern0pt}{\isacharprime}{\kern0pt}x{\isacharprime}{\kern0pt}{\isacharprime}{\kern0pt}{\isacharcomma}{\kern0pt}\ Exp\ {\isacharparenleft}{\kern0pt}Num\ {\isadigit{0}}{\isacharparenright}{\kern0pt}{\isacharparenright}{\kern0pt}{\isacharcomma}{\kern0pt}\ {\isacharparenleft}{\kern0pt}{\isacharprime}{\kern0pt}{\isacharprime}{\kern0pt}{\isachardollar}{\kern0pt}foo{\isacharcolon}{\kern0pt}{\isacharcolon}{\kern0pt}Param{\isacharprime}{\kern0pt}{\isacharprime}{\kern0pt}{\isacharcomma}{\kern0pt}\ Exp\ {\isacharparenleft}{\kern0pt}Num\ {\isadigit{0}}{\isacharparenright}{\kern0pt}{\isacharparenright}{\kern0pt}{\isacharbrackright}{\kern0pt}{\isasymrrangle}{\isacharparenright}{\kern0pt}\ {\isasymleadsto}\isanewline
\ \ \ State\ {\isasymllangle}fm\ {\isacharbrackleft}{\kern0pt}{\isacharparenleft}{\kern0pt}{\isacharprime}{\kern0pt}{\isacharprime}{\kern0pt}x{\isacharprime}{\kern0pt}{\isacharprime}{\kern0pt}{\isacharcomma}{\kern0pt}\ Exp\ {\isacharparenleft}{\kern0pt}Num\ {\isadigit{1}}{\isacharparenright}{\kern0pt}{\isacharparenright}{\kern0pt}{\isacharcomma}{\kern0pt}\ {\isacharparenleft}{\kern0pt}{\isacharprime}{\kern0pt}{\isacharprime}{\kern0pt}{\isachardollar}{\kern0pt}foo{\isacharcolon}{\kern0pt}{\isacharcolon}{\kern0pt}Param{\isacharprime}{\kern0pt}{\isacharprime}{\kern0pt}{\isacharcomma}{\kern0pt}\ Exp\ {\isacharparenleft}{\kern0pt}Num\ {\isadigit{0}}{\isacharparenright}{\kern0pt}{\isacharparenright}{\kern0pt}{\isacharbrackright}{\kern0pt}{\isasymrrangle}{\isacharparenright}{\kern0pt}\ {\isasymleadsto}\isanewline
\ \ State\ {\isasymllangle}fm\ {\isacharbrackleft}{\kern0pt}{\isacharparenleft}{\kern0pt}{\isacharprime}{\kern0pt}{\isacharprime}{\kern0pt}x{\isacharprime}{\kern0pt}{\isacharprime}{\kern0pt}{\isacharcomma}{\kern0pt}\ Exp\ {\isacharparenleft}{\kern0pt}Num\ {\isadigit{1}}{\isacharparenright}{\kern0pt}{\isacharparenright}{\kern0pt}{\isacharcomma}{\kern0pt}\ {\isacharparenleft}{\kern0pt}{\isacharprime}{\kern0pt}{\isacharprime}{\kern0pt}{\isachardollar}{\kern0pt}foo{\isacharcolon}{\kern0pt}{\isacharcolon}{\kern0pt}Param{\isacharprime}{\kern0pt}{\isacharprime}{\kern0pt}{\isacharcomma}{\kern0pt}\ Exp\ {\isacharparenleft}{\kern0pt}Num\ {\isadigit{2}}{\isacharparenright}{\kern0pt}{\isacharparenright}{\kern0pt}{\isacharbrackright}{\kern0pt}{\isasymrrangle}{\isacharcomma}{\kern0pt}\isanewline
\ \ {\isacharparenleft}{\kern0pt}{\isacharparenleft}{\kern0pt}{\isacharparenleft}{\kern0pt}{\isacharparenleft}{\kern0pt}{\isacharparenleft}{\kern0pt}{\isacharparenleft}{\kern0pt}{\isacharparenleft}{\kern0pt}{\isasymlangle}fm\ {\isacharbrackleft}{\kern0pt}{\isacharparenleft}{\kern0pt}{\isacharprime}{\kern0pt}{\isacharprime}{\kern0pt}x{\isacharprime}{\kern0pt}{\isacharprime}{\kern0pt}{\isacharcomma}{\kern0pt}\ Exp\ {\isacharparenleft}{\kern0pt}Num\ {\isadigit{0}}{\isacharparenright}{\kern0pt}{\isacharparenright}{\kern0pt}{\isacharbrackright}{\kern0pt}{\isasymrangle}\ {\isasymleadsto}\ State\ {\isasymllangle}fm\ {\isacharbrackleft}{\kern0pt}{\isacharparenleft}{\kern0pt}{\isacharprime}{\kern0pt}{\isacharprime}{\kern0pt}x{\isacharprime}{\kern0pt}{\isacharprime}{\kern0pt}{\isacharcomma}{\kern0pt}\ Exp\ {\isacharparenleft}{\kern0pt}Num\ {\isadigit{0}}{\isacharparenright}{\kern0pt}{\isacharparenright}{\kern0pt}{\isacharbrackright}{\kern0pt}{\isasymrrangle}{\isacharparenright}{\kern0pt}\ {\isasymleadsto}\isanewline
\ \ \ \ \ \ \ \ Event\ {\isasymllangle}invEv{\isacharcomma}{\kern0pt}{\isacharbrackleft}{\kern0pt}P\ {\isacharprime}{\kern0pt}{\isacharprime}{\kern0pt}foo{\isacharprime}{\kern0pt}{\isacharprime}{\kern0pt}{\isacharcomma}{\kern0pt}\ A\ {\isacharparenleft}{\kern0pt}Num\ {\isadigit{0}}{\isacharparenright}{\kern0pt}{\isacharbrackright}{\kern0pt}{\isasymrrangle}{\isacharparenright}{\kern0pt}\ {\isasymleadsto}\isanewline
\ \ \ \ \ \ \ State\ {\isasymllangle}fm\ {\isacharbrackleft}{\kern0pt}{\isacharparenleft}{\kern0pt}{\isacharprime}{\kern0pt}{\isacharprime}{\kern0pt}x{\isacharprime}{\kern0pt}{\isacharprime}{\kern0pt}{\isacharcomma}{\kern0pt}\ Exp\ {\isacharparenleft}{\kern0pt}Num\ {\isadigit{0}}{\isacharparenright}{\kern0pt}{\isacharparenright}{\kern0pt}{\isacharbrackright}{\kern0pt}{\isasymrrangle}{\isacharparenright}{\kern0pt}\ {\isasymleadsto}\isanewline
\ \ \ \ \ \ Event\ {\isasymllangle}invREv{\isacharcomma}{\kern0pt}{\isacharbrackleft}{\kern0pt}P\ {\isacharprime}{\kern0pt}{\isacharprime}{\kern0pt}foo{\isacharprime}{\kern0pt}{\isacharprime}{\kern0pt}{\isacharcomma}{\kern0pt}\ A\ {\isacharparenleft}{\kern0pt}Num\ {\isadigit{0}}{\isacharparenright}{\kern0pt}{\isacharbrackright}{\kern0pt}{\isasymrrangle}{\isacharparenright}{\kern0pt}\ {\isasymleadsto}\isanewline
\ \ \ \ \ State\ {\isasymllangle}fm\ {\isacharbrackleft}{\kern0pt}{\isacharparenleft}{\kern0pt}{\isacharprime}{\kern0pt}{\isacharprime}{\kern0pt}x{\isacharprime}{\kern0pt}{\isacharprime}{\kern0pt}{\isacharcomma}{\kern0pt}\ Exp\ {\isacharparenleft}{\kern0pt}Num\ {\isadigit{0}}{\isacharparenright}{\kern0pt}{\isacharparenright}{\kern0pt}{\isacharbrackright}{\kern0pt}{\isasymrrangle}{\isacharparenright}{\kern0pt}\ {\isasymleadsto}\isanewline
\ \ \ \ State\ {\isasymllangle}fm\ {\isacharbrackleft}{\kern0pt}{\isacharparenleft}{\kern0pt}{\isacharprime}{\kern0pt}{\isacharprime}{\kern0pt}x{\isacharprime}{\kern0pt}{\isacharprime}{\kern0pt}{\isacharcomma}{\kern0pt}\ Exp\ {\isacharparenleft}{\kern0pt}Num\ {\isadigit{0}}{\isacharparenright}{\kern0pt}{\isacharparenright}{\kern0pt}{\isacharcomma}{\kern0pt}\ {\isacharparenleft}{\kern0pt}{\isacharprime}{\kern0pt}{\isacharprime}{\kern0pt}{\isachardollar}{\kern0pt}foo{\isacharcolon}{\kern0pt}{\isacharcolon}{\kern0pt}Param{\isacharprime}{\kern0pt}{\isacharprime}{\kern0pt}{\isacharcomma}{\kern0pt}\ Exp\ {\isacharparenleft}{\kern0pt}Num\ {\isadigit{0}}{\isacharparenright}{\kern0pt}{\isacharparenright}{\kern0pt}{\isacharbrackright}{\kern0pt}{\isasymrrangle}{\isacharparenright}{\kern0pt}\ {\isasymleadsto}\isanewline
\ \ \ State\ {\isasymllangle}fm\ {\isacharbrackleft}{\kern0pt}{\isacharparenleft}{\kern0pt}{\isacharprime}{\kern0pt}{\isacharprime}{\kern0pt}x{\isacharprime}{\kern0pt}{\isacharprime}{\kern0pt}{\isacharcomma}{\kern0pt}\ Exp\ {\isacharparenleft}{\kern0pt}Num\ {\isadigit{0}}{\isacharparenright}{\kern0pt}{\isacharparenright}{\kern0pt}{\isacharcomma}{\kern0pt}\ {\isacharparenleft}{\kern0pt}{\isacharprime}{\kern0pt}{\isacharprime}{\kern0pt}{\isachardollar}{\kern0pt}foo{\isacharcolon}{\kern0pt}{\isacharcolon}{\kern0pt}Param{\isacharprime}{\kern0pt}{\isacharprime}{\kern0pt}{\isacharcomma}{\kern0pt}\ Exp\ {\isacharparenleft}{\kern0pt}Num\ {\isadigit{2}}{\isacharparenright}{\kern0pt}{\isacharparenright}{\kern0pt}{\isacharbrackright}{\kern0pt}{\isasymrrangle}{\isacharparenright}{\kern0pt}\ {\isasymleadsto}\isanewline
\ \ State\ {\isasymllangle}fm\ {\isacharbrackleft}{\kern0pt}{\isacharparenleft}{\kern0pt}{\isacharprime}{\kern0pt}{\isacharprime}{\kern0pt}x{\isacharprime}{\kern0pt}{\isacharprime}{\kern0pt}{\isacharcomma}{\kern0pt}\ Exp\ {\isacharparenleft}{\kern0pt}Num\ {\isadigit{1}}{\isacharparenright}{\kern0pt}{\isacharparenright}{\kern0pt}{\isacharcomma}{\kern0pt}\ {\isacharparenleft}{\kern0pt}{\isacharprime}{\kern0pt}{\isacharprime}{\kern0pt}{\isachardollar}{\kern0pt}foo{\isacharcolon}{\kern0pt}{\isacharcolon}{\kern0pt}Param{\isacharprime}{\kern0pt}{\isacharprime}{\kern0pt}{\isacharcomma}{\kern0pt}\ Exp\ {\isacharparenleft}{\kern0pt}Num\ {\isadigit{2}}{\isacharparenright}{\kern0pt}{\isacharparenright}{\kern0pt}{\isacharbrackright}{\kern0pt}{\isasymrrangle}{\isacharcomma}{\kern0pt}\isanewline
\ \ {\isacharparenleft}{\kern0pt}{\isacharparenleft}{\kern0pt}{\isacharparenleft}{\kern0pt}{\isacharparenleft}{\kern0pt}{\isacharparenleft}{\kern0pt}{\isacharparenleft}{\kern0pt}{\isacharparenleft}{\kern0pt}{\isasymlangle}fm\ {\isacharbrackleft}{\kern0pt}{\isacharparenleft}{\kern0pt}{\isacharprime}{\kern0pt}{\isacharprime}{\kern0pt}x{\isacharprime}{\kern0pt}{\isacharprime}{\kern0pt}{\isacharcomma}{\kern0pt}\ Exp\ {\isacharparenleft}{\kern0pt}Num\ {\isadigit{0}}{\isacharparenright}{\kern0pt}{\isacharparenright}{\kern0pt}{\isacharbrackright}{\kern0pt}{\isasymrangle}\ {\isasymleadsto}\ State\ {\isasymllangle}fm\ {\isacharbrackleft}{\kern0pt}{\isacharparenleft}{\kern0pt}{\isacharprime}{\kern0pt}{\isacharprime}{\kern0pt}x{\isacharprime}{\kern0pt}{\isacharprime}{\kern0pt}{\isacharcomma}{\kern0pt}\ Exp\ {\isacharparenleft}{\kern0pt}Num\ {\isadigit{0}}{\isacharparenright}{\kern0pt}{\isacharparenright}{\kern0pt}{\isacharbrackright}{\kern0pt}{\isasymrrangle}{\isacharparenright}{\kern0pt}\ {\isasymleadsto}\isanewline
\ \ \ \ \ \ \ \ Event\ {\isasymllangle}invEv{\isacharcomma}{\kern0pt}{\isacharbrackleft}{\kern0pt}P\ {\isacharprime}{\kern0pt}{\isacharprime}{\kern0pt}foo{\isacharprime}{\kern0pt}{\isacharprime}{\kern0pt}{\isacharcomma}{\kern0pt}\ A\ {\isacharparenleft}{\kern0pt}Num\ {\isadigit{0}}{\isacharparenright}{\kern0pt}{\isacharbrackright}{\kern0pt}{\isasymrrangle}{\isacharparenright}{\kern0pt}\ {\isasymleadsto}\isanewline
\ \ \ \ \ \ \ State\ {\isasymllangle}fm\ {\isacharbrackleft}{\kern0pt}{\isacharparenleft}{\kern0pt}{\isacharprime}{\kern0pt}{\isacharprime}{\kern0pt}x{\isacharprime}{\kern0pt}{\isacharprime}{\kern0pt}{\isacharcomma}{\kern0pt}\ Exp\ {\isacharparenleft}{\kern0pt}Num\ {\isadigit{0}}{\isacharparenright}{\kern0pt}{\isacharparenright}{\kern0pt}{\isacharbrackright}{\kern0pt}{\isasymrrangle}{\isacharparenright}{\kern0pt}\ {\isasymleadsto}\isanewline
\ \ \ \ \ \ State\ {\isasymllangle}fm\ {\isacharbrackleft}{\kern0pt}{\isacharparenleft}{\kern0pt}{\isacharprime}{\kern0pt}{\isacharprime}{\kern0pt}x{\isacharprime}{\kern0pt}{\isacharprime}{\kern0pt}{\isacharcomma}{\kern0pt}\ Exp\ {\isacharparenleft}{\kern0pt}Num\ {\isadigit{1}}{\isacharparenright}{\kern0pt}{\isacharparenright}{\kern0pt}{\isacharbrackright}{\kern0pt}{\isasymrrangle}{\isacharparenright}{\kern0pt}\ {\isasymleadsto}\isanewline
\ \ \ \ \ Event\ {\isasymllangle}invREv{\isacharcomma}{\kern0pt}{\isacharbrackleft}{\kern0pt}P\ {\isacharprime}{\kern0pt}{\isacharprime}{\kern0pt}foo{\isacharprime}{\kern0pt}{\isacharprime}{\kern0pt}{\isacharcomma}{\kern0pt}\ A\ {\isacharparenleft}{\kern0pt}Num\ {\isadigit{0}}{\isacharparenright}{\kern0pt}{\isacharbrackright}{\kern0pt}{\isasymrrangle}{\isacharparenright}{\kern0pt}\ {\isasymleadsto}\isanewline
\ \ \ \ State\ {\isasymllangle}fm\ {\isacharbrackleft}{\kern0pt}{\isacharparenleft}{\kern0pt}{\isacharprime}{\kern0pt}{\isacharprime}{\kern0pt}x{\isacharprime}{\kern0pt}{\isacharprime}{\kern0pt}{\isacharcomma}{\kern0pt}\ Exp\ {\isacharparenleft}{\kern0pt}Num\ {\isadigit{1}}{\isacharparenright}{\kern0pt}{\isacharparenright}{\kern0pt}{\isacharbrackright}{\kern0pt}{\isasymrrangle}{\isacharparenright}{\kern0pt}\ {\isasymleadsto}\isanewline
\ \ \ State\ {\isasymllangle}fm\ {\isacharbrackleft}{\kern0pt}{\isacharparenleft}{\kern0pt}{\isacharprime}{\kern0pt}{\isacharprime}{\kern0pt}x{\isacharprime}{\kern0pt}{\isacharprime}{\kern0pt}{\isacharcomma}{\kern0pt}\ Exp\ {\isacharparenleft}{\kern0pt}Num\ {\isadigit{1}}{\isacharparenright}{\kern0pt}{\isacharparenright}{\kern0pt}{\isacharcomma}{\kern0pt}\ {\isacharparenleft}{\kern0pt}{\isacharprime}{\kern0pt}{\isacharprime}{\kern0pt}{\isachardollar}{\kern0pt}foo{\isacharcolon}{\kern0pt}{\isacharcolon}{\kern0pt}Param{\isacharprime}{\kern0pt}{\isacharprime}{\kern0pt}{\isacharcomma}{\kern0pt}\ Exp\ {\isacharparenleft}{\kern0pt}Num\ {\isadigit{0}}{\isacharparenright}{\kern0pt}{\isacharparenright}{\kern0pt}{\isacharbrackright}{\kern0pt}{\isasymrrangle}{\isacharparenright}{\kern0pt}\ {\isasymleadsto}\isanewline
\ \ State\ {\isasymllangle}fm\ {\isacharbrackleft}{\kern0pt}{\isacharparenleft}{\kern0pt}{\isacharprime}{\kern0pt}{\isacharprime}{\kern0pt}x{\isacharprime}{\kern0pt}{\isacharprime}{\kern0pt}{\isacharcomma}{\kern0pt}\ Exp\ {\isacharparenleft}{\kern0pt}Num\ {\isadigit{1}}{\isacharparenright}{\kern0pt}{\isacharparenright}{\kern0pt}{\isacharcomma}{\kern0pt}\ {\isacharparenleft}{\kern0pt}{\isacharprime}{\kern0pt}{\isacharprime}{\kern0pt}{\isachardollar}{\kern0pt}foo{\isacharcolon}{\kern0pt}{\isacharcolon}{\kern0pt}Param{\isacharprime}{\kern0pt}{\isacharprime}{\kern0pt}{\isacharcomma}{\kern0pt}\ Exp\ {\isacharparenleft}{\kern0pt}Num\ {\isadigit{2}}{\isacharparenright}{\kern0pt}{\isacharparenright}{\kern0pt}{\isacharbrackright}{\kern0pt}{\isasymrrangle}{\isacharbraceright}{\kern0pt}{\isachardoublequoteclose}\isanewline
\isadelimproof
\ \ \ \ %
\endisadelimproof
\isatagproof
\isacommand{by}\isamarkupfalse%
\ {\isacharparenleft}{\kern0pt}simp\ add{\isacharcolon}{\kern0pt}\ WL\isactrlsub E{\isacharunderscore}{\kern0pt}derivation{\isacharunderscore}{\kern0pt}system\ WL{\isacharunderscore}{\kern0pt}ex\isactrlsub {\isadigit{2}}{\isacharunderscore}{\kern0pt}def{\isacharparenright}{\kern0pt}%
\endisatagproof
{\isafoldproof}%
\isadelimproof
\endisadelimproof
\begin{isamarkuptext}%
The intuition behind program \isa{WL{\isacharunderscore}{\kern0pt}ex\isactrlsub {\isadigit{3}}} is simple. It first receives an unknown 
  input variable from another system, and subsequentially increments it by one. \par
  As can be inferred below, the global trace introduces a new symbolic input variable 
  \isa{{\isachardollar}{\kern0pt}x{\isacharcolon}{\kern0pt}{\isacharcolon}{\kern0pt}Input}, which is then concretized with the standard value 0 during the trace 
  composition. We also insert an input event into the trace, which could (in further 
  extensions of this model) serve as a trace composition interaction~point.%
\end{isamarkuptext}\isamarkuptrue%
\ \ \isacommand{lemma}\isamarkupfalse%
\ {\isachardoublequoteopen}Tr\ WL{\isacharunderscore}{\kern0pt}ex\isactrlsub {\isadigit{3}}\ {\isacharparenleft}{\kern0pt}{\isasymsigma}\isactrlsub I\ WL{\isacharunderscore}{\kern0pt}ex\isactrlsub {\isadigit{3}}{\isacharparenright}{\kern0pt}\ {\isacharequal}{\kern0pt}\isanewline
\ \ \ \ \ \ \ \ \ \ {\isacharbraceleft}{\kern0pt}{\isacharparenleft}{\kern0pt}{\isacharparenleft}{\kern0pt}{\isacharparenleft}{\kern0pt}{\isasymlangle}fm\ {\isacharbrackleft}{\kern0pt}{\isacharparenleft}{\kern0pt}{\isacharprime}{\kern0pt}{\isacharprime}{\kern0pt}x{\isacharprime}{\kern0pt}{\isacharprime}{\kern0pt}{\isacharcomma}{\kern0pt}\ Exp\ {\isacharparenleft}{\kern0pt}Num\ {\isadigit{0}}{\isacharparenright}{\kern0pt}{\isacharparenright}{\kern0pt}{\isacharcomma}{\kern0pt}\ {\isacharparenleft}{\kern0pt}{\isacharprime}{\kern0pt}{\isacharprime}{\kern0pt}{\isachardollar}{\kern0pt}x{\isacharcolon}{\kern0pt}{\isacharcolon}{\kern0pt}Input{\isacharprime}{\kern0pt}{\isacharprime}{\kern0pt}{\isacharcomma}{\kern0pt}\ Exp\ {\isacharparenleft}{\kern0pt}Num\ {\isadigit{0}}{\isacharparenright}{\kern0pt}{\isacharparenright}{\kern0pt}{\isacharbrackright}{\kern0pt}{\isasymrangle}\ {\isasymleadsto}\isanewline
\ \ \ \ \ \ \ \ \ \ State\ {\isasymllangle}fm\ {\isacharbrackleft}{\kern0pt}{\isacharparenleft}{\kern0pt}{\isacharprime}{\kern0pt}{\isacharprime}{\kern0pt}x{\isacharprime}{\kern0pt}{\isacharprime}{\kern0pt}{\isacharcomma}{\kern0pt}\ Exp\ {\isacharparenleft}{\kern0pt}Num\ {\isadigit{0}}{\isacharparenright}{\kern0pt}{\isacharparenright}{\kern0pt}{\isacharcomma}{\kern0pt}\ {\isacharparenleft}{\kern0pt}{\isacharprime}{\kern0pt}{\isacharprime}{\kern0pt}{\isachardollar}{\kern0pt}x{\isacharcolon}{\kern0pt}{\isacharcolon}{\kern0pt}Input{\isacharprime}{\kern0pt}{\isacharprime}{\kern0pt}{\isacharcomma}{\kern0pt}\ Exp\ {\isacharparenleft}{\kern0pt}Num\ {\isadigit{0}}{\isacharparenright}{\kern0pt}{\isacharparenright}{\kern0pt}{\isacharbrackright}{\kern0pt}{\isasymrrangle}{\isacharparenright}{\kern0pt}\ {\isasymleadsto}\isanewline
\ \ \ \ \ \ \ \ \ \ Event\ {\isasymllangle}inpEv{\isacharcomma}{\kern0pt}{\isacharbrackleft}{\kern0pt}A\ {\isacharparenleft}{\kern0pt}Num\ {\isadigit{0}}{\isacharparenright}{\kern0pt}{\isacharbrackright}{\kern0pt}{\isasymrrangle}{\isacharparenright}{\kern0pt}\ {\isasymleadsto}\isanewline
\ \ \ \ \ \ \ \ \ \ State\ {\isasymllangle}fm\ {\isacharbrackleft}{\kern0pt}{\isacharparenleft}{\kern0pt}{\isacharprime}{\kern0pt}{\isacharprime}{\kern0pt}x{\isacharprime}{\kern0pt}{\isacharprime}{\kern0pt}{\isacharcomma}{\kern0pt}\ Exp\ {\isacharparenleft}{\kern0pt}Num\ {\isadigit{0}}{\isacharparenright}{\kern0pt}{\isacharparenright}{\kern0pt}{\isacharcomma}{\kern0pt}\ {\isacharparenleft}{\kern0pt}{\isacharprime}{\kern0pt}{\isacharprime}{\kern0pt}{\isachardollar}{\kern0pt}x{\isacharcolon}{\kern0pt}{\isacharcolon}{\kern0pt}Input{\isacharprime}{\kern0pt}{\isacharprime}{\kern0pt}{\isacharcomma}{\kern0pt}\ Exp\ {\isacharparenleft}{\kern0pt}Num\ {\isadigit{0}}{\isacharparenright}{\kern0pt}{\isacharparenright}{\kern0pt}{\isacharbrackright}{\kern0pt}{\isasymrrangle}{\isacharparenright}{\kern0pt}\ {\isasymleadsto}\isanewline
\ \ \ \ \ \ \ \ \ \ State\ {\isasymllangle}fm\ {\isacharbrackleft}{\kern0pt}{\isacharparenleft}{\kern0pt}{\isacharprime}{\kern0pt}{\isacharprime}{\kern0pt}x{\isacharprime}{\kern0pt}{\isacharprime}{\kern0pt}{\isacharcomma}{\kern0pt}\ Exp\ {\isacharparenleft}{\kern0pt}Num\ {\isadigit{1}}{\isacharparenright}{\kern0pt}{\isacharparenright}{\kern0pt}{\isacharcomma}{\kern0pt}\ {\isacharparenleft}{\kern0pt}{\isacharprime}{\kern0pt}{\isacharprime}{\kern0pt}{\isachardollar}{\kern0pt}x{\isacharcolon}{\kern0pt}{\isacharcolon}{\kern0pt}Input{\isacharprime}{\kern0pt}{\isacharprime}{\kern0pt}{\isacharcomma}{\kern0pt}\ Exp\ {\isacharparenleft}{\kern0pt}Num\ {\isadigit{0}}{\isacharparenright}{\kern0pt}{\isacharparenright}{\kern0pt}{\isacharbrackright}{\kern0pt}{\isasymrrangle}{\isacharbraceright}{\kern0pt}{\isachardoublequoteclose}\isanewline
\isadelimproof
\ \ \ \ %
\endisadelimproof
\isatagproof
\isacommand{by}\isamarkupfalse%
\ {\isacharparenleft}{\kern0pt}simp\ add{\isacharcolon}{\kern0pt}\ WL\isactrlsub E{\isacharunderscore}{\kern0pt}derivation{\isacharunderscore}{\kern0pt}system\ WL{\isacharunderscore}{\kern0pt}ex\isactrlsub {\isadigit{3}}{\isacharunderscore}{\kern0pt}def{\isacharparenright}{\kern0pt}%
\endisatagproof
{\isafoldproof}%
\isadelimproof
\endisadelimproof
\begin{isamarkuptext}%
Note that the trace derivation speed of our proof automation is much slower 
  than for the standard While Language. This is a direct result of the increased 
  language complexity. We therefore propose further speed-related optimizations 
  (i.e. additional simplifications) as an idea for further extensions of~this~model.%
\end{isamarkuptext}\isamarkuptrue%
\isadelimdocument
\endisadelimdocument
\isatagdocument
\isamarkupsubsubsection{Code Generation%
}
\isamarkuptrue%
\endisatagdocument
{\isafolddocument}%
\isadelimdocument
\endisadelimdocument
\begin{isamarkuptext}%
Due to our focus on the code generation, we can again use the \isa{value} keyword
  to execute code for the global trace construction of arbitrary programs, and
  output the corresponding results in the console. Note that the code execution itself 
  is very performant, and can therefore be used to compute the global traces for any 
  arbitrary program \isa{S} and arbitrary initial state \isa{{\isasymsigma}}. We again propose further work 
  on exports of this code to several other programming languages supported by Isabelle 
  (e.g. Haskell, Scala) as an idea for extending the work of this~thesis.%
\end{isamarkuptext}\isamarkuptrue%
\ \ \isacommand{value}\isamarkupfalse%
\ {\isachardoublequoteopen}Tr\ WL{\isacharunderscore}{\kern0pt}ex\isactrlsub {\isadigit{1}}\ {\isacharparenleft}{\kern0pt}{\isasymsigma}\isactrlsub I\ WL{\isacharunderscore}{\kern0pt}ex\isactrlsub {\isadigit{1}}{\isacharparenright}{\kern0pt}{\isachardoublequoteclose}\isanewline
\ \ \isacommand{value}\isamarkupfalse%
\ {\isachardoublequoteopen}Tr\ WL{\isacharunderscore}{\kern0pt}ex\isactrlsub {\isadigit{2}}\ {\isacharparenleft}{\kern0pt}{\isasymsigma}\isactrlsub I\ WL{\isacharunderscore}{\kern0pt}ex\isactrlsub {\isadigit{2}}{\isacharparenright}{\kern0pt}{\isachardoublequoteclose}\isanewline
\ \ \isacommand{value}\isamarkupfalse%
\ {\isachardoublequoteopen}Tr\ WL{\isacharunderscore}{\kern0pt}ex\isactrlsub {\isadigit{3}}\ {\isacharparenleft}{\kern0pt}{\isasymsigma}\isactrlsub I\ WL{\isacharunderscore}{\kern0pt}ex\isactrlsub {\isadigit{3}}{\isacharparenright}{\kern0pt}{\isachardoublequoteclose}%
\isadelimdocument
\endisadelimdocument
\isatagdocument
\isamarkupsubsection{Trace Equivalence%
}
\isamarkuptrue%
\endisatagdocument
{\isafolddocument}%
\isadelimdocument
\endisadelimdocument
\begin{isamarkuptext}%
Similar to \isa{WL}, we again propose a notion of equivalence between programs. 
  We call two programs \isa{S} and \isa{S{\isacharprime}{\kern0pt}} of $WL_{EXT}$ trace equivalent under a given initial 
  state \isa{{\isasymsigma}} iff \isa{S} and \isa{S{\isacharprime}{\kern0pt}} called in \isa{{\isasymsigma}} generate the exact same set of global 
  traces upon termination. Note that the formalization of the inductive predicate
  is identical to the definition for \isa{WL}. \par
  Remember that the \isa{code{\isacharunderscore}{\kern0pt}pred} keyword is used to generate code for the
  inductive~definition.%
\end{isamarkuptext}\isamarkuptrue%
\ \ \isacommand{inductive}\isamarkupfalse%
\isanewline
\ \ \ \ equivalent\ {\isacharcolon}{\kern0pt}{\isacharcolon}{\kern0pt}\ {\isachardoublequoteopen}program\ {\isasymRightarrow}\ program\ {\isasymRightarrow}\ {\isasymSigma}\ {\isasymRightarrow}\ bool{\isachardoublequoteclose}\ {\isacharparenleft}{\kern0pt}{\isachardoublequoteopen}{\isacharunderscore}{\kern0pt}\ {\isasymsim}\ {\isacharunderscore}{\kern0pt}\ {\isacharbrackleft}{\kern0pt}{\isacharunderscore}{\kern0pt}{\isacharbrackright}{\kern0pt}{\isachardoublequoteclose}\ {\isadigit{8}}{\isadigit{0}}{\isacharparenright}{\kern0pt}\ \isakeyword{where}\isanewline
\ \ \ \ {\isachardoublequoteopen}{\isasymlbrakk}\ Tr\ prog\ {\isasymsigma}\ {\isacharequal}{\kern0pt}\ Tr\ prog{\isacharprime}{\kern0pt}\ {\isasymsigma}\ {\isasymrbrakk}\ {\isasymLongrightarrow}\ prog\ {\isasymsim}\ prog{\isacharprime}{\kern0pt}\ {\isacharbrackleft}{\kern0pt}{\isasymsigma}{\isacharbrackright}{\kern0pt}{\isachardoublequoteclose}\isanewline
\ \ \isacommand{code{\isacharunderscore}{\kern0pt}pred}\isamarkupfalse%
\ equivalent%
\isadelimproof
\ %
\endisadelimproof
\isatagproof
\isacommand{{\isachardot}{\kern0pt}}\isamarkupfalse%
\endisatagproof
{\isafoldproof}%
\isadelimproof
\endisadelimproof
\begin{isamarkuptext}%
We again automatically generate inductive simplifications for our trace
  equivalence notion using the \isa{inductive{\isacharminus}{\kern0pt}simps} keyword, thereby adding them to the 
  Isabelle simplifier~equations. This guarantees that we can conduct proofs 
  inferring trace~equivalence.%
\end{isamarkuptext}\isamarkuptrue%
\ \ \isacommand{inductive{\isacharunderscore}{\kern0pt}simps}\isamarkupfalse%
\ tequivalence{\isacharcolon}{\kern0pt}\ {\isachardoublequoteopen}prog\ {\isasymsim}\ prog{\isacharprime}{\kern0pt}\ {\isacharbrackleft}{\kern0pt}{\isasymsigma}{\isacharbrackright}{\kern0pt}{\isachardoublequoteclose}%
\begin{isamarkuptext}%
Last but not least, we derive an example trace equivalence conclusion
  in Isabelle utilizing the inductive definition~above.%
\end{isamarkuptext}\isamarkuptrue%
\ \ \isacommand{lemma}\isamarkupfalse%
\ {\isachardoublequoteopen}Program\ {\isacharbrackleft}{\kern0pt}{\isacharbrackright}{\kern0pt}\ {\isasymlbrace}\ CO\ {\isacharprime}{\kern0pt}{\isacharprime}{\kern0pt}x{\isacharprime}{\kern0pt}{\isacharprime}{\kern0pt}\ {\isacharcolon}{\kern0pt}{\isacharequal}{\kern0pt}\ Num\ {\isadigit{1}}\ {\isasymparallel}\ {\isacharprime}{\kern0pt}{\isacharprime}{\kern0pt}x{\isacharprime}{\kern0pt}{\isacharprime}{\kern0pt}\ {\isacharcolon}{\kern0pt}{\isacharequal}{\kern0pt}\ Num\ {\isadigit{2}}\ OC\ {\isasymrbrace}\ {\isasymsim}\ \isanewline
\ \ \ \ \ \ \ \ \ \ \ \ \ \ Program\ {\isacharbrackleft}{\kern0pt}{\isacharbrackright}{\kern0pt}\ {\isasymlbrace}\ CO\ {\isacharprime}{\kern0pt}{\isacharprime}{\kern0pt}x{\isacharprime}{\kern0pt}{\isacharprime}{\kern0pt}\ {\isacharcolon}{\kern0pt}{\isacharequal}{\kern0pt}\ Num\ {\isadigit{2}}\ {\isasymparallel}\ {\isacharprime}{\kern0pt}{\isacharprime}{\kern0pt}x{\isacharprime}{\kern0pt}{\isacharprime}{\kern0pt}\ {\isacharcolon}{\kern0pt}{\isacharequal}{\kern0pt}\ Num\ {\isadigit{1}}\ OC\ {\isasymrbrace}\ {\isacharbrackleft}{\kern0pt}{\isasymcircle}{\isacharbrackright}{\kern0pt}{\isachardoublequoteclose}\isanewline
\isadelimproof
\ \ \ \ %
\endisadelimproof
\isatagproof
\isacommand{using}\isamarkupfalse%
\ tequivalence\ \isacommand{by}\isamarkupfalse%
\ {\isacharparenleft}{\kern0pt}simp\ add{\isacharcolon}{\kern0pt}\ WL\isactrlsub E{\isacharunderscore}{\kern0pt}derivation{\isacharunderscore}{\kern0pt}system{\isacharparenright}{\kern0pt}%
\endisatagproof
{\isafoldproof}%
\isadelimproof
\isanewline
\endisadelimproof
\isadelimtheory
\isanewline
\endisadelimtheory
\isatagtheory
\isacommand{end}\isamarkupfalse%
\endisatagtheory
{\isafoldtheory}%
\isadelimtheory
\endisadelimtheory
\end{isabellebody}%

%% file: ms.bbl
\begin{thebibliography}{4}
\providecommand{\natexlab}[1]{#1}
\providecommand{\url}[1]{\texttt{#1}}
\expandafter\ifx\csname urlstyle\endcsname\relax
  \providecommand{\doi}[1]{doi: #1}\else
  \providecommand{\doi}{doi: \begingroup \urlstyle{rm}\Url}\fi

\bibitem[Brookes(1996)]{BrookesA}
Stephen~D. Brookes.
\newblock Full {A}bstraction for a {S}hared-{V}ariable {P}arallel {L}anguage.
\newblock In \emph{Inf. Comput. 127, 2 (1996)}, pages 145--163, 1996.

\bibitem[Brookes(2002)]{BrookesB}
Stephen~D. Brookes.
\newblock Traces, {P}omsets, {F}airness and {F}ull {A}bstraction for
  {C}ommunicating {P}rocesses.
\newblock In \emph{Proc. 13th Intl. Conf. on Concurrency Theory (CONCUR 2002)
  (LNCS, Vol. 2421), Lubos Brim, Petr Jancar, Mojmir Kret\'insk\'y, and
  Anton\'in Kucera (Eds.)}, pages 466--482. Springer, Berlin Heidelberg, 2002.

\bibitem[Din et~al.()Din, Hähnle, Henrio, Johnsen, Pun, and
  Tarifa]{LAGCSemantics}
Crystal~Chang Din, Reiner Hähnle, Ludovic Henrio, Einar~Broch Johnsen, Violet
  Ka~I Pun, and S.~Lizeth~Tapia Tarifa.
\newblock Locally {A}bstract, {G}lobally {C}oncrete {S}emantics of {C}oncurrent
  {P}rogramming {L}anguages.
\newblock Draft dated June 2021.

\bibitem[Xia et~al.(2020)Xia, Zakowski, He, Hur, Malecha, Pierce, and
  Zdancewic]{Xia}
Li{-}yao Xia, Yannick Zakowski, Paul He, Chung-Kil Hur, Gregory Malecha,
  Benjamin~C. Pierce, and Steve Zdancewic.
\newblock Interaction trees: representing recursive and impure programs in
  {C}oq.
\newblock In \emph{Proceedings of the ACM on Programming Languages 4, POPL (Jan
  2020)}, pages 1--32, 2020.
\newblock \doi{https://doi.org/10.1145/3371119}.

\end{thebibliography}
